\documentclass{article}
\usepackage[utf8]{inputenc}
\usepackage{authblk}
\usepackage{parskip}
\usepackage{multirow}
\usepackage[toc,page]{appendix}
\usepackage{comment}
\usepackage[style=numeric-comp, sorting=none]{biblatex}
\usepackage{graphicx}
\usepackage{epsfig}
\usepackage{amsmath}
\usepackage{subcaption}
\graphicspath{ {./image/} }
\usepackage{amssymb}
\usepackage{amsfonts}
\usepackage{float}

\usepackage{hyperref}
\hypersetup{
    colorlinks=true,
    linkcolor=blue,
    filecolor=magenta,      
    urlcolor=cyan,
    pdftitle={Overleaf Example},
    pdfpagemode=FullScreen,
    }
\usepackage{geometry}
\begin{document}
\date{}
\title{Magnetic black holes in 4$D$ Einstein--Gauss--Bonnet massive 
gravity coupled to nonlinear electrodynamics}

\author[1]{Prosenjit Paul\thanks{prosenjitpaul629@gmail.com}}
\author[2,3]{S. I. Kruglov\thanks{serguei.krouglov@utoronto.ca}}
\affil[1]{Indian Institute Of Engineering Science and Technology (IIEST), Shibpur-711103, WB, India}
\affil[2]{Department of Physics, University of Toronto, \protect\\ 60 St. Georges St.,
Toronto, ON M5S 1A7, Canada} 
\affil[3]{Canadian Quantum Research Center, 204-3002 32 Ave Vernon, BC V1T 2L7, Canada.}

\maketitle
\begin{center}
\rule{17cm}{0.4mm}
\end{center}

\begin{abstract}
We investigate Einstein--Gauss--Bonnet (EGB) $4D$ massive gravity coupled to nonlinear 
electrodynamics (NED) in an Anti--de--Sitter ($AdS$) background and find 
an exact magnetically charged black hole solution. The metric function 
was analyzed for different values of massive gravity parameters. The 
first law of black hole thermodynamics and generalized Smarr formula 
were verified, where we treated the cosmological constant as thermodynamic 
pressure. We define vacuum polarization as the conjugate to NED parameter. To analyze the local stability of the black hole we compute specific heat. 
We investigated the Van der Waals-like/reentrant phase transition of 
the black holes and estimated the critical points. We observe small/large 
black hole (SBH/LBH) and small/intermediate/large black hole (SBH/IBH/LBH) 
phase transition. The Joule-–Thomson coefficient, inversion, and isenthalpic 
curves are discussed. Finally, the minimum inversion temperature and 
the corresponding event horizon radius are obtained using numerical techniques. 
\end{abstract}

\newpage
\begin{center}
\rule{17cm}{0.4mm}
\end{center}
\tableofcontents
\begin{center}
\rule{17cm}{0.4mm}
\end{center}

\section{Introduction}
In 1905, Einstein originally proposed special relativity. After that, 
Einstein developed his field equation, which basically describes 
gravitation as a curvature of spacetime caused by mass and energy. 
General Relativity(GR) is based on two fundamental principles. First, 
the Einstein Equivalence Principle ``\emph{In small enough regions of 
spacetime, the laws of physics reduced to those of special relativity; 
it is impossible to detect the existence of a gravitational field by 
means of local experiments}'' \cite{carroll2019spacetime}. Second, 
the Principle of general Covariance ``\emph{The laws of physics holds 
in the absence of gravity; i.e. when the metric tensor $g_{\mu \nu}$ 
equals to the Minkowski tensor $\eta_{\mu \nu}$ and laws of physics 
are covariant; i.e. it preserves its form under a general coordinate 
transformation $x \to x^{\prime}$}'' \cite{weinberg1972gravitation}. 
The central theme of GR is Einstein's field equation. These equations 
relate the curvature of spacetime to matter distribution within it. 
These equations can be written as

\begin{equation}
R_{\mu \nu} - \frac{1}{2} R g_{\mu \nu} + \Lambda g_{\mu \nu} = 8 \pi G T_{\mu \nu},
\end{equation}
where $R_{\mu \nu}$ is the Ricci tensor, $R$ is the scalar curvature, 
and $g_{\mu \nu}$ is the metric tensor that encodes the geometry of 
spacetime. $\Lambda$ is the cosmological constant and $G$ is the
gravitational constant. The stress-energy tensor $T_{\mu \nu}$ 
describes the matter and energy content of the system and can include 
various components such as mass, energy, pressure, and momentum. It is 
defined in terms of the energy-momentum tensor for each type of matter 
or field present in the system.

General relativity also predicts several phenomena that have been 
experimentally verified. One of the most well-known predictions of 
GR is the bending of light rays by massive objects. According to GR, 
the path of light is bent when it passes through a region of spacetime 
with a strong gravitational field. This was confirmed by observations 
during the solar eclipse of 1919. Other predictions of GR are 
black holes, Gravitational Lensing, Gravitational Waves, Gravitational Redshift, 
and Gravitational Time Dilation. In conclusion, GR is a highly successful theory of gravitation but it opens some 
questions, especially how to unify quantum mechanics with gravitation.

GR is not a full-fledged theory of quantum gravitation, therefore 
people try to modify GR in various ways. For example,  introducing 
a scalar field to the action of GR, known as Scalar-Tensor theories 
of gravity \cite{Barrabes:1997kk,Cai:1996pj,Capozziello:2005bu,Sotiriou:2006hs,Moffat:2005si,Faraoni:2007yn}, 
$f(R)$ gravity, here the gravitational action includes a general 
function of the scalar curvature 
\cite{Sotiriou:2008ve,Sotiriou:2008rp,Capozziello:2011et,Berry:2011pb,Liang:2017ahj,Gogoi:2020ypn}, 
Lovelock gravity \cite{Lovelock:1971yv,Lovelock:1972vz,Deruelle:1989fj} 
and Horava–Lifshitz gravity \cite{Horava:2009uw,Blas:2009yd,Sotiriou:2010wn,Wang:2017brl}.

Lovelock theories, as described in Refs. 
\cite{Lovelock:1971yv,Lovelock:1972vz}, represent a special 
class of higher-order gravity theories. These theories 
maintain essential properties such as diffeomorphism 
invariance, metricity, and second-order equations of 
motion. Lovelock's theories of gravity provide a rich 
framework in which the theory of Gauss--Bonnet gravity 
naturally emerges when considering higher-dimensional 
spacetimes. $4D$ Gauss--Bonnet gravity does not 
contribute to the dynamics of the theories. However, 
it becomes significant and contributes to the dynamics 
when the dimensions of spacetime exceed four. In recent 
research, Glavan and Lin \cite{glavan2020einstein} made a 
notable contribution by finding a solution to the 
Einstein--Gauss--Bonnet field equation in four dimensions. 
They achieved this by rescaling the Gauss--Bonnet coupling 
parameter $\alpha$ in a different manner, specifically 
by $\alpha/D-4$, where $D$ represents spacetime dimensions. 
The electrically charged $AdS$ black hole in $4D$  EGB gravity 
was found in Ref. \cite{Fernandes:2020rpa}. Black hole 
solutions in $4D$ or higher dimension EGB gravity studied 
in Refs. \cite{Hegde:2020xlv,Wei:2020poh,Wang:2020pmb,Singh:2020nwo,Singh:2020xju,EslamPanah:2020hoj,Ghosh:2020ijh,Singh:2021xbk,Upadhyay:2022axg,Godani:2022jwz,Paul:2023mlh,Myrzakulov:2023rkr,Kumar:2023cmo}.

Another theory of modified GR is Massive gravity. 
Massive gravity is a framework that attempts to modify 
GR by adding mass terms to the Einstein--Hilbert action. 
The main motivation for introducing mass to the graviton is 
to account for the accelerated expansion of the universe. 
The pioneering work of Fierz and Pauli led to the first 
formulation of massive gravity \cite{Fierz:1939ix,Fierz:1939zz}. 
However, when this theory is applied in a curved background, 
it encounters ghost instabilities \cite{Boulware:1972yco}. 
The recent formulation of the de Rham--Gabadadze--Tolley (dRGT) 
massive gravity theory \cite{deRham:2010ik,deRham:2010kj} 
solves the ghost instabilities and provides a consistent 
framework. Massive gravity theories can make distinct 
predictions regarding the properties of gravitational waves. 
The LIGO observational data sets a constraint on the mass of 
the graviton, indicating that its value must not exceed 
$m \le 1.2 \times 10^{-22}$ eV \cite{LIGOScientific:2016lio}. 
Black hole solutions \& phase transition in massive gravity 
are studied in Refs. \cite{Hendi:2015bna,Hendi:2016hbe,Upadhyay:2018vfu,Hendi:2018hdo,Singh:2020rnm,Upadhyay:2022axg,Paul:2023mlh,Singh:2024jgo}.

In 1933-1934 Born and Infeld first proposed the theory of 
NED \cite{Born:1934gh} to remove the singularity of a 
point-like charge. Due to quantum electron-positron one-loop 
interactions, Heisenberg and Euler derived a comprehensive 
effective action that accounted for the non-linear corrections 
to Maxwell's theory of electromagnetism \cite{Heisenberg:1936nmg}. 
Plebanski later further expanded the theory of NED in the context 
of special relativity by including a general function of the 
electromagnetic field invariants \cite{plebanski1966non}. 
The low-energy regime of heterotic string theory 
\cite{Natsuume:1994hd,Kats:2006xp,Cai:2008ph,Liu:2008kt,Anninos:2008sj} 
could recover the NED. One of the 
main motivation for NED is tackling the fundamental problem 
in cosmology, e.g. NED has been explored in the context of 
inflationary models, NED has been proposed as a candidate 
for describing dark energy or as an alternative explanation 
to the cosmological constant and NED describes singularity 
free or regular black hole solutions. The black hole solutions 
in GR with Born--Infeld (BI) NED have studied Refs. \cite{cataldo1999three,cai2004born,panahiyan2020nonlinearly,kruglov2017born,Yang:2020jno,Zhang:2021kha,Yerra:2022eov}. 
Apart from Born--Infeld \& Heisenberg--Euler electrodynamics, 
there exists a variety of NED models such as logarithmic 
\cite{Soleng:1995kn,Hendi:2012zz,Hendi:2013dwa,Hendi:2014mna,Kruglov:2019ybs,Kruglov:2022sxx,Kruglov:2023qed}, 
generalized logarithmic \cite{Kruglov:2014iqa}, 
double-logarithmic \cite{Gullu:2020ant}, exponential 
\cite{Hendi:2012zz,Hendi:2013dwa,Hendi:2014mna,Kruglov:2017fck}, 
power--Maxwell \cite{Hendi:2012um,Hendi:2016usw}, $\arcsin$ 
\cite{kruglov2015nonlinear,Kruglov:2016ezw,Kruglov:2018xzs,Kruglov:2019okd} 
\& some others NED \cite{Kruglov:2016ymq,kruglov2017nonlinear,Yang:2020jno,Kruglov:2021stm,Kruglov:2021btd,Kruglov:2021pdp,Kruglov:2021qzd,Kruglov:2021rqf,kruglov2022nonlinearly,}. 
In this paper, we consider the NED model proposed in \cite{kruglov2017nonlinear}. 
The reason to consider NED \cite{kruglov2017nonlinear} is its 
simplicity. The mass and metric functions are expressed in the 
form of simple elementary functions while in the Born–-Infeld model 
these functions are of the hypergeometric type.

Black holes are one of the fascinating objects of our observable universe. 
The work of Bekenstein and Hawking showed that black holes can be 
considered as a thermodynamic system with entropy \& temperature 
proportional to the area of the event horizon \& surface gravity 
at the horizon \cite{bekenstein2020black,bekenstein1973extraction,bekenstein1974generalized,bardeen1973four,hawking1975particle}. 
Hawking \& Page showed that a phase transition between 
Schwarzschild $AdS$ black hole and thermal $AdS$ space occurs, 
which is known as  Hawking–-Page transition in the literature 
\cite{Hawking:1982dh}. The fascinating concept of the $AdS/CFT$ 
correspondence provides an insightful explanation: the 
Hawking--Page phase transition can be interpreted as the 
gravitational dual of the confinement/deconfinement phase 
transition \cite{maldacena1999large,Witten:1998qj,gubser1998gauge,Aharony:1999ti}. 
As a result, this correspondence has served as a driving 
force behind the exploration of black holes and their 
thermodynamics in the $AdS$ spacetime. Recently, it was 
discovered that the cosmological constant played the 
role of thermodynamic pressure associated with the black 
hole \& thermodynamics volume is its conjugate variable 
\cite{kastor2009enthalpy,kastor2010smarr,dolan2011cosmological,dolan2011pressure}. 
In the extended phase space, the mass of the black hole 
is considered as enthalpy, rather than internal energy. 
Therefore, many interesting phenomena emerge, such as $P--v$ 
criticality \& Van der Waals like phase transition 
\cite{Kubiznak:2012wp,Hendi:2012um,Gunasekaran:2012dq,dehghani2014p,hennigar2015p,Xu:2015rfa,hendi2016extended,hansen2017universality,majhi2017pv,hendi2017geometrical,upadhyay2017p}, 
Joule--Thomson (J--T) expansion of the black holes \cite{Okcu:2016tgt,Okcu:2017qgo,Mo:2018rgq,Zhao:2018kpz,Nam:2020gud,kruglov2022nonlinearly,Kruglov:2022bhx,Kruglov:2022mde,Kruglov:2022sxx}, 
black holes as heat engines \cite{johnson2014holographic,belhaj2015heat,setare2015polytropic,johnson2016gauss,johnson2016born,bhamidipati2017heat,hennigar2017holographic,mo2017heat,hendi2018black} 
and reentrant phase transitions \cite{altamirano2013reentrant,frassino2014multiple,hennigar2015reentrant,NaveenaKumara:2020biu,Zhang:2020obn,Kruglov:2023ktg}.

Mass of the black hole treated as enthalpy, in extended phase 
space. Therefore, one might naturally apply the Joule--Thomson 
expansion technique to the charged $AdS$ black hole. The concept 
of Joule--Thomson expansion for a black hole in Einstein's 
gravity was first studied by Ökcü \& Aydıner in 
Ref. \cite{Okcu:2016tgt}. After that, Joule--Thomson expansion 
of $D$ dimension charged black hole, Kerr--$AdS$, Kerr--Newman--$AdS$, 
and black hole in massive gravity studied in Refs. 
\cite{Okcu:2017qgo,Mo:2018rgq,Zhao:2018kpz,Nam:2020gud}. 
The Joule--Thomson expansion of black hole in General Relativity 
coupled to NED (eq. \eqref{eq:2.4} and others) 
studied in Refs. \cite{kruglov2022nonlinearly,Kruglov:2022bhx,Kruglov:2022mde,Kruglov:2022sxx}. 
Joule--Thomson effects of $4D$ EGB gravity coupled to Maxwell/BI 
electrodynamics studied in Refs. \cite{Hegde:2020xlv,Zhang:2021kha}.

In this paper, we investigate magnetically charged black hole 
solutions in $4D$ Einstein--Gauss--Bonnet massive gravity 
coupled to NED. We obtain an exact black hole solution in 
extended phase space. We study the thermodynamics, phase transition 
\& Joule--Thomson expansion of the black hole in $4D$ EGB massive 
gravity, $4D$ EGB massless gravity \& $4D$ Einstein massive gravity.

The paper is organized as follows. In section \ref{sec:2}, we 
write $D$ dimensional EGB Massive Gravity 
action in $AdS$ space coupled to NED and find the metric function 
for magnetized black holes in $4D$. The effects of graviton mass 
and NED parameters on the horizon of the black holes are depicted. 
In section \ref{sec:3}, we study the Hawking temperature of the 
black hole in $4D$ EGB massive gravity, $4D$ EGB massless gravity, 
and massive Einstein gravity. The black hole in massive Einstein 
gravity undergoes Hawking--Page-like phase transition for some particular 
values of the parameters. We verified the first law of black hole 
thermodynamics and generalized Smarr formula in extended phase space. 
From the first law of black hole thermodynamics, we compute magnetic 
potential and vacuum polarization. The specific heat of the black holes 
is studied. In section \ref{sec:4}, we numerically obtain the critical 
radius, critical temperature, and critical pressure for the black hole 
in $4D$ EGB massive gravity,  $4D$ EGB massless gravity, and massive 
Einstein gravity. Furthermore, we analyzed the $G-T_{H}$ and $P-v$ plots. 
The reentrant phase transitions of the black holes are analyzed in section \ref{sec:RPT}. 
In section \ref{sec:5}, we investigate the Joule--Thomson adiabatic 
expansion of the black holes in $4D$ EGB massive gravity, $4D$ EGB 
massless gravity, and massive Einstein gravity. We analyzed the effects 
of massive gravity \& NED parameter on the constant mass curve and inverse 
curve. Furthermore, we numerically investigate the minimum horizon radius 
and corresponding inverse temperature. Finally, the Joule--Thomson coefficient 
as a function of horizon radius is studied.

\section{4D Einstein--Gauss--Bonnet Massive Gravity Coupled to NED}\label{sec:2}

The $D$ dimensional action for EGB massive gravity coupled
to NED in $AdS$ background given by 

\begin{equation}\label{eq:2.1}
S= \frac{1}{16 \pi} \int d^{D}x \sqrt{-g} \Biggr[R -2 \Lambda + \alpha \mathcal{G} +\mathcal{L}_{NED}     + m^2 \sum_{i=1}^{4} c_{i} \mathcal{U}_{i}(g,h)\Biggr],
\end{equation}

where $g$ is  determinant of the metric $g_{\mu \nu}$. We use units with $G=1$, 
$\Lambda$ is the negative cosmological constant. $R$ is Ricci scalar, $\alpha$ 
is Gauss--Bonnet coupling parameter, 
$\mathcal{G}= R_{\mu \nu \rho \sigma} R^{\mu \nu \rho \sigma} -4 R_{\mu \nu } 
R^{\mu \nu } + R^2$ is the Gauss--Bonnet term. $R_{\mu \nu \rho \sigma}$ is  
Riemann tensor, $R_{\mu \nu }$ is Ricci tensor and m is a parameter related 
to graviton mass, $c_i (i=1,2,3,4)$ is constant \footnote{``\emph{In order to 
have a self-consistent massive gravity theory, the coupling parameters $c_i$ 
might be required to be negative if the squared mass of the graviton is 
positive. However, in the $AdS$ spacetime, the coupling parameters $c_i$ can 
still take the positive values. This is because the fluctuations of the fields 
with the negative squared masses in the $AdS$ spacetime could still be stable 
if their squared masses obey the corresponding Breitenlohner–-Freedman bounds''} 
\cite{Nam:2020gud}.} \cite{Nam:2020gud} and $\mathcal{U}_i(g, h)$ are symmetric 
polynomials of eigenvalues of matrix $\mathcal{K}_{\nu}^{\mu}= \sqrt{g^{\mu 
\alpha} h_{\alpha \nu}}$ given by 
\begin{equation} \label{eq:2.2}
\begin{split}
\mathcal{U}_{1} & = \bigr[  \mathcal{K} \bigr],\\
\mathcal{U}_{2} & = \bigr[  \mathcal{K} \bigr]^{2} -  \bigr[  \mathcal{K}^2 \bigr],\\
\mathcal{U}_{3} & = \bigr[  \mathcal{K} \bigr]^{3} - 3\bigr[  \mathcal{K} \bigr] \bigr[  \mathcal{K}^2 \bigr] + 2  \bigr[  \mathcal{K}^{3} \bigr], \\
\mathcal{U}_{4} & = \bigr[\mathcal{K} \bigr]^{4} - 6\bigr[\mathcal{K}^2 \bigr] \bigr[  \mathcal{K}\bigr]^2 + 8 \bigr[\mathcal{K}^{3}\bigr] \bigr[\mathcal{K}\bigr] + 3 \bigr[\mathcal{K}^2\bigr]^2 -6 \bigr[\mathcal{K}^4\bigr].
\end{split}
\end{equation}
In $D=4$ dimensions, Gauss--Bonnet term does not contribute to the dynamics. 
Therefore, we rescale \cite{glavan2020einstein} the Gauss--Bonnet coupling parameter 
as $\alpha \to \alpha/(D-4)$, finally action takes the following form
\begin{equation}\label{eq:2.3}
S= \frac{1}{16 \pi} \int d^{D}x \sqrt{-g} \Biggr[R -2 \Lambda + \frac{\alpha}{D-4} \mathcal{G} +\mathcal{L}_{NED} + m^2 \sum_{i} c_{i} \mathcal{U}_{i}(g,h)    \Biggr].
\end{equation}

We use the NED Lagrangian proposed in \cite{kruglov2017nonlinear,Kruglov:2021stm,kruglov2022nonlinearly} 
\begin{equation}\label{eq:2.4}
    \mathcal{L}_{NED}=  -\frac{\mathcal{F}}{1+  \sqrt{2\beta \mathcal{F}}},
\end{equation}
where $\mathcal{F}=F_{\mu \nu}F^{\mu \nu}=2(B^2-E^2)$ and 
$F_{\mu \nu}= \partial_{\mu}{A_{\nu}}-\partial_{\nu}{A_{\mu}}$ is Maxwell's 
tensor\footnote{we use a different notation of $\mathcal{F}$ compared to 
\cite{kruglov2017nonlinear,kruglov2022nonlinearly,Kruglov:2021stm}} and 
$\beta$ is the positive coupling. Using the above NED Lagrangian action is given by 
\begin{equation}\label{eq:2.5}
S= \frac{1}{16 \pi} \int d^{D}x \sqrt{-g} \Biggr[R -2 \Lambda + \frac{\alpha}{D-4} \mathcal{G} -\frac{\mathcal{F}}{1+  \sqrt{2\beta \mathcal{F}}}    + m^2 \sum_{i} c_{i} \mathcal{U}_{i}(g,h)   \Biggr].
\end{equation}
Variation of the above action with respect to $A_{\mu}$ gives 
\begin{equation}\label{eq:2.6}
    \partial_{\mu} \Bigl( \sqrt{-g} \mathcal{L}_{\mathcal{F}} F^{\mu \nu} \Bigl)=0,
\end{equation}
where $\mathcal{L}_{\mathcal{F}} = \partial {\mathcal{L}}/\partial {\mathcal{F}}$. Using the Bianchi identities
\begin{equation}\label{eq:2.7}
    \nabla_{\mu}  \Tilde{F}^{\mu \nu}=0,
\end{equation}

where $\Tilde{F}^{\mu \nu}$ is the dual field-strength tensor, which is defined 
as $\Tilde{F}^{\mu \nu} = \frac{1}{2} \epsilon^{\mu \nu \gamma \delta} 
F_{\gamma \delta}$ and $\epsilon_{\mu \nu \gamma \delta}$ is the Levi--Civita 
symbol, totally antisymmetric with respect to all pairs of indices. Now 
we consider the static and spherical symmetric solution of the form 
\begin{equation}\label{eq:2.8}
    ds^{2}= - e^{2A(r)} dt^2 +  e^{2C(r)} dr^2 +r^2 d{\Omega}_{D-2}^2.
\end{equation}
Following the Ref. \cite{Cai:2014znn} we take the  reference metric as
\begin{equation}\label{eq:2.9}
    h_{\mu \nu}= \text{diag}\bigl( 0, 0, c^2h_{D-2}  \bigl),
\end{equation}
where $c$ is a dimensionless positive constant \cite{Cai:2014znn}. The reference 
metric $h_{\mu \nu}$ is a rank two symmetric tensor. Now substituting the equation 
\eqref{eq:2.9} into equation \eqref{eq:2.2} we obtain 

\begin{equation} \label{eq:2.10}
\begin{split}
\mathcal{U}_{1} & = \frac{(D-2)c}{r}, \\
\mathcal{U}_{2} & = \frac{(D-2)(D-3)c^{2}}{r^{2}},\\
\mathcal{U}_{3} & = \frac{(D-2)(D-3)(D-4)c^{3}}{r^{3}}, \\
\mathcal{U}_{4} & = \frac{(D-2)(D-3)(D-4)(D-4)c^{4}}{r^{4}}.
\end{split}
\end{equation}

In this paper, we only consider magnetically charged black holes. Hence, the field 
strength tensor component $F_{\theta \phi}$ is 
\begin{equation}\label{eq:2.11}
F_{\theta \phi}= \frac{Q_{m}}{r^{D-4}} \sin{\theta}_{D-3}  \biggr[  \prod_{i=1}^{D-4} \sin^{2}{\theta}_{i}  \biggr], 
\end{equation}
where $Q_m$ is the magnetic charge. Therefore, NED Lagrangian \eqref{eq:2.4} is
\begin{equation}\label{eq:2.12}
    \mathcal{L}_{NED}= -\frac{2 Q_{m}^{2}}{2 Q_{m} \sqrt{\beta} r^{D-2}+r^{2D-4}}.
\end{equation}
Finally, using above NED Lagrangian and equation \eqref{eq:2.8} we reduced the action \eqref{eq:2.5} as
\begin{equation*}
S=\frac{\Sigma_{D-2}}{16 \pi} \int dt dr (D-2)  e^{A+C}  \biggr[ r^{D-1} \psi \Bigl( 1+ \alpha (D-3) \psi\Bigl)^{\prime} + \frac{r^{D-1}}{l^2} - \frac{2 Q_m^2 r^{D-2} }{ (r^{2D-4}+2Q_m\sqrt{\beta} r^{D-2})(D-2)} +\frac{m^2cc_1r^{D-2}}{r}
\end{equation*}    
\begin{equation}\label{eq:2.13}
+\frac{m^2c^2c_2 (D-3) r^{D-2}}{r^2}+\frac{m^2c^3c_3(D-3)(D-4)r^{D-2}}{r^3}+\frac{m^2c^4c_4(D-3)(D-4)(D-5)r^{D-2}}{(D-2)r^4}   \biggr],
\end{equation}
where prime denotes differentiation with respect to radial coordinates and we used 
$\Lambda=-3/l^2$, where $l$ is $AdS$ radius. Finally, taking the limit $D \to 4$ and 
combining everything inside radial derivative, one can obtain
\begin{equation}\label{eq:2.14}
    S=\frac{\Sigma_{2}}{8 \pi} \int dt dr   e^{A+C}  \biggr[ r^{3} \psi \Bigl( 1+ \alpha \psi\Bigl) + \frac{r^{3}}{l^2} - \frac{ Q_m^2  \arctan{({r}/{k})}  }{k } + a + m^2 \Bigl\{ \frac{c_{1} c r^{2}}{2} + {c_{2}c^2 r}\Bigl\}   \biggr]^{\prime},
\end{equation}

where $\Sigma_{2}={2{\pi}^\frac{3}{2}}/{\Gamma\bigl( {\frac{3}{2}}\bigl)}$ is the area of two dimensional sphere, $\psi= r^{-2} \bigl( 1- e^{-2C} \bigl)$, $k=\sqrt{2 Q_m \sqrt{\beta}}$ and $a$ is an integration constant. From above equation, one can obtain the following solutions 
\begin{equation}\label{eq:2.15}
    e^{A+C}=1,
\end{equation} 
\begin{equation}\label{eq:2.16}
  \psi  \bigl( 1+ \alpha \psi\bigl) + \frac{1}{l^2} - \frac{Q_m^2 \arctan{({r}/{k})}  }{kr^3} +\frac{a}{r^3}+ \frac{m^2}{r^3} \Bigl( \frac{c_{1} c r^{2}}{2} + {c_{2}c^2 r}\Bigl) = \frac{8 \pi M}{\Sigma_{2}r^3},
\end{equation}
 where $ M$ is an integration constant and $M$ is related to the mass of the black hole. Therefore, the exact solution is
\begin{equation}\label{eq:2.17}
e^{2A}= e^{-2C}= 1+ \frac{r^2}{2\alpha} \Biggr[1 \pm \sqrt{1+4\alpha \biggl\{ \frac{(2M-a)}{r^3} +    \frac{Q_m^2 \arctan{(r/k)}}{k r^3}  - \frac{1}{l^2} - \frac{m^2}{2r^2}\Bigl( c c_{1} r + 2 c^2 c_{2}\Bigl)  \biggl\} }  \Biggr]. 
\end{equation}
We will discuss three different black hole solutions. The first solution is $4D$ EGB 
massive gravity black holes with NED. In the limit $\beta \to 0$, above equation must 
be reduced to $4D$ EGB massive gravity with Maxwell electrodynamics \cite{Paul:2023mlh} and one can obtain $a=\pi Q_m^2/2k$. Therefore
\begin{equation}\label{eq:2.18}
    e^{2A}= e^{-2C}= 1+ \frac{r^2}{2\alpha} \Biggr[1 \pm \sqrt{1+4\alpha \biggl\{ \frac{2M}{r^3} - \frac{Q_m^2}{kr^3} \Bigl( \frac{\pi}{2} - \arctan{(r/k)} \Bigl) - \frac{1}{l^2} - \frac{m^2}{2r^2}\Bigl( c c_{1} r + 2 c^2 c_{2}\Bigl)  \biggl\} }  \Biggr].
\end{equation}
The negative branch corresponds to the $4D$ magnetically charged $AdS$ EGB massive gravity black hole, whereas the $+ve$ branch does not lead to a physically meaningful solution because the positive sign in the mass term indicates graviton instabilities \cite{Upadhyay:2022axg}, so we only take the negative branch of equation \eqref{eq:2.18}. Using the trigonometric identity and limit $\beta \to 0$ above equation reduces to magnetically charged $AdS$ black hole in $4D$ EGB massive gravity \cite{Paul:2023mlh}
\begin{equation}\label{eq:2.19}
    e^{2A}= e^{-2C}= 1+ \frac{r^2}{2\alpha} \Biggr[1 - \sqrt{1+4\alpha \biggl\{ \frac{2M}{r^3} - \frac{Q_m^2}{r^4}  - \frac{1}{l^2} - \frac{m^2}{2r^2}\Bigl( c c_{1} r + 2 c^2 c_{2}\Bigl)  \biggl\} }  \Biggr].
\end{equation}
Taking further massless limit into above equation gives electrically charged $AdS$ black holes in Maxwell electrodynamics \cite{Fernandes:2020rpa}. The second solution is $4D$ EGB massless gravity black holes with NED. This solution can be obtained from equation \eqref{eq:2.18} by setting $m=0$
\begin{equation}\label{eq:2.20}
e^{2A}= e^{-2C}= 1+ \frac{r^2}{2\alpha} \Biggr[1 - \sqrt{1+4\alpha \biggl\{ \frac{2M}{r^3} - \frac{Q_m^2}{kr^3} \Bigl( \frac{\pi}{2} - \arctan{(r/k)} \Bigl) - \frac{1}{l^2}  \biggl\} }  \Biggr].
\end{equation}
In the limit $l \to \infty$ above solution reduces to \cite{Kruglov:2021stm}
\begin{equation}\label{eq:2.21}
e^{2A}= e^{-2C}= 1+ \frac{r^2}{2\alpha} \Biggr[1 - \sqrt{1+4\alpha \biggl\{ \frac{2M_0}{r^3} + \frac{Q_m^2 \arctan{(r/k)}}{kr^3}  \biggl\} }  \Biggr],
\end{equation}
where $2M_0=2M-a$. Taking further $\beta \to 0$ limit above equation gives 
charged black holes in Maxwell electrodynamics \cite{glavan2020einstein}. The third 
solution is $4D$  massive Einstein gravity black holes with NED. These solutions can 
be obtained from equation \eqref{eq:2.18}, in the limit $\alpha \to 0$
\begin{equation}\label{eq:2.22}
   e^{2A}= e^{-2C}= 1-\frac{2 M}{r}+\frac{a}{r } - \frac{Q_m^2 \arctan  ({r}/{k})}{rk} +\frac{r^{2}}{l^{2}}+m^{2} c_{2} c^{2}+\frac{m^{2} c_{1} r c}{2}.
\end{equation}
In the massless limit above solution reduces to black hole in NED, which was obtained in \cite{kruglov2022nonlinearly}
\begin{equation}\label{eq:2.23}
   e^{2A}= e^{-2C}= 1-\frac{2 M_0}{r} - \frac{Q_m^2 \arctan  ({r}/{k})}{rk} +\frac{r^{2}}{l^{2}},
\end{equation}
where we used $2M_0=2M-a$. Using trigonometric identity and limit $\beta \to 0$ equation \eqref{eq:2.22} reduces to Reissner-–Nordstrom $AdS$ black holes in massive gravity \cite{Cai:2014znn}
 \begin{equation}\label{eq:2.24}
     e^{2A}= e^{-2C}= 1-\frac{2 M}{r}+\frac{Q_{m}^{2} }{ r^2 }+\frac{r^{2}}{l^{2}}+m^{2} c_{2} c^{2}+\frac{m^{2} c_{1} c r}{2}.
 \end{equation}
Furthermore, if we take chargeless limit($M= Q_m \to 0$) or $r \to \infty$ into equation \eqref{eq:2.22} then vacuum solution comes out \cite{Cai:2014znn,Nam:2019zyk}
\begin{equation}\label{eq:2.25}
e^{2A}= e^{-2C}= 1+\frac{r^{2}}{l^{2}}+m^{2} c_{2} c^{2}+\frac{m^{2} c_{1} r c}{2}.
\end{equation}

\begin{figure}[H]
\centering
\subfloat[$\alpha=0.2$]{\includegraphics[width=.5\textwidth]{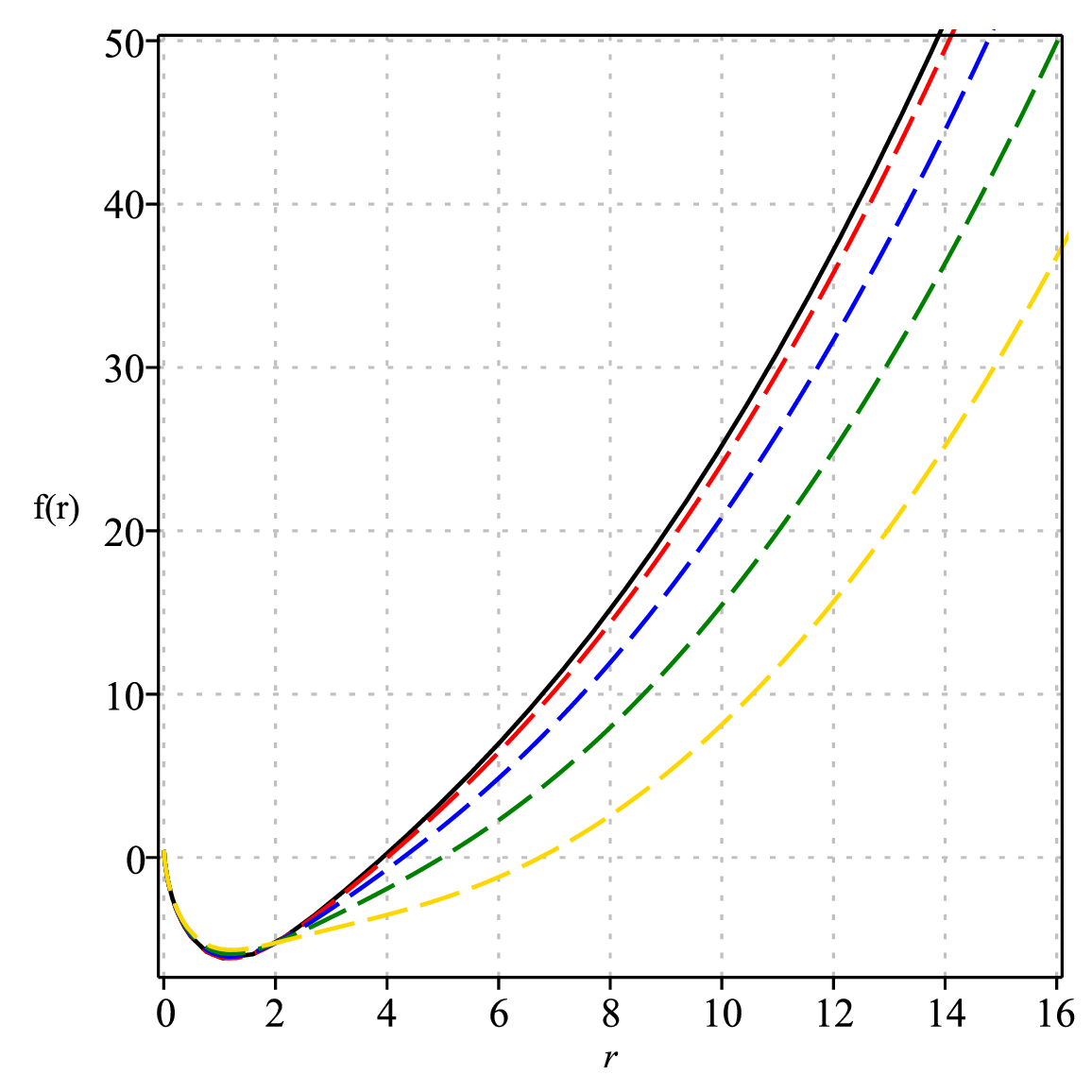}}\hfill
\subfloat[$\alpha=0.4$]{\includegraphics[width=.5\textwidth]{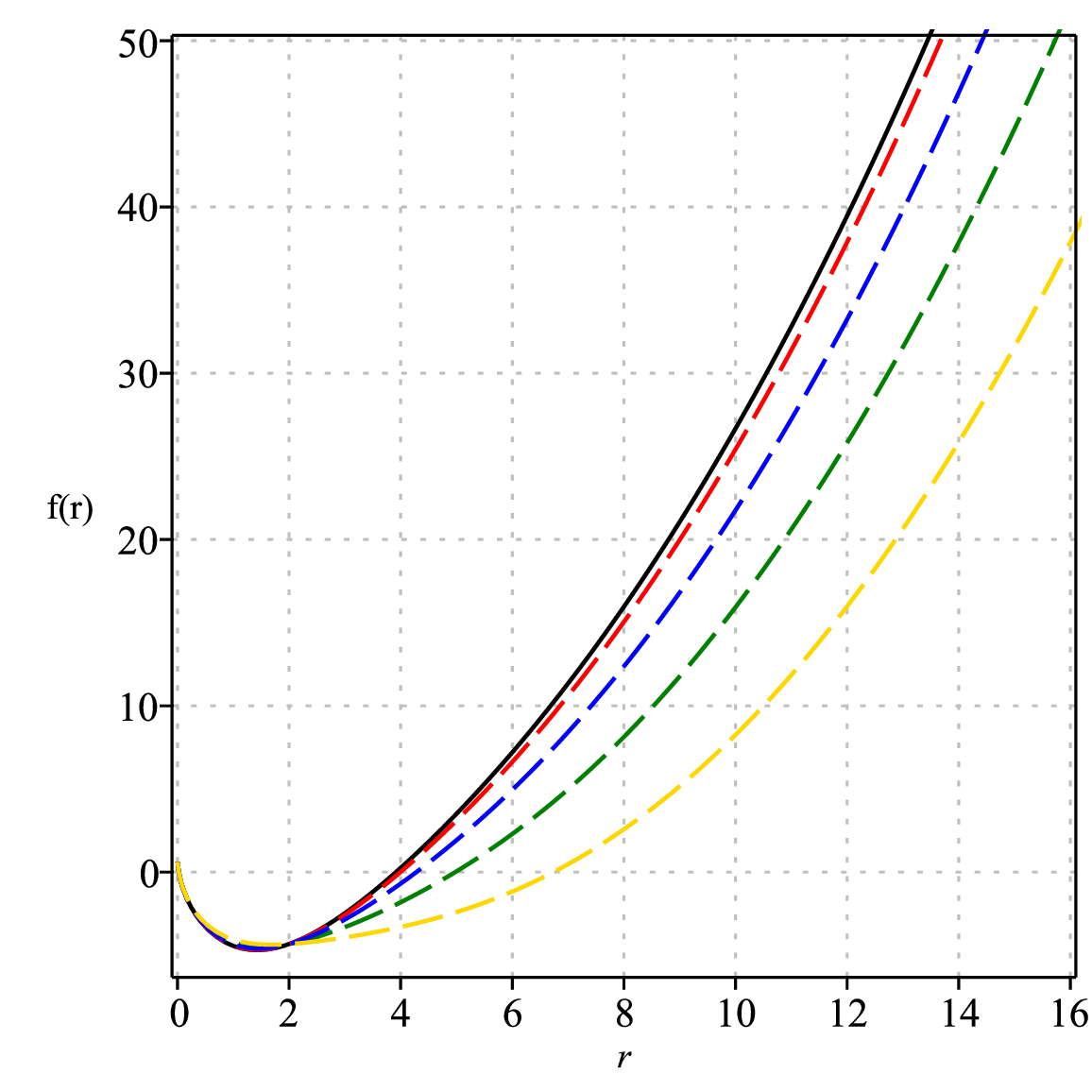}}\hfill
\caption{$m=0.0$ denoted by solid black line, $m=0.5$ denoted by red dash line with, $m=1.0$ denoted by blue dash line, $m=1.5$ denoted by green dash line and $m=2.0$ denoted by gold dash line  in EGB-NED with $M=10$, $Q_m=2$, $\beta=0.5$, $c=1$, $c_1=-1$, $c_2=1$ and $l=2$. }\label{fig:1}
\end{figure} 
\begin{figure}[H]
\centering
\subfloat[$\alpha=0.02$]{\includegraphics[width=.5\textwidth]{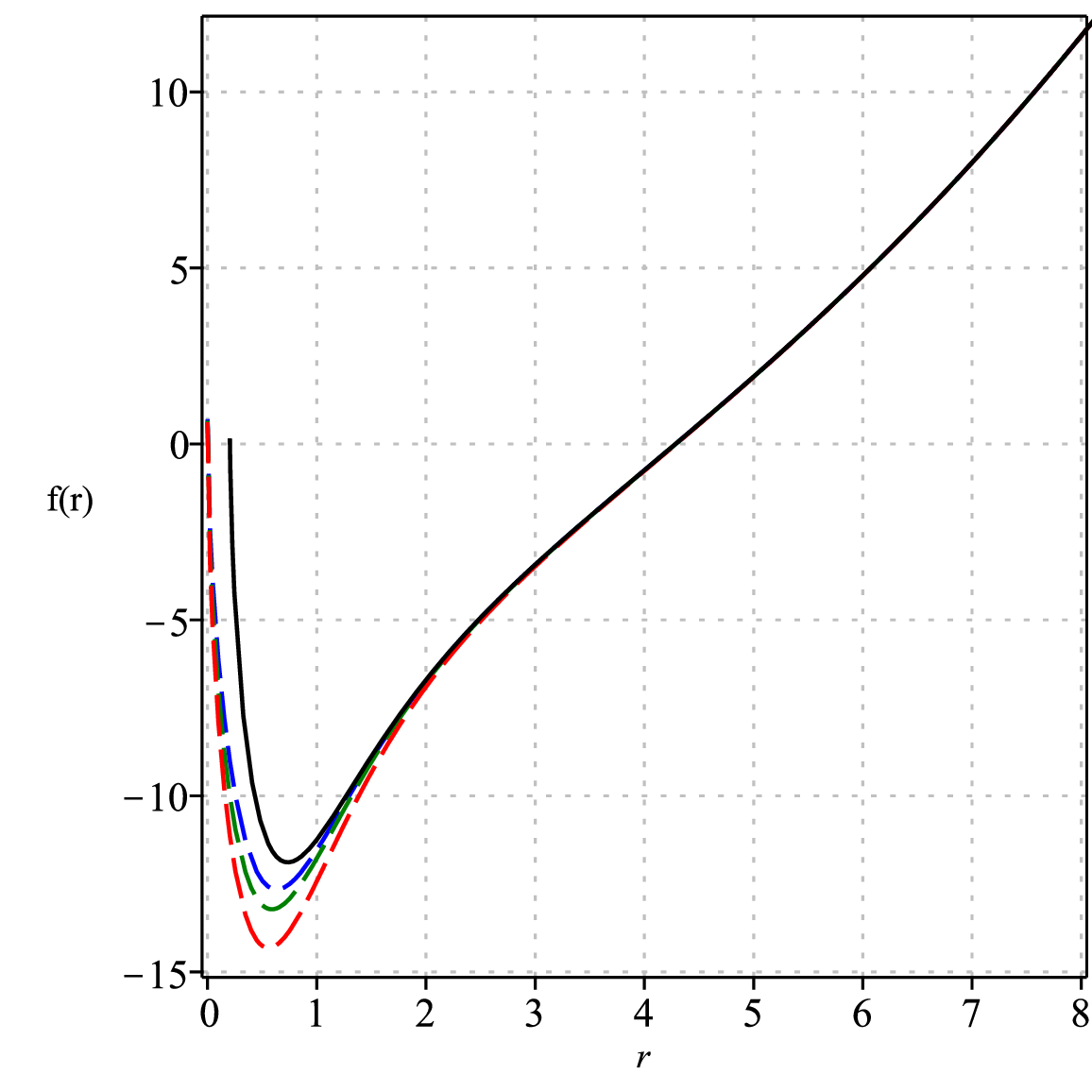}}\hfill
\subfloat[$\alpha=0.2$]{\includegraphics[width=.5\textwidth]{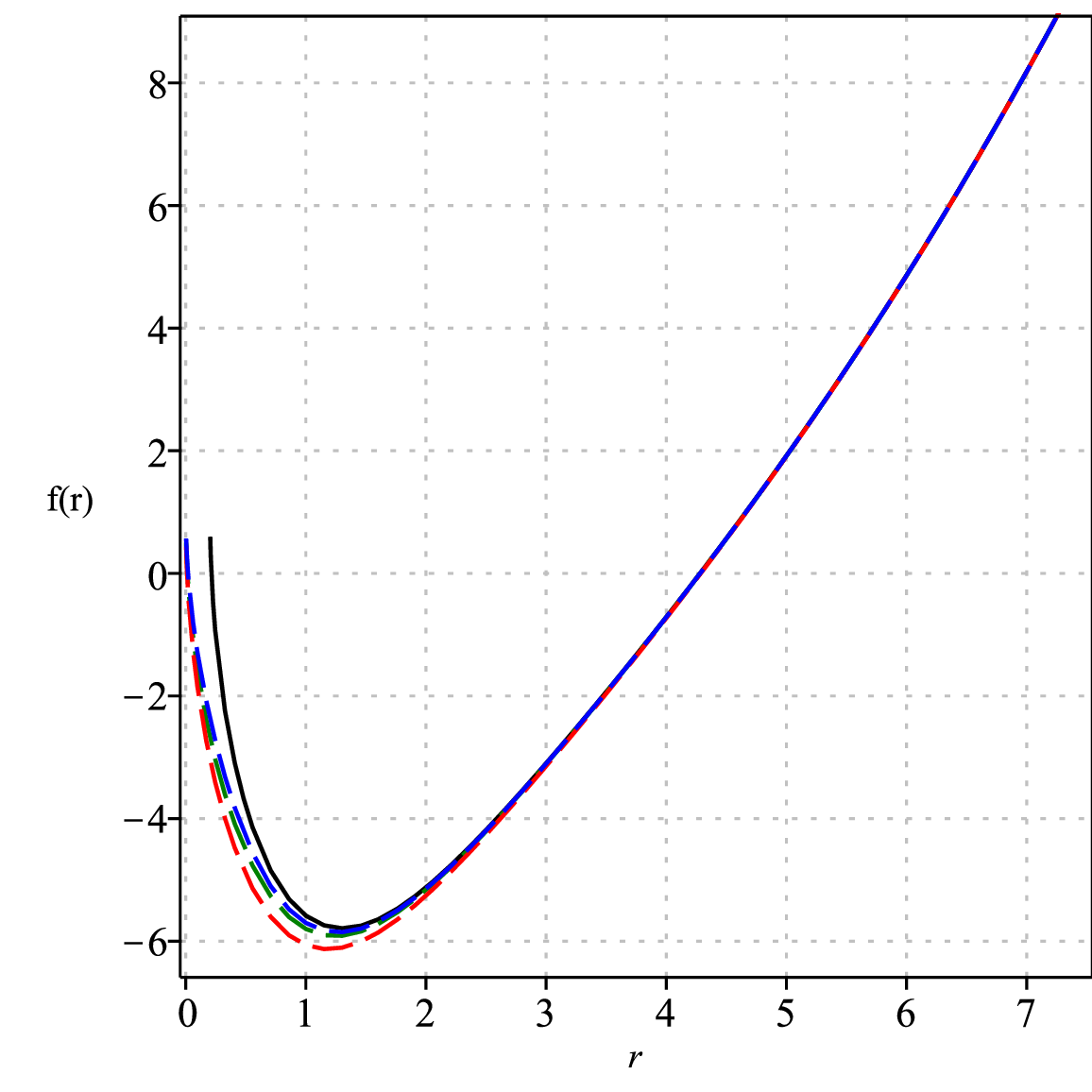}}\hfill
\caption{$\beta=0.0$ denoted by solid black line, $\beta=0.01$ denoted by blue dash line with, $\beta=0.05$ denoted by green dash line and  $\beta=1.0$ denoted by red dash line in EGB-NED with $M=10$, $Q_m=2$, $m=1.0$, $c=1$, $c_1=-1$, $c_2=1$ and $l=2$. }\label{fig:2}
\end{figure} 
\begin{figure}[H]
\centering
\subfloat[$m=0.0$ \& $Q_m=2$]{\includegraphics[width=.5\textwidth]{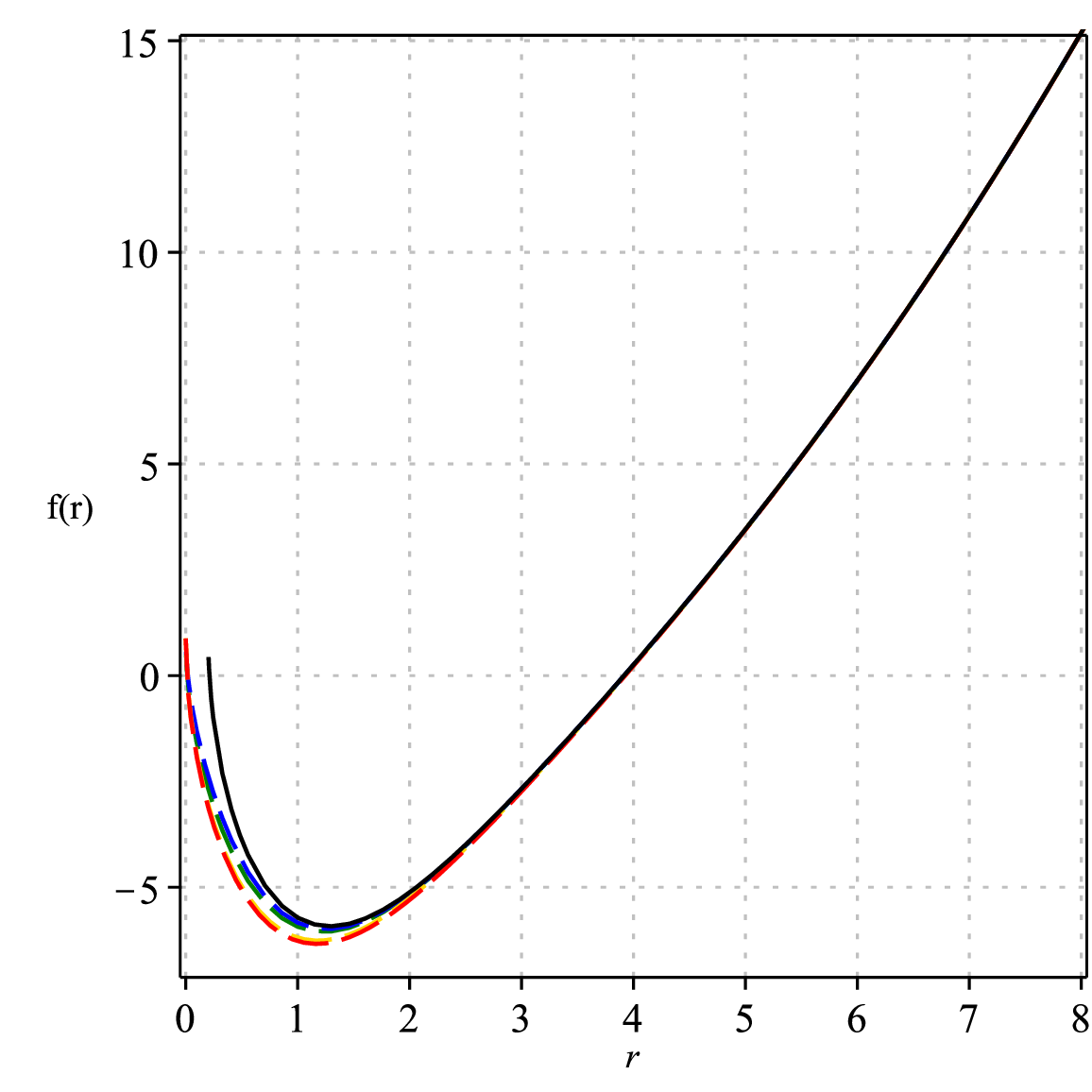}}\hfill
\subfloat[$m=1.0$ \& $\beta=0.5$]{\includegraphics[width=.5\textwidth]{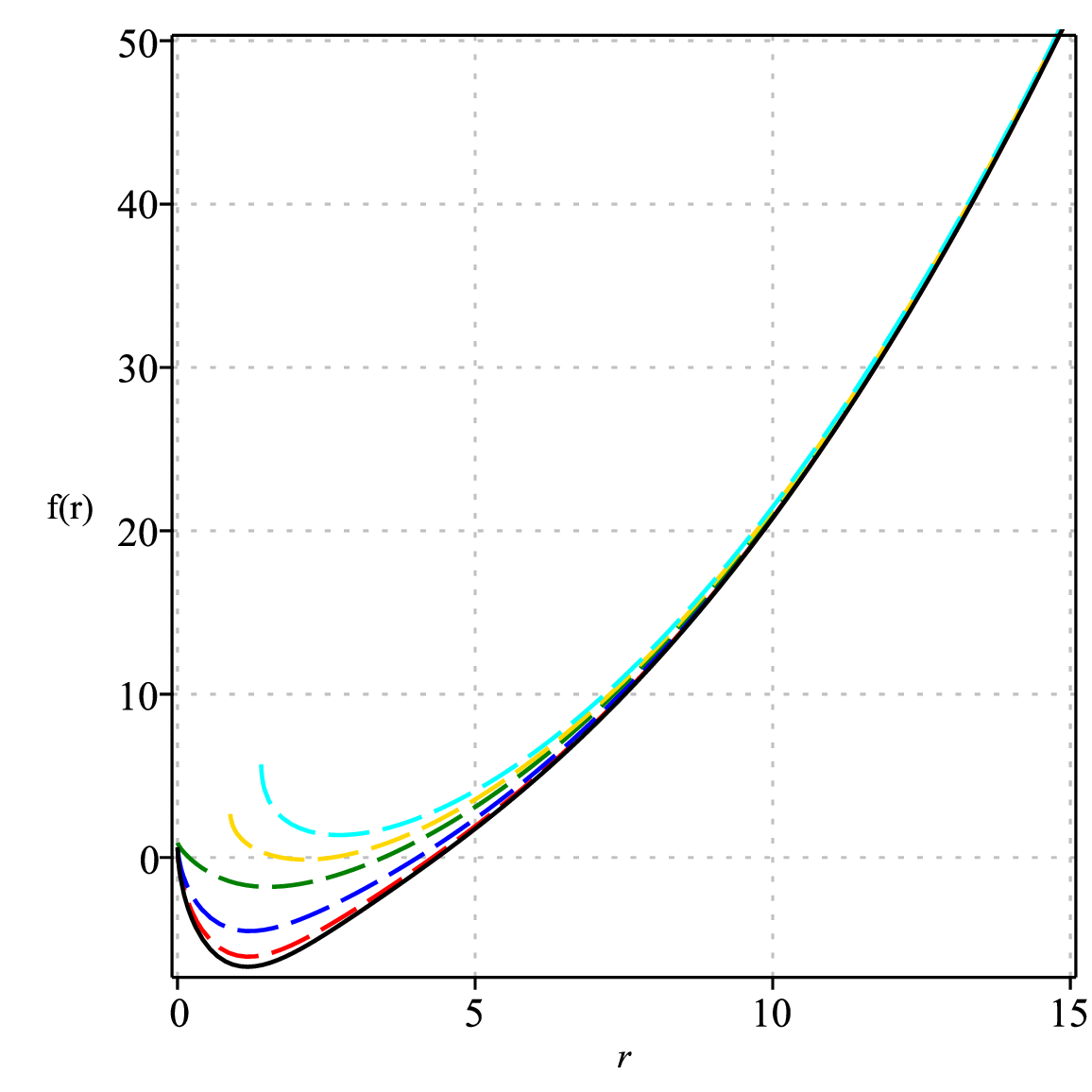}}\hfill
\caption{Left Panel: $\beta=0.0$ denoted by solid black line, $\beta=0.01$ denoted by blue dash line with, $\beta=0.05$ denoted by green dash line, $\beta=1.0$ denoted by gold dash line and $\beta=2.5$ denoted by red dash line in EGB-NED. Right Panel: $Q_m=0.0$ denoted by solid black line, $Q_m=2.0$ denoted by red dash line with, $Q_m=4.0$ denoted by blue dash line, $Q_m=6.0$ denoted by green dash line and $Q_m=7.0$ denoted by gold dash line and $Q_m=8.0$ denoted by cyan dash line in EGB-NED with $M=10$,  $\alpha=0.2$, $c=1$, $c_1=-1$, $c_2=1$ and $l=2$. }\label{fig:3}
\end{figure} 

In the above and below figures, we take $f(r)=e^{2A}$. In 
Fig. \ref{fig:1} and Fig. \ref{fig:2} we depicted the metric function 
for different values of $\alpha$, $\beta$ and graviton mass. Depending 
upon the graviton mass parameter black holes have one or two horizons. 
As the graviton mass increases, the position of the event horizon radius 
increases also keeping Gauss--Bonnet and NED parameters fixed. In Fig. 
\ref{fig:3} (b) the metric function is depicted for different values 
of charge and the black hole has two horizons, one is the Cauchy horizon 
and another is the black hole event horizon. For a critical value of 
charge $Q_m^{crit}=7.082$, black hole has only one degenerate horizon, 
and for $Q_m >Q_m^{crit}$ there is no horizon radius, and thus no black 
hole solution exists.

\begin{figure}[H]
\centering
\subfloat[$\beta=0.5$]{\includegraphics[width=.5\textwidth]{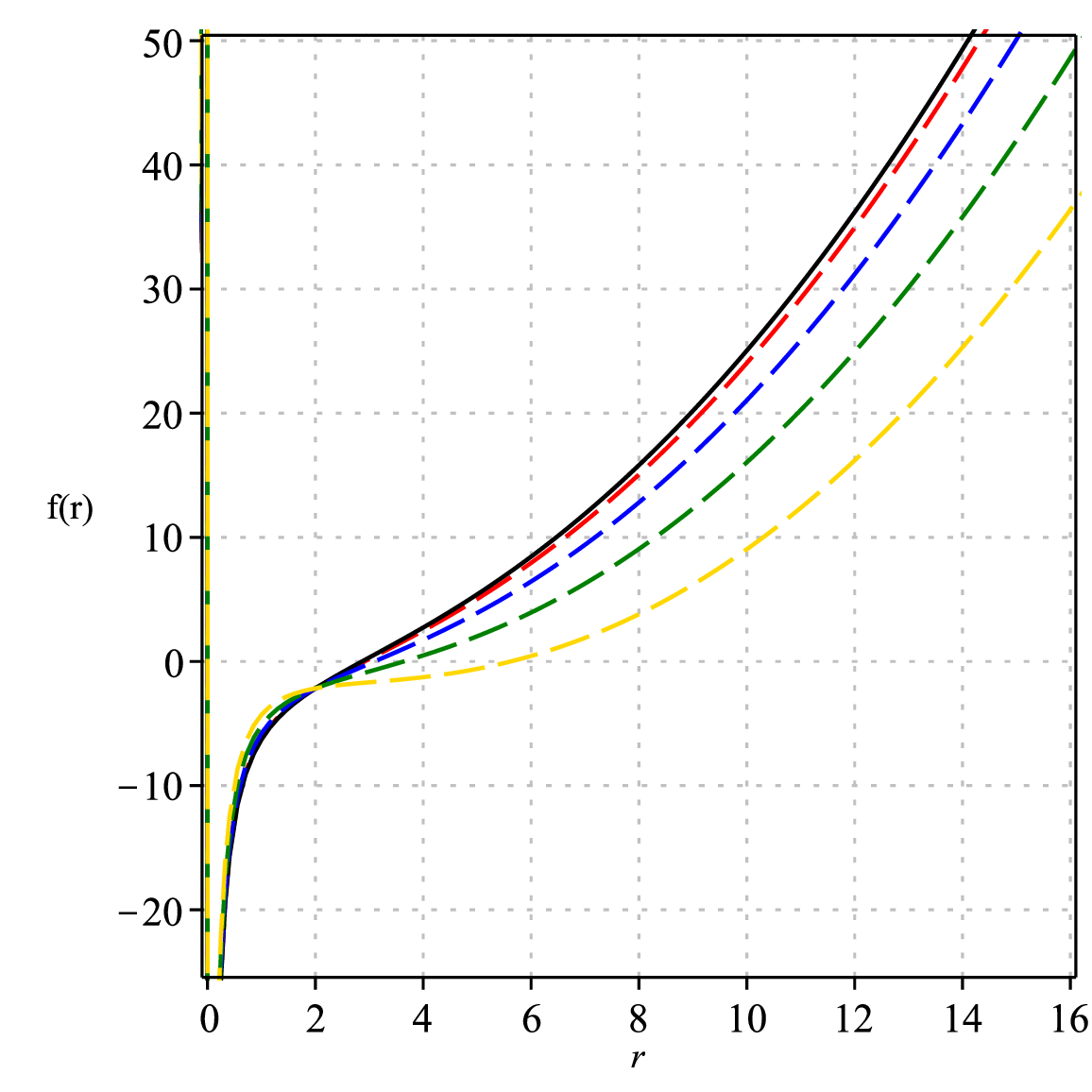}}\hfill
\subfloat[$m=1$]{\includegraphics[width=.5\textwidth]{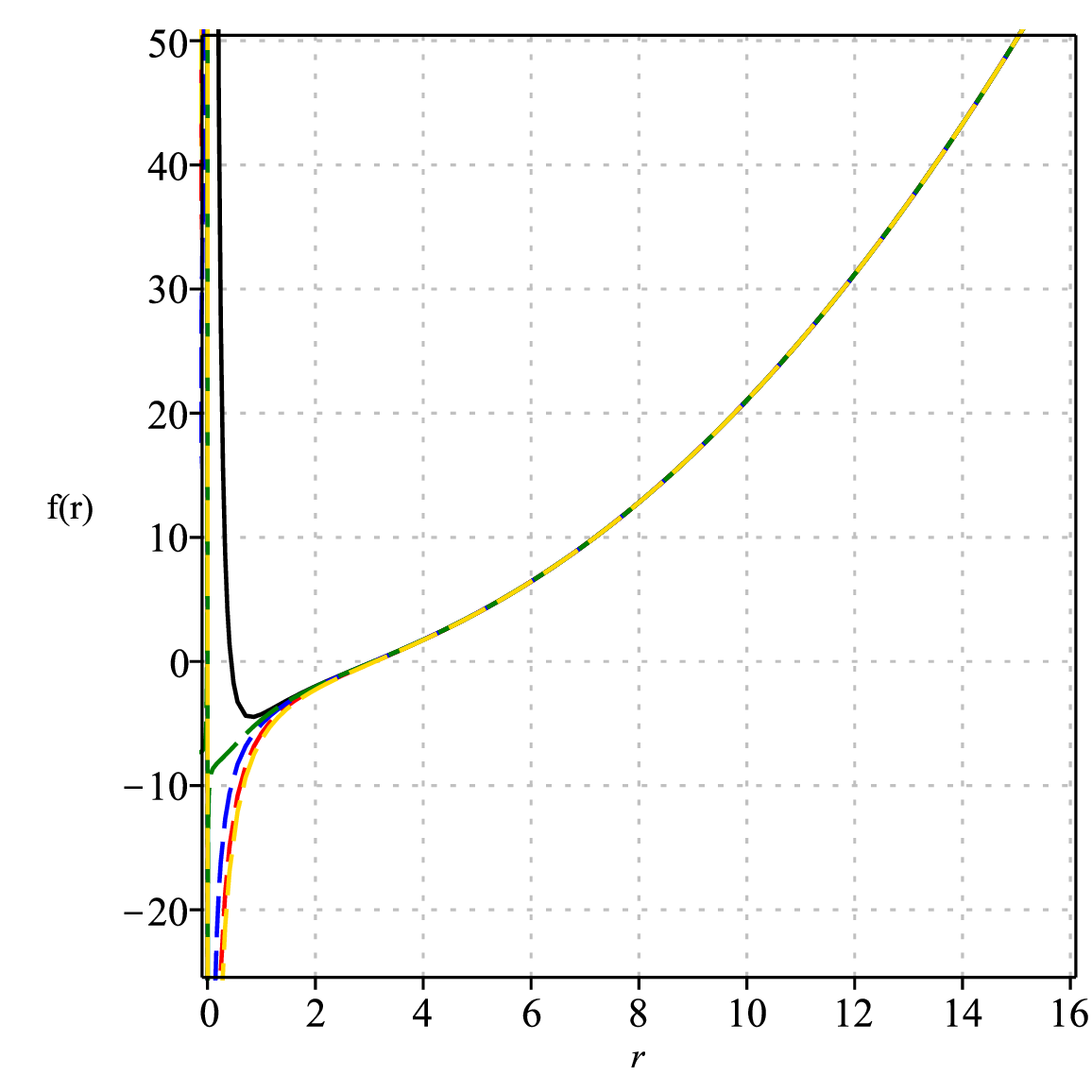}}\hfill
\caption{Left panel: $m=0.0$ denoted by solid black line, $m=0.5$ denoted by red dash line with, $m=1.0$ denoted by blue dash line, $m=1.5$ denoted by green dash line and $m=2.0$ denoted by gold dash line. Right panel: $\beta=0.0$ denoted by solid black line, $\beta=0.05$ denoted by red dash line with, $\beta=0.01$ denoted by green dash line and  $\beta=0.05$ denoted by blue dash line and $\beta=2.0$ denoted by gold dash line in GR-NED with $M=5$, $Q_m=2$,  $c=1$, $c_1=-1$, $c_2=1$ and $l=2$ }\label{fig:4}
\end{figure}
\begin{figure}[H]
    \centering
    \includegraphics[width=.6\textwidth]{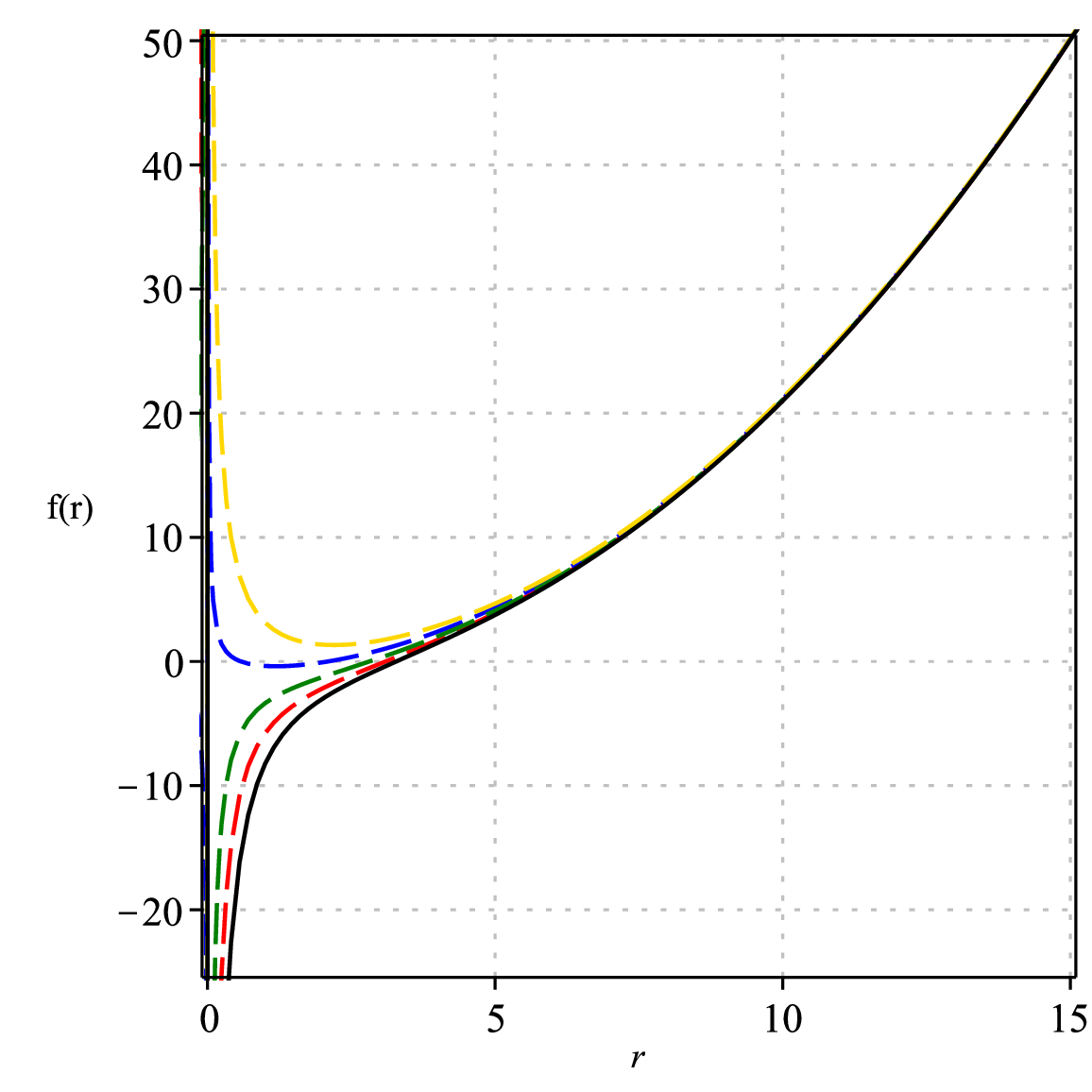}
    \caption{$Q_m=0.0$ denoted by solid black line, $Q_m=2.0$ denoted by red dash line with, $Q_m=3.0$ denoted by green dash line, $Q_m=4.0$ denoted by blue dash line and $Q_m=5.0$ denoted by gold dash line in GR-NED with $M=5$, $m=1.0$, $\beta=0.5$, $c=1$, $c_1=-1$, $c_2=1$ and $l=2$.}
    \label{fig:5}
\end{figure}

\begin{figure}[H]
    \centering
    \includegraphics[width=.6\textwidth]{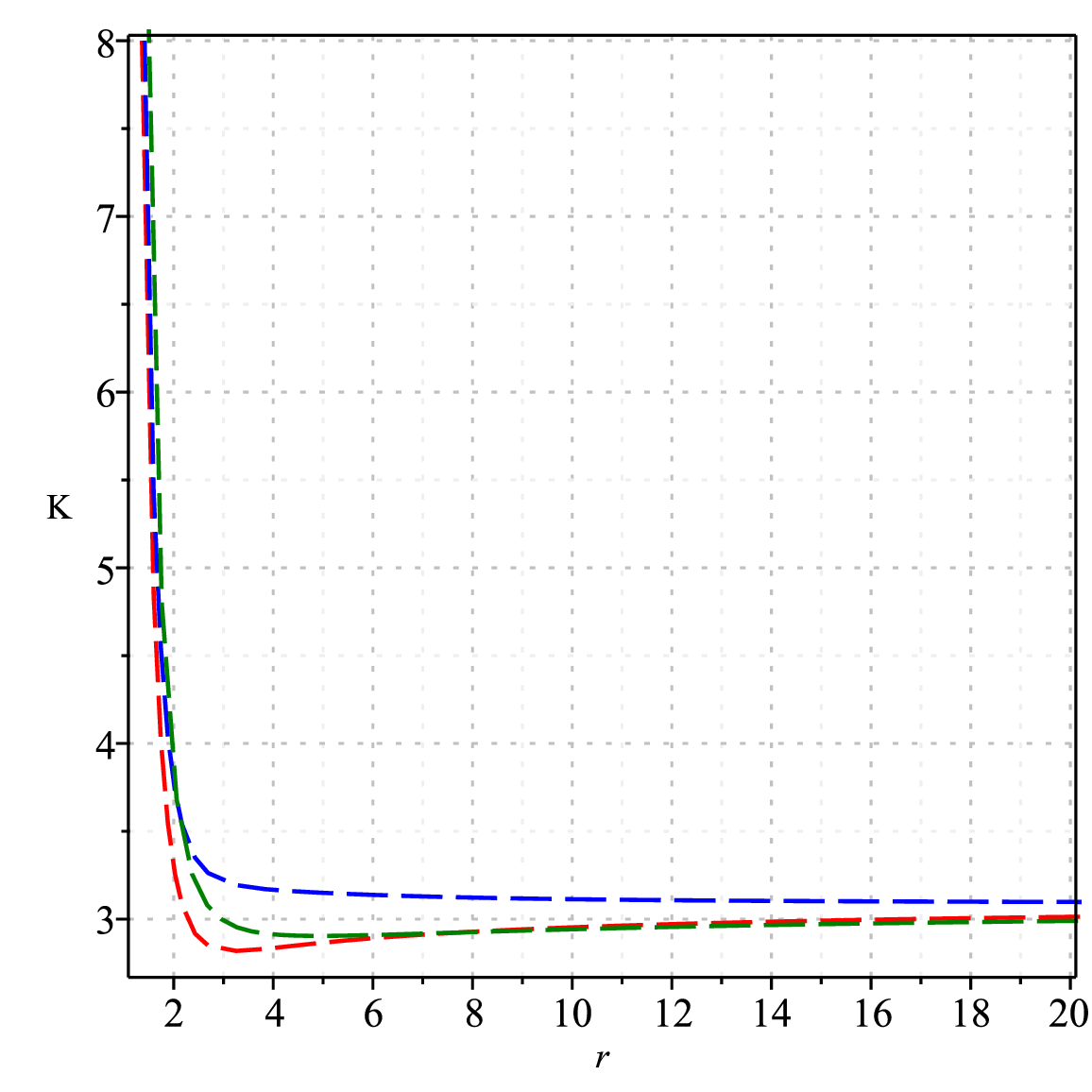}
    \caption{ $m=0.5$ denoted by red dash line in EGB--NED, $m=0.0$ denoted by blue dash line in EGB-NED and $m=0.5$ denoted by green dash line in GR-NED with $M=2$, $Q_m=0.6$, $\beta=0.5$, $\alpha=0.2$, $c=1$, $c_1=-1$, $c_2=1$ and $l=2$.}
    \label{fig:6}
\end{figure}

 In Fig. \ref{fig:4} (a), Fig. \ref{fig:4} (b), and \ref{fig:5} we 
 depicted the metric function of black hole in GR 
 coupled to NED for different values of graviton mass and NED parameter 
 $\beta$. The black hole has only one horizon and as we increase the 
 graviton mass the position of the horizon increases also. In Fig. 
 \ref{fig:5} metric function is depicted for different values of charge. 
 Depending upon the charge black hole has one or two horizons. For 
 $Q_m=4$ black hole has two horizons, $Q_m^{crit}=4.1681$ two horizons 
 coincide and there exists only one degenerate horizon. For 
 $Q_m>Q_m^{crit}$ there are no horizons and thus no black hole 
 solution exists. The Kretschmann scalar is plotted in Fig. \ref{fig:6}, 
 which shows that at $r \to 0$ Kretschmann scalar goes to infinity. 
 Therefore black holes have a true singularity at $r \to 0$, which is 
 hidden by the event horizon, and as $r \to \infty$ Kretschmann scalar 
 takes a constant positive value.

\section{Black Hole Thermodynamics}\label{sec:3}

In this section, we study the thermodynamics of black holes in extended-phase space. 
The cosmological constant is related to the pressure of the black holes 
\cite{Dolan:2010ha}. Thermodynamics of the black holes in GR coupled to 
Maxwell/BI electrodynamics was studied in 
Refs. \cite{Kubiznak:2012wp,Gunasekaran:2012dq}. Black hole thermodynamics in 
GR coupled to NED Lagrangian \eqref{eq:2.4} was studied in Ref. 
\cite{kruglov2022nonlinearly}. Thermodynamics of $4D$ EGB gravity black holes 
in Maxwell/BI was studied in Refs. 
\cite{Hegde:2020xlv,Yang:2020jno,Upadhyay:2022axg}. 
The physical mass of the black hole can be obtained from the relations $f(r_{+})=0$

\begin{equation}\label{eq:3.1}
M=\frac{r_{{+}}^{3}}{2}\Biggr[\frac{\alpha}{r_{{+}}^{4}}+\frac{1}{r_{{+}}^{2}}+\frac{Q_{m}^{2} (\frac{\pi}{2}-\arctan (\frac{r_{{+}}}{k}))}{k r_{{+}}^{3}}+\frac{1}{l^{2}}+\frac{m^{2} (2 c^{2} c_{2} +c c_{1} r_{{+}} )}{2 r_{{+}}^{2}}\Biggr].
\end{equation}
In the limit $\alpha \to 0$, above mass function reduces to the mass of massive Einstein gravity
\begin{equation}\label{eq:3.2}
M=\frac{r_{{+}}}{2}+\frac{Q_{m}^{2} \pi}{4 k}-\frac{Q_{m}^{2} \arctan  (\frac{r_{{+}}}{k})}{2 k}+\frac{r_{{+}}^{3}}{2 l^{2}}+\frac{r_{{+}} m^{2} c^{2} c_{2}}{2}+\frac{r_{{+}}^{2} m^{2} c c_{1}}{4}.
\end{equation}

If we take $\beta \to 0$ into equation \eqref{eq:3.1}, then the mass 
function reduces to the mass function of $4D$ EGB massive gravity with Maxwell 
electrodynamics \cite{Paul:2023mlh}. In the limit $m \to 0$, and $l=\infty$ 
equation \eqref{eq:3.1} is reduced to the mass function of black hole 
in EGB massless gravity coupled to NED, which was obtained in Ref. 
\cite{Kruglov:2021stm}. If one takes $\beta \to 0$ and $m=0$ into equation 
\eqref{eq:3.1}, then it reduces to $4D$ EGB massless 
gravity in Maxwell electrodynamics \cite{Fernandes:2020rpa}.  Furthermore, 
in the massless limit $m \to 0$ equation \eqref{eq:3.2} is reduced to the 
mass function of the black hole in GR coupled to NED 
\cite{kruglov2022nonlinearly}.

\begin{figure}[H]
\centering
\subfloat[$\alpha=0.2$]{\includegraphics[width=.5\textwidth]{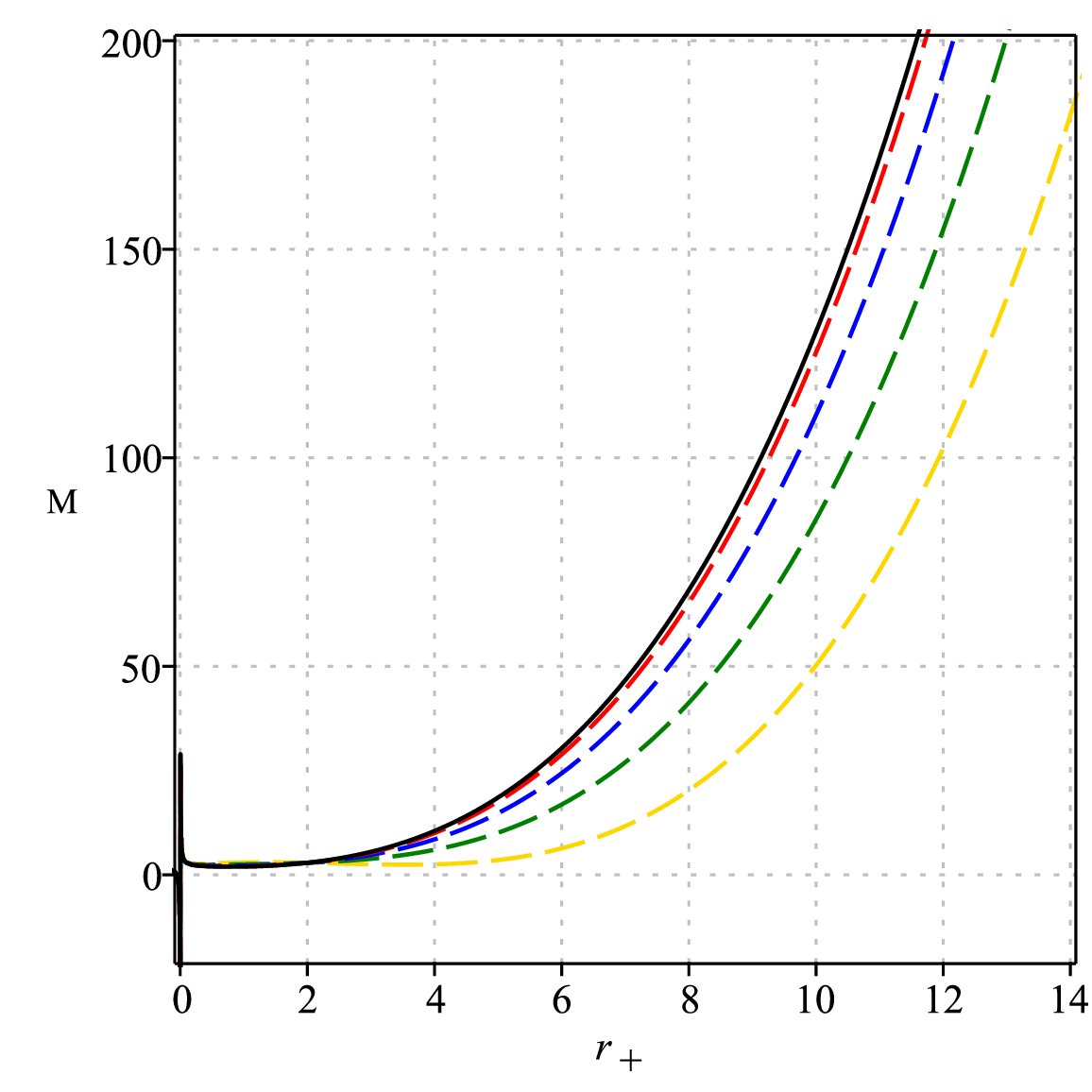}}\hfill
\subfloat[$\alpha=0.4$]{\includegraphics[width=.5\textwidth]{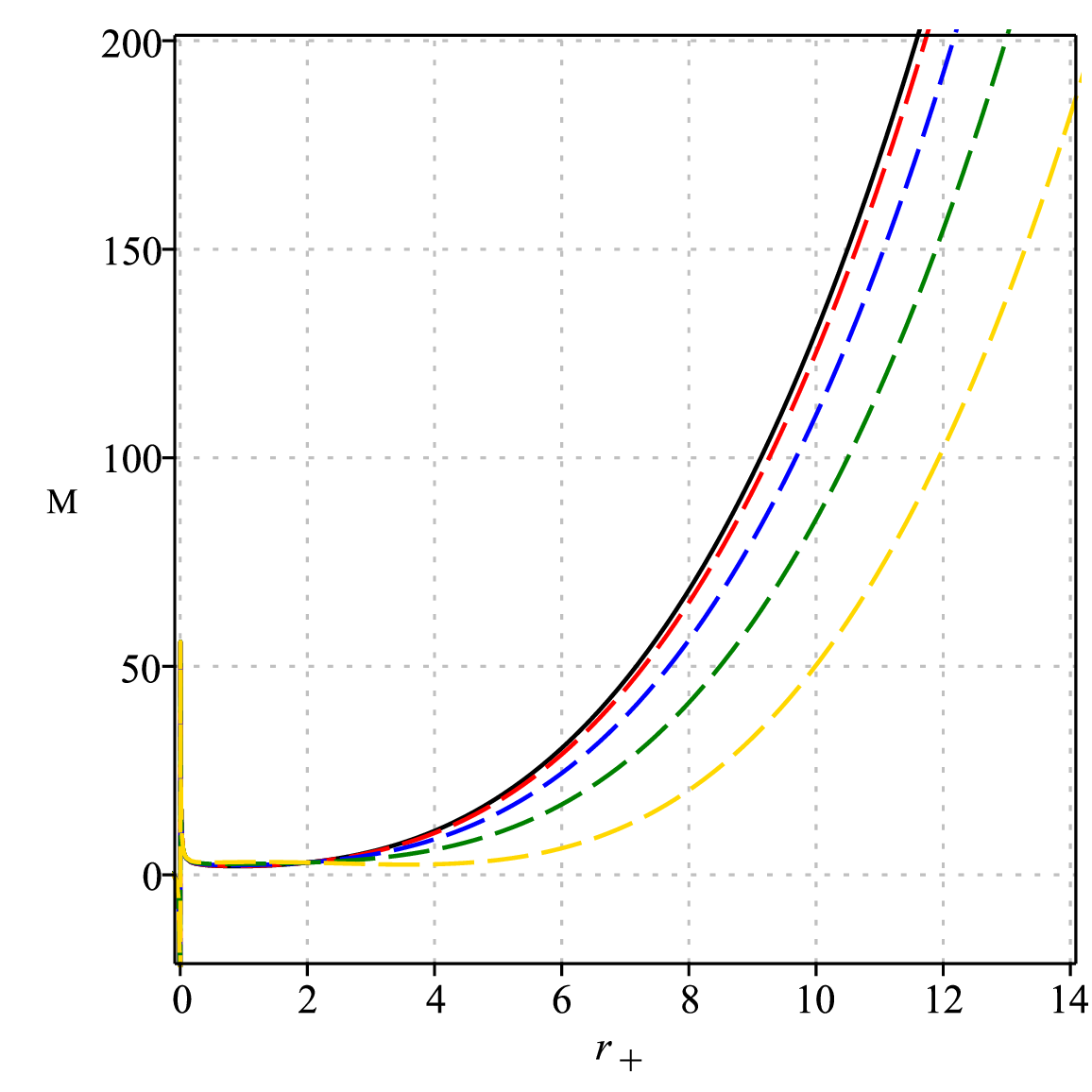}}\hfill
\caption{$m=0.0$ denoted by solid black line, $m=0.5$ denoted by red dash line with, $m=1.0$ denoted by blue dash line, $m=1.5$ denoted by green dash line and $m=2.0$ denoted by gold dash line  in EGB-NED with $Q_m=2$, $\beta=0.5$, $c=1$, $c_1=-1$, $c_2=1$ and $l=2$. }\label{fig:7}
\end{figure}

\begin{figure}[H]
\centering
\subfloat[$\alpha=0.02$]{\includegraphics[width=.5\textwidth]{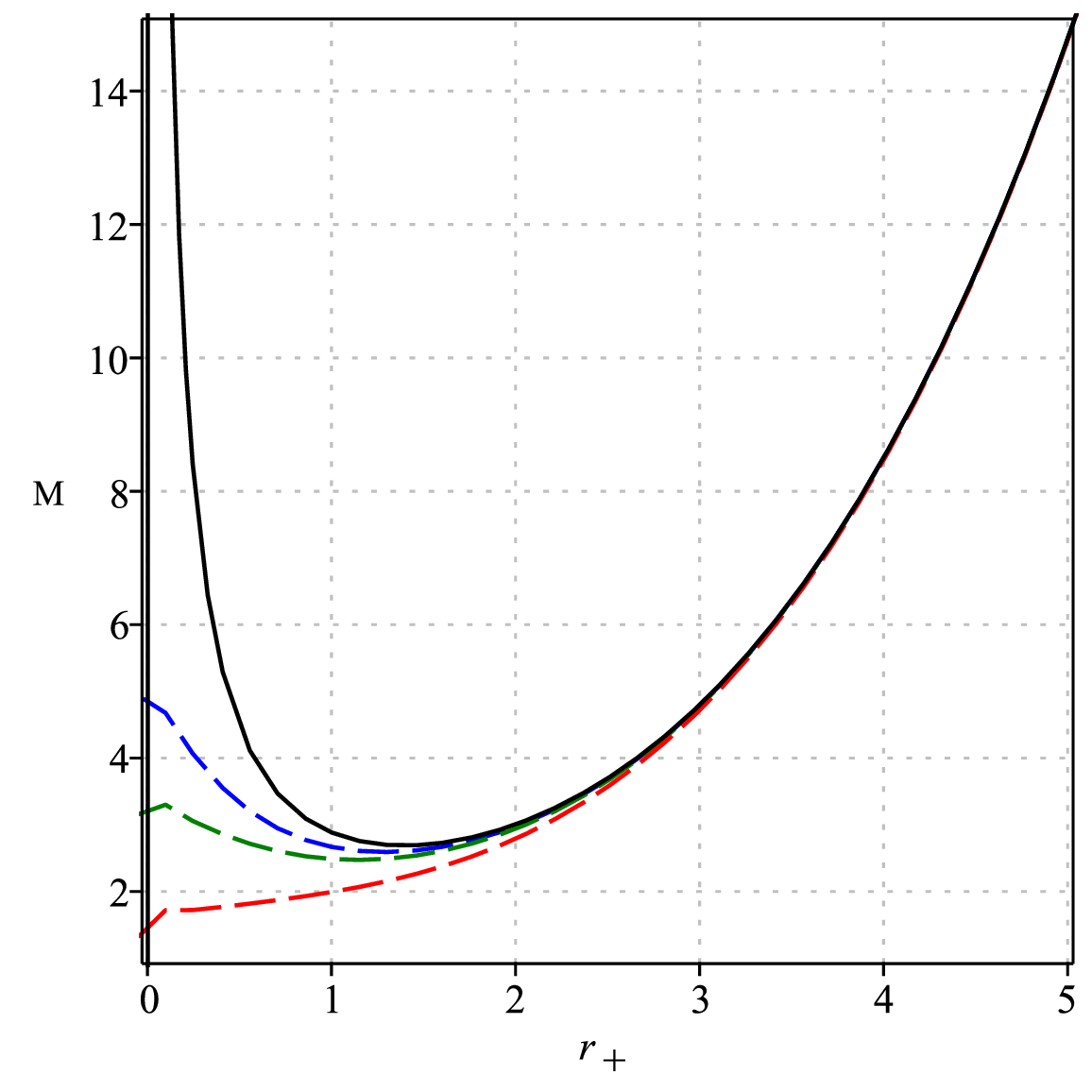}}\hfill
\subfloat[$\alpha=0.2$]{\includegraphics[width=.5\textwidth]{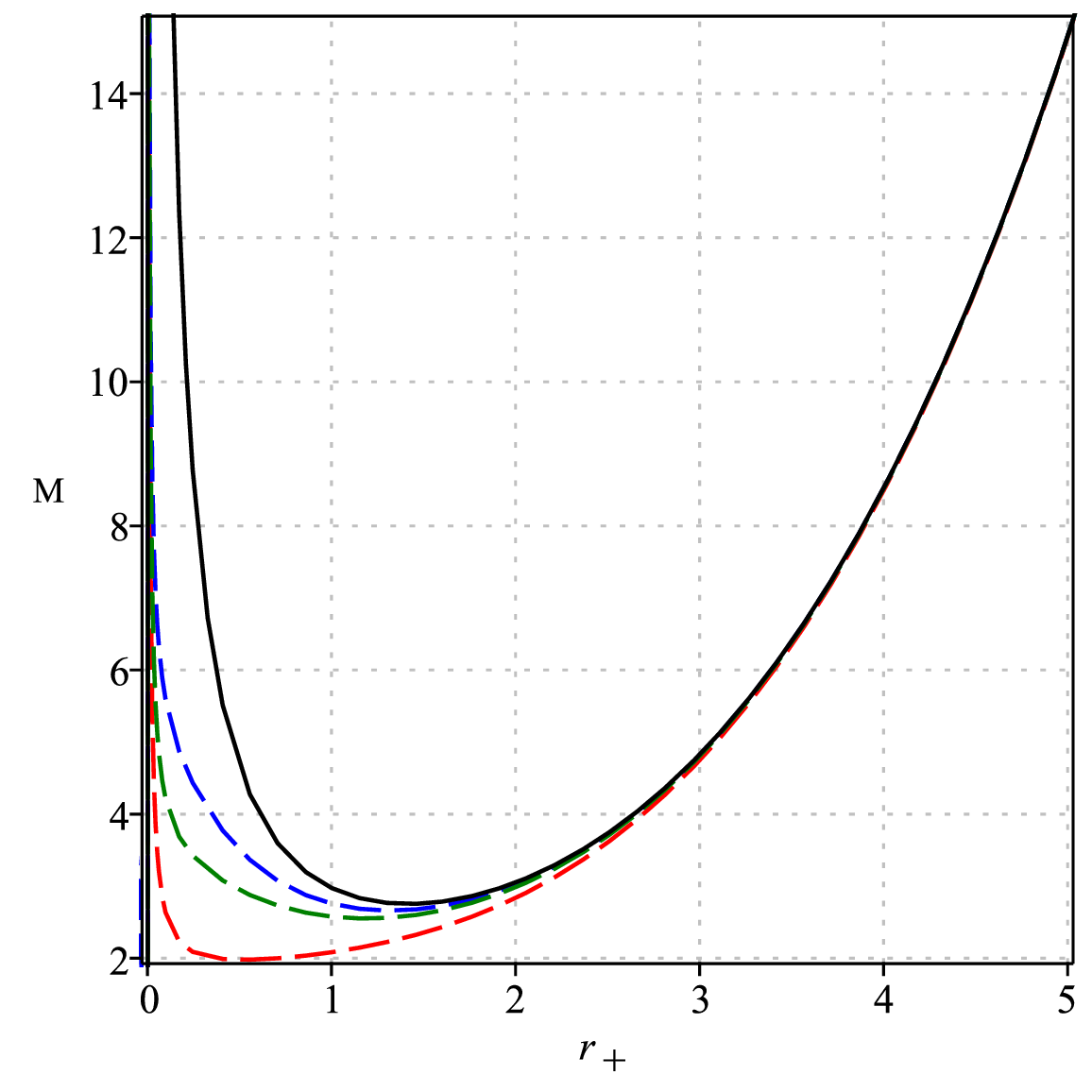}}\hfill
\caption{$\beta=0.0$ denoted by solid black line, $\beta=0.01$ denoted by blue dash line with, $\beta=0.05$ denoted by green dash line and  $\beta=1.0$ denoted by red dash line in EGB-NED with  $Q_m=2$, $m=1.0$, $c=1$, $c_1=-1$, $c_2=1$ and $l=2$. }\label{fig:8}
\end{figure}

The effects of graviton mass and Gauss--Bonnet coupling parameter on the 
physical mass of the black hole in NED are shown in Fig. \ref{fig:7} (a) and 
(b). There is a minimum horizon radius 
$r_{GB}^{min}$, when $r_{+}<r_{GB}^{min}$ graviton mass does not 
have any effects on the black hole mass but when $r_{+}>r_{GB}^{min}$ 
graviton mass slowly decreases black hole mass function, i.e. larger black 
hole ($r_{+}>>r_{GB}^{min}$) in massive gravity coupled to NED have smaller 
mass compared to a larger black hole ($r_{+}>>r_{GB}^{min}$) in massless gravity 
coupled to NED. The effects of nonlinear parameter $\beta$ on the black hole 
mass are shown in Fig. \ref{fig:8} (a) and (b). The effects are clearly visible 
for smaller-sized black holes and for larger-sized black holes, the NED 
parameter does not have any effect on the mass function. For small-sized 
black hole mass function increases as the $\beta$ value decreases. 

\begin{figure}[H]
\centering
\subfloat[$m=0.0$ \&  $Q_m=2$]{\includegraphics[width=.5\textwidth]{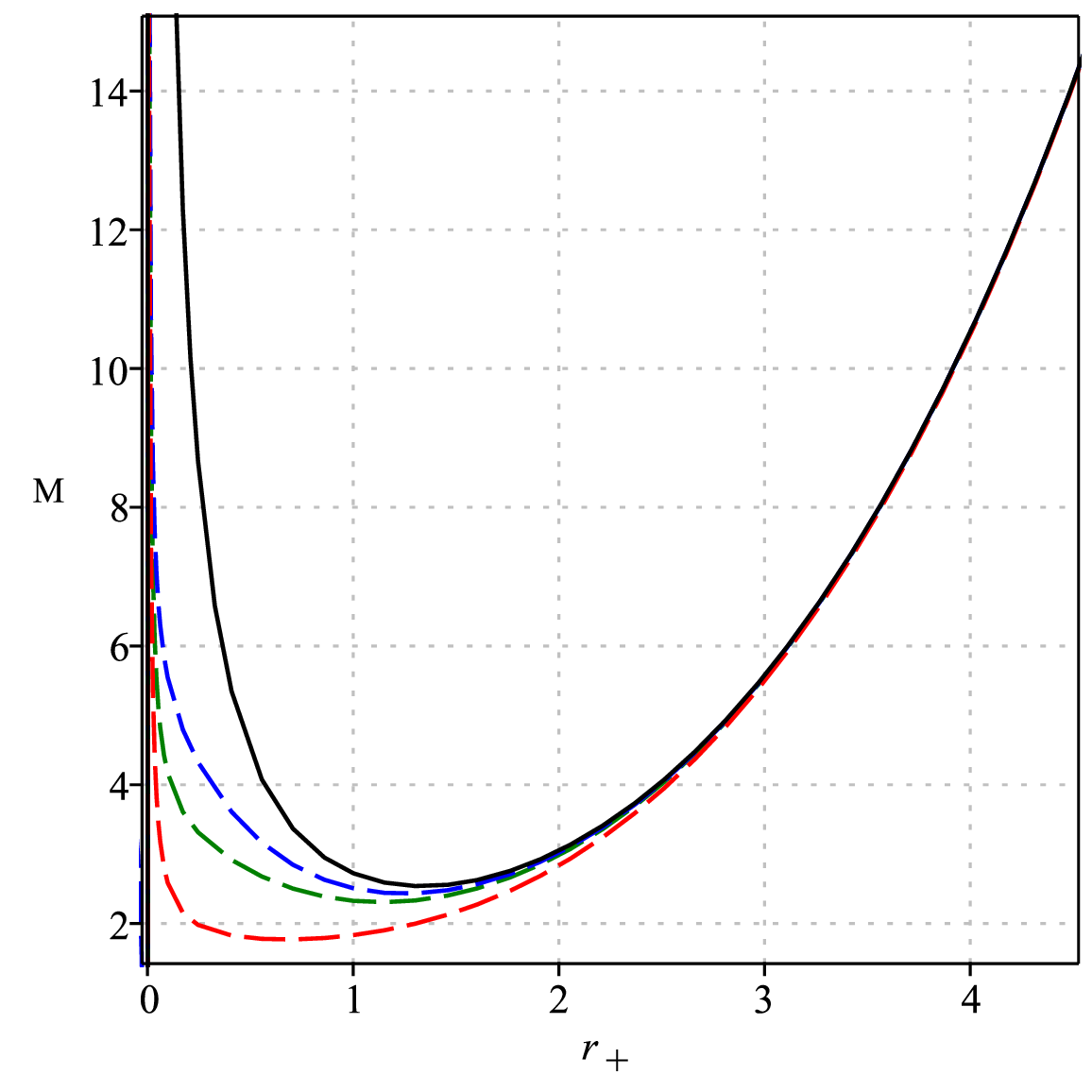}}\hfill
\subfloat[$m=1.0$ \&  $\beta=0.5$]{\includegraphics[width=.5\textwidth]{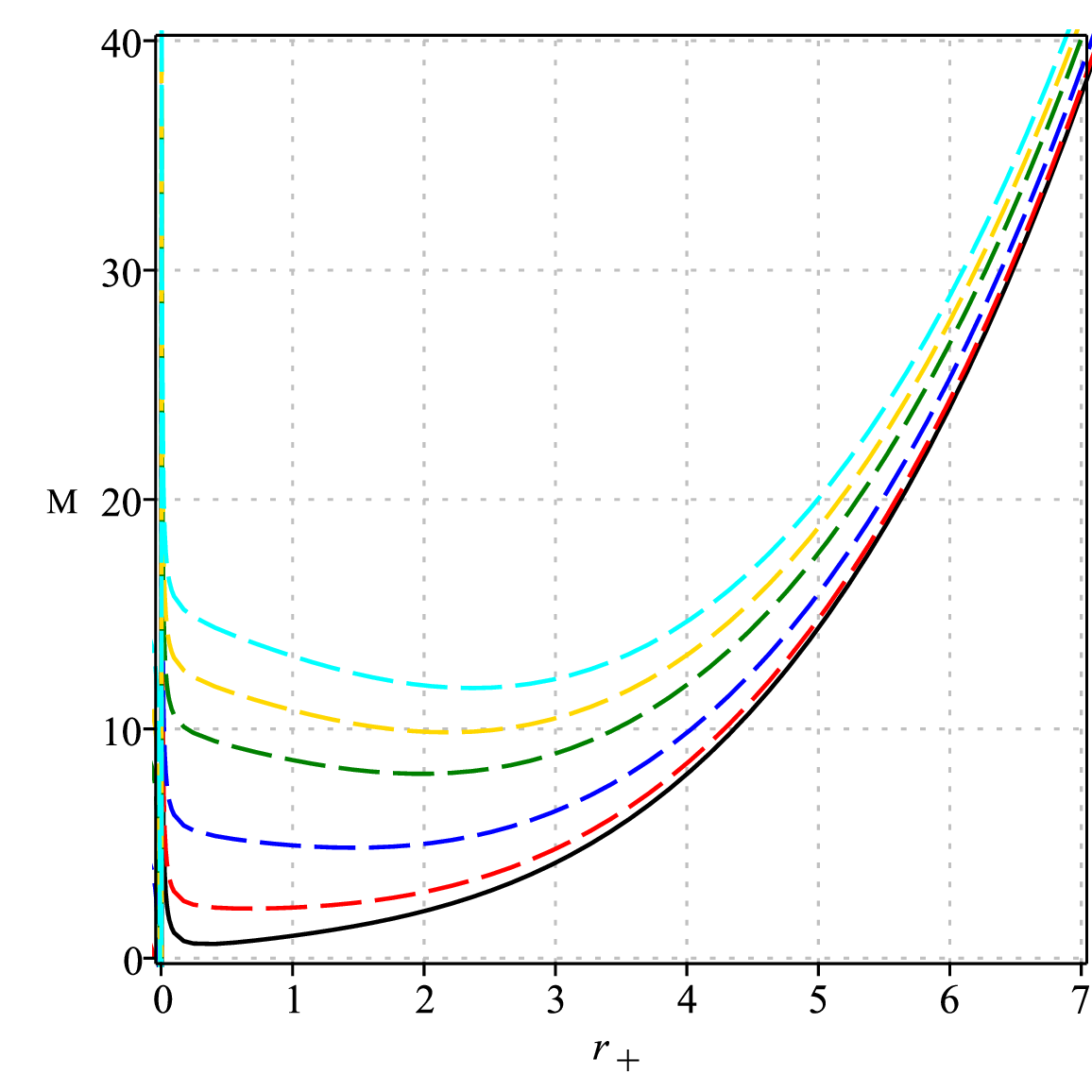}}\hfill
\caption{Left Panel: $\beta=0.0$ denoted by solid black line, $\beta=0.01$ denoted by blue dash line with, $\beta=0.05$ denoted by green dash line, $\beta=1.0$ denoted by red dash line in EGB-NED. Right Panel: $Q_m=0.0$ denoted by solid black line, $Q_m=2.0$ denoted by red dash line with, $Q_m=4.0$ denoted by blue dash line, $Q_m=6.0$ denoted by green dash line and $Q_m=7.0$ denoted by gold dash line and $Q_m=8.0$ denoted by cyan dash line in EGB-NED with $Q_m=2$, $\alpha=0.2$, $c=1$, $c_1=-1$, $c_2=1$ and $l=2$.}\label{fig:9}
\end{figure}

\begin{figure}[H]
\centering
\subfloat[$\beta=0.5$]{\includegraphics[width=.5\textwidth]{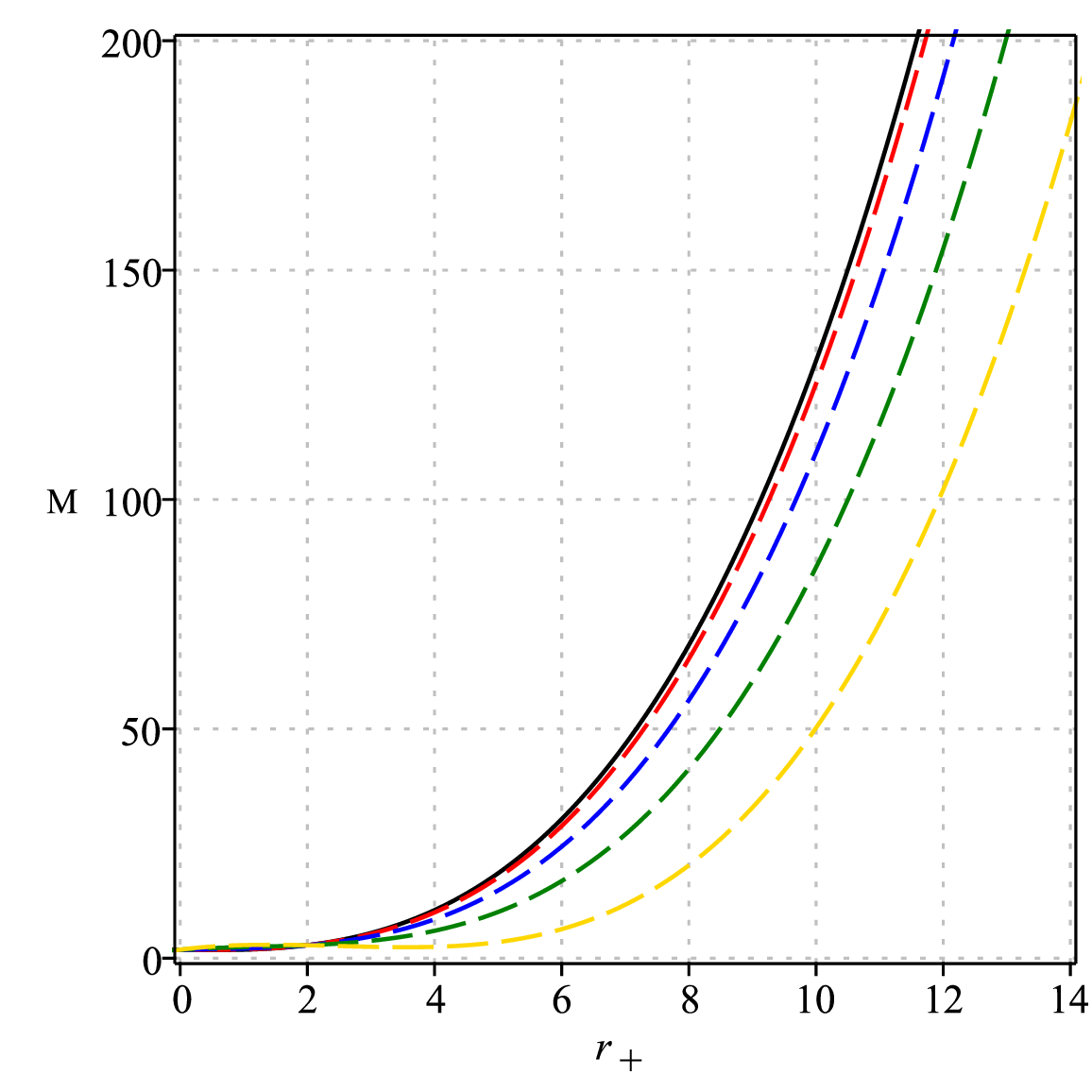}}\hfill
\subfloat[$m=1$]{\includegraphics[width=.5\textwidth]{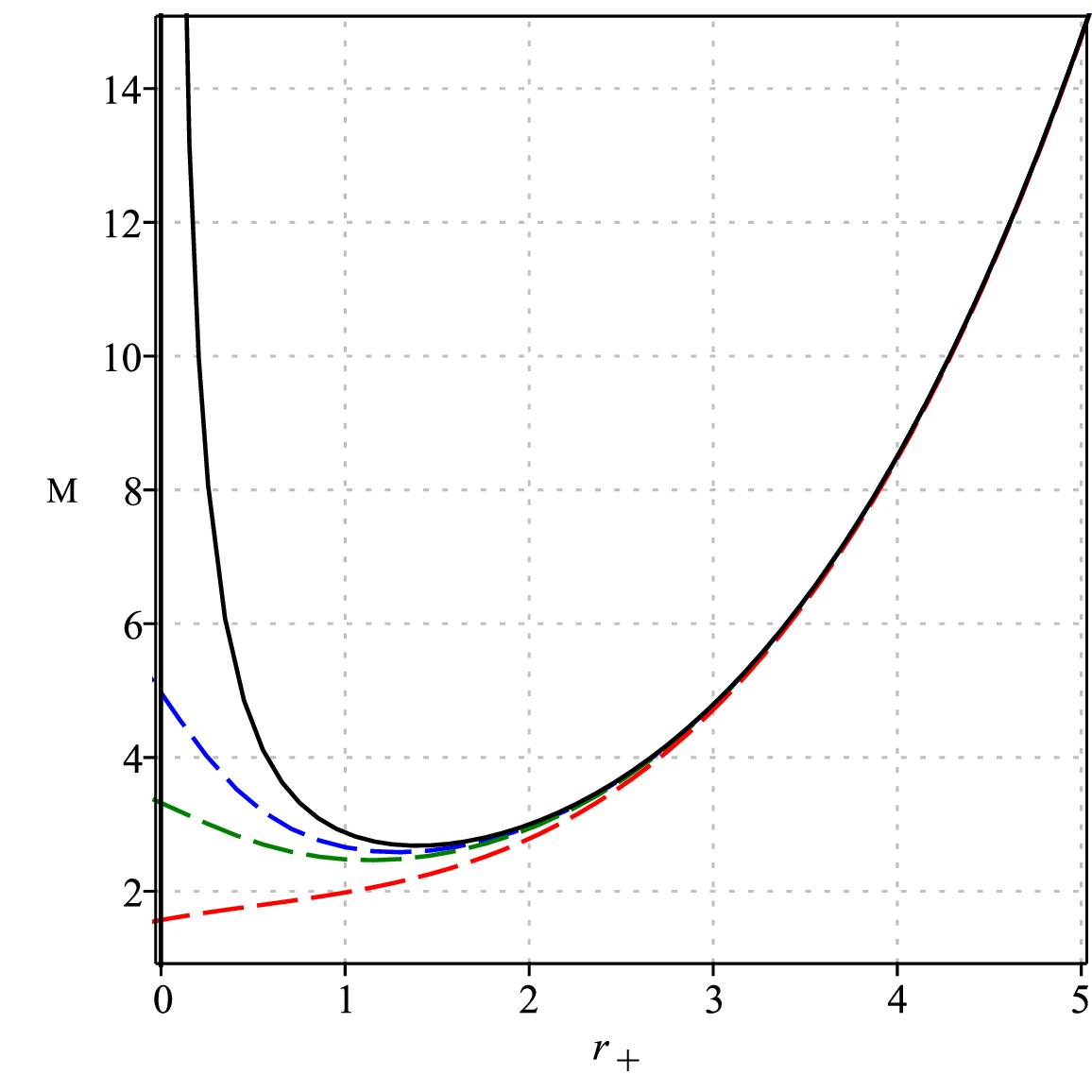}}\hfill
\caption{Left panel: $m=0.0$ denoted by solid black line, $m=0.5$ denoted by red dash line with, $m=1.0$ denoted by blue dash line, $m=1.5$ denoted by green dash line and $m=2.0$ denoted by gold dash line. Right panel: $\beta=0.0$ denoted by solid black line, $\beta=0.01$ denoted by blue dash line, $\beta=0.05$ denoted by green dash line and  $\beta=1.0$ denoted by red dash line. GR-NED with $Q_m=2$,  $c=1$, $c_1=-1$, $c_2=1$ and $l=2$ }\label{fig:10}
\end{figure}

\begin{figure}[H]
    \centering
    \includegraphics[width=.6\textwidth]{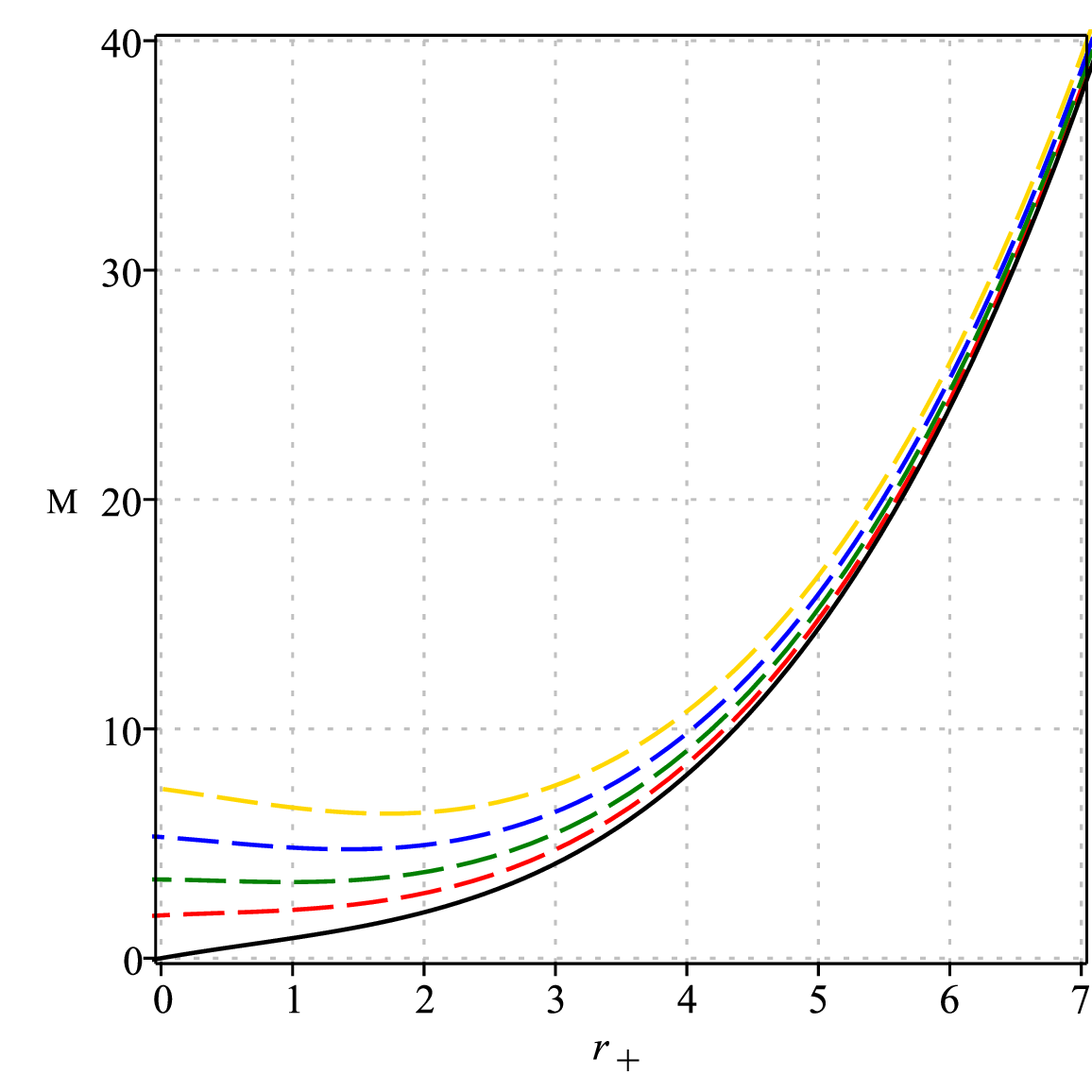}
    \caption{$Q_m=0.0$ denoted by solid black line, $Q_m=2.0$ denoted by red dash line with, $Q_m=3.0$ denoted by green dash line, $Q_m=4.0$ denoted by blue dash line and $Q_m=5.0$ denoted by gold dash line in GR-NED with $m=1.0$, $\beta=0.5$, $c=1$, $c_1=-1$, $c_2=1$ and $l=2$.}
    \label{fig:11}
\end{figure}
A similar kind of behaviour is shown (Fig. \ref{fig:10} \& 
Fig. \ref{fig:11}) for black holes in massive GR 
coupled to NED, $r_{GR}^{min}$ is the minimum horizon radius of the 
black hole in GR.

The Hawking temperature of the black hole is defined as 
\begin{equation}\label{eq:3.3}
    T_{H} = \frac{f^{\prime}(r) }{4 \pi} \bigg\vert_{r=r_{+}},
\end{equation}
where prime denotes differentiation with respect to r. Therefore, one can obtain Hawking temperature as 
\begin{equation*}
T_{H}= \frac{1}{ 4 \pi  r_{{+}} l^{2} (r_{{+}}^{2}+2 \alpha )  (k^{2}+r_{{+}}^{2})}\Biggr[ 3 r_{{+}}^{6}+c c_{1} l^{2} m^{2} r_{{+}}^{5}+ \bigl( (c^{2} c_{2} m^{2}+1) l^{2}+3 k^{2}\bigl) r_{{+}}^{4}+c c_{1} k^{2} l^{2} m^{2} r_{{+}}^{3} 
\end{equation*}
\begin{equation}\label{eq:3.4}
+ \bigl( (c^{2} c_{2} m^{2}+1) k^{2}-Q_{m}^{2}-\alpha \bigl) l^{2} r_{{+}}^{2}-\alpha k^{2} l^{2} \Biggr].
\end{equation}

\begin{figure}[H]
\centering
\subfloat[$\alpha=0.2$]{\includegraphics[width=.5\textwidth]{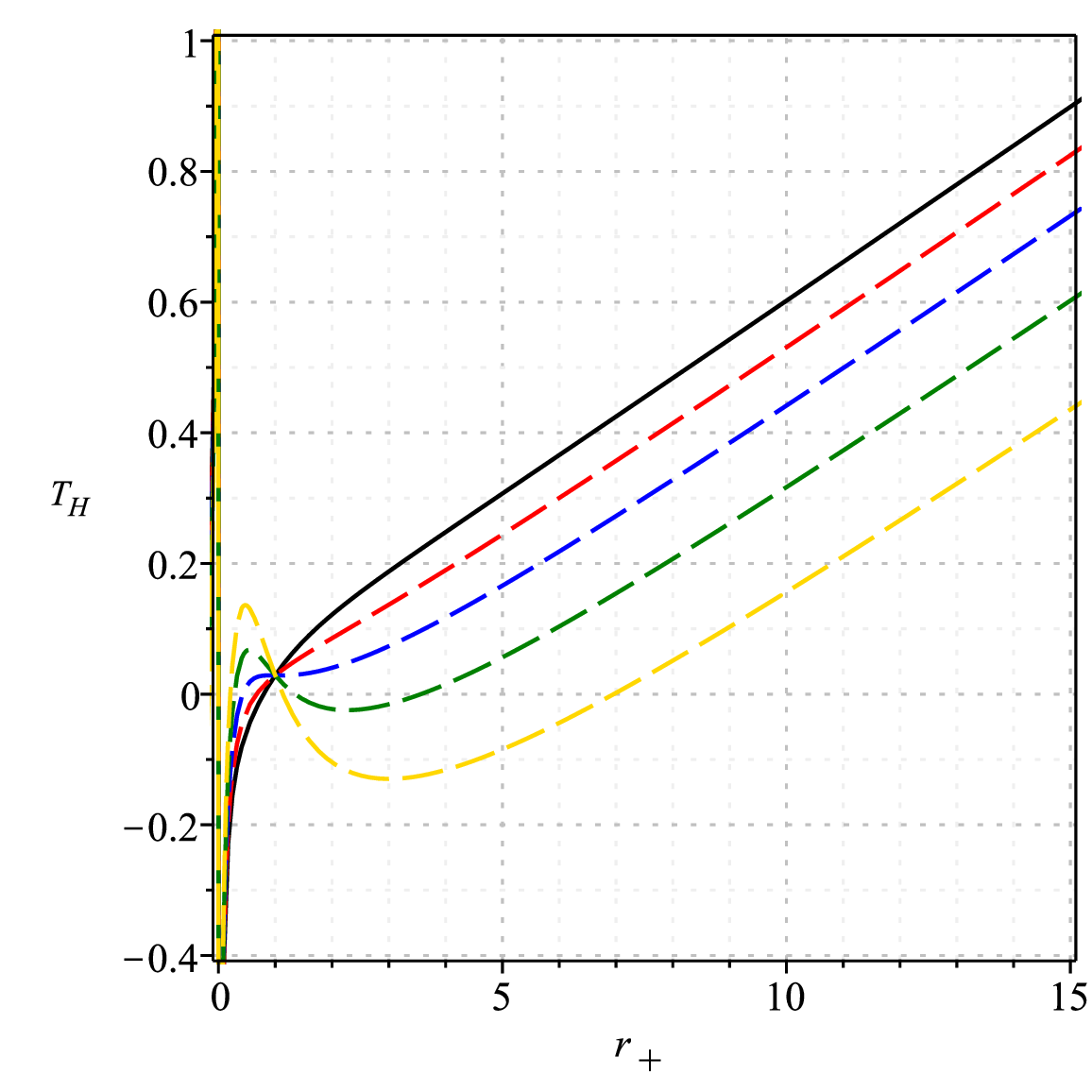}}\hfill
\subfloat[$\alpha=0.4$]{\includegraphics[width=.5\textwidth]{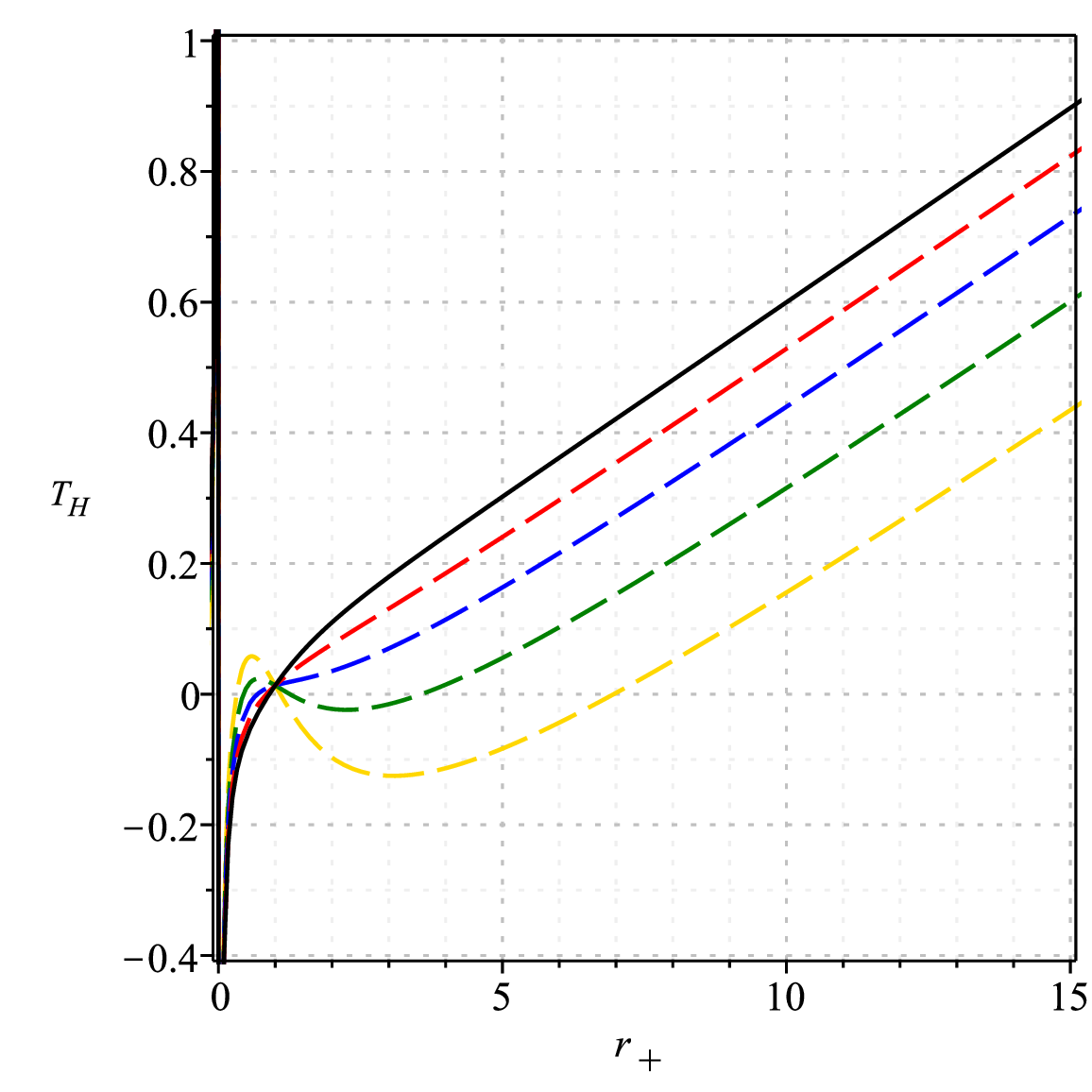}}\hfill
\caption{$m=0.0$ denoted by solid black line, $m=0.5$ denoted by red dash line with, $m=1.0$ denoted by blue dash line, $m=1.5$ denoted by green dash line and $m=2.0$ denoted by gold dash line  in EGB-NED with $Q_m=2$, $\beta=0.5$, $c=1$, $c_1=-1$, $c_2=1$ and $l=2$. }\label{fig:12}
\end{figure}

\begin{figure}[H]
\centering
\subfloat[$\alpha=0.04$]{\includegraphics[width=.5\textwidth]{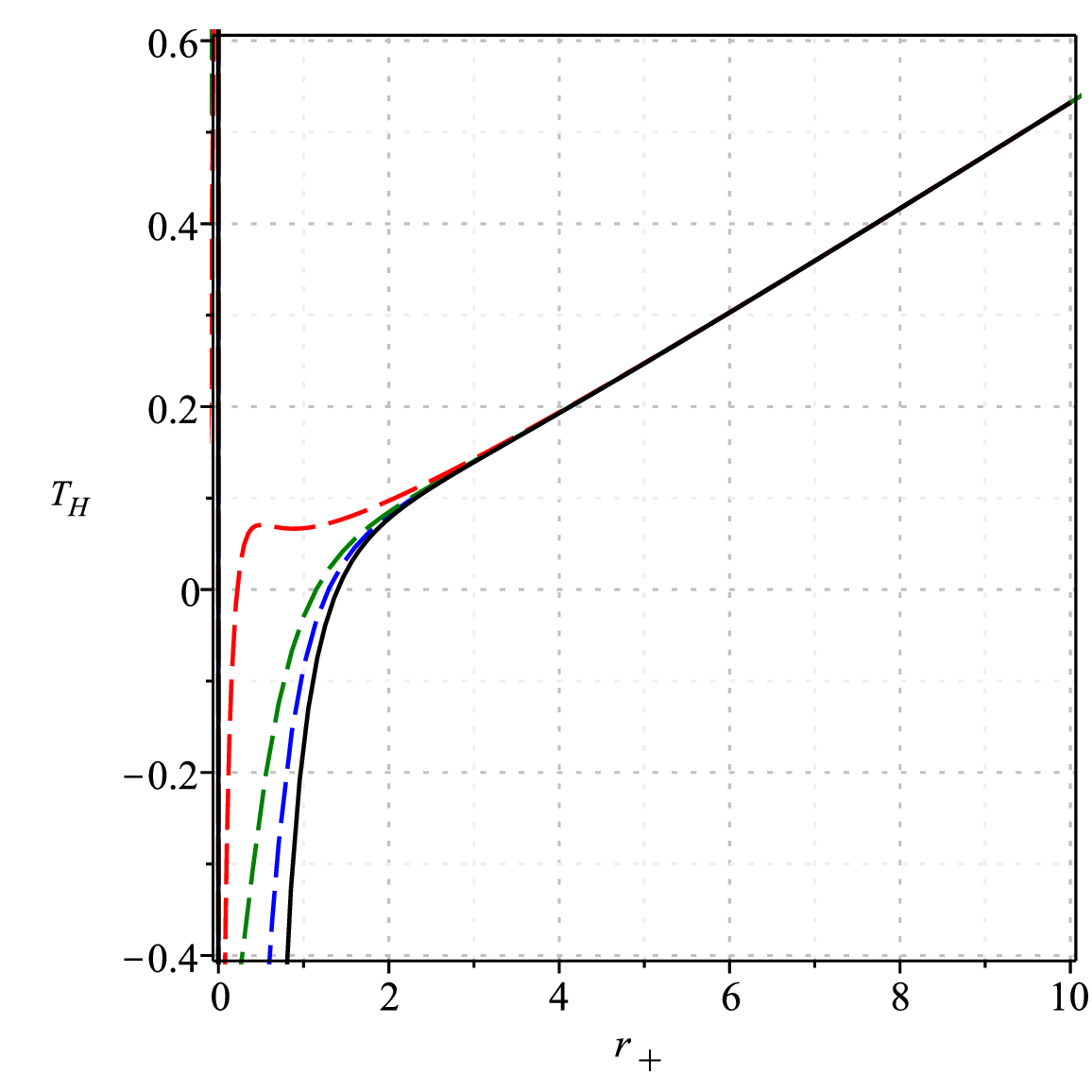}}\hfill
\subfloat[$\alpha=0.4$]{\includegraphics[width=.5\textwidth]{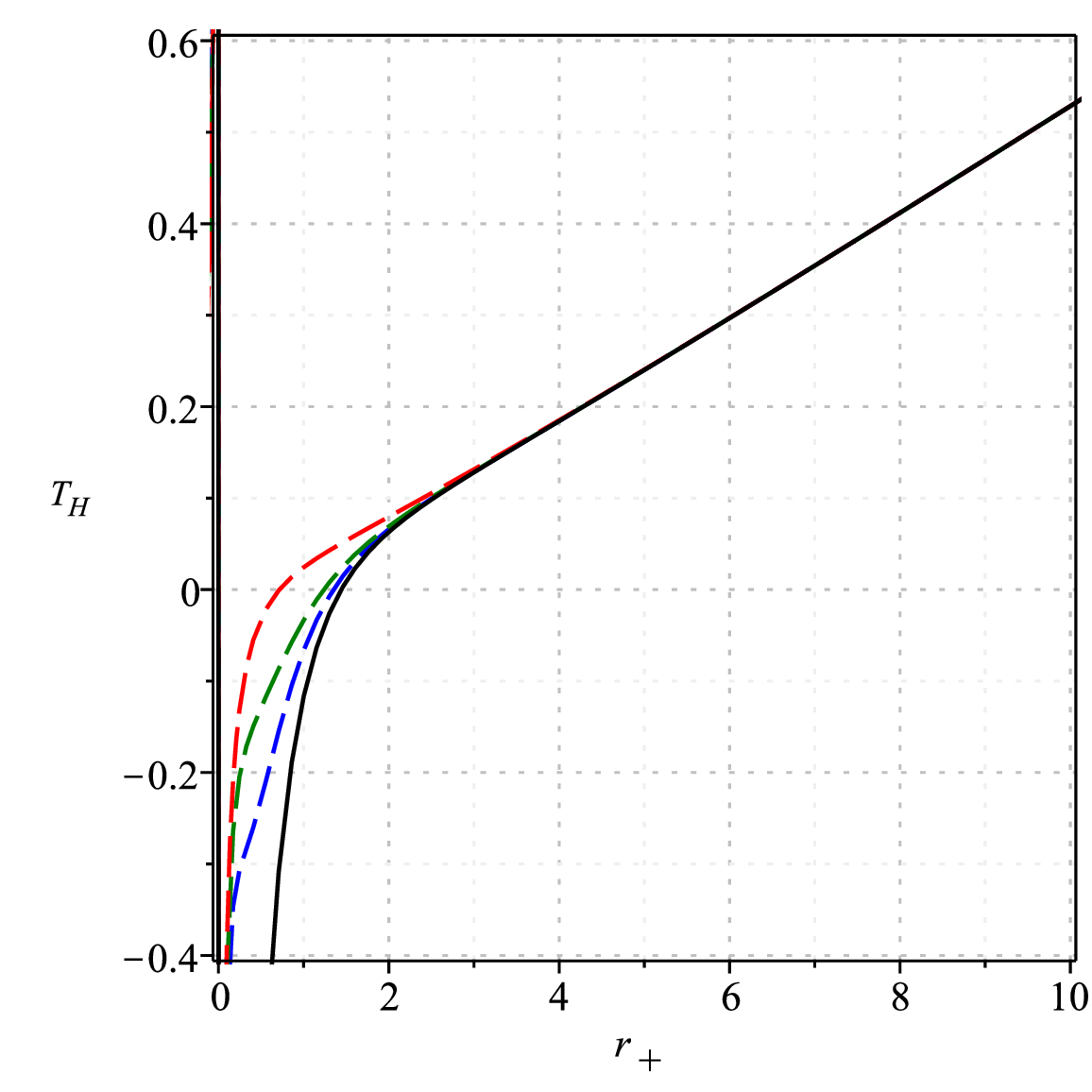}}\hfill
\caption{$\beta=0.0$ denoted by solid black line, $\beta=0.01$ denoted by blue dash line with, $\beta=0.05$ denoted by green dash line and  $\beta=1.0$ denoted by red dash line in EGB-NED with  $Q_m=2$, $m=1.0$, $c=1$, $c_1=-1$, $c_2=1$ and $l=2$. }\label{fig:13}
\end{figure}

\begin{figure}[H]
\centering
\subfloat[$m=0.0$ \&  $Q_m=2$]{\includegraphics[width=.5\textwidth]{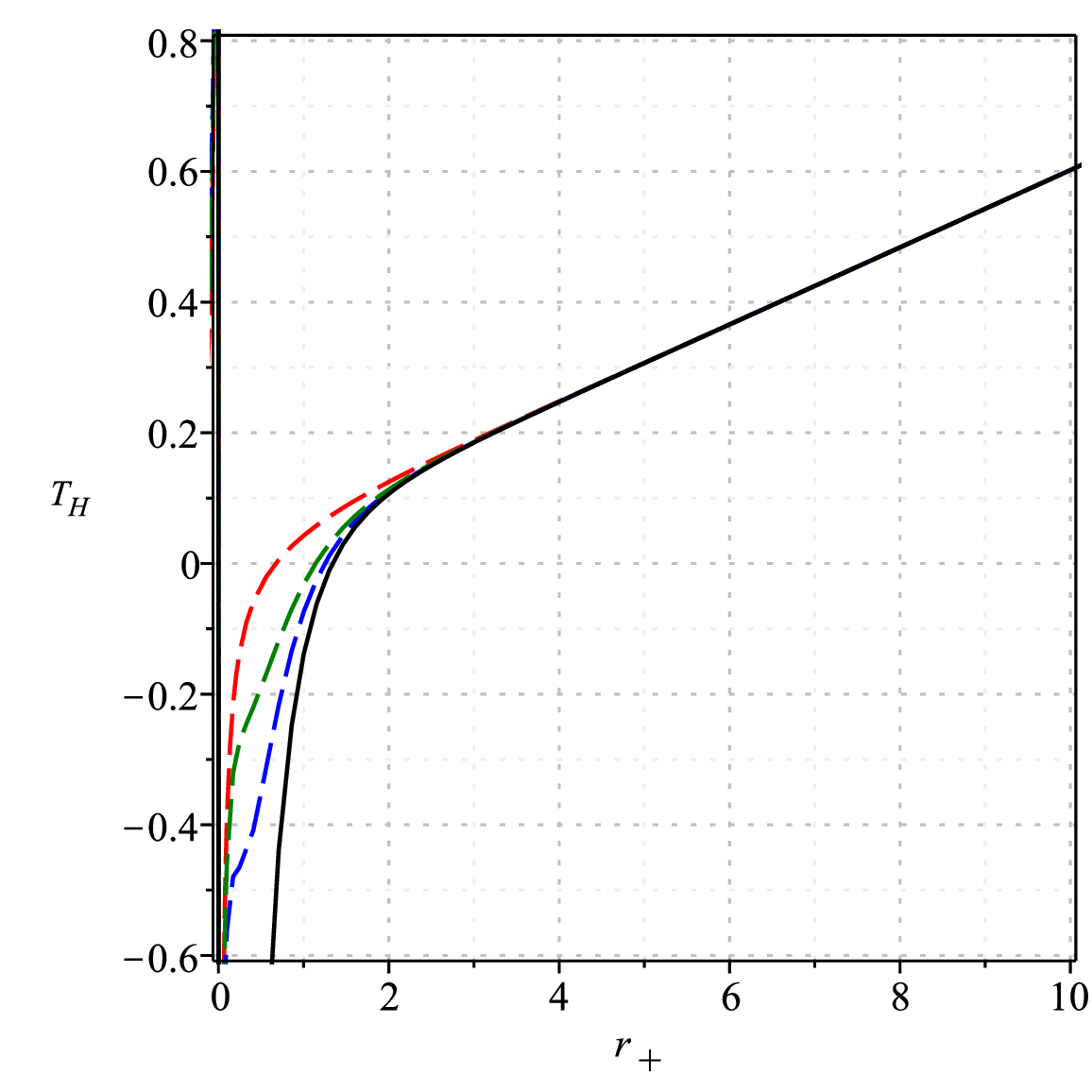}}\hfill
\subfloat[$m=1.0$ \&  $\beta=0.5$]{\includegraphics[width=.5\textwidth]{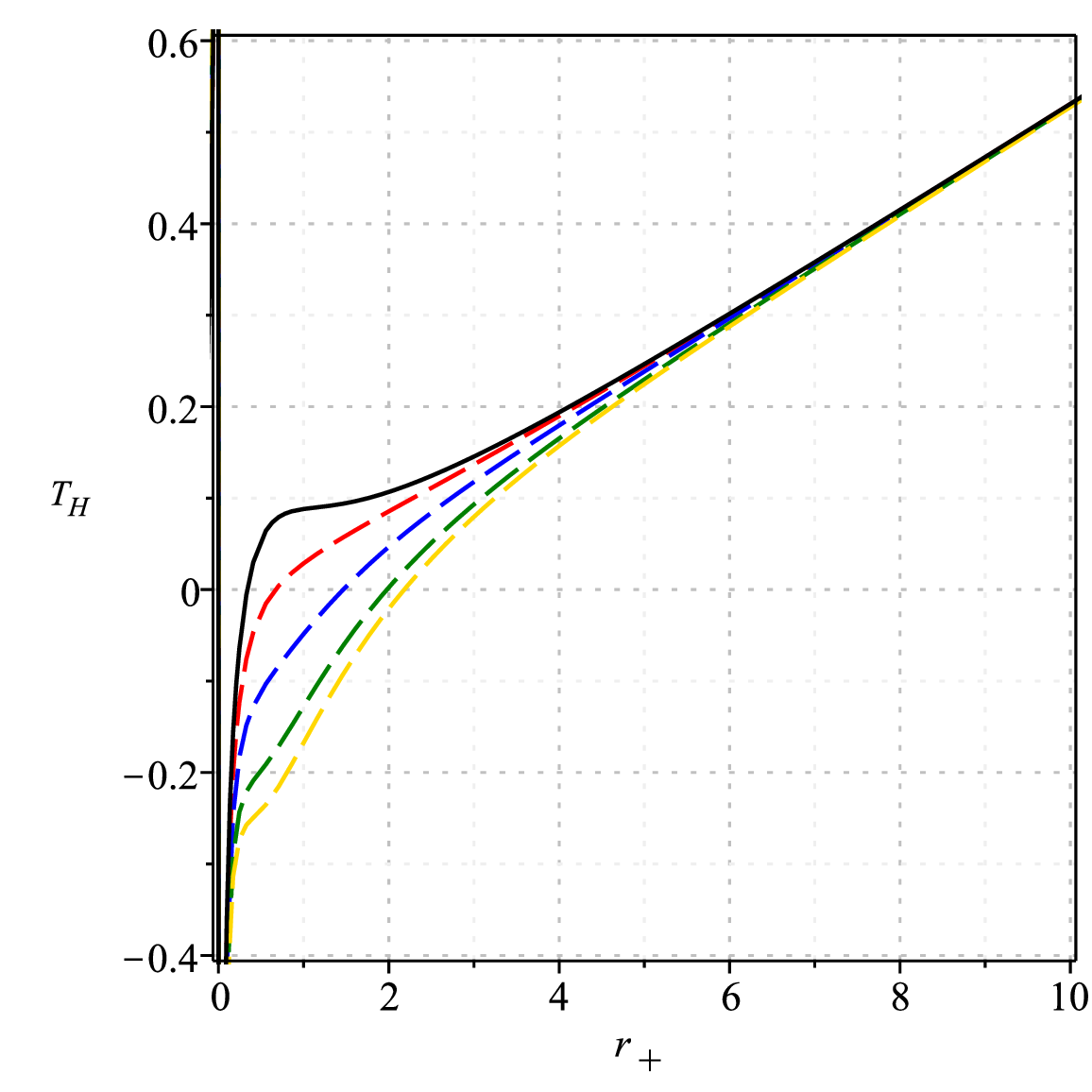}}\hfill
\caption{Left Panel: $\beta=0.0$ denoted by solid black line, $\beta=0.01$ denoted by blue dash line with, $\beta=0.05$ denoted by green dash line, $\beta=1.0$ denoted by red dash line in EGB-NED. Right Panel: $Q_m=0.0$ denoted by solid black line, $Q_m=2.0$ denoted by red dash line with, $Q_m=4.0$ denoted by blue dash line, $Q_m=6.0$ denoted by green dash line and $Q_m=7.0$ denoted by gold dash line in EGB-NED with $\alpha=0.2$, $c=1$, $c_1=-1$, $c_2=1$ and $l=2$.}\label{fig:14}
\end{figure}

In the limit $\alpha \to 0$, above equation is reduced into Hawking 
temperature of black hole in massive Einstein gravity coupled to NED
\begin{equation}\label{eq:3.5}
T_{H}=\frac{1}{4 \pi  r_{{+}} l^{2} (k^{2}+r_{{+}}^{2})} \Biggr[  \Bigl(c m^{2} c_{1} r_{{+}}^{3}+(c^{2} c_{2} m^{2}+1) r_{{+}}^{2} +c k^{2} m^{2} c_{1} r_{{+}} +(c^{2} c_{2} m^{2}+1) k^{2}-Q_{m}^{2} \Bigl) l^{2} +3 r_{{+}}^{4} +3 k^{2} r_{{+}}^{2} \Biggr].
\end{equation}
In the limit $m \to 0$, and $l=\infty$, equation \eqref{eq:3.4} is reduced into Hawking temperature of EGB massless gravity coupled to NED, which was obtained in Ref. \cite{Kruglov:2021stm}
\begin{equation}\label{eq:3.6}
T_{H}=\frac{r_{{+}}^{4}+(-Q_{m}^{2}+k^{2}-\alpha ) r_{{+}}^{2}-\alpha k^{2}}{4 \pi  r_{{+}} (r_{{+}}^{2}+2 \alpha ) (k^{2}+r_{{+}}^{2})}.
\end{equation}
Taking $\beta \to 0$ limit into equation \eqref{eq:3.4} one can obtain Hawking temperature of $4D$ EGB massive gravity black holes in Maxwell electrodynamics \cite{Paul:2023mlh}
\begin{equation}\label{eq:3.7}
T_{H}=\frac{(c m^{2} c_{1} r_{{+}}^{3}+(c^{2} c_{2} m^{2}+1) r_{{+}}^{2}-Q_{m}^{2}-\alpha ) l^{2}+3 r_{{+}}^{4}}{4 \pi  r_{{+}} (r_{{+}}^{2}+2 \alpha ) l^{2}}.
\end{equation}
Furthermore, if one takes massless limit into the equation 
\eqref{eq:3.7} then Hawking temperature of $4D$ EGB gravity black 
holes in Maxwell electrodynamics are obtained \cite{Fernandes:2020rpa}. 
In the massless limit equation \eqref{eq:3.5} reduces to Hawking 
temperature of the black hole in GR coupled to NED 
\cite{kruglov2022nonlinearly}. In Fig. \ref{fig:12} Hawking temperature 
is plotted for different values of graviton mass. For the higher value 
of graviton mass Hawking temperature attains local maxima at $r_{+}^{a}$ 
and local minima at $r_{+}^{b}$, where $r_{+}^b > r_{+}^a$. These local 
maxima and minima slowly disappear as graviton mass decreases. The 
maximum and minimum values of the Hawking temperature increase as we 
increase the graviton mass. In Fig. \ref{fig:13}, Fig. \ref{fig:14}(a) 
and Fig. \ref{fig:14}(b) Hawking temperature of $4D$ EGB black holes is 
plotted for different values of $\beta$ in massive gravity, massless 
gravity, and for different values of magnetic charge in massive gravity.

The behaviour of Hawking temperature in equation \eqref{eq:3.6} is shown in \cite{Kruglov:2021stm}. For a particular value of horizon radius Hawking temperature attains maxima and after the maximum point, it is decreasing functions of horizon radius. 

\begin{figure}[H]
\centering
\subfloat[$\beta=0.5$]{\includegraphics[width=.5\textwidth]{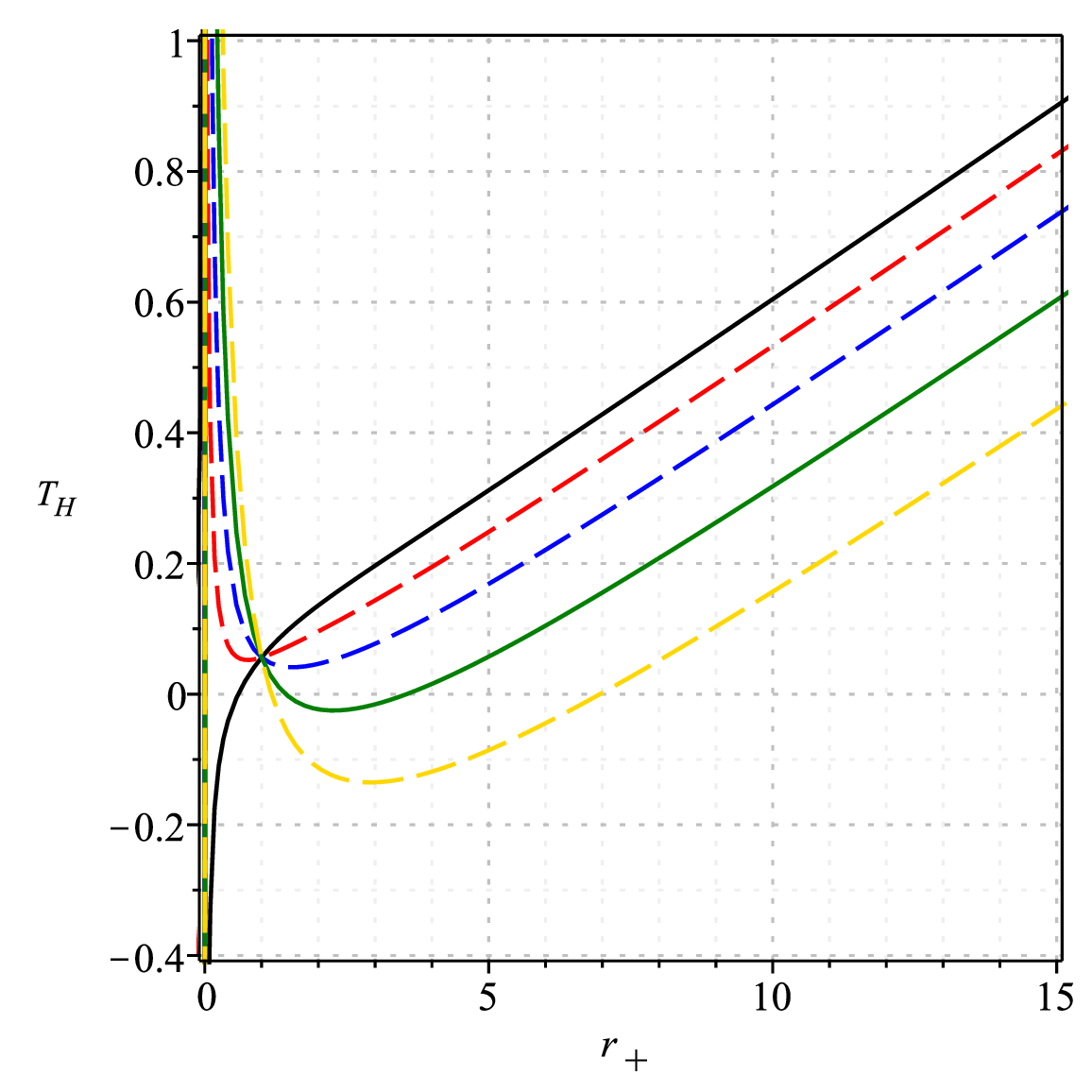}}\hfill
\subfloat[$m=1$]{\includegraphics[width=.5\textwidth]{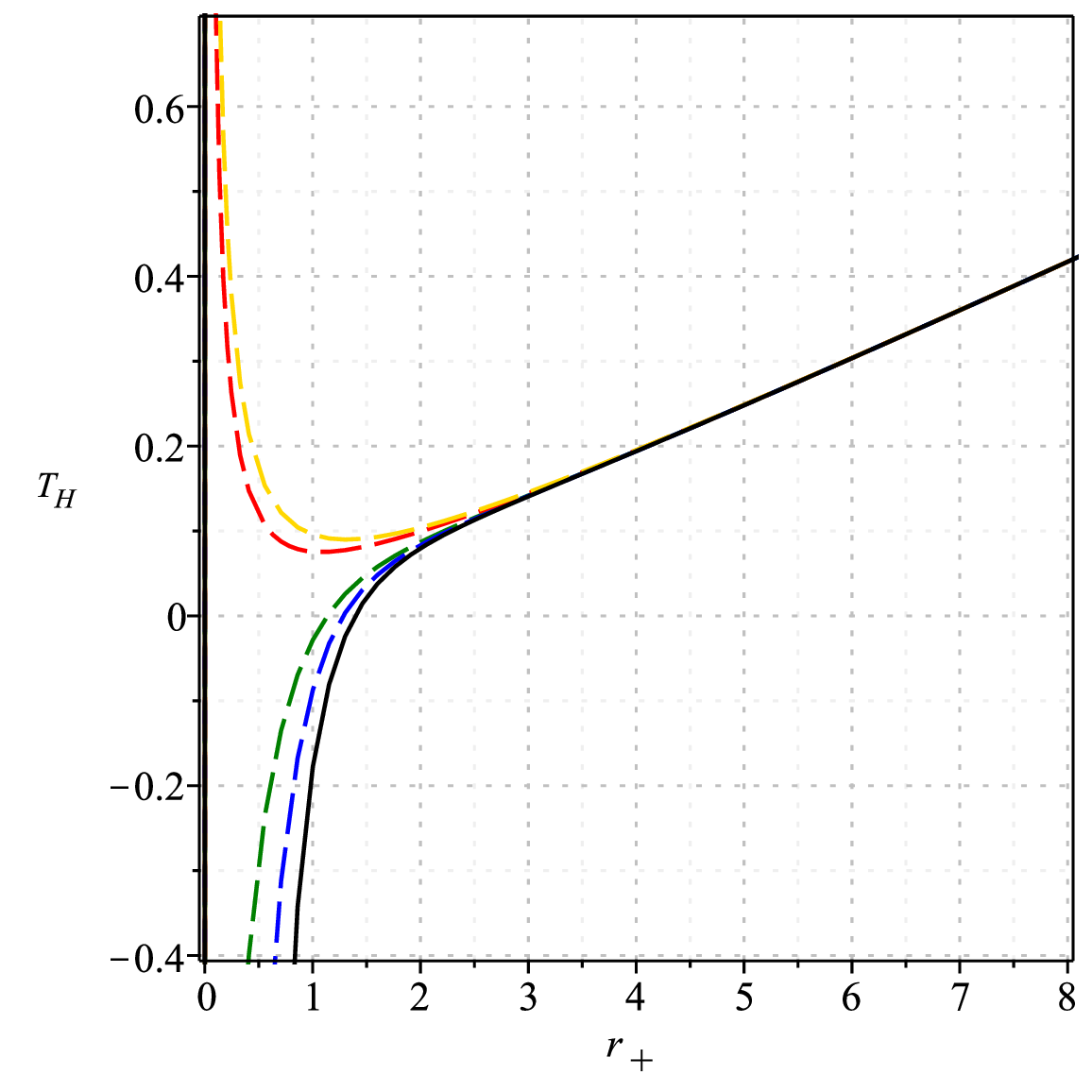}}\hfill
\caption{Left panel: $m=0.0$ denoted by solid black line, $m=1.0$ denoted by red dash line with, $m=1.5$ denoted by blue dash line, $m=2.0$ denoted by green dash line and $m=2.5$ denoted by gold dash line. Right panel: $\beta=0.0$ denoted by solid black line, $\beta=0.01$ denoted by blue dash line, $\beta=0.05$ denoted by green dash line and  $\beta=1.0$ denoted by red dash line and $\beta=2.5$ denoted by gold dash line. GR-NED with $Q_m=2$,  $c=1$, $c_1=-1$, $c_2=1$ and $l=2$ }\label{fig:15}
\end{figure}

\begin{figure}[H]
    \centering
    \includegraphics[width=.6\textwidth]{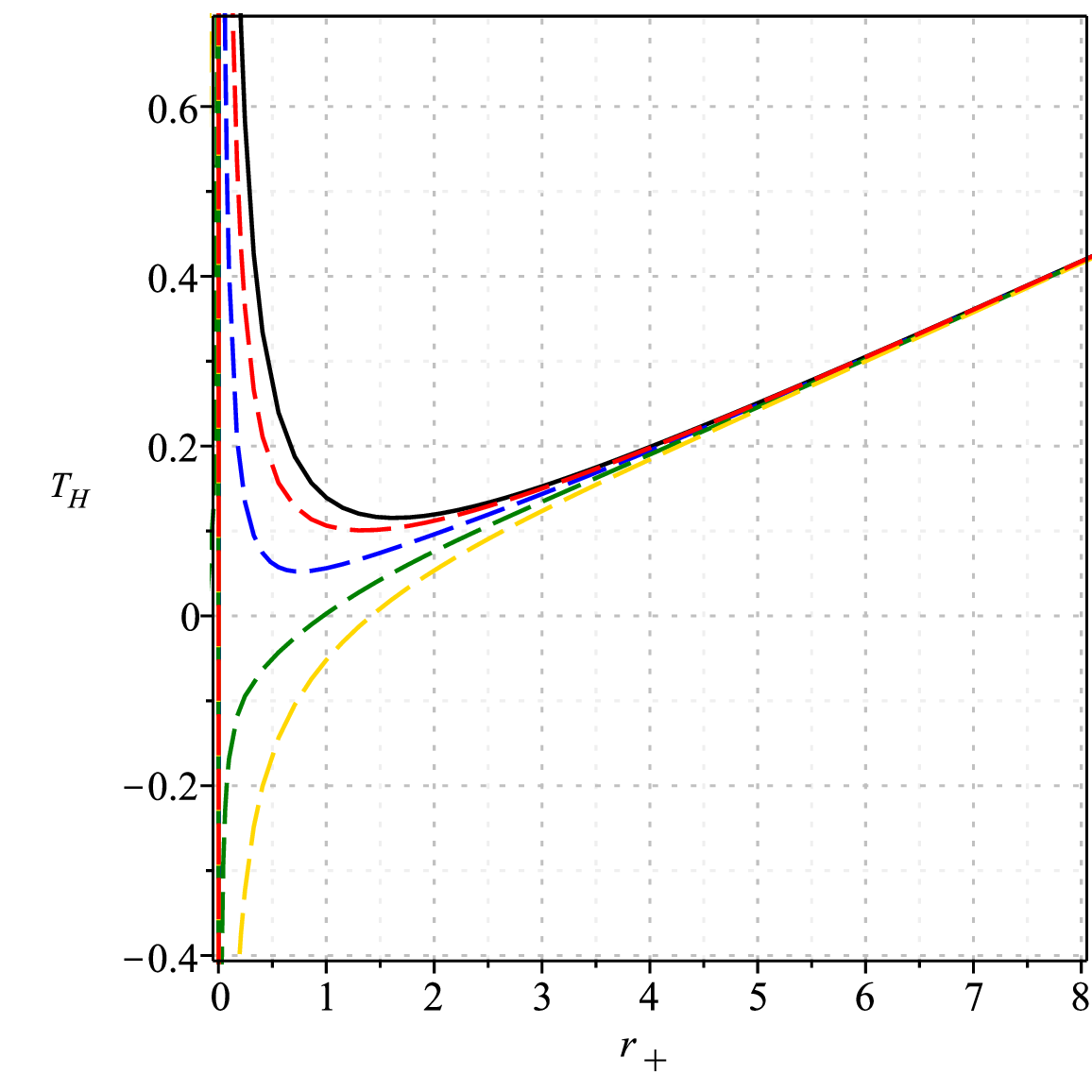}
    \caption{$Q_m=0.0$ denoted by solid black line, $Q_m=1.0$ denoted by red dash line with, $Q_m=2.0$ denoted by blue dash line, $Q_m=3.0$ denoted by green dash line and $Q_m=4.0$ denoted by gold dash line in GR-NED with $m=1.0$, $\beta=0.5$, $c=1$, $c_1=-1$, $c_2=1$ and $l=2$.}
    \label{fig:16}
\end{figure}

The Hawking temperature of the black hole in massive GR coupled to NED is shown in Fig \ref{fig:15} and Fig. 
\ref{fig:16}. In Fig. \ref{fig:15}(a), Fig. \ref{fig:15}(b) and 
Fig. \ref{fig:16} temperature is plotted for different values of 
graviton mass, NED parameter, and magnetic charge of the black hole. 
There is a minimum temperature($T_{GR}^{min}$) at some critical value 
of horizon radius(say, $r_{0}$) which is positive for $m=1.0$, $m=1.5$ 
and for  $\beta=1.0$, $\beta=2.5$ \& for $Q_m=0,1,2$. The minimum temperature divides the black holes into small and large. Above the minimum temperature, small and large black holes coexist at all temperatures. This behavior of the temperature is very similar to the temperature of the Schwarzschild--$AdS$ black hole \cite{Hawking:1982dh}. The equation for critical horizon radius at which $T_{GR}^{min}$ occurs can be obtain from equation \eqref{eq:3.5}

\begin{equation}\label{eq:3.8}
    \frac{\partial T_{H}}{\partial r_{+}}  \bigg\vert_{r_{+}=r_{0}} =0,
\end{equation}

\begin{equation}\label{eq:3.9}
3 r_{0}^{6}+\Bigl( 6 k^{2} -l^{2}  -c^{2} c_{2} l^{2} m^{2} \Bigl) r_{0}^{4} + \Bigl( 3 k^{4} r-2 k^{2} l^{2}  +3 Q_{m}^{2} l^{2}  -2 c^{2} c_{2} k^{2} l^{2} m^{2} \Bigl) r_{0}^{2} -c^{2} c_{2} k^{4} l^{2} m^{2}+Q_{m}^{2} k^{2} l^{2}-k^{4} l^{2}= 0.
\end{equation}

In table \ref{table:1} we estimated the minimum Hawking temperature $T_{GR}^{min}$ and critical horizon radius $r_0$ at which $T_{GR}^{min}$ occurs. 

\begin{table}[H]
\centering
\begin{tabular}{ |p{1.5cm}|p{1.5cm}|p{1.5cm}| } 
\hline
\multicolumn{3}{|c|}{\textbf{Fig. \ref{fig:15}(a)}} \\
\hline
\textbf{m} & \textbf{$r_{0}$} & \textbf{$T_{GR}^{min}$} \\ [0.5ex]  
\hline
1.0 & 0.7554 & 0.0522  \\ \hline
1.5 & 1.5402 & 0.0759  \\ \hline
2.0 & 2.2656 & -0.0251  \\ \hline
2.5 & 2.9134 & -0.1351  \\ \hline
\multicolumn{3}{|c|}{\textbf{Fig. \ref{fig:15}(b)}} \\
\hline
\textbf{$\beta$} & \textbf{$r_{0}$} & \textbf{$T_{GR}^{min}$} \\ [0.5ex]  
\hline
1.0 & 1.0808 & 0.0751  \\ \hline
2.5 & 1.3115 & 0.0898  \\ \hline
\multicolumn{3}{|c|}{\textbf{Fig. \ref{fig:16}}} \\
\hline
\textbf{$Q_m$} & \textbf{$r_{0}$} & \textbf{$T_{GR}^{min}$}  \\ [0.5ex]  
\hline
0.0 & 1.6330 & 0.1153  \\ \hline
1.0 & 1.3360 & 0.1006  \\ \hline
2.0 & 0.7554 & 0.0522  \\ [1ex] 
\hline
\end{tabular}
\caption{}
\label{table:1}
\end{table}

The entropy of the black hole defined as
\begin{equation}\label{eq:3.10}
    S= \int \frac{dM}{T_{H}}.
\end{equation}
Using equation \eqref{eq:3.1} and equation \eqref{eq:3.4} one can obtain entropy as
\begin{equation}\label{eq:3.11}
    S= \pi  r_{{+}}^{2}+4 \pi  \alpha  \ln(r_{{+}} ) +S_{0},
\end{equation}
where $S_0$ is an integration constant. Next, we will derive the first law of black 
hole thermodynamics. We take the potential $\mathcal{A}$, $\mathcal{C}_1$ and 
$\mathcal{C}_2$ corresponding to Gauss--Bonnet coupling parameter($\alpha$), 
constant $c_1$ and $c_2$.  The first law of black hole thermodynamics in extended 
phase space and Smarr formula \cite{Kubiznak:2012wp,Gunasekaran:2012dq,Xu:2015rfa,kruglov2022nonlinearly} are given by
\begin{equation}\label{eq:3.12}
    dM= T_{H} dS + V dP + \Phi_{m} dQ_m + \mathcal{A} d\alpha + \mathcal{C}_{1} dc_1 + \mathcal{C}_{2} dc_2 + \mathcal{B} d\beta,
\end{equation}
\begin{equation}\label{eq:3.13}
    M=2T_{H}S-2PV+2\mathcal{A}\alpha+\Phi_{m}Q_m +2\beta \mathcal{B} -\mathcal{C}_1 c_1.
\end{equation}
From the first law, one can obtain the following quantities
\begin{equation}\label{eq:3.14}
\Phi_m(r_{+})= \frac{ Q_{m} r_{+}}{4 r_{+}^{2}+8 Q_{m} \sqrt{\beta}} +\frac{3 \sqrt{2Q_{m}}  \pi}{16 \beta^{{1}/{4}}}    -\frac{3 \sqrt{2Q_{m}}  \arctan({r_{+}}/{k})}{8 \beta^{{1}/{4}}},
\end{equation}
\begin{equation}\label{eq:3.15}
\mathcal{B}(r_{+})= \frac{r_{+}^3}{2}\biggr[\frac{Q_{m}^{2}}{4 r_{+}^{2} \beta  \bigl(2 Q_{m} \sqrt{\beta}+r^{2}\bigl)}-\frac{\sqrt{2} Q_{m}^{{3}/{2}}  \bigl(\pi -2 \arctan({r_{+}}/{k})\bigl)}{16 \beta^{{5}/{4}} r_{+}^{3}} \biggr],
\end{equation}
\begin{equation}\label{eq:3.16}
 V=\frac{4}{3}\pi r_{+}^3,
\end{equation}
\begin{equation}\label{eq:3.17}
    \mathcal{A}=\frac{1}{2r_{+}},
\end{equation}
\begin{equation}\label{eq:3.18}
    \mathcal{C}_1=\frac{m^2cr_{+}^2}{4},
\end{equation}
\begin{equation}\label{eq:3.19}
    \mathcal{C}_2=\frac{m^2c^2r_{+}}{4}.
\end{equation}

\begin{figure}[H]
\centering
\subfloat[$Q_m=1$]{\includegraphics[width=.5\textwidth]{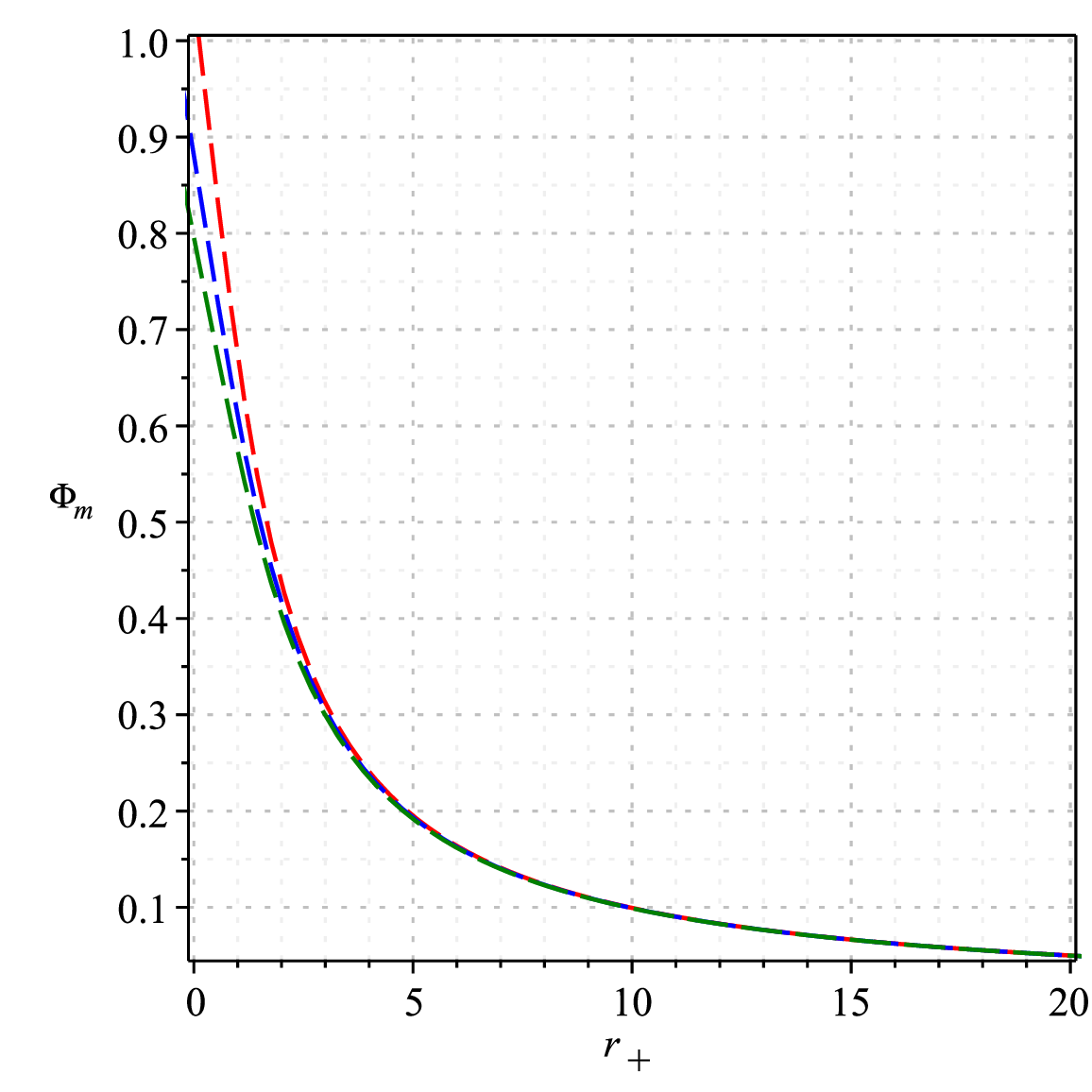}}\hfill
\subfloat[$Q_m=1$]{\includegraphics[width=.5\textwidth]{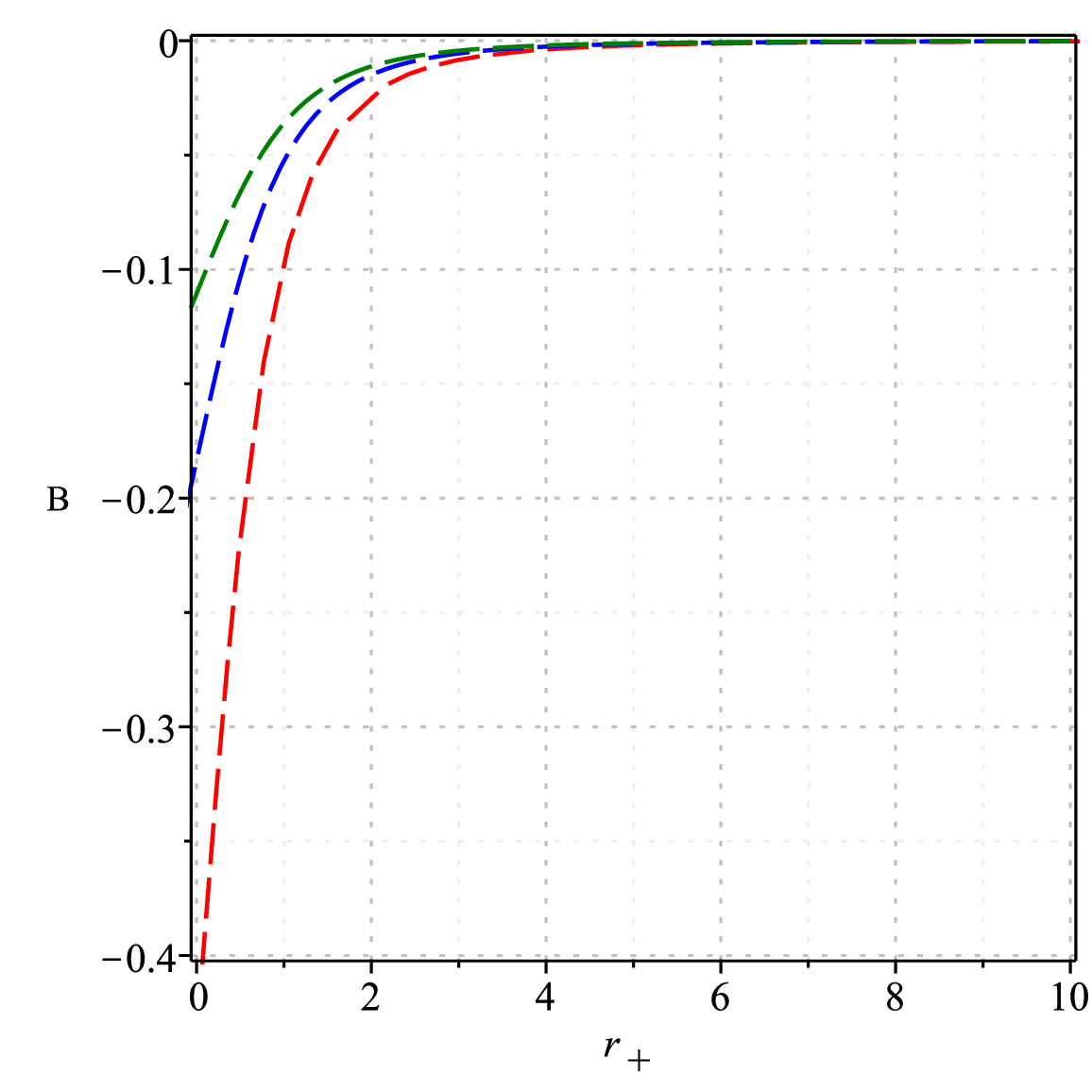}}\hfill
\caption{$\beta=0.4$ denoted by dash red line, $\beta=0.8$ denoted by blue dash  line with and $\beta=1.2$ denoted by green dash line.}\label{fig:17}
\end{figure}

In Fig. \ref{fig:17}(a) and Fig. \ref{fig:17}(b) magnetic potential 
($\Phi_m$) and  vacuum polarization ($\mathcal{B}$) are shown. At $r_{+}=0$ 
magnetic potential takes constant positive value 
($\Phi_m(0)=3\pi \sqrt{2Q_m}/16\beta^{1/4}$) and at 
$r_{+} \to \infty$ it is zero. At $r_{+}=0$ vacuum polarization takes 
constant negative value and at $r_{+} \to \infty$ it is zero.

Next, we investigate local stability of the black holes. The local thermodynamical stability depends on the specific heat of the black hole, a positive $C>0$ implies that black holes are locally stable and a negative specific heat implies black holes are locally unstable. The specific heat is defined as

\begin{equation}\label{eq:3.20}
    C=\frac{\partial M}{\partial T_{H}} =\frac {\partial M/\partial r_{+}}{\partial T_{H}/\partial r_{+}}.
\end{equation}
Substituting mass and temperature from equation \eqref{eq:3.1} and equation \eqref{eq:3.4} into above equation, we obtain specific heat of the $4D$ EGB massive gravity black hole in NED

\begin{equation*}
\frac{\partial M}{\partial r_+}=\frac{1}{2 r_{{+}}^{2} (k^{2}+r_{{+}}^{2}) l^{2}}\Biggr[ 3 r_{{+}}^{6}+c c_{1} l^{2} m^{2} r_{{+}}^{5}+((c^{2} c_{2} m^{2}+1) l^{2}+3 k^{2}) r_{{+}}^{4}+c c_{1} k^{2} l^{2} m^{2} r_{{+}}^{3}
\end{equation*}
\begin{equation}\label{eq:3.21}
+\biggl( (c^{2} c_{2} m^{2}+1) k^{2}-Q_{m}^{2}-\alpha \biggl) l^{2} r_{{+}}^{2}-\alpha  k^{2} l^{2}\Biggr],
\end{equation}

\begin{equation*}
\frac{\partial T_H}{\partial r_+}=\frac{1}{16 (\frac{r_{{+}}^{2}}{2}+\alpha )^{2} r_{{+}}^{2} l^{2} \pi  (k^{2}+r_{{+}}^{2})^{2}}\biggr[ 3 r_{{+}}^{10}+\bigl((-c^{2} c_{2} m^{2}-1) l^{2}+6 k^{2}+18 \alpha \bigl) r_{{+}}^{8}+4 \alpha  c c_{1} l^{2} m^{2} r_{{+}}^{7}
\end{equation*}
\begin{equation*}
+\biggl( \Bigl( (-2 c^{2} c_{2} m^{2}-2) k^{2}+2 c^{2} \alpha  m^{2} c_{2} +3 Q_{m}^{2}+5 \alpha \Bigl) l^{2}+3 k^{4}+36 \alpha  k^{2}\biggl) r_{{+}}^{6}+8 \alpha  c c_{1} k^{2} l^{2} m^{2} r_{{+}}^{5}
\end{equation*}
\begin{equation*}
+\biggl( \Bigl( (-c^{2} c_{2} m^{2}-1) k^{4}+(4 c^{2} \alpha  m^{2} c_{2} +Q_{m}^{2}+10 \alpha ) k^{2}+2 \alpha  Q_{m}^{2}+2 \alpha^{2}\Bigl) l^{2}+18 \alpha  k^{4}\biggl) r_{{+}}^{4}+4 \alpha  c c_{1} k^{4} l^{2} m^{2} r_{{+}}^{3}
\end{equation*}
\begin{equation}
+2 k^{2} \alpha  l^{2} \Bigl( (c^{2} c_{2} m^{2}+{5}/{2}) k^{2}-Q_{m}^{2}+2 \alpha \Bigl) r_{{+}}^{2}+2 \alpha^{2} k^{4} l^{2} \biggr].
\end{equation}

In the limit $m \to 0$ and $l=\infty$ above equation is reduced to specific heat of $4D$ EGB massless gravity coupled to NED \cite{Kruglov:2021stm} 
\begin{equation}\label{eq:3.23}
\frac{\partial M}{\partial r_+}=\frac{r_{{+}}^{4}+(-Q_{m}^{2}+k^{2}-\alpha ) r_{{+}}^{2}-\alpha  k^{2}}{2 (k^{2}+r_{{+}}^{2}) r_{{+}}^{2}},
\end{equation}
\begin{equation*}
\frac{\partial T_H}{\partial r_+}=\frac{1}{4 (r_{{+}}^{2}+2 \alpha )^{2} r_{{+}}^{2} \pi  (k^{2}+r_{{+}}^{2})^{2}}\Biggr[ -r_{{+}}^{8}+(3 Q_{m}^{2}-2 k^{2}+5 \alpha ) r_{{+}}^{6}+(2 \alpha^{2}+(2 Q_{m}^{2}+10 k^{2}) \alpha -k^{4}+k^{2} Q_{m}^{2}) r_{{+}}^{4}
\end{equation*}
\begin{equation}
+5 (k^{2}-{2 Q_{m}^{2}}/{5}+{4 \alpha}/{5}) \alpha  k^{2} r_{{+}}^{2}+2 \alpha^{2} k^{4}\Biggr].
\end{equation}

\begin{figure}[H]
\centering
\subfloat[$\alpha=0.2$]{\includegraphics[width=.5\textwidth]{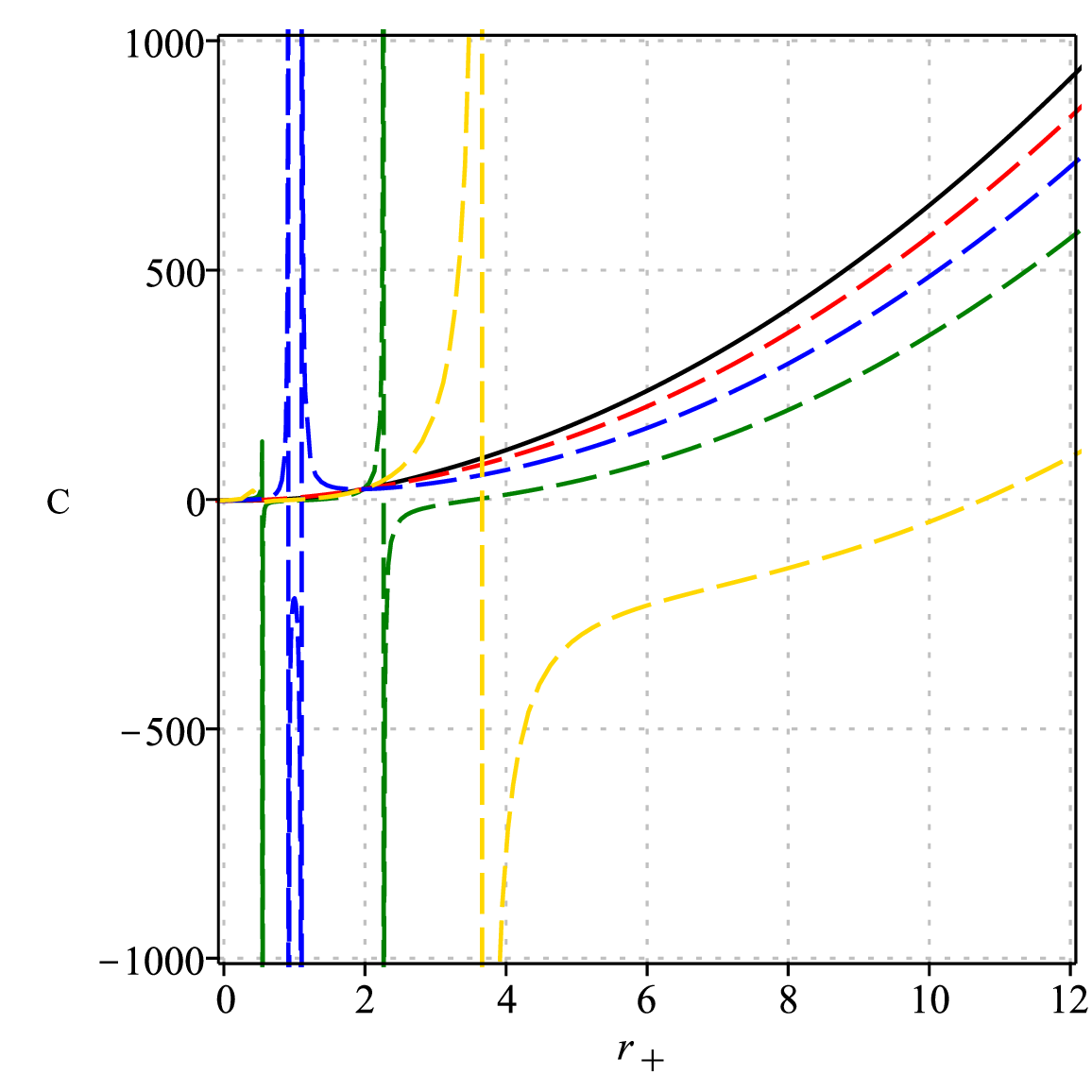}}\hfill
\subfloat[$\alpha=0.4$]{\includegraphics[width=.5\textwidth]{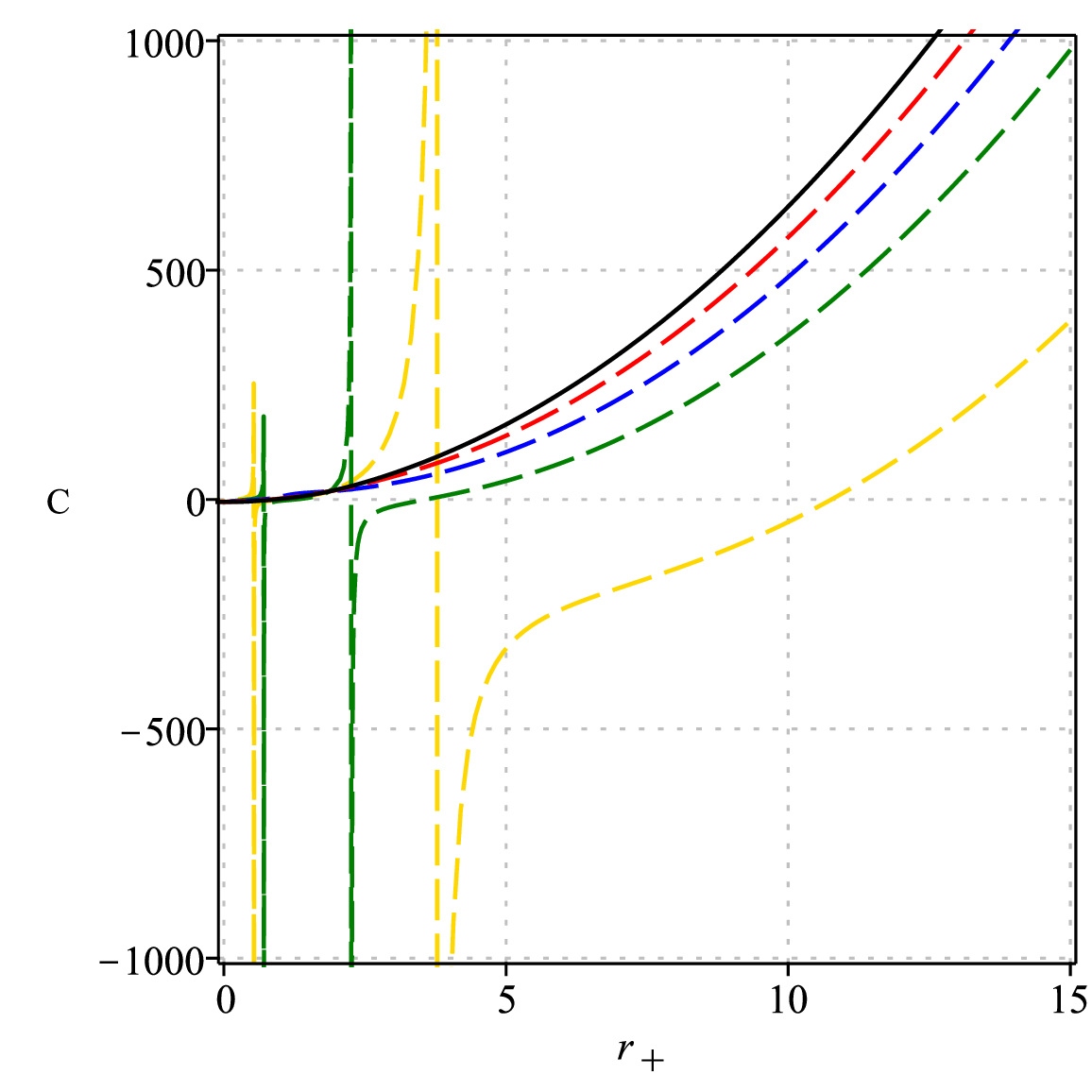}}\hfill
\caption{$m=0.0$ denoted by solid black line, $m=1.0$ denoted by red dash line with, $m=1.5$ denoted by blue dash line, $m=2.0$ denoted by green dash line and $m=3.0$ denoted by gold dash line  in EGB-NED with $Q_m=2$, $\beta=0.5$, $c=1$, $c_1=-1$, $c_2=1$ and $l=2$. }\label{fig:18}
\end{figure} 

\begin{figure}[H]
\centering
\subfloat[$\alpha=0.02$]{\includegraphics[width=.5\textwidth]{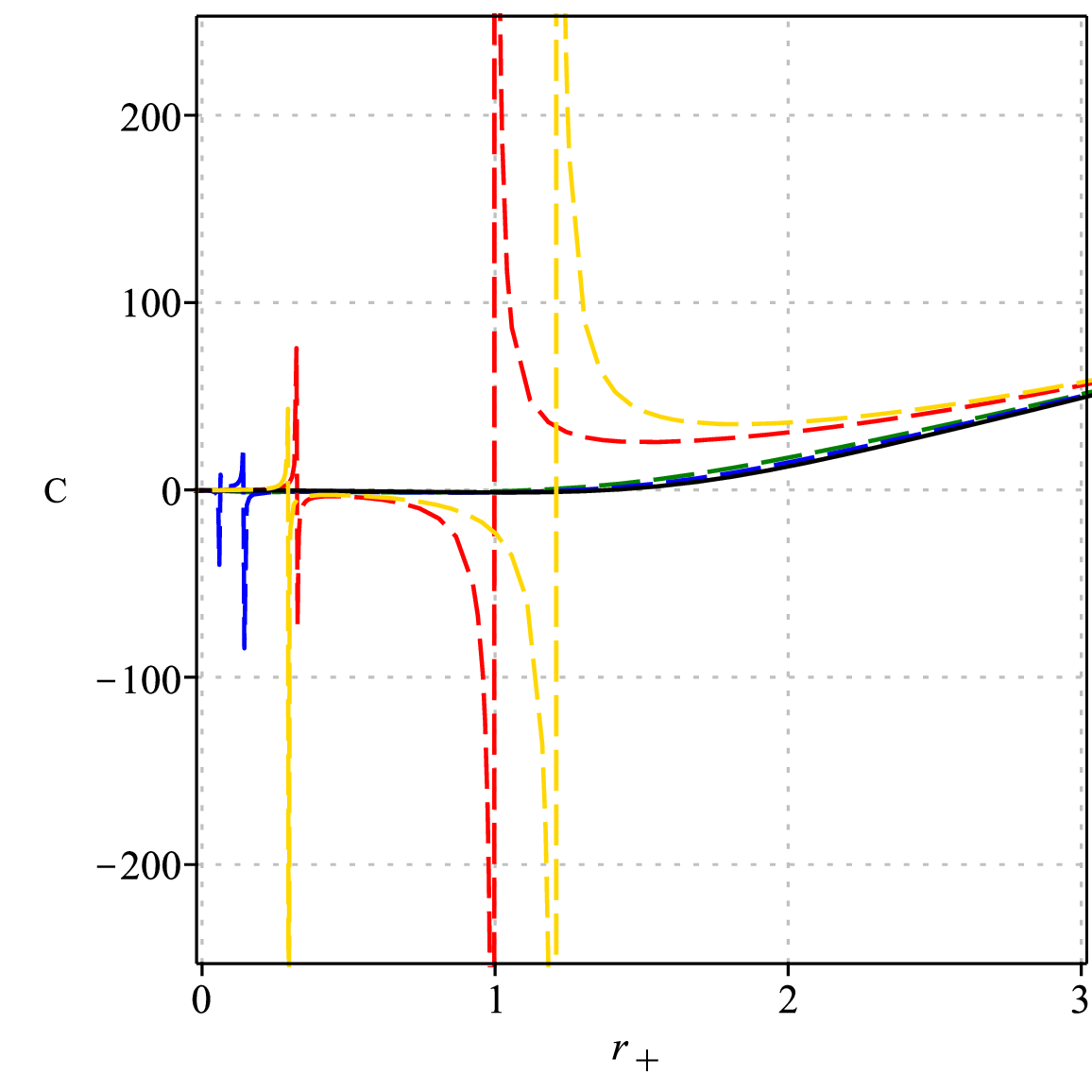}}\hfill
\subfloat[$\alpha=0.2$]{\includegraphics[width=.5\textwidth]{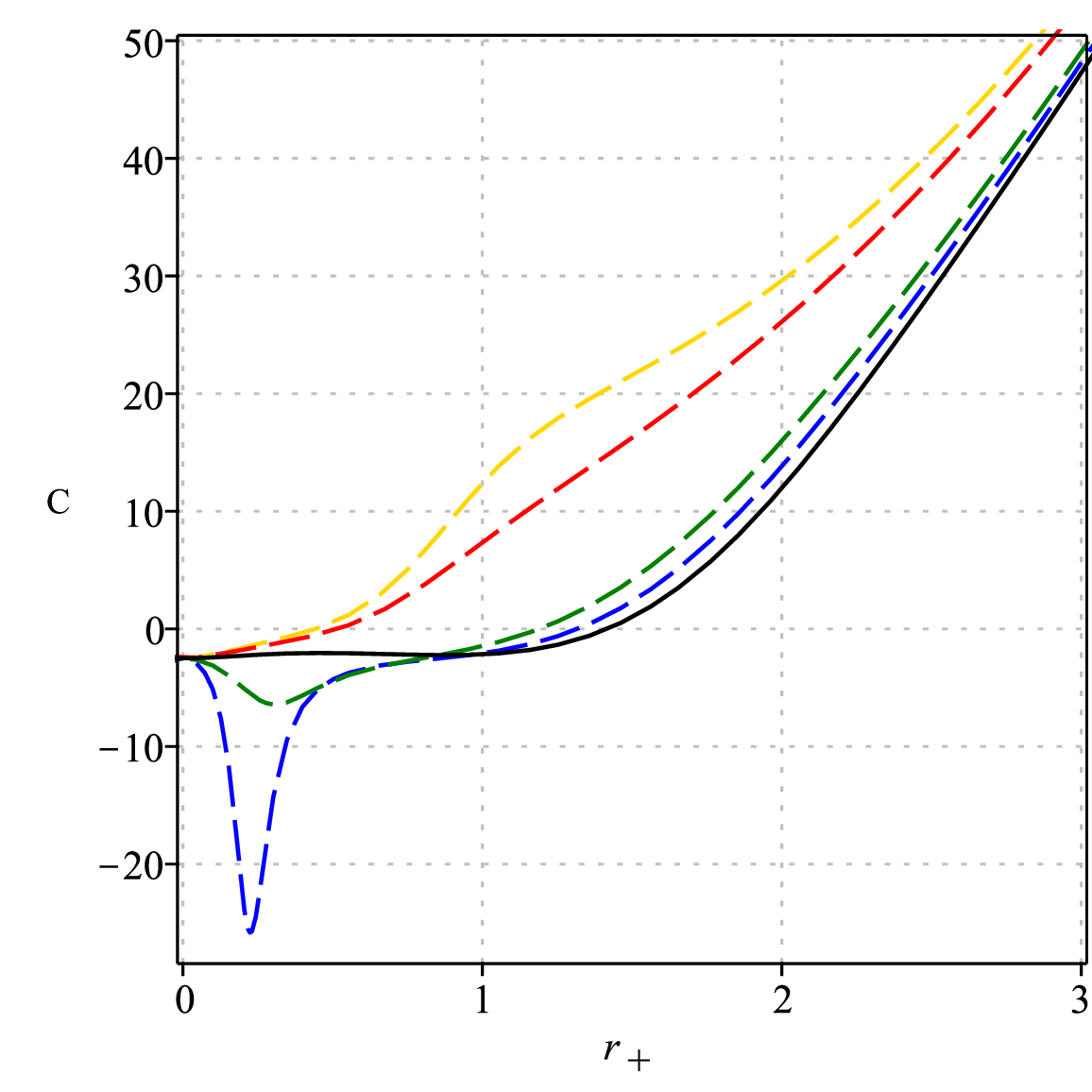}}\hfill
\caption{$\beta=0.0$ denoted by solid black line, $\beta=0.01$ denoted by blue dash line with, $\beta=0.05$ denoted by green dash line,  $\beta=1.0$ denoted by red dash line and $\beta=2.0$ denoted by gold dash line in EGB-NED with  $Q_m=2$, $m=1.0$, $c=1$, $c_1=-1$, $c_2=1$ and $l=2$. }\label{fig:19}
\end{figure} 

\begin{figure}[H]
\centering
\subfloat[$m=0.0$ \&  $Q_m=2$]{\includegraphics[width=.5\textwidth]{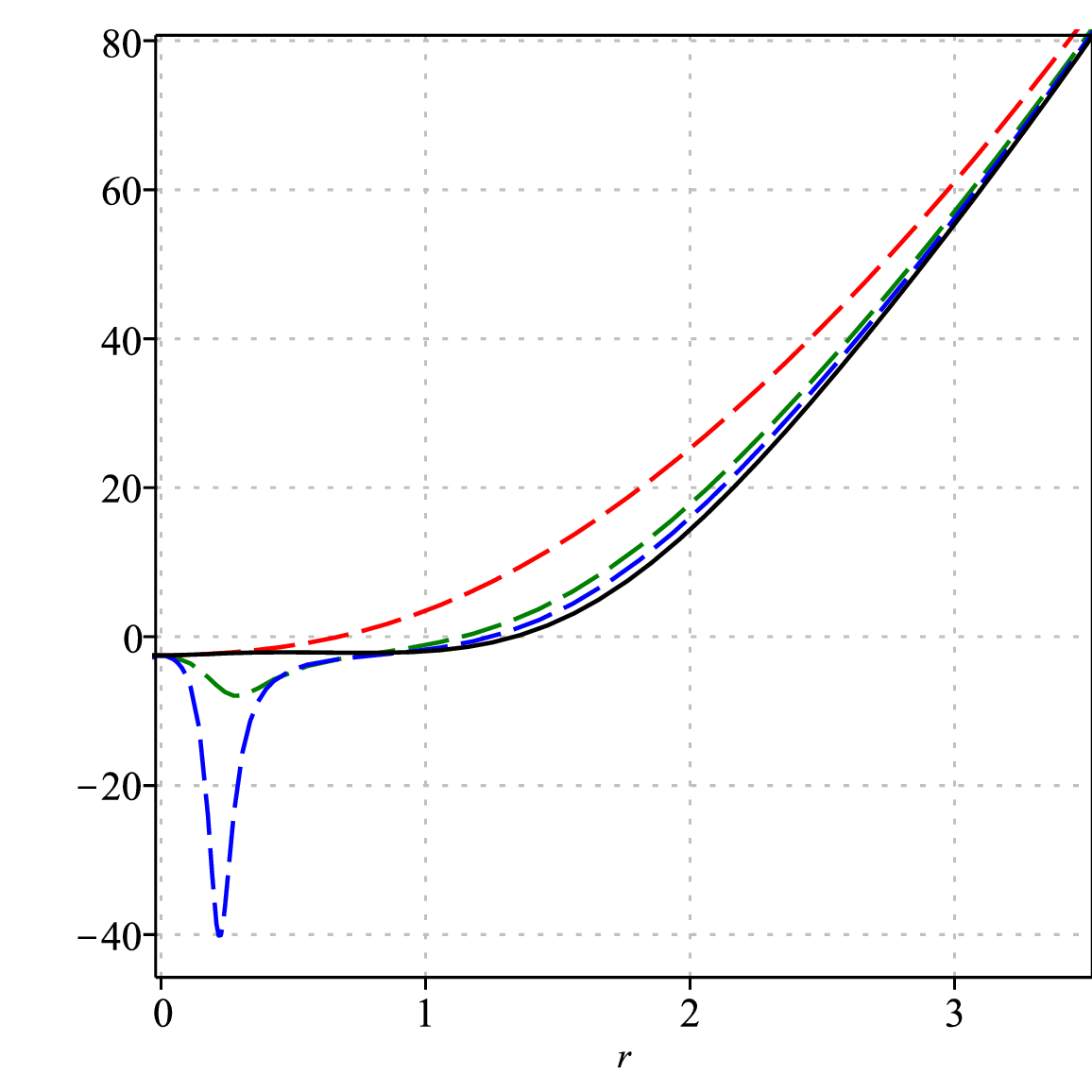}}\hfill
\subfloat[$m=1.0$ \&  $\beta=0.5$]{\includegraphics[width=.5\textwidth]{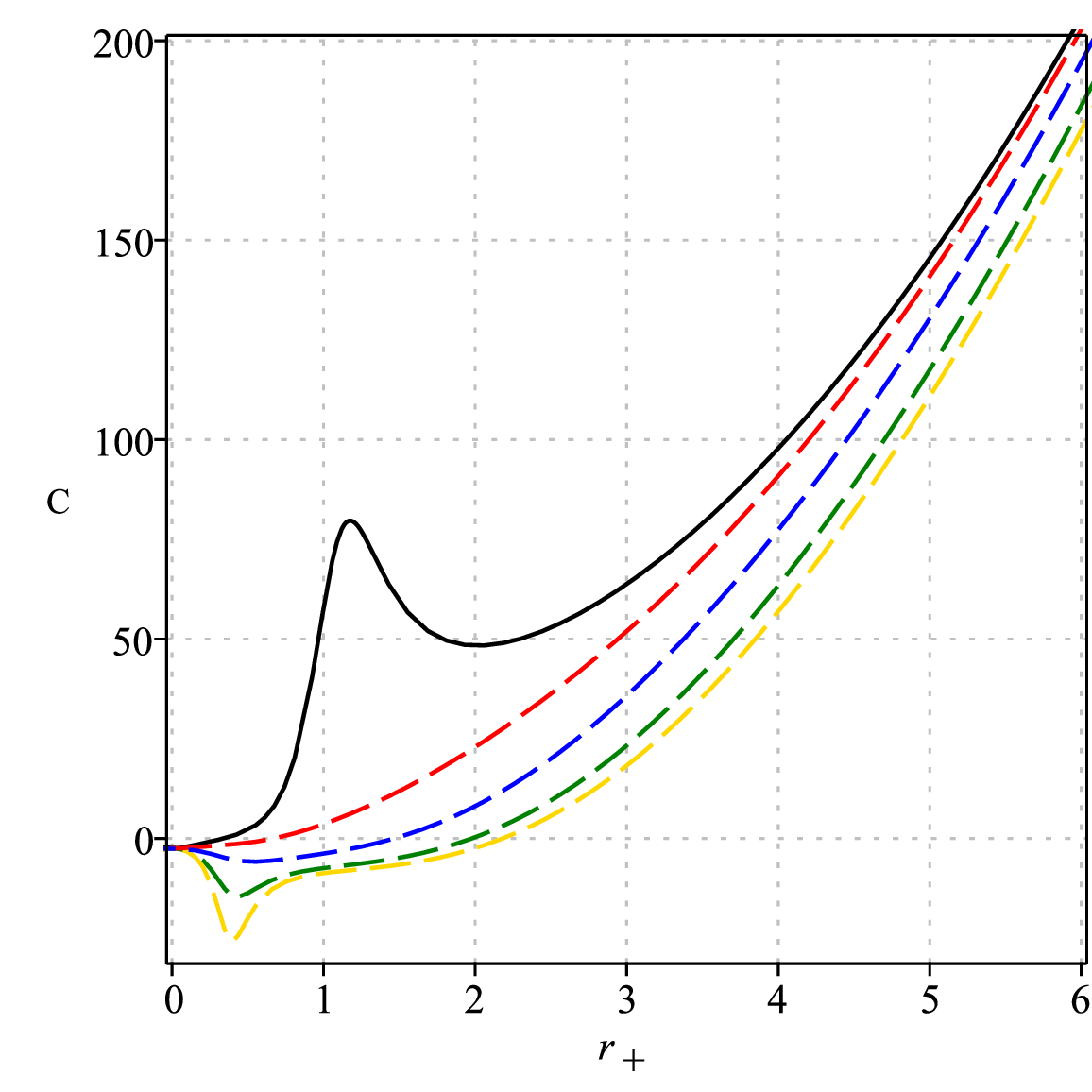}}\hfill
\caption{Left Panel: $\beta=0.0$ denoted by solid black line, $\beta=0.01$ denoted by blue dash line with, $\beta=0.05$ denoted by green dash line, $\beta=1.0$ denoted by red dash line in EGB-NED. Right Panel: $Q_m=0.0$ denoted by solid black line, $Q_m=2.0$ denoted by red dash line with, $Q_m=4.0$ denoted by blue dash line, $Q_m=6.0$ denoted by green dash line and $Q_m=7.0$ denoted by gold dash line in EGB-NED with $\alpha=0.2$, $c=1$, $c_1=-1$, $c_2=1$ and $l=2$.}\label{fig:20}
\end{figure}

In Fig. \ref{fig:18} we plot the specific heat of $4D$ EGB massive 
gravity black hole in NED for different values of graviton mass and 
Gauss--Bonnet coupling parameter. For $m=0,1.0$ specific heat is positive 
definite and continuous function of $r_{+}$. This implies no phase 
transition occurs and black holes are locally stable.  Clearly specific 
heat is singular at $r_{GB}^{a}$ and $r_{GB}^{b}$ with 
$r_{GB}^{a} < r_{GB}^{b}$ for $m=1.5,2.0$ and $2.5$. Two second-order 
phase transition occurs at $r_{GB}^{a}$ and $r_{GB}^{b}$, where Hawking 
temperature attains local maxima and minima (Fig. \ref{fig:12}). 
These two diverging points separate three regions. First region, 
$r_{+} < r_{GB}^{a}$ here specific heat is negative. Second region 
$r_{GB}^a < r_{+}<r_{GB}^b$, between two diverging points specific 
heat is also negative. Third region, $r_{+} > r_{GB}^b$ in this 
region specific heat is positive. Therefore, black holes with 
$r_{+} > r_{GB}^b$ are only thermodynamically locally stable. 
In Fig. \ref{fig:19} specific heat is plotted for different values 
of NED parameter $\beta$. Fig. \ref{fig:19}(a) shows similar 
behaviour as Fig. \ref{fig:18}. In Fig. \ref{fig:19}(b) specific 
heat is only positive when $r_{+}$ is greater than a critical horizon 
radius.

In the limit $\alpha \to 0$ equation \eqref{eq:3.20} is reduced to specific 
heat of black hole in $4D$ massive Einstein's gravity coupled to NED

\begin{equation*}
\frac{\partial M}{\partial r_+}=\frac{1}{2 r_{{+}}^{2} (k^{2}+r_{{+}}^{2}) l^{2}}\Biggr[ 3 r_{{+}}^{6}+c c_{1} l^{2} m^{2} r_{{+}}^{5}+((c^{2} c_{2} m^{2}+1) l^{2}+3 k^{2}) r_{{+}}^{4}+c c_{1} k^{2} l^{2} m^{2} r_{{+}}^{3}
\end{equation*}
\begin{equation}\label{eq:3.25}
+\biggl( (c^{2} c_{2} m^{2}+1) k^{2}-Q_{m}^{2} \biggl) l^{2} r_{{+}}^{2}\Biggr],
\end{equation}

\begin{equation*}
\frac{\partial T_H}{\partial r_+}=\frac{1}{8  r_{{+}}^{4} l^{2} \pi  (k^{2}+r_{{+}}^{2})^{2}}\biggr[ 3 r_{{+}}^{10}+\bigl((-c^{2} c_{2} m^{2}-1) l^{2}+6 k^{2}\bigl) r_{{+}}^{8}+\biggl( \Bigl( (-2 c^{2} c_{2} m^{2}-2) k^{2} +3 Q_{m}^{2} \Bigl) l^{2}+3 k^{4}\biggl) r_{{+}}^{6}
\end{equation*}
\begin{equation}
+\biggl( \Bigl( (-c^{2} c_{2} m^{2}-1) k^{4}+Q_{m}^{2} k^{2}\Bigl) l^{2}\biggl) r_{{+}}^{4} \biggr].
\end{equation}

In the limit $m \to 0$, above equations is reduced to specific heat of $4D$ 
massless Einstein's gravity coupled to NED \cite{kruglov2022nonlinearly}
\begin{equation}
\frac{\partial M}{\partial r_+}= \frac{3 r_{{+}}^{4}+(3 k^{2}+l^{2}) r_{{+}}^{2}+l^{2} (-Q_{m}^{2}+k^{2})}{2 (k^{2}+r_{{+}}^{2}) l^{2}},    
\end{equation}
\begin{equation}
\frac{\partial T_H}{\partial r_+}=\frac{3 r_{{+}}^{6}+(6 k^{2}-l^{2}) r_{{+}}^{4}+(3 Q_{m}^{2} l^{2}+3 k^{4}-2 k^{2} l^{2}) r_{{+}}^{2}-k^{4} l^{2}+k^{2} Q_{m}^{2} l^{2}}{4 r_{{+}}^{2} \pi  (k^{2}+r_{{+}}^{2})^{2} l^{2}}.
\end{equation}

\begin{figure}[H]
\centering
\subfloat[$\beta=0.5$]{\includegraphics[width=.5\textwidth]{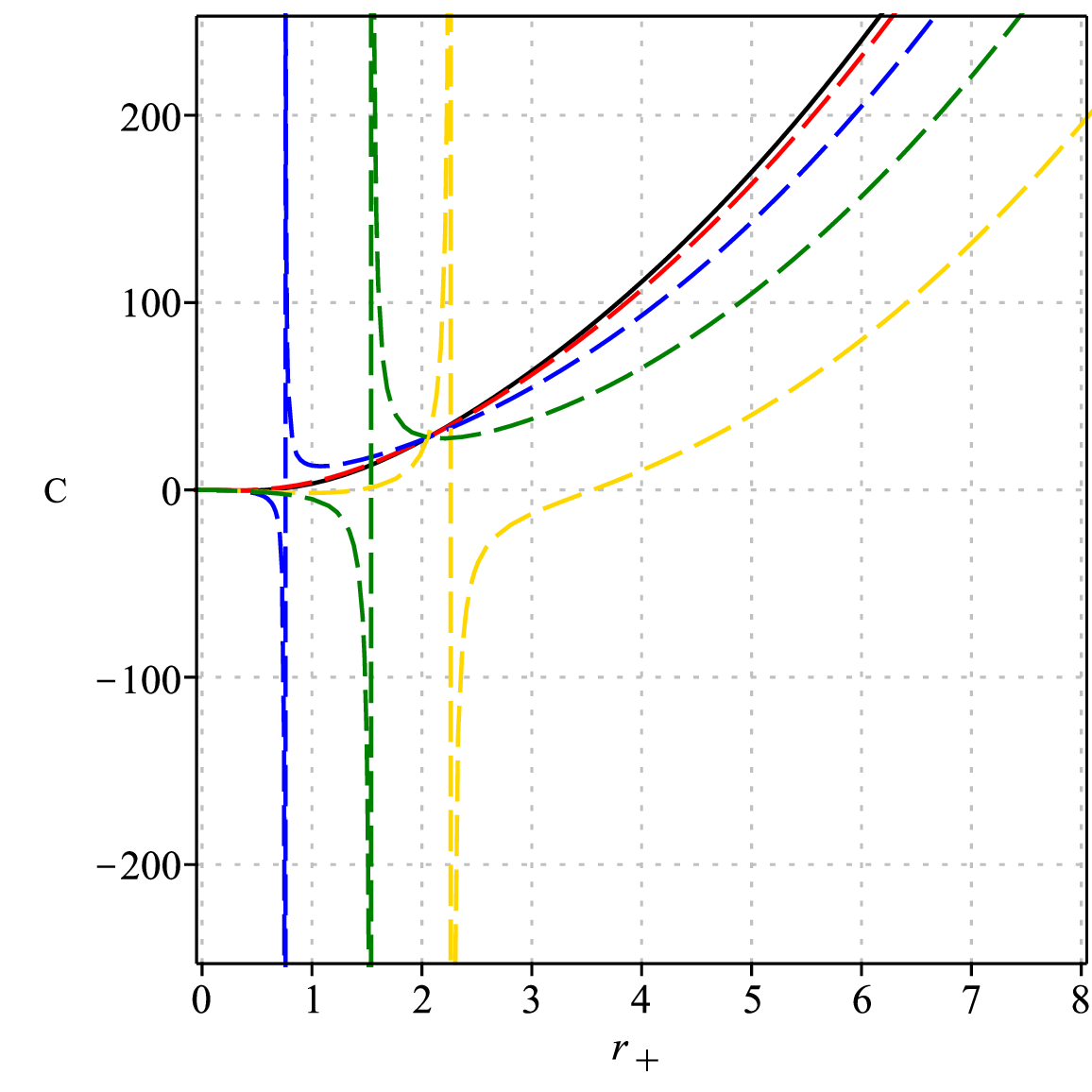}}\hfill
\subfloat[$m=1$]{\includegraphics[width=.5\textwidth]{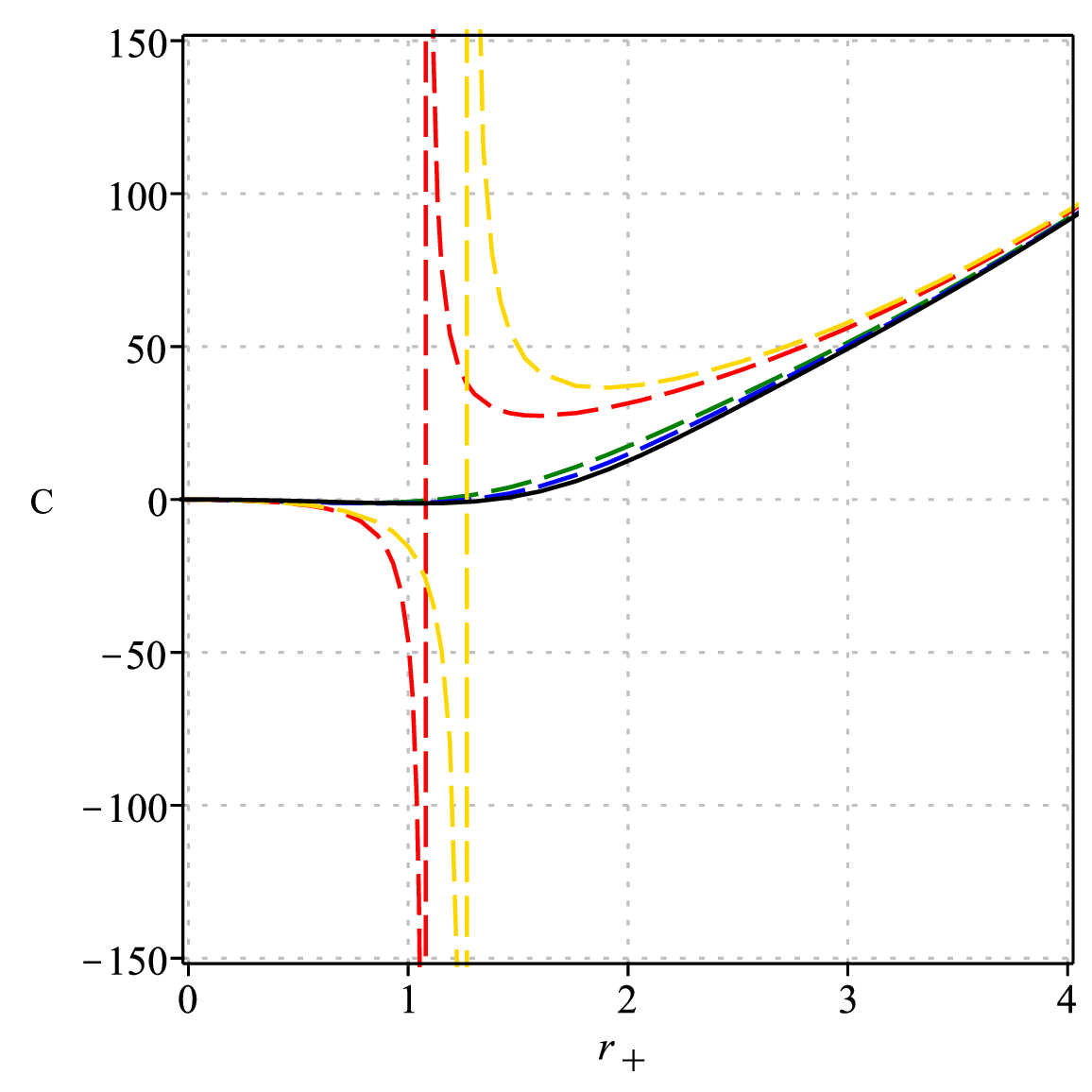}}\hfill
\caption{Left panel: $m=0.0$ denoted by solid black line, $m=0.5$ denoted by red dash line with, $m=1.0$ denoted by blue dash line, $m=1.5$ denoted by green dash line and $m=2.0$ denoted by gold dash line. Right panel: $\beta=0.0$ denoted by solid black line, $\beta=0.01$ denoted by blue dash line, $\beta=0.05$ denoted by green dash line and  $\beta=1.0$ denoted by red dash line and $\beta=2.0$ denoted by gold dash line. GR-NED with $Q_m=2$,  $c=1$, $c_1=-1$, $c_2=1$ and $l=2$ }\label{fig:21}
\end{figure}

\begin{figure}[H]
    \centering
    \includegraphics[width=.6\textwidth]{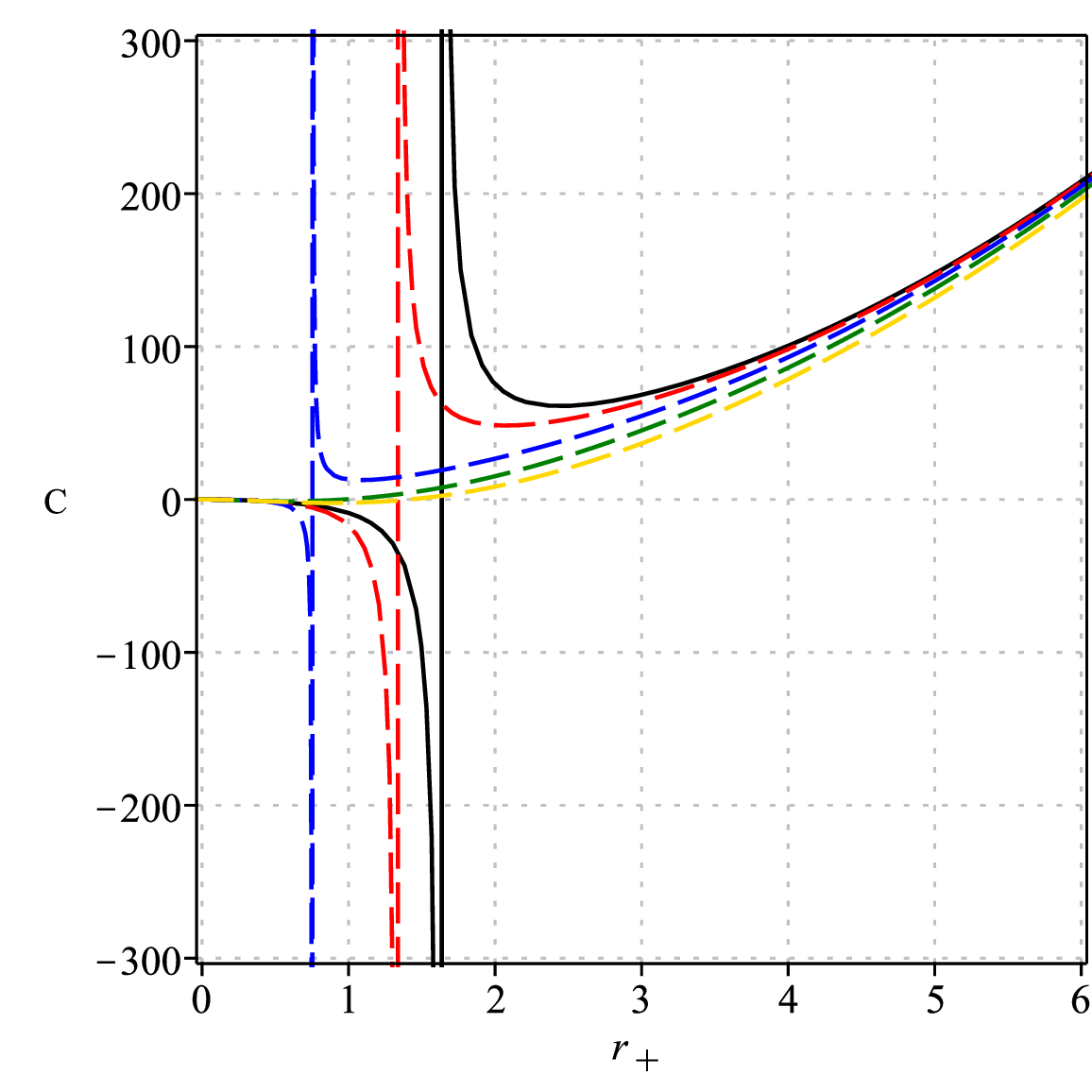}
    \caption{$Q_m=0.0$ denoted by solid black line, $Q_m=1.0$ denoted by red dash line with, $Q_m=2.0$ denoted by blue dash line, $Q_m=3.0$ denoted by green dash line and $Q_m=4.0$ denoted by gold dash line in GR-NED with $m=1.0$, $\beta=0.5$, $c=1$, $c_1=-1$, $c_2=1$ and $l=2$.}
    \label{fig:22}
\end{figure}

In Figs. \ref{fig:21} and \ref{fig:22} specific heat of black hole in 
massive GR coupled to NED are plotted. In Fig. 
\ref{fig:21}(a) specific heat is shown for different values of 
graviton mass. For $m=0$ and $0.5$, specific heat is negative for 
smaller-sized black holes, it attains zero at some critical values 
of horizon radius and finally increases function of horizon radius. 
Clearly specific heat is discontinuous for $m=1,1.5$ and $2$, i.e. a 
second order phase transition occurs for such black hole at $r_{0}$. 
The singular point of specific heat is where Hawking temperature 
(Figs. \ref{fig:15} and \ref{fig:16}) attains a minimum ($T_{GR}^{min}$). 
The singular point at $r_{0}$ separates specific heat into two regions. 
The first region, $r_{+} < r_0$ where specific heat is negative, i.e. 
black hole with horizon radius $r_{+} < r_0$ is thermodynamically 
unstable. The second region, $r_{+} > r_0$ where specific heat is 
positive, i.e. black hole with horizon radius $r_{+} > r_0$ is 
thermodynamically stable. A similar kind of behaviour is shown in 
Figs. \ref{fig:21}(b) and \ref{fig:22}.
 
Next, we study the global stability of $4D$ EGB black holes in NED. Gibbs free energy is defined as
\begin{equation}\label{eq:3.30}
    G=M-T_{H}S.
\end{equation}
Using equations \eqref{eq:3.1}, \eqref{eq:3.4} and \eqref{eq:3.10} we obtain
\begin{equation*}
G=\frac{r_{{+}}}{2}+\frac{Q_{m}^{2} \pi}{4 k}-\frac{Q_{m}^{2} \arctan  ({r_{{+}}}/{k})}{2 k}+\frac{r_{{+}} m^{2} c^{2} c_{2}}{2}+\frac{r_{{+}}^{2} m^{2} c c_{1}}{4}+\frac{\alpha}{2 r_{{+}}}+\frac{r_{{+}}^{3}}{2 l^{2}}-\frac{(\alpha  \ln(r_{{+}} )+\frac{r_{{+}}^{2}}{4})}{{r_{{+}} (r_{{+}}^{2}+2 \alpha ) l^{2} (k^{2}+r_{{+}}^{2})}} \biggr[ 3 r_{{+}}^{6}
\end{equation*}
\begin{equation}\label{eq:3.31}
+c c_{1} l^{2} m^{2} r_{{+}}^{5}+((c^{2} c_{2} m^{2}+1) l^{2}+3 k^{2}) r_{{+}}^{4}+c c_{1} k^{2} l^{2} m^{2} r_{{+}}^{3}+((c^{2} c_{2} m^{2}+1) k^{2}-Q_{m}^{2}-\alpha ) l^{2} r_{{+}}^{2}-\alpha  k^{2} l^{2} \biggr].
\end{equation}

In the limit $\alpha \to 0$, above equation reduced to Gibbs free energy of 
black hole in massive GR coupled to NED
\begin{equation*}
G=-\frac{r_{+} m^{2} c_{2} k^{2} c^{2}}{4 (k^{2}+ r_{+}^{2})}-\frac{r_{+}^{3} m^{2} c_{2} c^{2}}{4 (k^{2}+ r_{+}^{2})}-\frac{r_{+}^{2} c c_{1} k^{2} m^{2}}{4 (k^{2}+r_{+}^{2})}-\frac{r_{+}^{4} c c_{1} m^{2}}{4 (k^{2}+r_{+}^{2})}+\frac{r_{+} Q_{m}^{2}}{4 (k^{2}+r_{+}^{2})}-\frac{r_{+} k^{2}}{4( k^{2}+ r_{+}^{2})}-\frac{3 r_{+}^{3} k^{2}}{4 l^{2} (k^{2}+r_{+}^{2})}
\end{equation*}
\begin{equation}\label{eq:3.32}
-\frac{r_{+}^{3}}{4 (k^{2}+ r_{+}^{2})}-\frac{3 r_{+}^{5}}{4 l^{2} (k^{2}+r_{+}^{2})}+\frac{r_{+}}{2}-\frac{Q_{m}^{2} \arctan  ({r_{+}}/{k})}{2 k}+\frac{r_{+}^{3}}{2 l^{2}}+\frac{r_{+} m^{2} c^{2} c_{2}}{2}+\frac{c m^{2} c_{1} r_{+}^{2}}{4}+\frac{\pi  Q_{m}^{2}}{4 k}.
\end{equation}

\begin{figure}[H]
\centering
\subfloat[$\alpha=0.2$]{\includegraphics[width=.5\textwidth]{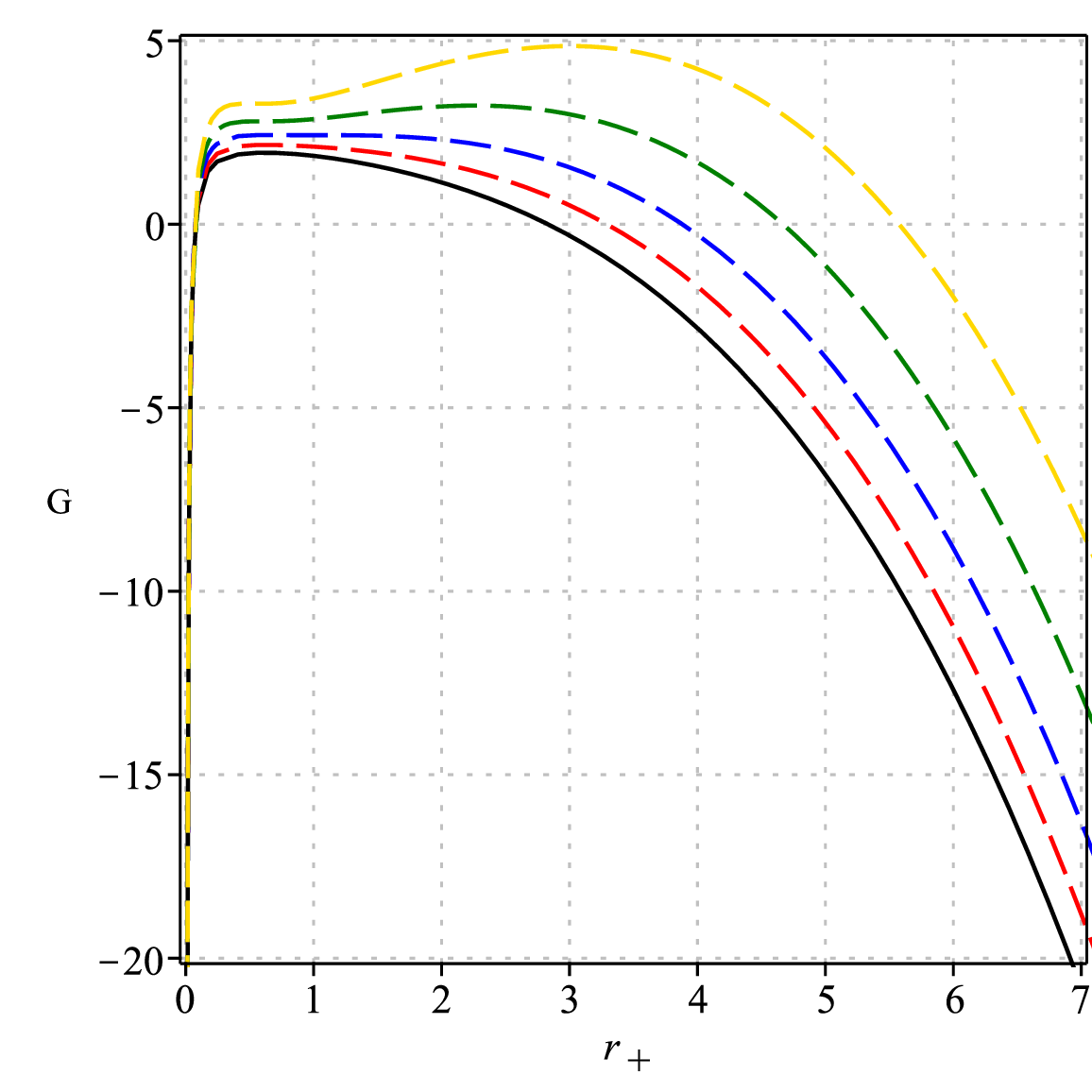}}\hfill
\subfloat[$\alpha=0.4$]{\includegraphics[width=.5\textwidth]{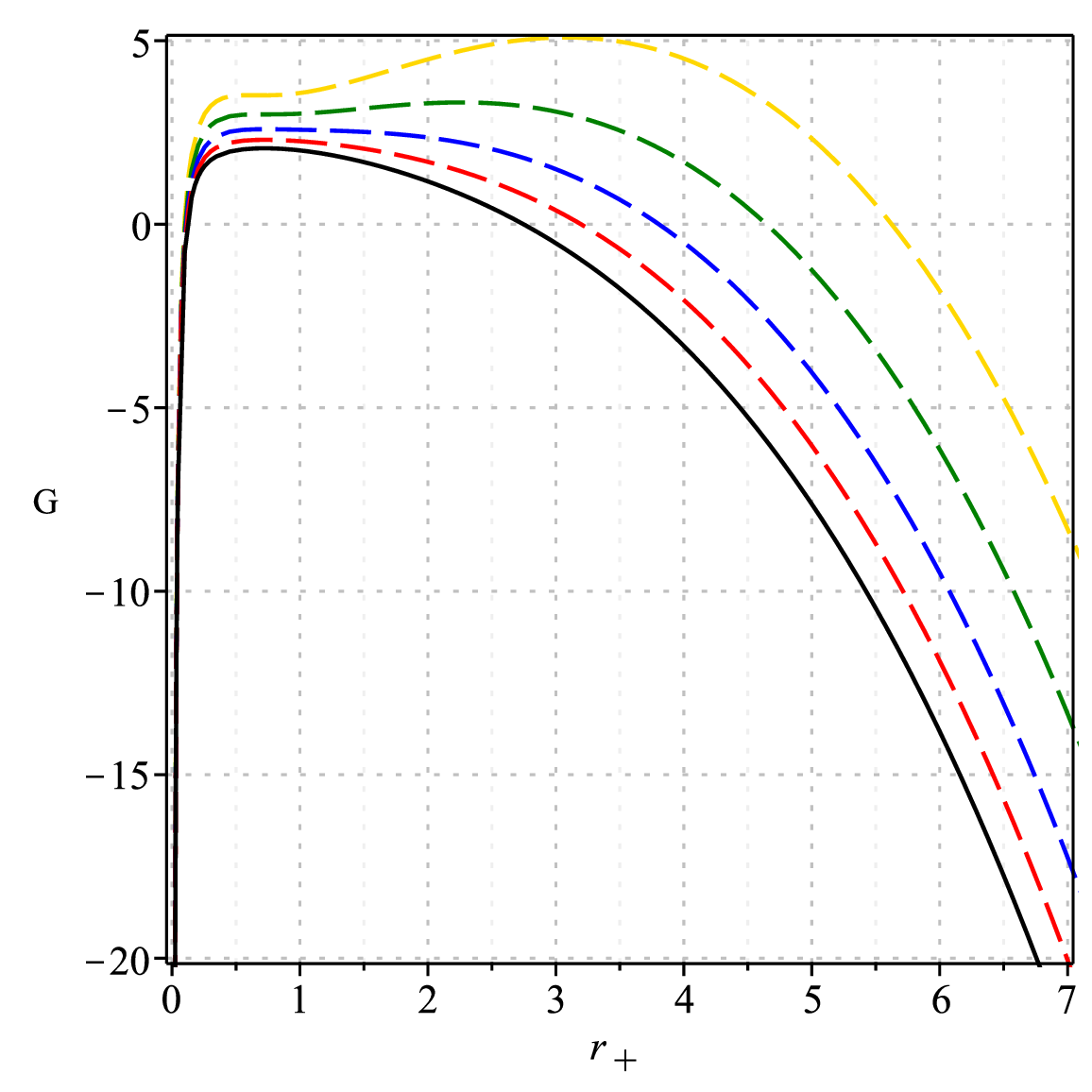}}\hfill
\caption{$m=0.0$ is denoted by solid black line, $m=1.0$ is denoted by red 
dash line with, $m=1.5$ is denoted by blue dash line, $m=2.0$ is denoted by 
green dash line and $m=3.0$ is denoted by gold dash line  in EGB-NED with 
$Q_m=2$, $\beta=0.5$, $c=1$, $c_1=-1$, $c_2=1$ and $l=2$. }\label{fig:23}
\end{figure} 

\begin{figure}[H]
\centering
\subfloat[$\alpha=0.02$]{\includegraphics[width=.5\textwidth]{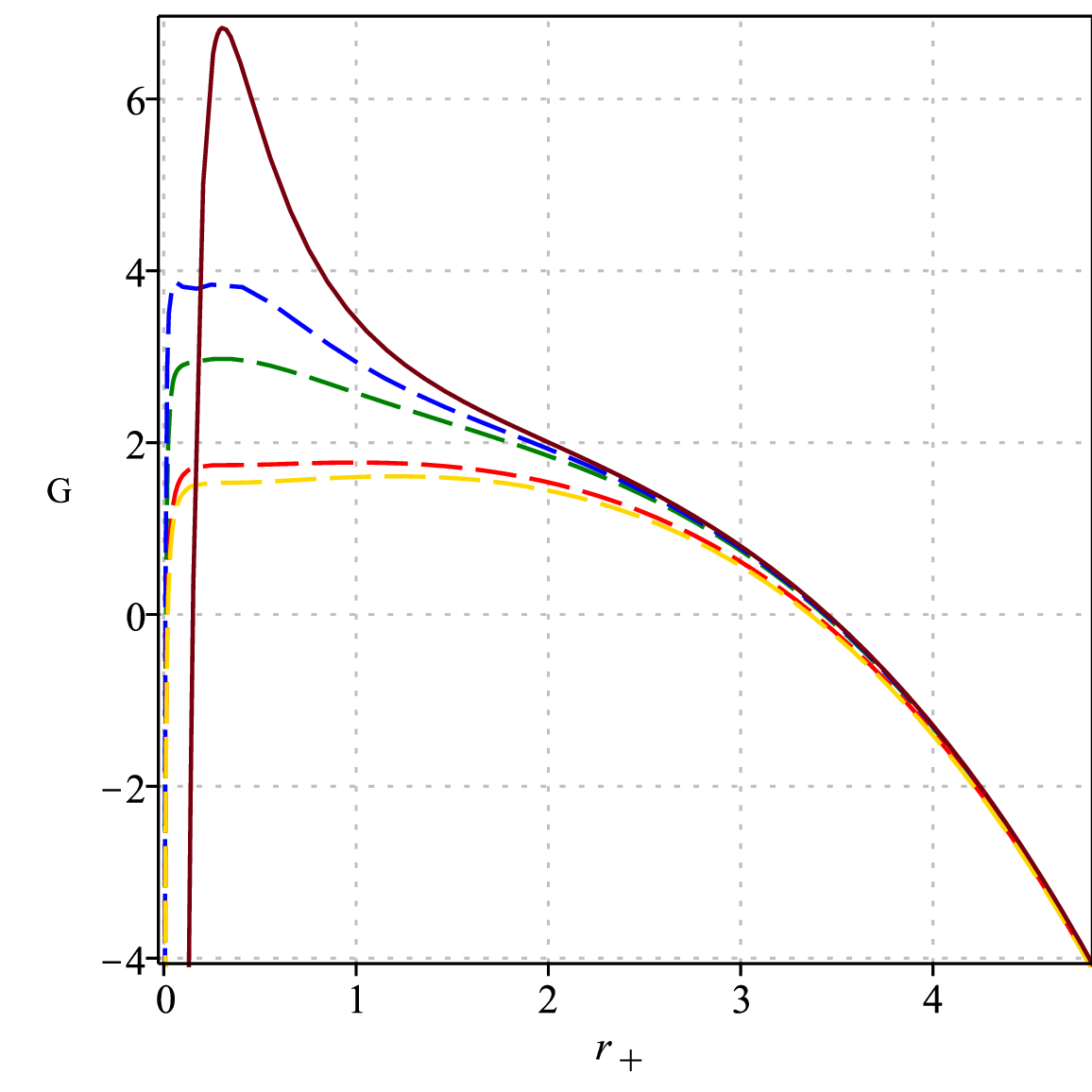}}\hfill
\subfloat[$\alpha=0.2$]{\includegraphics[width=.5\textwidth]{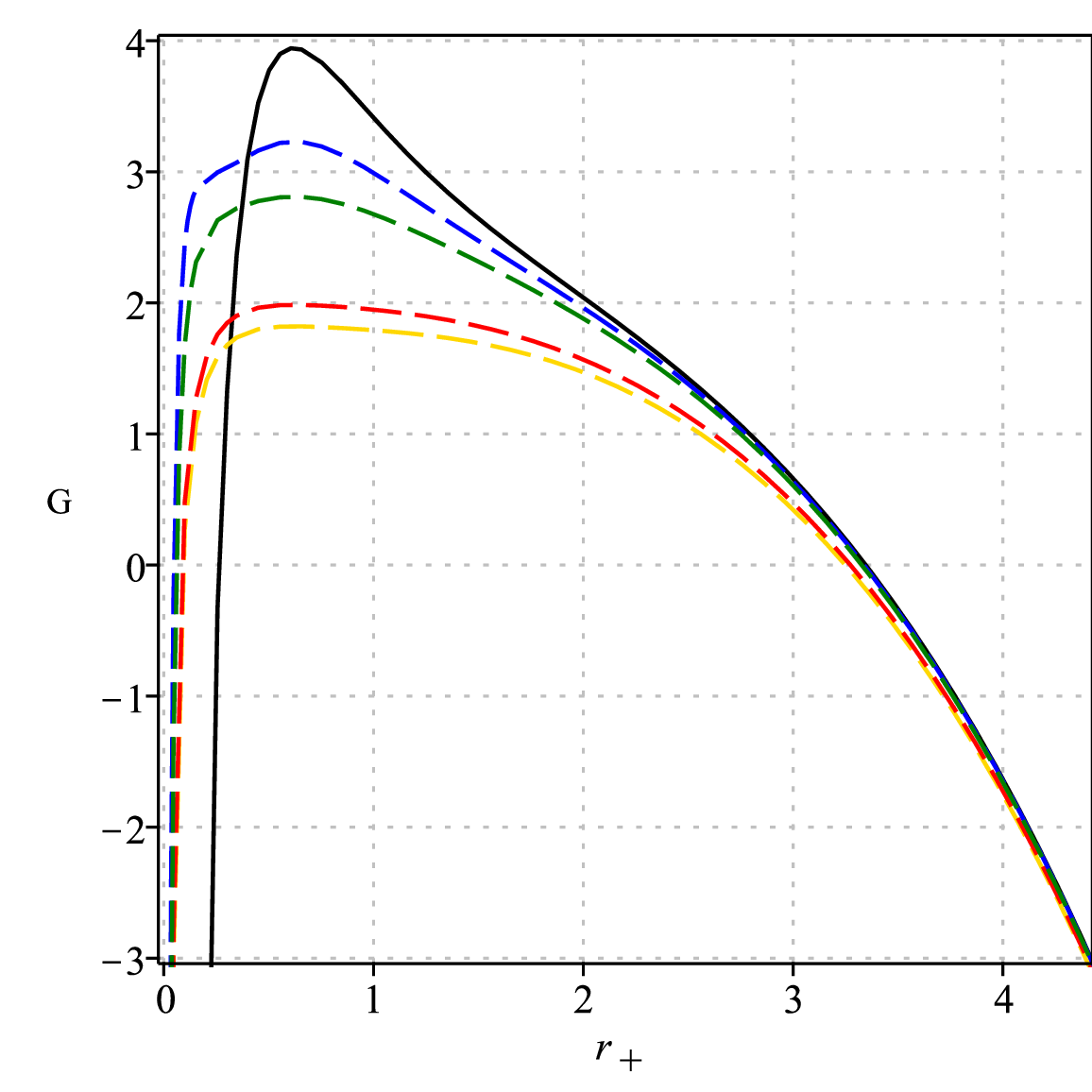}}\hfill
\caption{$\beta=0.0$ is denoted by solid black line, $\beta=0.01$ is denoted 
by blue dash line with, $\beta=0.05$ is denoted by green dash line,  
$\beta=1.0$ denoted by red dash line and $\beta=2.0$ is denoted by gold dash 
line in EGB-NED with  $Q_m=2$, $m=1.0$, $c=1$, $c_1=-1$, $c_2=1$ and $l=2$. }\label{fig:24}
\end{figure} 

\begin{figure}[H]
\centering
\subfloat[$m=0.0$ \&  $Q_m=2$]{\includegraphics[width=.5\textwidth]{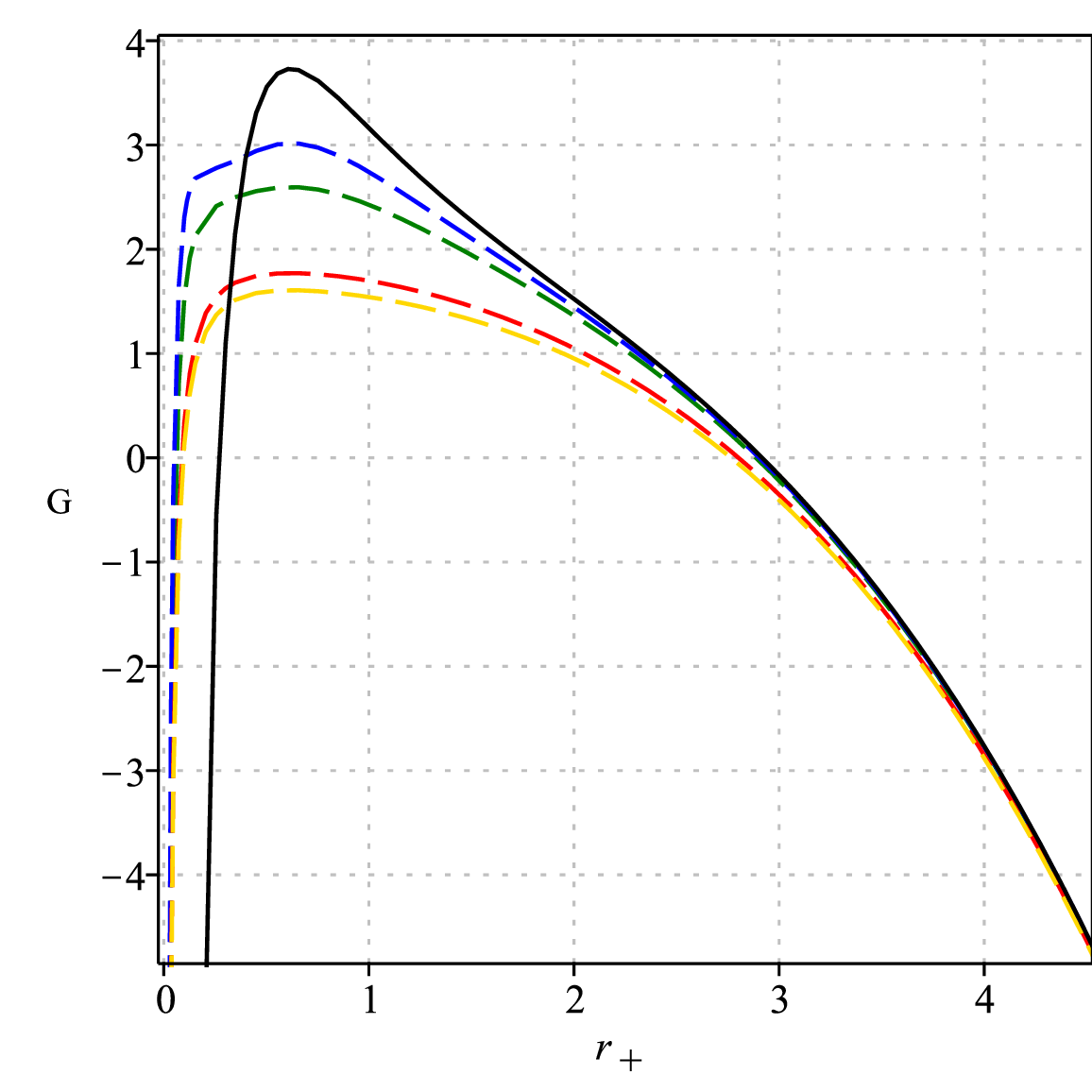}}\hfill
\subfloat[$m=1.0$ \&  $\beta=0.5$]{\includegraphics[width=.5\textwidth]{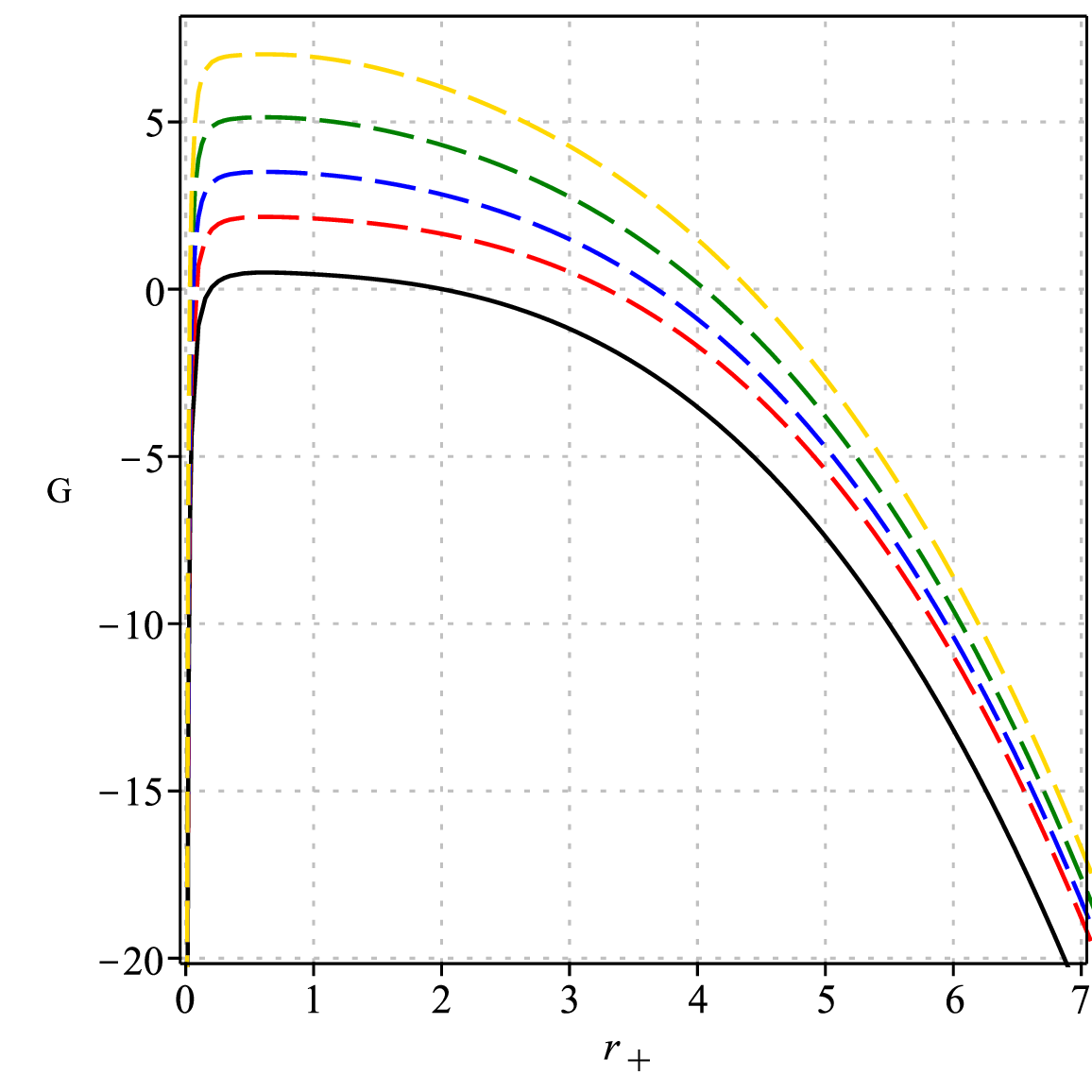}}\hfill
\caption{Left Panel: $\beta=0.0$ is denoted by solid black line, $\beta=0.01$ 
is denoted by blue dash line with, $\beta=0.05$ is denoted by green dash 
line, $\beta=1.0$ is denoted by red dash line in EGB-NED. Right Panel: 
$Q_m=0.0$ is denoted by solid black line, $Q_m=2.0$ is denoted by red dash 
line with, $Q_m=4.0$ is denoted by blue dash line, $Q_m=6.0$ is denoted by 
green dash line and $Q_m=7.0$ is denoted by gold dash line in EGB-NED with 
$\alpha=0.2$, $c=1$, $c_1=-1$, $c_2=1$ and $l=2$.}\label{fig:25}
\end{figure}

\begin{figure}[H]
\centering
\subfloat[$\beta=0.5$]{\includegraphics[width=.5\textwidth]{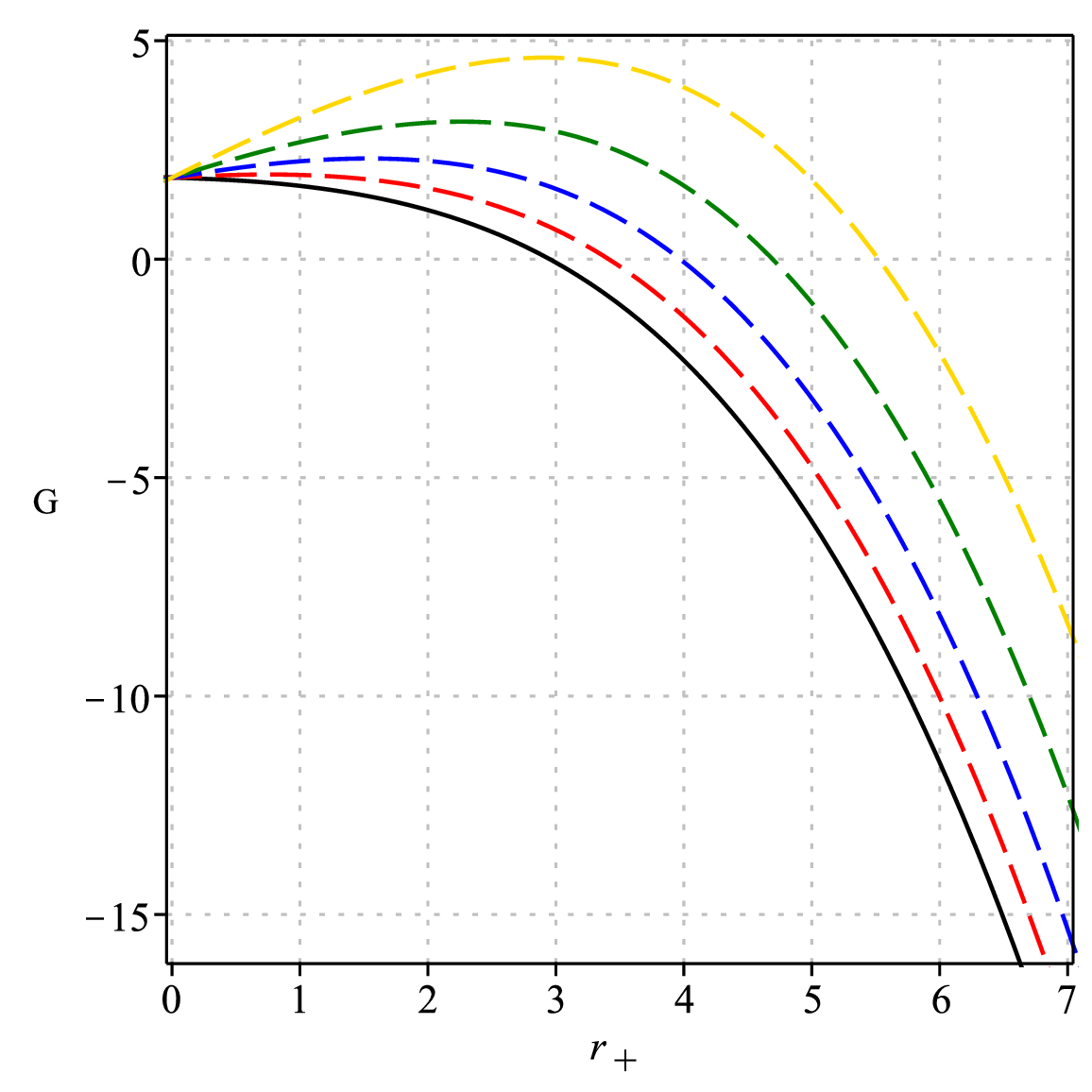}}\hfill
\subfloat[$m=1$]{\includegraphics[width=.5\textwidth]{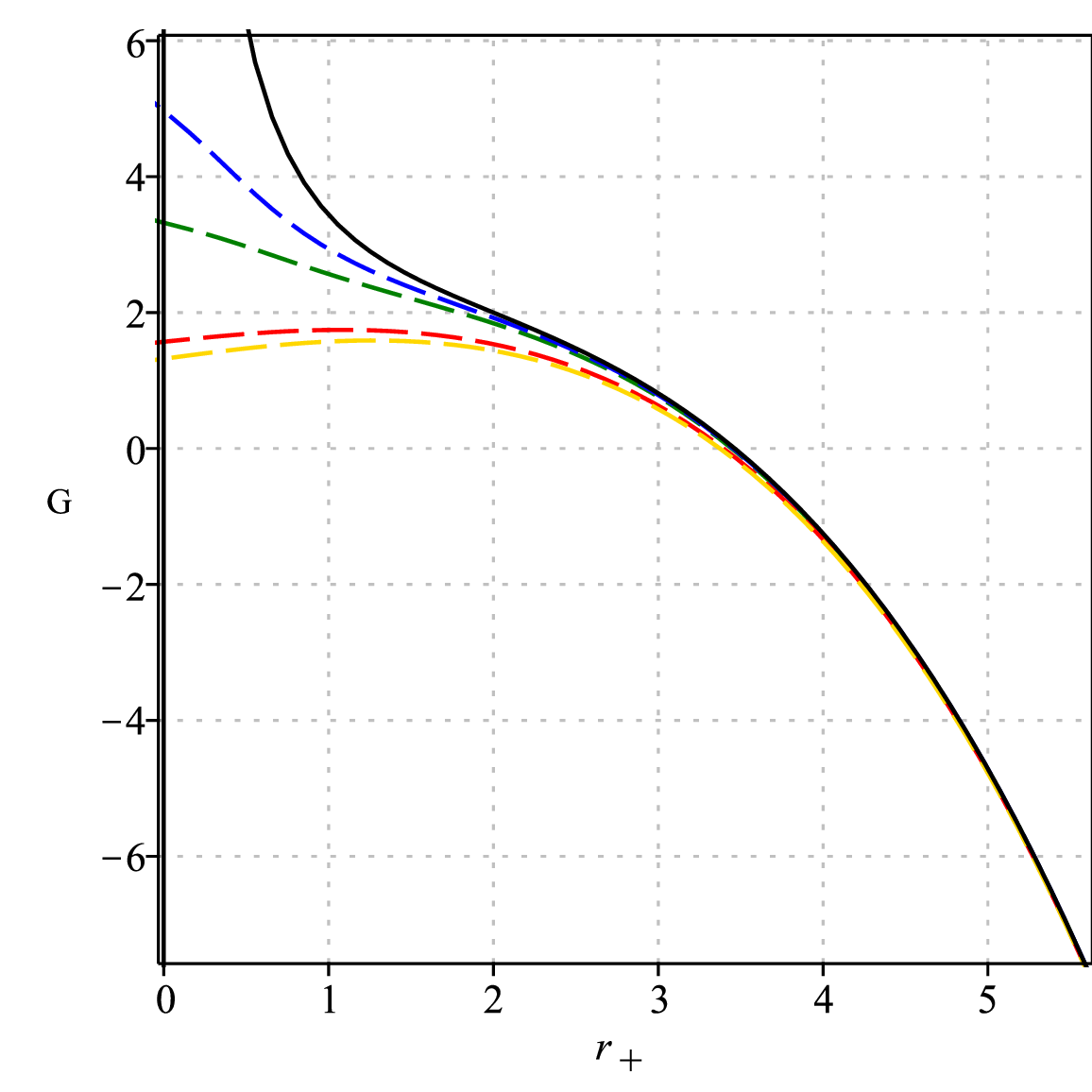}}\hfill
\caption{Left panel: $m=0.0$ is denoted by solid black line, $m=0.5$ is 
denoted by red dash line with, $m=1.0$ is denoted by blue dash line, $m=1.5$ is 
denoted by green dash line and $m=2.0$ is denoted by gold dash line. Right 
panel: $\beta=0.0$ is denoted by solid black line, $\beta=0.01$ is denoted 
by blue dash line, $\beta=0.05$ is denoted by green dash line and 
$\beta=1.0$ is denoted by red dash line and $\beta=2.0$ is denoted by gold 
dash line. GR-NED with $Q_m=2$,  $c=1$, $c_1=-1$, $c_2=1$ and $l=2$ }\label{fig:26}
\end{figure}

\begin{figure}[H]
    \centering
    \includegraphics[width=.6\textwidth]{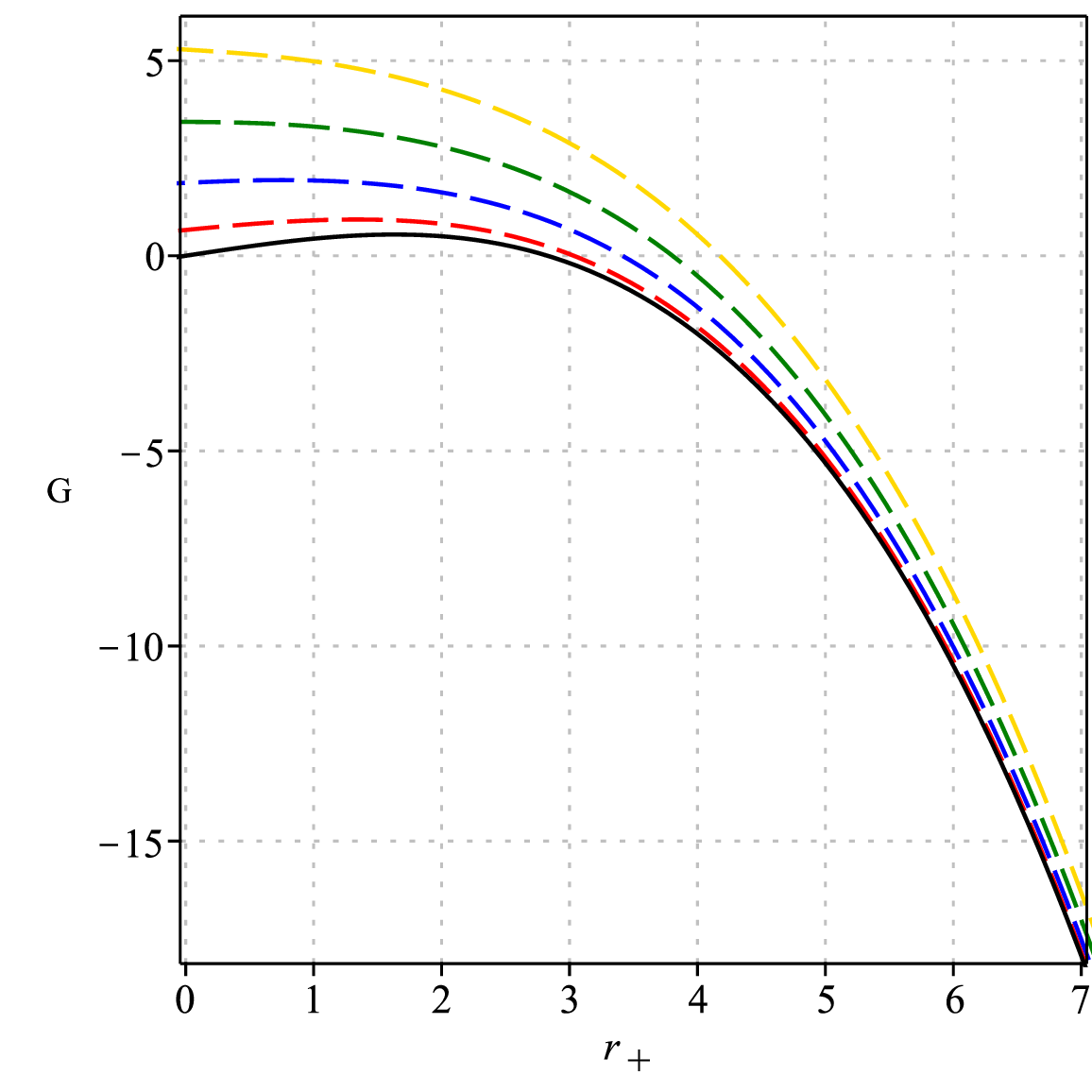}
    \caption{$Q_m=0.0$ is denoted by solid black line, $Q_m=1.0$ is denoted by 
    red dash line with, $Q_m=2.0$ is denoted by blue dash line, $Q_m=3.0$ is 
    denoted by green dash line and $Q_m=4.0$ is denoted by gold dash line 
    in GR-NED with $m=1.0$, $\beta=0.5$, $\alpha=0.0$, $c=1$, $c_1=-1$, $c_2=1$ and $l=2$.}
    \label{fig:27}
\end{figure}
\section{Van--der Waals Like Phase Transition}\label{sec:4}
In this section, we study the Van der Waals-like behaviour of a $4D$ EGB black hole. 
The cosmological constant is related to the black hole pressure by the relations 
$P=-\Lambda/8\pi=3/8\pi l^2$. The phase transition of black holes in GR/EGB coupled 
to Maxwell/BI electrodynamics was studied in Refs. \cite{Kubiznak:2012wp,Gunasekaran:2012dq,Hegde:2020xlv,Zhang:2020obn}, 
GR coupled to NED (in eq. \eqref{eq:2.4}) was studied in 
\cite{kruglov2022nonlinearly}. Phase transition of black hole in massive 
GR/EGB coupled to Maxwell electrodynamics was studied in Refs. 
\cite{Cai:2014znn,Fernando:2016qhq,Hendi:2015pda,Zou:2016sab,Zou:2017juz,Hendi:2016yof,Upadhyay:2022axg}. 

The Van der Waals equation of state of a fluid is given by
\begin{equation}\label{eq:4.1}
    \Bigl( P+\frac{a}{v^2}\Bigl)\bigl(v-b \bigl) =RT,
\end{equation}
where $a$ and $b$ represent interaction between the fluid molecules and size of the molecules. $v$ is the specific volume of the fluid molecules.
From the above equation one can obtain pressure as 
\begin{equation}\label{eq:4.2}
    P=\frac{RT}{v-b}- \frac{a}{v^2}.
\end{equation}

\subsection{Black Holes in 4D EGB Massive gravity coupled to NED}\label{sec:4.1}
The pressure of the black hole in $4D$ EGB massive gravity coupled to NED can be obtained from Hawking temperature \eqref{eq:3.4} as
\begin{equation*}
P=\frac{1}{16 \pi  k^{2} v^{3}+4 \pi  v^{5}} \Bigr[ (-c c_{1} m^{2}+4 \pi  T_{H} ) v^{4}+(-2 c^{2} c_{2} m^{2}-2) v^{3}+\Bigl( (-4 c c_{1} m^{2}+16 \pi  T_{H} ) k^{2}+32 \pi  T_{H} \alpha \Bigl) v^{2}
\end{equation*}
\begin{equation}\label{eq:4.3}
+\Bigl((-8 c^{2} c_{2} m^{2}-8) k^{2}+8 Q_{m}^{2}+8 \alpha \Bigl) v +128 \pi  T_{H} \alpha  k^{2} \Bigr] +\frac{8 \alpha  k^{2}}{\pi  v^{4} (4 k^{2}+v^{2})},
\end{equation}
where we take $v=2r_+$. At critical points
\begin{equation}\label{eq:4.4}
    \biggl( \frac{\partial P}{\partial v}\biggl)_{T_c, v_c} = 0 = \biggl( \frac{\partial^{2} P}{\partial v^{2}} \biggl)_{T_c, v_c}.
\end{equation}
Therefore, using above equation and equation \eqref{eq:4.3} one can obtain equations for critical volume, critical temperature and pressure as

\begin{equation*}
\Bigr[ -(c^{2} c_{2} m^{2}+1) v_{c}^{10}+12 \alpha  c c_{1} m^{2} v_{c}^{9}+\Bigl( -12( c^{2} c_{2} m^{2}+1) k^{2}+24 m^{2} c_{2} c^{2} \alpha +24 Q_{m}^{2}+48 \alpha \Bigl) v_{c}^{8} +144 \alpha  c c_{1} k^{2} m^{2} v_{c}^{7}
\end{equation*}
\begin{equation*}
+\Bigl( (-48 c^{2} c_{2} m^{2}-48) k^{4}+(288 m^{2} c_{2} c^{2} \alpha +48 Q_{m}^{2}+576 \alpha ) k^{2}+192 \alpha  (Q_{m}^{2}+\alpha )\Bigl) v_{c}^{6}+576 \alpha  c c_{1} k^{4} m^{2} v_{c}^{5}
\end{equation*}
\begin{equation*}
-64 \Bigl( (c^{2} c_{2} m^{2}+1) k^{4}+(-18 m^{2} c_{2} c^{2} \alpha -Q_{m}^{2}-36 \alpha ) k^{2}+18 \alpha  Q_{m}^{2}-36 \alpha^{2}\Bigl) k^{2} v_{c}^{4}+768 \alpha  c c_{1} k^{6} m^{2} v_{c}^{3}
\end{equation*}
\begin{equation}\label{eq:4.5}
+1536 \alpha  k^{4} \Bigl( (c^{2} c_{2} m^{2}+2) k^{2}-Q_{m}^{2}+6 \alpha \Bigl)v_{c}^{2}+12288 \alpha^{2} k^{6} \Bigr]=0,
\end{equation}

\begin{equation*}
T_{c} = \frac{1}{64 \pi  v_{c} (k^{2}+\frac{v_{c}^{2}}{4})^{2} (v_{c}^{2}+24 \alpha )} \biggr[ c c_{1} m^{2} v_{c}^{7}+(4 c^{2} c_{2} m^{2}+4) v_{c}^{6}+8 c c_{1} k^{2} m^{2} v_{c}^{5}+\Bigl( (32 c^{2} c_{2} m^{2}+32) k^{2}-32 Q_{m}^{2}-32 \alpha \Bigl) v_{c}^{4}
\end{equation*}
\begin{equation}\label{eq:4.6}
+16 c c_{1} k^{4} m^{2} v_{c}^{3}+64 k^{2} \Bigl( (c^{2} c_{2} m^{2}+1) k^{2}-Q_{m}^{2}-4 \alpha \Bigl) v_{c}^{2}-512 \alpha  k^{4}\biggr],
\end{equation}

\begin{equation*}
P_{c} = \frac{1}{32 v_{c}^{4} (v_{c}^{2}+24 \alpha ) (k^{2}+\frac{v_{c}^{2}}{4})^{2} \pi}\biggr[ (c^{2} c_{2} m^{2}+1) v_{c}^{8}-8 \alpha  c c_{1} m^{2} v_{c}^{7}+\Bigl( (8 c^{2} c_{2} m^{2}+8) k^{2}-8 m^{2} c_{2} c^{2} \alpha -12 Q_{m}^{2}-20 \alpha \Bigl) v_{c}^{6}
\end{equation*}
\begin{equation*}
-64 \alpha  c c_{1} k^{2} m^{2} v_{c}^{5}+ \Bigl( (16 c^{2} c_{2} m^{2}+16) k^{4}+(-64 m^{2} c_{2} c^{2} \alpha -16 Q_{m}^{2}-160 \alpha ) k^{2}-32 \alpha  Q_{m}^{2}-32 \alpha^{2}\Bigl) v_{c}^{4}-128 \alpha  c c_{1} k^{4} m^{2} v_{c}^{3}
\end{equation*}
\begin{equation}\label{eq:4.7}
-128 \alpha  k^{2} \Bigl( (c^{2} c_{2} m^{2}+{5}/{2}) k^{2}-Q_{m}^{2}+2 \alpha \Bigl) v_{c}^{2}-512 \alpha^{2} k^{4}\biggr].
\end{equation}

In the limit $m \to 0$, and $\alpha \to 0$, above equations are reduced to 
critical temperature and pressure of $4D$ massless GR coupled to NED 
\cite{kruglov2022nonlinearly}

\begin{equation}\label{eq:4.8}
  \Bigl( v_{c}^{2} +4k^2 \Bigl)^3 - 8 Q_m^2 \Bigl( 3v_c^4 +6k^2v_c^2+8k^4 \Bigl) =0,  
\end{equation}
\begin{equation}
T_{c} = \frac{1}{ \pi  v_{c}  }  - \frac{8 Q_m^2 \bigl( v_c^2 +2k^2\bigl)}{\pi v_c (4 k^{2}+{v_{c}^{2}})^{2}}, 
\end{equation}
\begin{equation}
P_{c} = \frac{1}{2 \pi  v_{c}^2  }  - \frac{2 Q_m^2 \bigl( 3v_c^2 +4k^2\bigl)}{\pi v_c^2 ({4} k^{2}+{v_{c}^{2}})^{2}}. 
\end{equation}

Equation \eqref{eq:4.5} for critical volume can not be solved analytically, 
so we use numerical techniques to obtain critical points as shown in the below 
table. In tables \ref{table:2}, \ref{table:3} and \ref{table:4} we estimate 
critical volume ($v_c$), critical pressure ($P_c$), critical temperature ($T_c$) 
and $\rho_c$ for different values of graviton mass $m$, NED parameter $\beta$ 
and GB coupling parameter $\alpha$. In table \ref{table:2} critical 
parameters are estimated, as graviton mass increases critical volume and 
temperature decreases, similarly critical pressure and $\rho_c$ increase as 
graviton mass increases from zero. The effects of NED parameter $\beta$ is 
shown in table \ref{table:3}, keeping the graviton mass and GB
coupling parameter fixed. As NED parameter increases critical volume and 
$\rho_c$ decrease, similarly critical pressure and temperature increase as 
NED parameter increases. The effect of GB parameter $\alpha$ is shown in 
table \ref{table:4}, keeping the graviton mass and NED coupling parameter 
fixed. Table \ref{table:4} shows the opposite behaviour of table \ref{table:3}.
\begin{table}[H]
\centering
\begin{tabular}{ |p{1.5cm}|p{1.5cm}|p{1.5cm}|p{1.5cm}|p{1.5cm}| } 
 \hline
 m & ${v_c}$ & $P_{c}$ & $T_{c}$ & $\rho_{c}$   \\ [0.5ex]  \hline
 \hline
 0.0 & 8.9469 & 0.0008 & 0.0218 & 0.3283 \\ \hline
 0.1 & 8.8749 & 0.0008 & 0.0214 & 0.3317 \\ \hline
 0.2 & 8.6646 & 0.0009 & 0.0202 & 0.3860 \\ \hline
 0.3 & 8.3315 & 0.0010 & 0.0182 & 0.4577 \\ \hline
 0.4 & 7.8969 & 0.0012 & 0.0156 & 0.6074 \\ \hline
 0.5 & 7.3838 & 0.0014 & 0.0124 & 0.8336 \\ [1ex] 
 \hline
\end{tabular}
\caption{Values of critical volume($T_{c}$), critical pressure($P_{c}$), critical temperature($T_{c}$) and $\rho_{c}=P_{c}v_{c}/T_{c}$ for different graviton mass with $Q_m=2$, $\alpha=0.2$, $\beta=0.5$, $c=1$ $c_{1}=-1$ and $c_{2}=1$.}
\label{table:2}
\end{table}

\begin{table}[H]
\centering
\begin{tabular}{ |p{1.5cm}|p{1.5cm}|p{1.5cm}|p{1.5cm}|p{1.5cm}| } 
 \hline
 $\beta$ & ${v_c}$ & $P_{c}$ & $T_{c}$ & $\rho_{c}$   \\ [0.5ex]  \hline
 \hline
 0.4 & 1.1811 & 0.0103 & 0.0064 & 1.9008 \\ \hline
 0.8 & 0.9416 & 0.0598 & 0.1030 & 0.5466 \\ \hline
 1.2 & 0.9034 & 0.0830 & 0.1479 & 0.5069 \\ \hline
 1.6 & 0.8865 & 0.0970 & 0.1750 & 0.4913 \\ \hline
 2.0 & 0.8766 & 0.1065 & 0.1936 & 0.4822 \\\hline
 2.4 & 0.8700 & 0.1135 & 0.2073 & 0.4763 \\        [1ex] 
 \hline
\end{tabular}
\caption{Values of critical volume($T_{c}$), critical pressure($P_{c}$), critical temperature($T_{c}$) 
and $\rho_{c}=P_{c}v_{c}/T_{c}$ for different GB coupling parameter with $Q=2$, $\alpha=0.02$, $m=1$, $c=1$ $c_{1}=-1$ and $c_{2}=1$.}
\label{table:3}
\end{table}

\begin{table}[H]
\centering
\begin{tabular}{ |p{1.5cm}|p{1.5cm}|p{1.5cm}|p{1.5cm}|p{1.5cm}| } 
\hline
$\alpha$ & ${v_c}$ & $P_{c}$ & $T_{c}$ & $\rho_{c}$   \\ [0.5ex]  \hline
\hline
0.00 & 4.0461 & 0.0043 & 0.0435 & 0.4062 \\ \hline
0.01 & 4.1224 & 0.0041 & 0.0424 & 0.4051 \\  \hline
0.05 & 4.4454 & 0.0036 & 0.0393 & 0.4072 \\  \hline
0.10 & 4.7788 & 0.0031 & 0.0361 & 0.4103 \\ \hline
0.40 & 6.2224 & 0.0018 & 0.0260 & 0.4307 \\ \hline
0.80 & 7.5720 & 0.0012 & 0.0200 & 0.4543 \\        [1ex] 
 \hline
\end{tabular}
\caption{Values of critical volume($T_{c}$), critical pressure($P_{c}$), 
critical temperature($T_{c}$) and $\rho_{c}=P_{c}v_{c}/T_{c}$ for different GB coupling parameter with $Q=1$, $\beta=0.1$, $m=0.2$, $c=1$ $c_{1}=-2$ and $c_{2}=0.75$.}
\label{table:4}
\end{table}

\begin{figure}[H]
\centering
\subfloat[$m=0.1$]{\includegraphics[width=.5\textwidth]{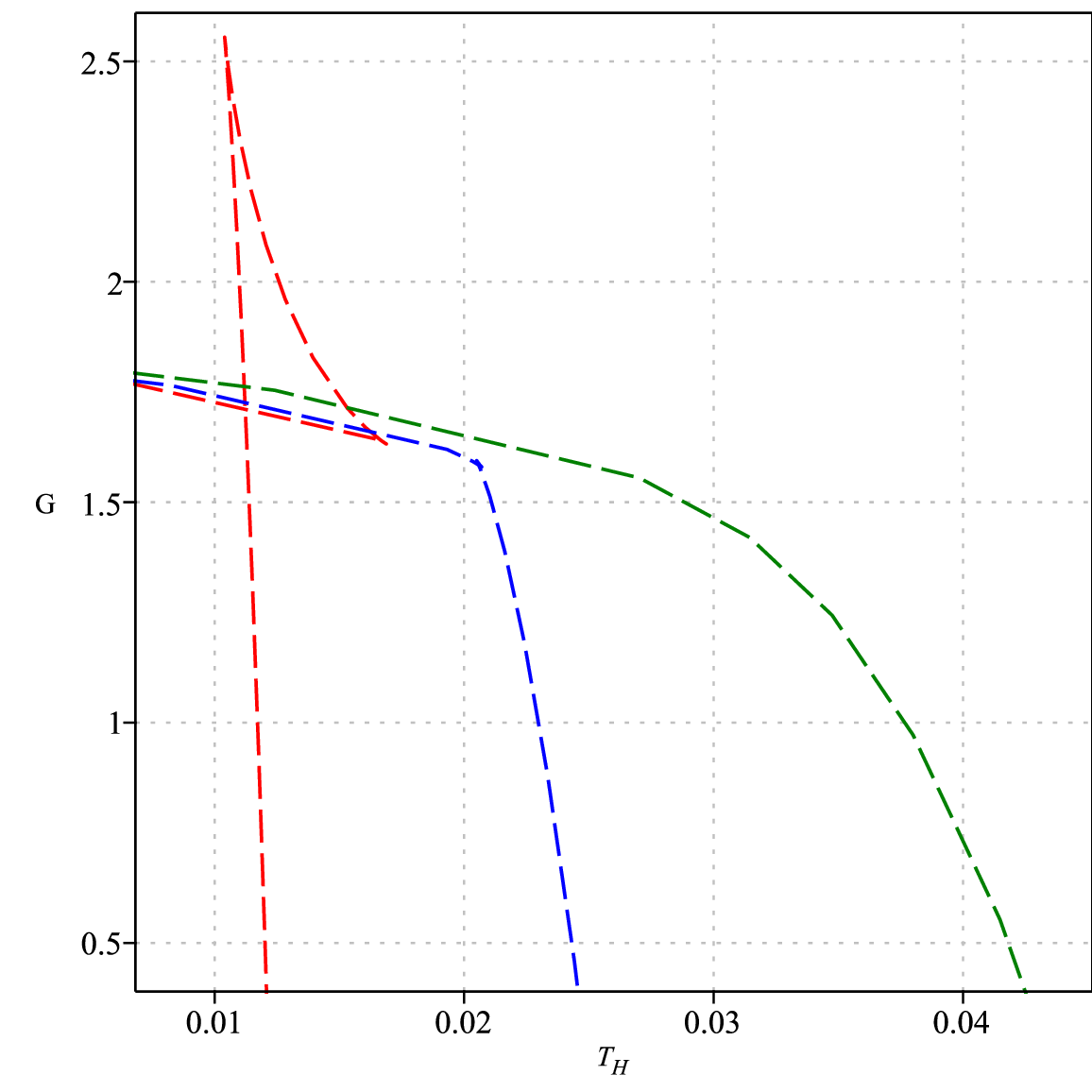}}\hfill
\subfloat[$m=0.5$]{\includegraphics[width=.5\textwidth]{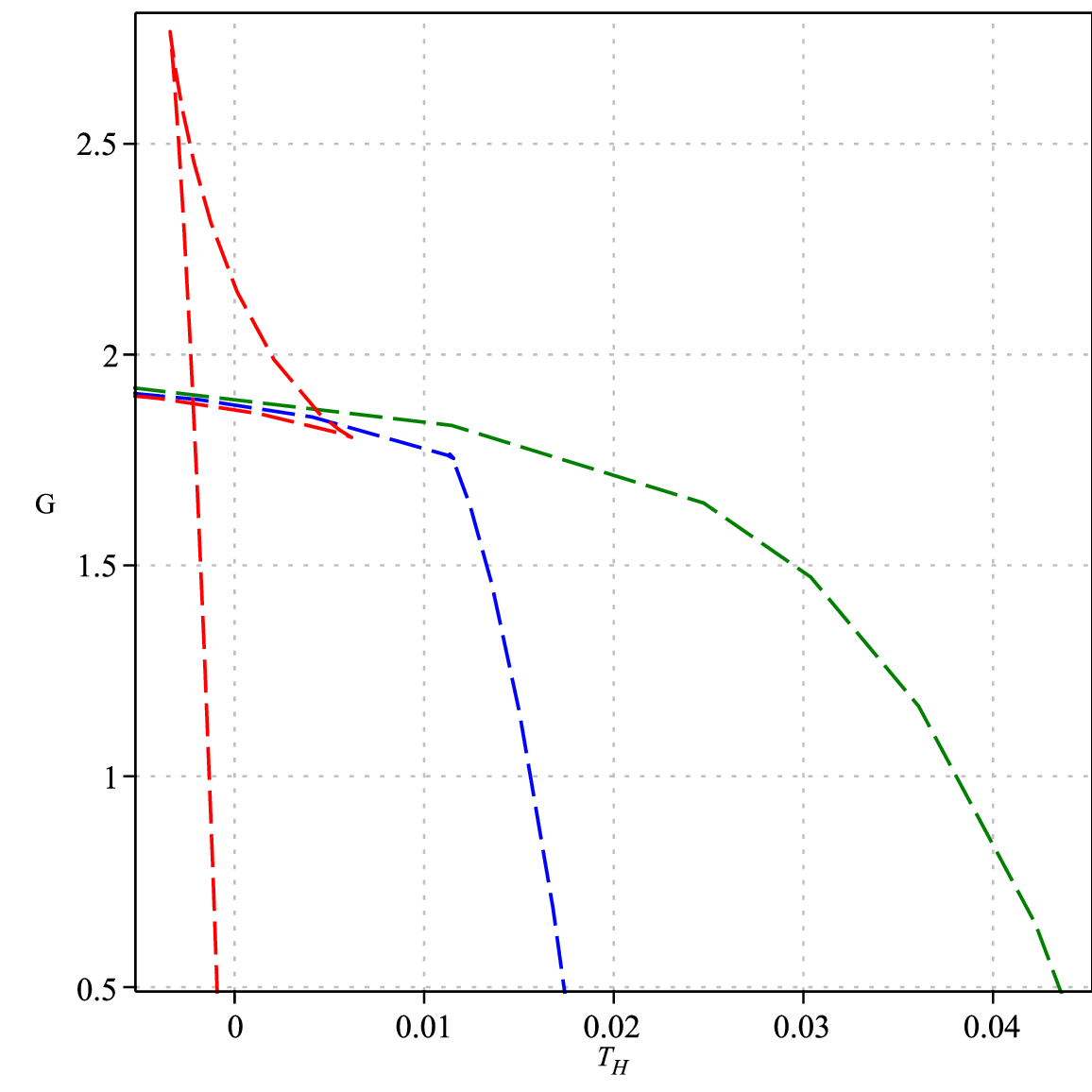}}\hfill
\caption{Red dash line is denoted $P=0.25P_c$, blue dash line is denoted $P=P_c$ and 
green dash line is denoted $P=3P_c$ with $Q_m=2$, $\alpha=0.2$, $\beta=0.5$, $c=1$, $c_1=-1$ and $c_2=1$. }\label{fig:28}
\end{figure}

We plot the $G-T_{H}$ diagram for different values of graviton mass 
$m$, NED parameter $\beta$, GB coupling parameter and critical 
pressure. In Fig. \ref{fig:28} $G-T_{H}$ diagram is depicted for 
different values of pressure and graviton mass. When $P<P_c$ Gibbs 
free energy shows swallow tail (triangular shape) behaviour, which 
clearly indicates that black hole undergoes a first order phase 
transition, i.e. when $P<P_c$ a phase transition occurs between \textbf{SBH} and 
\textbf{LBH}. From the 
point $T_{H}=0$ to the intersection point of the red dash curve a 
\textbf{SBH} branch is preferred and beyond the point of intersection, 
a \textbf{LBH} is preferred. Hence, at the point of intersection a 
transition between \textbf{SBH} and \textbf{LBH} occurs. The horizon 
radius of the \textbf{SBH} and \textbf{LBH} is different therefore, 
there is a discontinuity in the black hole horizon radius at the point 
of intersection for $P<P_c$. Finally, one can say that entropy of the 
black hole is discontinuous at the point of intersection as entropy 
depends on the horizon radius.

\begin{figure}[H]
\centering
\subfloat[$\beta=0.8$]{\includegraphics[width=.5\textwidth]{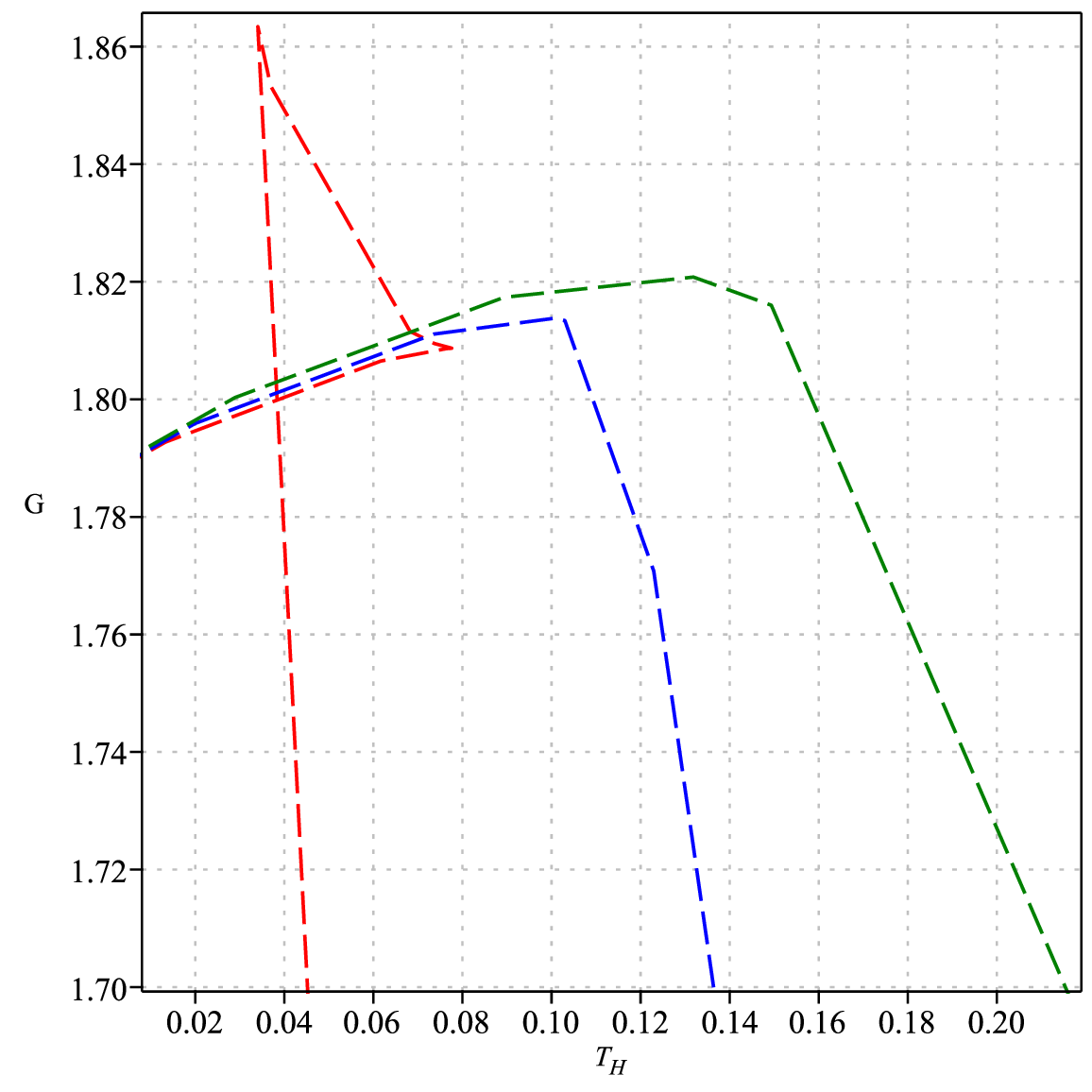}}\hfill
\subfloat[$\beta=1.2$]{\includegraphics[width=.5\textwidth]{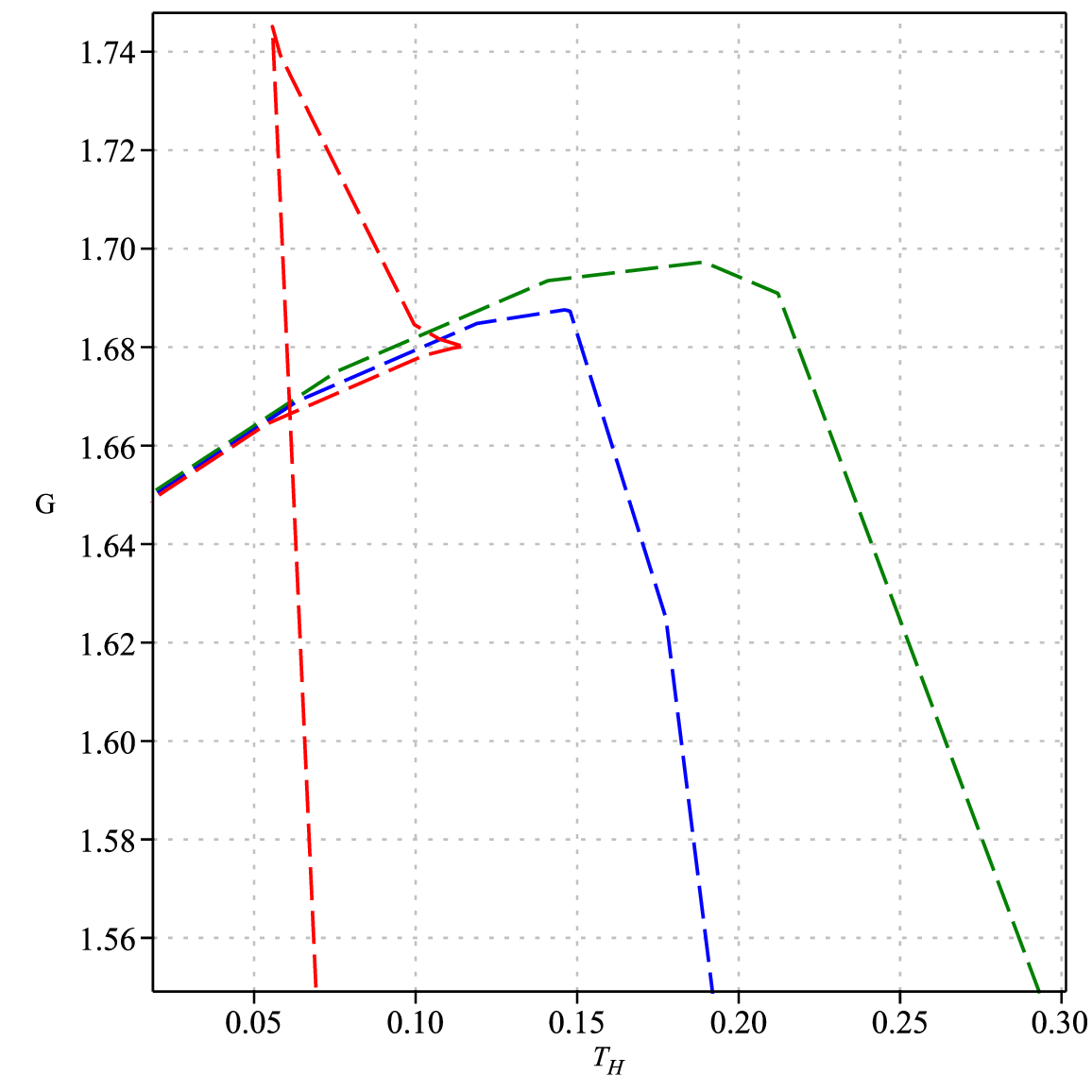}}\hfill
\caption{Red dash line is denoted $P=0.25P_c$, blue dash line is denoted $P=P_c$ and green dash line is 
denoted $P=2P_c$ with $Q_m=2$, $\alpha=0.02$, $m=1$, $c=1$, $c_1=-1$ and $c_2=1$.}\label{fig:29}
\end{figure} 

\begin{figure}[H]
\centering
\subfloat[$\alpha=0.4$]{\includegraphics[width=.5\textwidth]{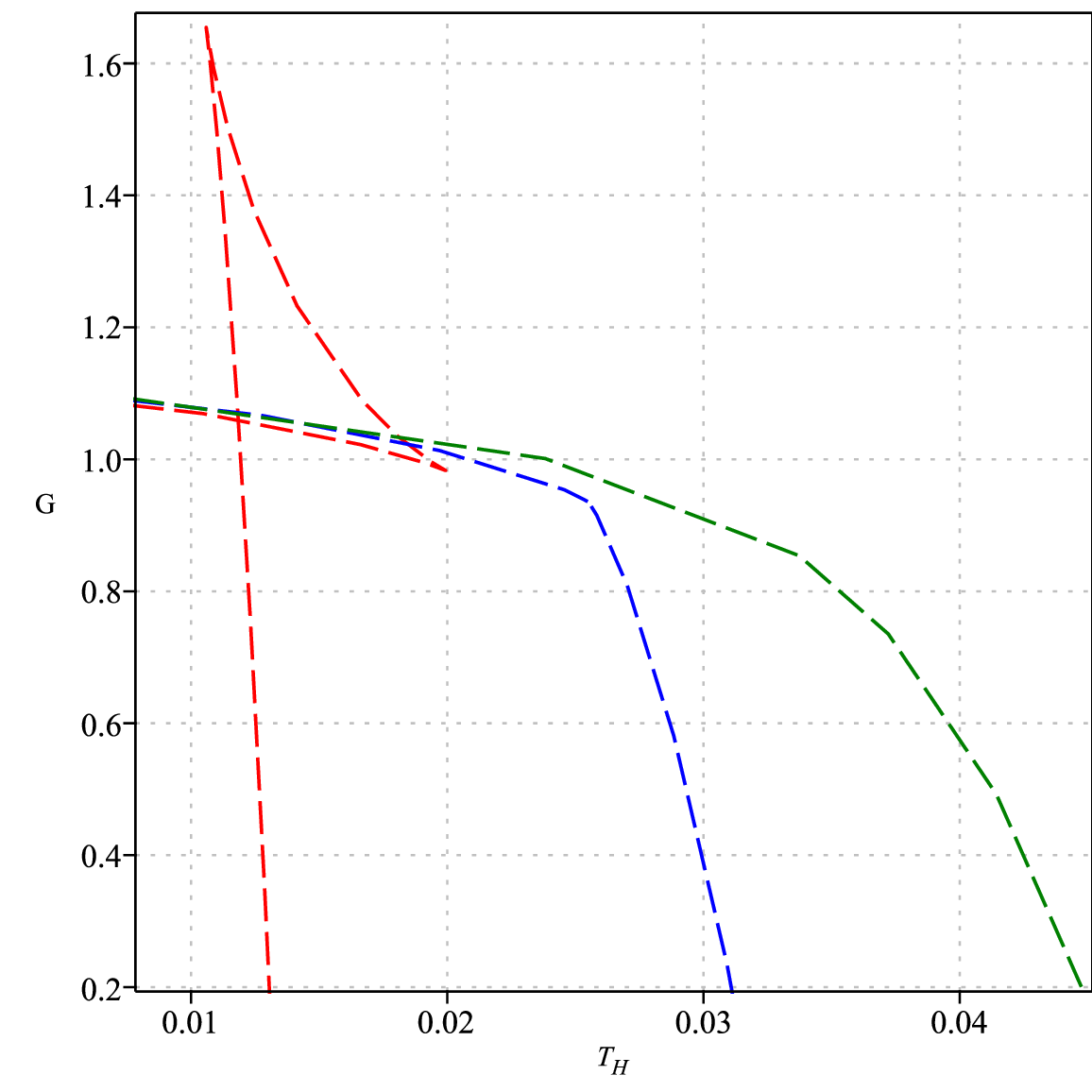}}\hfill
\subfloat[$\alpha=0.8$]{\includegraphics[width=.5\textwidth]{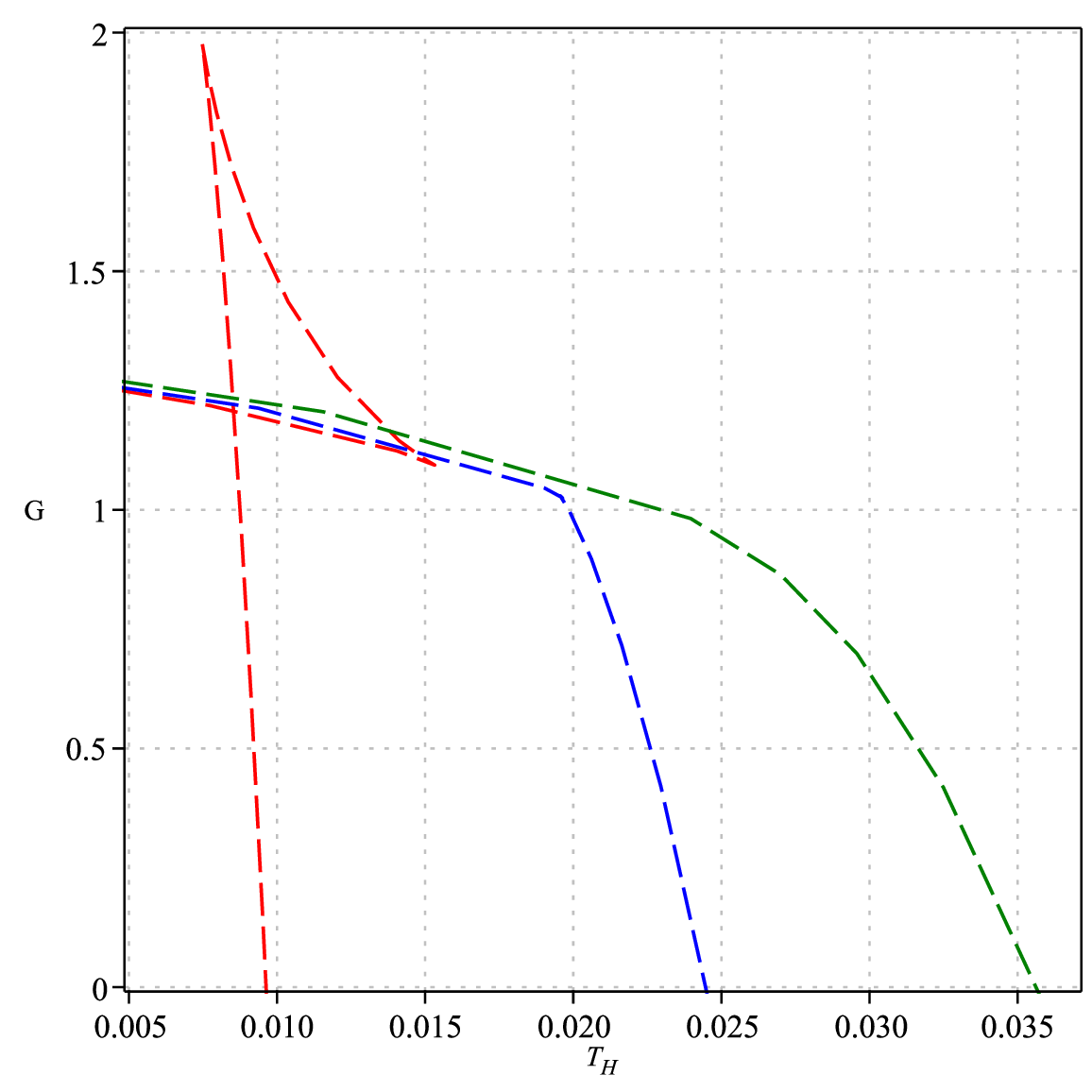}}\hfill
\caption{ Red dash line is denoted $P=0.25P_c$, blue dash line is denoted $P=P_c$ and green dash line is 
denoted $P=2P_c$ with $Q_m=1$, $\beta=0.1$, $m=0.2$, $c=1$, $c_1=-2$ and $c_2=0.75$.}\label{fig:30}
\end{figure}
\begin{figure}[H]
\centering
\subfloat[$\beta=1.2$]{\includegraphics[width=.5\textwidth]{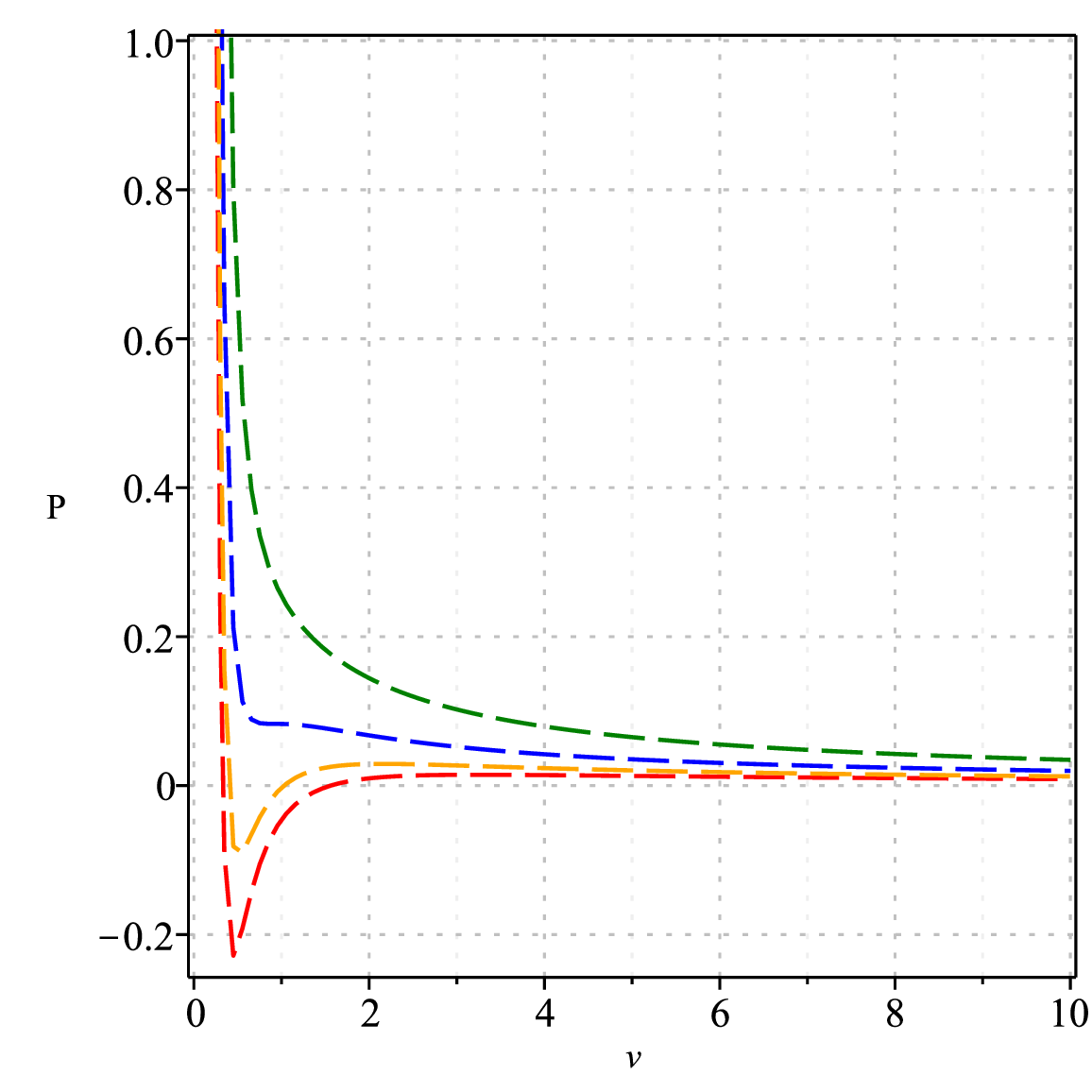}}\hfill
\subfloat[$\beta=2.0$]{\includegraphics[width=.5\textwidth]{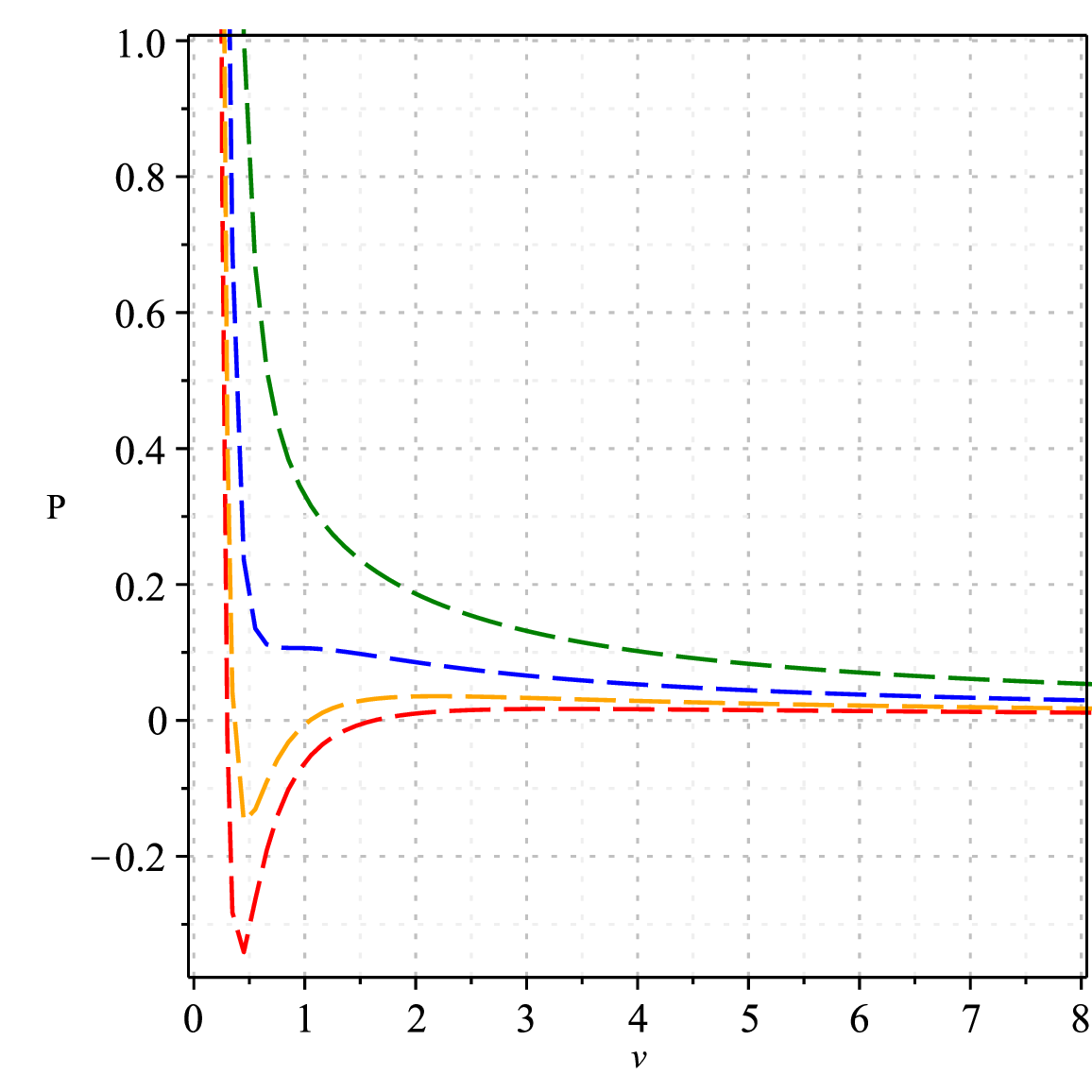}}\hfill
\caption{Red dash line is denoted $T_{H}=0.25T_c$, orange dash line is denoted $T_{H}=0.5T_c$, blue dash 
line is denoted $T_{H}=T_c$ and green dash line is denoted $T_{H}=2T_c$ with $Q_m=2$, $\alpha=0.02$, $m=1.0$, $c=1$, $c_1=-1$ and $c_2=1$.}\label{fig:31}
\end{figure}

 
Swallow tail (triangular shape) behaviour disappears at critical point 
$P=P_c$. For $P>P_c$, no phase transition occurs. Similar behaviour is 
shown in Fig. \ref{fig:29} \& Fig. \ref{fig:30} for different values of 
NED and GB parameters. The $P-v$ diagram is depicted in 
Fig. \ref{fig:31} for different values of NED parameters. The isotherm 
undergoes liquid-gas like phase transition and inflection point is 
present at critical temperature $T_{H}=T_c$. For $T>T_c$ no phase 
transition occurs. 

\subsection{Black Holes in 4D EGB Massless gravity coupled to NED}\label{sec:4.2}

The pressure of the black hole in $4D$ EGB massless gravity coupled to NED can be obtained from Hawking temperature \eqref{eq:3.4} by taking the massless limit $m \to 0$ 
\begin{equation*}
P=\frac{1}{4 \pi v^3 (4 \pi  k^{2} +  v^{2})} \Bigr[ 4 \pi  T_{H}  v^{4}-2 v^{3}+\Bigl( +16 \pi  T_{H}  k^{2}+32 \pi  T_{H} \alpha \Bigl) v^{2}
\end{equation*}
\begin{equation}\label{eq:4.11}
+\Bigl( -8k^{2}+8 Q_{m}^{2}+8 \alpha \Bigl) v +128 \pi  T_{H} \alpha  k^{2} \Bigr] +\frac{8 \alpha  k^{2}}{\pi  v^{4} (4 k^{2}+v^{2})},
\end{equation}
where we take $v=2r_+$. At critical points
\begin{equation}\label{eq:4.12}
    \biggl( \frac{\partial P}{\partial v}\biggl)_{T_c, v_c} = 0 = \biggl( \frac{\partial^{2} P}{\partial v^{2}} \biggl)_{T_c, v_c}.
\end{equation}
Therefore, using above equation and equation \eqref{eq:4.11} one can obtain equations for critical volume, critical temperature and pressure as

\begin{equation*}
\Bigr[ - v_{c}^{10}+\Bigl( -12 k^{2} +24 Q_{m}^{2}+48 \alpha \Bigl) v_{c}^{8} 
+\Bigl( -48 k^{4}+( 48 Q_{m}^{2}+576 \alpha ) k^{2}+192 \alpha  (Q_{m}^{2}+\alpha )\Bigl) v_{c}^{6}
\end{equation*}
\begin{equation}\label{eq:4.13}
-64 \Bigl( k^{4}-( Q_{m}^{2}+36 \alpha ) k^{2}+18 \alpha  Q_{m}^{2}-36 \alpha^{2}\Bigl) k^{2} v_{c}^{4} +1536 \alpha  k^{4} \Bigl( 2 k^{2}-Q_{m}^{2}+6 \alpha \Bigl)v_{c}^{2}+12288 \alpha^{2} k^{6} \Bigr]=0,
\end{equation}

\begin{equation}\label{eq:4.14}
T_{c} = \frac{1}{64 \pi  v_{c} (k^{2}+\frac{v_{c}^{2}}{4})^{2} (v_{c}^{2}+24 \alpha )} \biggr[ 4 v_{c}^{6}+\Bigl( 32 k^{2}-32 Q_{m}^{2}-32 \alpha \Bigl) v_{c}^{4}
+64 k^{2} \Bigl( k^{2}-Q_{m}^{2}-4 \alpha \Bigl) v_{c}^{2}-512 \alpha  k^{4}\biggr],
\end{equation}

\begin{equation*}
P_{c} = \frac{1}{32 v_{c}^{4} (v_{c}^{2}+24 \alpha ) (k^{2}+\frac{v_{c}^{2}}{4})^{2} \pi}\biggr[  v_{c}^{8}+\Bigl( 8 k^{2} -12 Q_{m}^{2}-20 \alpha \Bigl) v_{c}^{6}
+ \Bigl( 16 k^{4}+( -16 Q_{m}^{2}-160 \alpha ) k^{2}-32 \alpha  Q_{m}^{2}
\end{equation*}
\begin{equation}\label{eq:4.15}
-32 \alpha^{2}\Bigl) v_{c}^{4}-128 \alpha  k^{2} \Bigl( ({5}/{2}) k^{2}-Q_{m}^{2}+2 \alpha \Bigl) v_{c}^{2}-512 \alpha^{2} k^{4}\biggr].
\end{equation}

\begin{table}[H]
\centering
\begin{tabular}{ |p{1.5cm}|p{1.5cm}|p{1.5cm}|p{1.5cm}|p{1.5cm}| } 
 \hline
 $\beta$ & ${v_c}$ & $P_{c}$ & $T_{c}$ & $\rho_{c}$   \\ [0.5ex]  \hline
 \hline
 0.0 & 5.9092 & 0.0021 & 0.0343 & 0.3617 \\ \hline
 0.2 & 5.3453 & 0.0024 & 0.0361 & 0.3553 \\ \hline
 0.4 & 5.1026 & 0.0025 & 0.0369 & 0.3457 \\ \hline
 0.6 & 4.9163 & 0.0026 & 0.0376 & 0.3399 \\ \hline
 0.8 & 4.7614 & 0.0027 & 0.0382 & 0.3365 \\ \hline
 1.0 & 4.6285 & 0.0028 & 0.0387 & 0.3348 \\  [1ex]
 \hline
\end{tabular}
\caption{Values of critical volume($T_{c}$), critical pressure($P_{c}$), 
critical temperature($T_{c}$) and $\rho_{c}=P_{c}v_{c}/T_{c}$ for different 
GB coupling parameter with $Q=1$, $\alpha=0.2$, $m=0$.}
\label{table:5}
\end{table}

\begin{table}[H]
\centering
\begin{tabular}{ |p{1.5cm}|p{1.5cm}|p{1.5cm}|p{1.5cm}|p{1.5cm}| } 
\hline
$\alpha$ & ${v_c}$ & $P_{c}$ & $T_{c}$ & $\rho_{c}$   \\ [0.5ex]  \hline
\hline
0.00 & 3.6916 & 0.0046 & 0.0503 & 0.3421 \\ \hline 
0.01 & 3.8223 & 0.0042 & 0.0489 & 0.3282 \\  \hline 
0.05 & 4.2508 & 0.0036 & 0.0448 & 0.3415 \\  \hline
0.10 & 4.6735 &0.0031 & 0.0411 & 0.3525 \\ \hline
0.20 & 5.3453 & 0.0029 & 0.0361 & 0.3553 \\   [1ex]    
\hline
\end{tabular}
\caption{Values of critical volume($T_{c}$), critical pressure($P_{c}$), critical temperature($T_{c}$) 
and $\rho_{c}=P_{c}v_{c}/T_{c}$ for different GB coupling parameter with $Q=1$, $\beta=0.2$ and $m=0.0$.}
\label{table:6}
\end{table}

Similar to subsection \ref{sec:4.1}, we numerically solve Eqs. 
\eqref{eq:4.13}, \eqref{eq:4.14} and \eqref{eq:4.15} to obtain critical 
points. The critical points are estimated in tables \ref{table:5} and 
\ref{table:6} for different values of NED parameter $\beta$ and 
GB coupling parameter $\alpha$. In table \ref{table:5}, 
critical parameters are shown for different values of $\beta$. As the 
NED parameter rises, the critical volume and $\rho_c$ decrease, while at 
the same time, the critical pressure and temperature increase. In table 
\ref{table:6}, we illustrate the effects of the GB parameter 
$\alpha$ on the critical parameters, while the NED parameter held constant. 
As the GB parameter rises, the critical volume and $\rho_c$ 
increase, while at the same time, the critical pressure and temperature 
decrease.

\begin{figure}[H]
\centering
\subfloat[$\beta=0.4$]{\includegraphics[width=.5\textwidth]{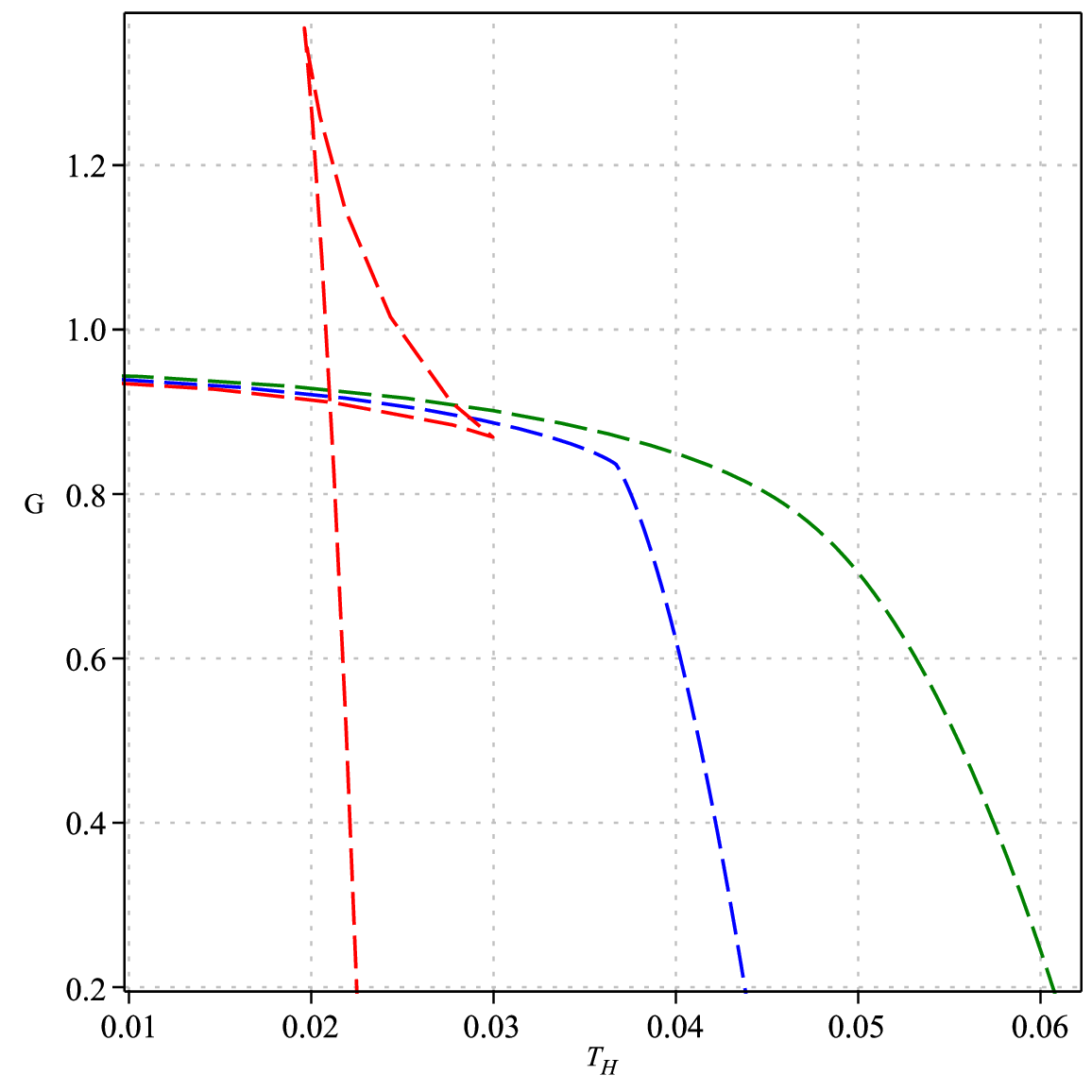}}\hfill
\subfloat[$\beta=0.8$]{\includegraphics[width=.5\textwidth]{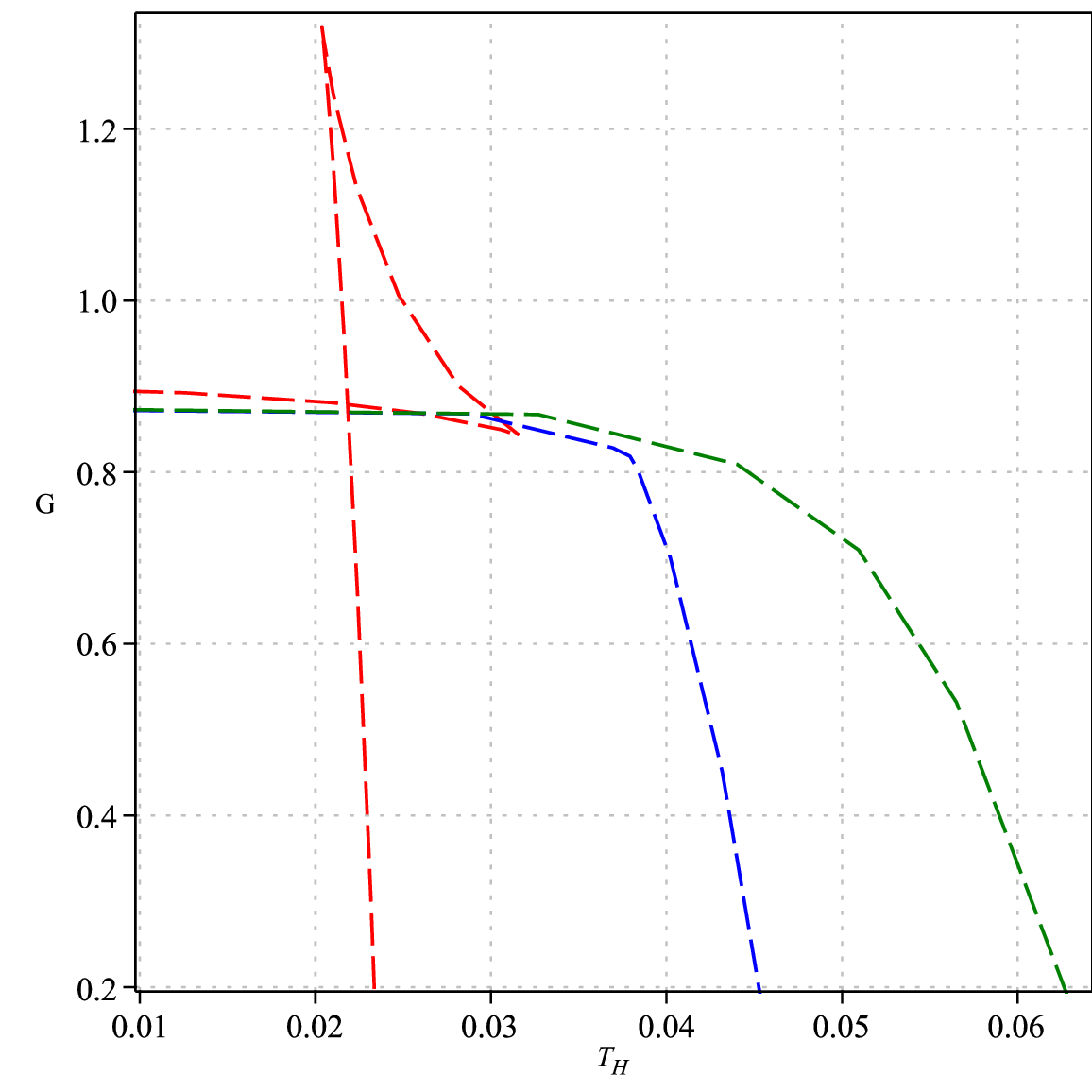}}\hfill
\caption{Red dash line is denoted by $P=0.25P_c$, blue dash line is denoted by $P=P_c$ and green dash line is 
denoted by $P=2P_c$ with $Q_m=1$, $\alpha=0.2$ and $m=0.0$.}\label{fig:32}
\end{figure} 

\begin{figure}[H]
\centering
\subfloat[$\alpha=0.01$]{\includegraphics[width=.5\textwidth]{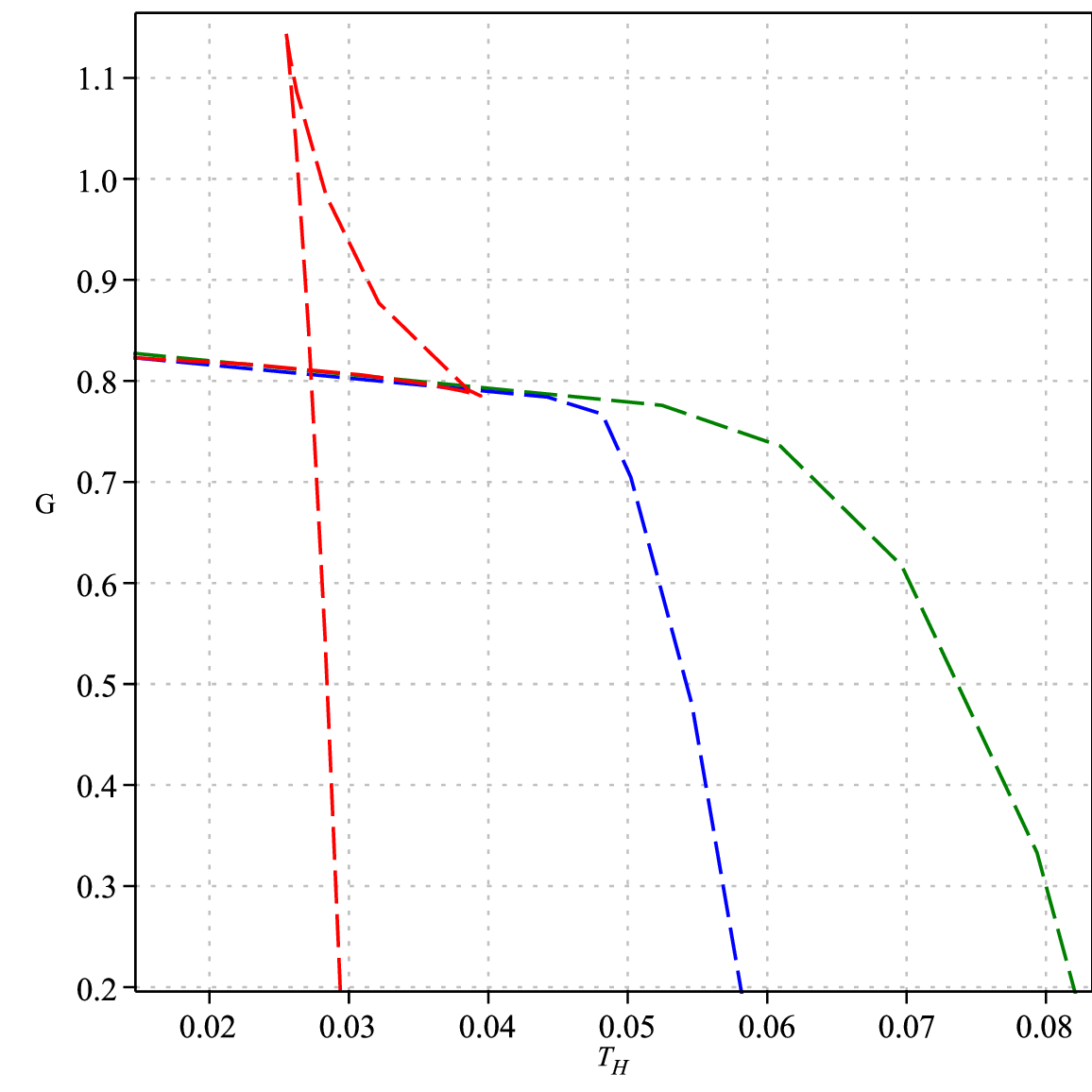}}\hfill
\subfloat[$\alpha=0.1$]{\includegraphics[width=.5\textwidth]{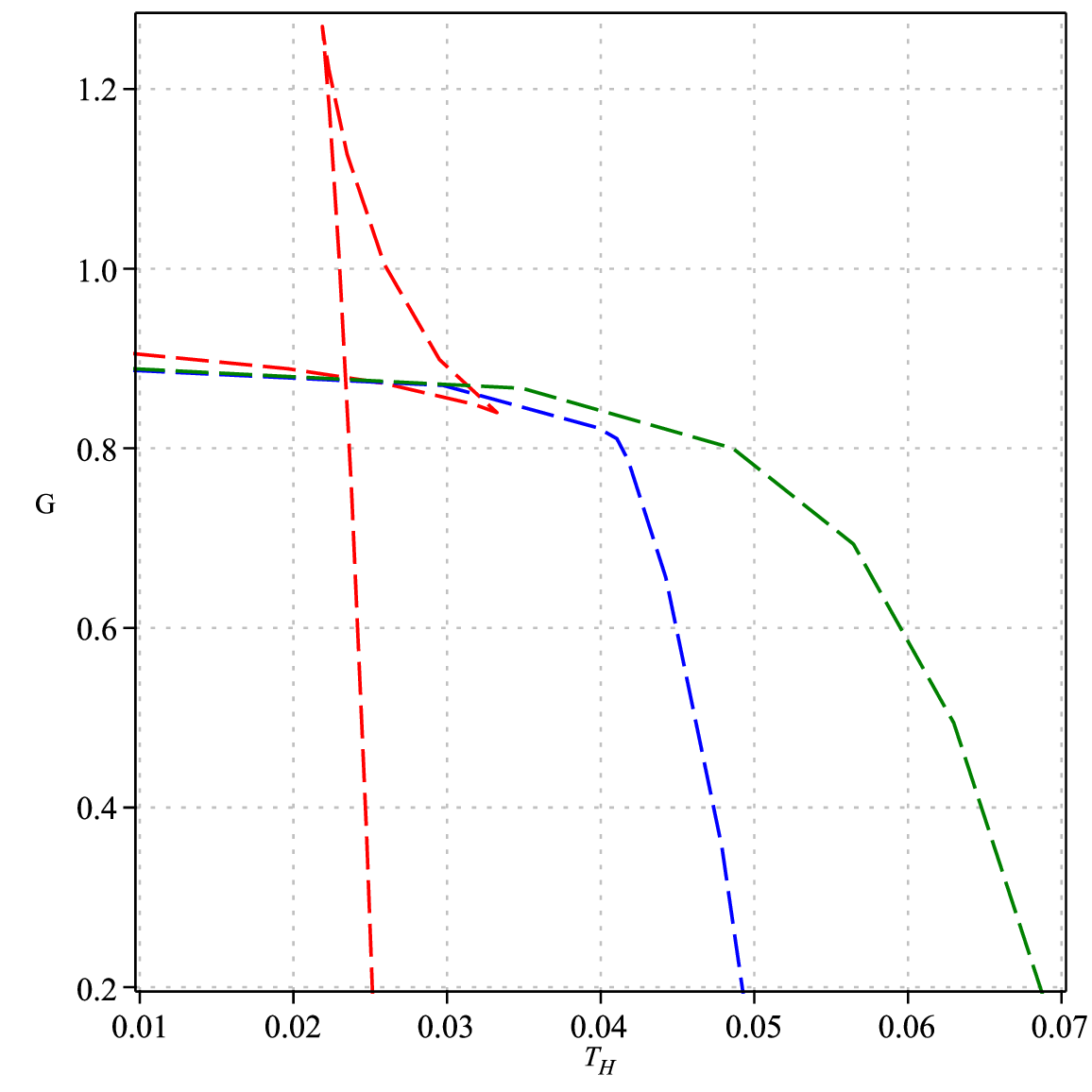}}\hfill
\caption{ Red dash line is denoted by $P=0.25P_c$, blue dash line is denoted by $P=P_c$ and green dash line is 
denoted by $P=2P_c$ with $Q_m=1$, $\beta=0.2$ and $m=0.0$.}\label{fig:33}
\end{figure} 
\pagebreak 

The $G-T_{H}$ diagram is shown for EGB massless gravity, by varying two key parameters: the 
NED parameter $\beta$ and the GB coupling parameter. This diagram is illustrated in Fig. 
\ref{fig:32} and \ref{fig:33}. In Fig. \ref{fig:32}, $G-T_{H}$ diagram is plotted for different values 
of NED parameter $\beta$. When the pressure ($P$) remains below a critical value denoted as $P_c$, the 
Gibbs free energy curve exhibits a characteristic swallowtail or triangular pattern. This behaviour 
signifies a first-order phase transition of the black holes. When $P<P_c$, a transition between two 
distinct types of black holes: the smaller ones (\textbf{SBH}) and the larger ones (\textbf{LBH}). As 
we move from the point $T_{H}=0$ towards the intersection point with the red dashed curve, a \textbf{SBH} 
branch is observed. Beyond this intersection point, the \textbf{LBH} branch becomes more favoured. This 
infers that at the intersection, a transition between the two types of black holes occurs. A discontinuity 
in the black hole's horizon radius occurs, specifically at the intersection point for $P<P_c$. This 
variation in horizon radius consequently leads to a discontinuity in the black hole's entropy at the 
same intersection point. This outcome emerges due to the dependency of entropy on the horizon radius. 
However, as the pressure reaches the critical point $P=P_c$, the swallowtail pattern disappears, 
signifying the absence of the phase transition. When the pressure surpasses $P_c$, there are no further 
phase transitions. A similar behaviour is displayed in Fig. \ref{fig:33} but with different values of the 
GB parameters.

\begin{figure}[H]
\centering
\subfloat[$\beta=1.0$]{\includegraphics[width=.5\textwidth]{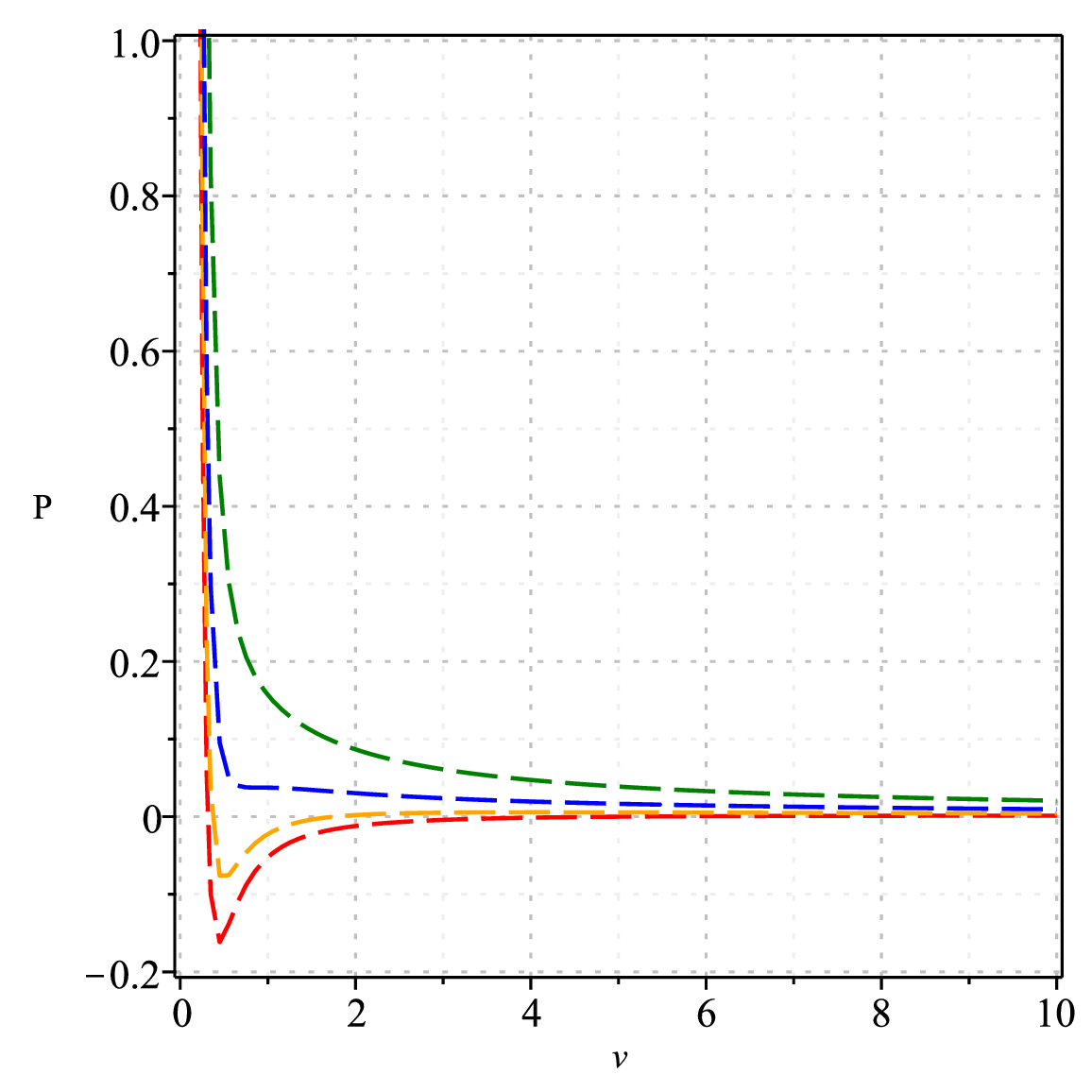}}\hfill
\subfloat[$\beta=1.5$]{\includegraphics[width=.5\textwidth]{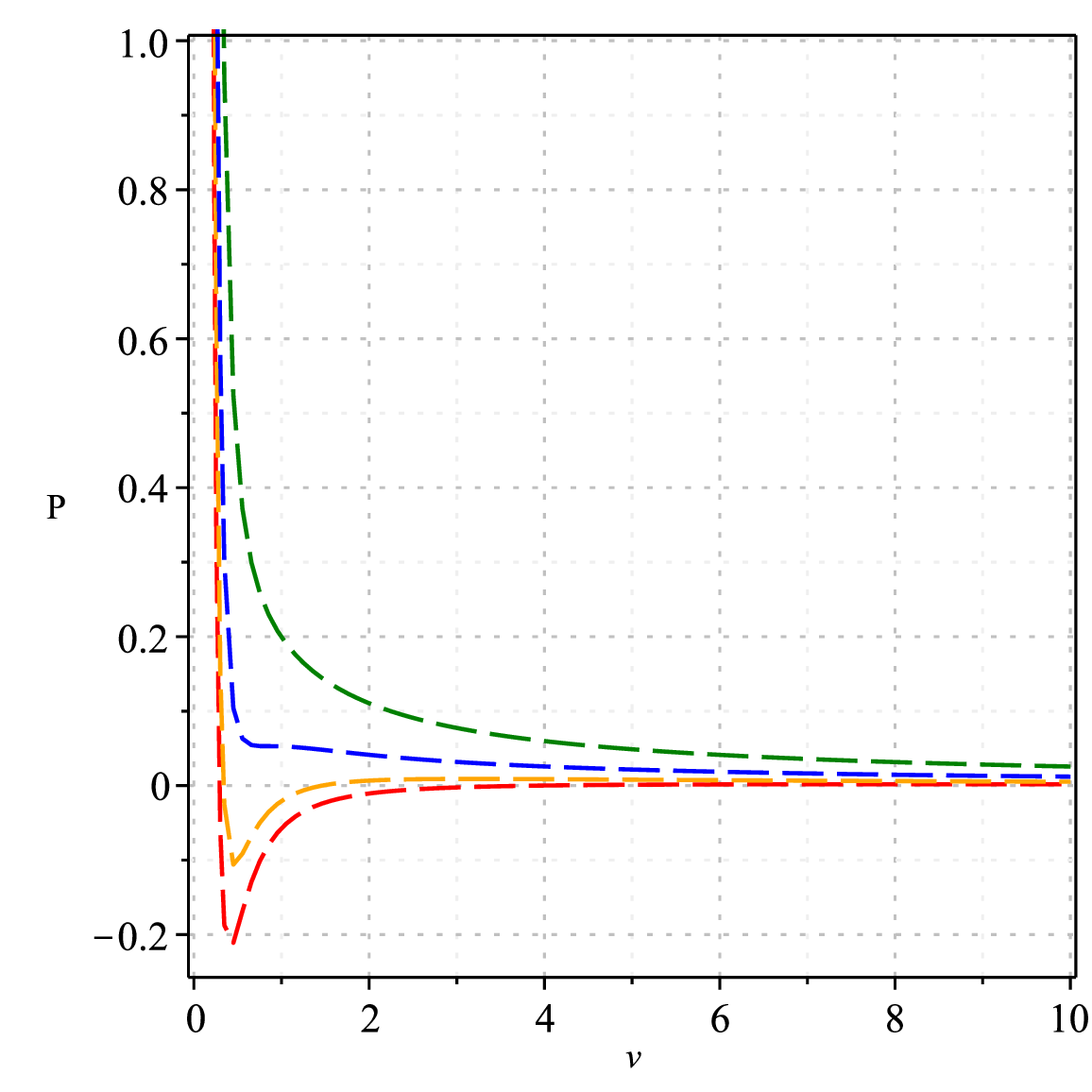}}\hfill
\caption{Red dash line is denoted by $T_{H}=0.25T_c$, orange dash line is 
denoted by $T_{H}=0.5T_c$, blue dash line is denoted by $T_{H}=T_c$ 
and green dash line is denoted by $T_{H}=2T_c$ with $Q_m=1$, $\alpha=0.01$ and $m=0.0$.}\label{fig:34}
\end{figure}

Additionally, a $P--v$ diagram is featured in Fig. \ref{fig:34} for different values of the NED parameter. 
Here, the isotherm showcases a liquid-gas type phase transition and is characterized by an inflection point occurring at the critical temperature $T_{H}=T_c$. It is crucial to note that for temperatures exceeding $T_c$, no phase transition takes place. This implies a clear distinction between the behaviour above and below the critical temperature.

\subsection{Black Holes in 4D Massive Einstein gravity coupled to NED}\label{sec:4.3}
The pressure of the black hole in $4D$ Einstein massive gravity coupled to NED can be obtained from Hawking temperature \eqref{eq:3.5} as
\begin{equation*}
P=\frac{1}{16 \pi  k^{2} v^{3}+4 \pi  v^{5}} \Bigr[ (-c c_{1} m^{2}+4 \pi  T_{H} ) v^{4}+(-2 c^{2} c_{2} m^{2}-2) v^{3}+ (-4 c c_{1} m^{2}+16 \pi  T_{H} ) k^{2} v^{2}
\end{equation*}
\begin{equation}\label{eq:4.16}
+\Bigl((-8 c^{2} c_{2} m^{2}-8) k^{2}+8 Q_{m}^{2} \Bigl) v  \Bigr],
\end{equation}
where we take $v=2r_+$. At critical points
\begin{equation}\label{eq:4.17}
    \biggl( \frac{\partial P}{\partial v}\biggl)_{T_c, v_c} = 0 = \biggl( \frac{\partial^{2} P}{\partial v^{2}} \biggl)_{T_c, v_c}.
\end{equation}
Therefore, using above equation and equation \eqref{eq:4.13} one can obtain equations for critical volume, critical temperature and pressure as

\begin{equation*}
\Bigr[ -(c^{2} c_{2} m^{2}+1) v_{c}^{6}+\Bigl( -12( c^{2} c_{2} m^{2}+1) k^{2} +24 Q_{m}^{2} \Bigl) v_{c}^{4} 
\end{equation*}
\begin{equation}\label{eq:4.18}
+48\Bigl( -( c^{2} c_{2} m^{2}+1) k^{4}+ Q_{m}^{2} k^{2} \Bigl) v_{c}^{2}-64 \Bigl( (c^{2} c_{2} m^{2}+1) k^{4} -Q_{m}^{2} k^{2}\Bigl) k^{2}  \Bigr]=0,
\end{equation}

\begin{equation*}
T_{c} = \frac{1}{64 \pi  v_{c}^3 (k^{2}+\frac{v_{c}^{2}}{4})^{2} } \biggr[ c c_{1} m^{2} v_{c}^{7}+(4 c^{2} c_{2} m^{2}+4) v_{c}^{6}+8 c c_{1} k^{2} m^{2} v_{c}^{5}+ 32 \Bigl( ( c^{2} c_{2} m^{2}+1) k^{2}- Q_{m}^{2} \Bigl) v_{c}^{4}
\end{equation*}
\begin{equation}\label{eq:4.19}
+16 c c_{1} k^{4} m^{2} v_{c}^{3}+64 k^{2} \Bigl( (c^{2} c_{2} m^{2}+1) k^{2}-Q_{m}^{2} \Bigl) v_{c}^{2}\biggr],
\end{equation}

\begin{equation}\label{eq:4.20}
P_{c} = \frac{1}{32 \pi v_{c}^{6} (k^{2}+\frac{v_{c}^{2}}{4})^{2} }\biggr[ (c^{2} c_{2} m^{2}+1) v_{c}^{8}+\Bigl( (8 c^{2} c_{2} m^{2}+8) k^{2} -12 Q_{m}^{2} \Bigl) v_{c}^{6} +16 \Bigl( ( c^{2} c_{2} m^{2}+1) k^{4} - Q_{m}^{2}  k^{2} \Bigl) v_{c}^{4}\biggr].
\end{equation}

In the massless limit, $m \to 0$ above equations are reduced to critical temperature and pressure of $4D$ massless GR coupled to NED \cite{kruglov2022nonlinearly}

\begin{equation}\label{eq:4.21}
  \Bigl( v_{c}^{2} +4k^2 \Bigl)^3 - 8 Q_m^2 \Bigl( 3v_c^4 +6k^2v_c^2+8k^4 \Bigl) =0,  
\end{equation}
\begin{equation}
T_{c} = \frac{1}{ \pi  v_{c}  }  - \frac{8 Q_m^2 \bigl( v_c^2 +2k^2\bigl)}{\pi v_c (4 k^{2}+{v_{c}^{2}})^{2}}, 
\end{equation}
\begin{equation}
P_{c} = \frac{1}{2 \pi  v_{c}^2  }  - \frac{2 Q_m^2 \bigl( 3v_c^2 +4k^2\bigl)}{\pi v_c^2 ({4} k^{2}+{v_{c}^{2}})^{2}}. 
\end{equation}

\begin{table}[H]
\centering
\begin{tabular}{ |p{1.5cm}|p{1.5cm}|p{1.5cm}|p{1.5cm}|p{1.5cm}| } 
 \hline
 $m$ & ${v_c}$ & $P_{c}$ & $T_{c}$ & $\rho_{c}$   \\ [0.5ex]  \hline
 \hline
 0.0 & 4.1346 & 0.0041 & 0.0476 & 0.3561 \\ \hline
 0.2 & 4.0173 & 0.0044 & 0.0475 & 0.3721 \\ \hline
 0.4 & 3.6936 & 0.0056 & 0.0479 & 0.4318 \\ \hline
 0.6 & 3.2222 & 0.0084 & 0.0504 & 0.5370 \\ \hline
 0.8 & 2.6330 & 0.0136 & 0.0587 & 0.6100 \\ \hline
 0.9 & 2.2622 & 0.0182 & 0.0673 & 0.6117 \\  [1ex]
 \hline
\end{tabular}
\caption{Values of critical volume($T_{c}$), critical pressure($P_{c}$), 
critical temperature($T_{c}$) and $\rho_{c}=P_{c}v_{c}/T_{c}$ for different 
GB coupling parameter with $Q=1$, $\alpha = 0.0$, $\beta=0.1$, 
$c=1$ $c_{1}=-1$ and $c_{2}=1$.}
\label{table:7}
\end{table}

\begin{table}[H]
\centering
\begin{tabular}{ |p{1.5cm}|p{1.5cm}|p{1.5cm}|p{1.5cm}|p{1.5cm}| } 
\hline
$\beta$ & ${v_c}$ & $P_{c}$ & $T_{c}$ & $\rho_{c}$   \\ [0.5ex]  \hline
\hline
0.0 & 5.8496 & 0.0023 & 0.0358 & 0.3758 \\ \hline 
0.1 & 5.1066 & 0.0040 & 0.0456 & 0.4479 \\  \hline
0.2 & 4.7125 & 0.0052 & 0.0514 & 0.4767 \\  \hline
0.3 & 4.3416 & 0.0066 & 0.0574 & 0.4992 \\  \hline
0.4 & 3.9387 & 0.0085 & 0.0645 & 0.5190 \\  \hline
0.5 & 3.4013 & 0.0120 & 0.0757 & 0.5391 \\  [1ex]
\hline
\end{tabular}
\caption{Values of critical volume($T_{c}$), critical pressure($P_{c}$), 
critical temperature($T_{c}$) and $\rho_{c}=P_{c}v_{c}/T_{c}$ for different
 GB coupling parameter with $Q=1.2$, $\alpha=0.0$, $m=0.1$, $c=1$, $c_1=-1$ and 
 $c_2=1$.}
\label{table:8}
\end{table}

Similar to subsection \ref{sec:4.1} and \ref{sec:4.2}, we numerically 
solved equation \eqref{eq:4.18} for critical volume and estimated critical 
points in table \ref{table:7} and \ref{table:8}. The effects of graviton 
mass on the critical parameters are shown in table \ref{table:7}. As mass of the 
graviton increases from zero critical volume decreases. Similarly, critical 
pressure, temperature and $\rho_c$ increase as graviton mass increases from 
zero. The effects of NED parameter $\beta$ (table \ref{table:8}) is similar 
as table \ref{table:7}.

\begin{figure}[H]
\centering
\subfloat[$m=0.4$]{\includegraphics[width=.5\textwidth]{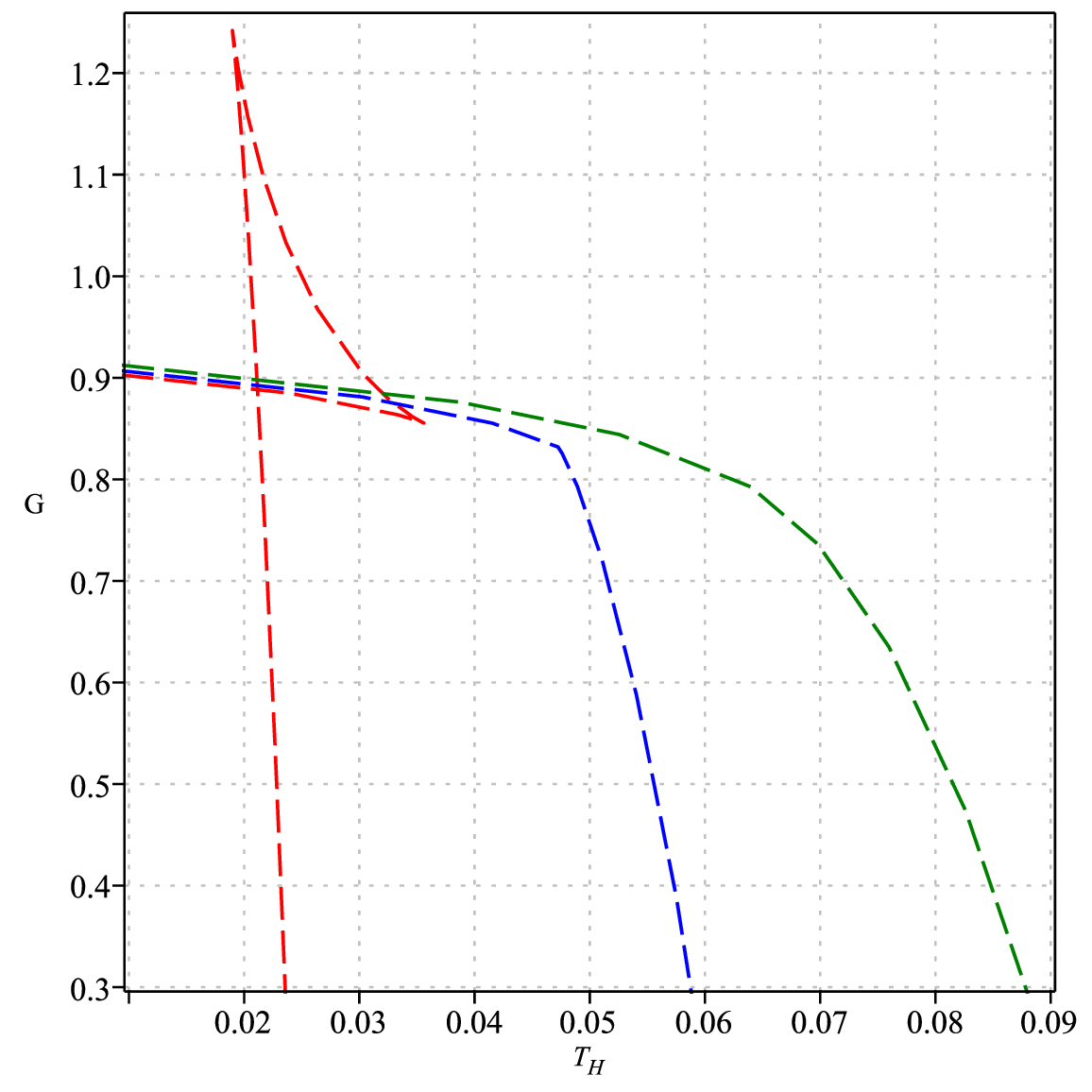}}\hfill
\subfloat[$m=0.6$]{\includegraphics[width=.5\textwidth]{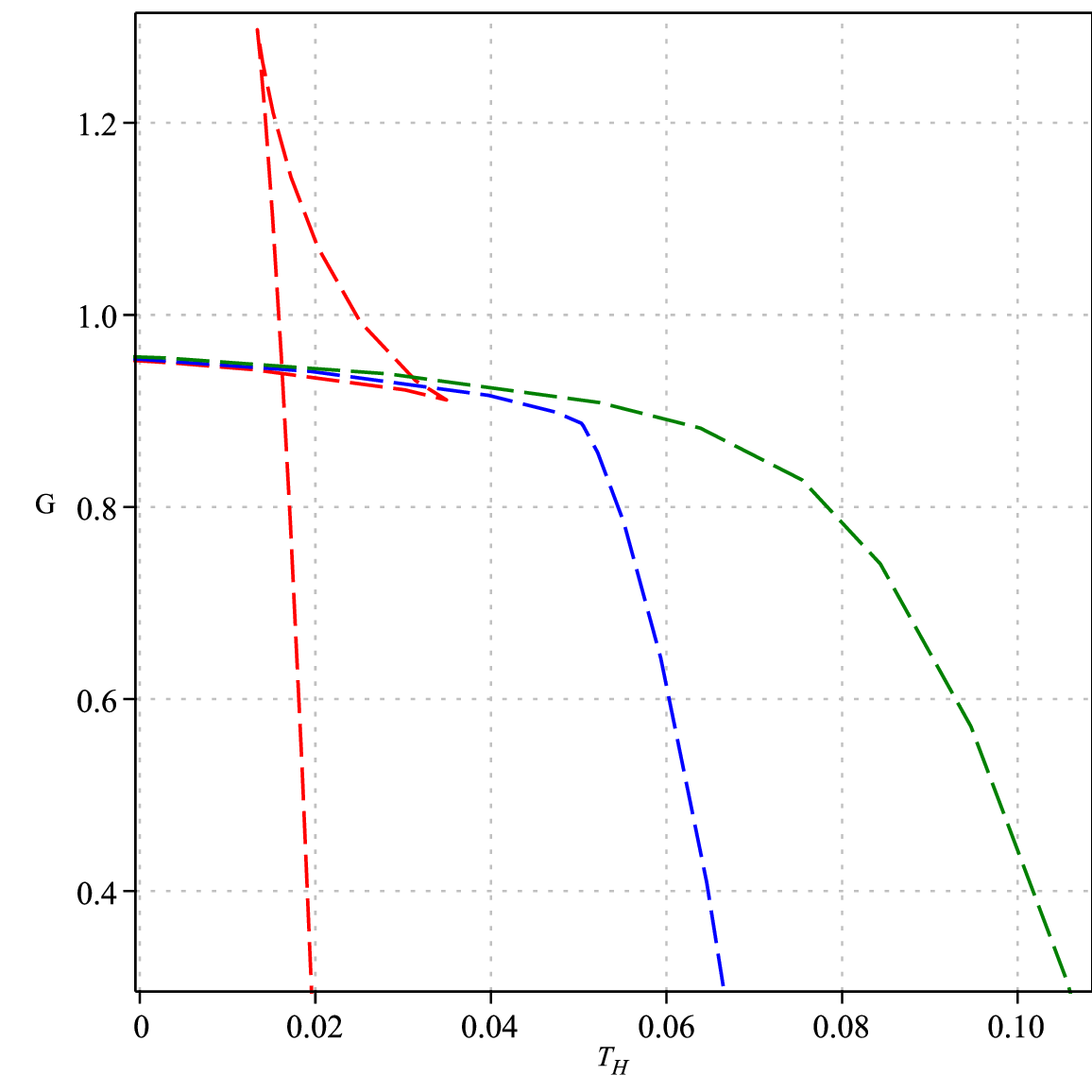}}\hfill
\caption{Red dash line is denoted by $P=0.25P_c$, blue dash line is denoted by $P=P_c$ and green dash line is 
denoted by $P=2P_c$ with $Q_m=1$, $\beta=0.1$, $c=1$, $c_1=-1$ and $c_2=1$.}\label{fig:35}
\end{figure}

\begin{figure}[H]
\centering
\subfloat[$\beta=0.1$]{\includegraphics[width=.5\textwidth]{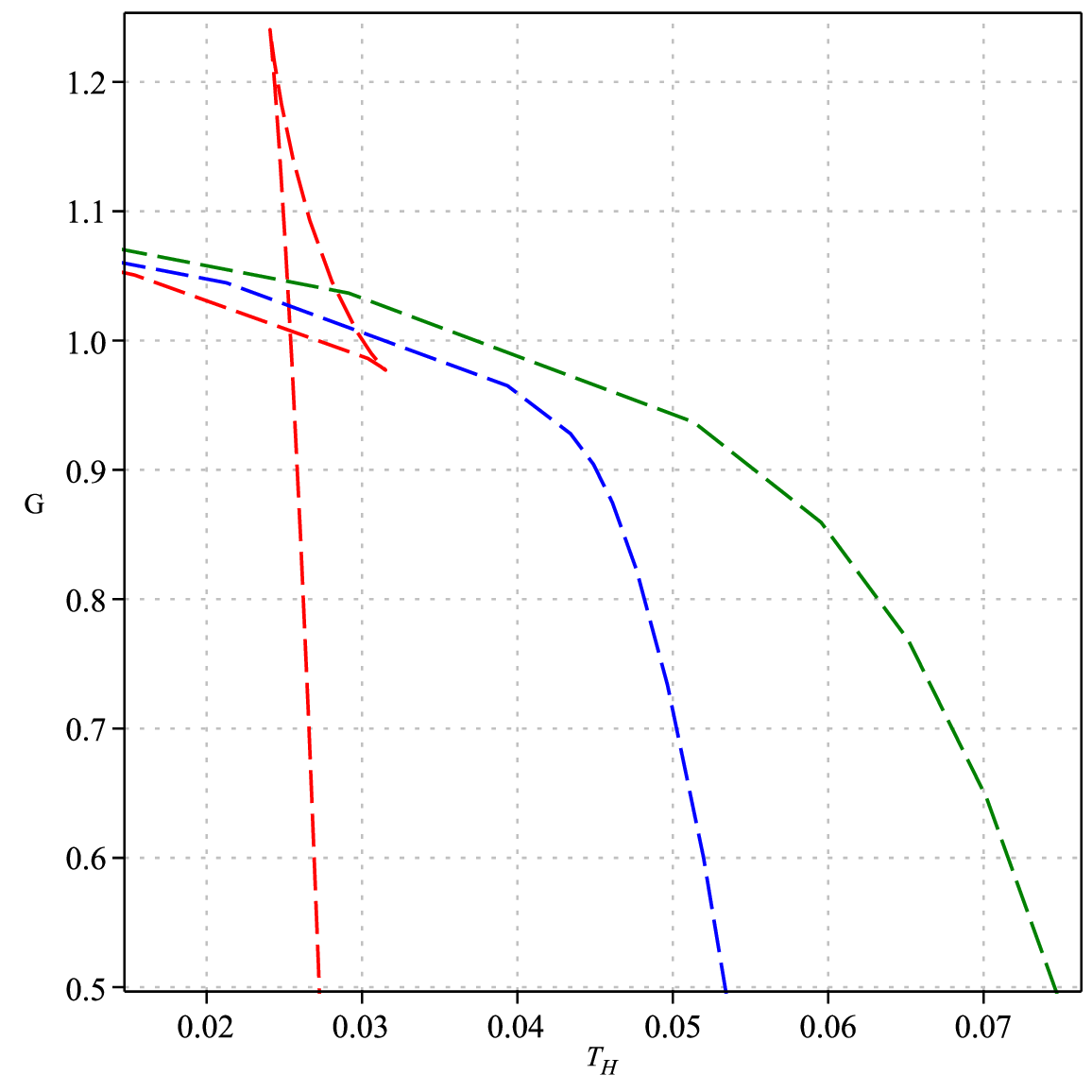}}\hfill
\subfloat[$\beta=0.4$]{\includegraphics[width=.5\textwidth]{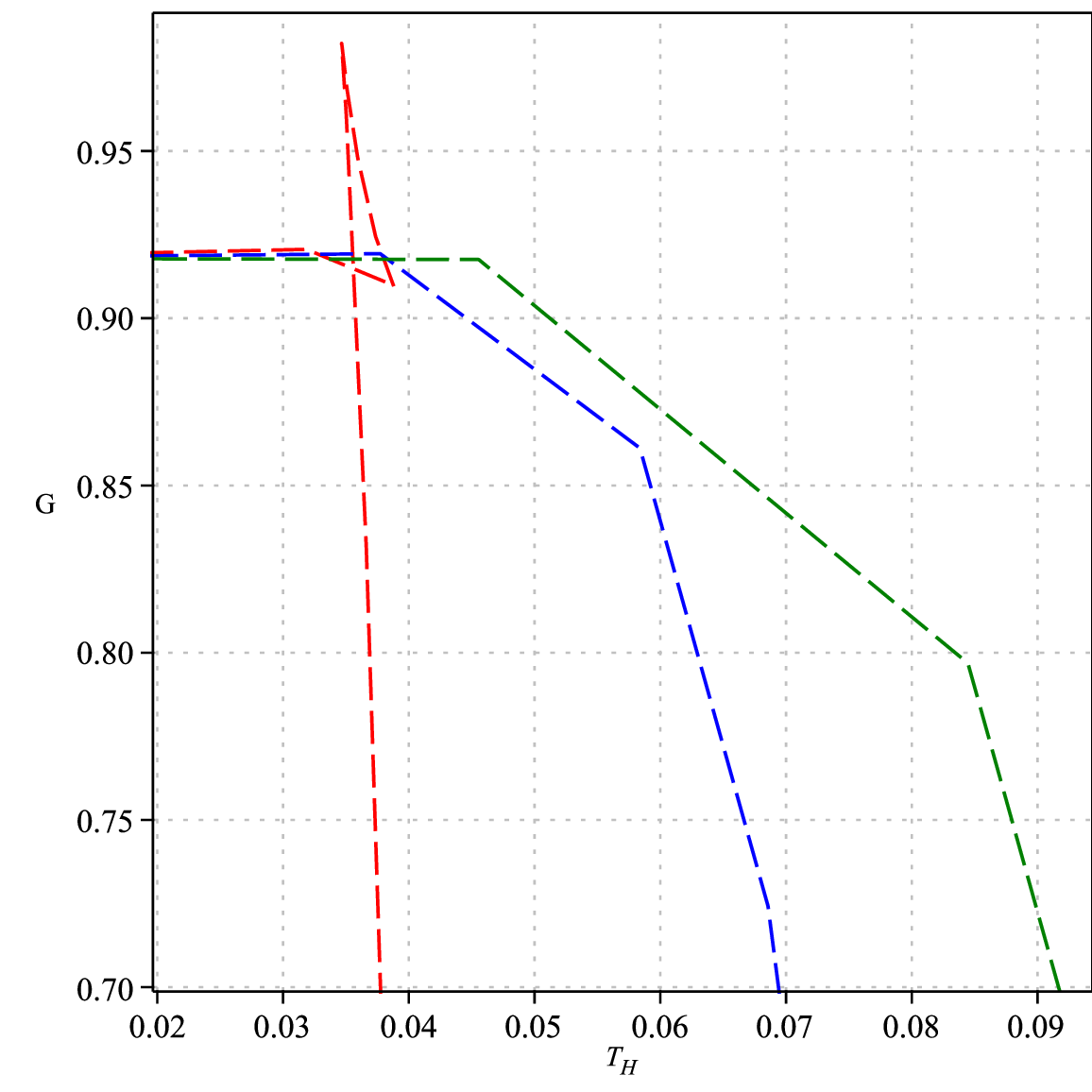}}\hfill
\caption{ Red dash line is denoted by $P=0.25P_c$, blue dash line is denoted by $P=P_c$ and green dash line 
is denoted by $P=2P_c$ with $Q_m=1.2$, $m=0.1$, $c=1$, $c_1=-1$ and $c_2=1$.}\label{fig:36}
\end{figure}

\begin{figure}[H]
\centering
\subfloat[$m=0.4$]{\includegraphics[width=.5\textwidth]{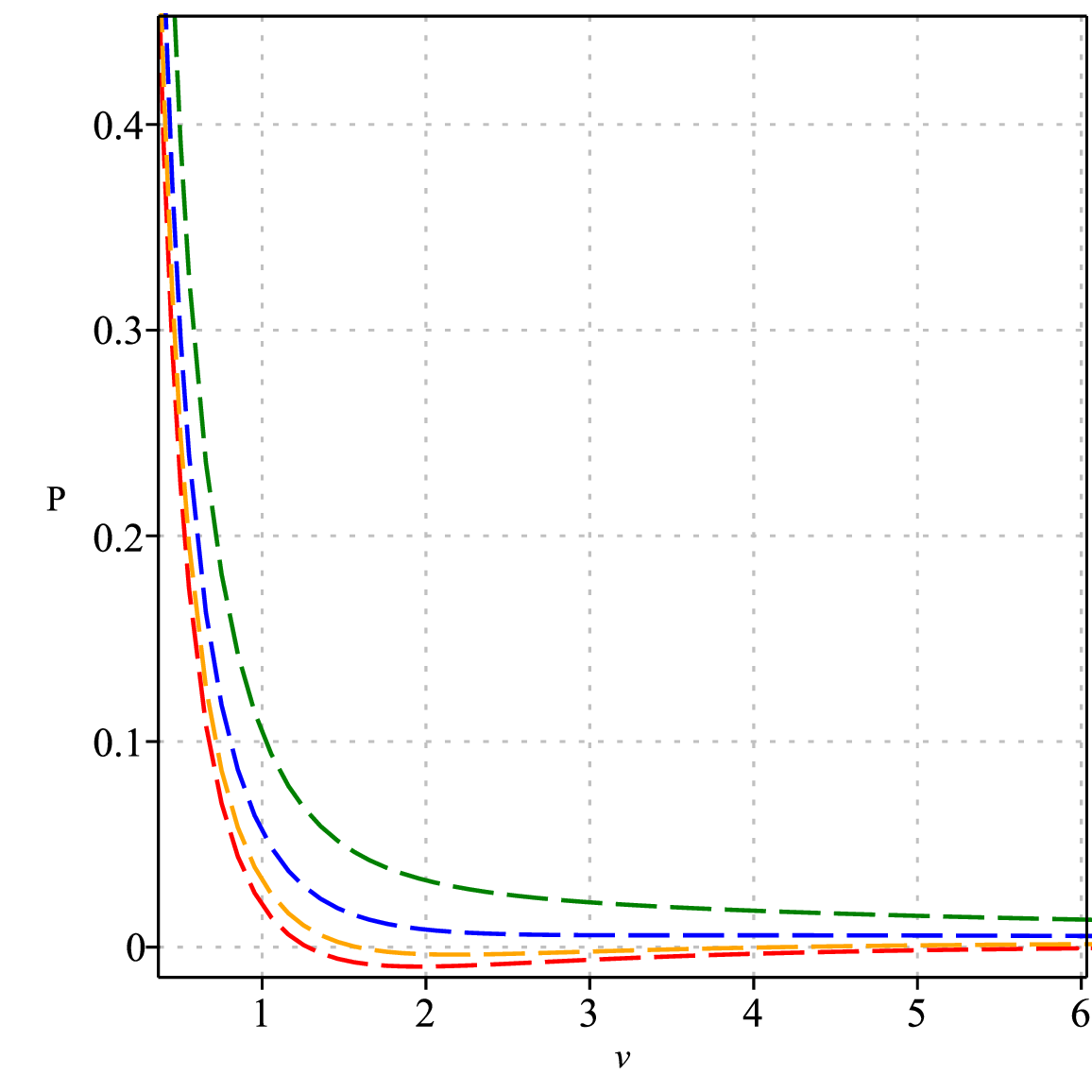}}\hfill
\subfloat[$m=0.8$]{\includegraphics[width=.5\textwidth]{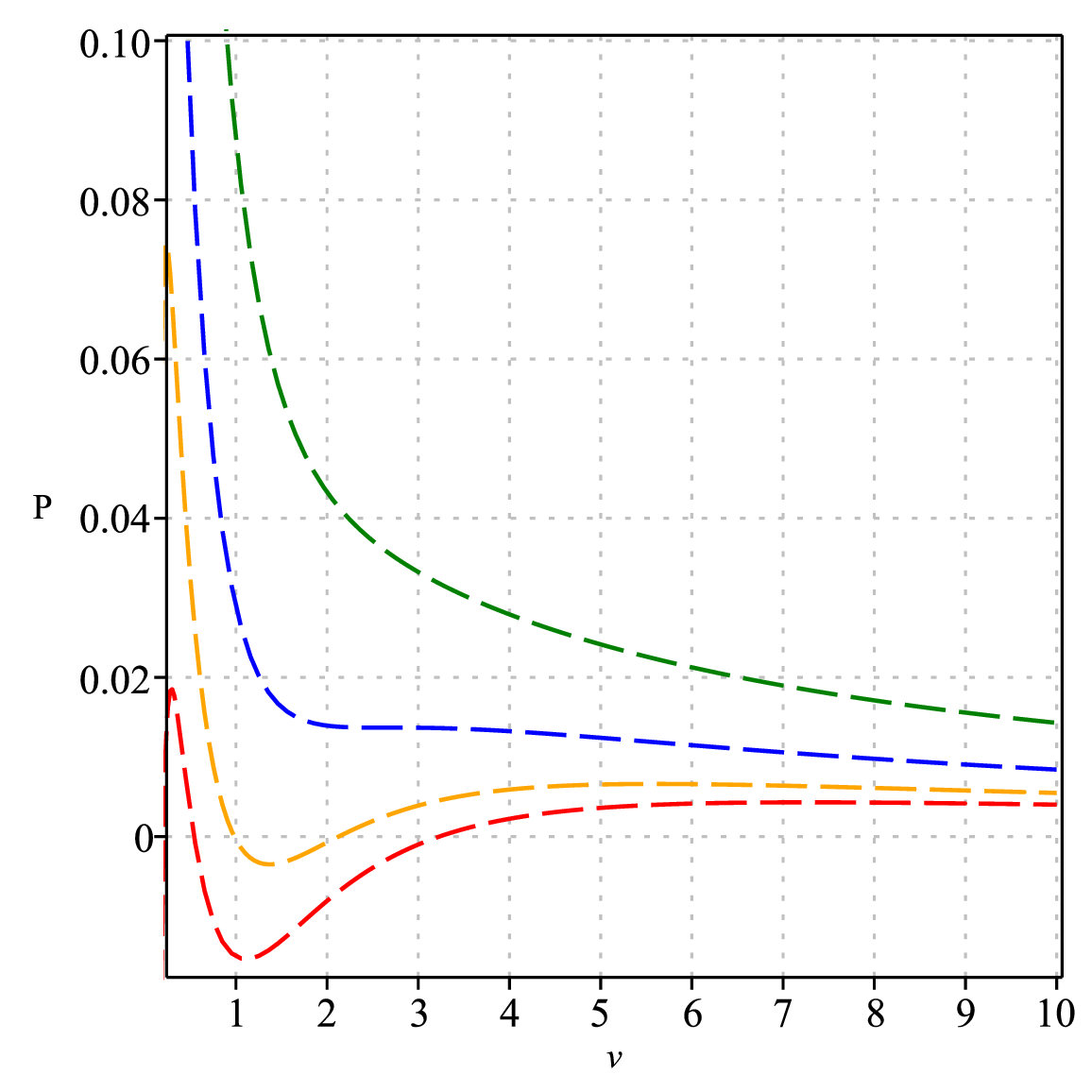}}\hfill
\caption{Red dash line is denoted by $T=0.25T_c$, orange dash line is denoted by $T=0.5T_c$, blue dash line is 
denoted by $T=T_c$ and green dash line is denoted by $T=2T_c$ with $Q_m=1$, $\beta=0.1$, $c=1$, $c_1=-1$ and $c_2=1$.}\label{fig:37}
\end{figure}

The $G-T_{H}$ diagram Fig. \ref{fig:35} \& Fig. \ref{fig:36} show swallow 
tail behaviour for $P<P_c$, i.e. a first-order phase transition occurs for 
such black hole. A \textbf{SBH} is preferred from $T_{H}=0$ to the 
intersection point for the red dash curve and \textbf{LBH} is preferred 
beyond the point of intersection. For $P=P_c$ and $P>P_c$ Swallowtail-like 
behaviour disappears. The $P-v$ diagram in Fig. \ref{fig:37} shows a 
liquid-gas-like phase transition.

\section{Reentrant Phase Transitions}\label{sec:RPT}
Black holes are the thermodynamic system that exhibits a rich phase structure  
(van der Waals-like and/or reentrant phase transitions) in $AdS$ background. 
A reentrant phase transition is defined as when a system undergoes more than one 
phase transition, such that thermodynamic variables of the system change 
monotonically but the initial and final state of the system is the same. This 
type of phenomenon was first observed in a nicotine/water mixture thermodynamic 
system. In this mixture, if we increase the temperature for a fixed percentage 
of nicotine then the system exhibits a reentrant phase transition \cite{hudson1904gegenseitige}. The initial 
and final state is a homogenous mixture of nicotine/water, with a distinct
nicotine/water intermediate state. This type of phase transition 
is observed for different physical systems \cite{narayanan1994reentrant}.

In black hole thermodynamics, RPT was first observed for black holes in $4D$
Einstein gravity coupled to BI electrodynamics \cite{Gunasekaran:2012dq}.  
For this black hole a \textbf{LBH}--\textbf{IBH}--\textbf{LBH} phase transition 
observed. But for higher dimensional in Einstein gravity coupled to BI 
electrodynamics RPT was completely absent. Surprisingly single and multi-spinning
black holes in higher dimension $(d\ge6$) anti-de Sitter/de Sitter space show RPT 
\cite{altamirano2013reentrant,Altamirano:2013uqa,Altamirano:2014tva,Kubiznak:2015bya}.
The RPT of black holes in higher-order theories of gravity 
studied in Refs. \cite{Frassino:2014pha,Wei:2014hba,Hennigar:2015wxa,Zhang:2020obn}.

\subsection{Black Holes in \texorpdfstring{$4D$}{TEXT} Einstein
 gravity coupled to NED}\label{sec:RPT1}

In this subsection, we study the RPT of black holes in $4D$ Einstein gravity 
coupled to NED. The equation for critical radius is \ref{eq:4.21}
\begin{equation}\label{eq:1}
\Bigl( v_{c}^{2} +4k^2 \Bigl)^3 - 8 Q_m^2 \Bigl( 3v_c^4 +6k^2v_c^2+8k^4 \Bigl) =0. 
\end{equation}
Putting $x=v_c^2+4k^2$ into the above equation we obtain
\begin{equation}\label{eq:2}
    x^3-24Q_m^2x^2+144Q^2k^2x-256Q_m^2k^4=0.
\end{equation}
In order to satisfy $v_c \geq 0$, we must have
\begin{equation}\label{eq:3}
\lvert x \rvert \geq 4k^2, 
\quad\text{or}\quad
\lvert x \rvert \geq 8\sqrt{\beta}Q_m.
\end{equation}

Next, we will find the solutions of equation \eqref{eq:2} in terms of 
trigonometric functions. Three real roots of equation \eqref{eq:2} occurs 
when the discriminant is
\begin{equation}\label{eq:4}
\Delta= 442368Q_m^4k^4(Q_m^2-k^2)(5Q_m^2-4k^2) < 0.
\end{equation}
When $\Delta > 0$ only one root is real and for $\Delta=0$ the equation 
\eqref{eq:2} has either one or two real solutions. From condition $\Delta < 0$ 
we obtain
\begin{equation}\label{eq:5}
    \frac{Q_m}{2}=\sqrt{\beta_0} < \sqrt{\beta} < \sqrt{\beta_2}= \frac{5Q_m}{8}.
\end{equation}
To find the solutions of equation \eqref{eq:2}, we will use the Tschirnhaus 
transformation method. Putting $x=t+B$ into equation \eqref{eq:2}

\begin{equation}\label{eq:6}
    t^3 +pt +q=0,
\end{equation}
where we set coefficients of $t^2$ equal to zero \& $B=8Q_m^2$. 
Finally the solutions of equations \eqref{eq:2} is 
\begin{equation}\label{eq:7}
    x_{j}=2 \sqrt{\frac{-p}{3}} \cos{\Biggr[\frac{1}{3} \arccos{\biggl(\frac{3q}{2p}\sqrt{\frac{-3}{p} } \biggl)}-\frac{2\pi j}{3} \Biggr]},
\end{equation}
where $j=0$, $1$ \& $2$. The condition in equation \eqref{eq:3} was satisfied 
for $x_0$ and $x_1$ only, $x_2$ does not satisfy condition \eqref{eq:3}. 
Therefore we have two real critical points. The constants $p$ and $q$ are given by
\begin{align}\label{eq:8}
p &=  3B^2 -48BQ_m^2 +144Q_m^2 k^2, \\ 
q &=  B^3 -24B^2Q_m^2 +144 B Q_m^2 k^2 -256 Q_m^2 k^4.
\end{align}
$\beta < \beta_{0}$ admits only one real critical point. For $\beta > \beta_2$ 
no critical points occur. Finally, the critical radius $v_c$ can be written as
\begin{equation}\label{eq:10}
v_c=\sqrt{x-4k^2} \text{,}  \:    x=
x_0 \; \& \; x_1, \: \text{where} \;  \beta_0  < \beta < \beta_2,
\end{equation}
\begin{equation}\label{eq:11}
T_c = \frac{-8 Q_{m}^{2} (2k^{2}+ v_{c}^{2})+(4 k^{2}+v_{c}^2)^{2}}{\pi  (4 k^{2}+v_{c}^{2})^{2} v_{c}},
\end{equation}
\begin{equation}\label{eq:12}
P_c = \frac{-4 Q_{m}^{2} (4k^{2}+3  v_{c}^{2})+(4 k^{2}+v_{c}^2)^{2}}{2 (4 k^{2}+v_{c}^{2})^{2} \pi  v_{c}^{2}}.
\end{equation}
For two critical pressures to be positive, we must have
\begin{equation}\label{eq:13}
 \sqrt{\beta} > \sqrt{\beta_1} = \frac{9Q_m}{16}.
\end{equation}

\begin{figure}[H]
\centering
\subfloat[ Two critical points at positive pressures.]{\includegraphics[width=.5\textwidth]{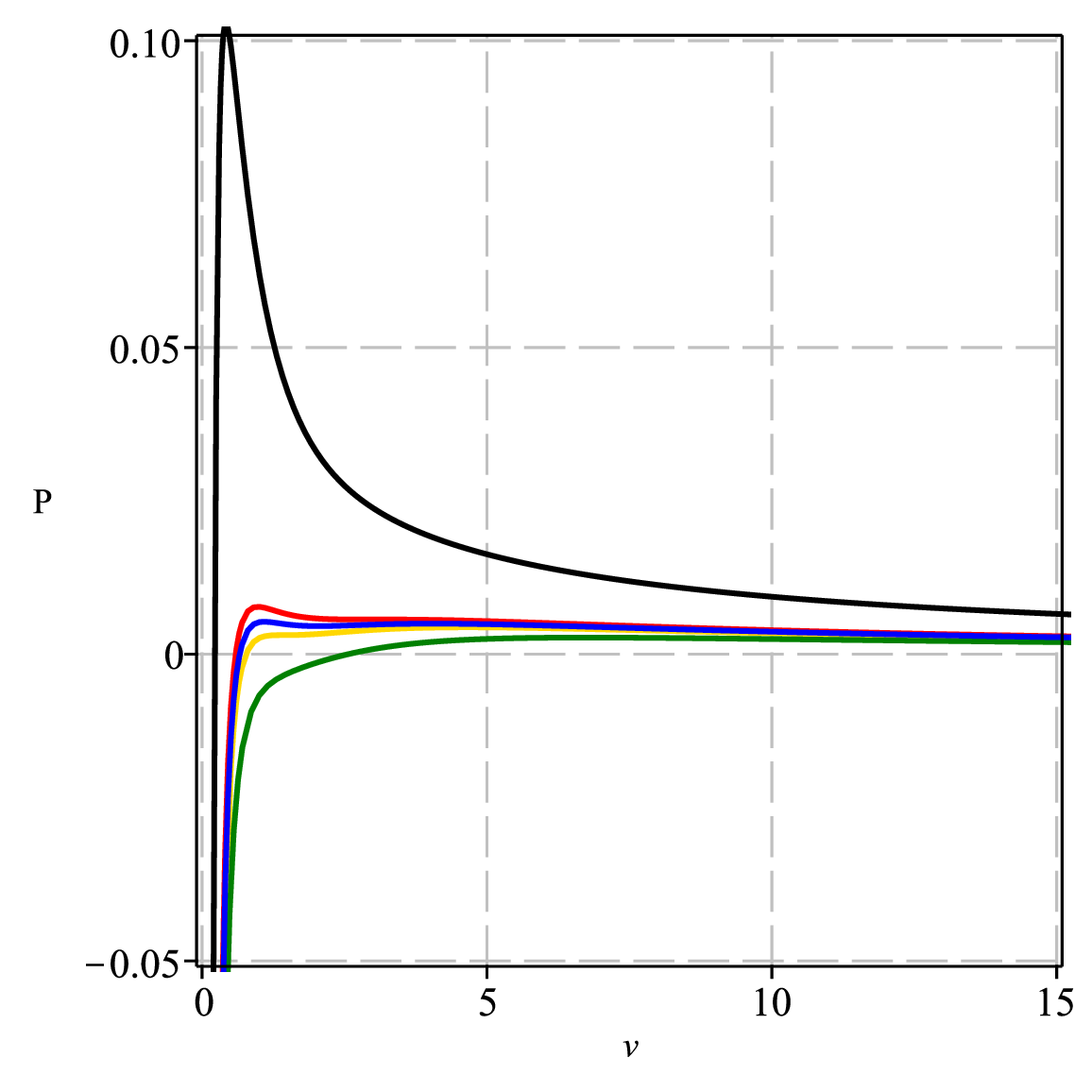}}\hfill
\subfloat[ Two critical points, one at positive pressure, the
other at negative pressure.]{\includegraphics[width=.5\textwidth]{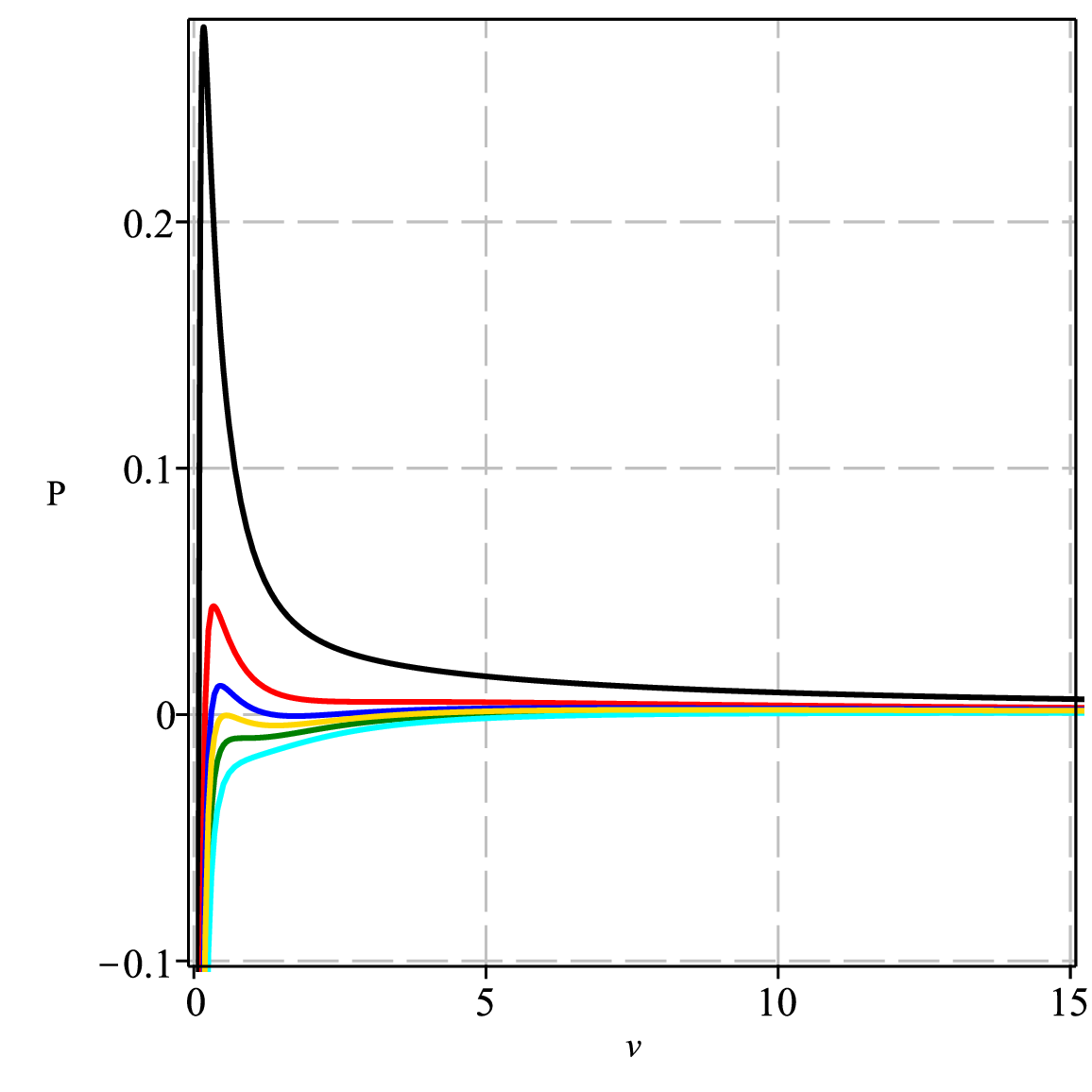}}\hfill
\caption{Left Pannel : Black line denotes $T=2T_{cp2}$,  Red line denotes $T=T_{cp2}$, blue line denotes $T_{cp1}<T=0.0520<T_{cp2}$, Gold line denotes $T=T_{cp1}$ \& green line denotes $T=0.0400<T_{cp1}$. Right Pannel : Black line denotes $T=2T_{cp2}$,  Red line denotes $T=T_{cp2}$, blue line denotes $T_{cp1}<T=0.0400<T_{cp2}$, gold line denotes $T=0.0340>T_{cp1}$, green line denotes $T=T_{cp1}$ \& cyan line denotes $T=0.0200<T_{cp1}$.}\label{fig:RPT1}
\end{figure} 

Excluding the range of $\beta$ from $\beta_0$ to $\beta_1$, we can say that two 
critical points are real and positive with two positive pressures for 
$\beta_1 < \beta <  \beta_2$\footnote{This condition on $\beta$ will not match 
with table 1 of Ref. \cite{kruglov2022nonlinearly} because we chose a different 
form of NED Lagrangian compared to Ref. \cite{kruglov2022nonlinearly}, which is 
different by some constant factor. The condition on $\beta$ for NED Lagrangian in Ref. 
\cite{kruglov2022nonlinearly} see appendices (\ref{appendicex}).}. For $\beta_0 < \beta <  \beta_1$ 
\footnotemark[3] we have two real critical points with one critical pressure 
being negative. When $\beta < \beta_0 $ we have a van der Waals-like phase 
transition, i.e. only one critical point is real and positive.

\begin{figure}[H]
     \centering
     \begin{subfigure}[b]{0.3\textwidth}
         \centering
         \includegraphics[width=\textwidth]{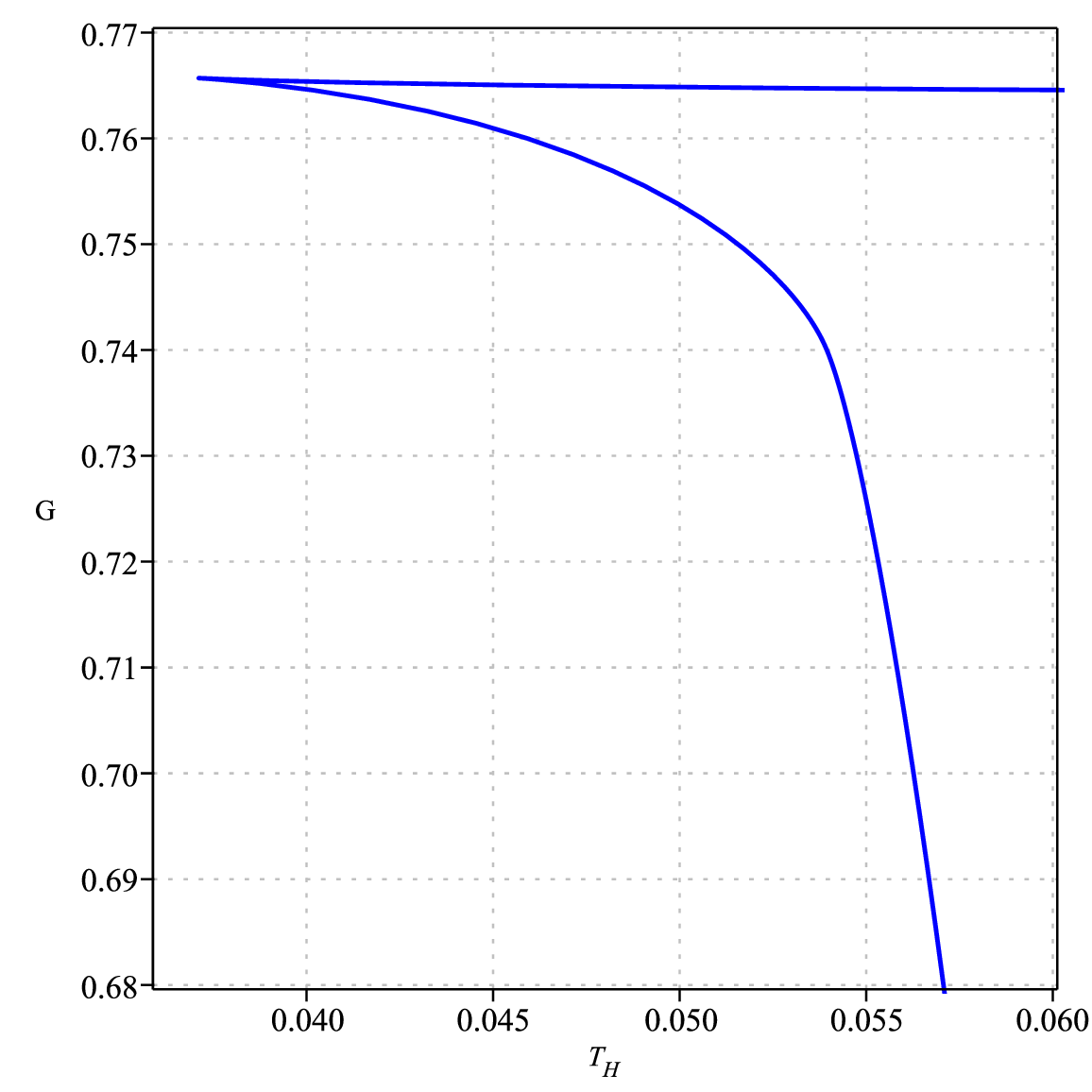}
         \caption{$P=0.0056>P_{cp2}$}
         \label{fig:5.1(a)}
     \end{subfigure}
     \hfill
     \begin{subfigure}[b]{0.3\textwidth}
         \centering
         \includegraphics[width=\textwidth]{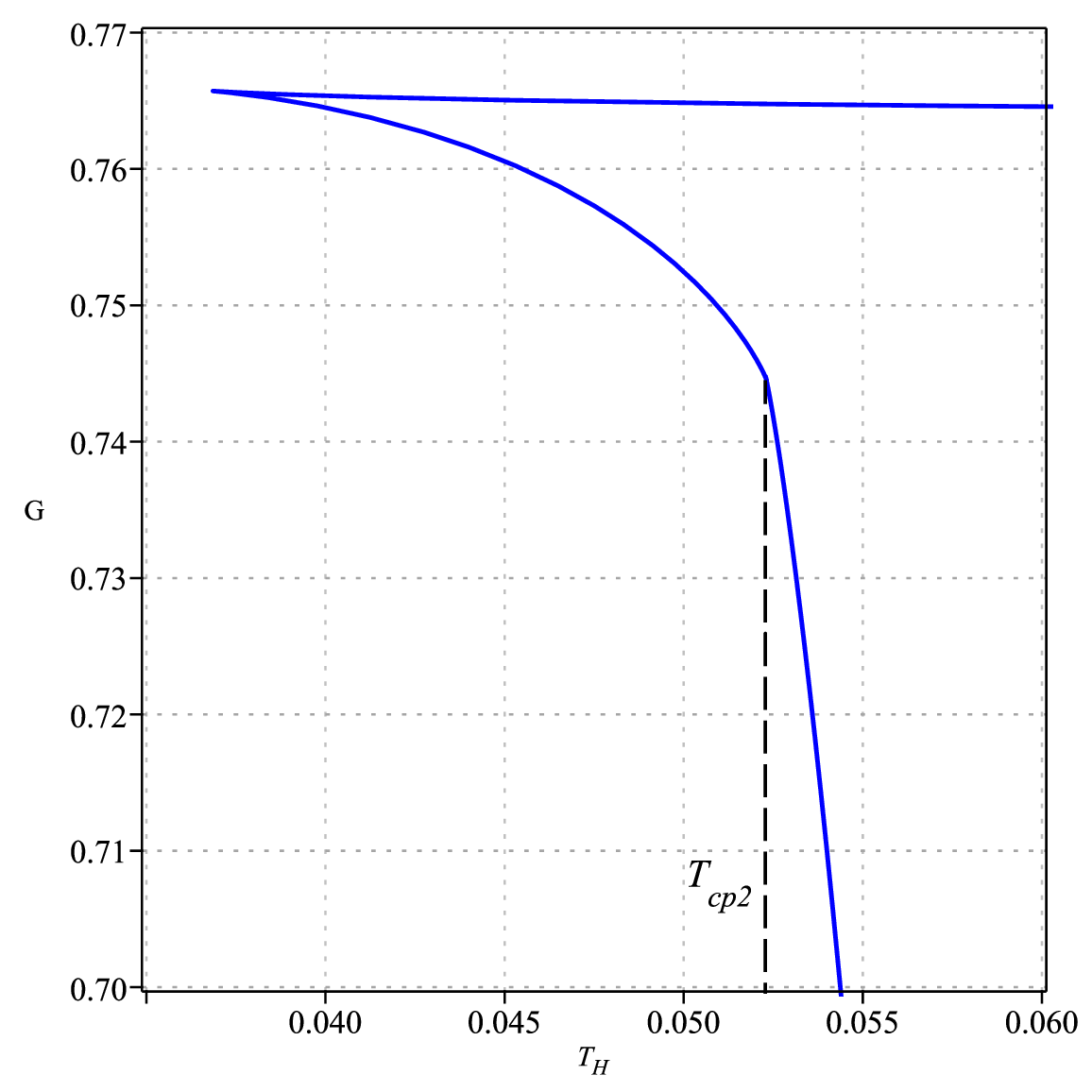}
         \caption{$P=0.0051=P_{cp2}$}
         \label{fig:5.1(b)}
     \end{subfigure}
     \hfill
     \begin{subfigure}[b]{0.3\textwidth}
         \centering
         \includegraphics[width=\textwidth]{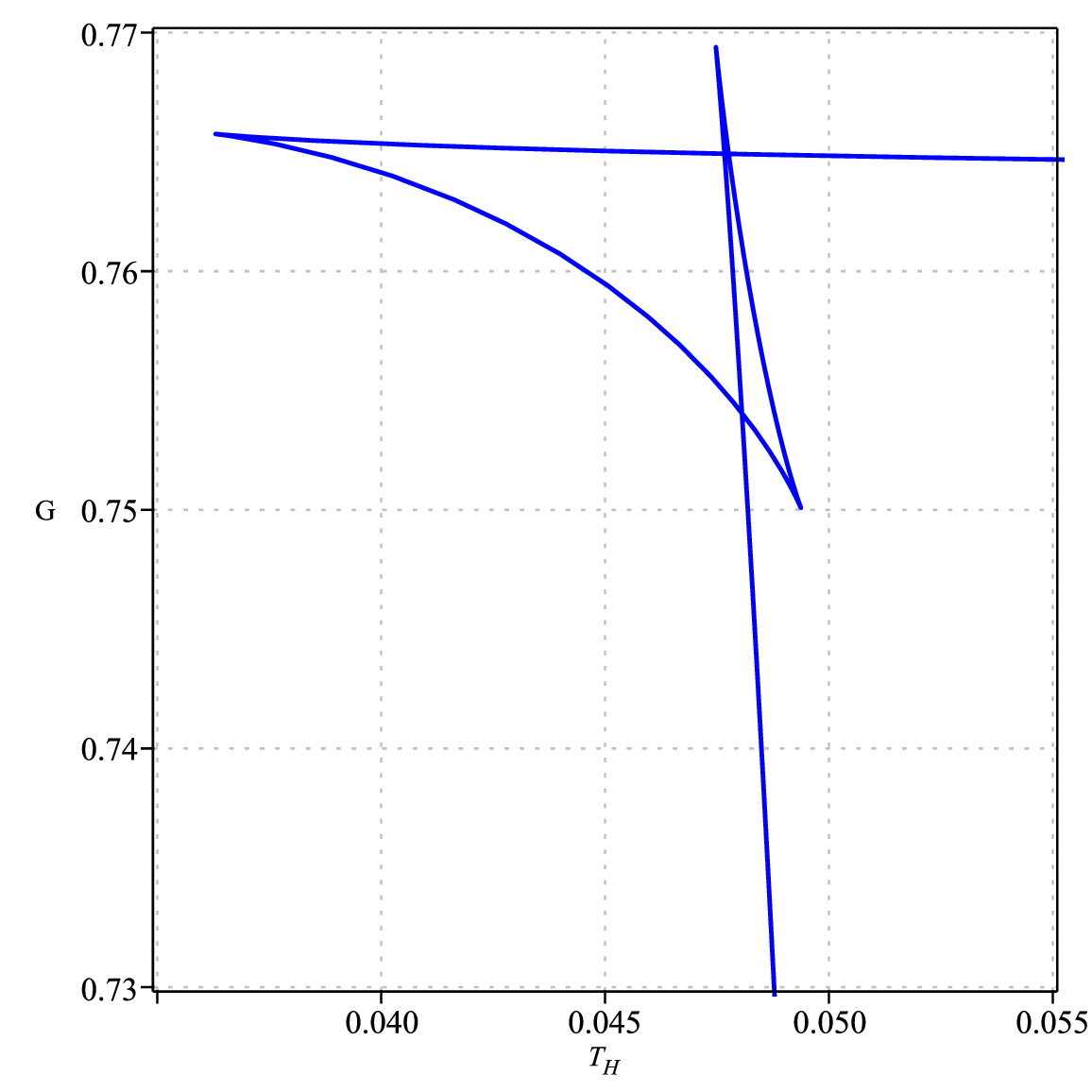}
         \caption{$P=0.0040<P_{cp2}$}
         \label{fig:5.1(c)}
     \end{subfigure}
     \hfill
     \begin{subfigure}[b]{0.3\textwidth}
         \centering
         \includegraphics[width=\textwidth]{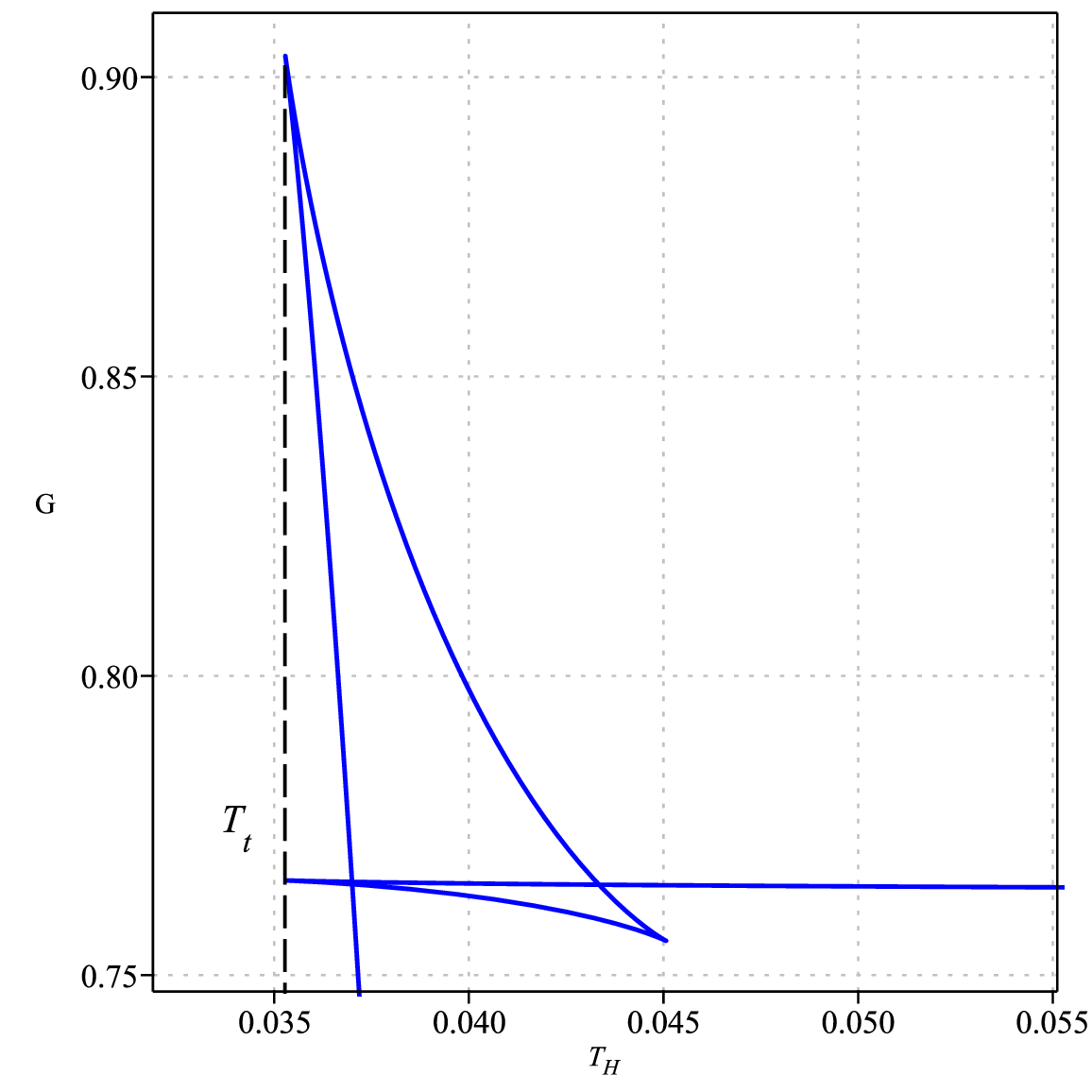}
         \caption{$P=0.00207=P_{t}$}
         \label{fig:5.1(d)}
     \end{subfigure}
     \hfill
     \begin{subfigure}[b]{0.3\textwidth}
         \centering
         \includegraphics[width=\textwidth]{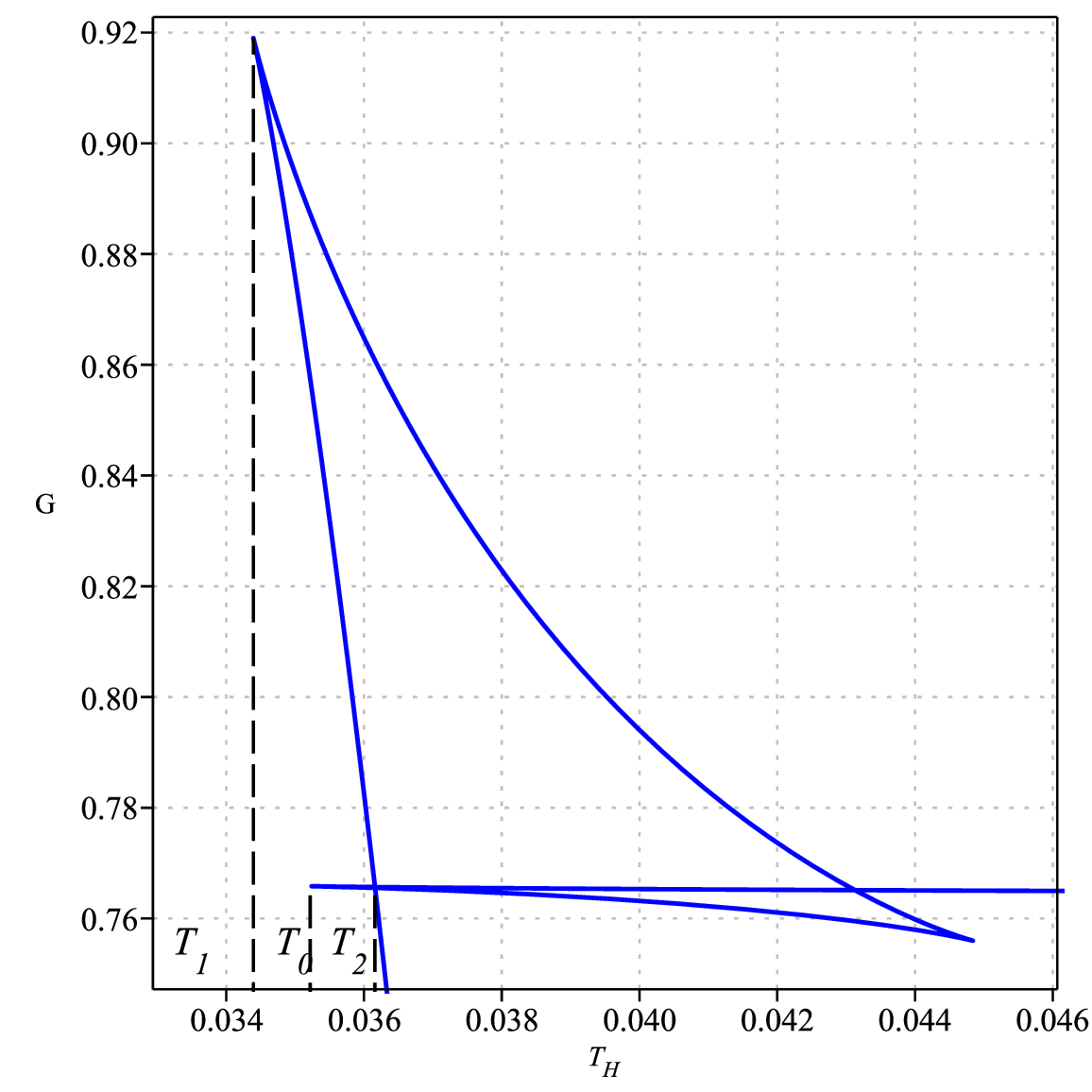}
         \caption{$P_{z}<P=0.00196<P_{t}$}
         \label{fig:5.1(e)}
     \end{subfigure}
     \hfill
     \begin{subfigure}[b]{0.3\textwidth}
         \centering
         \includegraphics[width=\textwidth]{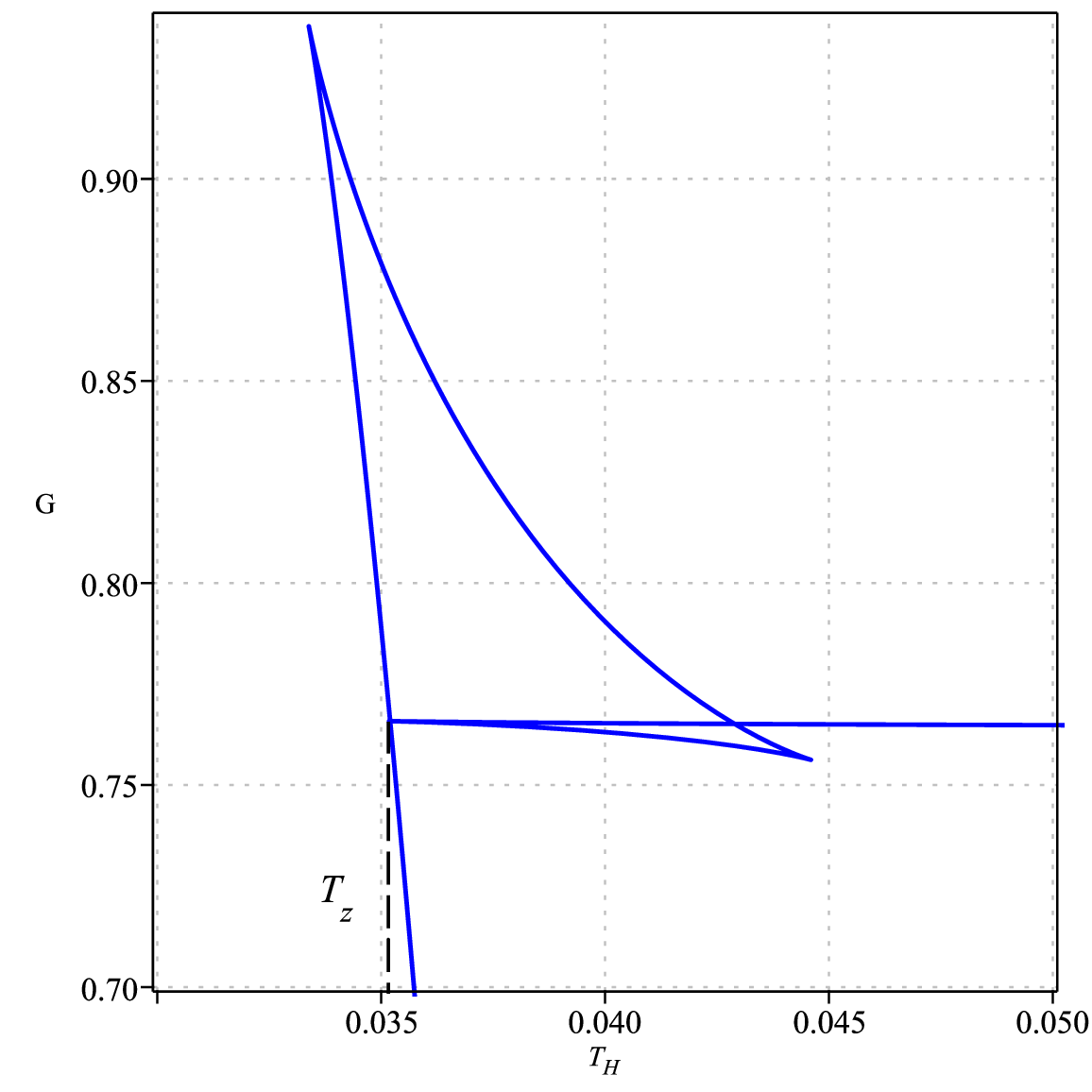}
         \caption{$P=0.00184=P_{z}$}
         \label{fig:5.1(f)}
     \end{subfigure}
      \hfill
     \begin{subfigure}[b]{0.3\textwidth}
         \centering
         \includegraphics[width=\textwidth]{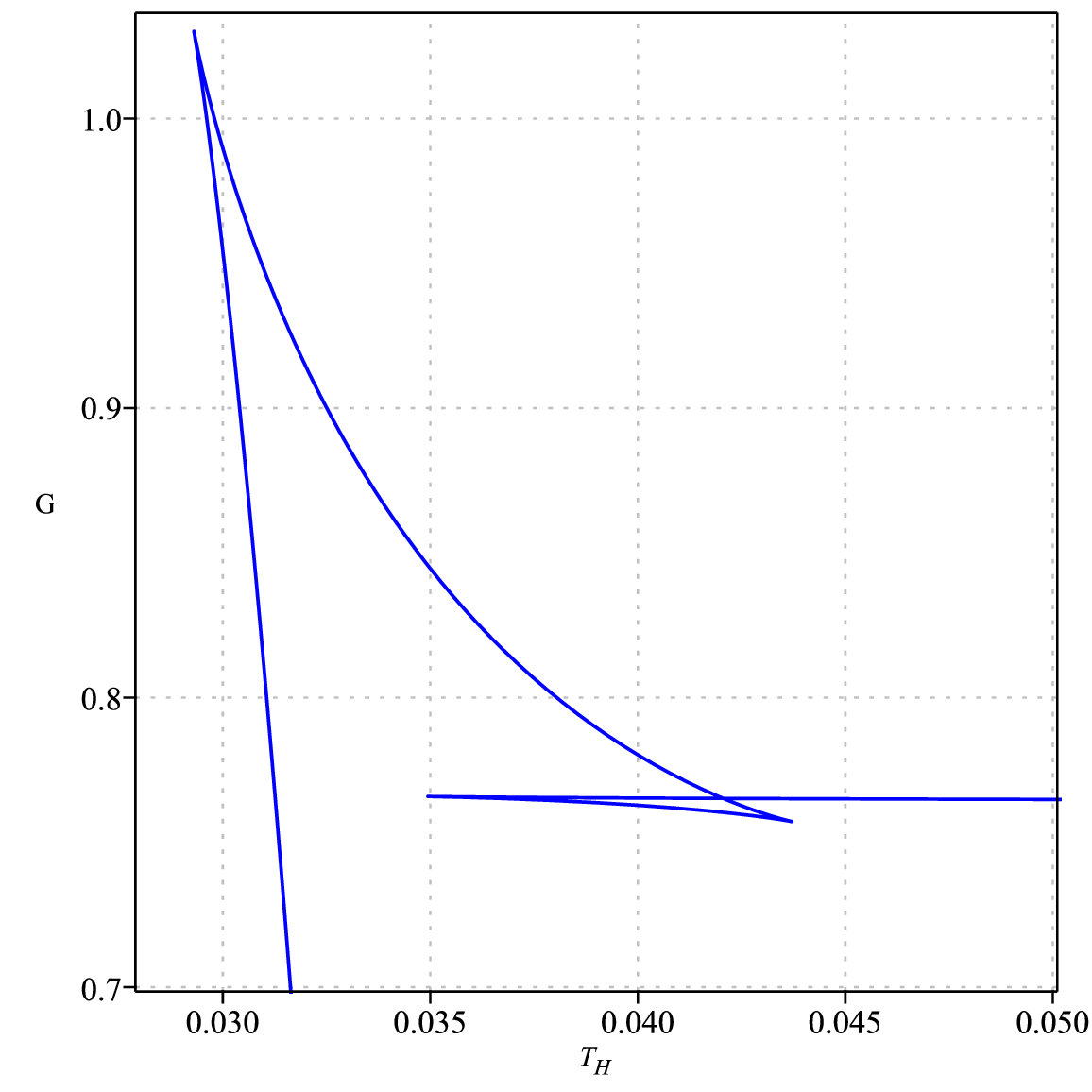}
         \caption{$P=0.0014<P_{z}$}
         \label{fig:5.1(g)}
     \end{subfigure}
      \hfill
     \begin{subfigure}[b]{0.3\textwidth}
         \centering
         \includegraphics[width=\textwidth]{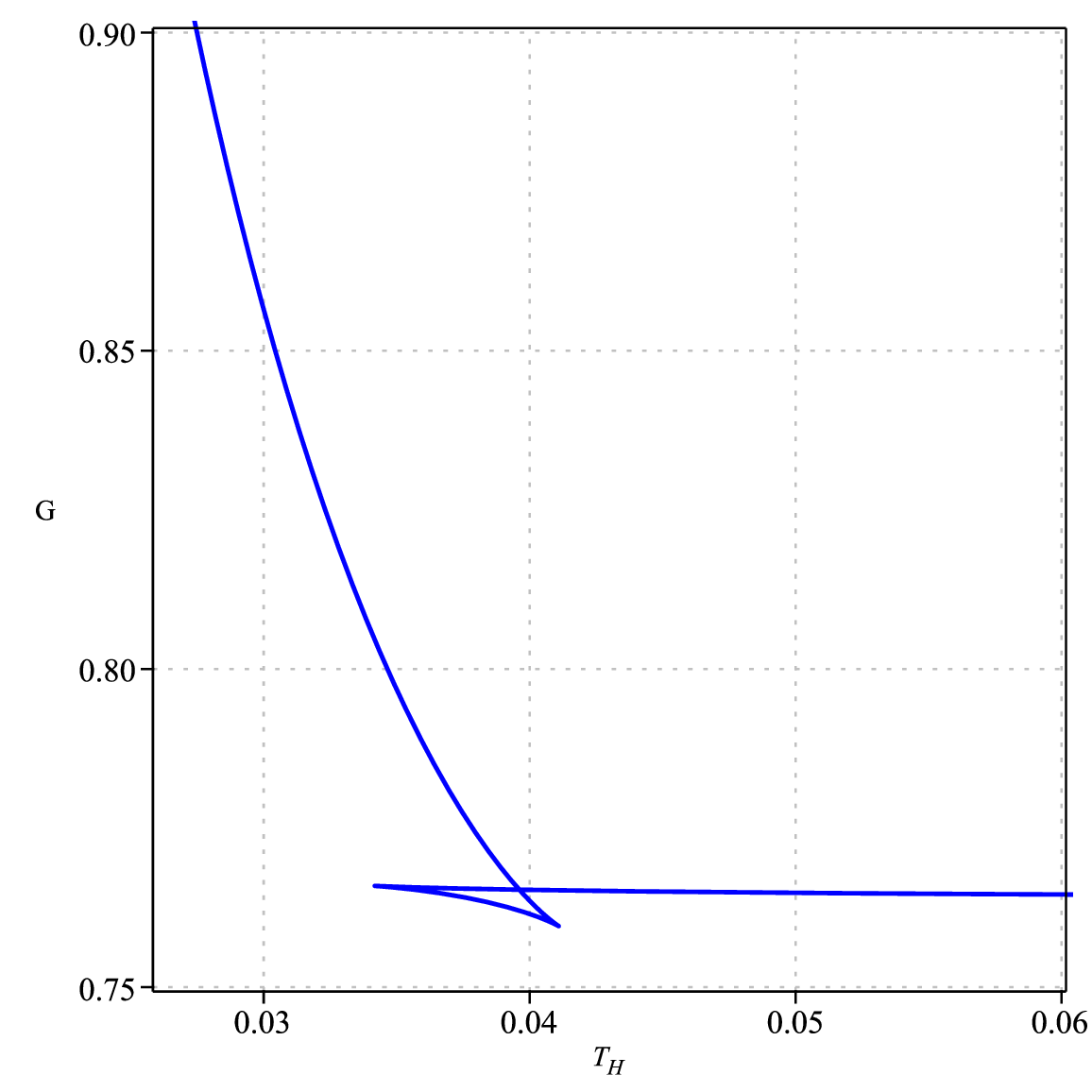}
         \caption{$P=0$}
         \label{fig:5.1(h)}
     \end{subfigure}
        \caption{Two critical points, one at positive critical pressure and
other at negative critical pressure.}
        \label{fig:RPT2}
\end{figure}

\begin{figure}[H]
     \centering
     \begin{subfigure}[b]{0.3\textwidth}
         \centering
         \includegraphics[width=\textwidth]{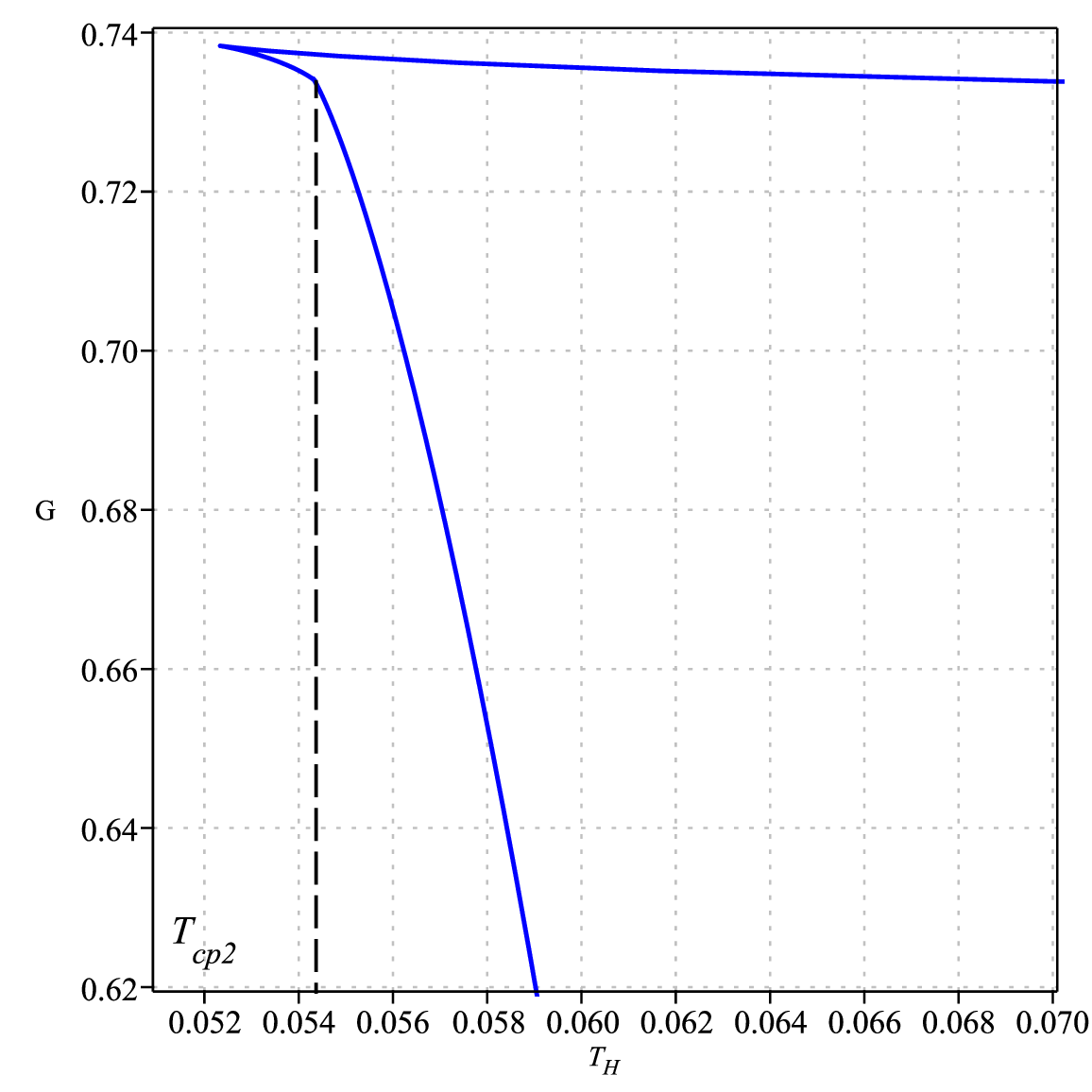}
         \caption{$P=0.0056=P_{cp2}$}
         \label{fig:5.1(i)}
     \end{subfigure}
     \hfill
     \begin{subfigure}[b]{0.3\textwidth}
         \centering
         \includegraphics[width=\textwidth]{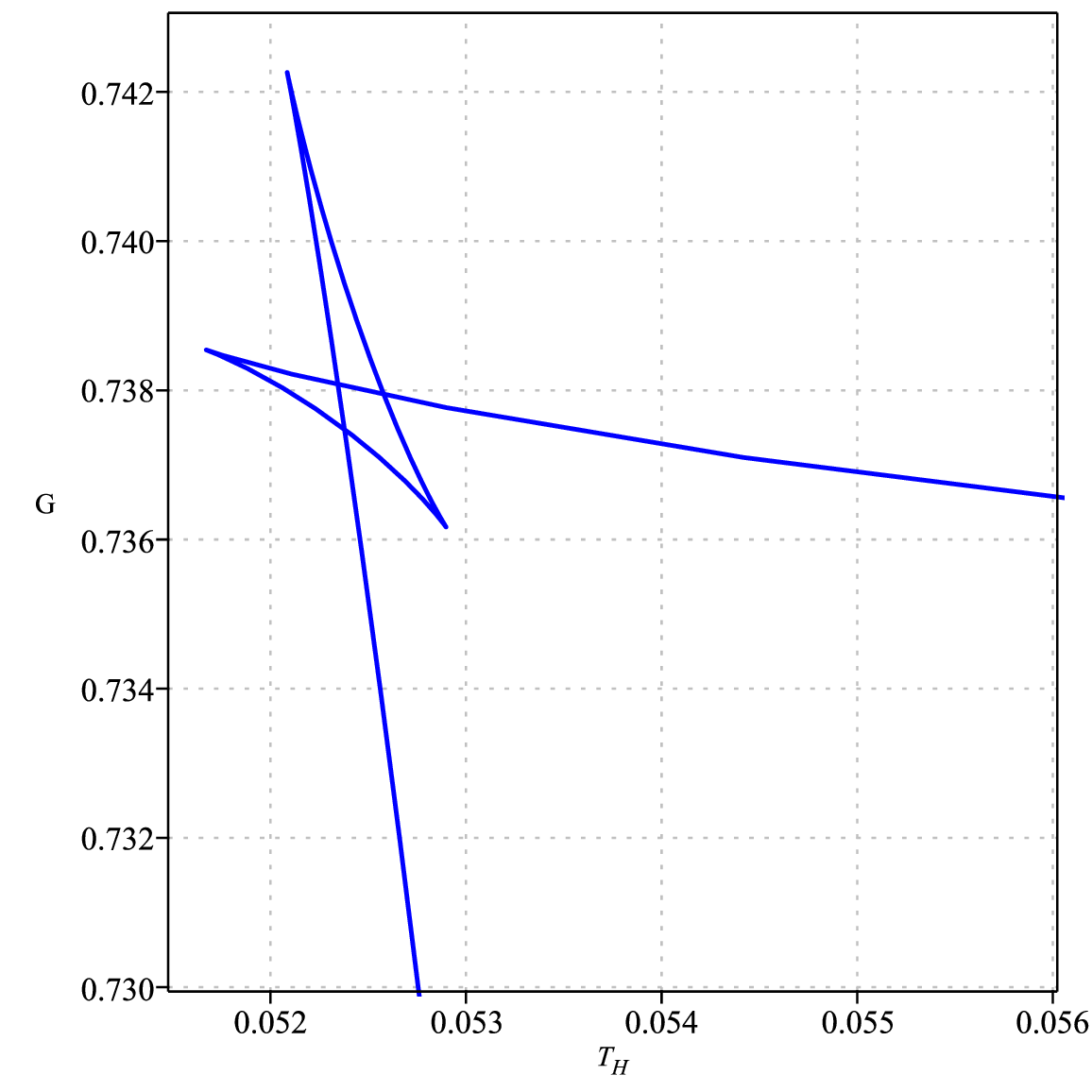}
         \caption{$P=0.0050<P_{cp2}$}
         \label{fig:5.1(j)}
     \end{subfigure}
     \hfill
     \begin{subfigure}[b]{0.3\textwidth}
         \centering
         \includegraphics[width=\textwidth]{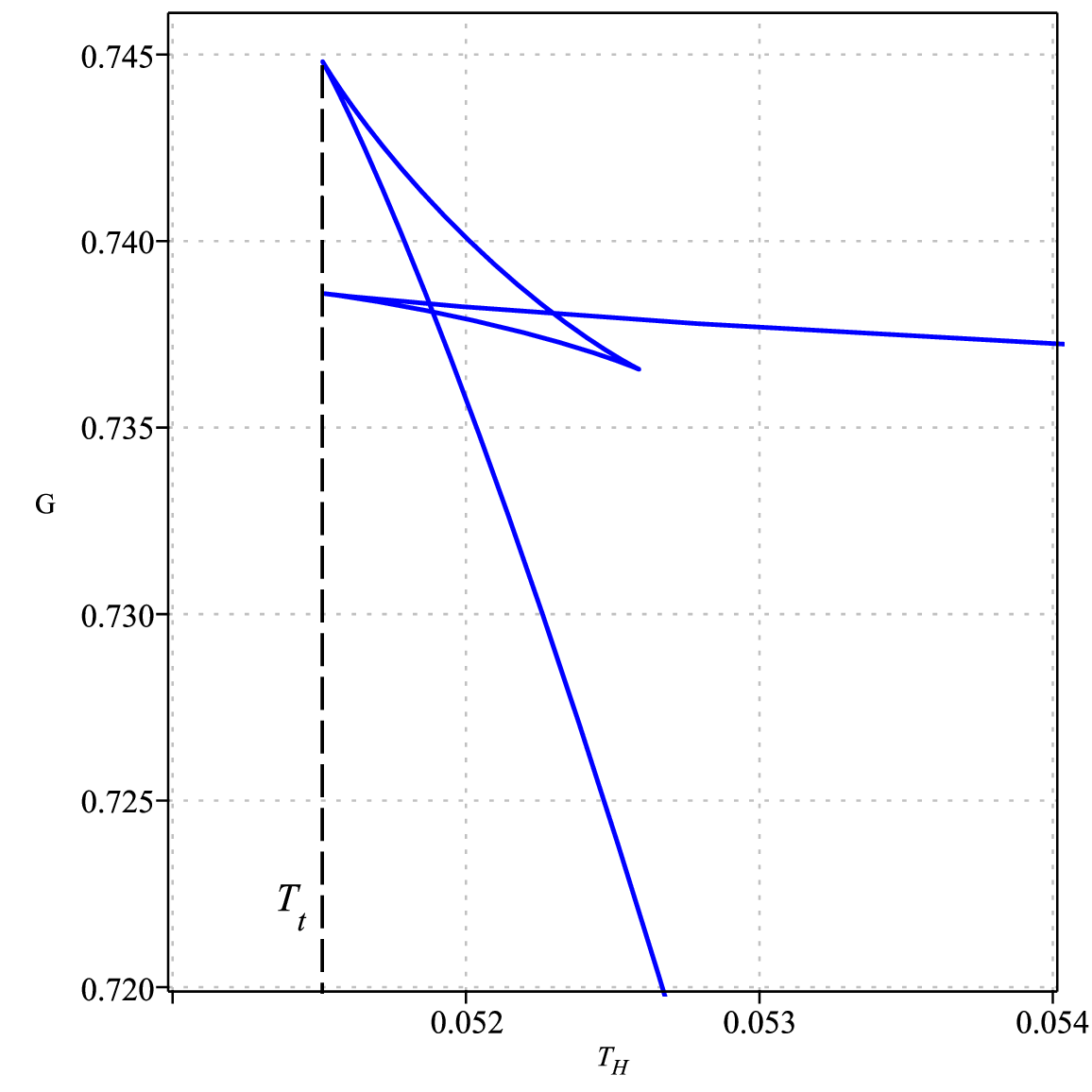}
         \caption{$P=0.00486=P_{t}$}
         \label{fig:5.1(k)}
     \end{subfigure}
     \hfill
     \begin{subfigure}[b]{0.3\textwidth}
         \centering
         \includegraphics[width=\textwidth]{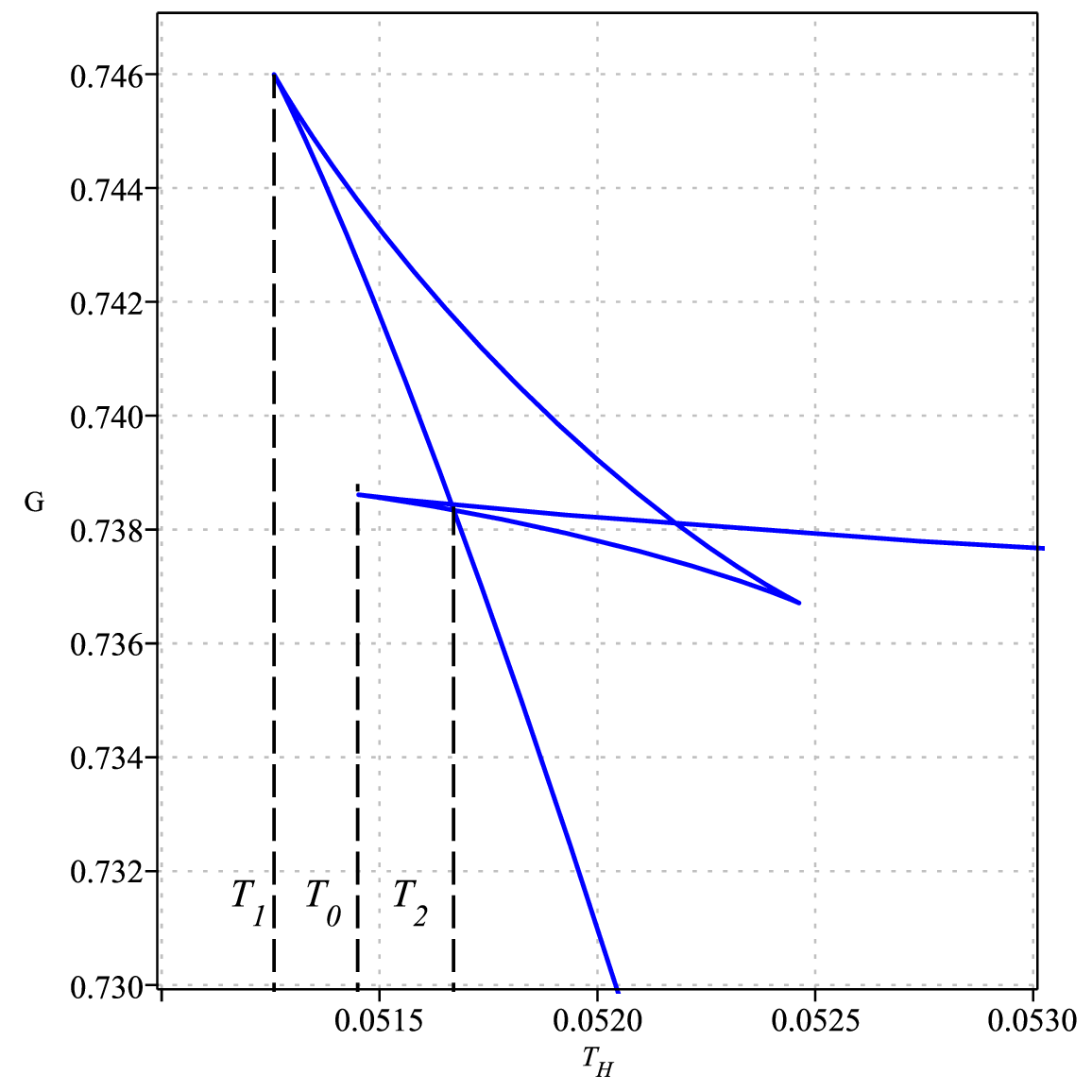}
         \caption{$P_{z}<P=0.00480<P_{t}$}
         \label{fig:5.1(l)}
     \end{subfigure}
     \hfill
     \begin{subfigure}[b]{0.3\textwidth}
         \centering
         \includegraphics[width=\textwidth]{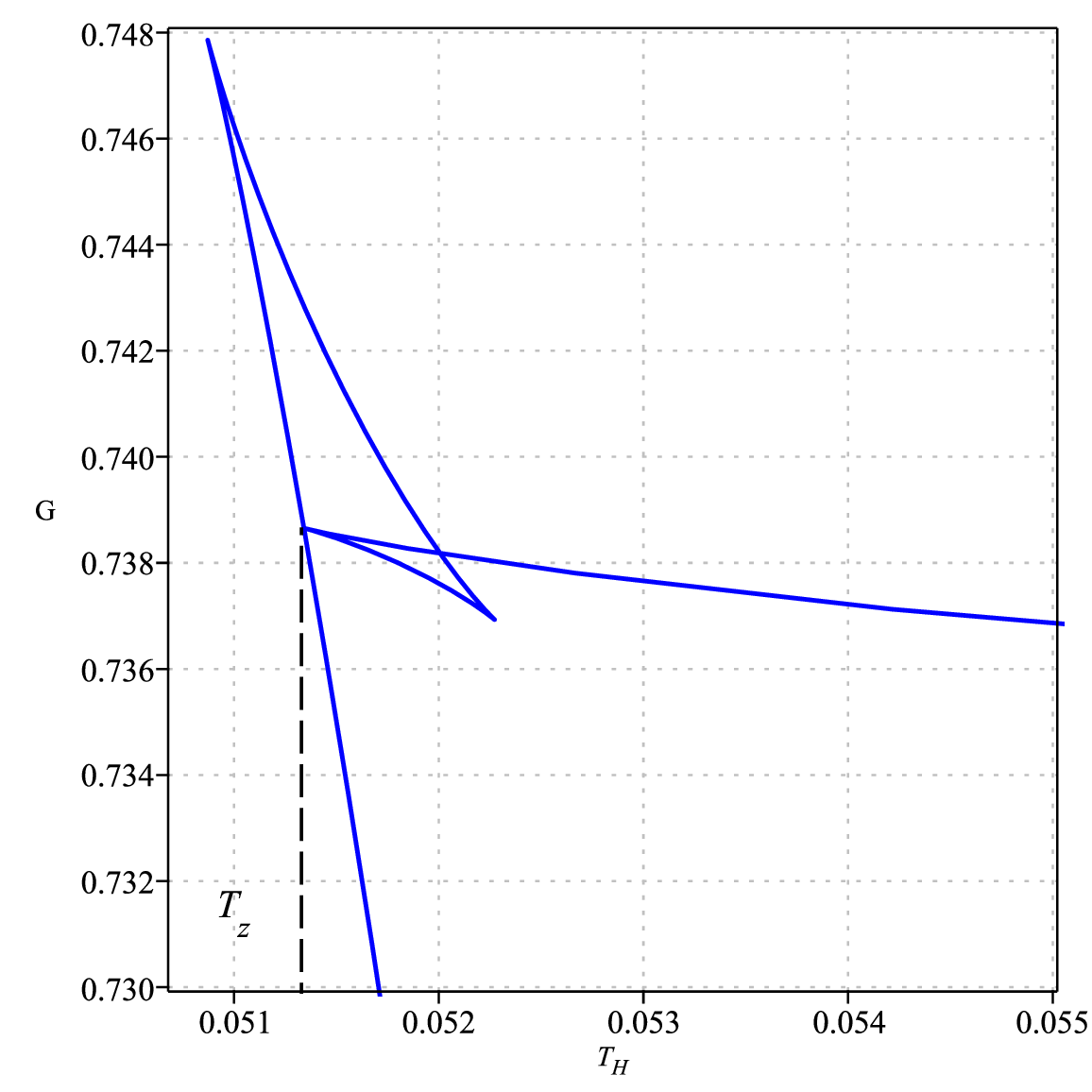}
         \caption{$P=0.00471=P_{z}$}
         \label{fig:5.1(m)}
     \end{subfigure}
     \hfill
     \begin{subfigure}[b]{0.3\textwidth}
         \centering
         \includegraphics[width=\textwidth]{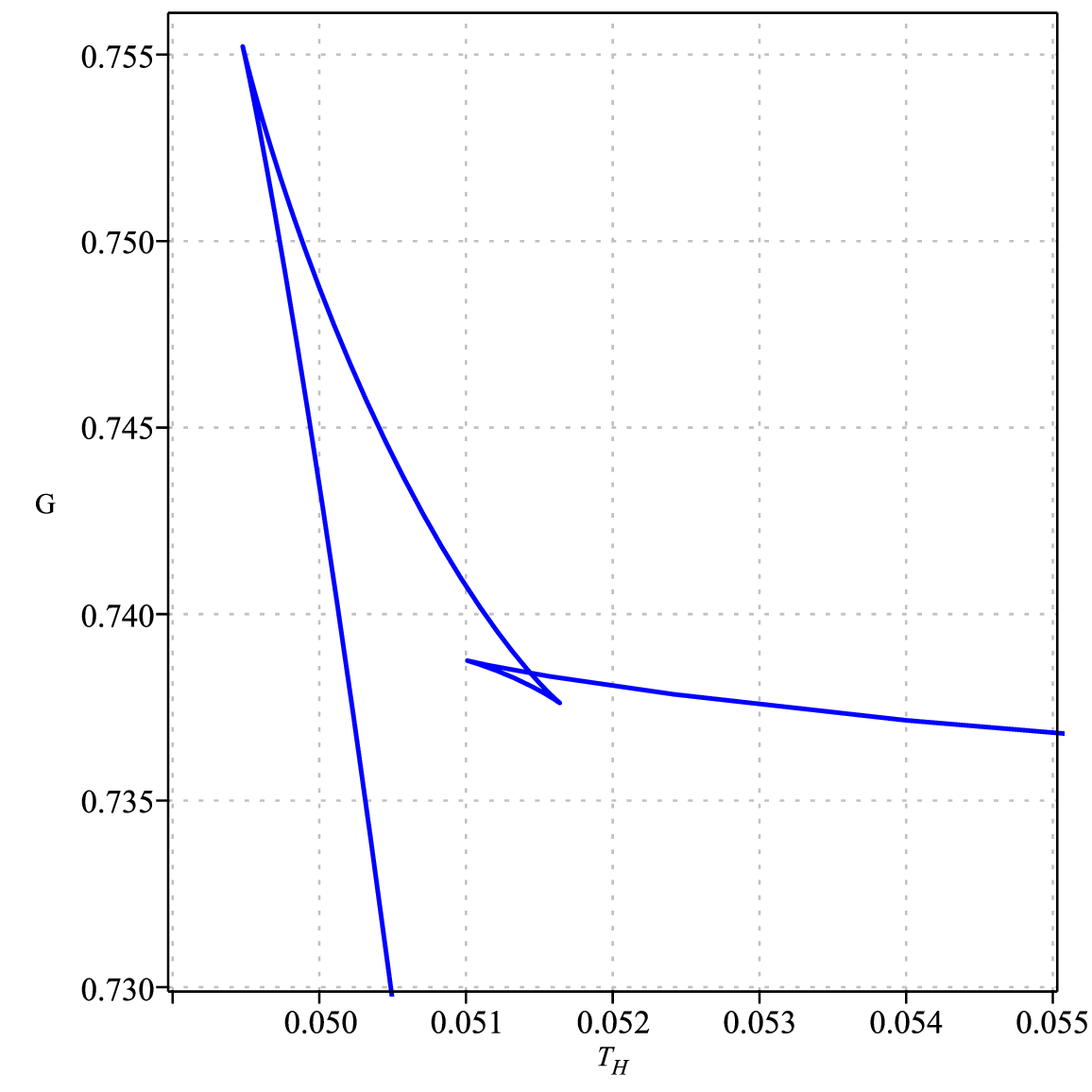}
         \caption{$P_{cp1}<P=0.0044<P_{z}$}
         \label{fig:5.1(n)}
     \end{subfigure}
      \hfill
     \begin{subfigure}[b]{0.3\textwidth}
         \centering
         \includegraphics[width=\textwidth]{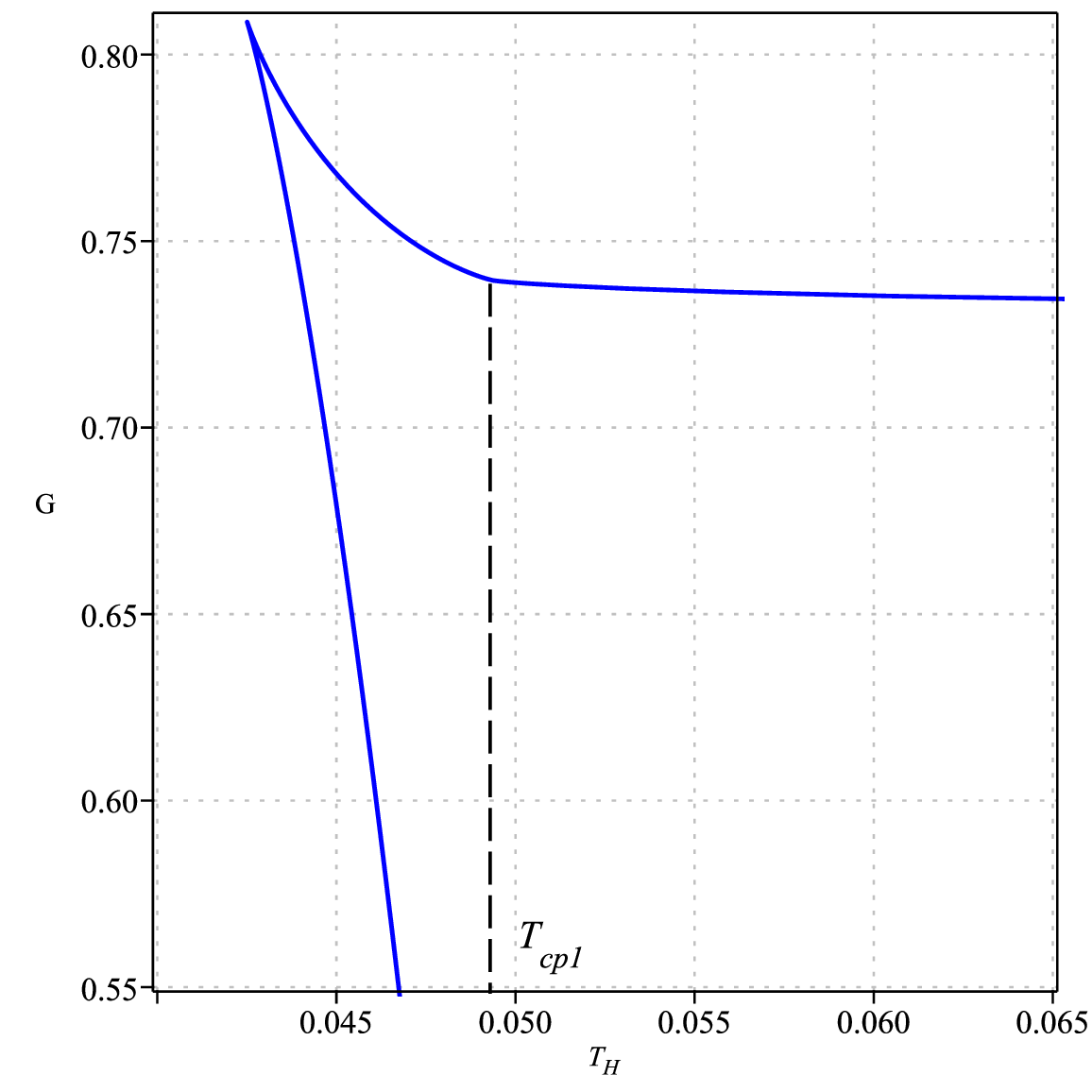}
         \caption{$P=0.0031=P_{cp1}$}
         \label{fig:5.1(o)}
     \end{subfigure}
      \hfill
     \begin{subfigure}[b]{0.3\textwidth}
         \centering
         \includegraphics[width=\textwidth]{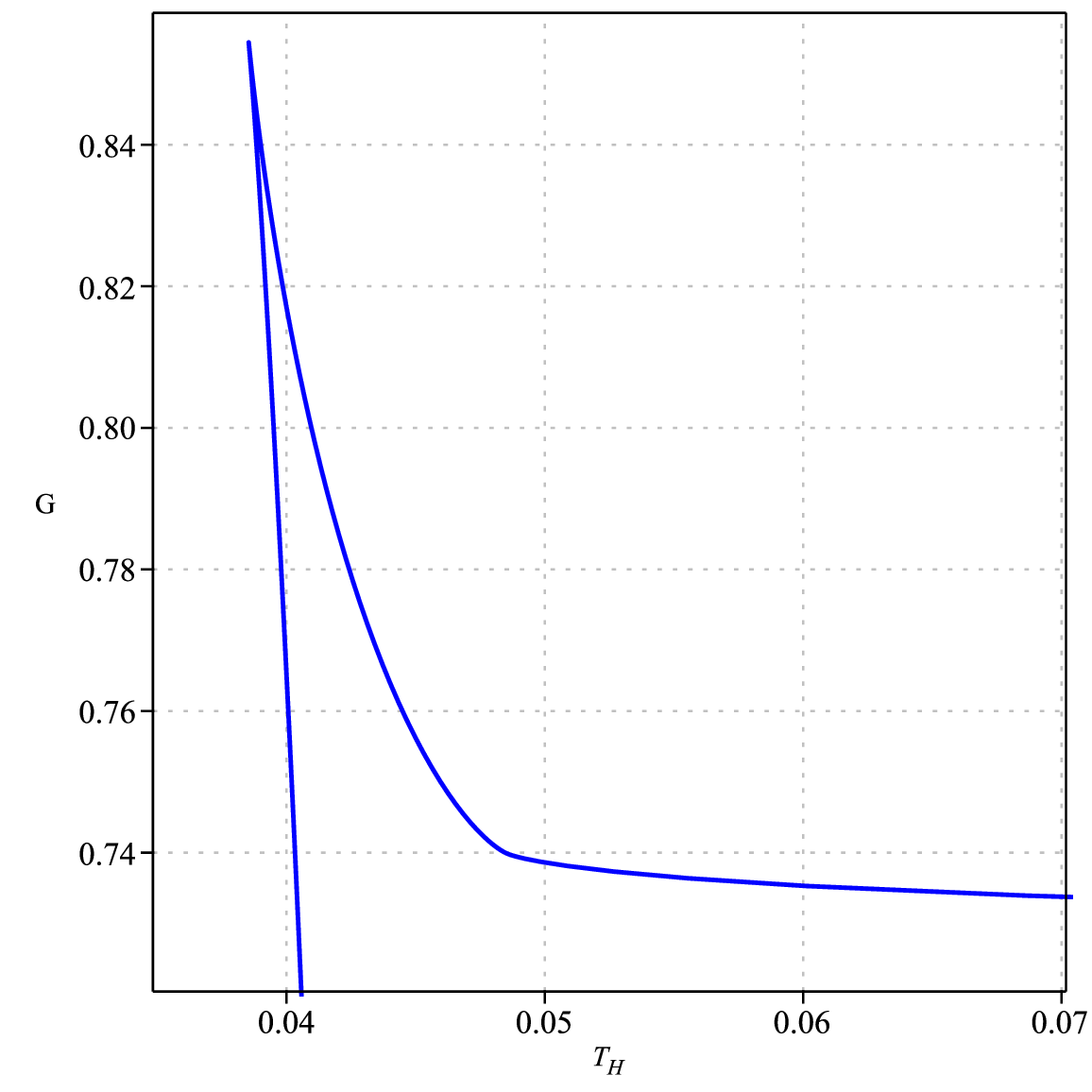}
         \caption{$P=0.0025<P_{cp1}$}
         \label{fig:5.1(p)}
     \end{subfigure}
        \caption{Two critical points with two positive critical 
        pressures.}
        \label{fig:RPT3}
\end{figure}

The behaviour of the $G-T_{H}$ diagram is shown in Figs. \ref{fig:RPT2} 
and \ref{fig:RPT3}, when one critical point is at positive pressure, 
the other at negative pressure, and two critical points at positive pressure. 
For $P<P_{cp2}$ a shallow tall behaviour occurs, which indicates a first-order 
phase transition Fig. \ref{fig:5.1(c)}. If we decrease the pressure then it 
continues till $P_{t}$, which is shown in Fig. \ref{fig:5.1(d)}. Therefore, 
the first-order phase transition occurs for $T \in (T_{t} , T_{cp2})$ 
and $P \in (P_{t} , P_{cp2})$. Once again, if we further decrease the pressure 
from $P_{t}$ then a  zeroth-order phase transition occurs Fig. \ref{fig:5.1(e)}.
 This zeroth-order phase transition occurs for a range of pressure 
 $P \in (P_{z} , P_{t})$ and temperature $T \in (T_{1} , T_{2})$. If we fixed 
 the pressure then a \textbf{LBH} is preferred between the temperature 
 $T \in (T_{1}, T_{0})$.

At temperature $T_{0}$ system goes to an intermediate 
black hole (\textbf{IBH}) phase. If we further increase the temperature 
from $T_{0}$ a transition from \textbf{IBH} to \textbf{LBH} occurs 
at $T_{2}$, which is known as first-order phase transition. Therefore, 
the system undergoes a \textbf{LBH}--\textbf{IBH}--\textbf{LBH} phase 
transition, as the initial and final phases are the same, this is called 
the reentrant phase transition. The phenomenon of reentrant phase transition 
disappears at $P=P_{z}$ in Fig. \ref{fig:5.1(f)}. A similar kind of behaviour 
is shown in Fig. \ref{fig:RPT3}, when two critical points are real with two 
real positive pressures.

\begin{table}[H]
\begin{center}
\begin{tabular}{ |c|c|c|c|c| } 
\hline
Case & CP & CP1 & CP2 \\
\hline
\multirow{3}{10em}{$\beta_1 < \beta=0.34 < \beta_2$} & $v_c$ & 1.4336 & 2.9587 \\ 
& $T_c$ & 0.0494 & 0.0545 \\ 
& $P_c$ & 0.0031 & 0.0056 \\ 
\hline
\multirow{3}{10em}{$\beta_0 < \beta=0.28 < \beta_1 $} & $v_c$ & 0.8870 & 3.3160 \\ 
& $T_c$ &  0.0280 & 0.0525 \\ 
& $P_c$ & -0.0094 & 0.0051 \\ 
\hline
\multirow{3}{10em}{$0.20=\beta < \beta_0  $} & $v_c$ & $--$ & 3.6916 \\ 
& $T_c$ & $--$  & 0.0503 \\ 
& $P_c$ & $--$  & 0.0046 \\ 
\hline
\end{tabular}
\end{center}
\caption{With $Q_m=1$.}
\label{ta}
\end{table}

\subsection{Black Holes in \texorpdfstring{$4D$}{TEXT} massive Einstein
 gravity coupled to NED}\label{sec:RPT2}

The equation for critical radius is given in equation \ref{eq:4.18}
\begin{equation}\label{eq:14}
a \Bigl( v_{c}^{2} +4k^2 \Bigl)^3 - 8 Q_m^2 \Bigl( 3v_c^4 +6k^2v_c^2+8k^4 \Bigl) =0,
\end{equation}
where we take $a=1+c^2c_2m^2$. Putting $x=v_c^2+4k^2$ into the above equation we obtain
\begin{equation}\label{eq:15}
    ax^3-24Q_m^2x^2+144Q^2k^2x-256Q_m^2k^4=0.
\end{equation}
In order to satisfy $v_c \geq 0$, we must have
\begin{equation}\label{eq:16}
\lvert x \rvert \geq 4k^2, 
\quad\text{or}\quad
\lvert x \rvert \geq 8\sqrt{\beta}Q_m.
\end{equation}

Three real roots of equation \eqref{eq:15} occur when the discriminant is
\begin{equation}\label{eq:17}
\Delta= 442368Q_m^4k^4(Q_m^2-ak^2)(5Q_m^2-4ak^2) < 0.
\end{equation}
When $\Delta > 0$ only one root is real and for $\Delta=0$ the equation 
\eqref{eq:15} has either one or two real solutions. 
From condition $\Delta < 0$ we obtain
\begin{equation}\label{eq:18}
    \frac{Q_m}{2\sqrt{\beta}}={a_0} < a < {a_2}= \frac{5Q_m}{8\sqrt{\beta}}.
\end{equation}
To find the solutions of equation \eqref{eq:15}, we will use the 
Tschirnhaus transformation method. Putting $x=t+B$ into equation \eqref{eq:15}

\begin{equation}\label{eq:19}
    t^3 +pt +q=0,
\end{equation}
where we set coefficients of $t^2$ equal to zero \& $B={8Q_m^2}/{a}$. 
Finally the solutions of equations \eqref{eq:15} is 
\begin{equation}\label{eq:20}
    x_{j}=2 \sqrt{\frac{-p}{3}} \cos{\Biggr[\frac{1}{3} \arccos{\biggl(\frac{3q}{2p}\sqrt{\frac{-3}{p} } \biggl)}-\frac{2\pi j}{3} \Biggr]},
\end{equation}
where $j=0$, $1$ \& $2$. The condition in equation \eqref{eq:16} was satisfied for $x_0$ and $x_1$ only, $x_2$ does not satisfy condition \eqref{eq:16}. Therefore we have two critical points. The constant $p$ and $q$ is given by
\begin{align}\label{eq:21}
p &=  3B^2 -\frac{48BQ_m^2}{a} +\frac{144Q_m^2 k^2}{a}, \\ 
q &=  B^3 -\frac{24B^2Q_m^2}{a} +\frac{144 B Q_m^2 k^2}{a} -\frac{256 Q_m^2 k^4}{a}.
\end{align}

\begin{figure}[H]
     \centering
     \begin{subfigure}[b]{0.3\textwidth}
         \centering
         \includegraphics[width=\textwidth]{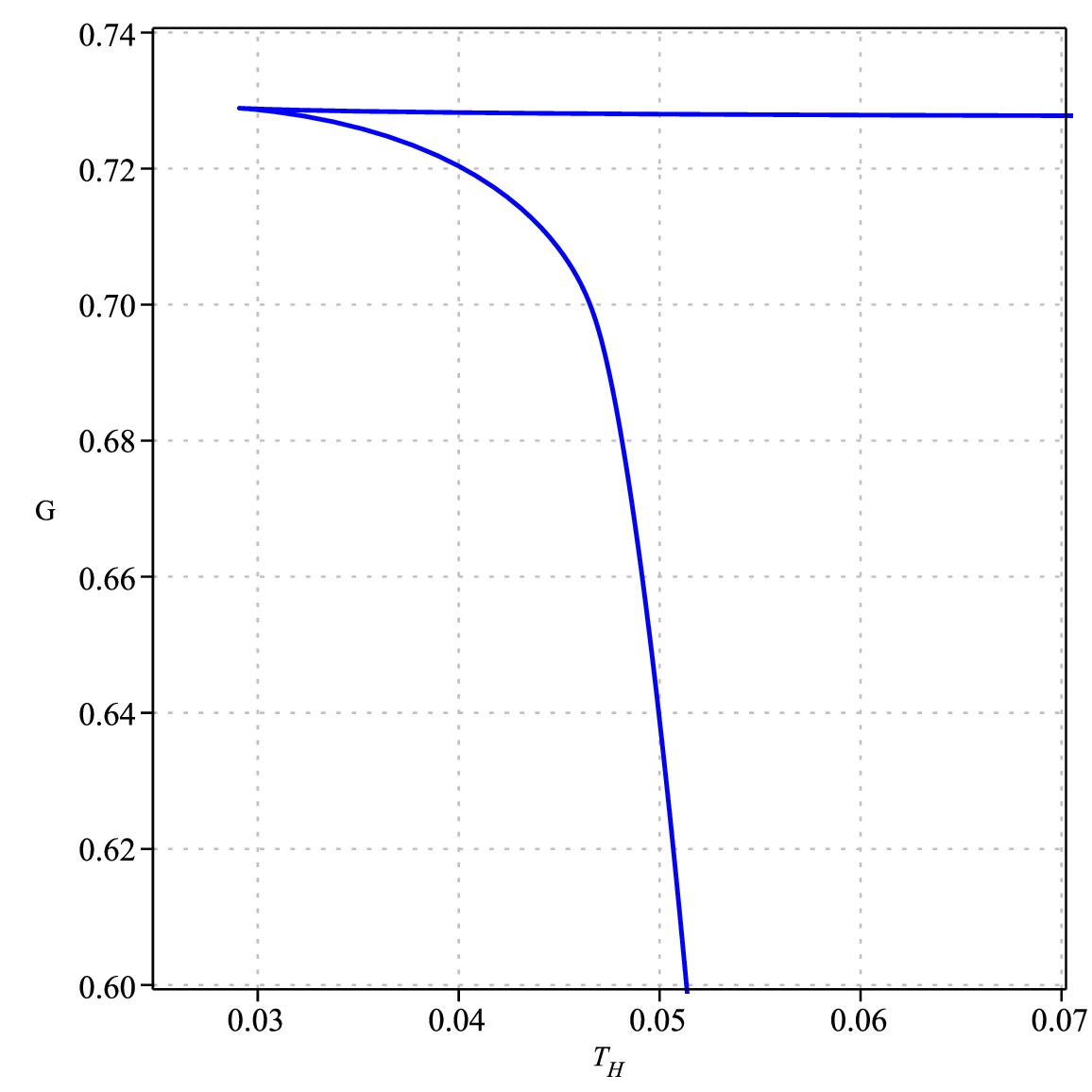}
         \caption{$P=0.0050>P_{cp2}$}
         \label{fig:5.2(a)}
     \end{subfigure}
     \hfill
     \begin{subfigure}[b]{0.3\textwidth}
         \centering
         \includegraphics[width=\textwidth]{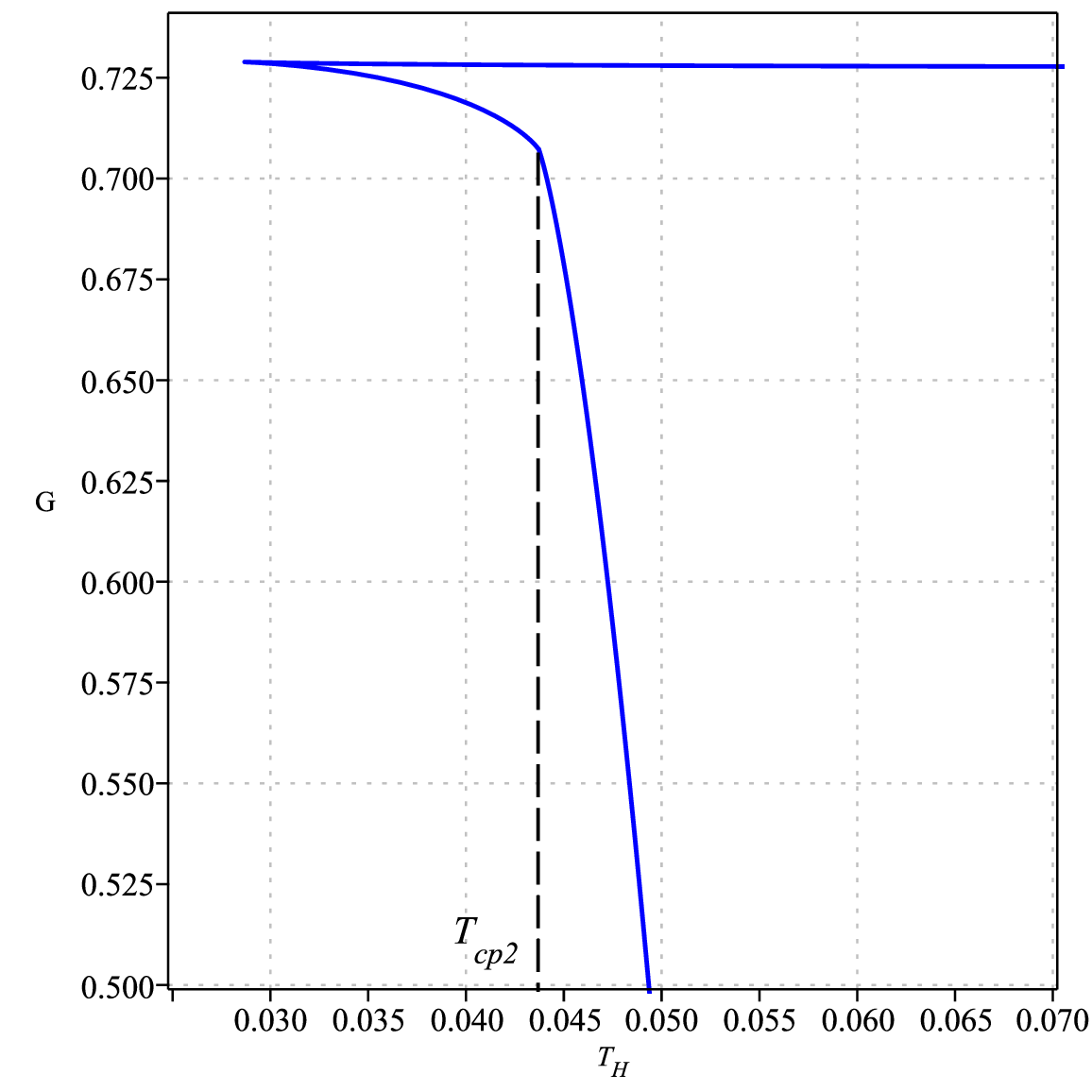}
         \caption{$P=0.0041=P_{cp2}$}
         \label{fig:5.2(b)}
     \end{subfigure}
     \hfill
     \begin{subfigure}[b]{0.3\textwidth}
         \centering
         \includegraphics[width=\textwidth]{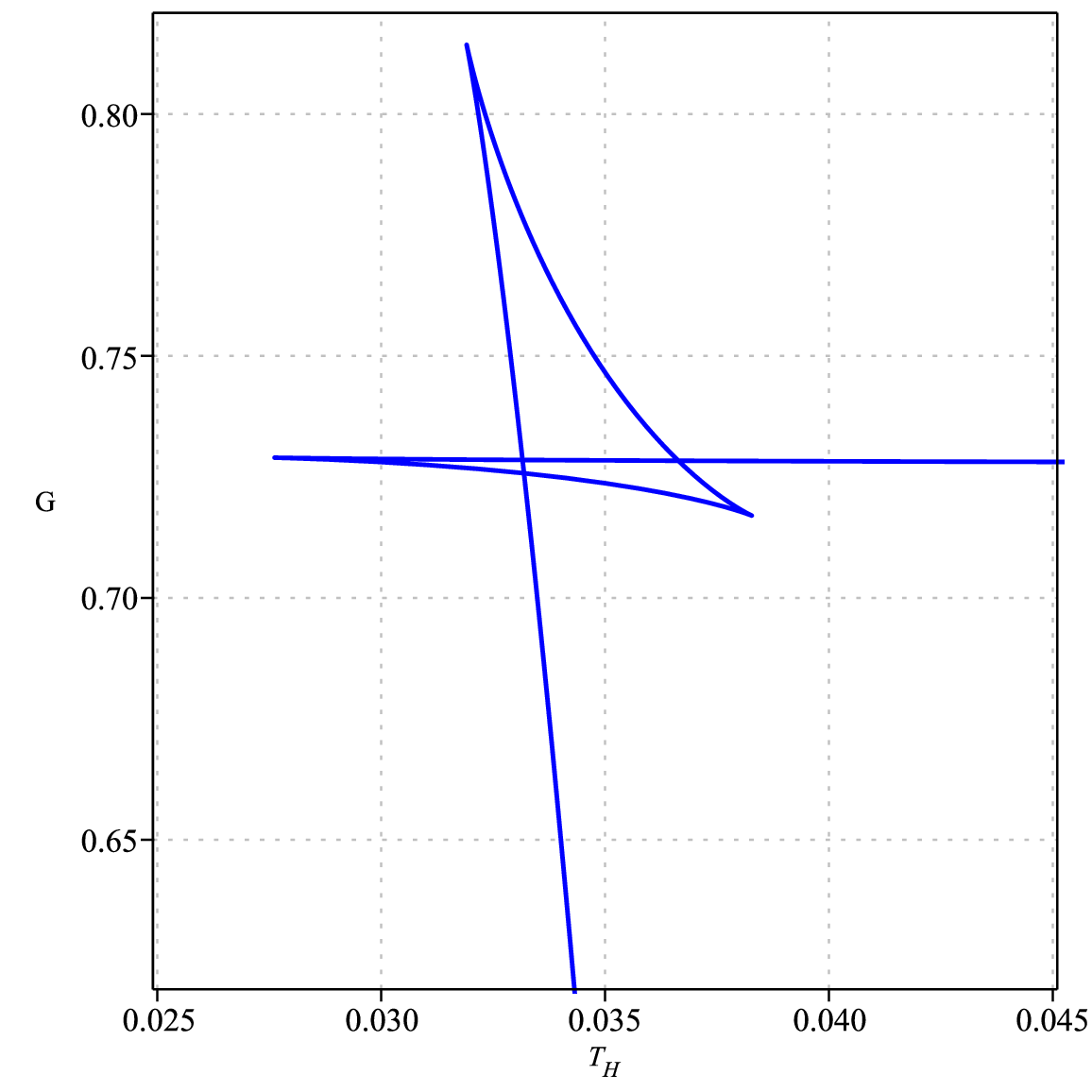}
         \caption{$P=0.0020<P_{cp2}$}
         \label{fig:5.2(c)}
     \end{subfigure}
     \hfill
     \begin{subfigure}[b]{0.3\textwidth}
         \centering
         \includegraphics[width=\textwidth]{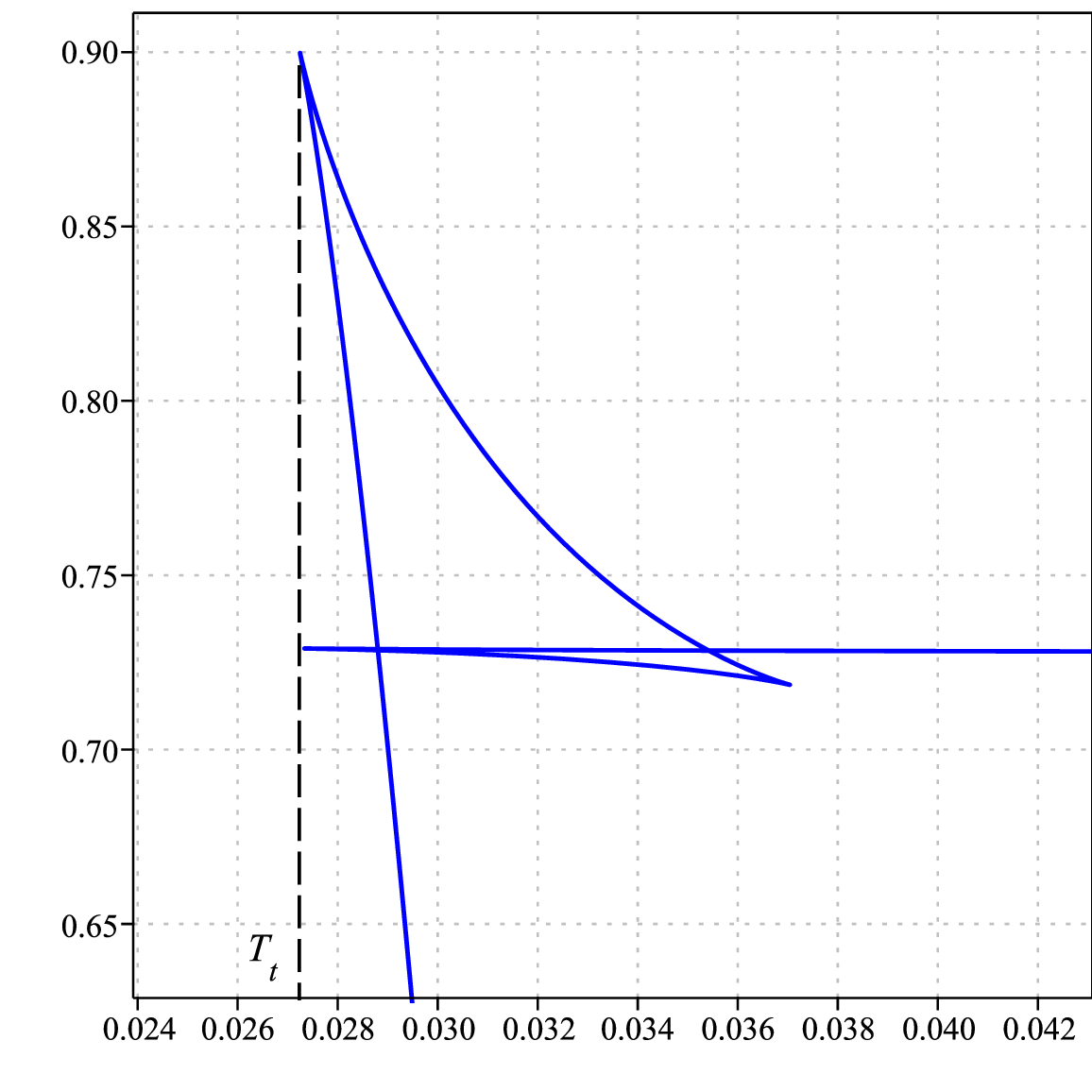}
         \caption{$P=0.00144=P_{t}$}
         \label{fig:5.2(d)}
     \end{subfigure}
     \hfill
     \begin{subfigure}[b]{0.3\textwidth}
         \centering
         \includegraphics[width=\textwidth]{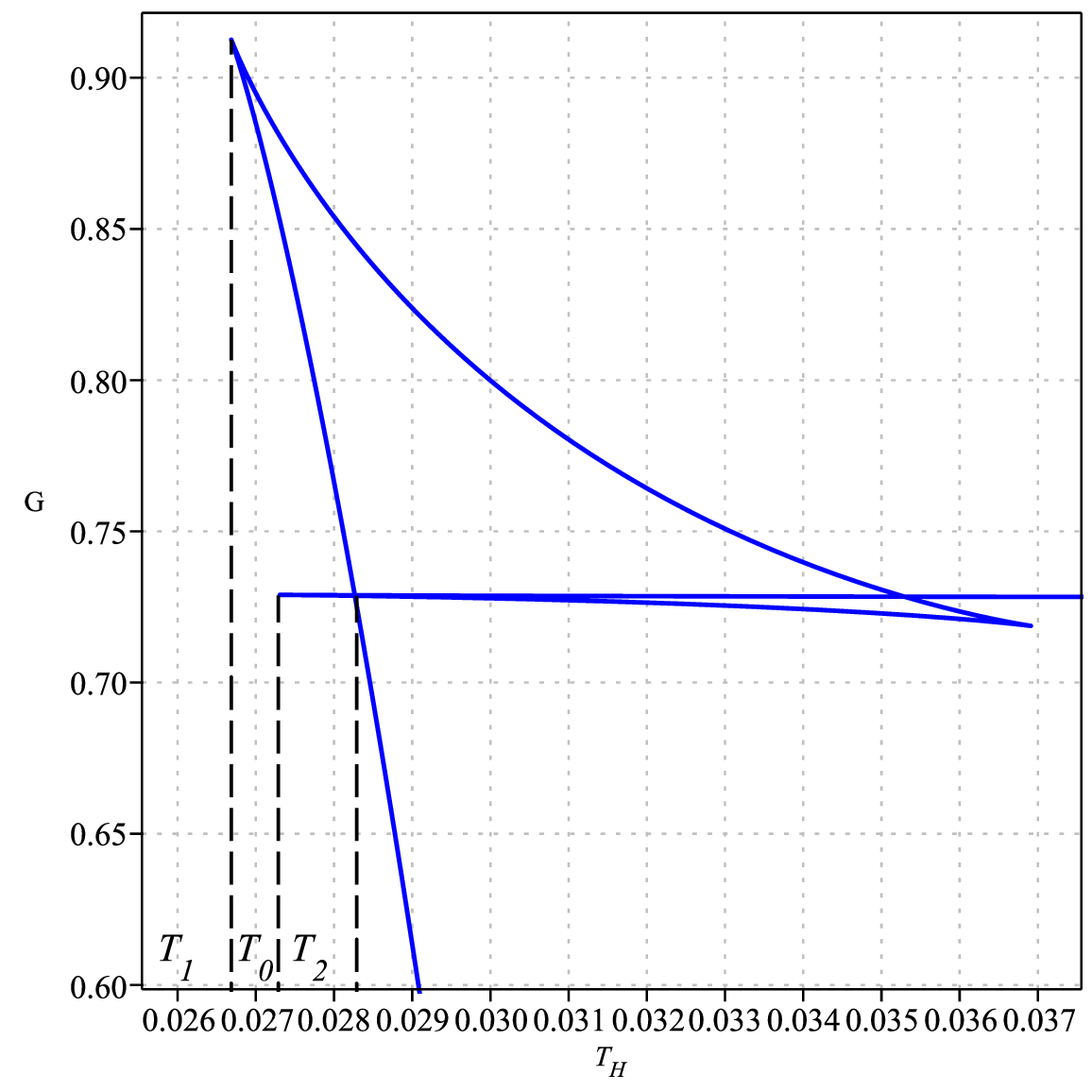}
         \caption{$P_{z}<P=0.00138<P_{t}$}
         \label{fig:5.2(e)}
     \end{subfigure}
     \hfill
     \begin{subfigure}[b]{0.3\textwidth}
         \centering
         \includegraphics[width=\textwidth]{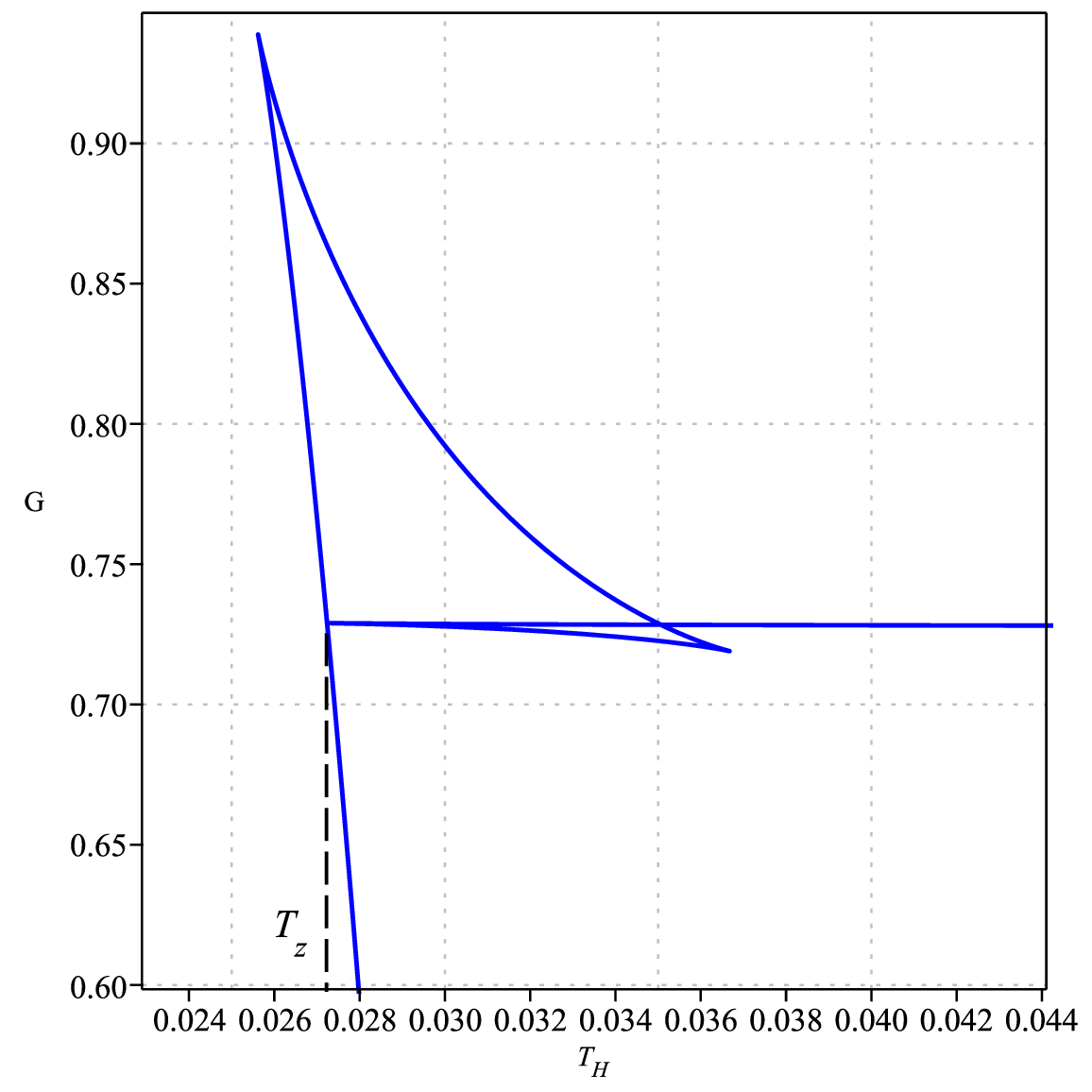}
         \caption{$P=0.00127=P_{z}$}
         \label{fig:5.2(f)}
     \end{subfigure}
      \hfill
     \begin{subfigure}[b]{0.3\textwidth}
      
         \includegraphics[width=\textwidth]{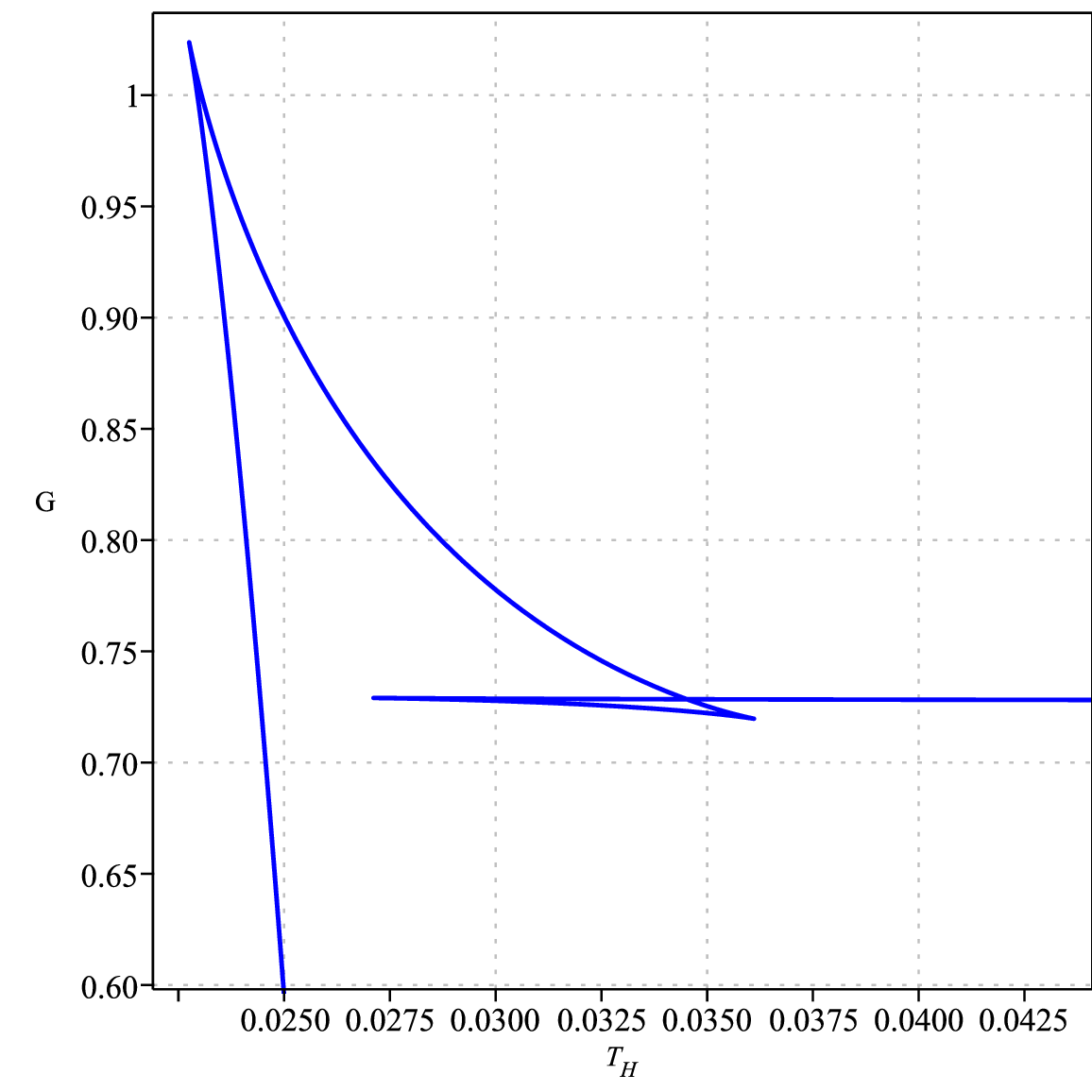}
         \caption{$P=0.0010<P_{z}$}
         \label{fig:5.2(g)}
     \end{subfigure}
      \hfill
     \begin{subfigure}[b]{0.3\textwidth}
         
         \includegraphics[width=\textwidth]{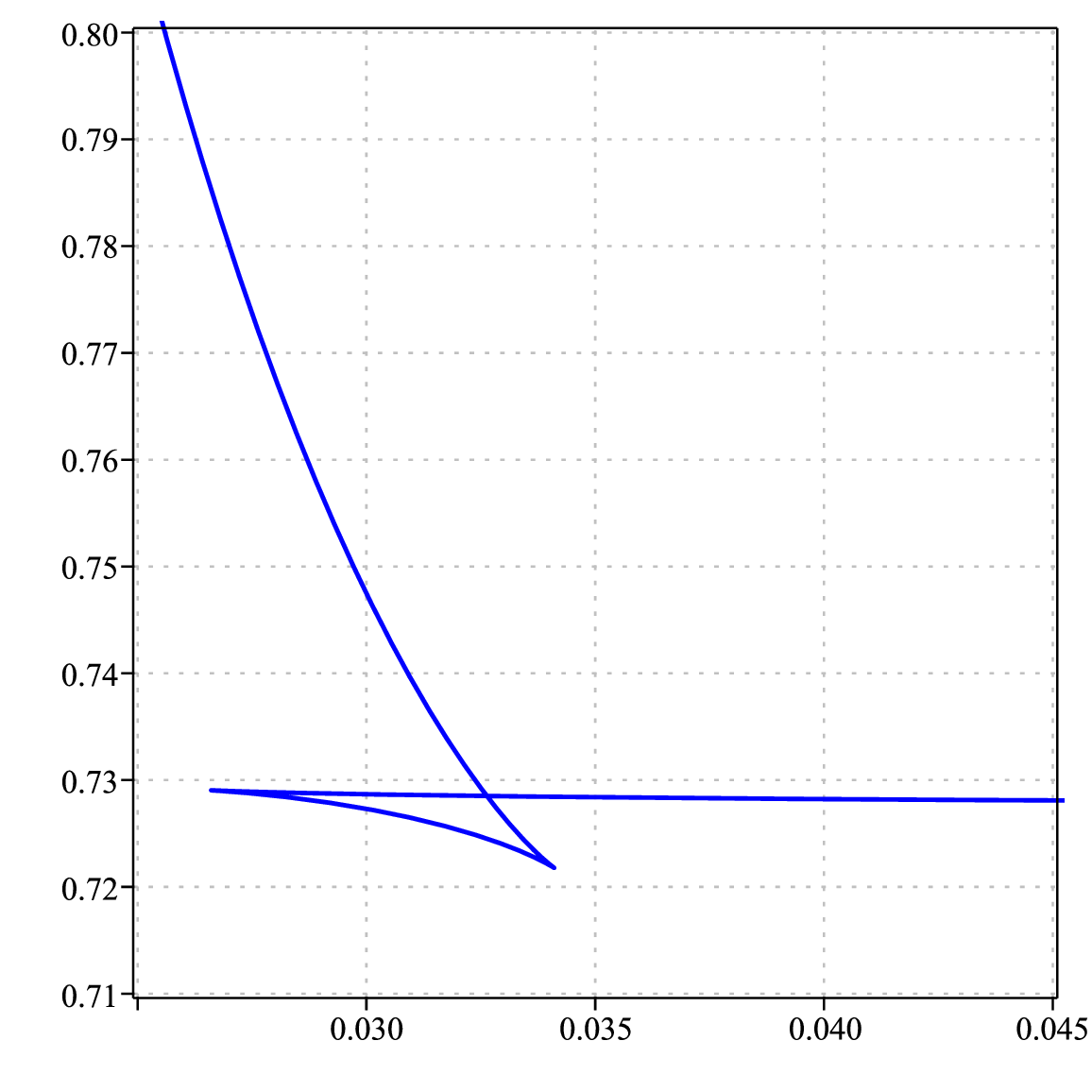}
         \caption{$P=0$}
         \label{fig:5.2(h)}
     \end{subfigure}
        \caption{Two critical points with one pressure is positive and one pressure is negative.}
        \label{fig:RPT5}
\end{figure}

\begin{figure}[H]
     \centering
     \begin{subfigure}[b]{0.3\textwidth}
         \centering
         \includegraphics[width=\textwidth]{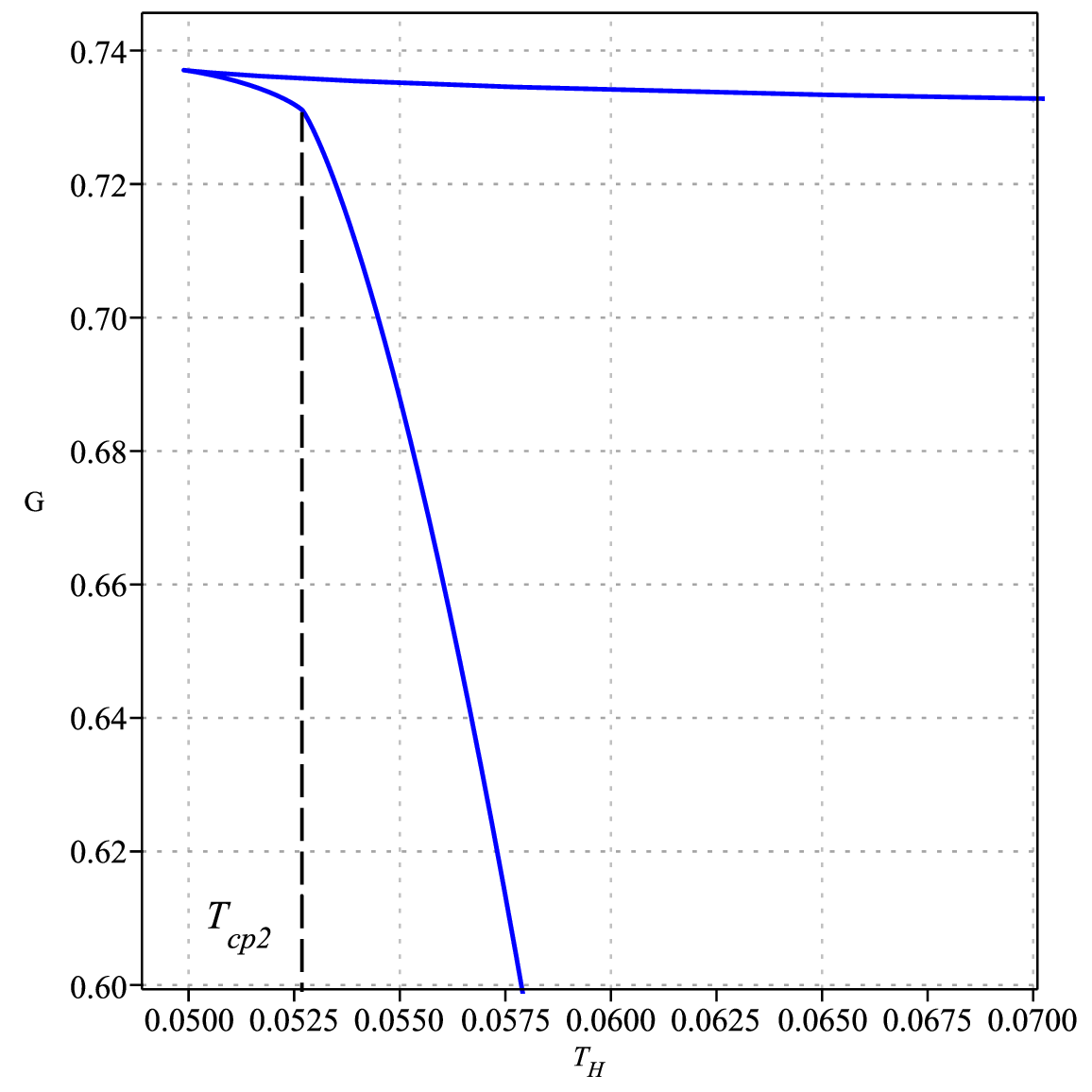}
         \caption{$P=0.0055=P_{cp2}$}
         \label{fig:5.2(i)}
     \end{subfigure}
     \hfill
     \begin{subfigure}[b]{0.3\textwidth}
         \centering
         \includegraphics[width=\textwidth]{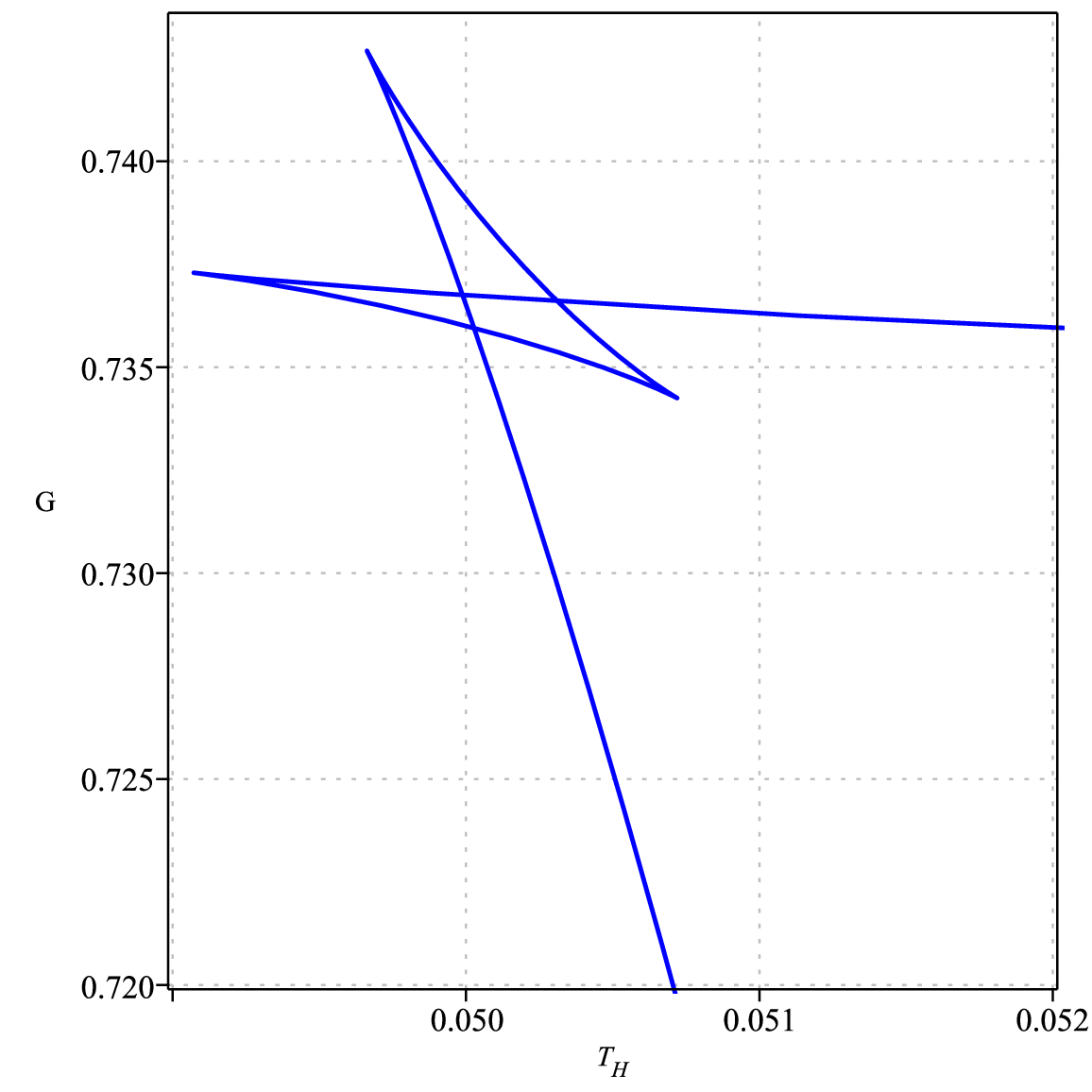}
         \caption{$P=0.0047<P_{cp2}$}
         \label{fig:5.2(j)}
     \end{subfigure}
     \hfill
     \begin{subfigure}[b]{0.3\textwidth}
         \centering
         \includegraphics[width=\textwidth]{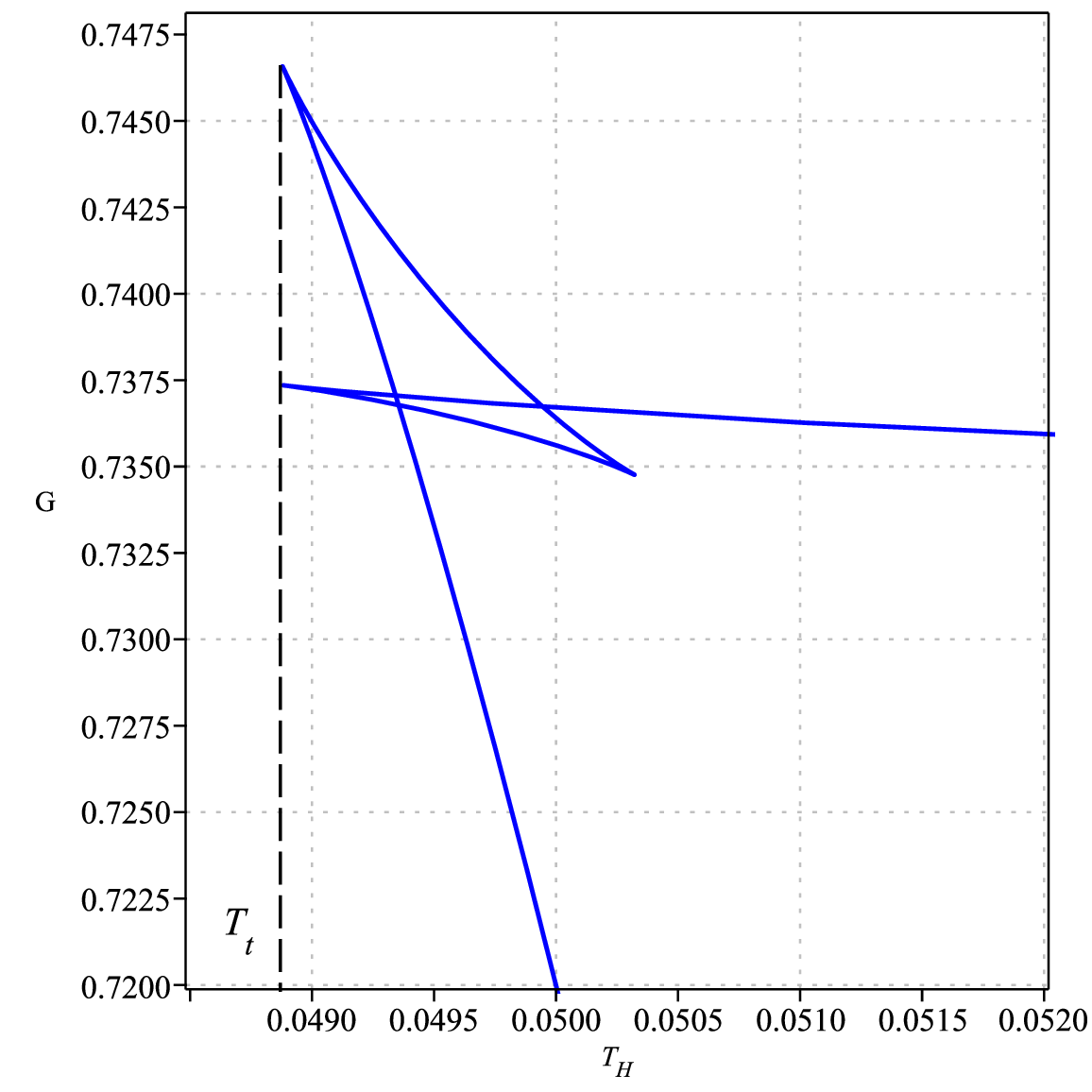}
         \caption{$P=0.00452=P_{t}$}
         \label{fig:5.2(k)}
     \end{subfigure}
     \hfill
     \begin{subfigure}[b]{0.3\textwidth}
         \centering
         \includegraphics[width=\textwidth]{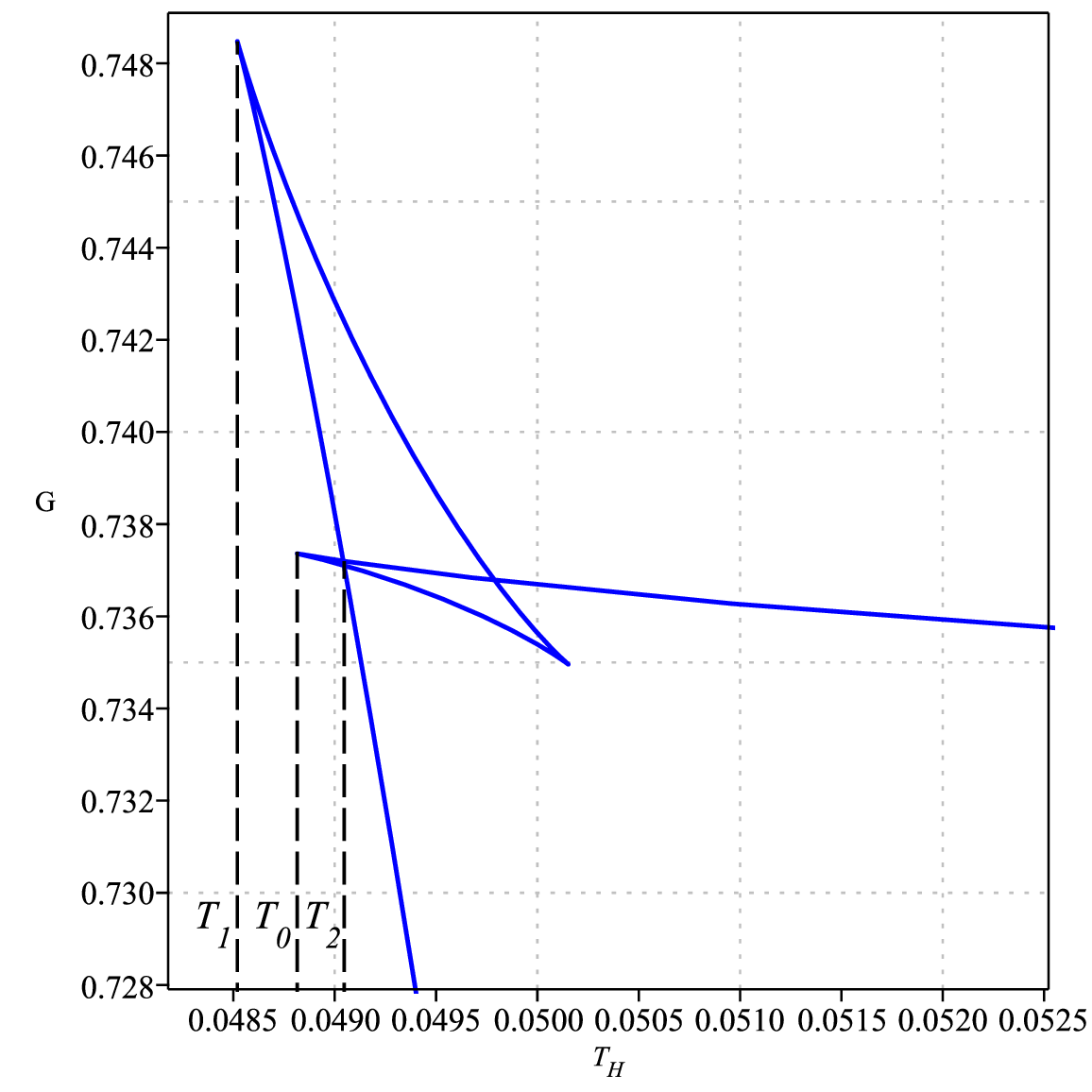}
         \caption{$P_{z}<P=0.00444<P_{t}$}
         \label{fig:5.2(l)}
     \end{subfigure}
     \hfill
     \begin{subfigure}[b]{0.3\textwidth}
         \centering
         \includegraphics[width=\textwidth]{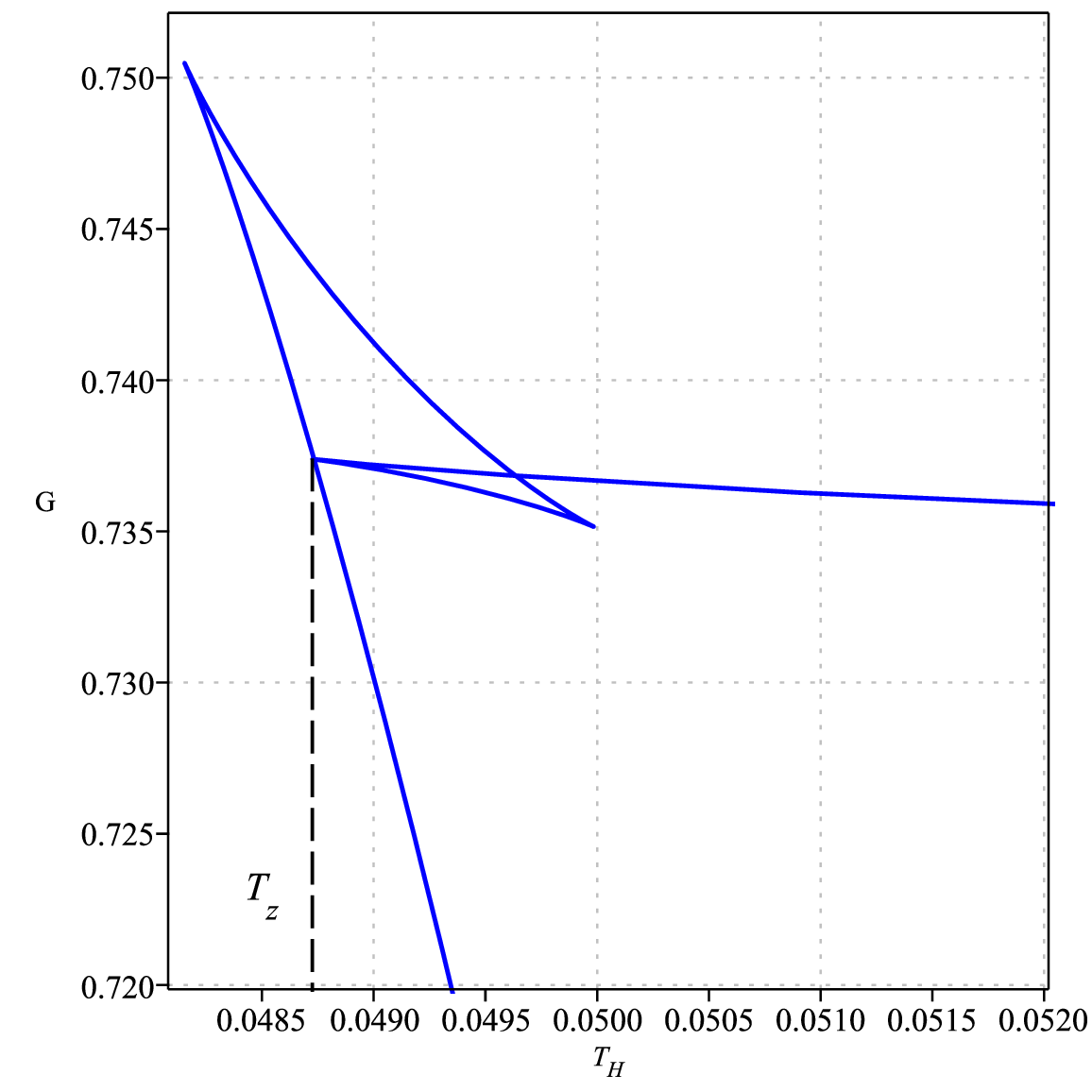}
         \caption{$P=0.00436=P_{z}$}
         \label{fig:5.2(m)}
     \end{subfigure}
     \hfill
     \begin{subfigure}[b]{0.3\textwidth}
         \centering
         \includegraphics[width=\textwidth]{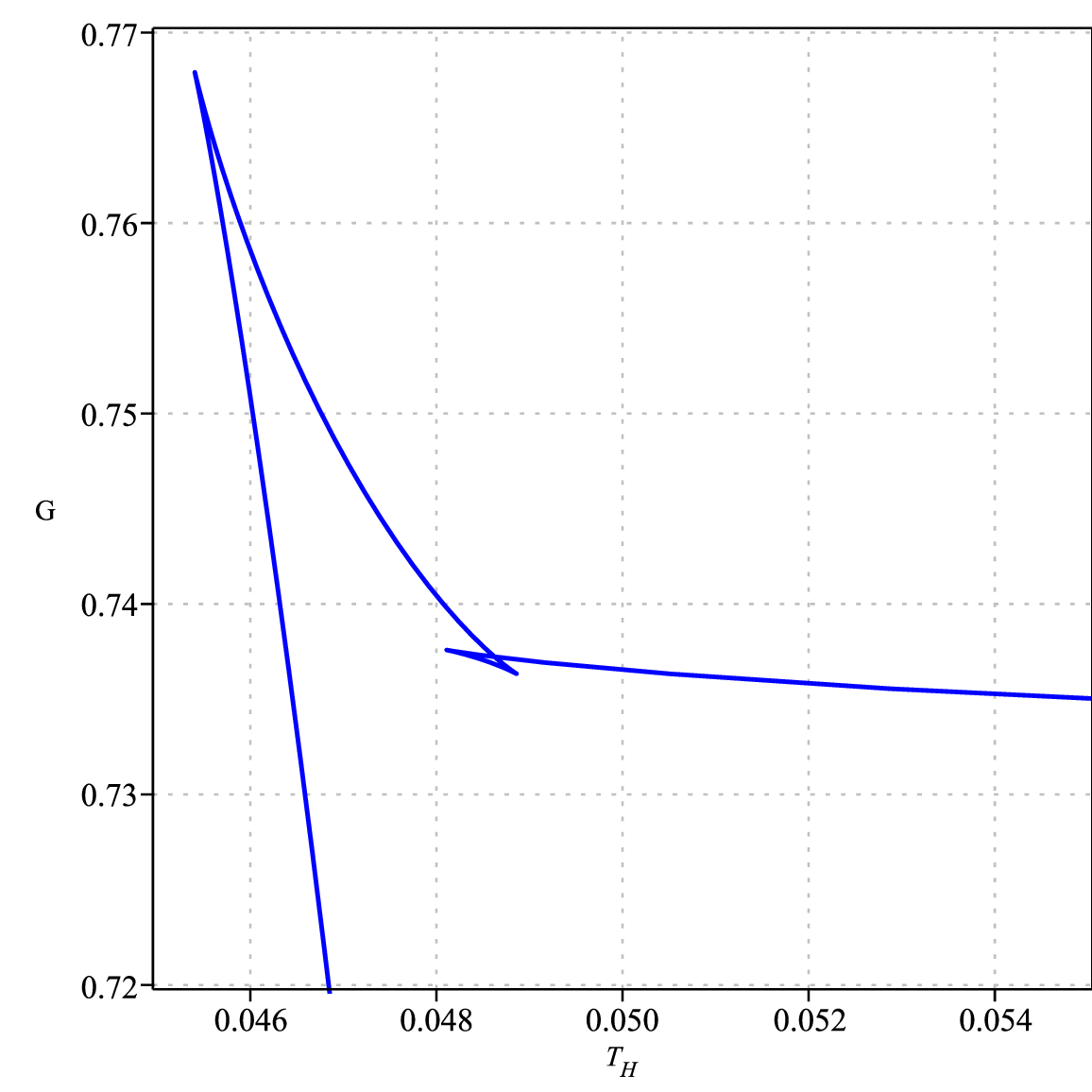}
         \caption{$P_{cp1}<P=0.0038<P_{z}$}
         \label{fig:5.2(n)}
     \end{subfigure}
      \hfill
     \begin{subfigure}[b]{0.3\textwidth}
         \centering
         \includegraphics[width=\textwidth]{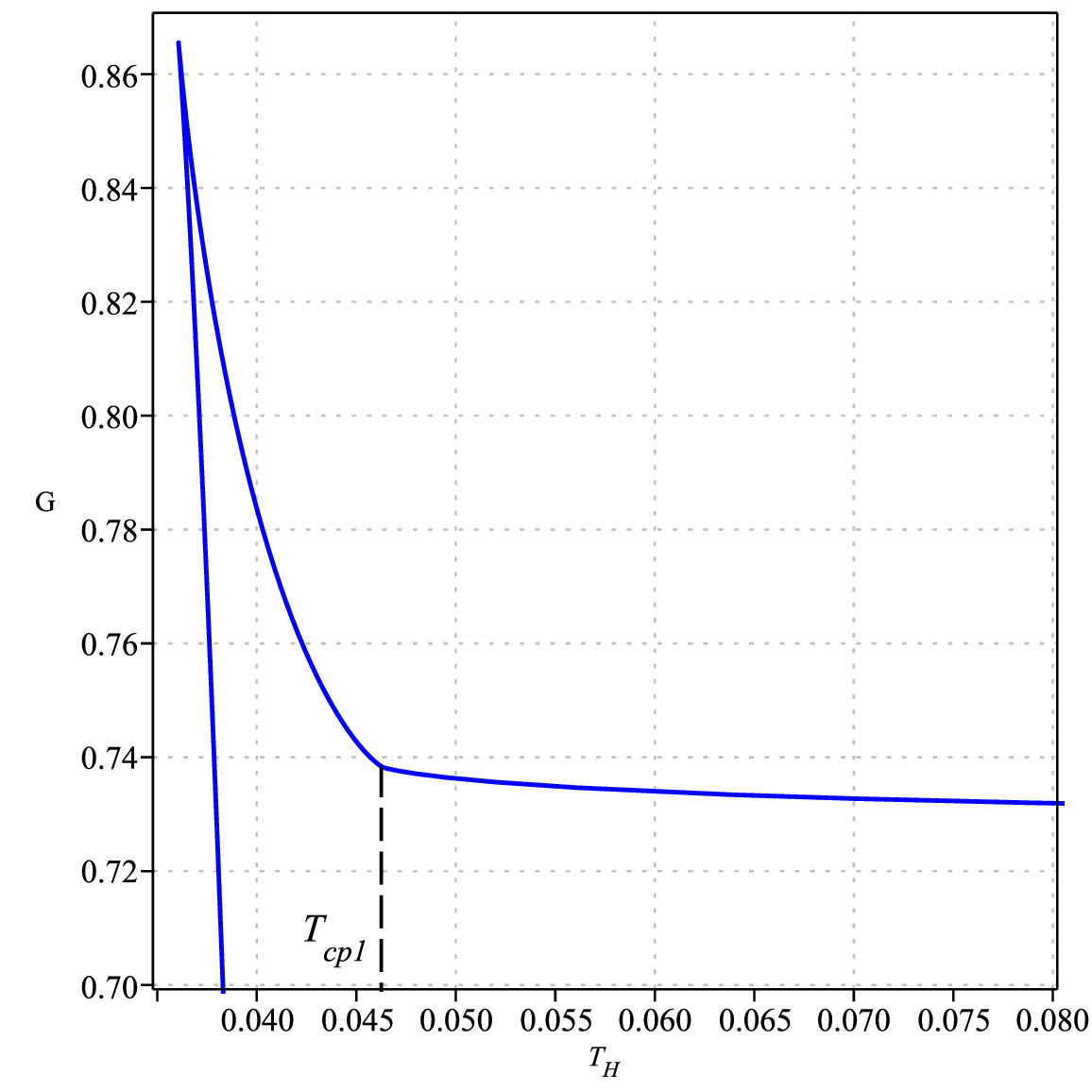}
         \caption{$P=0.0023=P_{cp1}$}
         \label{fig:5.2(0)}
     \end{subfigure}
      \hfill
     \begin{subfigure}[b]{0.3\textwidth}
         \centering
         \includegraphics[width=\textwidth]{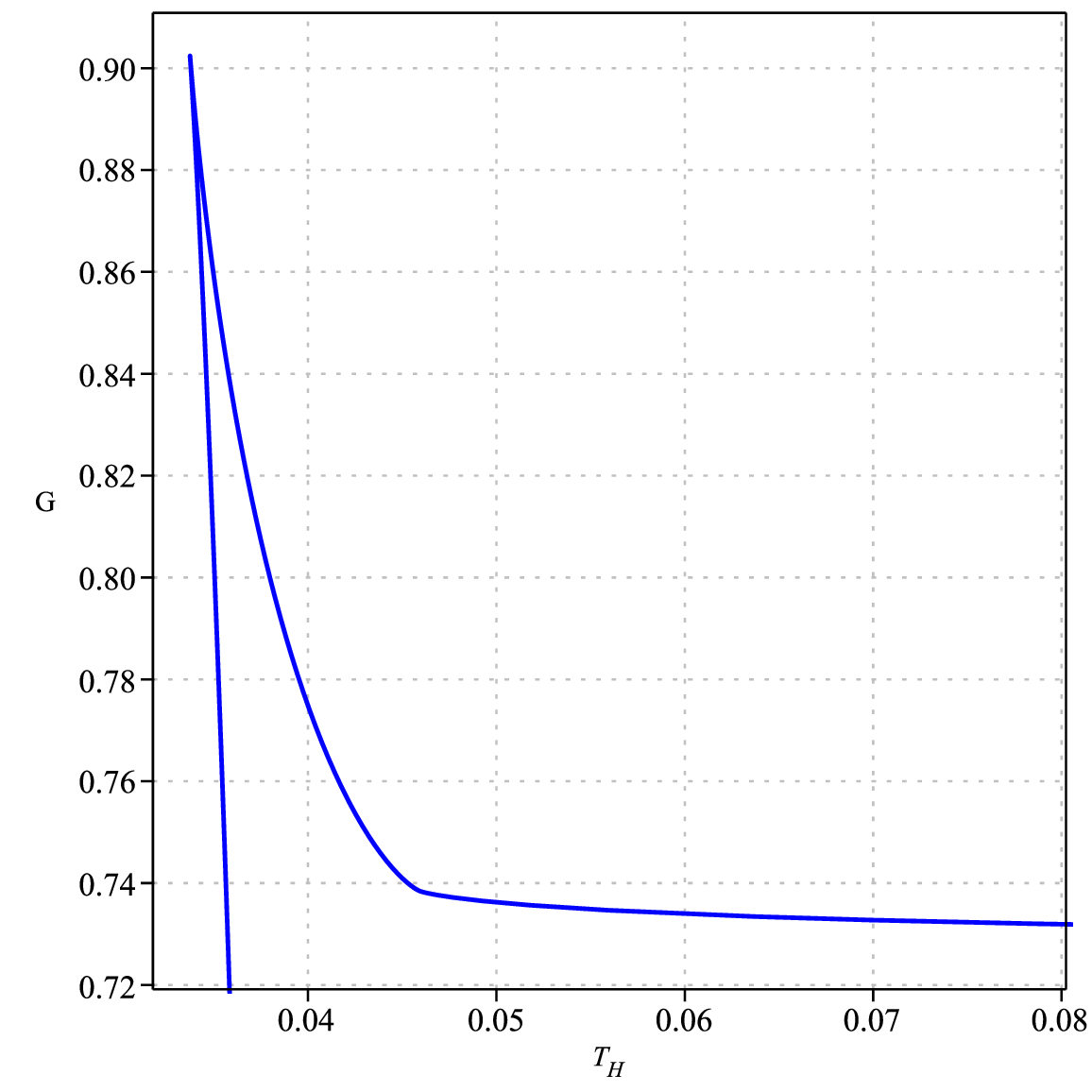}
         \caption{$P=0.0020<P_{cp1}$}
         \label{fig:5.2(p)}
     \end{subfigure}
        \caption{Two critical points with two positive critical pressures.}
        \label{fig:RPT6}
\end{figure}

$a < a_{0}$ admits only one real solution. For $a > a_2$ no critical points occur. Finally, the critical radius $v_c$ can be written as
\begin{equation}\label{eq:22}
v_c=\sqrt{x-4k^2} \text{,}  \:    x=
x_0 \; \& \; x_1, \: \text{where} \;  a_0  < a < a_2,
\end{equation}
\begin{equation}\label{eq:23}
T_c = \frac{(4 k^{2}+v_{c}^{2})^{2} (c c_{1} m^{2} v_{c} +4 a )
-64 Q_{m}^{2} k^{2}-32 Q_{m}^{2} v_{c}^{2}}{4 \pi  v_{c} (4 k^{2}+v_{c}^{2})^{2}},
\end{equation}
\begin{equation}\label{eq:24}
P_c = \frac{a (4 k^{2}+v_{c}^{2})^{2}-16 Q_{m}^{2} k^{2}-12 Q_{m}^{2} v_{c}^{2}}{2 (4 k^{2}+v_{c}^{2})^{2} \pi  v_{c}^{2}}.
\end{equation}
For two critical pressures to be positive, we must have
\begin{equation}\label{eq:25}
 a > {a_1} = \frac{9Q_m}{16\sqrt{\beta}}.
\end{equation}
Excluding the range of $a$ from $a_0$ to $a_1$, we can say that critical points 
are positive and real within the range  $a_1 < a <  a_2$. For $a_0 < a <  a_0$,
two positive and real critical points are obtained with one critical pressure being 
negative. The $G-T_{H}$ diagram is depicted in Figs. \ref{fig:RPT5} and \ref{fig:RPT6}, 
which is the same as the behaviour of $G-T_{H}$ diagram in \ref{sec:RPT1}, i.e., massive 
gravity does not change phase structure of the black holes. The massive gravity
affects critical points. Under the effects of massive gravity critical points 
$(P_{t}, T_{t})$ and $(P_{z}, T_{z})$ take lower values compared to massless Einstein 
gravity.

\begin{figure}[H]
\centering
\subfloat[Two pressures are positive]{\includegraphics[width=.5\textwidth]{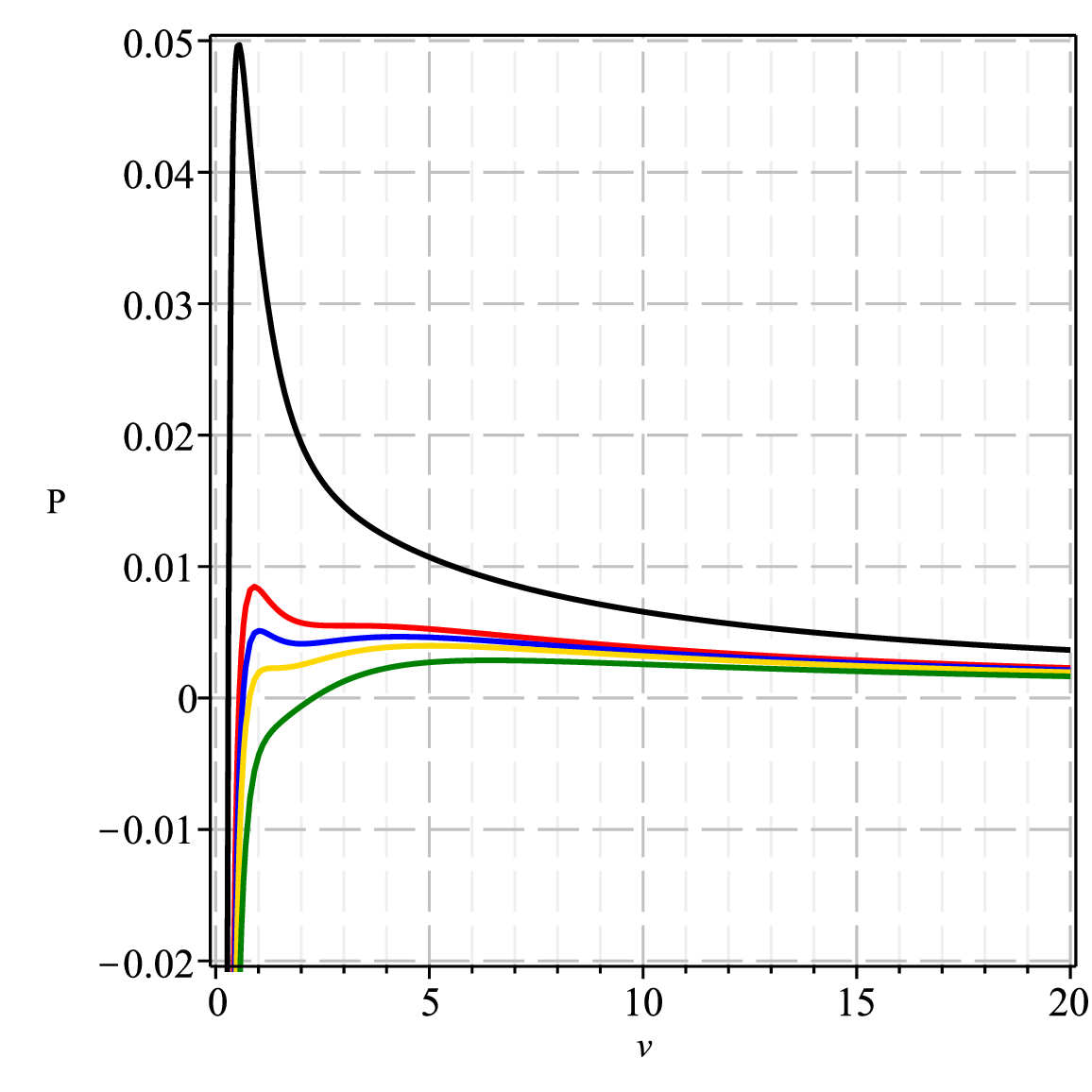}}\hfill
\subfloat[ One pressure is negative]{\includegraphics[width=.5\textwidth]{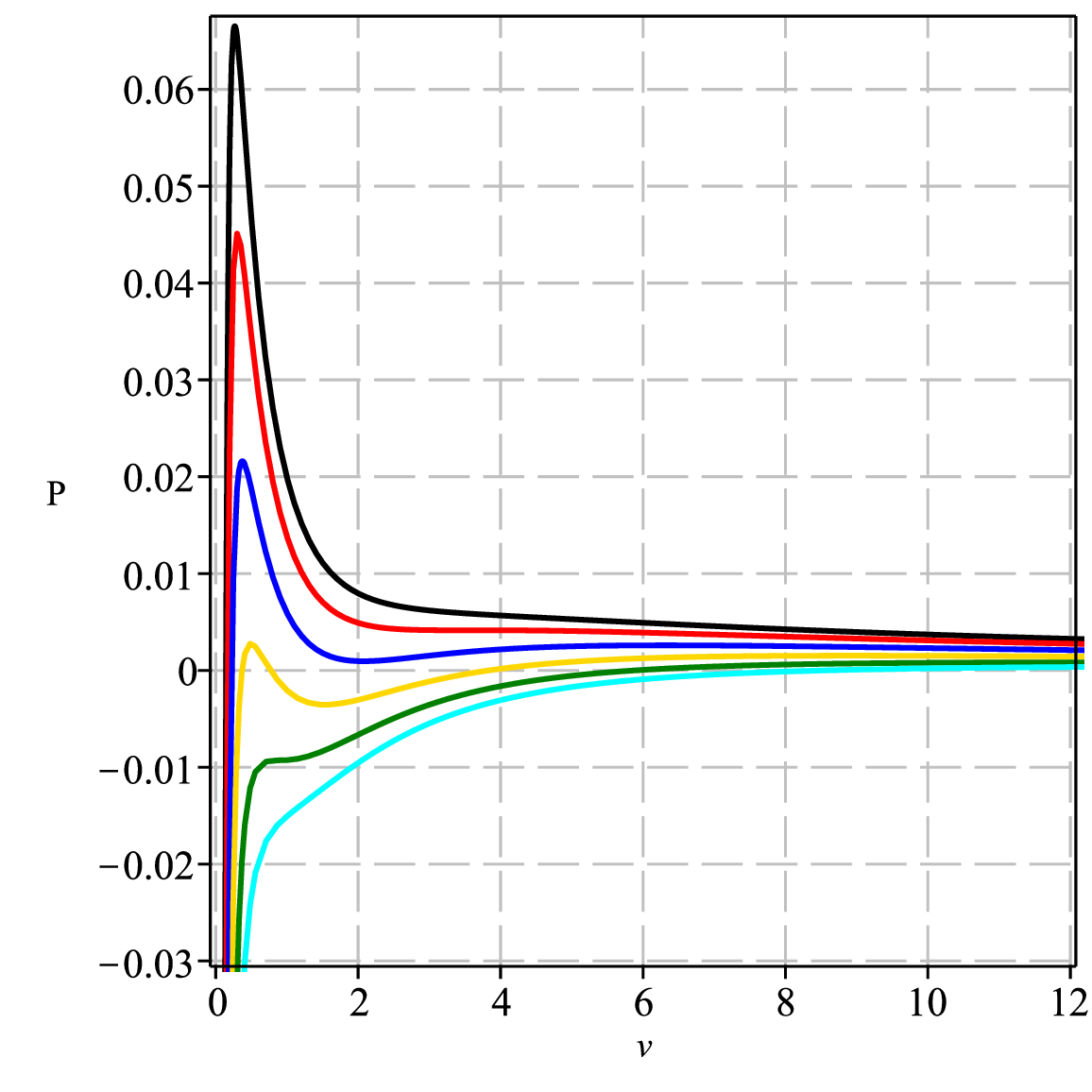}}\hfill
\caption{Left Panel : Black line denotes $T=0.0800>T_{cp2}$,  Red line denotes 
$T=T_{cp2}$, Blue line denotes $T_{cp1}<T=0.0495<T_{cp2}$, Gold line denotes 
$T=T_{cp1}$ \& Green line denotes $T=0.0400<T_{cp1}$. Right Panel :  
Black line denotes $T=0.0500>T_{cp2}$,  Red line denotes $T=T_{cp2}$, 
Blue line denotes $T_{cp1}<T=0.0360<T_{cp2}$, Gold line denotes $T=0.0280<T_{cp2}$, 
Green line denotes $T=T_{cp1}$ \& Cyan line denotes $T=0.0150<T_{cp1}$.}\label{fig:RPT4}
\end{figure}

\begin{table}[H]
\begin{center}
\begin{tabular}{ |c|c|c|c|c| } 
\hline
Case & CP & CP1 & CP2 \\
\hline
\multirow{3}{10em}{$a_1 \leq a=0.99 \leq a_2$} & $v_c$ & 1.3793 & 3.0212 \\ 
& $T_c$ & 0.0463 & 0.0527 \\ 
& $P_c$ & 0.0023 & 0.0055 \\ 
\hline
\multirow{3}{10em}{$a_0 \leq a=0.90 \leq a_1$} & $v_c$ & 0.8817 & 3.5202 \\ 
& $T_c$ &  0.0208 & 0.0439 \\ 
& $P_c$ & -0.0092 & 0.0041 \\ 
\hline
\multirow{3}{10em}{$0.80=a \leq a_0  $} & $v_c$ & $--$ & 4.0398 \\ 
& $T_c$ & $--$  & 0.0355 \\ 
& $P_c$ & $--$  & 0.0030 \\ 
\hline
\end{tabular}
\end{center}
\caption{With $Q_m=1$, $c=1$, $c_1=-1$, $m=0.1$ and $\beta=0.34$}
\label{ta1}
\end{table}

\subsection{Black Holes in \texorpdfstring{$4D$}{TEXT}  EGB massless gravity coupled to NED}\label{sec:RPT3}
In this subsection, we study the RPT of a black hole in massless EGB gravity. The equation
for critical radius is given in equation \ref{eq:4.13}

\begin{equation*}
    \Bigr[ - v_{c}^{10}+\Bigl( -12 k^{2} +24 Q_{m}^{2}+48 \alpha \Bigl) v_{c}^{8} 
    +\Bigl( -48 k^{4}+( 48 Q_{m}^{2}+576 \alpha ) k^{2}+192 \alpha  (Q_{m}^{2}+\alpha )\Bigl) v_{c}^{6}
    \end{equation*}
    \begin{equation}\label{eq:27}
    -64 \Bigl( k^{4}-( Q_{m}^{2}+36 \alpha ) k^{2}+18 \alpha  Q_{m}^{2}-36 \alpha^{2}\Bigl) k^{2} v_{c}^{4} +1536 \alpha  k^{4} \Bigl( 2 k^{2}-Q_{m}^{2}+6 \alpha \Bigl)v_{c}^{2}+12288 \alpha^{2} k^{6} \Bigr]=0.
\end{equation}

Inspired by the rich phase structure of the black hole in Einstein 
gravity coupled to NED (see section \ref{sec:RPT1}), here we consider
a small contribution $(\alpha=0.0001)$ of the GB coupling 
parameter and we set $Q_{m}=1$. In the range, 
$\beta \in (\beta_1, \beta_2)$ equation \ref{eq:27} has three real and positive 
solutions with three real positive critical pressures. We take 
$\beta=0.34$ and obtained three real positive values of the 
critical parameters in table \ref{ta2}. In the range, 
$\beta \in (\beta_0, \beta_1)$ equation \ref{eq:27} has three real and positive 
solutions with two real positive critical pressures and one negative
critical pressure. We take $\beta=0.28$ and obtain the critical 
parameters in table \ref{ta2}.

\begin{table}[H]
\begin{center}
\begin{tabular}{ |c|c|c|c|c|c| } 
\hline
Case & CP & CP1 & CP2 & CP3 \\
\hline
\multirow{3}{10em}{$\beta_1 \leq \beta=0.34 \leq \beta_2$} & $v_c$ & 0.1411 & 2.9622 & 1.4242  \\ 
& $T_c$ & 0.2060 & 0.0545 & 0.0493  \\ 
& $P_c$ & 0.5111 & 0.0056 & 0.0030  \\ 
\hline
\multirow{3}{10em}{$\beta_0 \leq \beta=0.28 \leq \beta_1 $} & $v_c$ & 0.2168 & 3.3181 & 0.8665 \\ 
& $T_c$ & 0.0533 & 0.0525 & 0.0276  \\ 
& $P_c$ & 0.0573 & 0.0051 & -0.0098  \\ 
\hline
\multirow{3}{10em}{$\beta =0.42 \geq \beta_2  $} & $v_c$ & 0.1163 & $--$ & $--$ \\ 
& $T_c$ & 0.3936 & $--$  & $--$ \\ 
& $P_c$ & 1.2206 & $--$  & $--$ \\ 
\hline
\end{tabular}
\end{center}
\caption{With $Q_m=1$ \& $\alpha=0.0001$.}
\label{ta2}
\end{table}

\begin{figure}[H]
\centering
\subfloat[One pressure is negative]{\includegraphics[width=.5\textwidth]{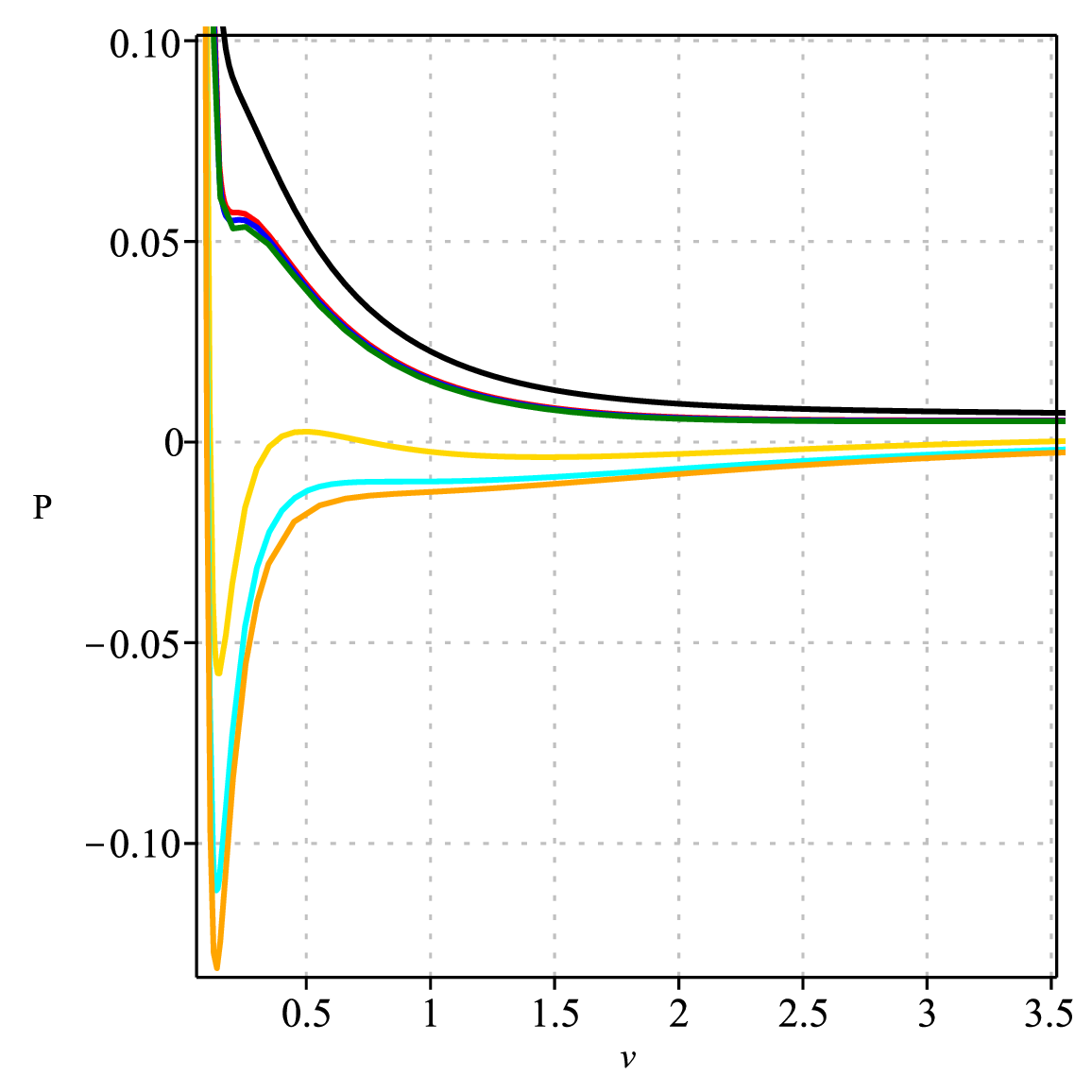}}\hfill
\subfloat[Three pressure are positive]{\includegraphics[width=.5\textwidth]{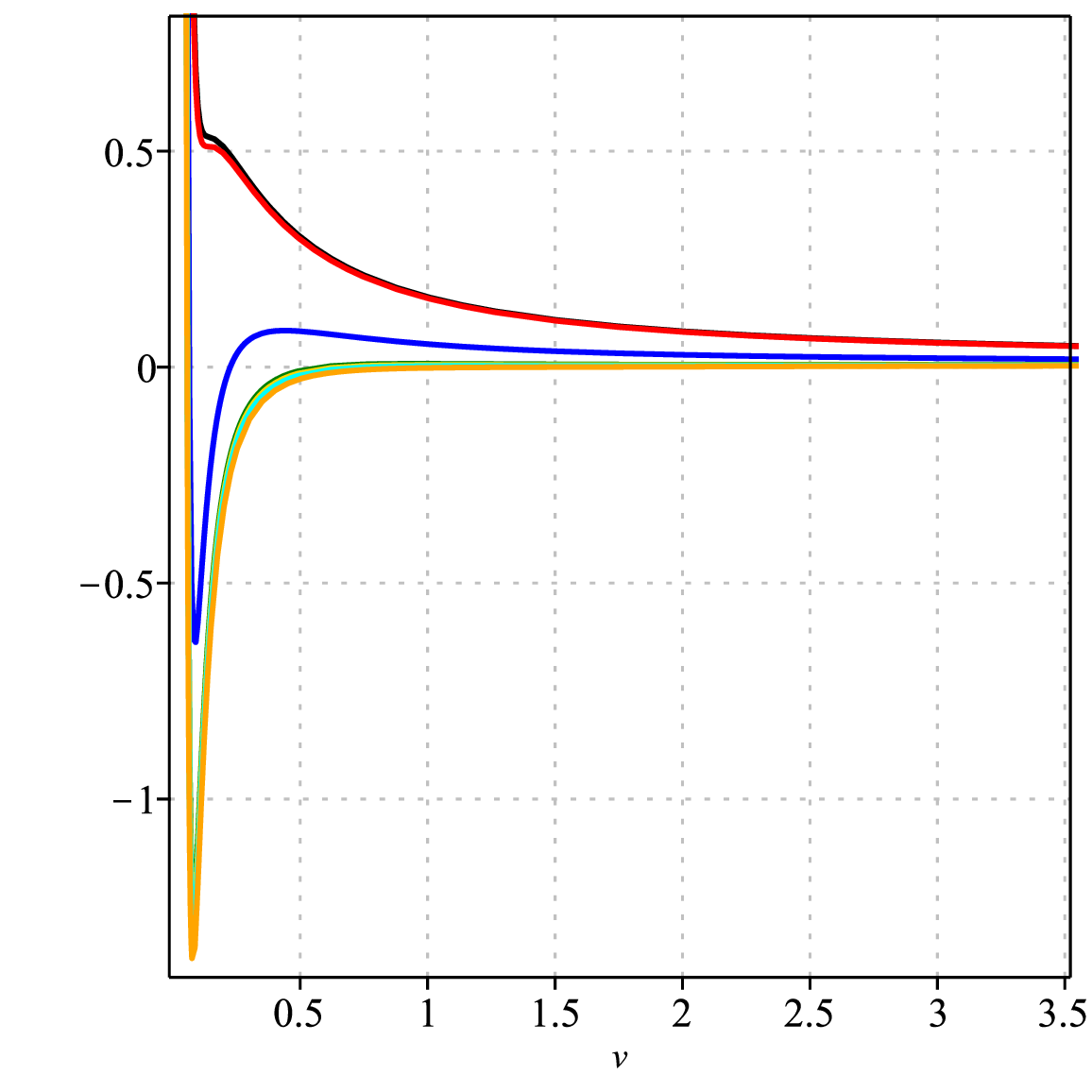}}\hfill
\caption{Left Panel : Black line denotes $T=0.0600>T_{cp1}$,  Red line 
denotes $T=0.0573=T_{cp2}$, Blue line denotes $T_{cp2}<T=0.0529<T_{cp1}$, 
Green line denotes $T=0.0525=T_{cp2}$, Gold line denotes 
$T_{cp3}<T=0.0350<T_{cp2}$, Cyan line denotes $T=0.0350=T_{}cp3$ \& 
Orange line denotes $T=0.0250<T_{cp3}$. Right Panel :  Black line denotes 
$T=0.2090>T_{cp1}$, Red line denotes $T=0.2060=T_{cp1}$, Blue line denotes 
$T_{cp2}<T=0.1000<T_{cp1}$, Green line denotes $T=0.0545=T_{cp2}$, Gold 
line denotes $T_{cp3}<T=0.0518<T_{cp2}$, Cyan line denotes 
$T=0.0493=T_{cp3}$ \& Orange line denotes $T=0.0450<T_{cp3}$.}\label{fig:RPT7}
\end{figure}

\begin{figure}[H]
     \centering
     \begin{subfigure}[b]{0.3\textwidth}
         \centering
         \includegraphics[width=\textwidth]{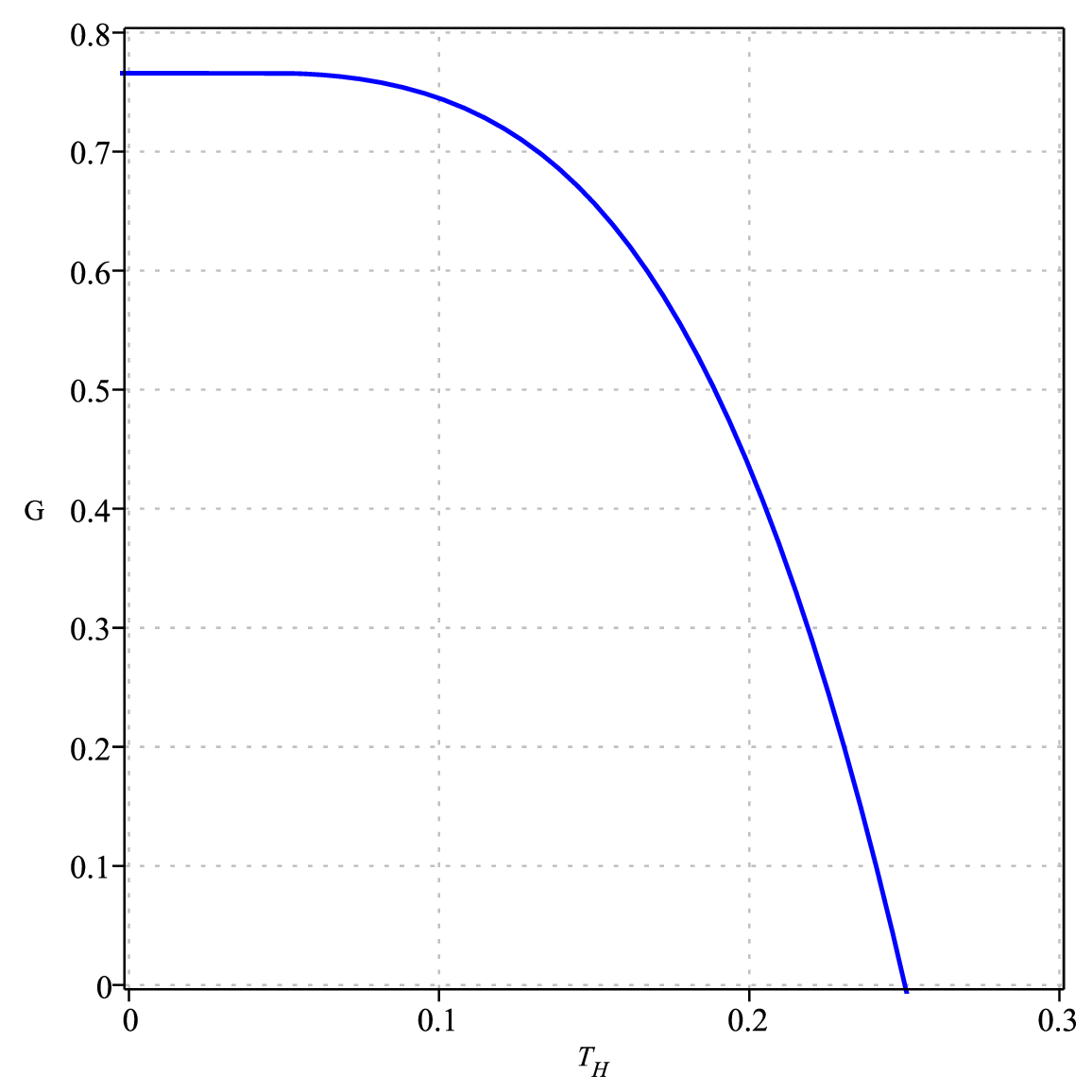}
         \caption{$P=0.0573=P_{cp1}$}
         \label{fig:5.3(a)}
     \end{subfigure}
     \hfill
     \begin{subfigure}[b]{0.3\textwidth}
         \centering
         \includegraphics[width=\textwidth]{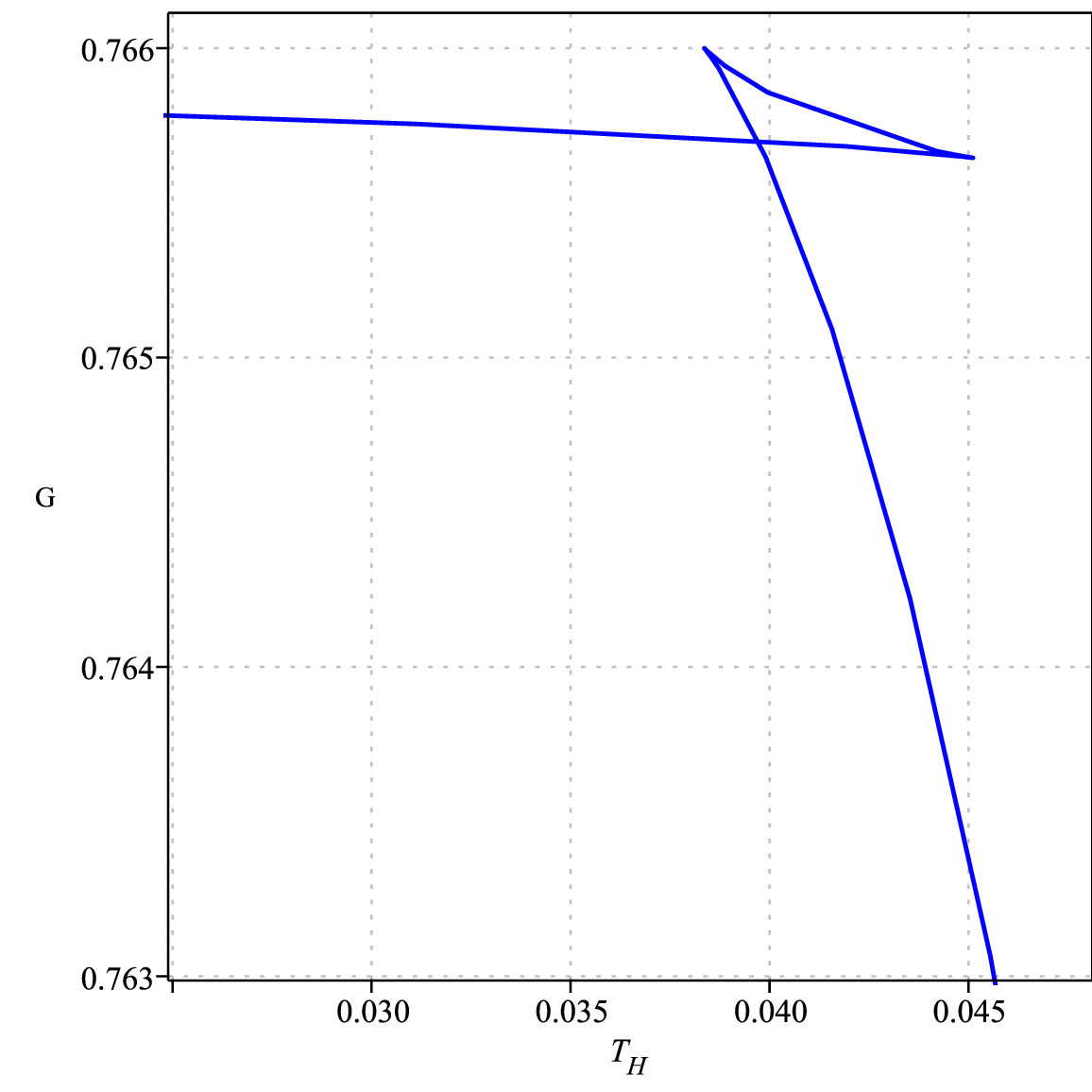}
         \caption{$P=0.0100<P_{cp1}$}
         \label{fig:5.3(b)}
     \end{subfigure}
     \hfill
     \begin{subfigure}[b]{0.3\textwidth}
         \centering
         \includegraphics[width=\textwidth]{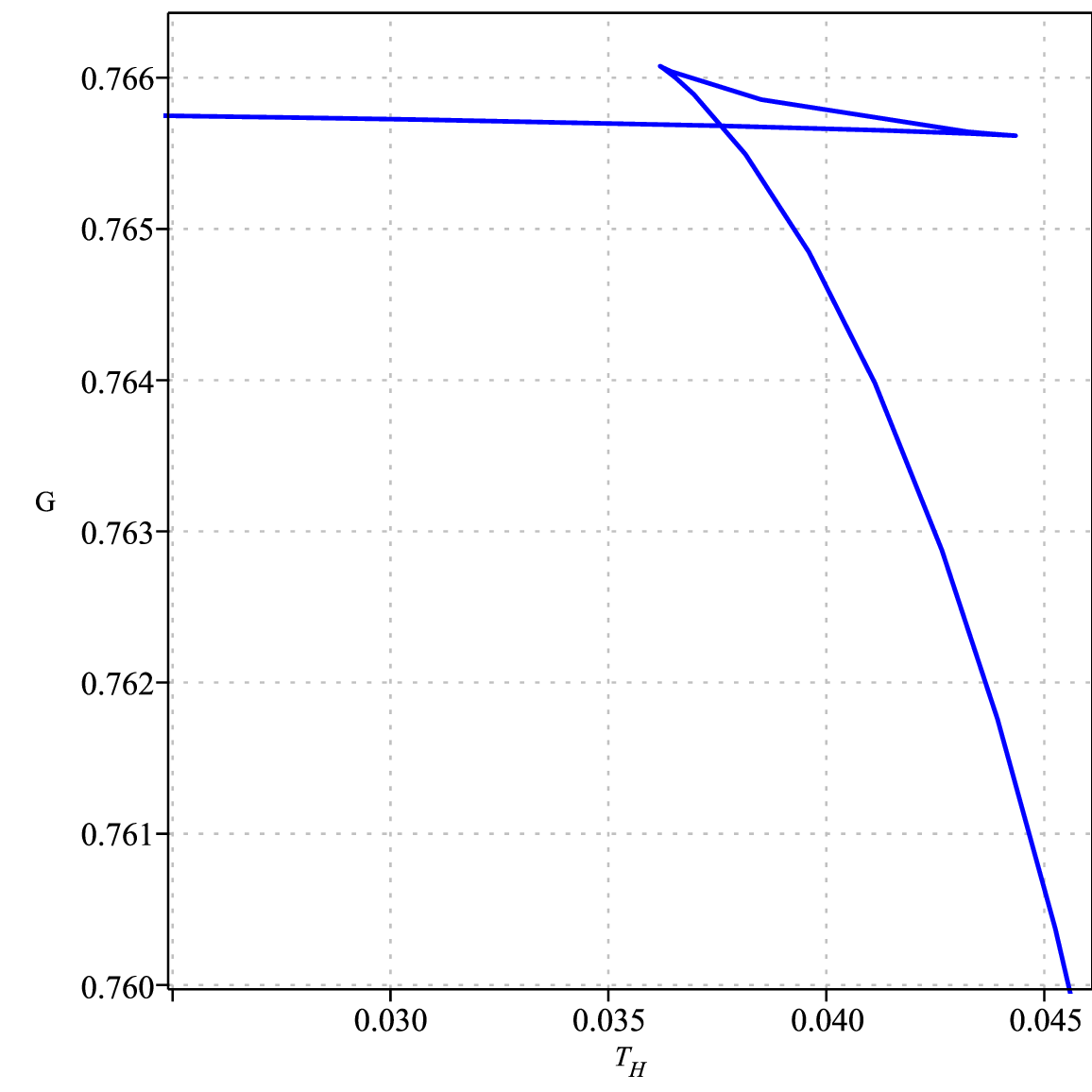}
         \caption{$P=0.0051=P_{cp2}$}
         \label{fig:5.3(c)}
     \end{subfigure}
     \hfill
     \begin{subfigure}[b]{0.3\textwidth}
         \centering
         \includegraphics[width=\textwidth]{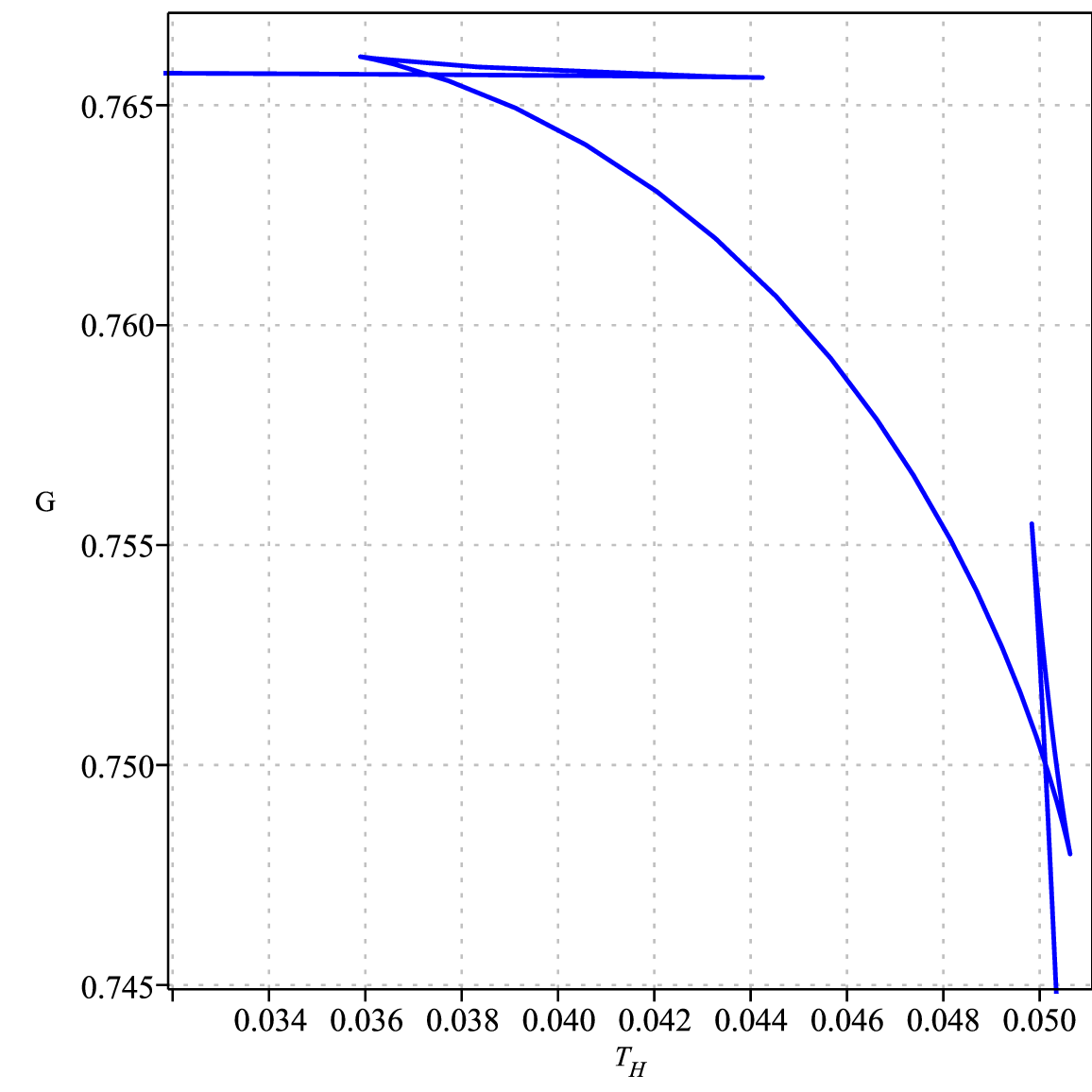}
         \caption{$P_{t}<P=0.0045<P_{cp2}$}
         \label{fig:5.3(d)}
     \end{subfigure}
     \hfill
     \begin{subfigure}[b]{0.3\textwidth}
         \centering
         \includegraphics[width=\textwidth]{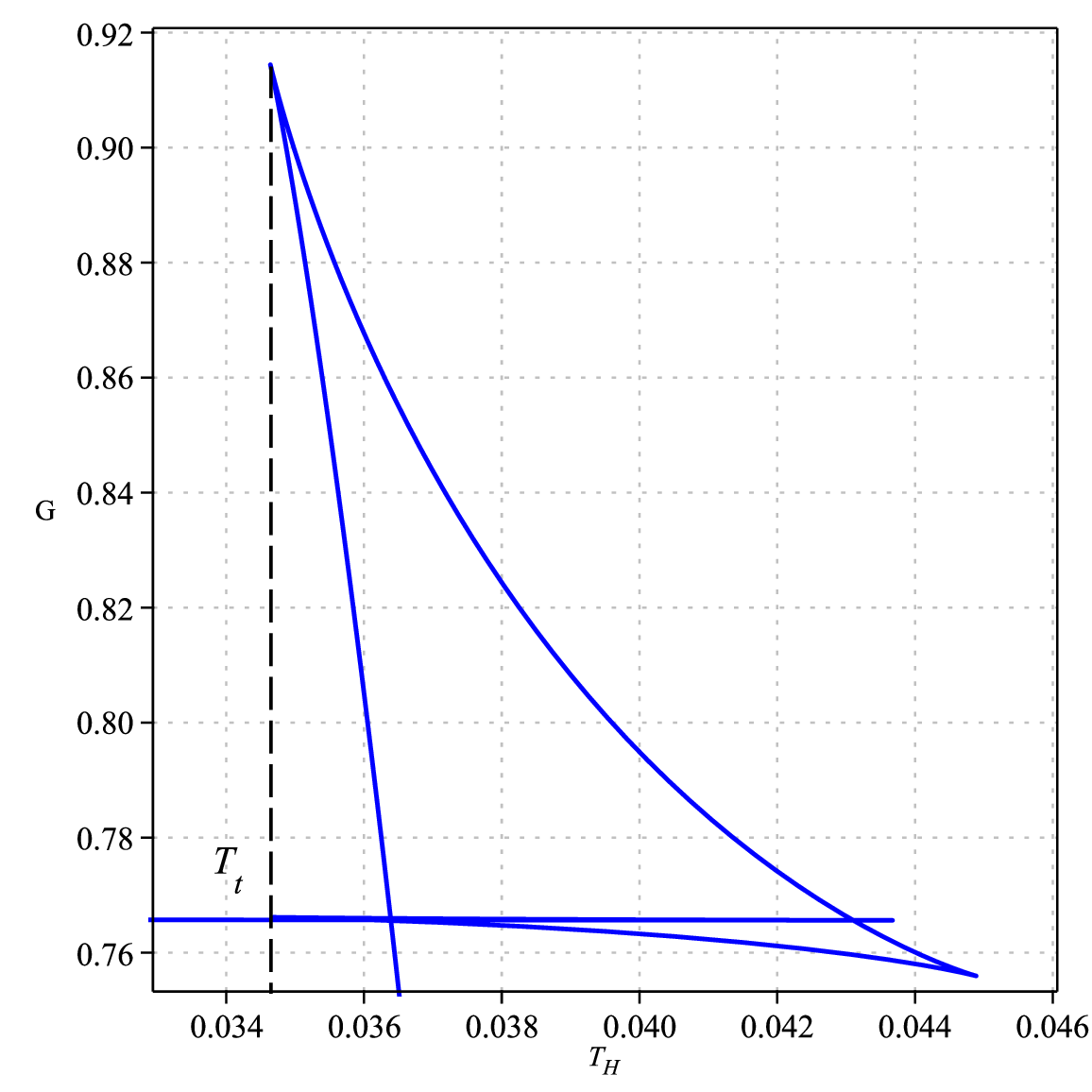}
         \caption{$P=0.00199=P_{t}$}
         \label{fig:5.3(e)}
     \end{subfigure}
     \hfill
     \begin{subfigure}[b]{0.3\textwidth}
         \centering
         \includegraphics[width=\textwidth]{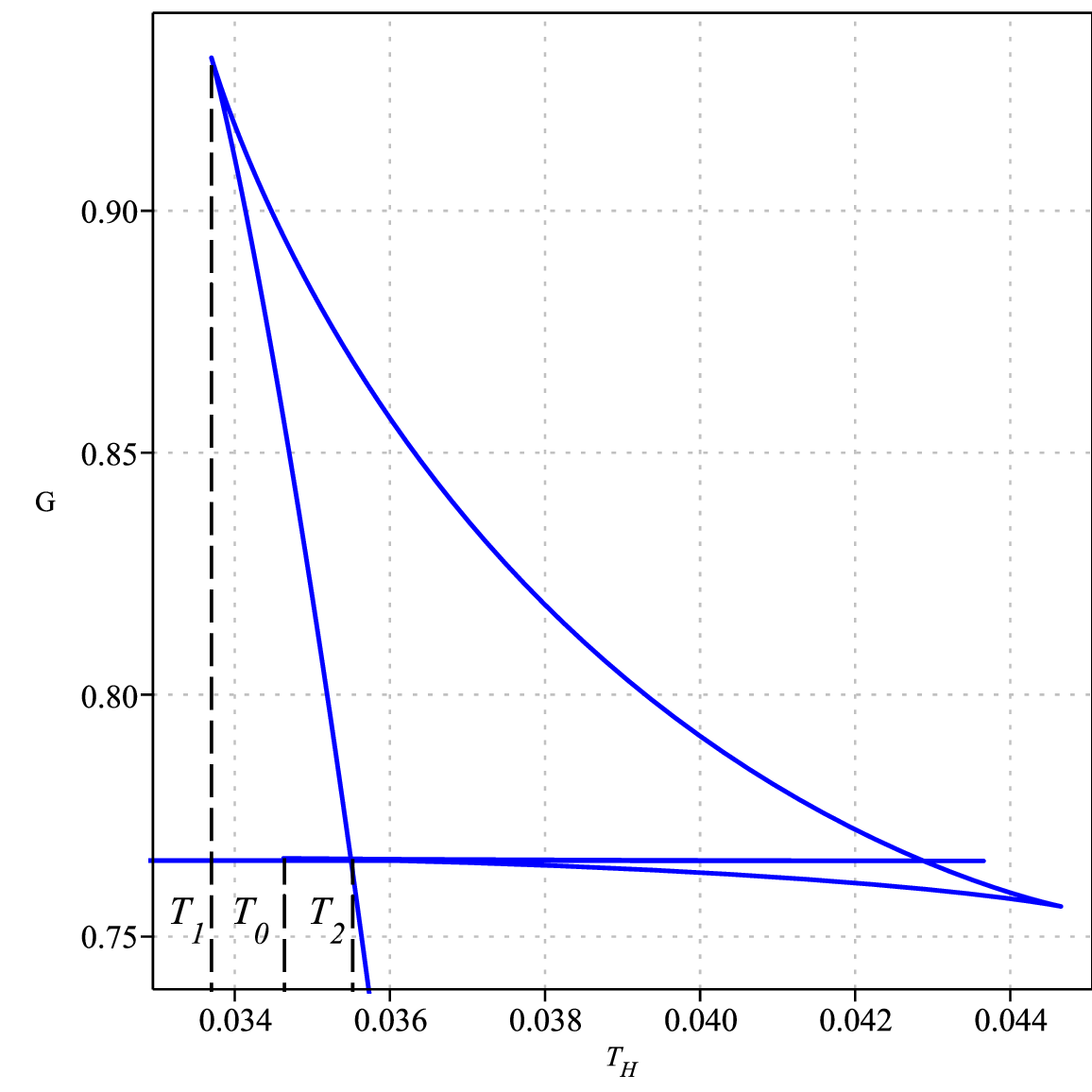}
         \caption{$P_{z}<P=0.001877<P_{t}$}
         \label{fig:5.3(f)}
     \end{subfigure}
      \hfill
     \begin{subfigure}[b]{0.3\textwidth}
         \centering
         \includegraphics[width=\textwidth]{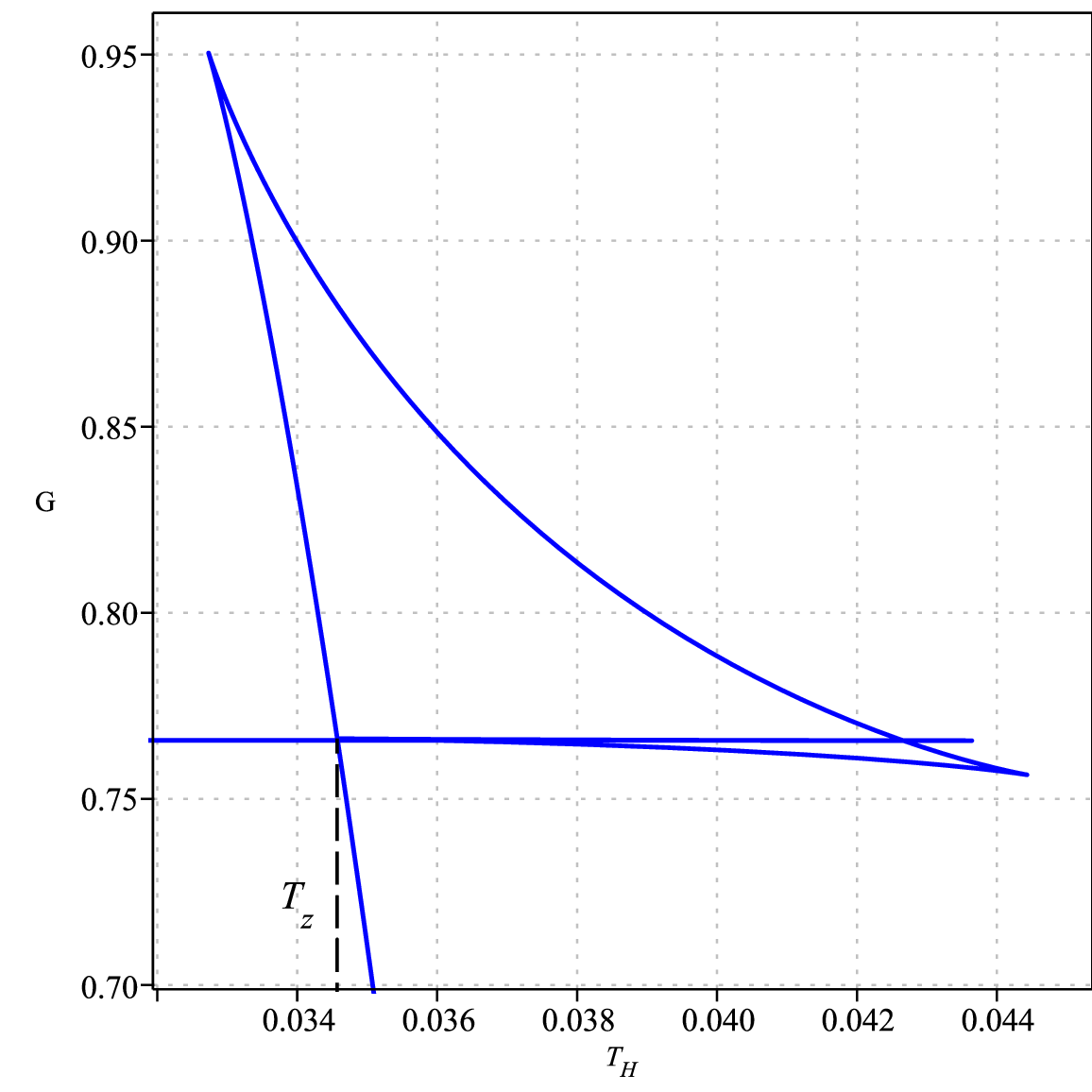}
         \caption{$P=0.001765=P_{z}$}
         \label{fig:5.3(g)}
     \end{subfigure}
      \hfill
     \begin{subfigure}[b]{0.3\textwidth}
         \centering
         \includegraphics[width=\textwidth]{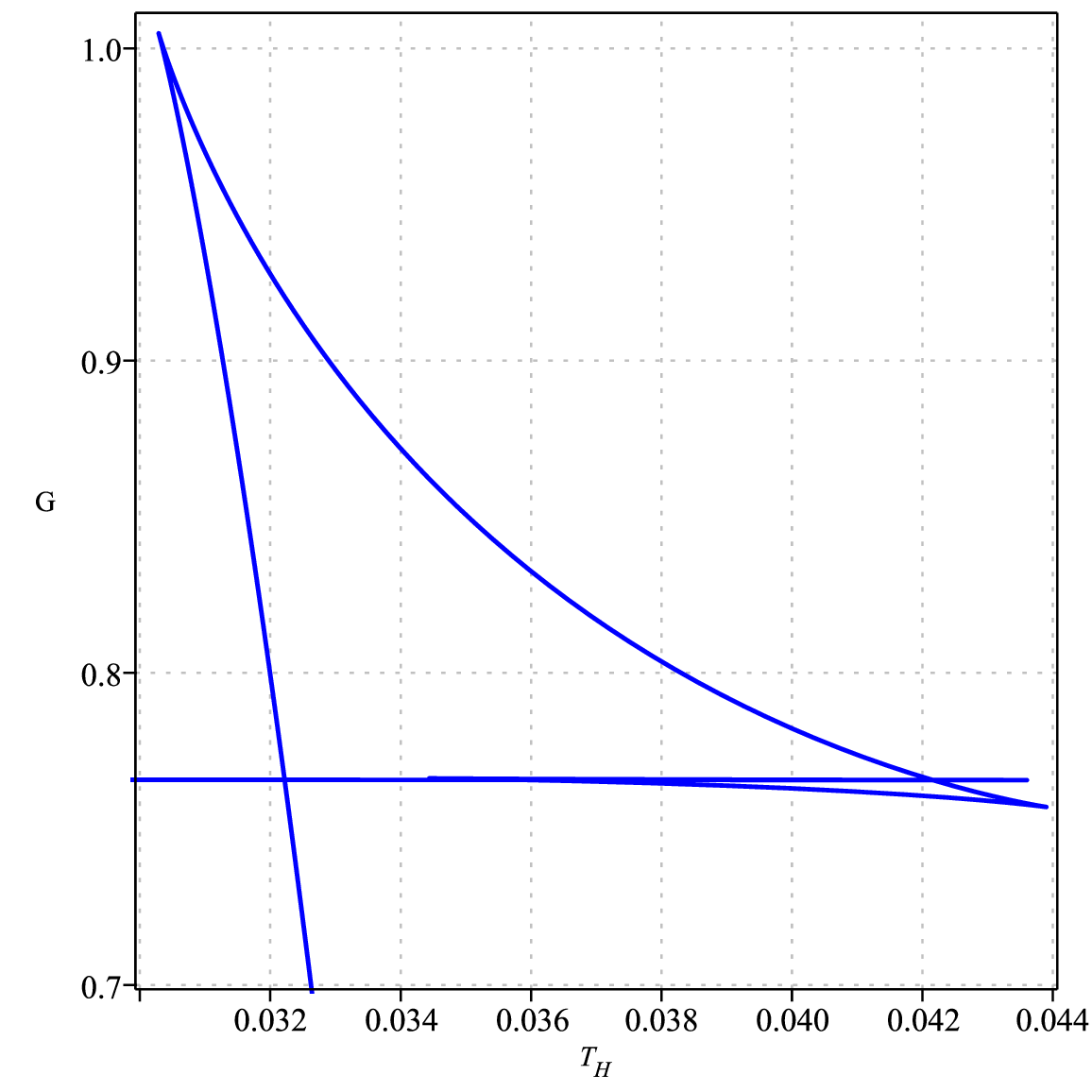}
         \caption{$P=0.0015<P_{z}$}
         \label{fig:5.3(h)}
     \end{subfigure}
        \caption{Three critical points, two at positive pressures and one at negative pressure.}
        \label{fig:RPT8}
\end{figure}

\begin{figure}[H]
     \centering
     \begin{subfigure}[b]{0.3\textwidth}
         \centering
         \includegraphics[width=\textwidth]{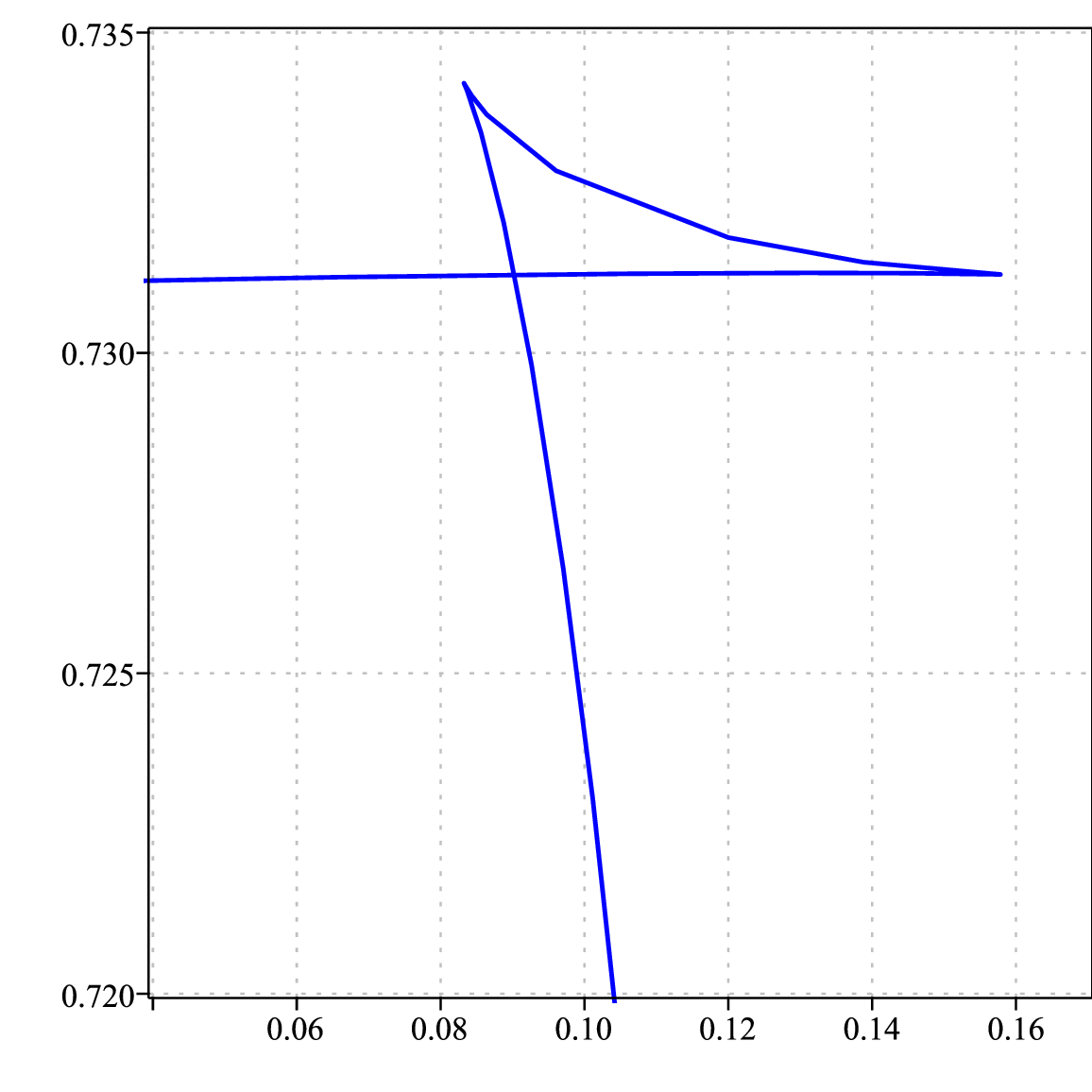}
         \caption{$P=0.0500<P_{cp1}$}
         \label{fig:5.3(i)}
     \end{subfigure}
     \hfill
     \begin{subfigure}[b]{0.3\textwidth}
         \centering
         \includegraphics[width=\textwidth]{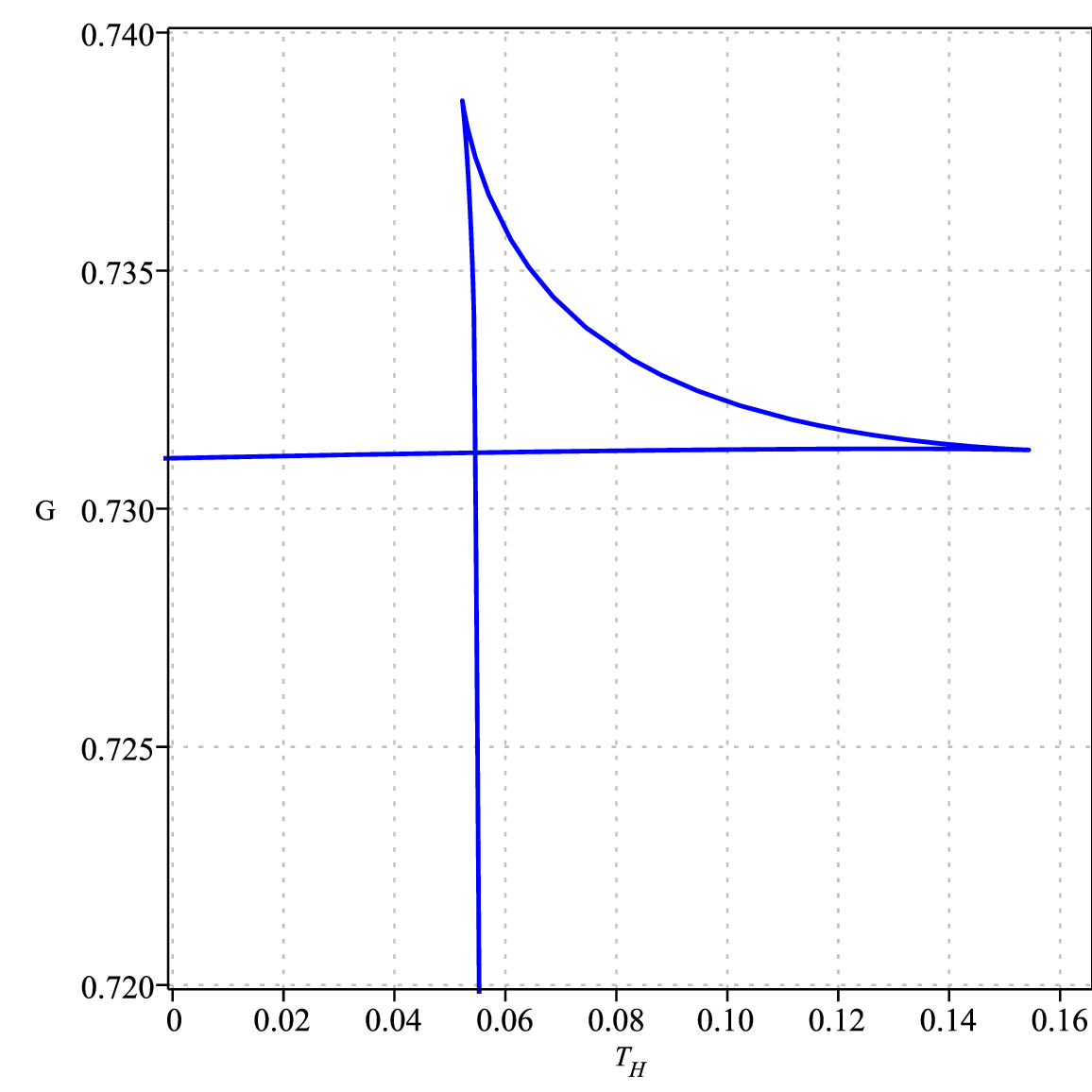}
         \caption{$P=0.0056=P_{cp2}$}
         \label{fig:5.3(j)}
     \end{subfigure}
     \hfill
     \begin{subfigure}[b]{0.3\textwidth}
         \centering
         \includegraphics[width=\textwidth]{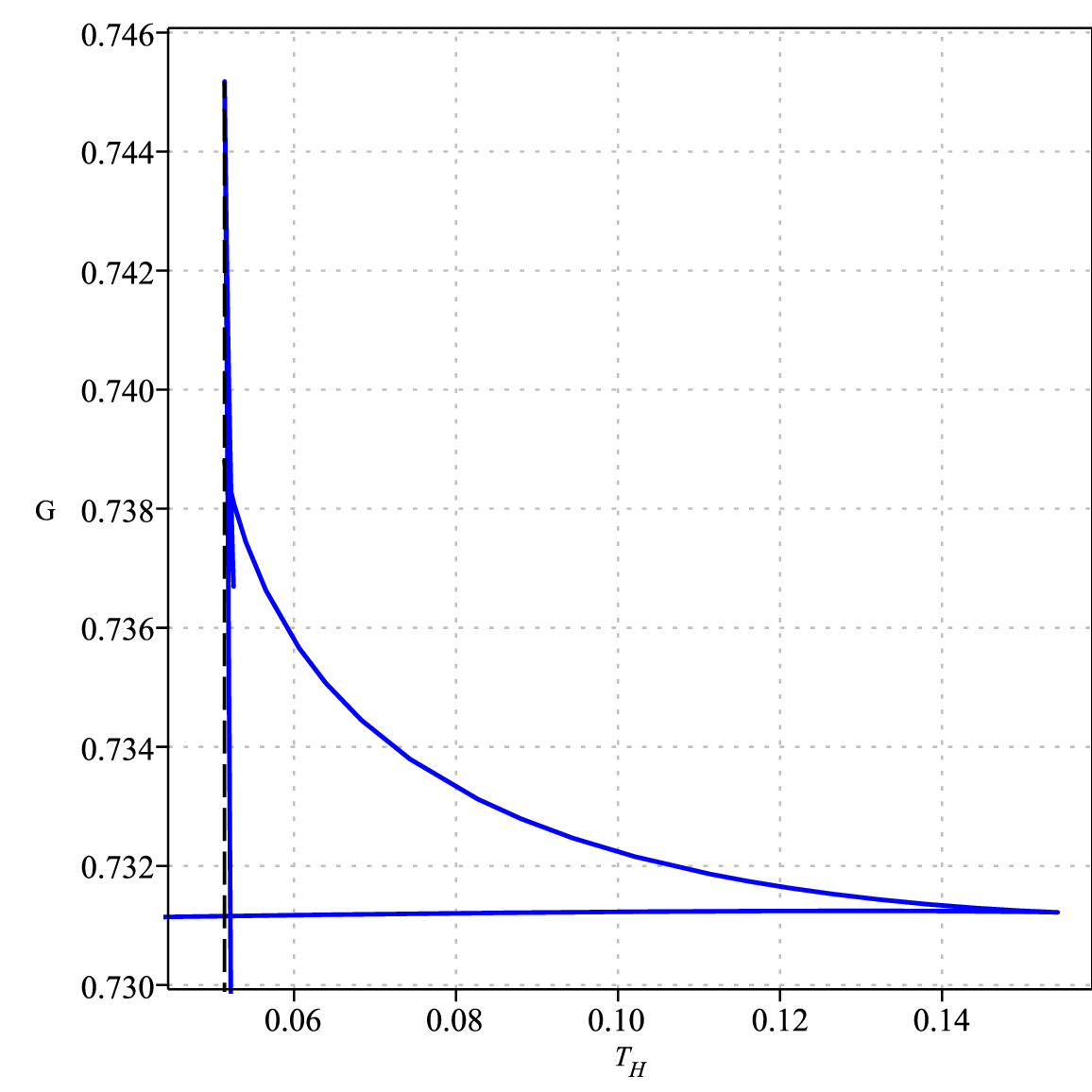}
         \caption{$P=0.00484=P_{t}$}
         \label{fig:5.3(k)}
     \end{subfigure}
     \hfill
     \begin{subfigure}[b]{0.3\textwidth}
         \centering
         \includegraphics[width=\textwidth]{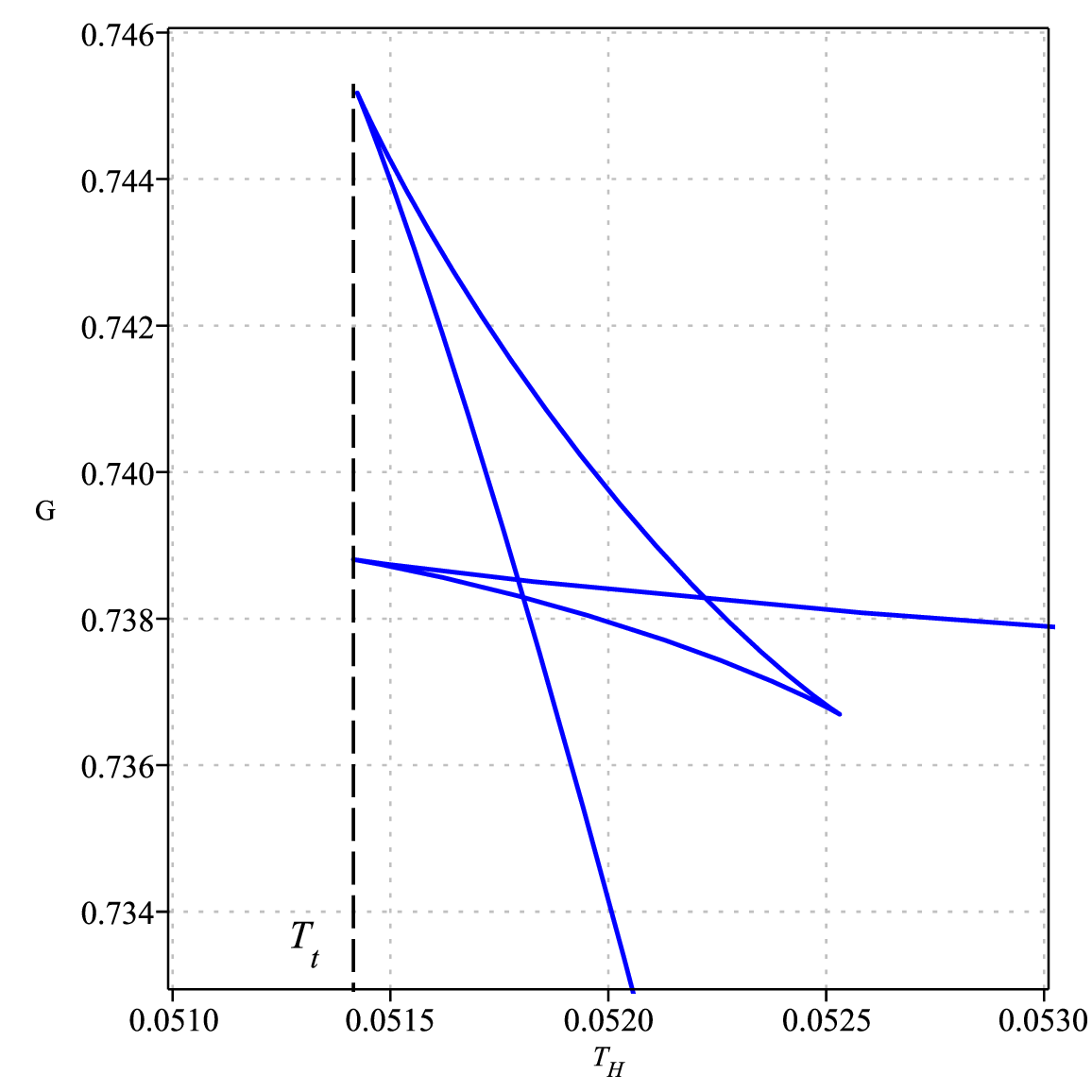}
         \caption{small scale of Fig. \ref{fig:5.3(k)}}
         \label{fig:5.3(l)}
     \end{subfigure}
     \hfill
     \begin{subfigure}[b]{0.3\textwidth}
         \centering
         \includegraphics[width=\textwidth]{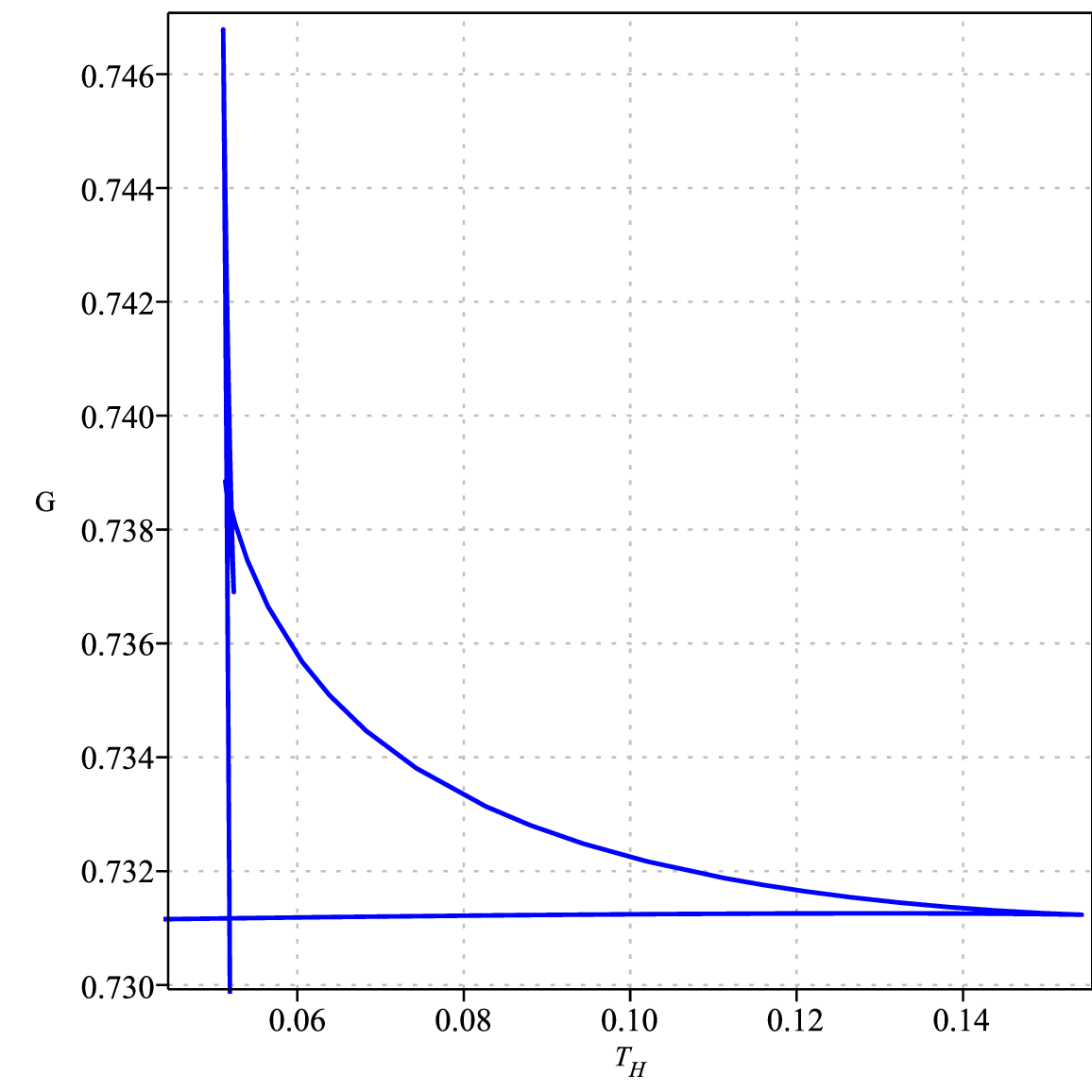}
         \caption{$P_{z}<P=0.00476<P_{t}$}
         \label{fig:5.3(m)}
     \end{subfigure}
     \hfill
     \begin{subfigure}[b]{0.3\textwidth}
         \centering
         \includegraphics[width=\textwidth]{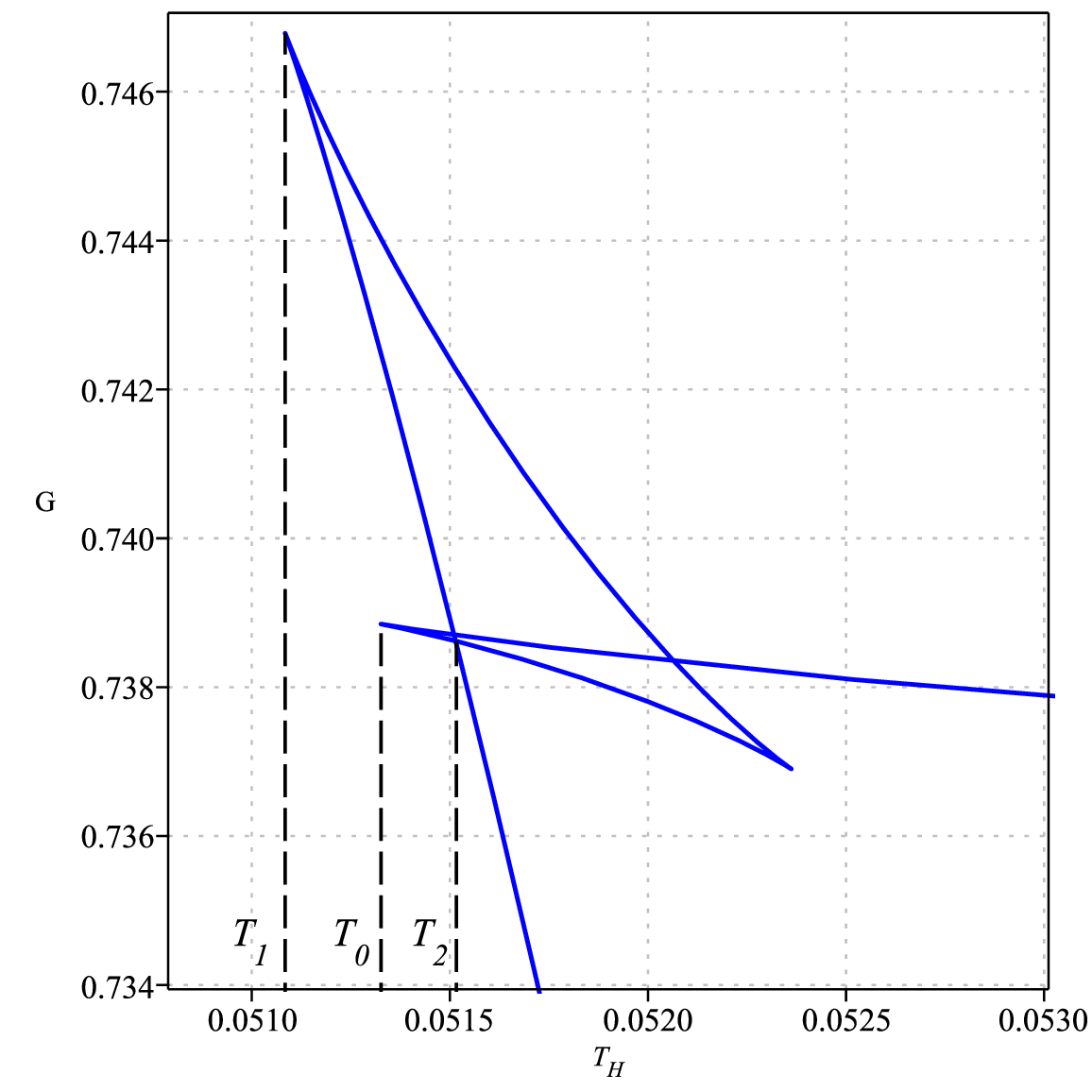}
         \caption{small scale of Fig. \ref{fig:5.3(m)}}
         \label{fig:5.3(n)}
     \end{subfigure}
      \hfill
     \begin{subfigure}[b]{0.3\textwidth}
         \centering
         \includegraphics[width=\textwidth]{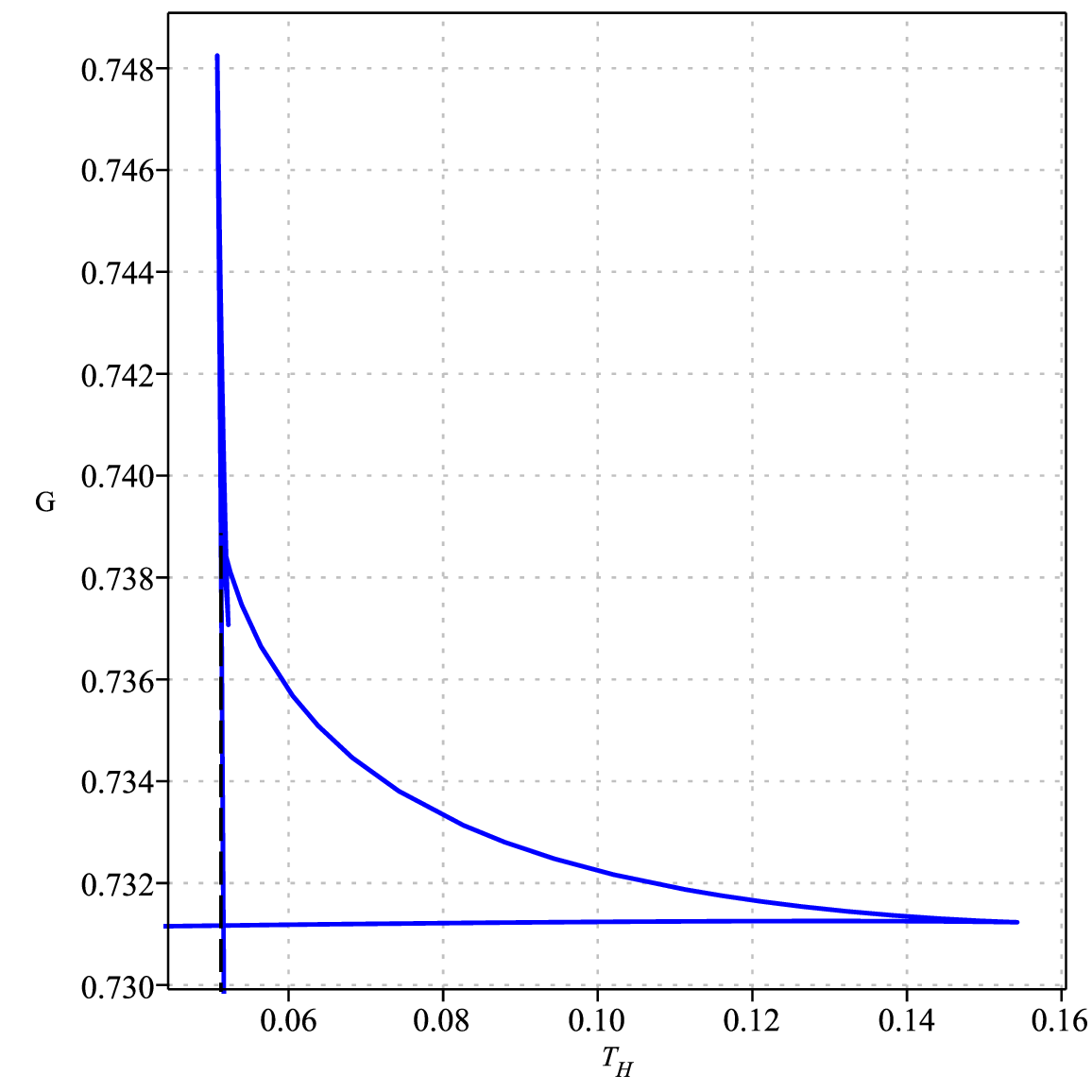}
         \caption{$P=0.00469=P_{z}$}
         \label{fig:5.3(o)}
     \end{subfigure}
      \hfill
     \begin{subfigure}[b]{0.3\textwidth}
         \centering
         \includegraphics[width=\textwidth]{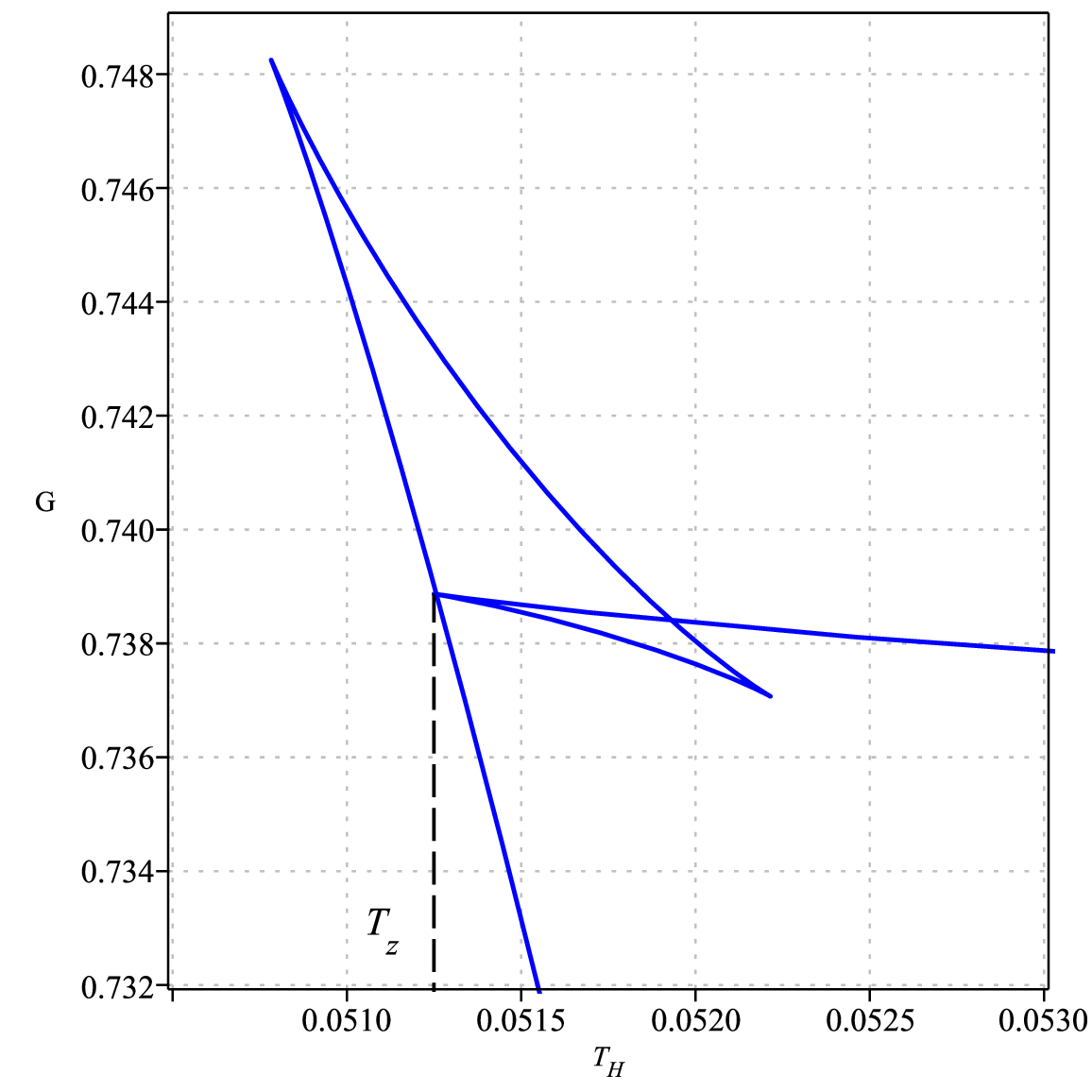}
         \caption{small scale of Fig. \ref{fig:5.3(o)}}
         \label{fig:5.3(p)}
     \end{subfigure}
     \begin{subfigure}[b]{0.3\textwidth}
         \centering
         \includegraphics[width=\textwidth]{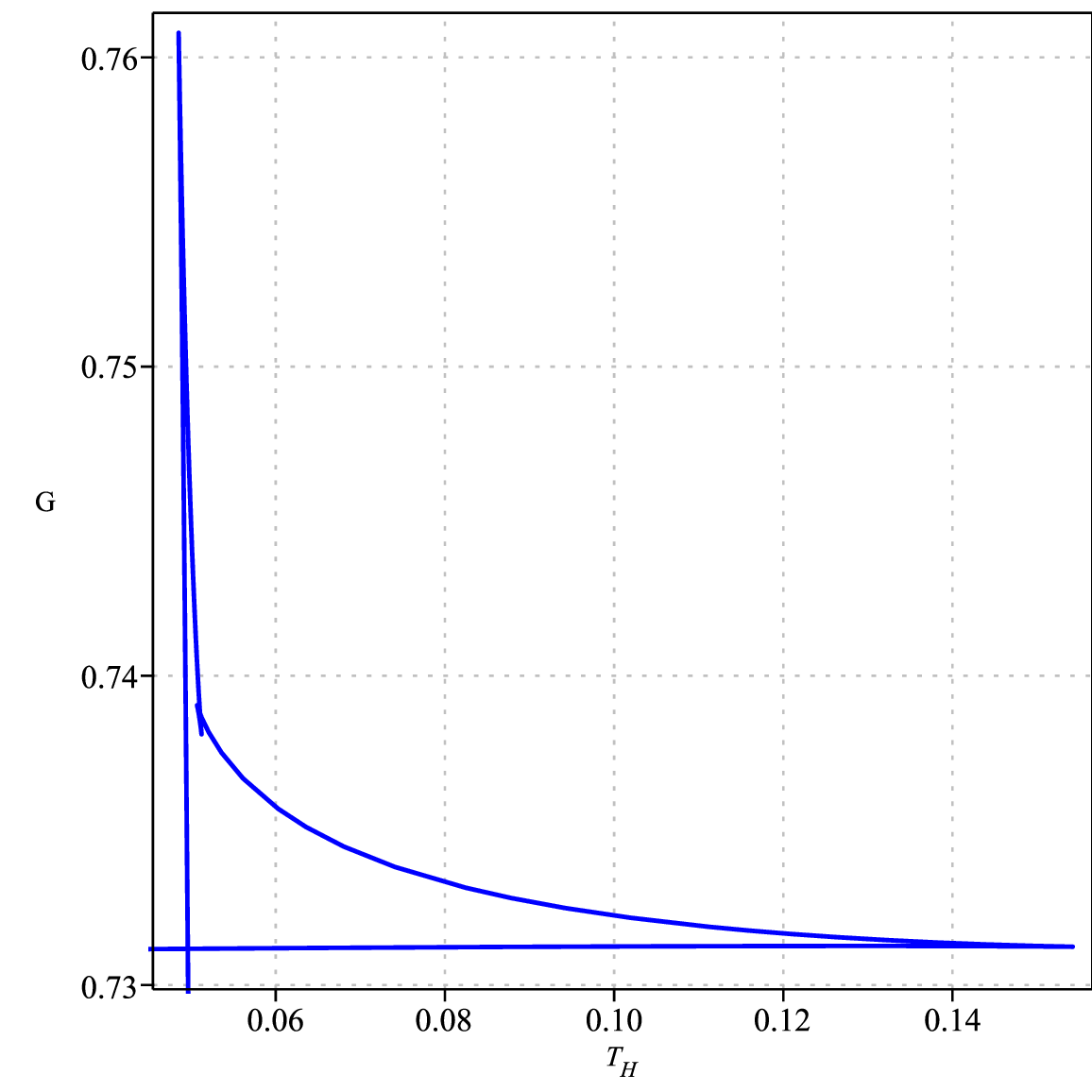}
         \caption{$P_{cp3}<P=0.0042<P_{z}$}
         \label{fig:5.3(q)}
     \end{subfigure}
     \begin{subfigure}[b]{0.3\textwidth}
         \centering
         \includegraphics[width=\textwidth]{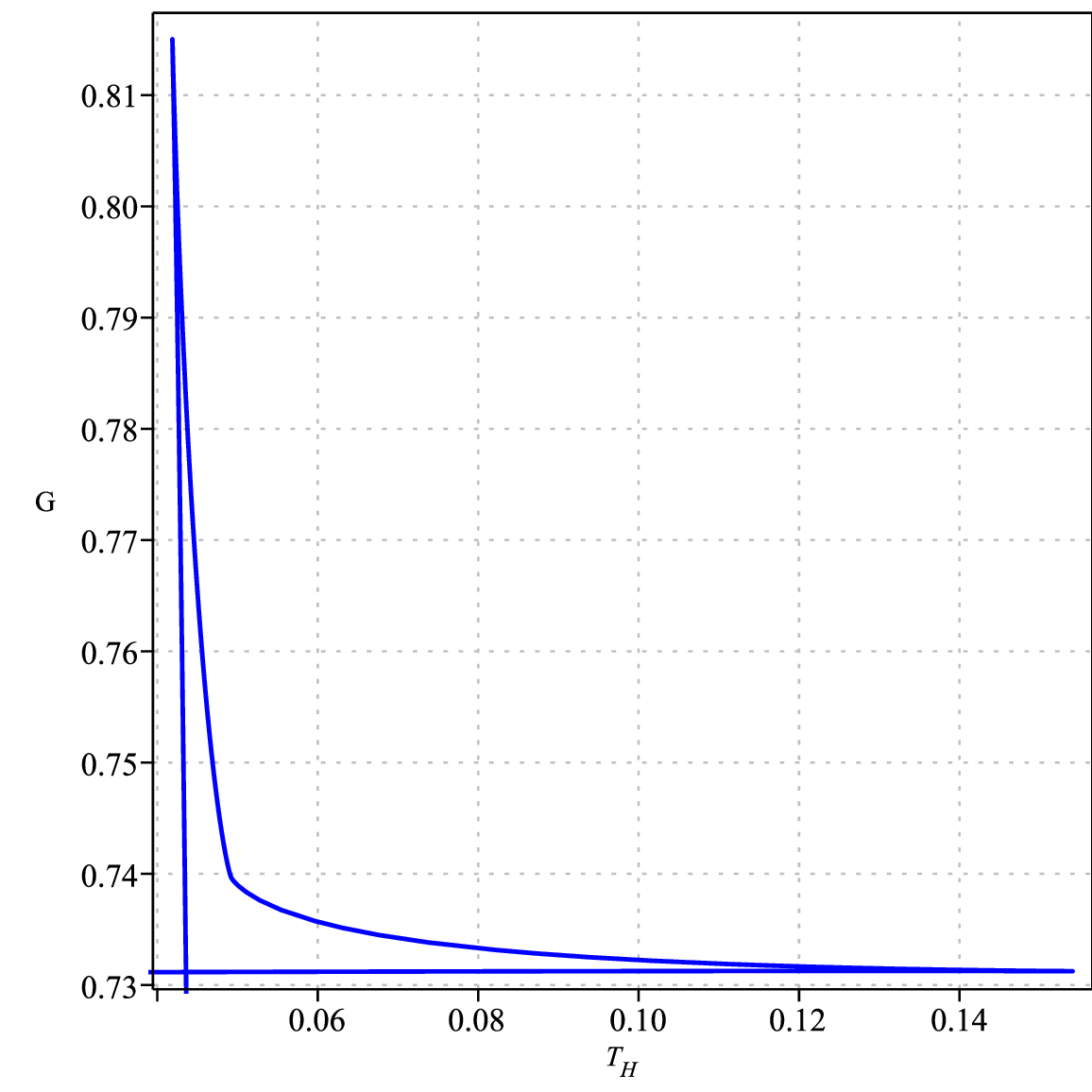}
         \caption{$P=0.0030=P_{cp3}$}
         \label{fig:5.3(r)}
     \end{subfigure}
      \begin{subfigure}[b]{0.3\textwidth}
         \centering
         \includegraphics[width=\textwidth]{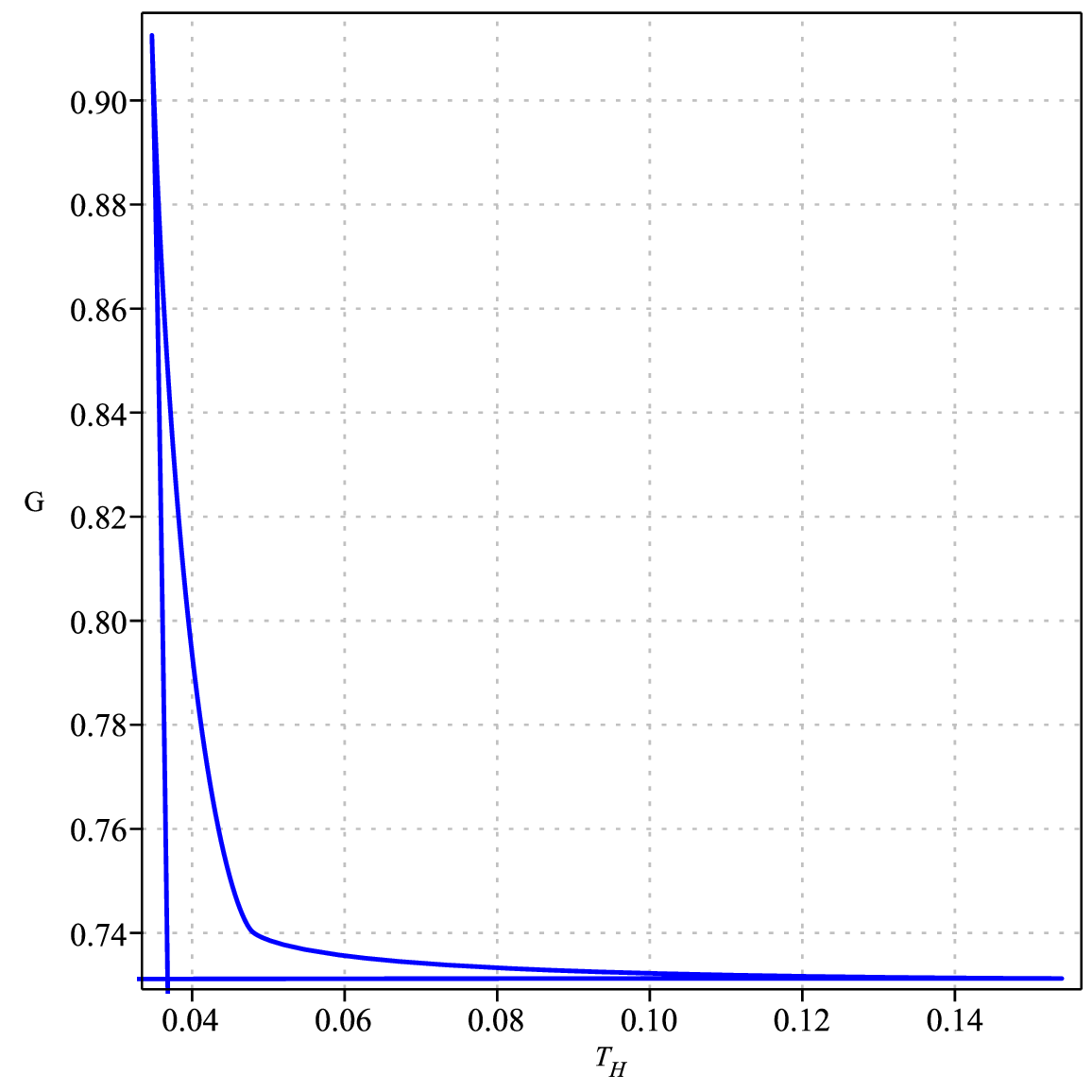}
         \caption{$P=0.0020<P_{cp3}$}
         \label{fig:5.3(s)}
     \end{subfigure}
        \caption{Three critical points with three positive pressures.}
        \label{fig:RPT9}
\end{figure}

The Gibbs free energy is illustrated in Figs. \ref{fig:RPT8} and 
\ref{fig:RPT9} for two positive critical pressures and three positive 
critical pressures. First, we discuss Fig. \ref{fig:RPT8}, when two 
critical pressures are positive and another is negative. For $P=P_{cp1}$ 
no phase transition occurs. For $P<P_{cp1}$ and $P=P_{cp2}$  
swallow tail behaviour appears, which is shown in 
Figs. \ref{fig:5.3(b)} and \ref{fig:5.3(c)}. For a range of pressure 
$P_{t} < P < P_{cp2}$, two swallow tails occur, which indicates 
\textbf{SBH}--\textbf{IBH} and \textbf{IBH}--\textbf{LBH} phase 
transitions in Fig. \ref{fig:5.3(d)}. At $P=P_{t}$ this behaviour 
disappears in Fig. \ref{fig:5.3(e)}. If we further decrease the 
pressure then a \textbf{SBH}--\textbf{IBH}--\textbf{LBH} phase 
transition happens in Fig. \ref{fig:5.3(f)}. This phase transition 
occurs for a range of pressure $P \in (P_{z},P_{t})$ and temperature 
$T \in (T_{1},T_{2})$. Finally, a \textbf{SBH}--\textbf{LBH} 
phase transition occurs for $P \leq P_{z}$ in Figs. \ref{fig:5.3(g)} 
and \ref{fig:5.3(h)}.

The Gibbs free energy is illustrated in Fig. \ref{fig:RPT9} for 
three positive critical pressures. At $P=P_{cp1}$, no phase transition 
occurs. The first order phase transition occurs for $P<P_{cp1}$ and 
it continues till $P=P_{t}$. In the range of pressure 
$P \in (P_{z}, P_{t})$ a zeroth-order phase transition occurs, 
which is similar to the phase transition studied in Fig. \ref{fig:RPT2}.

\subsection{Black Holes in \texorpdfstring{$4D$}{TEXT} EGB massive gravity 
coupled to NED}\label{sec:RPT4}

Inspired by the rich phase structure of the black hole in massive Einstein 
gravity coupled to NED (see section \ref{sec:RPT2}), here we consider
a small contribution $(\alpha=0.0001)$ of the GB coupling 
parameter and we set $Q_{m}=1$. In the range, 
$a \in (a_1, a_2)$ equation \ref{eq:4.5} has three real and positive 
solutions with three real positive critical pressures. We take 
$a=0.99$ and obtain three real positive values of the 
critical parameters in table \ref{ta3}. In the range, 
$a \in (a_0, a_1)$ equation \ref{eq:4.5} has three real and positive 
solutions with two real positive critical pressures and one negative
critical pressure. We take $a=0.90$ and obtain the critical 
parameters in table \ref{ta3}.

\begin{figure}[H]
     \centering
     \begin{subfigure}[b]{0.3\textwidth}
         \centering
         \includegraphics[width=\textwidth]{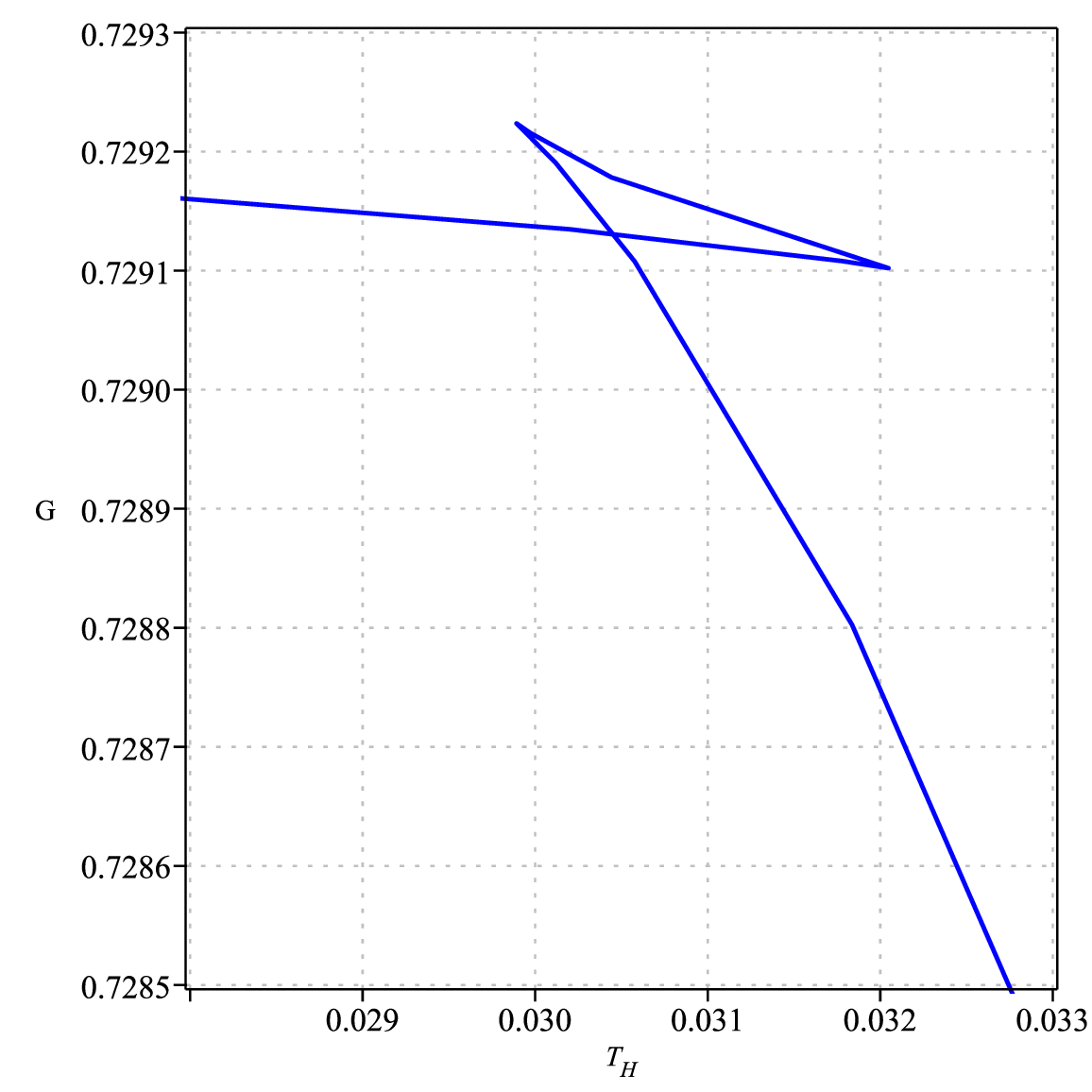}
         \caption{$P=0.0090<P_{cp1}$}
         \label{fig:5.4(a)}
     \end{subfigure}
     \hfill
     \begin{subfigure}[b]{0.3\textwidth}
         \centering
         \includegraphics[width=\textwidth]{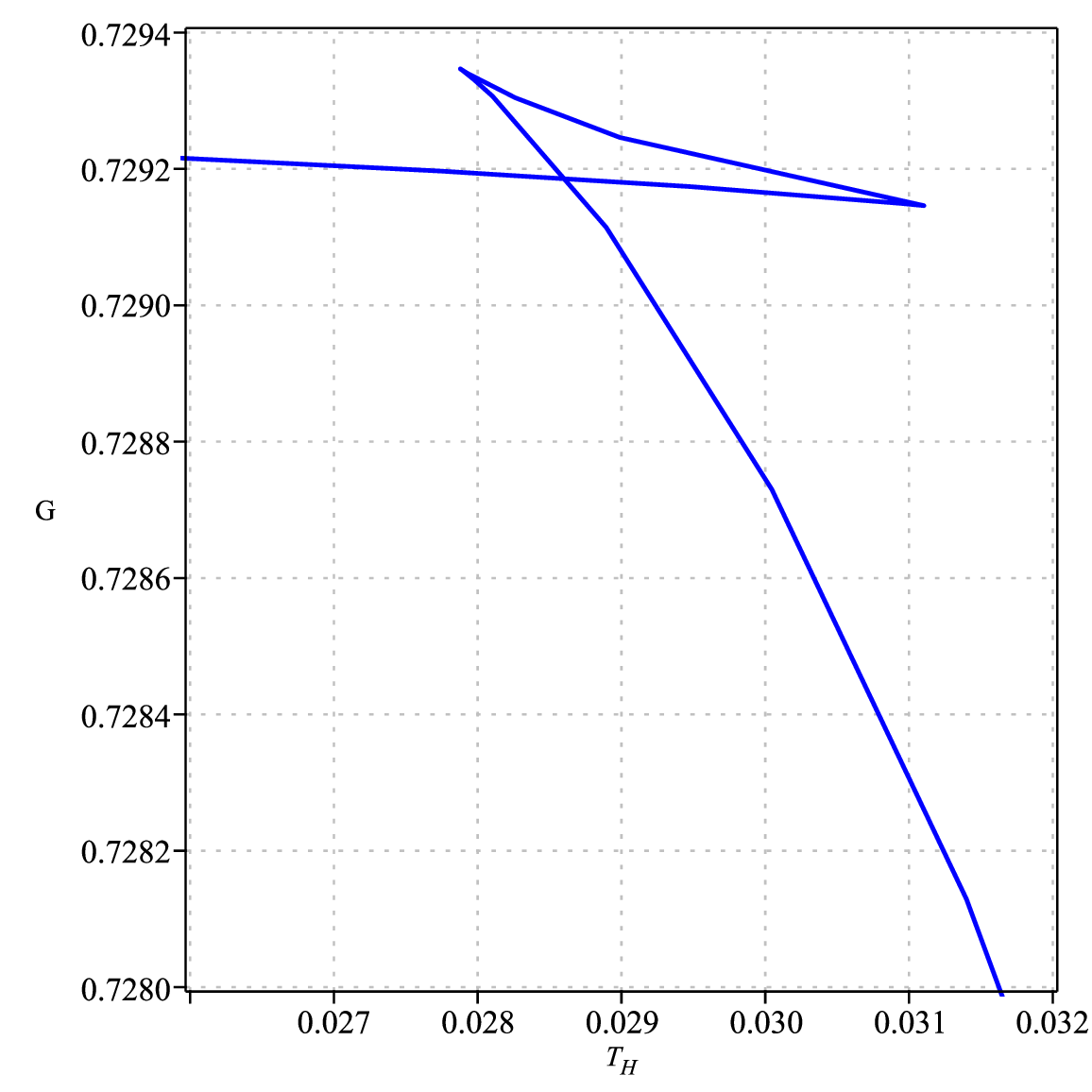}
         \caption{$P=0.0041=P_{cp2}$}
         \label{fig:5.4(b)}
     \end{subfigure}
     \hfill
     \begin{subfigure}[b]{0.3\textwidth}
         \centering
         \includegraphics[width=\textwidth]{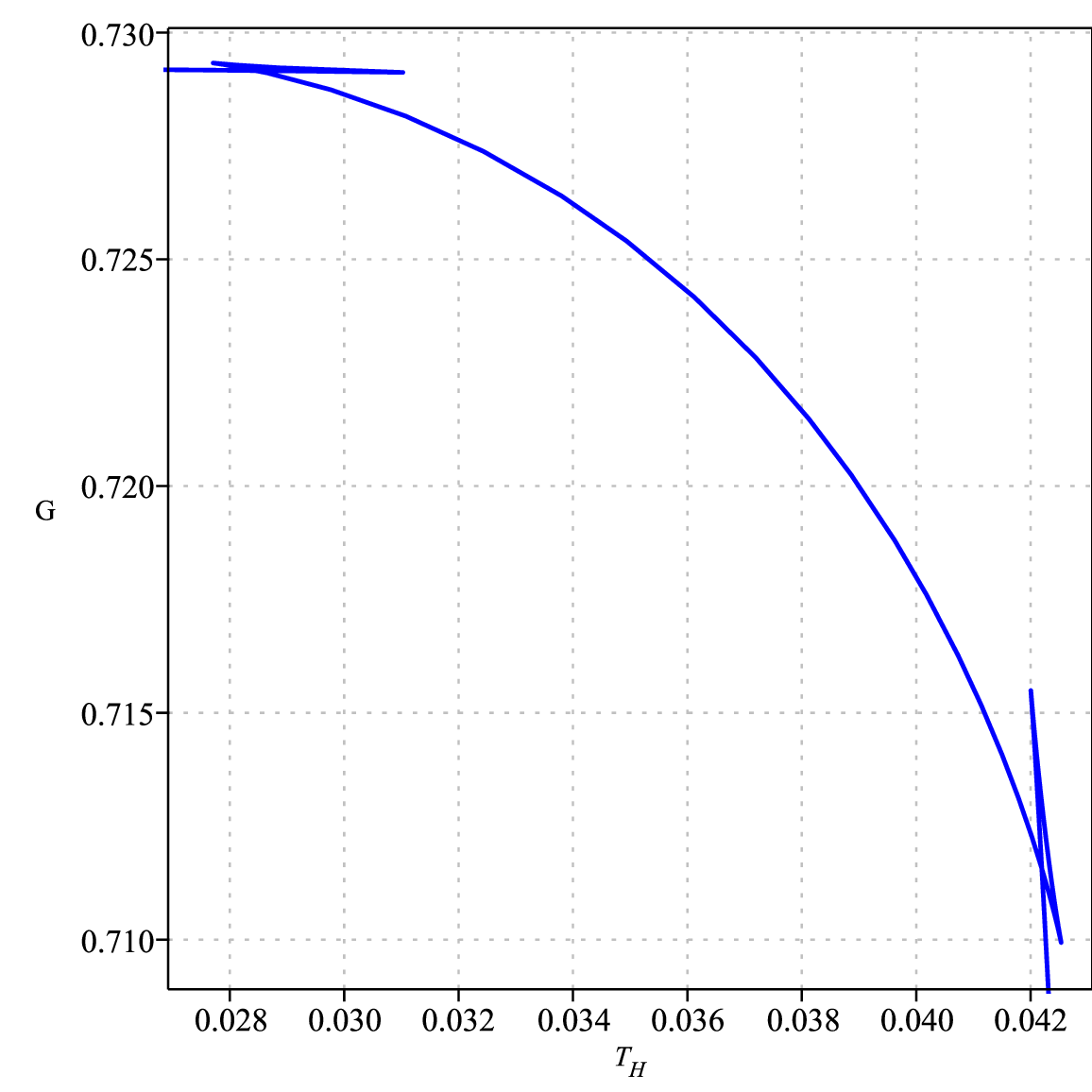}
         \caption{$P=0.0037<P_{cp2}$}
         \label{fig:5.4(c)}
     \end{subfigure}
     \hfill
     \begin{subfigure}[b]{0.3\textwidth}
         \centering
         \includegraphics[width=\textwidth]{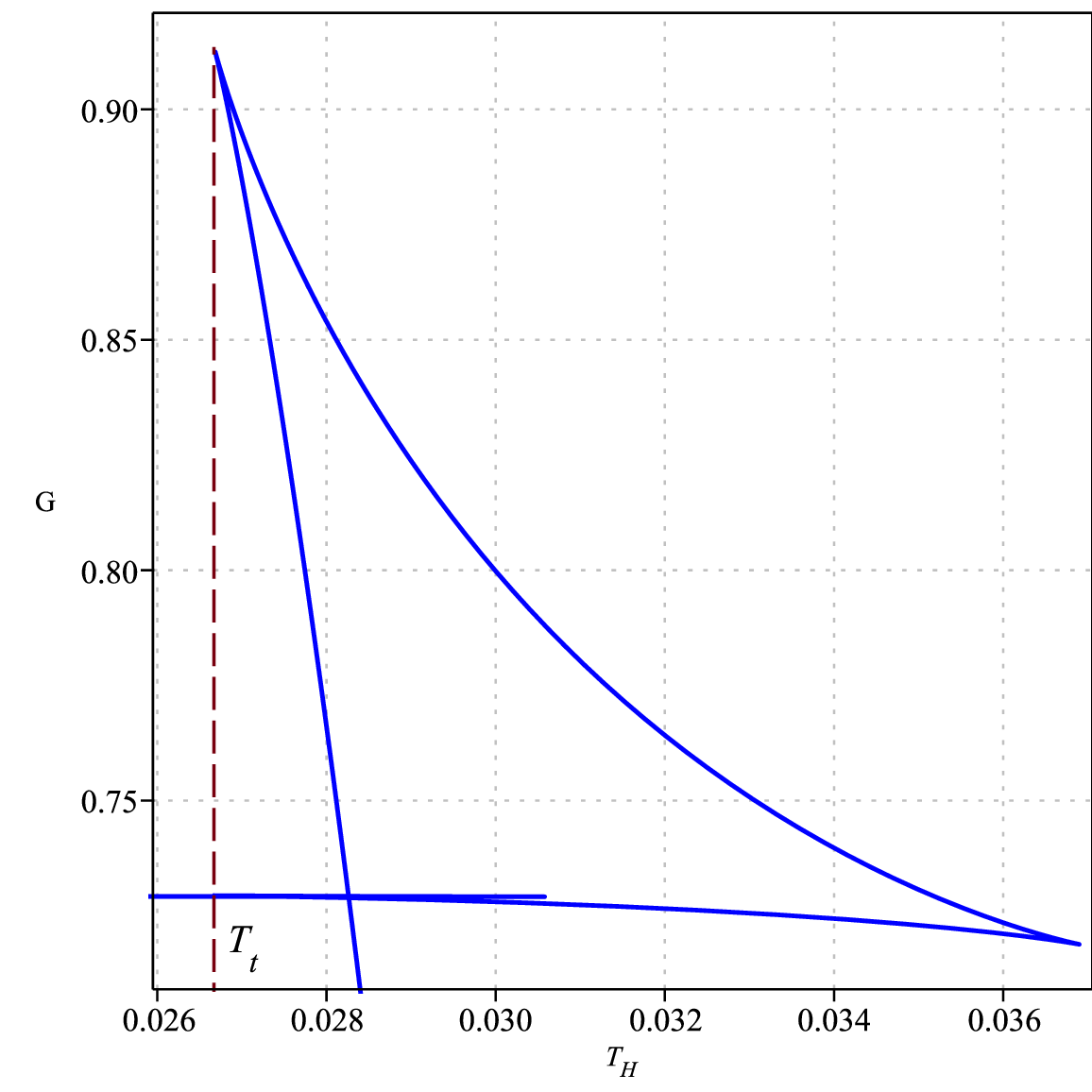}
         \caption{$P=0.00138=P_{t}$}
         \label{fig:5.4(d)}
     \end{subfigure}
     \hfill
     \begin{subfigure}[b]{0.3\textwidth}
         \centering
         \includegraphics[width=\textwidth]{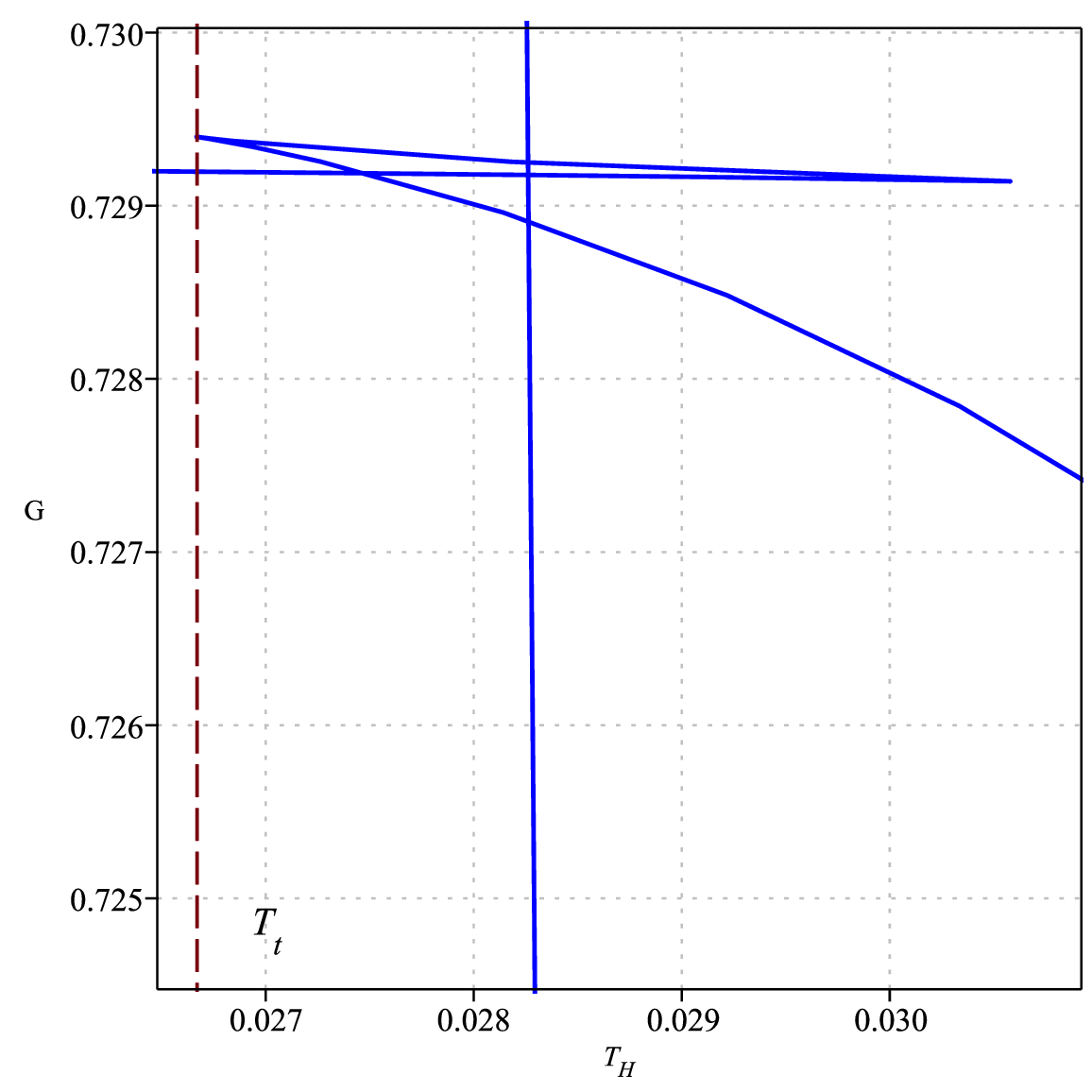}
         \caption{small scale of Fig. \ref{fig:5.4(d)}}
         \label{fig:5.4(e)}
     \end{subfigure}
     \hfill
     \begin{subfigure}[b]{0.3\textwidth}
         \centering
         \includegraphics[width=\textwidth]{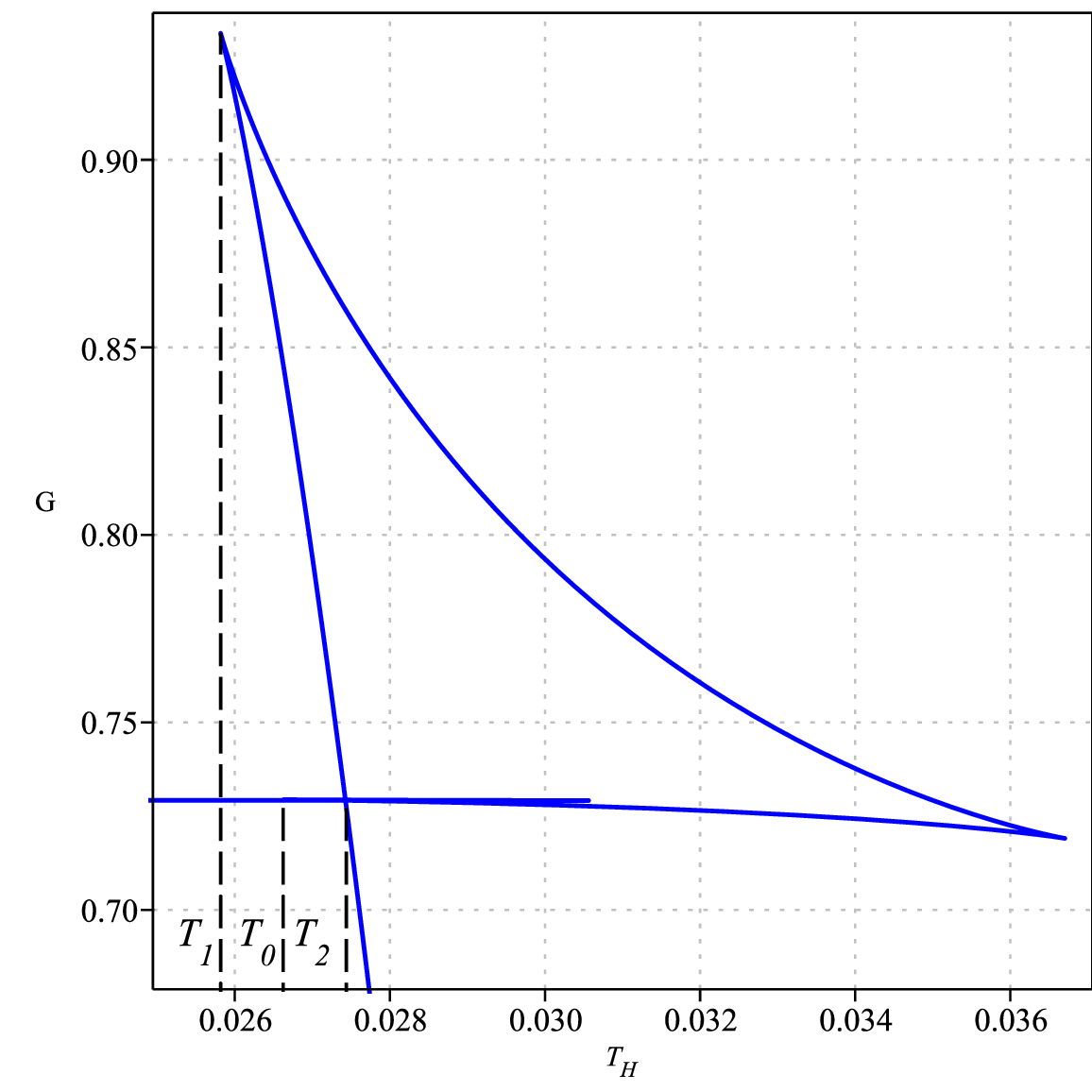}
         \caption{$P_{z}<P=0.00129<P_{t}$}
         \label{fig:5.4(f)}
     \end{subfigure}
      \hfill
     \begin{subfigure}[b]{0.3\textwidth}
         \centering
         \includegraphics[width=\textwidth]{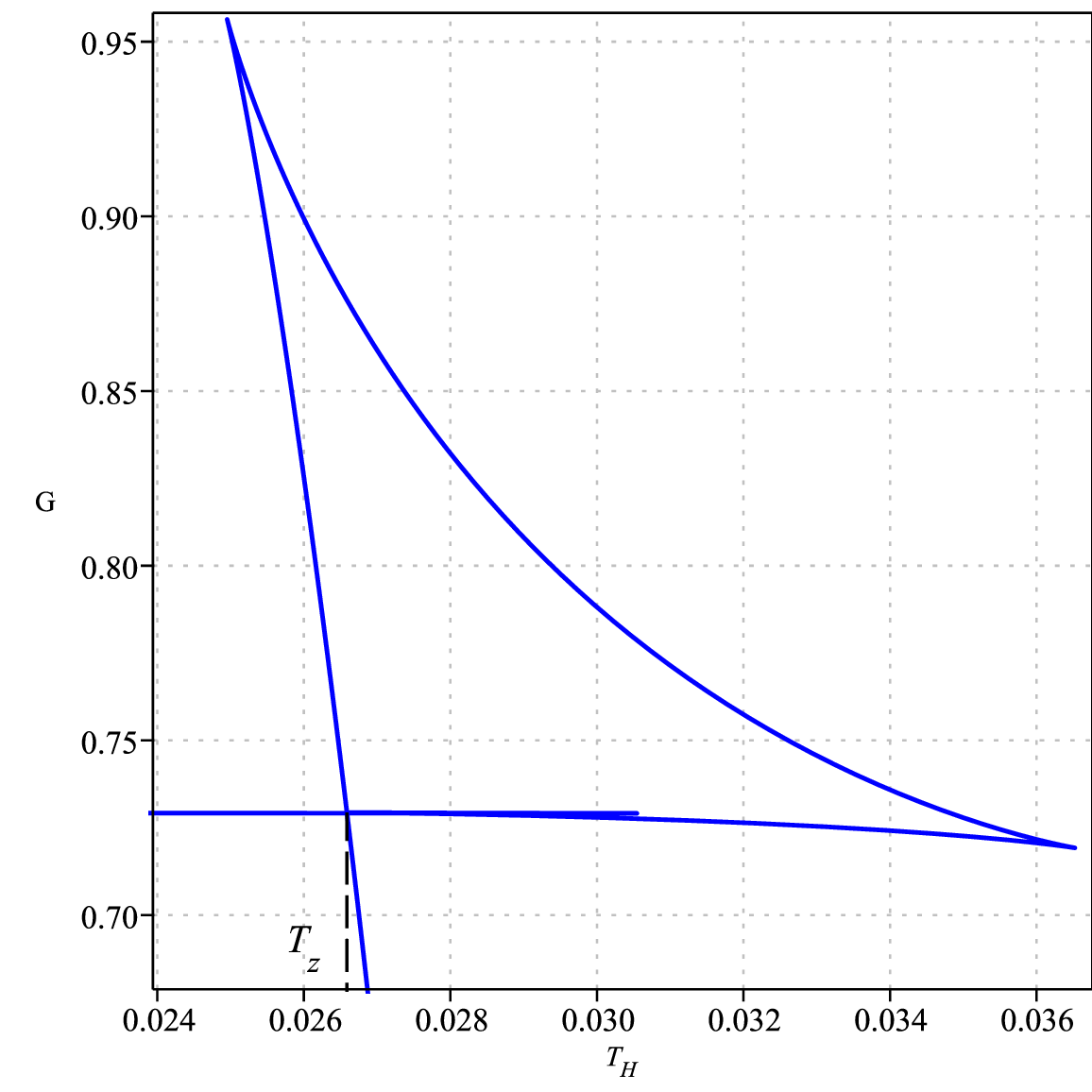}
         \caption{$P=0.001204=P_{z}$}
         \label{fig:5.4(h)}
     \end{subfigure}
          \begin{subfigure}[b]{0.3\textwidth}
         \centering
         \includegraphics[width=\textwidth]{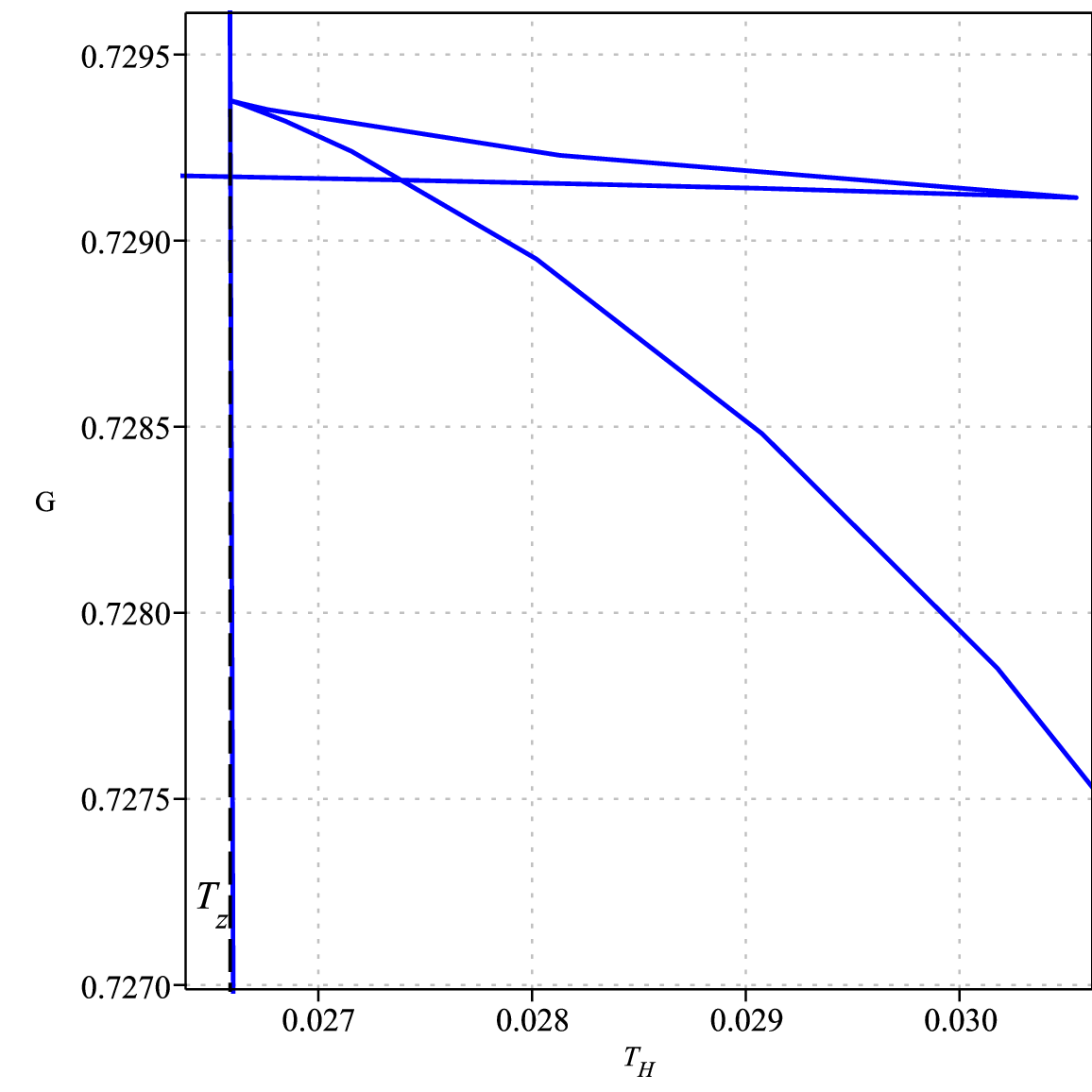}
         \caption{small scale of Fig. \ref{fig:5.4(h)}}
         \label{fig:5.4(i)}
     \end{subfigure}
          \begin{subfigure}[b]{0.3\textwidth}
         \centering
         \includegraphics[width=\textwidth]{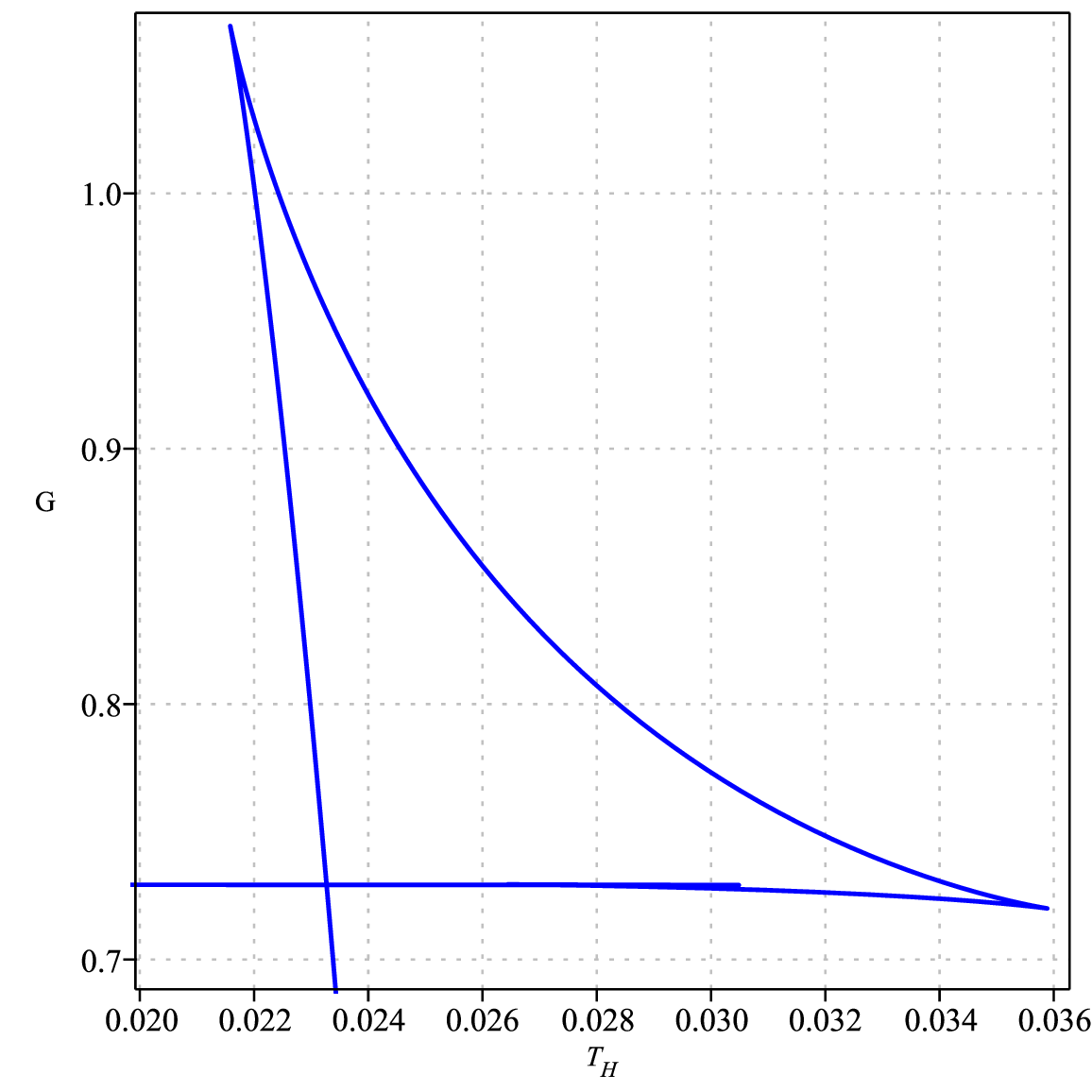}
         \caption{$P=0.0009<P_{z}$}
         \label{fig:5.4(j)}
     \end{subfigure}
          \begin{subfigure}[b]{0.3\textwidth}
         \centering
         \includegraphics[width=\textwidth]{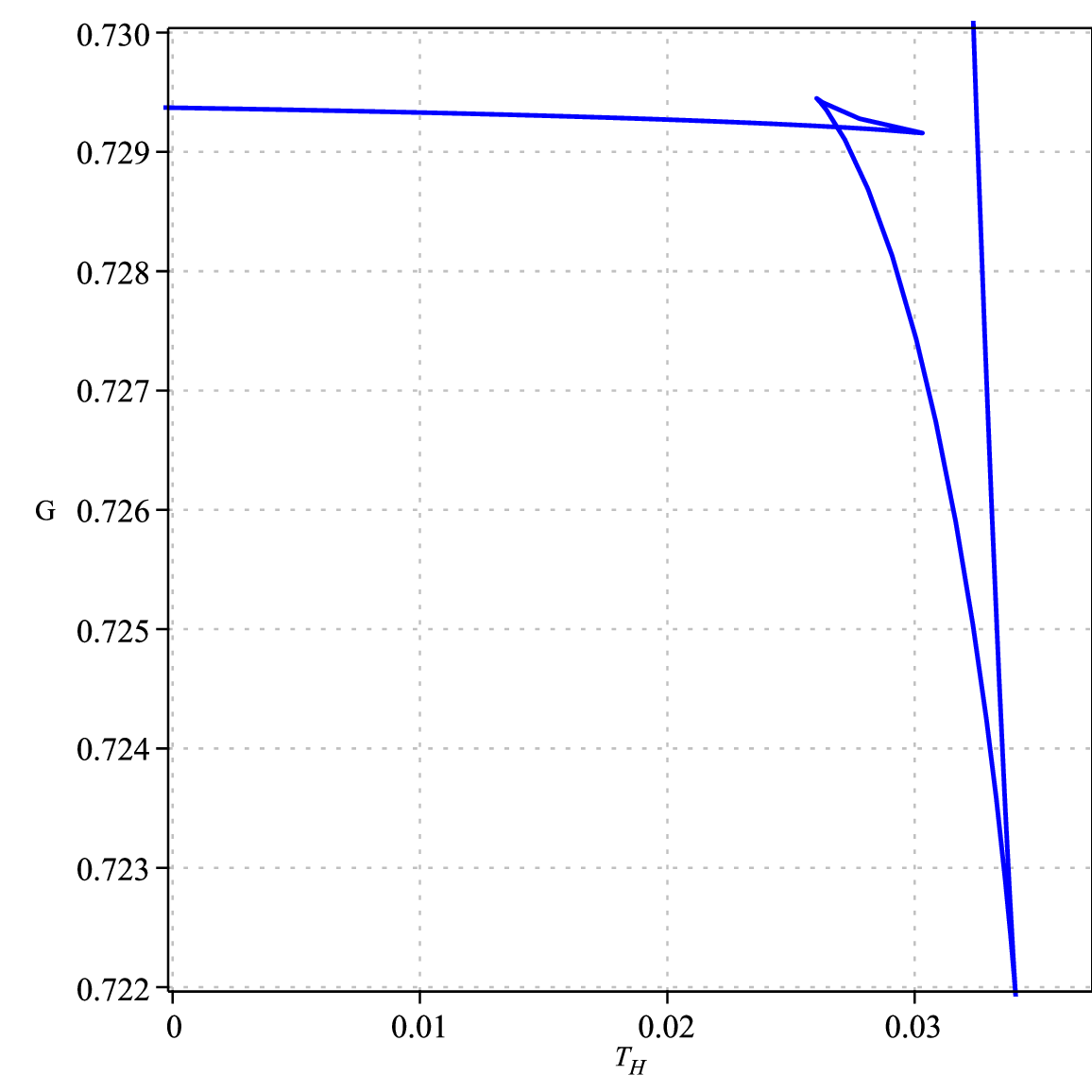}
         \caption{$P=0.0$}
         \label{fig:5.4(k)}
     \end{subfigure}
        \caption{Three critical points, two at positive pressures and one at negative pressure.}
        \label{fig:RPT11}
\end{figure}

\begin{figure}[H]
     \centering
     \begin{subfigure}[b]{0.3\textwidth}
         \centering
         \includegraphics[width=\textwidth]{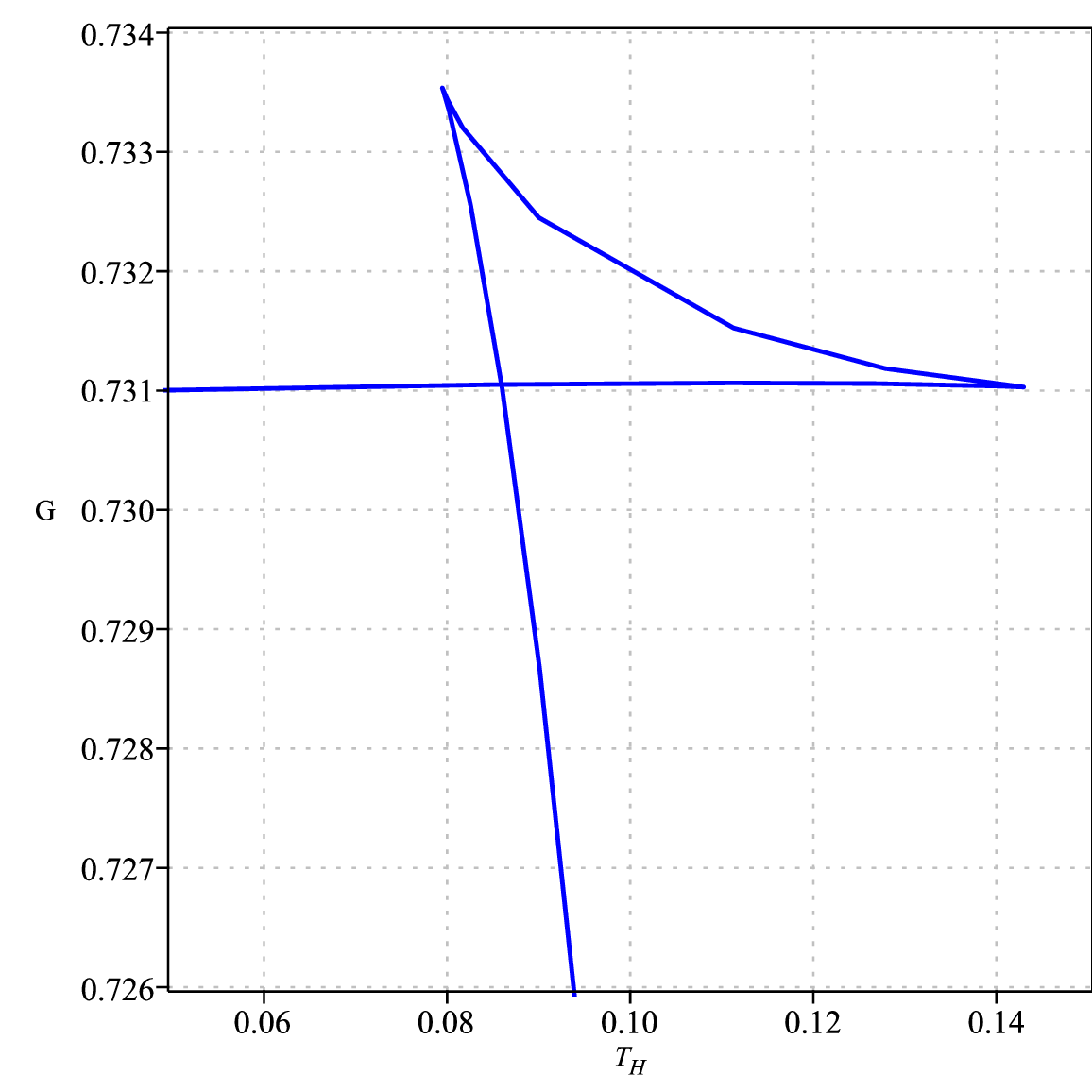}
         \caption{$P=0.0500<P_{cp1}$}
         \label{fig:5.4(l)}
     \end{subfigure}
     \hfill
     \begin{subfigure}[b]{0.3\textwidth}
         \centering
         \includegraphics[width=\textwidth]{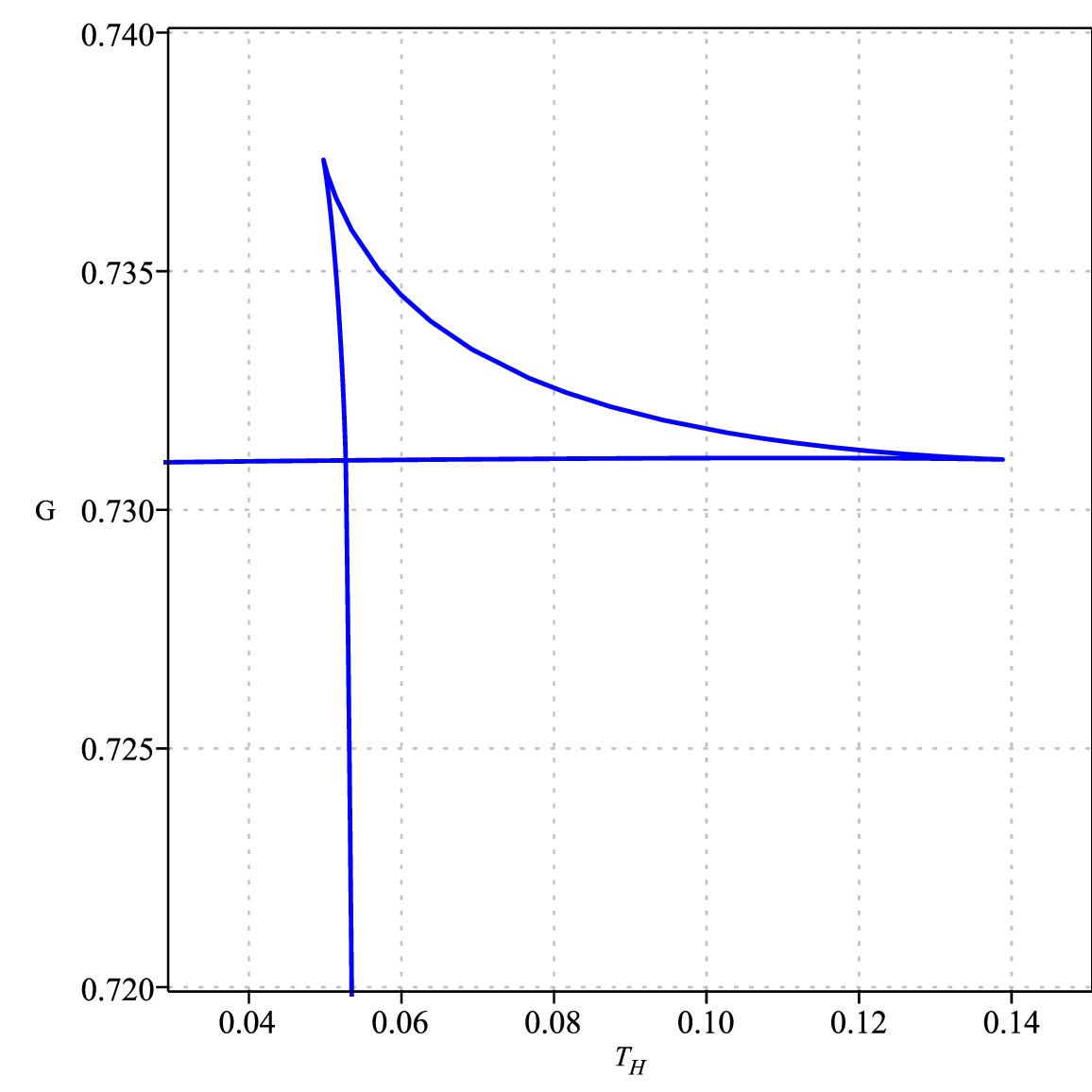}
         \caption{$P=0.0055=P_{cp2}$}
         \label{fig:5.4(m)}
     \end{subfigure}
     \hfill
     \begin{subfigure}[b]{0.3\textwidth}
         \centering
         \includegraphics[width=\textwidth]{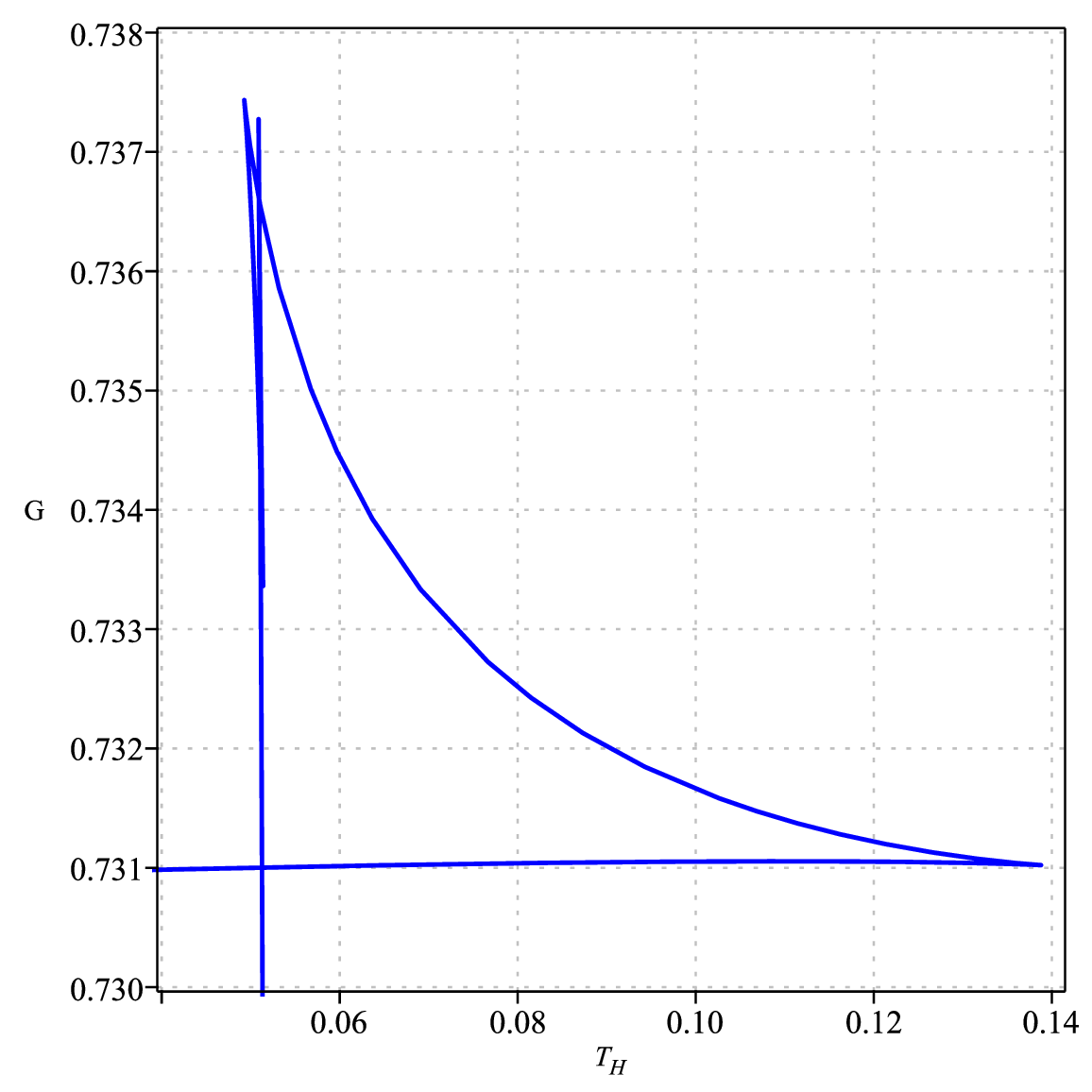}
         \caption{$P=0.0050<P_{cp2}$}
         \label{fig:5.4(n)}
     \end{subfigure}
     \hfill
     \begin{subfigure}[b]{0.3\textwidth}
         \centering
         \includegraphics[width=\textwidth]{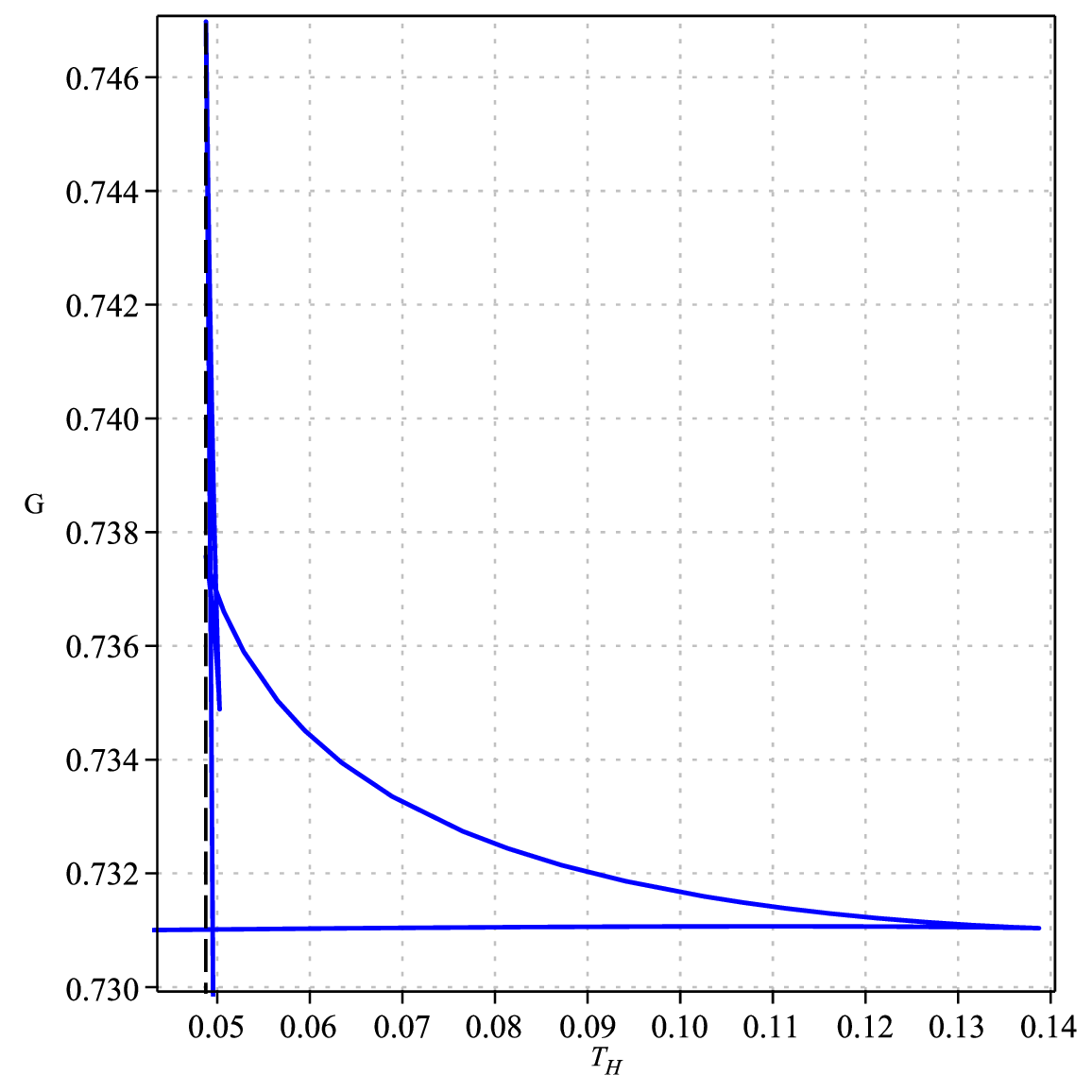}
         \caption{$P=0.0045=P_{t}$}
         \label{fig:5.4(o)}
     \end{subfigure}
     \hfill
     \begin{subfigure}[b]{0.3\textwidth}
         \centering
         \includegraphics[width=\textwidth]{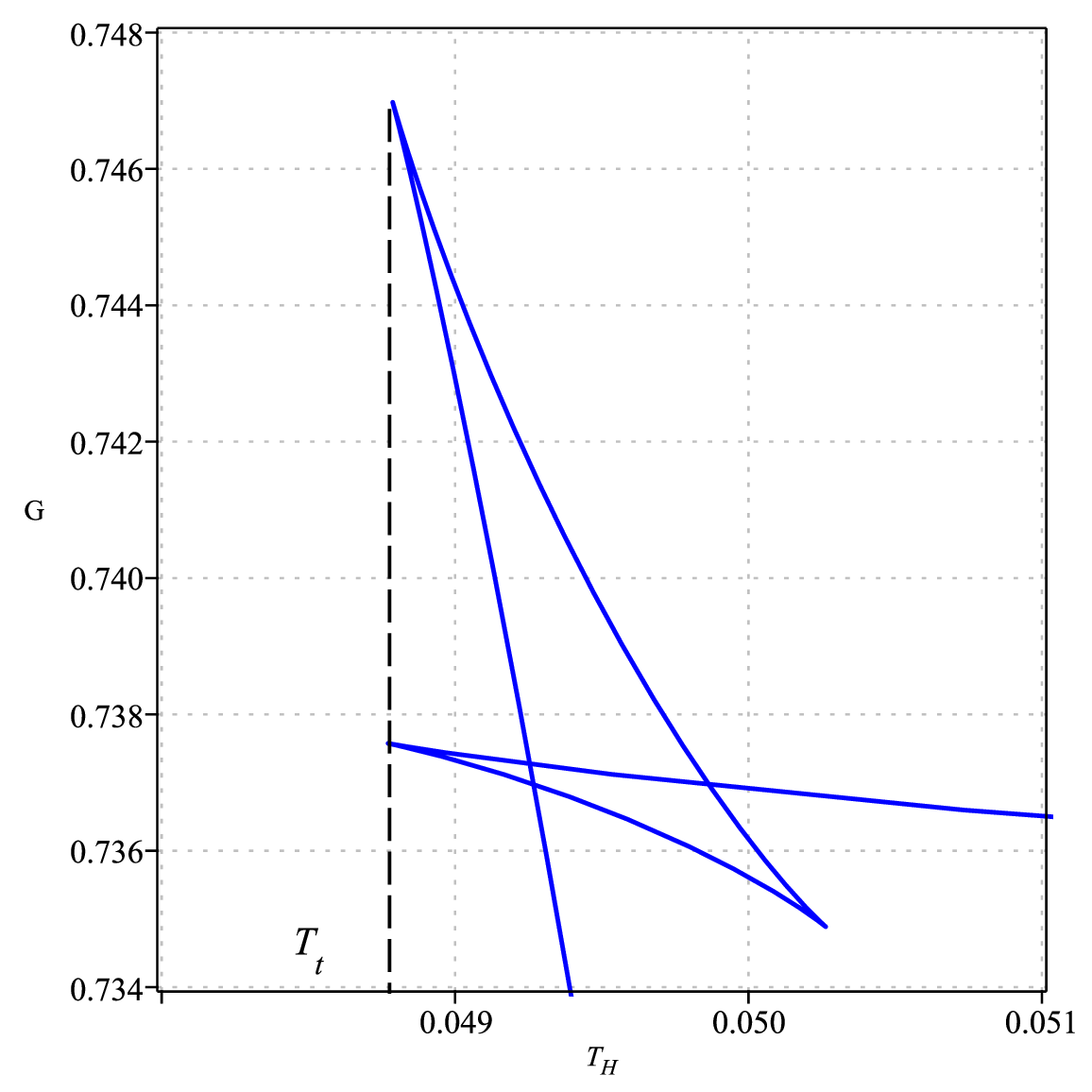}
         \caption{small scale of Fig. \ref{fig:5.4(o)}}
         \label{fig:5.4(p)}
     \end{subfigure}
     \hfill
     \begin{subfigure}[b]{0.3\textwidth}
         \centering
         \includegraphics[width=\textwidth]{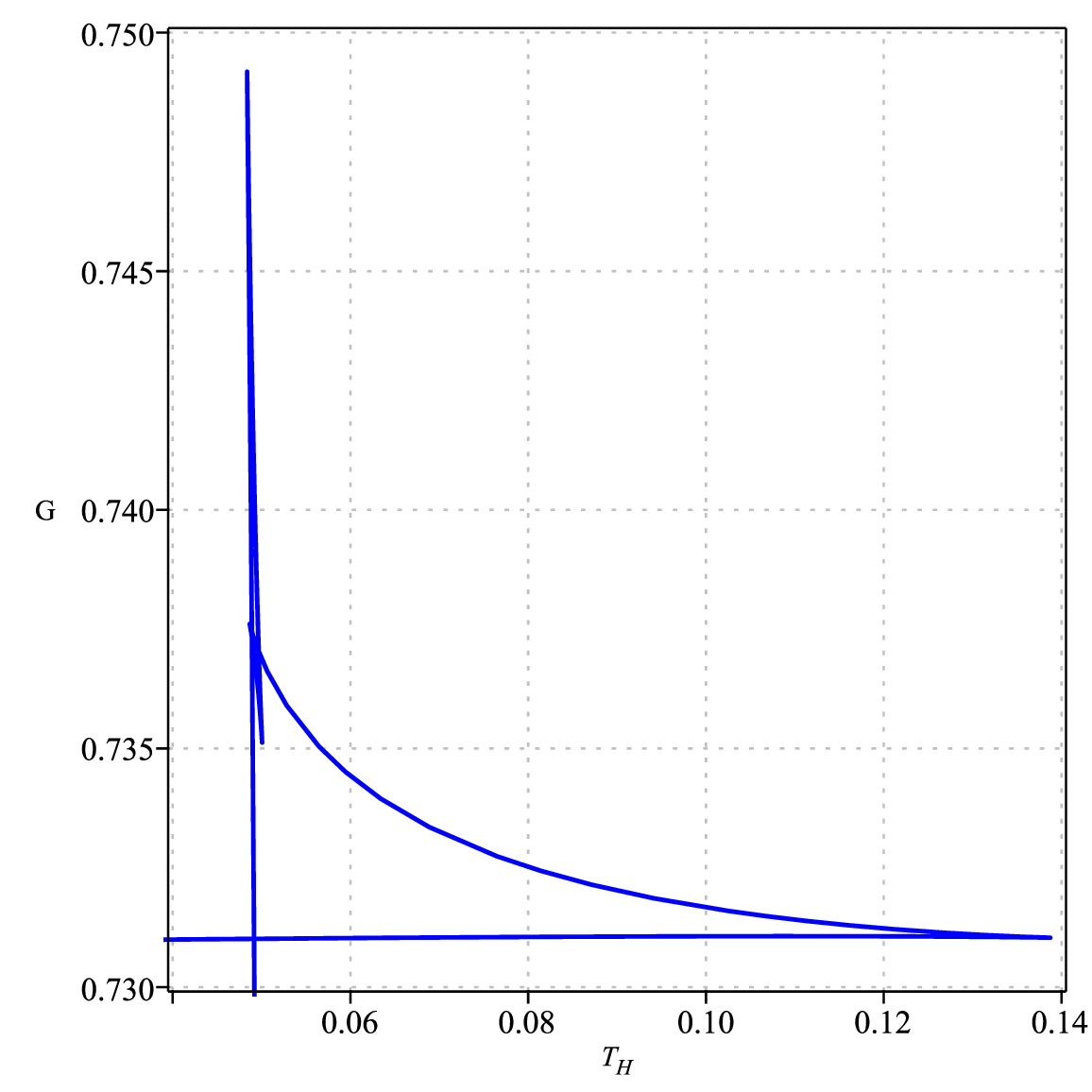}
         \caption{$P_{z}<P=0.00441<P_{t}$}
         \label{fig:5.4(q)}
     \end{subfigure}
      \hfill
     \begin{subfigure}[b]{0.3\textwidth}
         \centering
         \includegraphics[width=\textwidth]{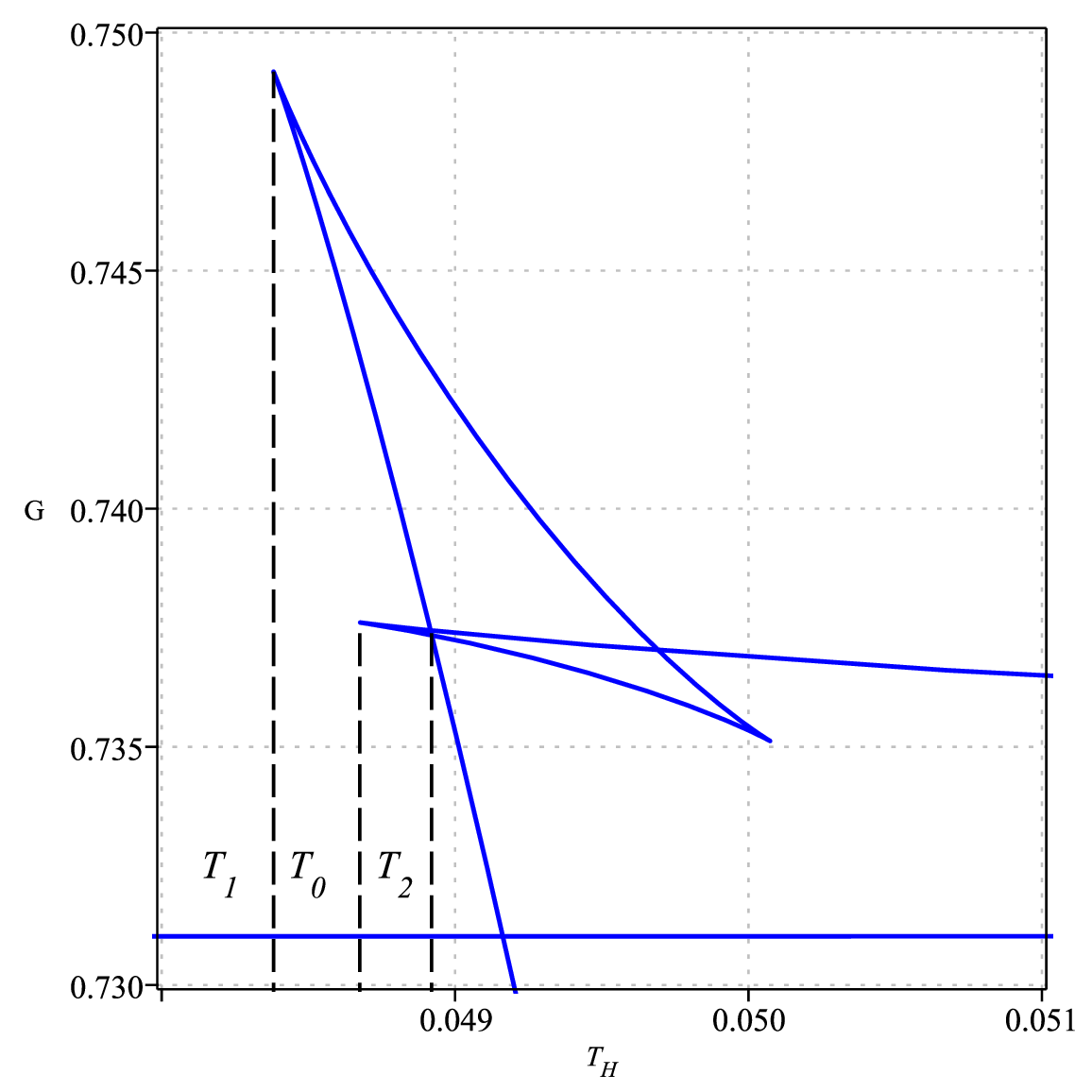}
         \caption{small scale of Fig. \ref{fig:5.4(q)}}
         \label{fig:5.4(r)}
     \end{subfigure}
      \hfill
     \begin{subfigure}[b]{0.3\textwidth}
         \centering
         \includegraphics[width=\textwidth]{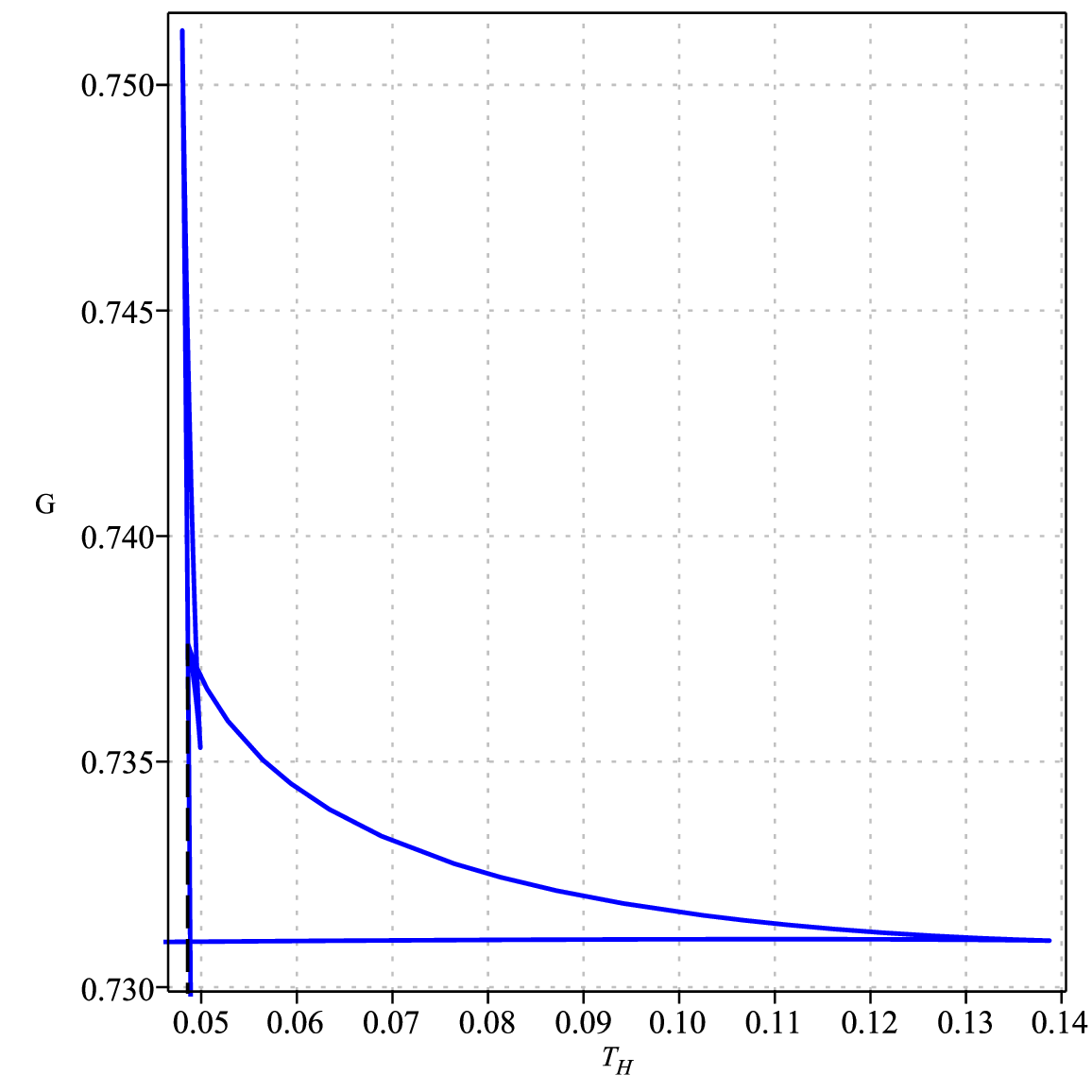}
         \caption{$P=0.00433=P_{z}$}
         \label{fig:5.4(s)}
     \end{subfigure}
     \begin{subfigure}[b]{0.3\textwidth}
         \centering
         \includegraphics[width=\textwidth]{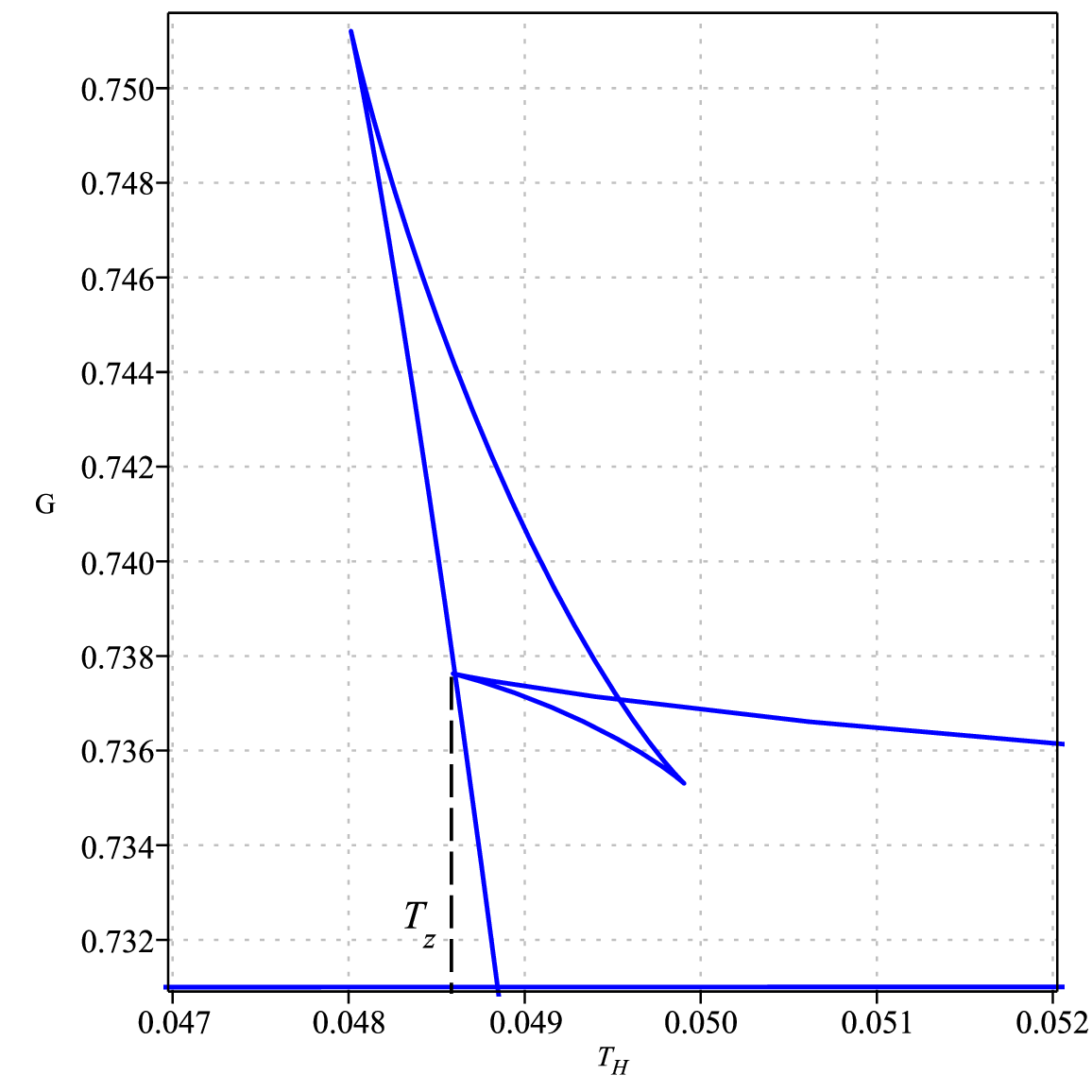}
         \caption{small scale of Fig. \ref{fig:5.4(s)}}
         \label{fig:5.4(t)}
     \end{subfigure}
     \begin{subfigure}[b]{0.3\textwidth}
         \centering
         \includegraphics[width=\textwidth]{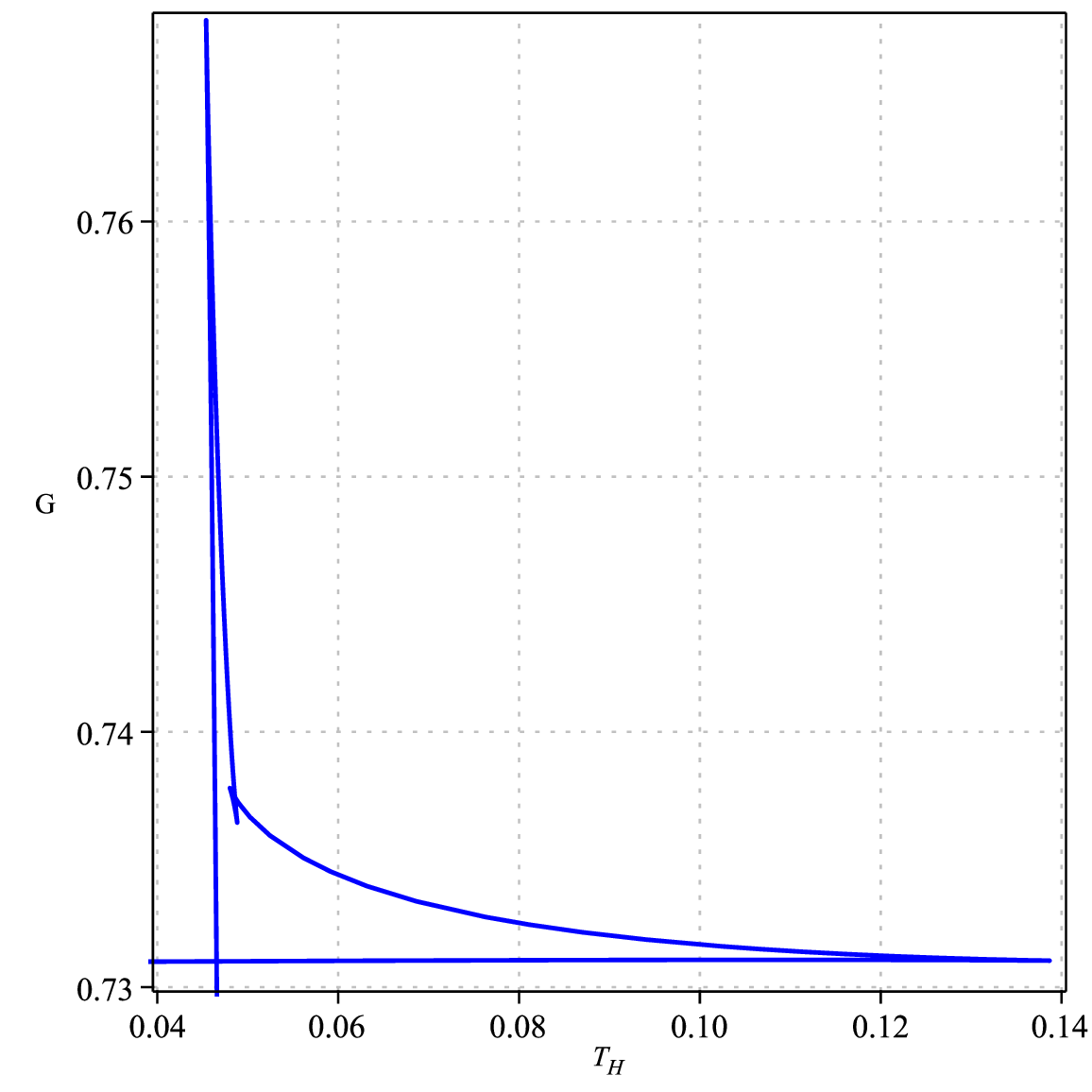}
         \caption{$P_{cp3}<P=0.0038<P_{z}$}
         \label{fig:5.4(u)}
     \end{subfigure}
    \begin{subfigure}[b]{0.3\textwidth}
         \centering
         \includegraphics[width=\textwidth]{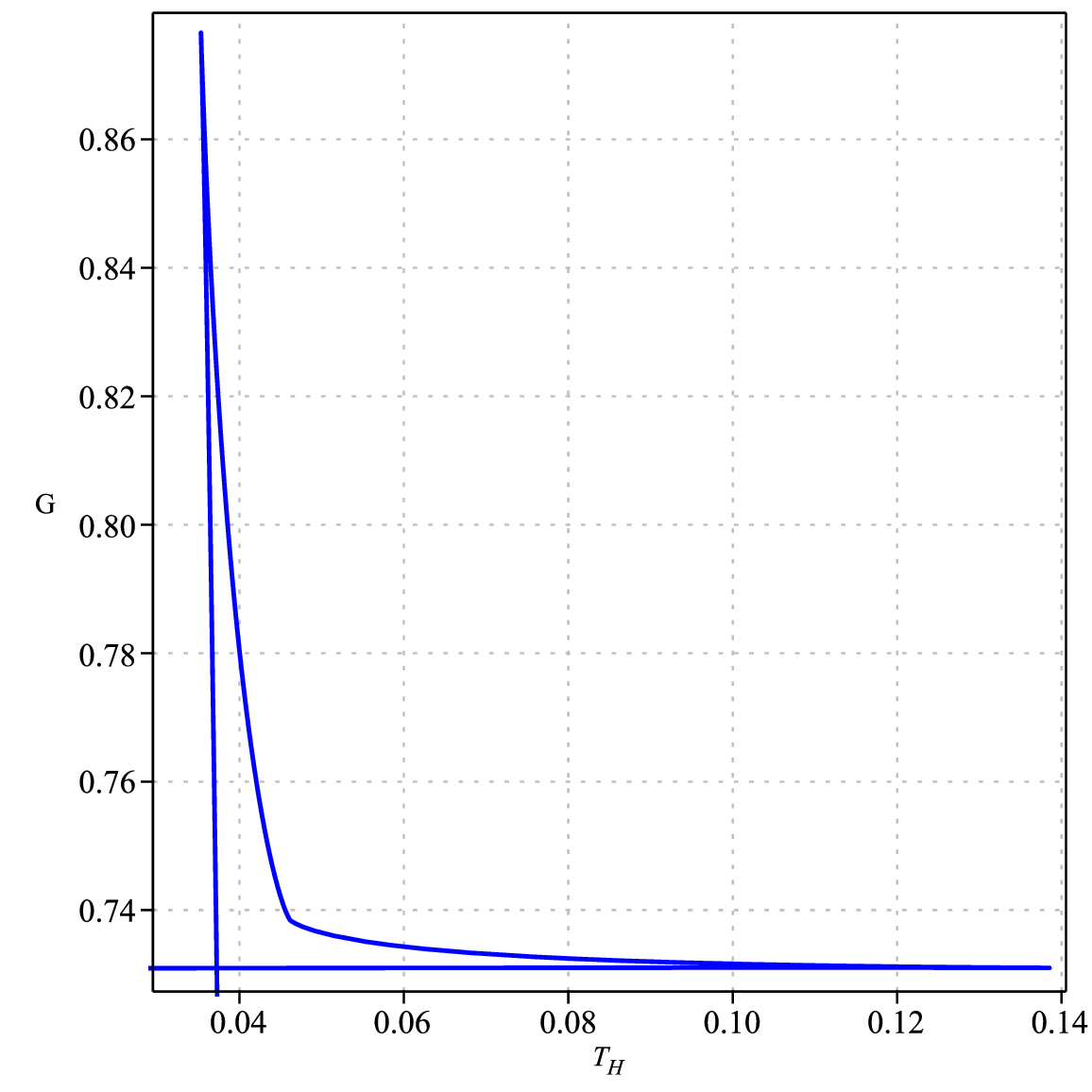}
         \caption{$P=0.0022=P_{cp3}$}
         \label{fig:5.4(v)}
     \end{subfigure}
    \begin{subfigure}[b]{0.3\textwidth}
         \centering
         \includegraphics[width=\textwidth]{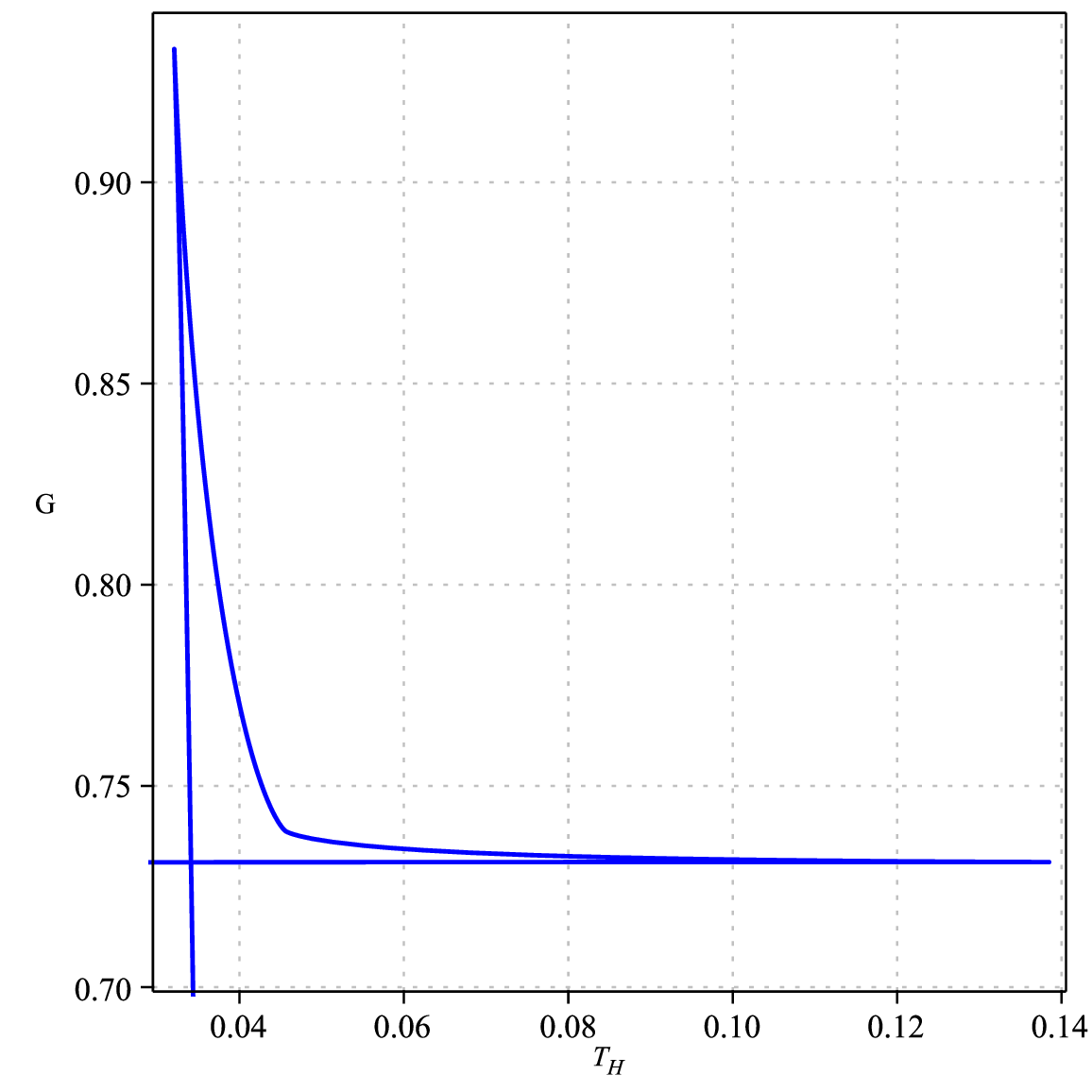}
         \caption{$P=0.0018<P_{cp3}$}
         \label{fig:5.4(w)}
     \end{subfigure}
        \caption{Three critical points, with three positive pressures.}
        \label{fig:RPT12}
\end{figure}

The $G-T_{H}$ diagram is shown in Figs. \ref{fig:RPT11}
and \ref{fig:RPT12}, which is the same as the behaviour of $G-T_{H}$ diagram in \ref{sec:RPT3}, 
i.e., massive gravity does not change the phase structure of the black holes. The massive gravity
affects critical points. Under the effects of massive gravity critical points 
$(P_{t}, T_{t})$ and $(P_{z}, T_{z})$ take lower values compared to massless EGB 
gravity. 

\begin{table}[H]
\begin{center}
\begin{tabular}{ |c|c|c|c|c|c| } 
\hline
Case & CP & CP1 & CP2 & CP3 \\
\hline
\multirow{3}{10em}{$a_1 \leq a=0.99 \leq a_2$} & $v_c$ & 0.1455 & 3.0244 & 1.3698   \\ 
& $T_c$ & 0.1854 & 0.0527 & 0.0462   \\ 
& $P_c$ & 0.4448 & 0.0055 & 0.0022   \\ 
\hline
\multirow{3}{10em}{$a_0 \leq a=0.90 \leq a_1 $} & $v_c$ & 0.2452 & 3.5222 & 0.8563\\ 
& $T_c$ &  0.0358 & 0.0439 & 0.0204 \\ 
& $P_c$ &  0.0276 & 0.0041 & -0.0095 \\ 
\hline
\multirow{3}{10em}{$a=1.2 \geq a_2  $} & $v_c$ & 0.0998 & $--$ & $--$ \\ 
& $T_c$ & 0.6732 & $--$  & $--$ \\ 
& $P_c$ & 2.4350 & $--$  & $--$ \\ 
\hline
\end{tabular}
\end{center}
\caption{With $Q_m=1$, $\beta=0.34$, $c=1$, $c_1=-1$, $m=0.1$ \& $\alpha=0.0001$.}
\label{ta3}
\end{table}

\begin{figure}[H]
\centering
\subfloat[One pressure is negative]{\includegraphics[width=.5\textwidth]{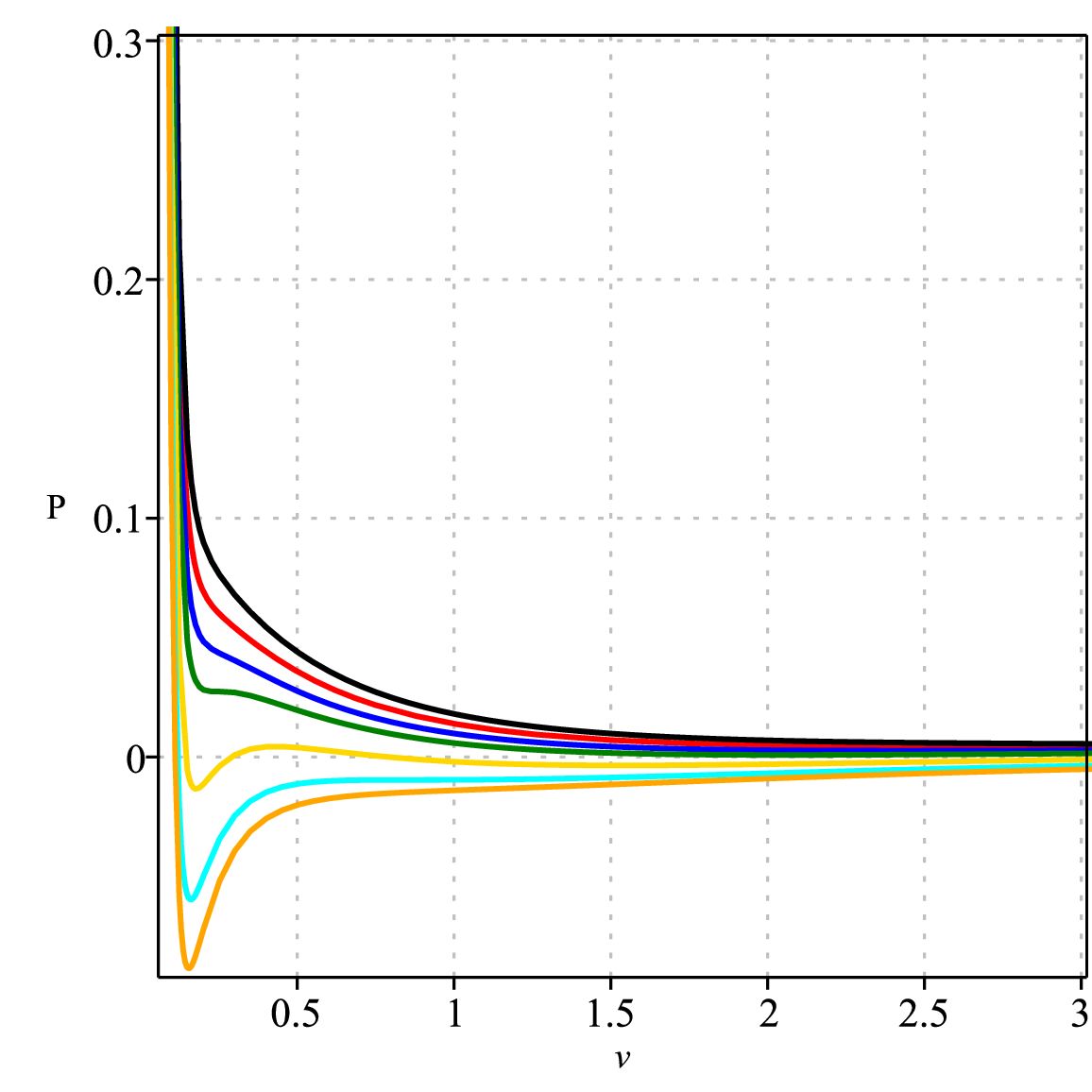}}\hfill
\subfloat[Three pressures are positive]{\includegraphics[width=.5\textwidth]{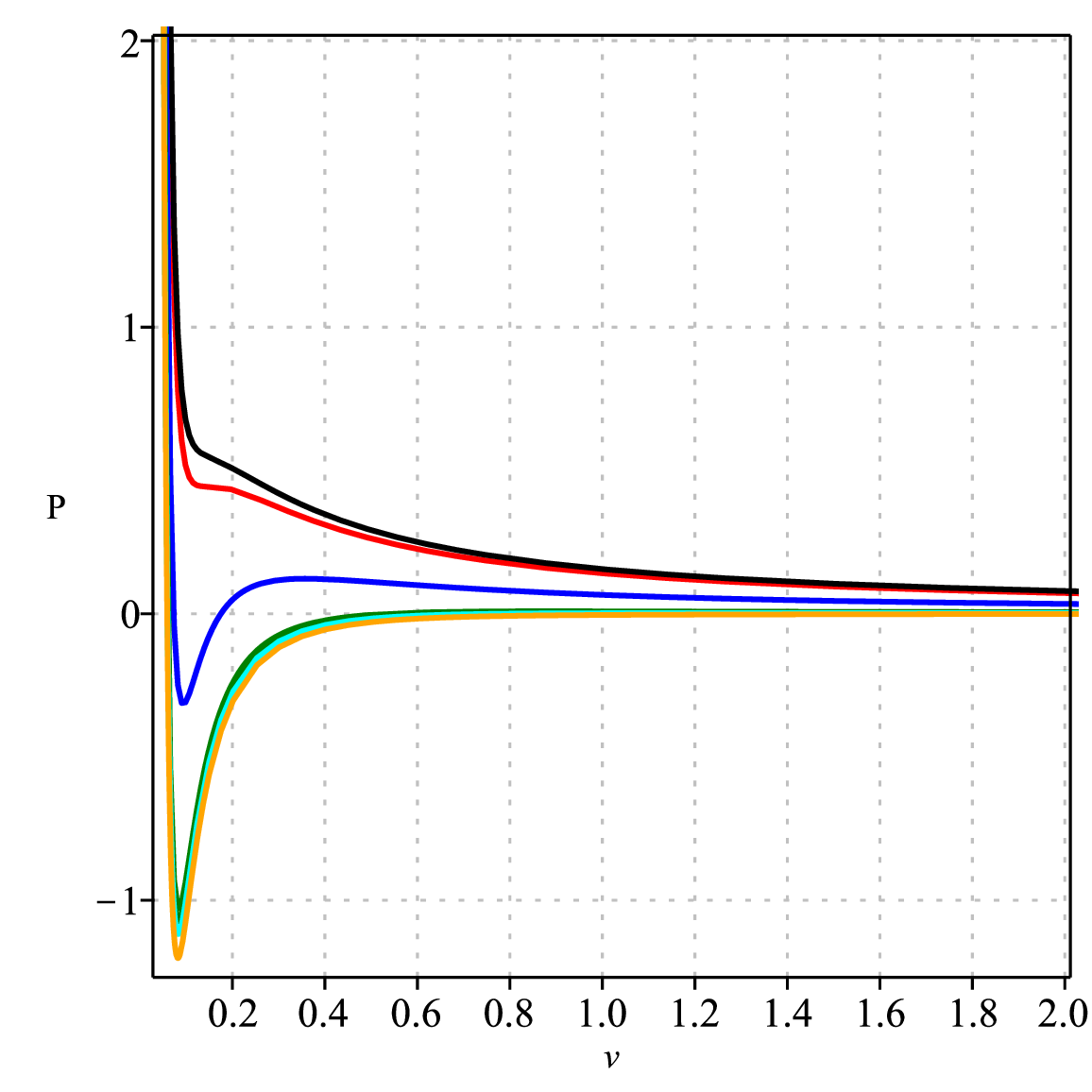}}\hfill
\caption{Left Panel : Black line denotes $T=0.0480>T_{cp2}$,  Red line denotes $T=0.0439=T_{cp2}$, 
Blue line denotes $T_{cp1}<T=0.0398<T_{cp2}$, Green line denotes $T=0.0358=T_{cp1}$, 
Gold line denotes $T_{cp3}<T=0.0280<T_{cp1}$, Cyan line denotes $T=0.0204=T_{cp3}$ \& Orange line denotes $T=0.0160<T_{cp3}$. Right Panel :  Black line denotes $T=0.2000>T_{cp1}$, Red line denotes $T=0.1854=T_{cp1}$, Blue line denotes $T_{cp2}<T=0.1100<T_{cp1}$, Green line denotes $T=0.0527=T_{cp2}$, Gold line denotes $T_{cp3}<T=0.0495<T_{cp2}$, Cyan line denotes $T=0.0462=T_{cp3}$ \& Orange line denotes $T=0.0400<T_{cp3}$.}\label{fig:RPT10}
\end{figure}

\section{Joule--Thomson Expansion}\label{sec:5}
In this section, we discuss Joule--Thomson expansion of black holes in $4D$ EGB massive/massless 
gravity and massive Einstein gravity. The Joule--Thomson expansion is a fundamental concept in 
thermodynamics that describes the cooling or heating effect of a gas when it is allowed to expand 
or contract through a porous plug or valve at constant enthalpy. The concept of Joule--Thomson 
expansion for a black hole was first studied by Ökcü \& Aydıner in Ref. \cite{Okcu:2016tgt}. 
After that, Joule--Thomson expansion of $D$ dimension charged black hole, Kerr--$AdS$, Kerr--Newman--$AdS$ 
and black holes in massive gravity studied in Refs. 
\cite{Okcu:2017qgo,Mo:2018rgq,Zhao:2018kpz,Nam:2020gud}. The Joule--Thomson expansion of black hole 
in GR coupled to NED (eq. \eqref{eq:2.4} and others) studied in Refs. 
\cite{kruglov2022nonlinearly,Kruglov:2022bhx,Kruglov:2022mde,Kruglov:2022sxx}. Joule--Thomson 
effects of $4D$ EGB gravity coupled to Maxwell/BI electrodynamics were studied in Refs. 
\cite{Hegde:2020xlv,Zhang:2021kha}. The Joule--Thomson coefficient ($\mu$) is a measure of the cooling 
or heating effect of the expansion and is defined as the rate of change of temperature with respect to 
pressure at constant enthalpy. The Joule--Thomson thermodynamic coefficient is given by
\begin{equation}\label{eq:5.1}
    \mu_{J}= \biggl( \frac{\partial{T}}{\partial{P}} \biggl)_{M}= \frac{1}{C_{P}} \biggr[  T \Bigl(\frac{\partial{V}}{\partial{T}}\Bigl)_P - V \biggr]=\frac{(\partial{T}/\partial{r_{+}})_M}{(\partial{P}/\partial{r_{+}})_M}.
\end{equation} 
If $\mu$ is positive, this corresponds to the cooling phase, while if $\mu$ is negative, this corresponds to the heating phase. From above equation we obtain inverse temperature ($\mu_{J}=0$) as
\begin{equation}\label{eq:5.2}
    T_{i}= V \biggl( \frac{\partial{T}}{\partial{V}} \biggl)_{P} = \frac{r_{+}}{3} \biggl( \frac{\partial{T}}{\partial{r_{+}}} \biggl)_{P}.
\end{equation}
  
\subsection{Black Holes in 4D EGB Massive gravity coupled to NED}
From Hawking temperature \eqref{eq:3.4} one can obtain black hole equation of state 
\begin{equation}\label{eq:5.3}
T_{H}=\frac{8 P \pi  r_{+}^{6}+c c_{1} m^{2} r_{+}^{5}+(c^{2} c_{2} m^{2}+8 P \pi  k^{2}+1) r_{+}^{4}+c c_{1} k^{2} m^{2} r_{+}^{3}+((c^{2} c_{2} m^{2}+1) k^{2}-Q_{m}^{2}-\alpha ) r_{+}^{2}-\alpha  k^{2}}{8 r_{+} (\frac{r_{+}^{2}}{2}+\alpha ) (k^{2}+r_{+}^{2}) \pi}.
\end{equation}
From the mass function pressure of the black hole is obtained as 
\begin{equation}\label{eq:5.4}
P=\frac{1}{{\pi  r_{+}^{4} k}}\biggr[-\frac{3 c^{2} c_{2} k m^{2} r_{+}^{2}}{8}-\frac{3 c c_{1} k m^{2} r_{+}^{3}}{16}+\frac{3 Q_{m}^{2} r_{+} \arctan (\frac{r_{+}}{k})}{8}-\frac{3 Q_{m}^{2} r_{+} \pi}{16}+\frac{3 M r_{+} k}{4}-\frac{3 k r_{+}^{2}}{8}-\frac{3 \alpha  k}{8} \biggr].
\end{equation}
From equation \eqref{eq:5.3} and equation \eqref{eq:5.2} we obtain inverse pressure as 
\begin{equation*}
P_{i}=\frac{1}{16 \pi  r_{+}^{6} (k^{2}+r_{+}^{2})^2} \biggr[  -4 c^{2} c_{2} k^{4} m^{2} r_{+}^{4}-8 c^{2} c_{2} k^{2} m^{2} r_{+}^{6}-4 c^{2} c_{2} m^{2} r_{+}^{8}-3 c c_{1} k^{4} m^{2} r_{+}^{5}-6 c c_{1} k^{2} m^{2} r_{+}^{7}
\end{equation*}
\begin{equation*}
-3 c c_{1} m^{2} r_{+}^{9}-4 \alpha  c^{2} c_{2} k^{4} m^{2} r_{+}^{2}-8 \alpha  c^{2} c_{2} k^{2} m^{2} r_{+}^{4}-4 \alpha  c^{2} c_{2} m^{2} r_{+}^{6}-2 \alpha  c c_{1} k^{4} m^{2} r_{+}^{3}-4 \alpha  c c_{1} k^{2} m^{2} r_{+}^{5}-2 \alpha  c c_{1} m^{2} r_{+}^{7}
\end{equation*}
\begin{equation*}
+4 Q_{m}^{2} k^{2} r_{+}^{4}+6 Q_{m}^{2} r_{+}^{6}-4 k^{4} r_{+}^{4}-8 k^{2} r_{+}^{6}-4 r_{+}^{8}+4 Q_{m}^{2} \alpha  k^{2} r_{+}^{2}+8 Q_{m}^{2} \alpha  r_{+}^{4}+2 \alpha  k^{4} r_{+}^{2}+4 \alpha  k^{2} r_{+}^{4}+2 \alpha  r_{+}^{6}+8 \alpha^{2} k^{4}
\end{equation*}
\begin{equation}\label{eq:5.5}
+16 \alpha^{2} k^{2} r_{+}^{2}+8 \alpha^{2} r_{+}^{4} \biggr].   
\end{equation}

Using equation \eqref{eq:5.3} and equation \eqref{eq:5.5} we obtain inverse temperature 
\begin{equation*}
T_i=\frac{1}{8 \pi  (k^{2}+r_{+}^{2})^{2} r_{+}^{3}} \biggr[  -c c_{1} m^{2} r_{+}^{7}+(-2 c^{2} c_{2} m^{2}-2) r_{+}^{6}-2 c c_{1} k^{2} m^{2} r_{+}^{5}+((-4 c^{2} c_{2} m^{2}-4) k^{2}+4 Q_{m}^{2}+4 \alpha ) r_{+}^{4}
\end{equation*}
\begin{equation}\label{eq:5.6}    
-c c_{1} k^{4} m^{2} r_{+}^{3}-2 ((c^{2} c_{2} m^{2}+1) k^{2}-Q_{m}^{2}-4 \alpha ) k^{2} r_{+}^{2}+4 \alpha  k^{4} \biggr].
\end{equation}
The expression for Joule-–Thomson coefficient looks cumbersome, so we will not present it here. If one 
takes $\beta \to 0$ limit to equations \eqref{eq:5.5} \& \eqref{eq:5.6} then inverse pressure \& 
temperature of electrically charged $AdS$ black holes in $4D$ EGB massive gravity are recovered 
\cite{Paul:2023mlh}. In the limit $m \to 0$, and $\alpha \to 0$, above equations \eqref{eq:5.5} \& 
\eqref{eq:5.6} are reduced to the inverse pressure \& temperature of black hole in $4D$ massless gravity 
couples to NED \cite{kruglov2022nonlinearly}
\begin{equation}\label{eq:5.7}
P_i=\frac{1}{16 r_{{+}}^{2} \pi  (k^{2}+r_{{+}}^{2})^{2}} \biggr[ 
4 Q_{m}^{2} k^{2}+6 Q_{m}^{2} r_{{+}}^{2}-4 k^{4}-8 k^{2} r_{{+}}^{2}-4 r_{{+}}^{4} \biggr],     
\end{equation}
\begin{equation}\label{eq:5.8}
T_i= \frac{1}{8 (k^{2}+r_{{+}}^{2})^{2} \pi  r_{{+}}}  \biggr[ -2 r_{{+}}^{4}+(-4 k^{2}+4 Q_{m}^{2}) r_{{+}}^{2}
 -2 k^{4}+2 Q_{m}^{2} k^{2} \biggr].
\end{equation}

\begin{figure}[H]
\centering
\subfloat[$\alpha=0.2$]{\includegraphics[width=.5\textwidth]{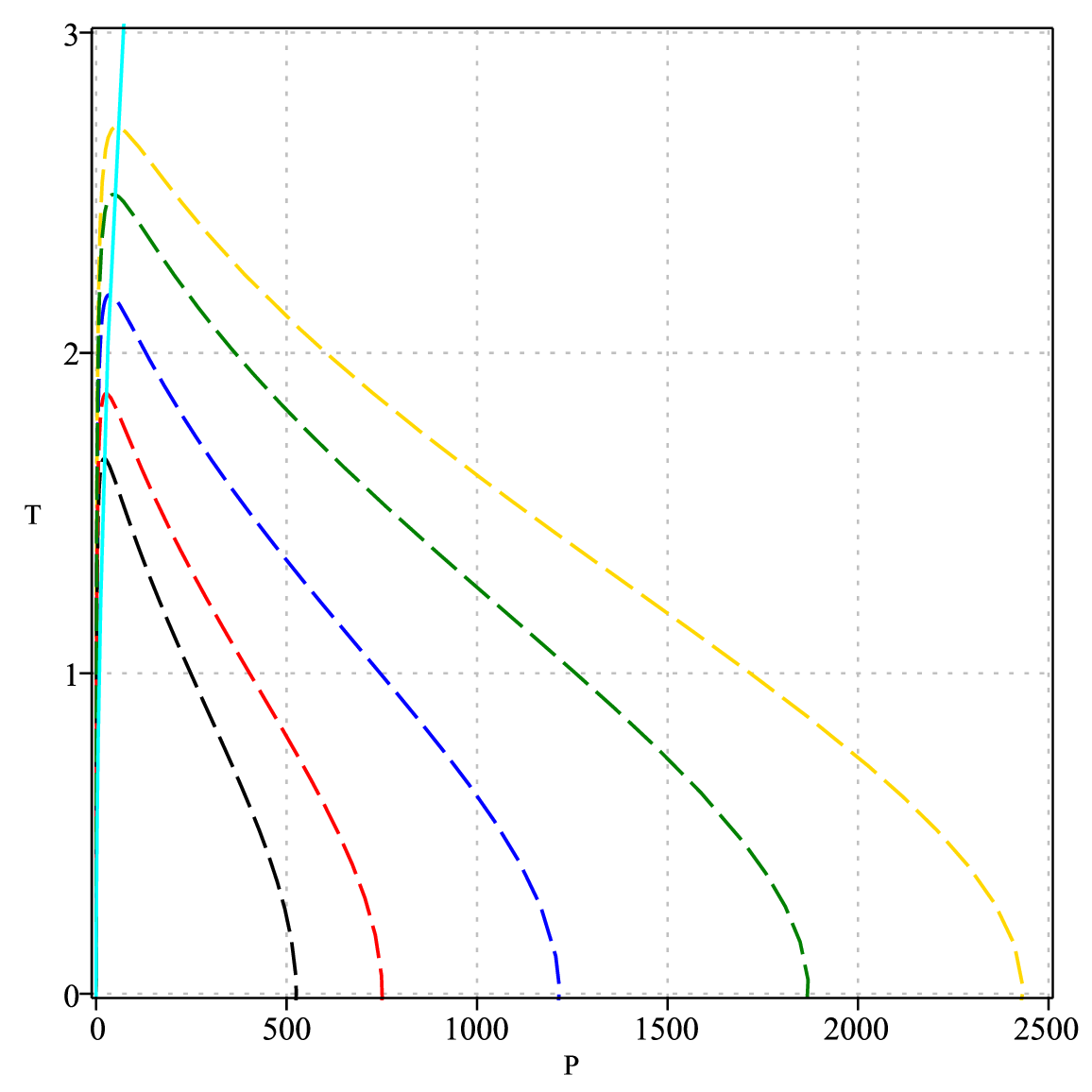}}\hfill
\subfloat[$\alpha=0.6$]{\includegraphics[width=.5\textwidth]{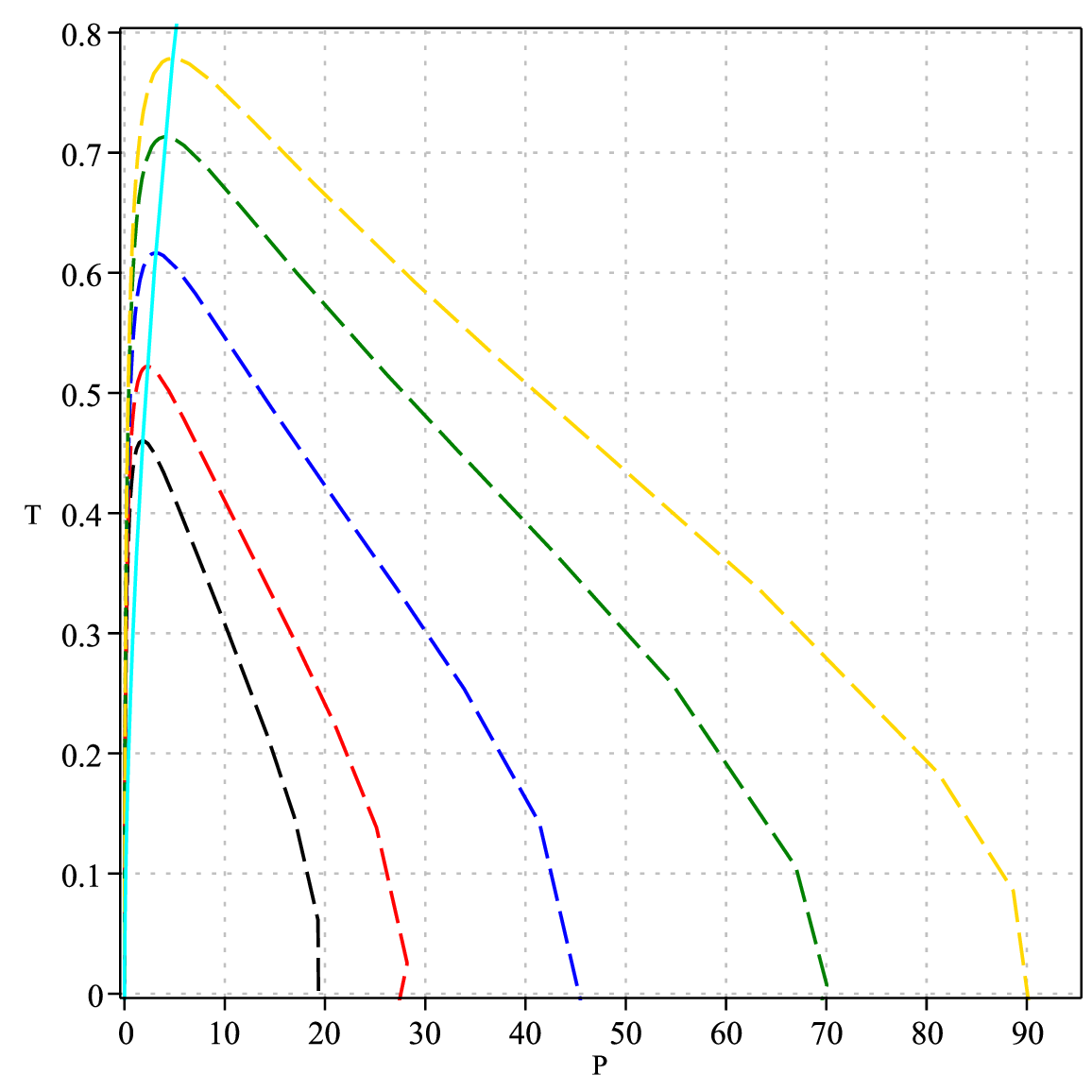}}\hfill
\caption{Black dash line denotes $M=4$, red dash line denotes $M=4.2$, blue dash line denoted $M=4.5$, green dash line denoted $M=4.8$, gold dash line denoted $M=5.0$ and solid cyan line denotes inverse curve with $Q_m=2$, $\beta=0.5$, $m=0.5$, $c=1$, $c_1=-1$ and $c_2=1$.}\label{fig:38}
\end{figure}

\begin{figure}[H]
\centering
\subfloat[$\beta=0.5$]{\includegraphics[width=.5\textwidth]{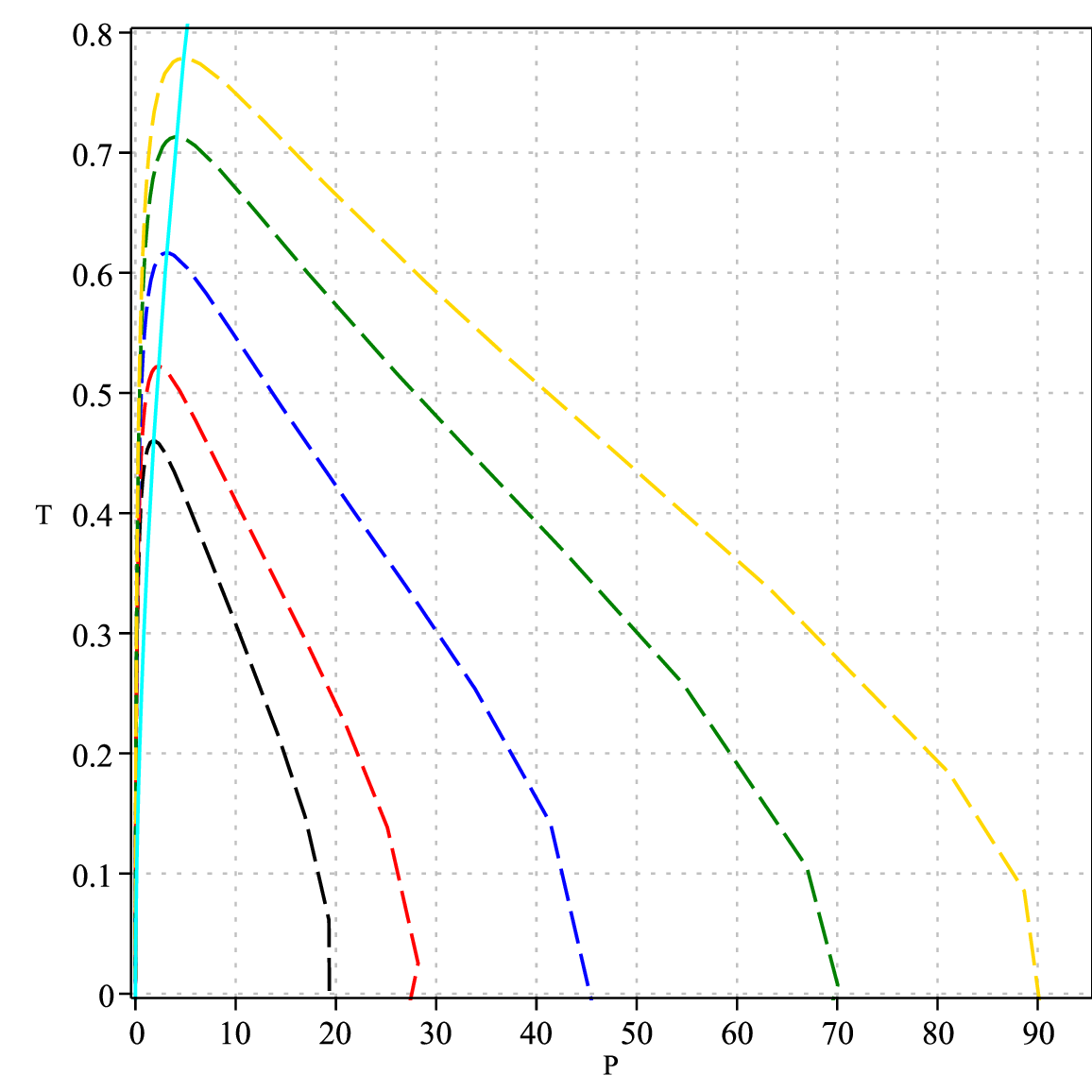}}\hfill
\subfloat[$\beta=1.0$]{\includegraphics[width=.5\textwidth]{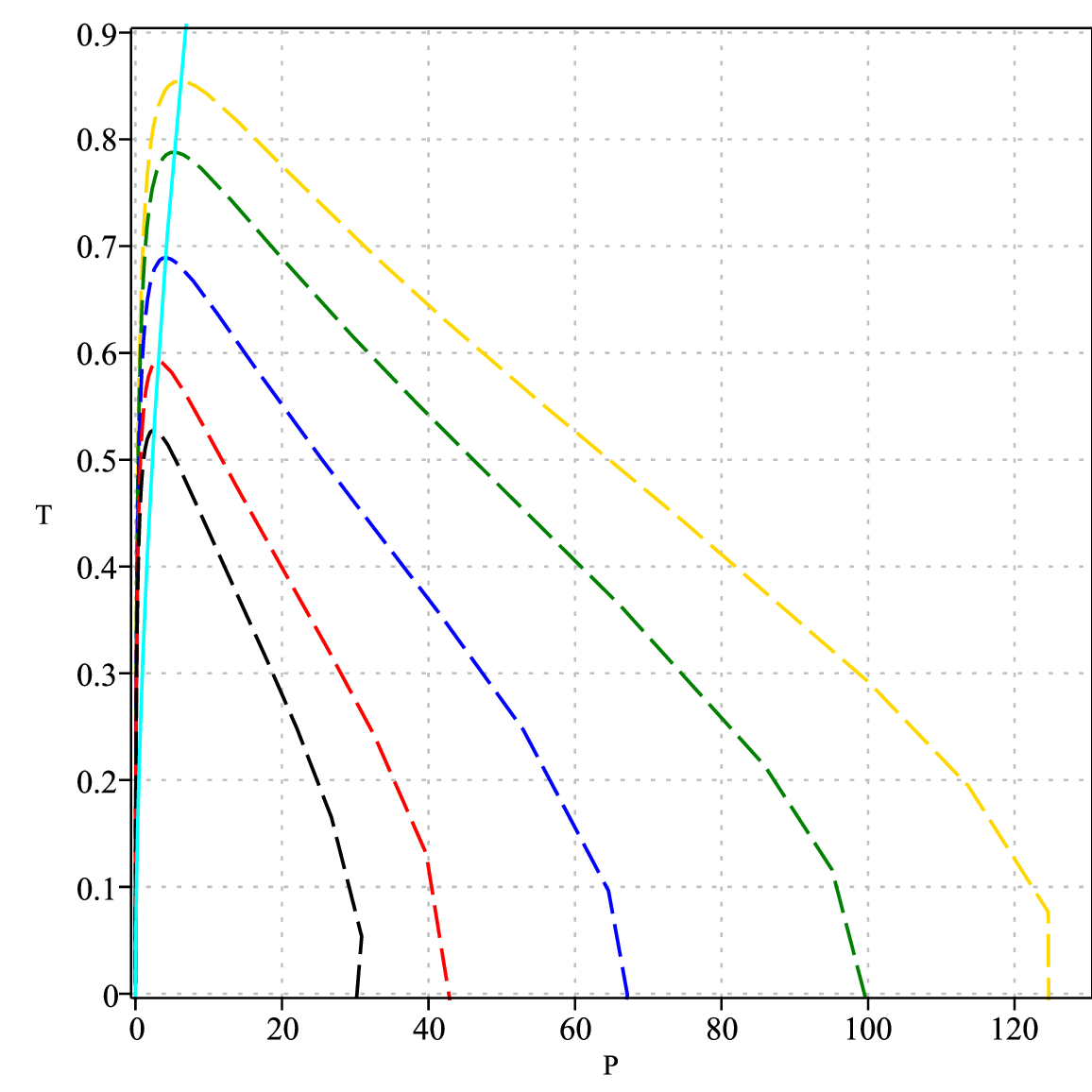}}\hfill
\caption{Black dash line denotes $M=4$, red dash line denotes $M=4.2$, blue dash line denoted $M=4.5$, green dash line denoted $M=4.8$, gold dash line denoted $M=5.0$ and solid cyan line denotes inverse curve with $Q_m=2$, $\beta=0.5$, $m=0.5$, $c=1$, $c_1=-1$ and $c_2=1$.}\label{fig:39}
\end{figure}

\begin{figure}[H]
\centering
\subfloat[$\alpha=0.2$]{\includegraphics[width=.5\textwidth]{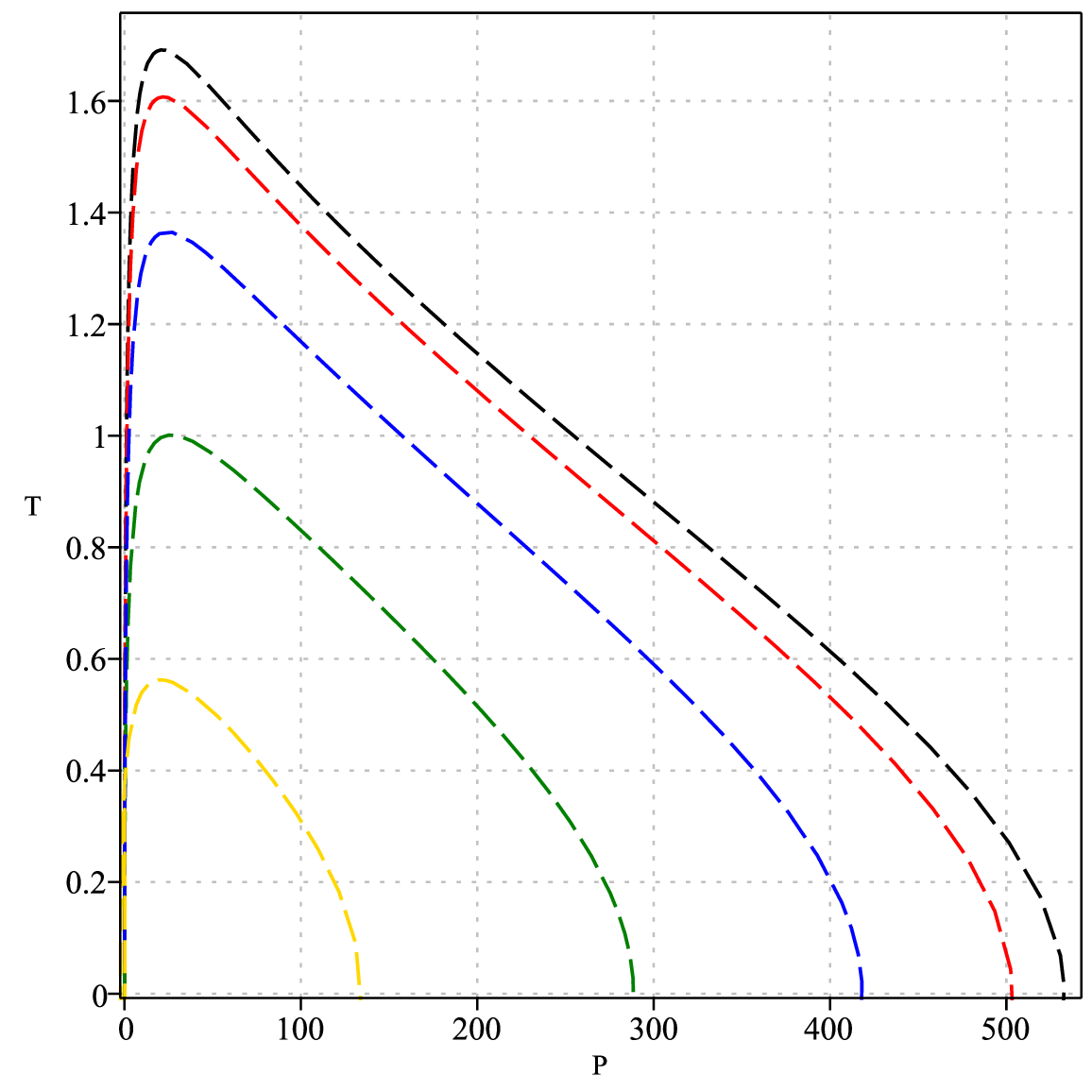}}\hfill
\subfloat[$\alpha=0.4$]{\includegraphics[width=.5\textwidth]{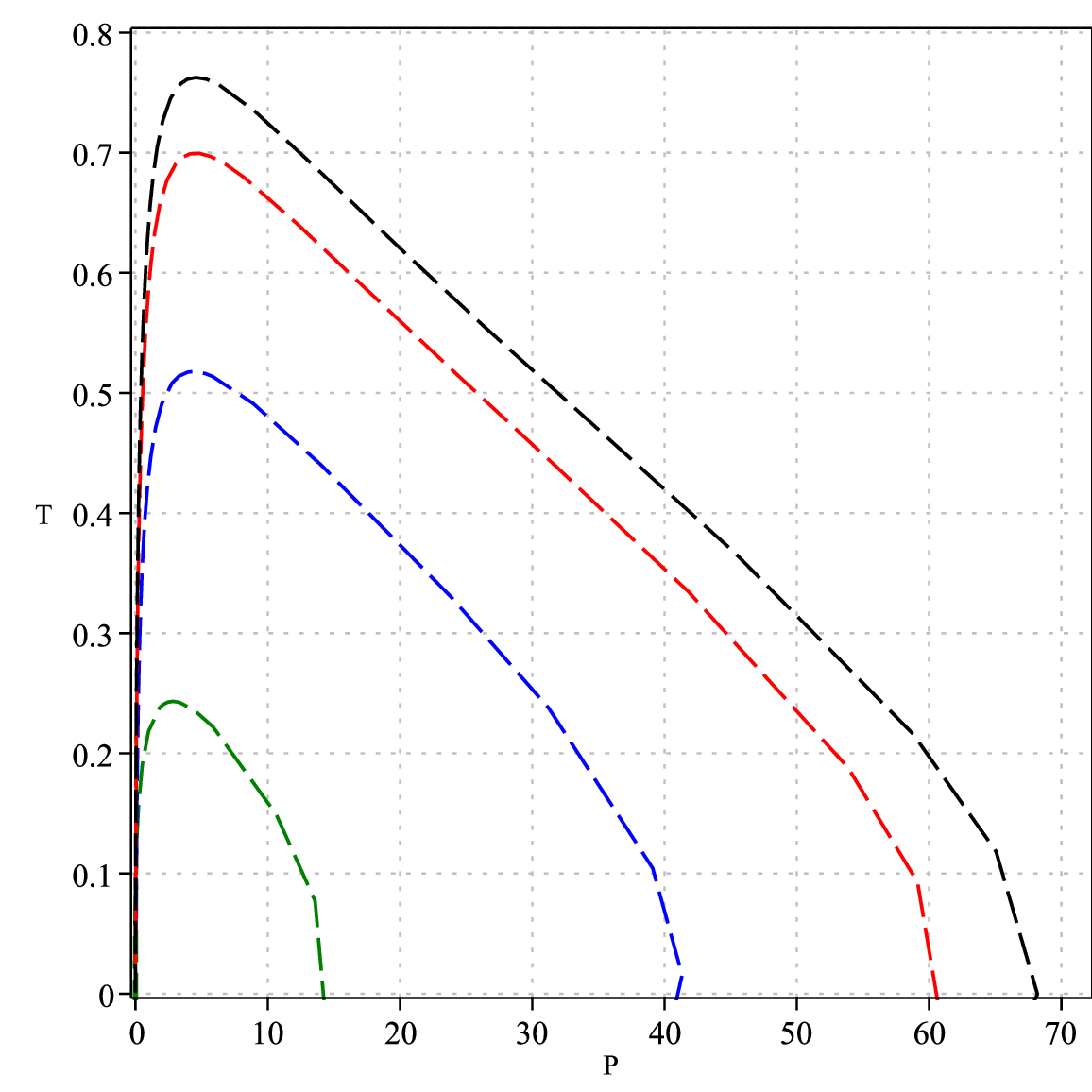}}\hfill
\caption{Black dash line denotes $m=0$, red dash line denotes $m=1.0$, blue dash line denoted $m=2.0$, green dash line denoted $m=3.0$ and gold dash line denoted $m=4.0$ with $Q_m=2$, $\beta=0.5$, $M=4$, $c=1$, $c_1=-1$ and $c_2=1$.}\label{fig:40}
\end{figure}

\begin{figure}[H]
\centering
\subfloat[$\beta=0.5$]{\includegraphics[width=.5\textwidth]{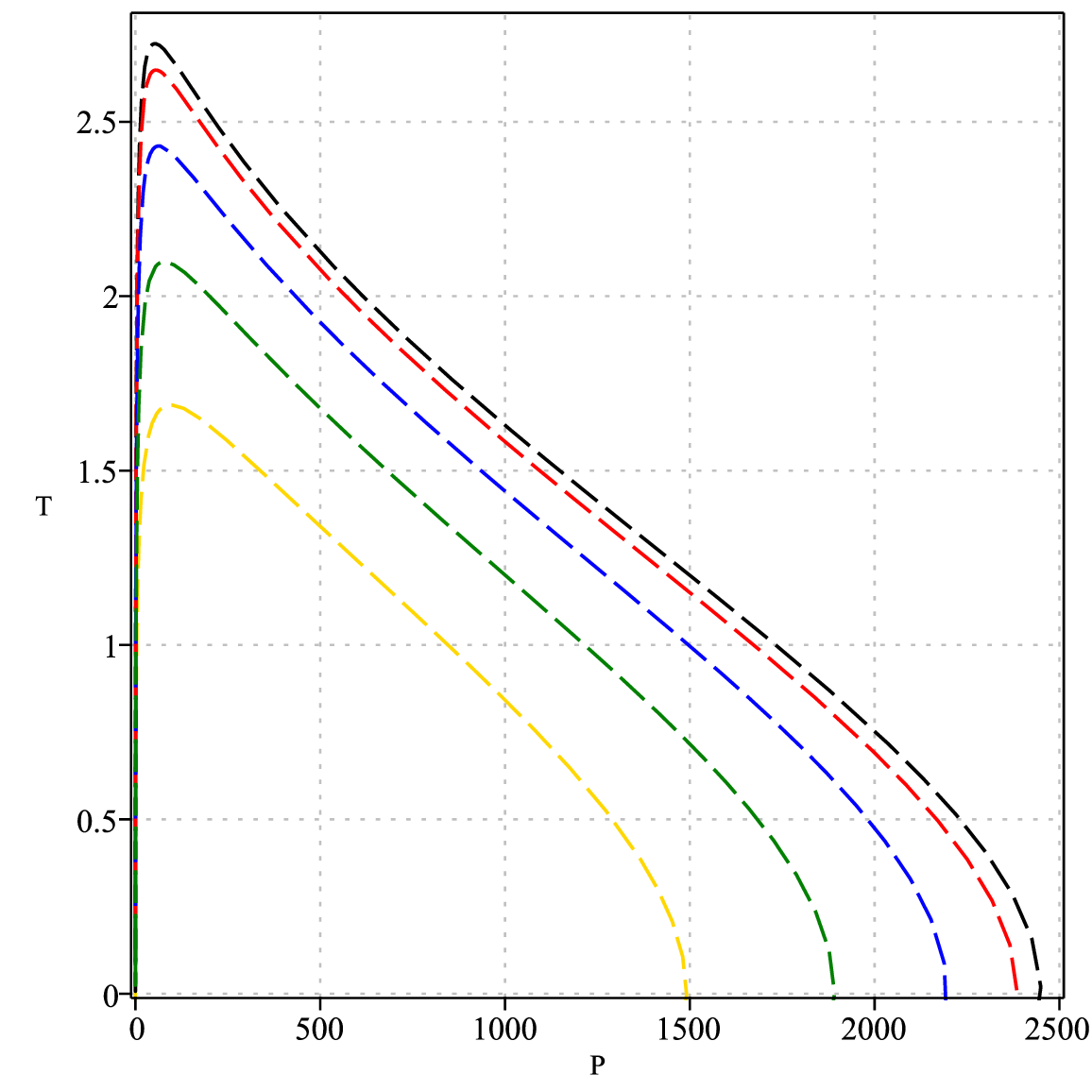}}\hfill
\subfloat[$\beta=1.0$]{\includegraphics[width=.5\textwidth]{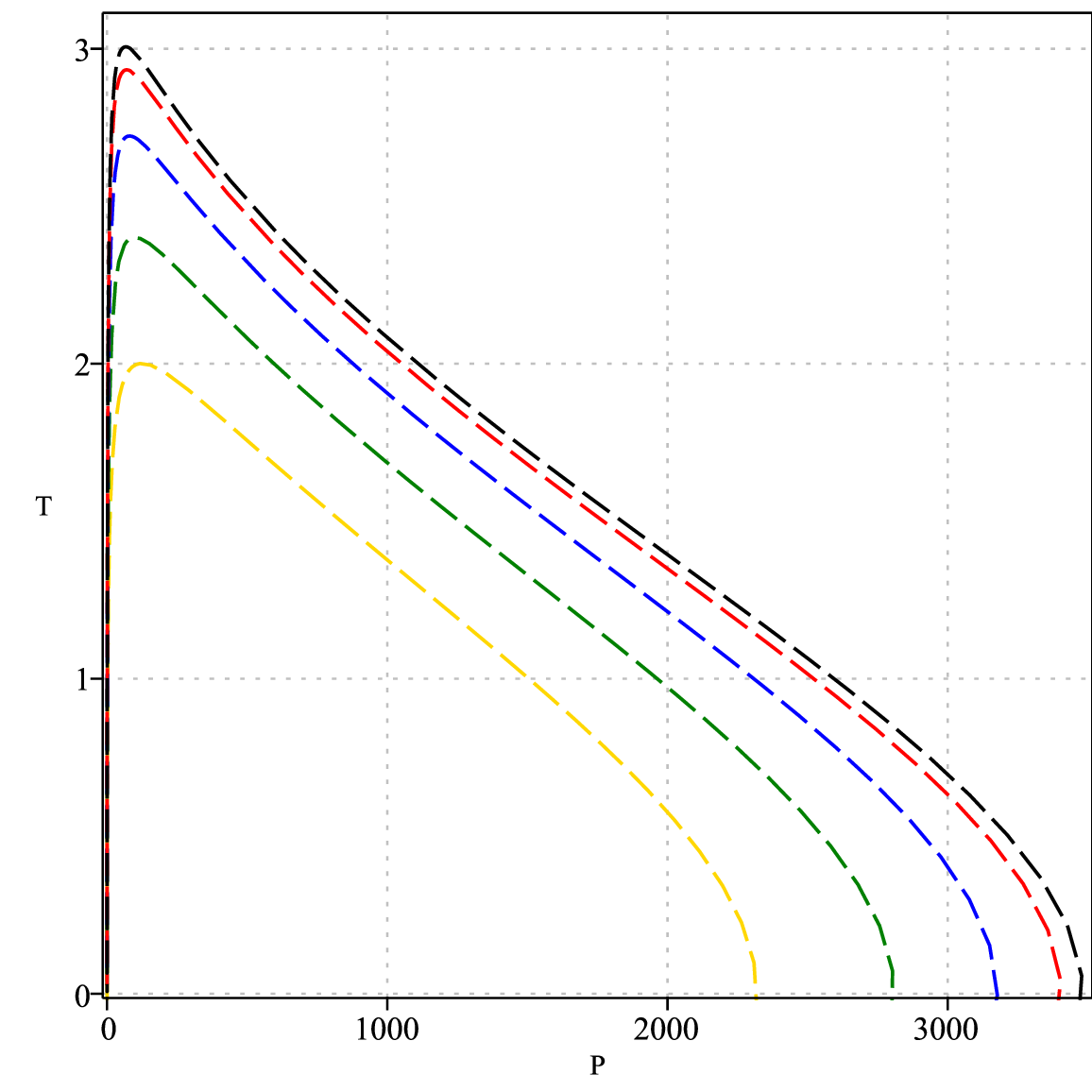}}\hfill
\caption{Black dash line denotes $m=0$, red dash line denotes $m=1.0$, blue dash line denoted $m=2.0$, green dash line denoted $m=3.0$ and gold dash line denoted $m=4.0$ with $Q_m=2$, $\alpha=0.2$, $M=5$, $c=1$, $c_1=-1$ and $c_2=1$.}\label{fig:41}
\end{figure}

\begin{figure}[H]
\centering
\subfloat[$c_1=-1$ \& $c_2=-1$]{\includegraphics[width=.5\textwidth]{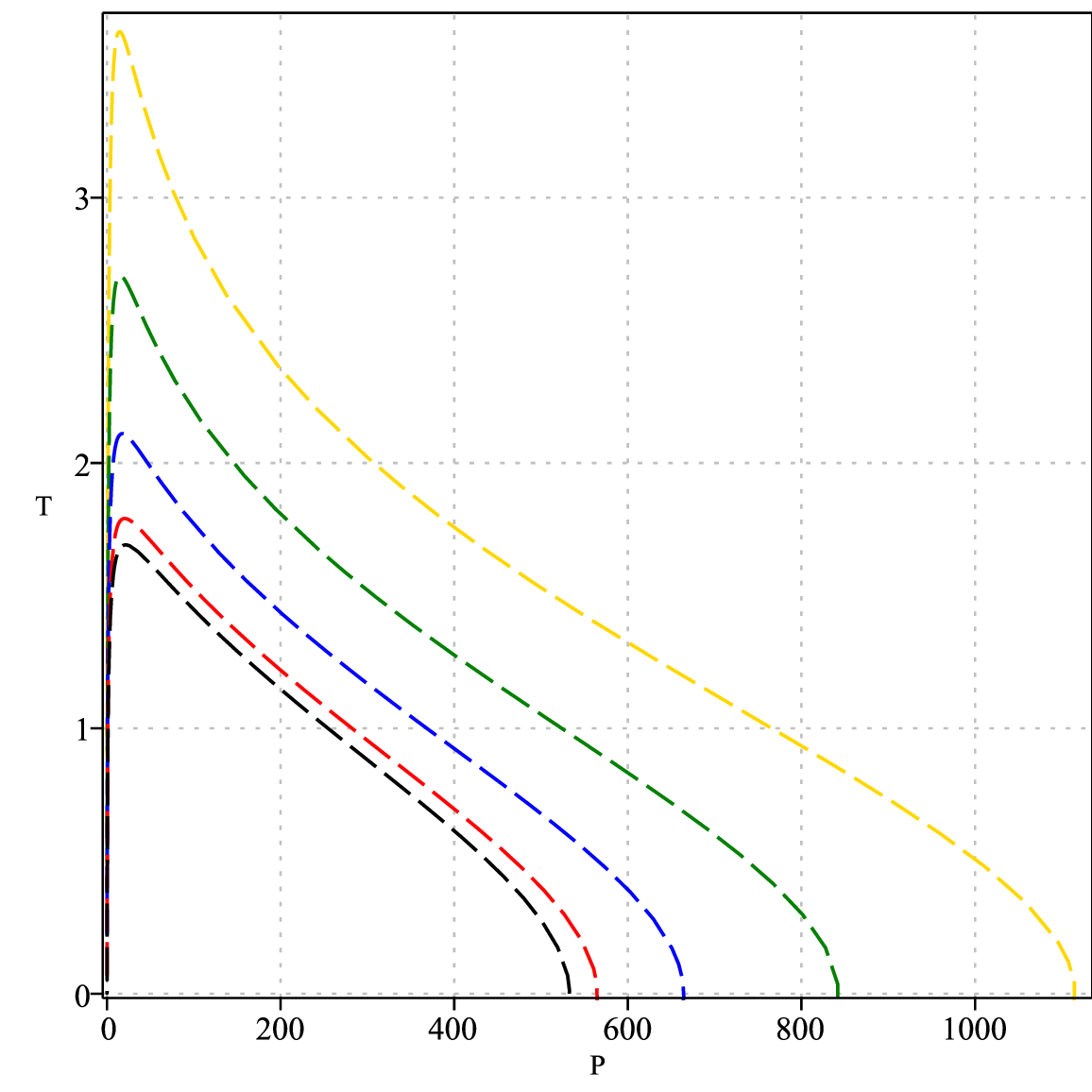}}\hfill
\subfloat[$c_1=-1$ \& $c_2=1$]{\includegraphics[width=.5\textwidth]{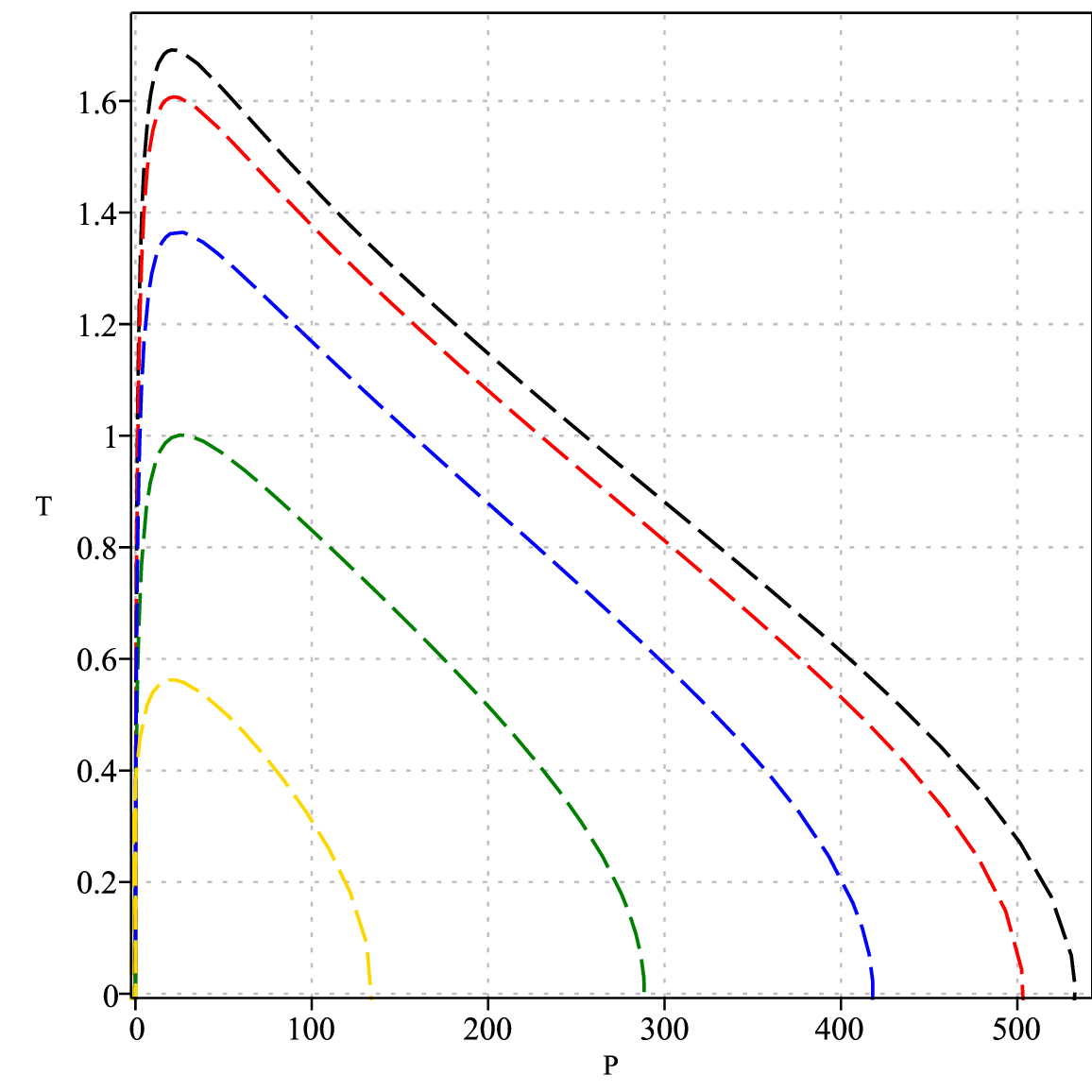}}\hfill
\subfloat[$c_1=1$ \& $c_2=-1$]{\includegraphics[width=.5\textwidth]{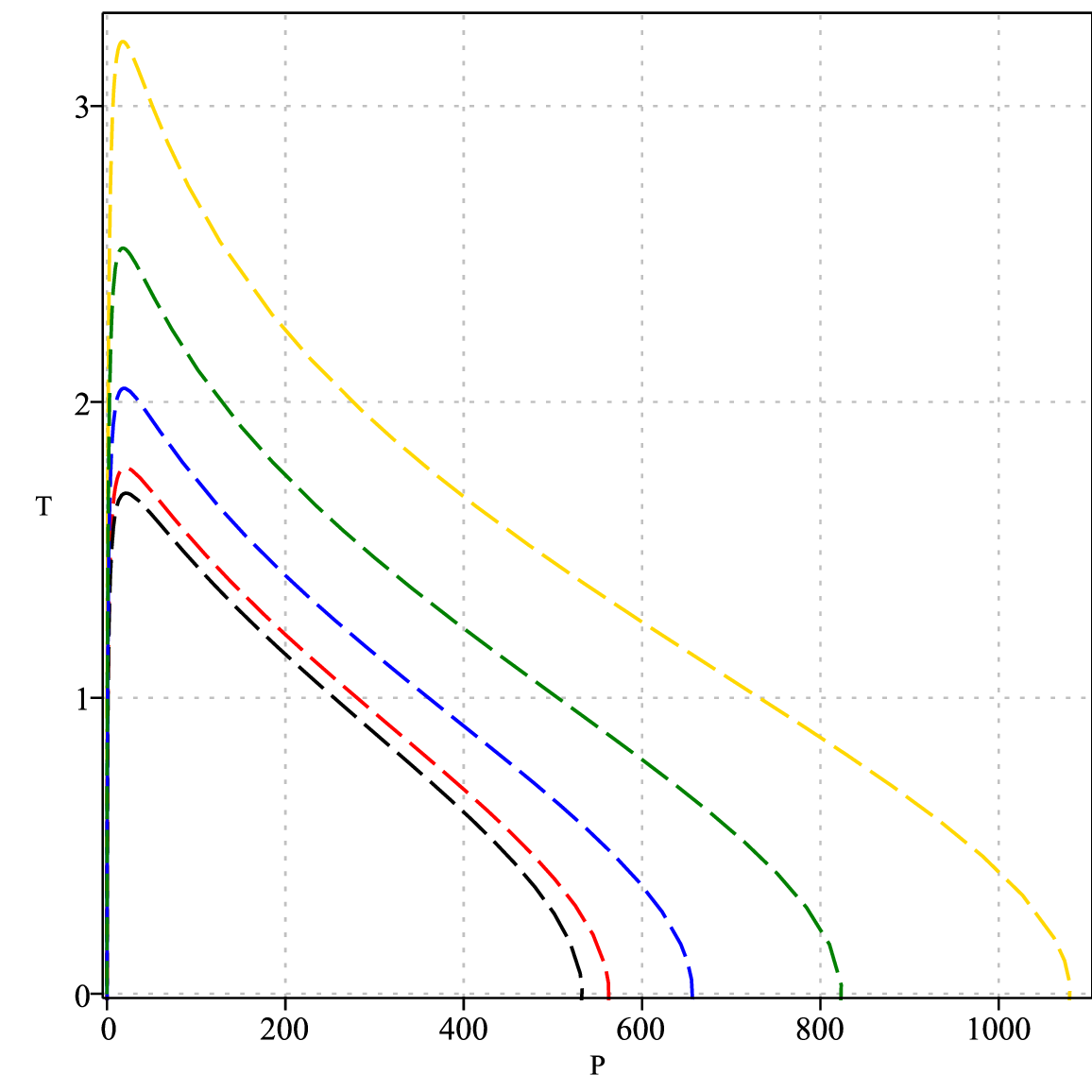}}\hfill
\subfloat[$c_1=1$ \& $c_2=1$]{\includegraphics[width=.5\textwidth]{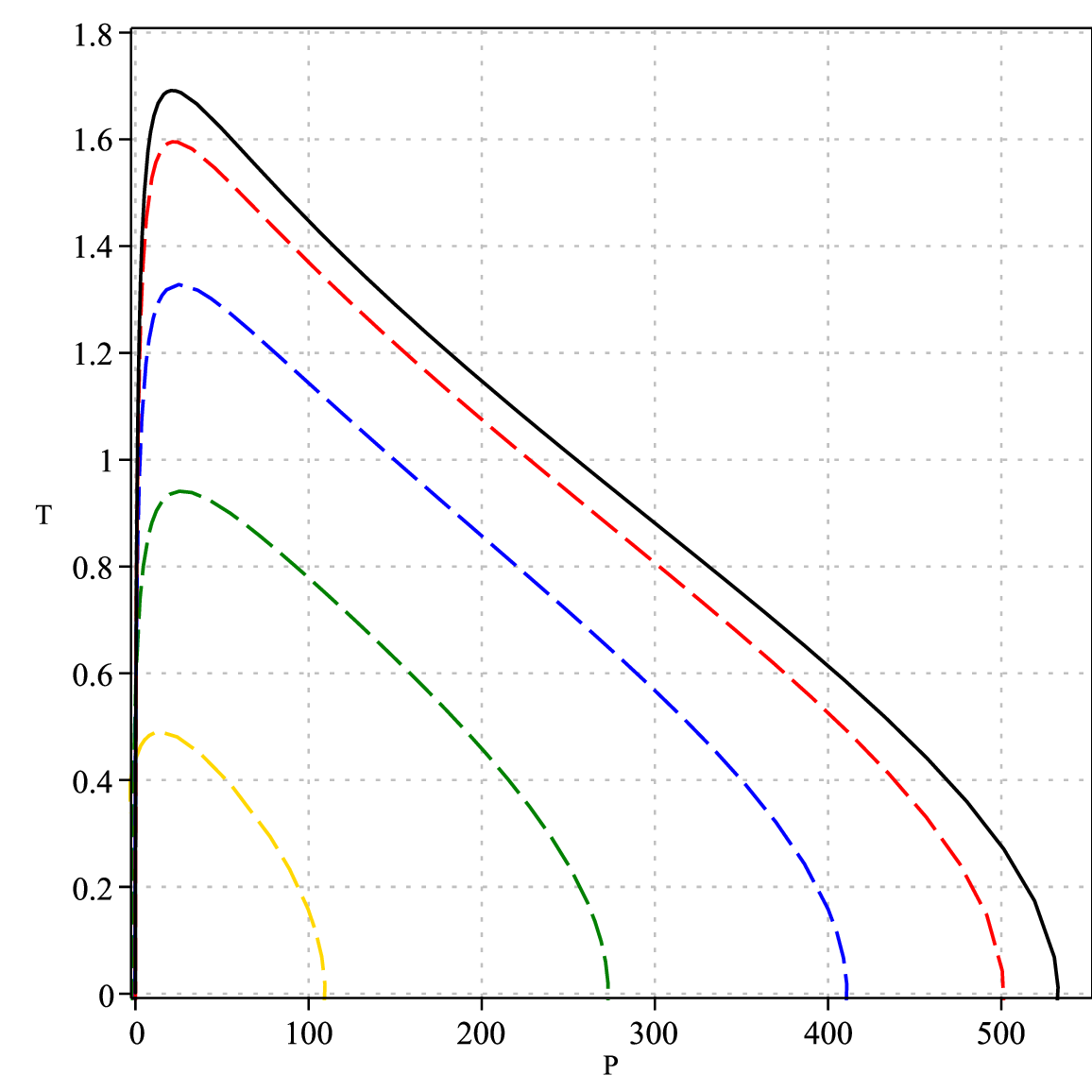}}\hfill
\caption{Black dash line denotes $m=0$, red dash line denotes $m=1.0$, blue dash line denoted $m=2.0$, green dash line denoted $m=3.0$ and gold dash line denoted $m=4.0$ with $Q_m=2$, $\alpha=0.2$, $\beta=0.5$, $M=4$ and $c=1$.}\label{fig:42}
\end{figure}

\begin{figure}[H]
\centering
\subfloat[$c_2=0$]{\includegraphics[width=.5\textwidth]{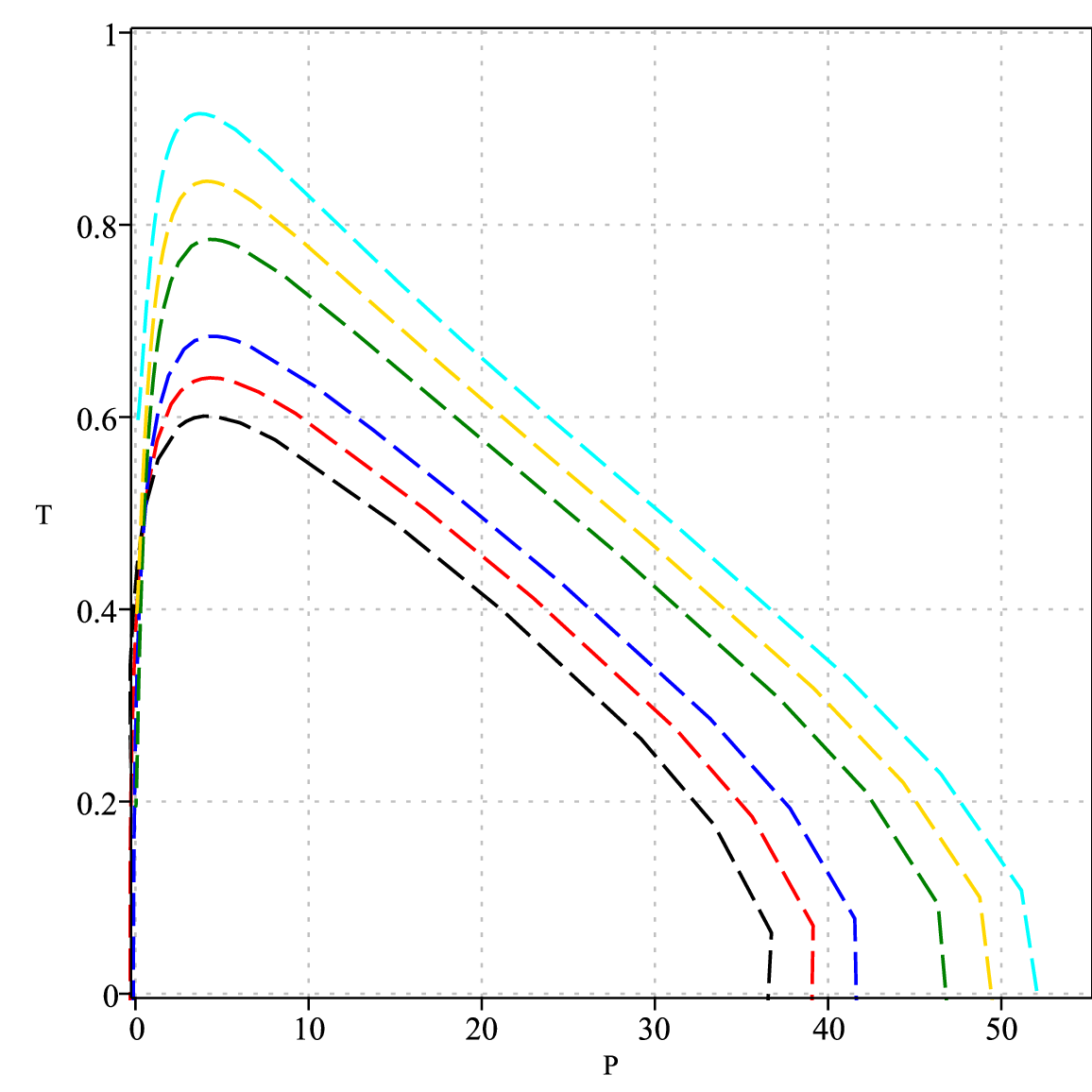}}\hfill
\subfloat[$c_1=0$]{\includegraphics[width=.5\textwidth]{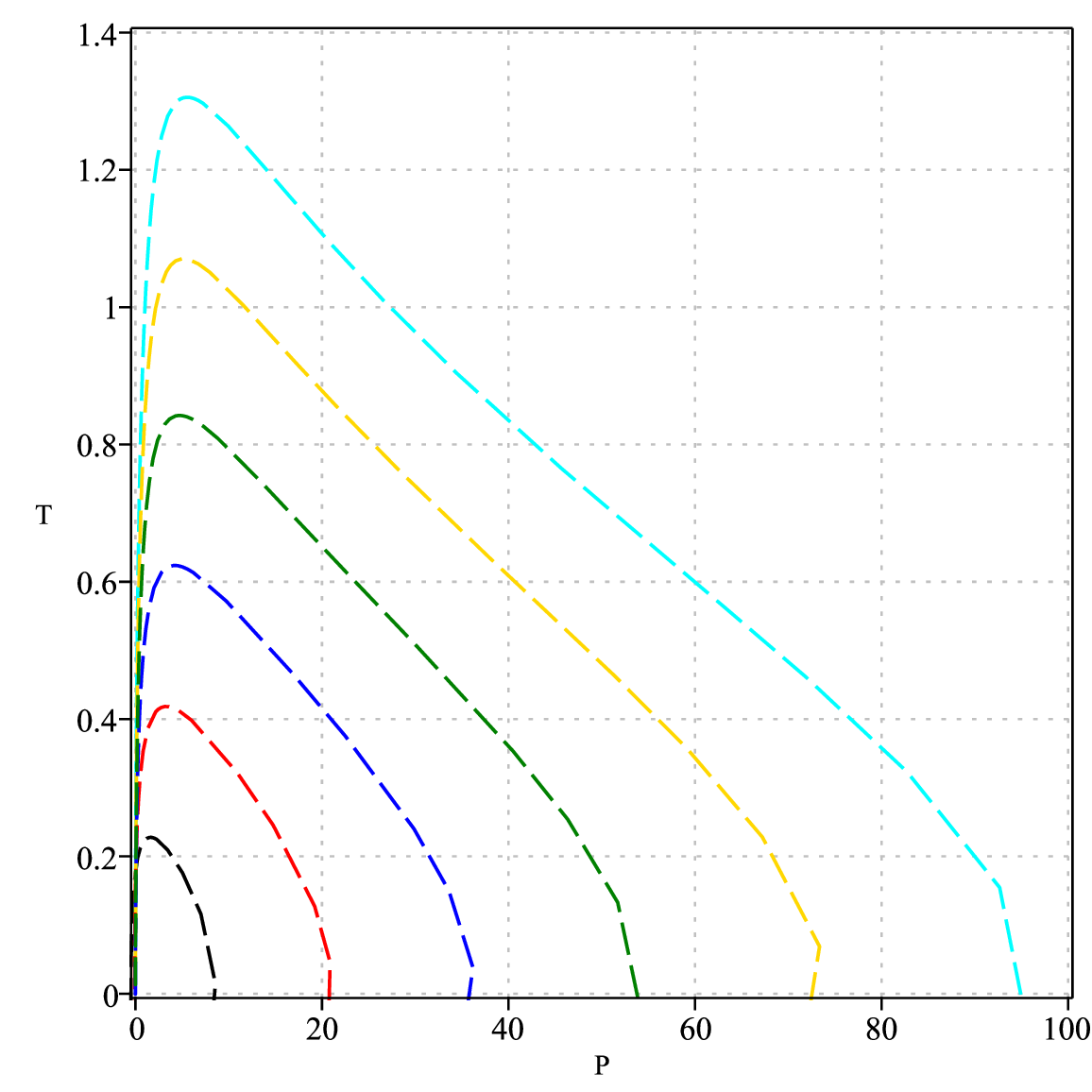}}\hfill
\caption{$Q_m=2$, $\alpha=0.2$, $\beta=0.5$, $M=3$, $m=1$ and $c=1$. 
Left panel: cyan dash line denotes $c_1=-15$, gold dash line denotes 
$c_1=-10$, green dash line denotes $c_1=-5$, blue dash line denotes 
$c_1=5$, red dash line denotes $c_1=10$ and black dash line denotes 
$c_1=15$. Right panel: cyan dash line denotes $c_2=-5$, gold dash line 
denotes $c_2=-5$, green dash line denotes $c_2=-1$, blue dash line 
denotes $c_2=1$, red dash line denotes $c_2=3$ and black dash line 
denotes $c_2=5$.}\label{fig:43}
\end{figure}

\begin{figure}[H]
\centering
\subfloat[$\beta=0.5$]{\includegraphics[width=.5\textwidth]{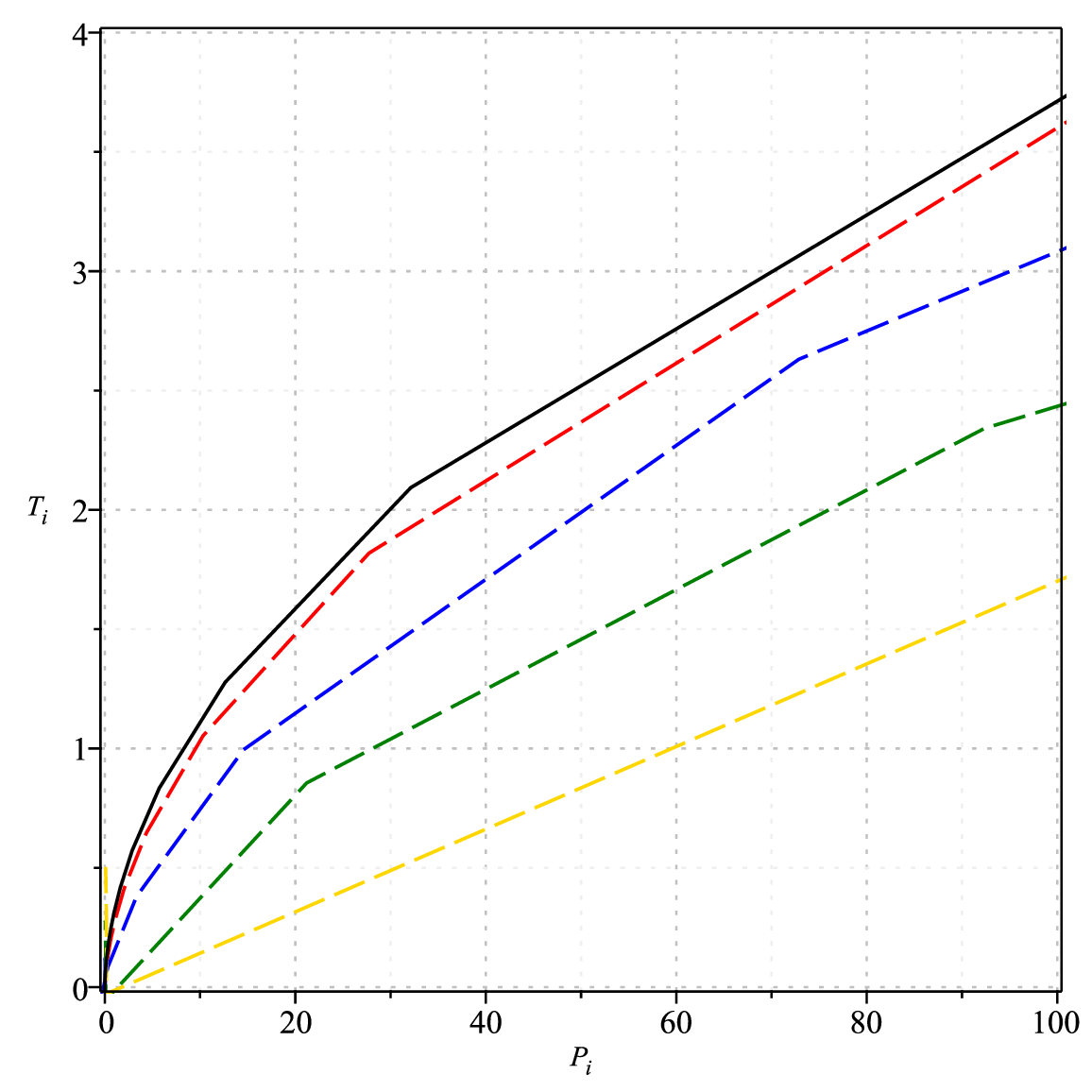}}\hfill
\subfloat[$m=0.5$]{\includegraphics[width=.5\textwidth]{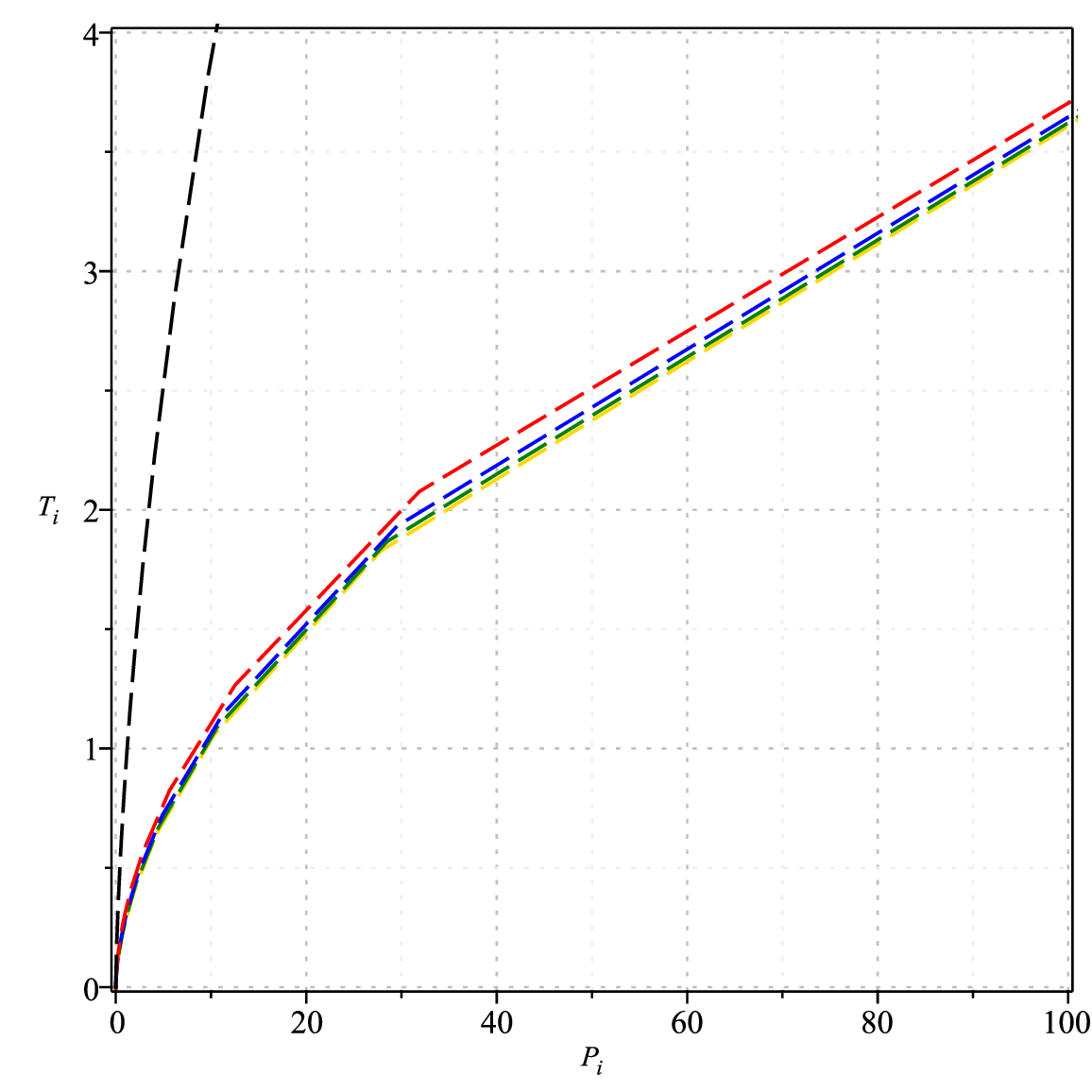}}\hfill
\caption{$Q_m=2$, $\alpha=0.2$, $\beta=0.5$, $c=1$, $c_1=-1$ and 
$c_2=1$. Left panel: black dash line denotes $m=0$, red dash line denotes $m=1.0$, blue dash line denoted $m=2.0$, green dash line denoted $m=3.0$ and gold dash line denoted $m=4.0$. Right panel: black dash line denotes $\beta=0$, red dash line denotes $\beta=0.4$, blue dash line denotes $\beta=0.8$, green dash line denotes $\beta=1.2$ and gold dash line denotes $\beta=1.6$.}\label{fig:44}
\end{figure}

\begin{figure}[H]
\centering
\subfloat[$c_2=0$]{\includegraphics[width=.5\textwidth]{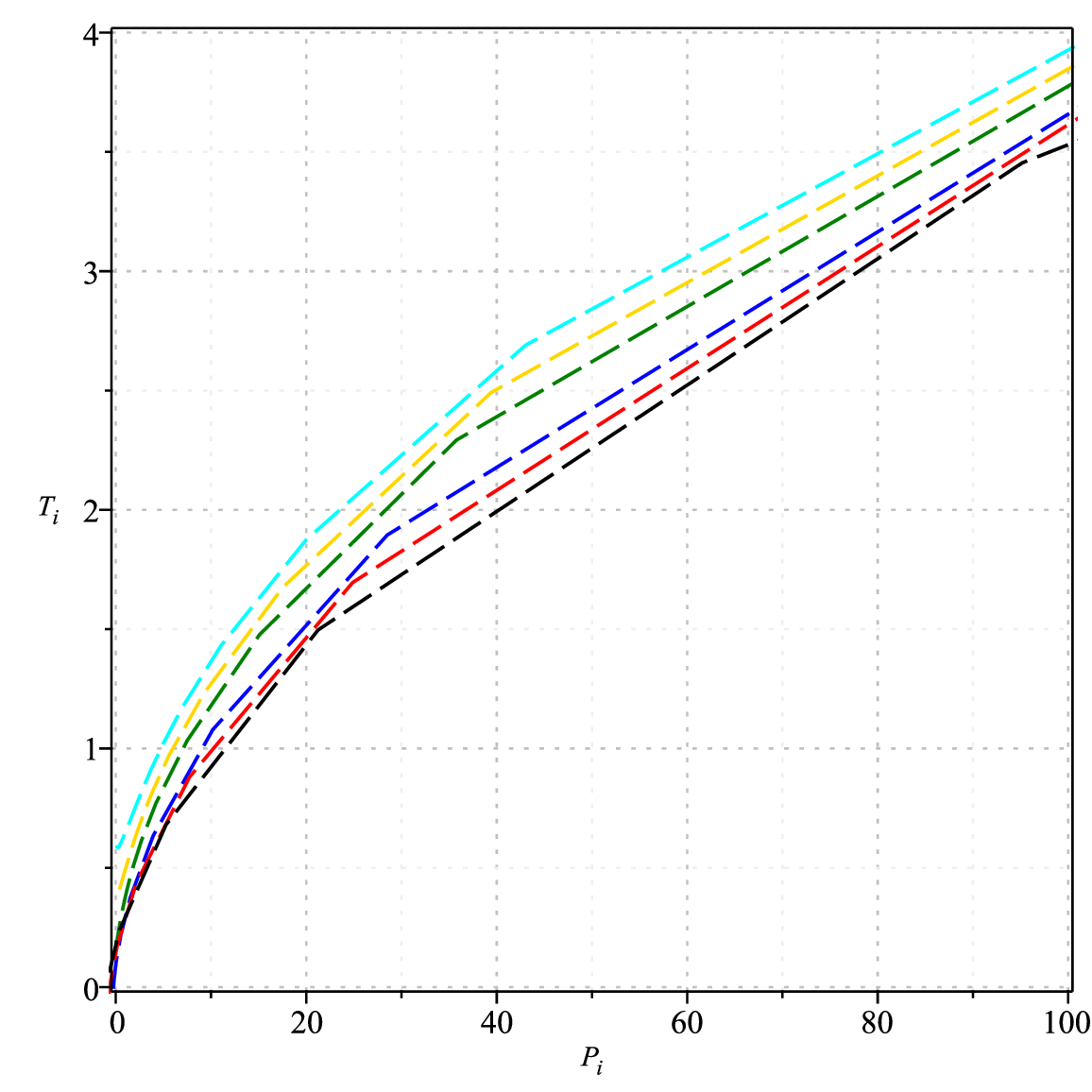}}\hfill
\subfloat[$c_1=0$]{\includegraphics[width=.5\textwidth]{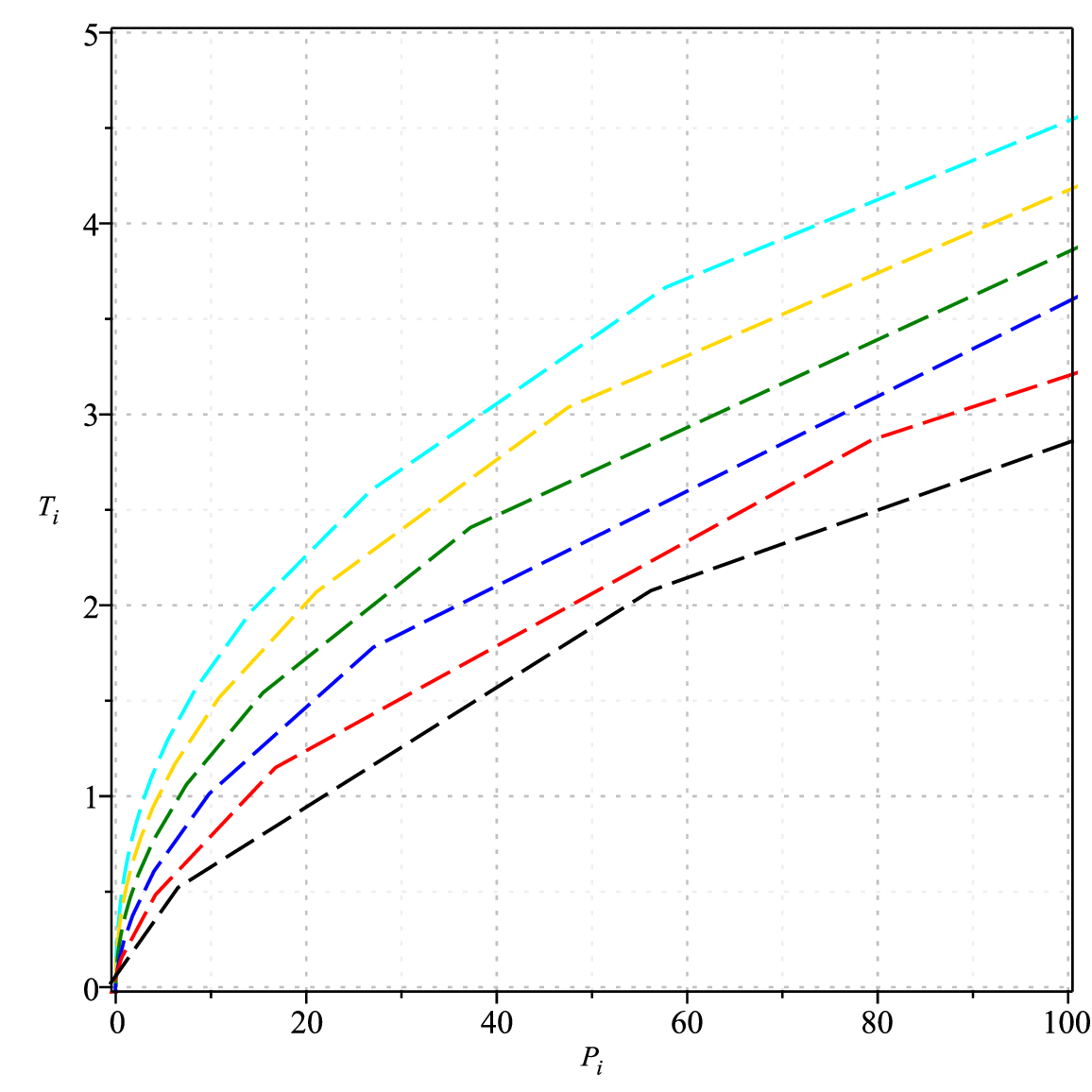}}\hfill
\caption{$Q_m=2$, $\alpha=0.2$, $\beta=0.5$, $M=3$, $m=1$ and 
$c=1$. Left panel: cyan dash line denotes $c_1=-15$, gold dash 
line denotes $c_1=-10$, green dash line denotes $c_1=-5$, blue 
dash line denotes $c_1=5$, red dash line denotes $c_1=10$ and 
black dash line denotes $c_1=15$. Right panel: cyan dash line denotes $c_2=-5$, gold dash line denotes $c_2=-5$, green dash line denotes $c_2=-1$, blue dash line denotes $c_2=1$, red dash line denotes $c_2=3$ and black dash line denotes $c_2=5$.}\label{fig:45}
\end{figure}

By setting $P_{i}=0$ into equation \eqref{eq:5.5}, one can obtain minimum event horizon radius as 
\begin{equation*}
(-4 c^{2} c_{2} m^{2}-4) (r_{{+}}^{min})^{8}-6 c_{1} c m^{2} (k^{2}+\frac{\alpha}{3}) (r_{{+}}^{min})^{7}+((-8 c^{2} c_{2} m^{2}-8) k^{2}-4 \alpha  c^{2} c_{2} m^{2}+6 Q_{m}^{2}+2 \alpha ) (r_{{+}}^{min})^{6}
\end{equation*}
\begin{equation*}
-3 c_{1} c m^{2} (k^{2}+{4 \alpha}/{3}) k^{2} (r_{{+}}^{min})^{5}+((-4 c^{2} c_{2} m^{2}-4) k^{4}+(-8 \alpha  c^{2} c_{2} m^{2}+4 Q_{m}^{2}+4 \alpha ) k^{2}+8 Q_{m}^{2} \alpha +8 \alpha^{2}) (r_{{+}}^{min})^{4}
\end{equation*}
\begin{equation}
-2 \alpha  c c_{1} k^{4} m^{2} (r_{{+}}^{min})^{3}-4 \alpha  ((c^{2} c_{2} m^{2}-{1}/{2}) k^{2}-Q_{m}^{2}-4 \alpha ) k^{2} (r_{{+}}^{min})^{2}+8 \alpha^{2} k^{4} -3 c c_{1} m^{2} (r_{{+}}^{min})^{9}=0.
\end{equation}

We numerically solve above equation and obtain the minimum event horizon radius. Using above equation and equation \eqref{eq:5.6} one can find out the minimum inverse temperature, which is listed in table \ref{table:9} \& \ref{table:10} for different values of graviton mass, NED parameter and Gauss-Bonnet coupling parameter.

\begin{table}[H]
\centering
\begin{tabular}{ |p{1.5cm}|p{1.5cm}|p{1.5cm}| } 
\hline
\multicolumn{3}{|c|}{$\alpha=0.2$ \& $\beta=0.5$ } \\
\hline
\textbf{m} & \textbf{$r_{+}^{min}$} & \textbf{$T_{i}^{min}$}  \\ [0.5ex]  
\hline
0.0 & 1.688 & 0.009  \\ \hline
0.1 & 1.693 & 0.009  \\ \hline
0.2 & 1.706 & 0.008  \\ \hline
0.3 & 1.733 & 0.007  \\ \hline
0.4 & 1.779 & 0.005  \\ \hline 
\end{tabular}
\caption{$Q_m=2$, $c_1=1$, $c_1=-1$ and $c_2=1$}
\label{table:9}
\end{table}

\begin{table}[H]
\centering
\begin{tabular}{ |p{1.5cm}|p{1.5cm}|p{1.5cm}| } 
\hline
\multicolumn{3}{|c|}{$m=0.1$ \& $\alpha=0.2$} \\
\hline
\textbf{$\beta$} & \textbf{$r_{+}^{min}$} & \textbf{$T_{i}^{min}$} \\ [0.5ex]  
\hline
0.0 & 2.533 & 0.0097  \\ \hline
0.1 & 2.191 & 0.0096  \\ \hline
0.4 & 1.794 & 0.0092 \\ \hline
0.8 & 1.437 & 0.0088 \\ \hline
1.0 & 1.302 & 0.0087 \\ \hline
\multicolumn{3}{|c|}{$m=0.1$ \& $\beta=0.5$} \\
\hline
\textbf{$\alpha$} & \textbf{$r_{+}^{min}$} & \textbf{$T_{i}^{min}$}  \\ [0.5ex]  
\hline
0.0 & 1.415 & 0.0094  \\ \hline
0.1 & 1.573 & 0.0093 \\ \hline
0.2 & 1.693 & 0.0091  \\ \hline
0.3 & 1.792 & 0.0088  \\ \hline
0.4 & 1.880 & 0.0086 \\ [1ex] 
\hline
\end{tabular}
\caption{$Q_m=2$, $c_1=1$, $c_1=-1$ and $c_2=1$}
\label{table:10}
\end{table}

\begin{figure}[H]
\centering
\subfloat[$c_2=0$]{\includegraphics[width=.5\textwidth]{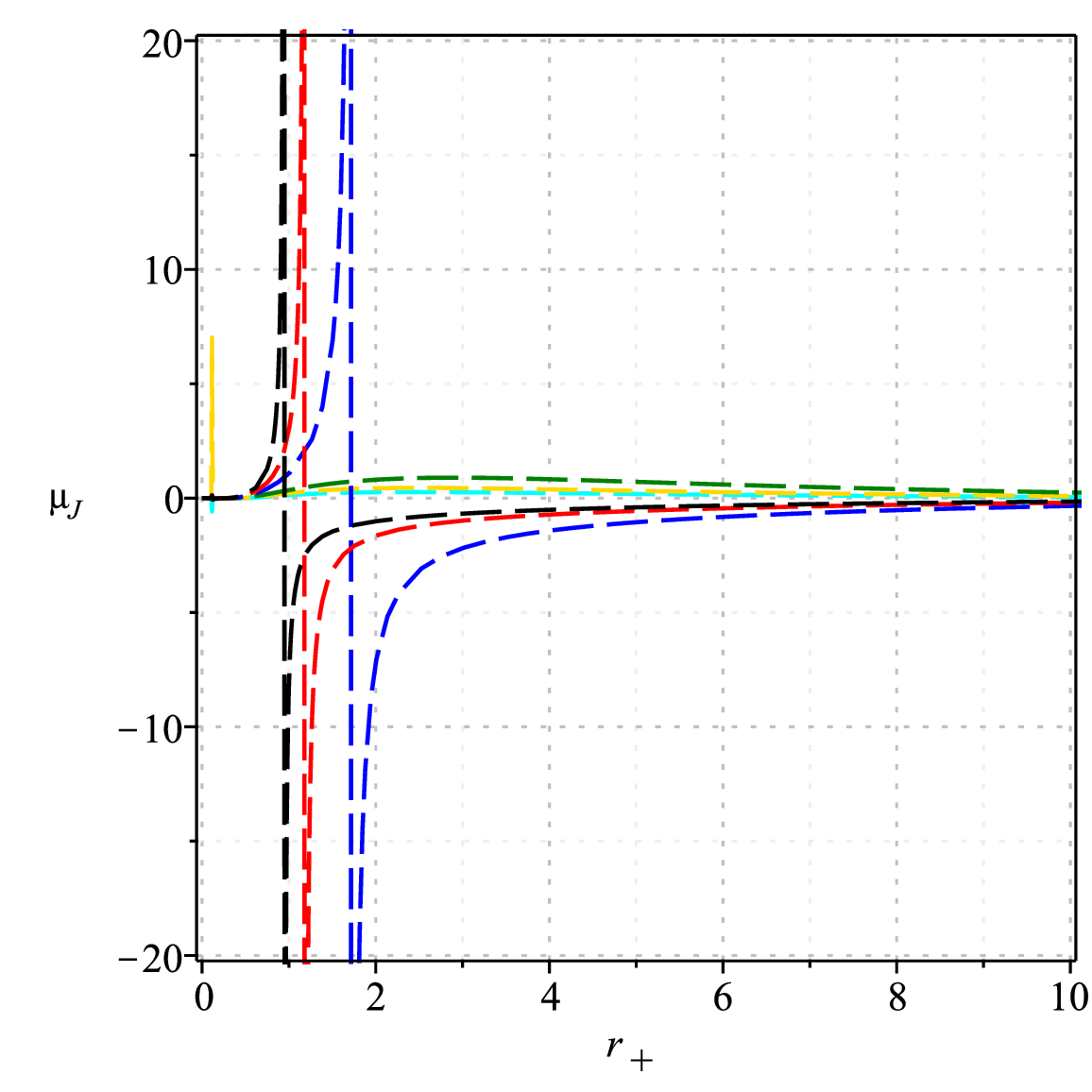}}\hfill
\subfloat[$c_1=0$]{\includegraphics[width=.5\textwidth]{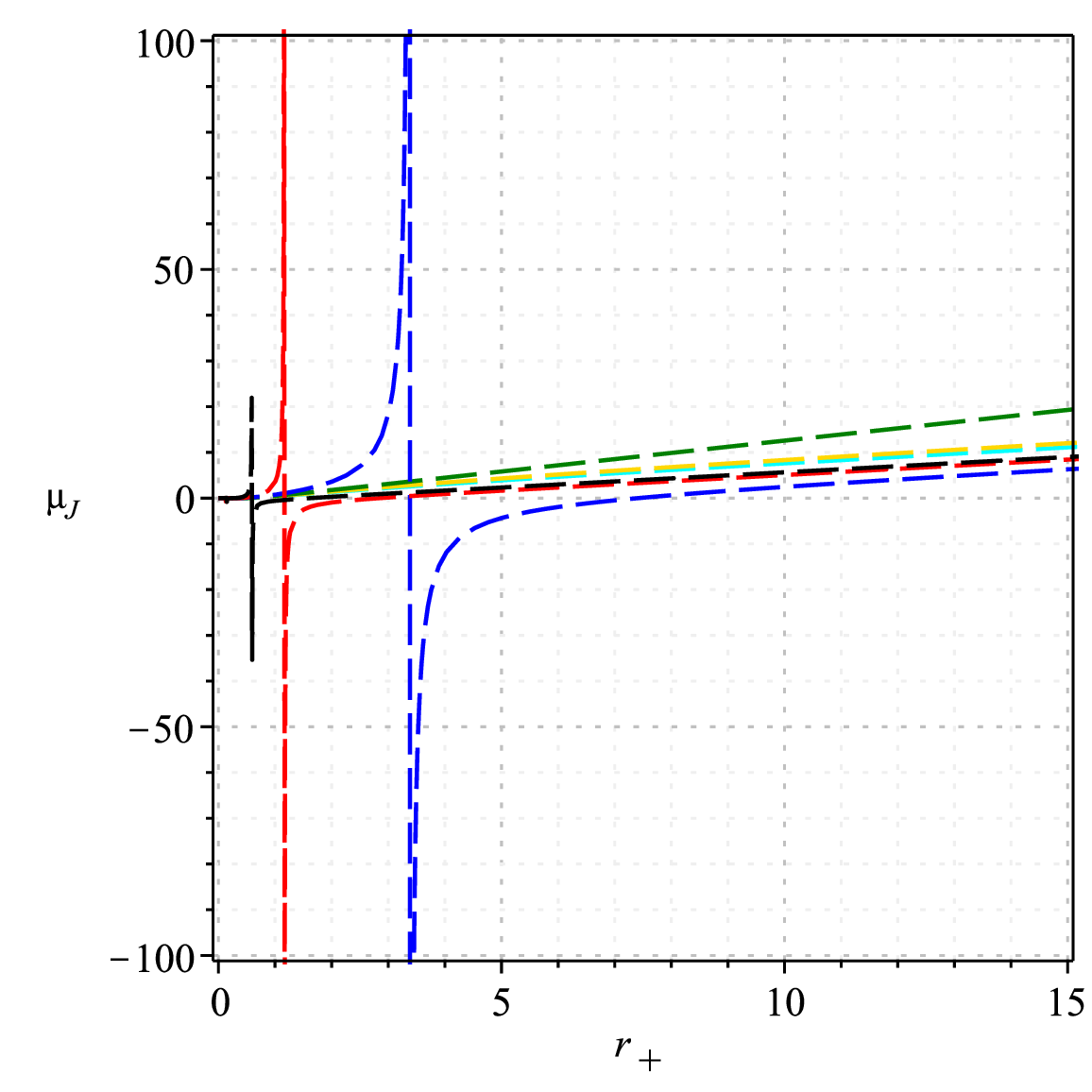}}\hfill
\subfloat[$c_2=0$]{\includegraphics[width=.3\textwidth]{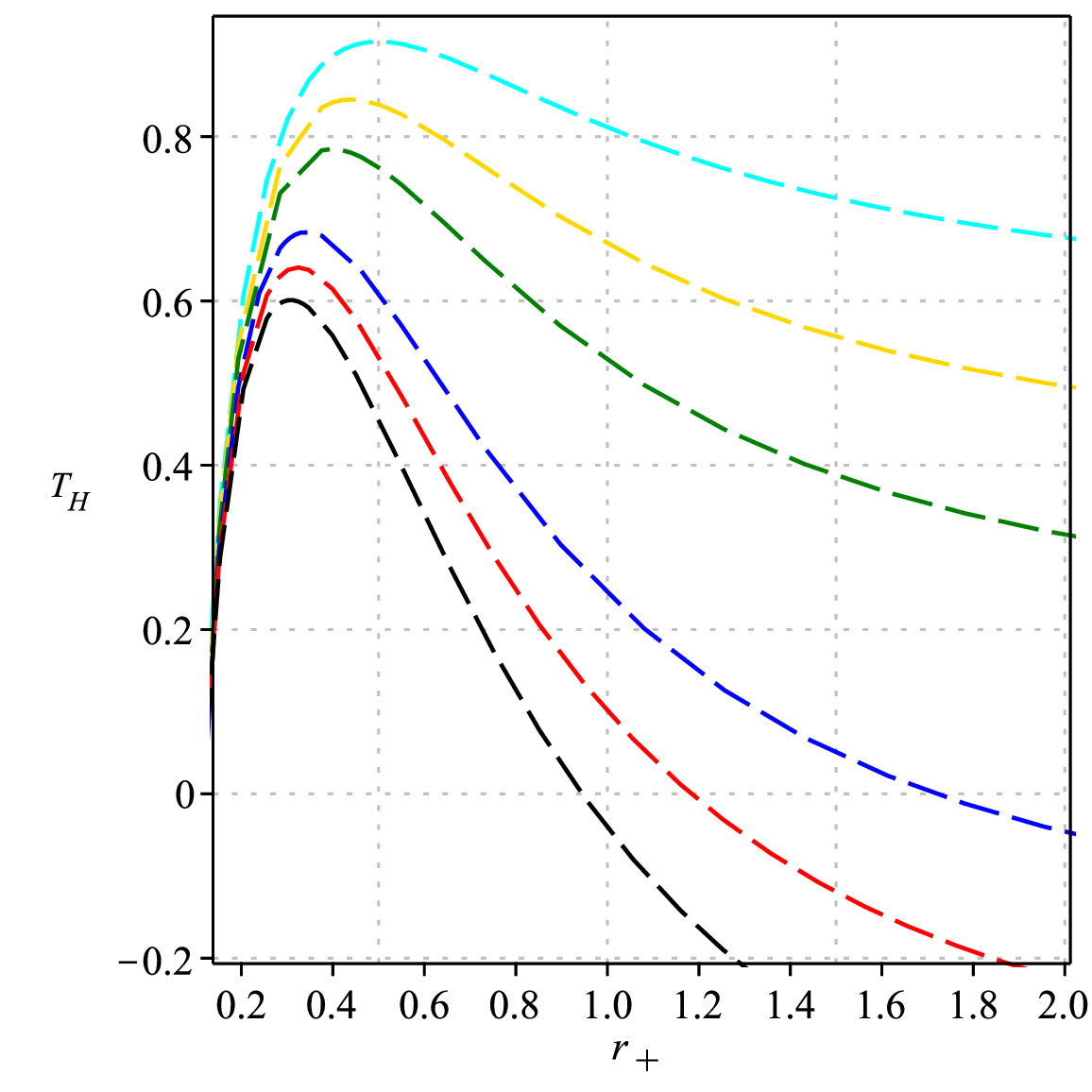}}\hfill
\subfloat[$c_1=0$]{\includegraphics[width=.3\textwidth]{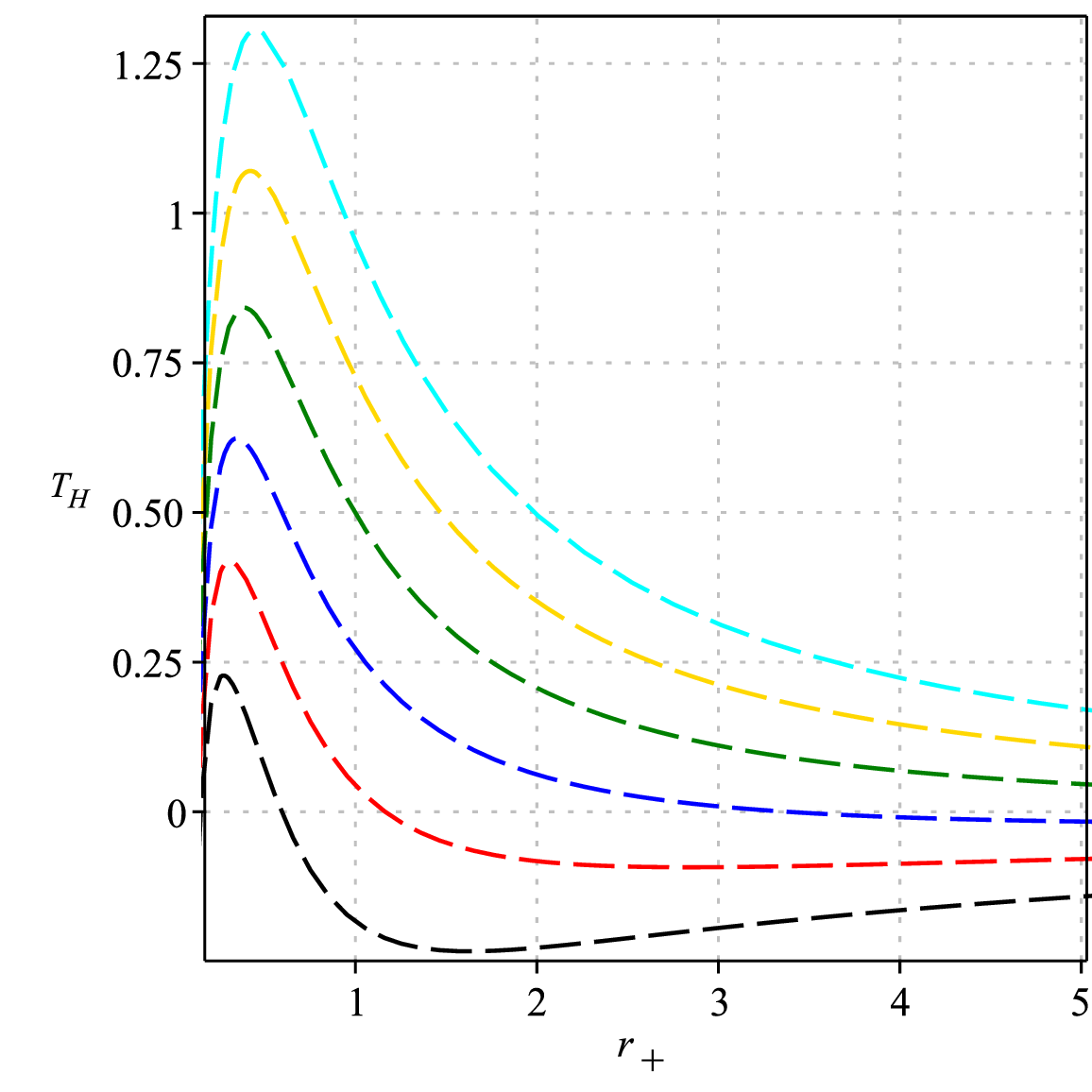}}\hfill
\caption{$Q_m=2$, $\alpha=0.2$, $\beta=0.5$, $M=3$, $m=1$ and $c=1$. Left panel: 
cyan dash line denotes $c_1=-15$, gold dash line denotes $c_1=-10$, green dash 
line denotes $c_1=-5$, blue dash line denotes $c_1=5$, red dash line denotes 
$c_1=10$ and black dash line denotes $c_1=15$. Right panel: cyan dash line denotes $c_2=-5$, gold dash line denotes $c_2=-3$, green dash line denotes $c_2=-1$, blue dash line denotes $c_2=1$, red dash line denotes $c_2=3$ and black dash line denotes $c_2=5$.}
\label{fig:46}
\end{figure}

In Fig. \ref{fig:46}(a) - \ref{fig:46}(b) we plot the Joule--Thomson coefficient 
of $4D$ EGB massive gravity black hole for different values of $c_{1}$ $(c_{2}=0)$ 
and $c_{2}$ $(c_{1}=0)$. For some particular values of horizon radius Joule--Thomson 
coefficient $\mu_{J}$ diverges. At the diverging points Hawking temperature goes to 
zero, i.e., $T_{H}=0$ which is shown in Figs. \ref{fig:46}(c) and \ref{fig:46}(d). 
Before the diverging points $\mu_{J}$ are positive, which indicates a cooling phase 
and at the diverging points a phase transition occurs for $\mu_{J}$. Joule--Thomson 
 coefficients change its phase from cooling to heating phase $\mu_{J} < 0$. 
Finally, at large values of $r_{+}$ Joule--Thomson coefficient $\mu_{J}=0$, which 
is known as the inverse phenomenon and the black hole goes to the cooling phase from 
heating phase.

\begin{figure}[H]
\centering
\subfloat[$\beta=0.5$ \& $\alpha=0.2$]{\includegraphics[width=.5\textwidth]{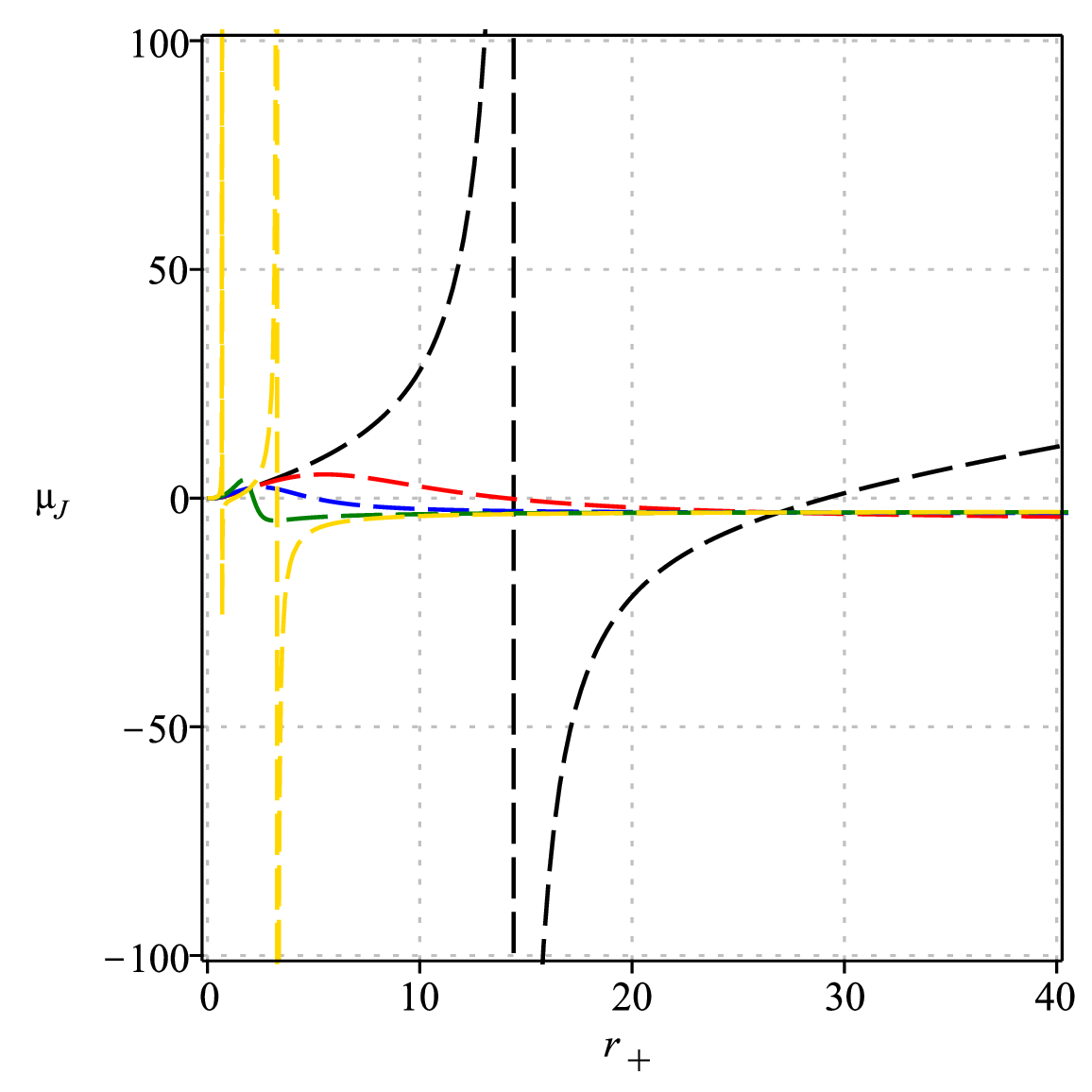}}\hfill
\subfloat[$\beta=1.0$ \& $\alpha=0.2$]{\includegraphics[width=.5\textwidth]{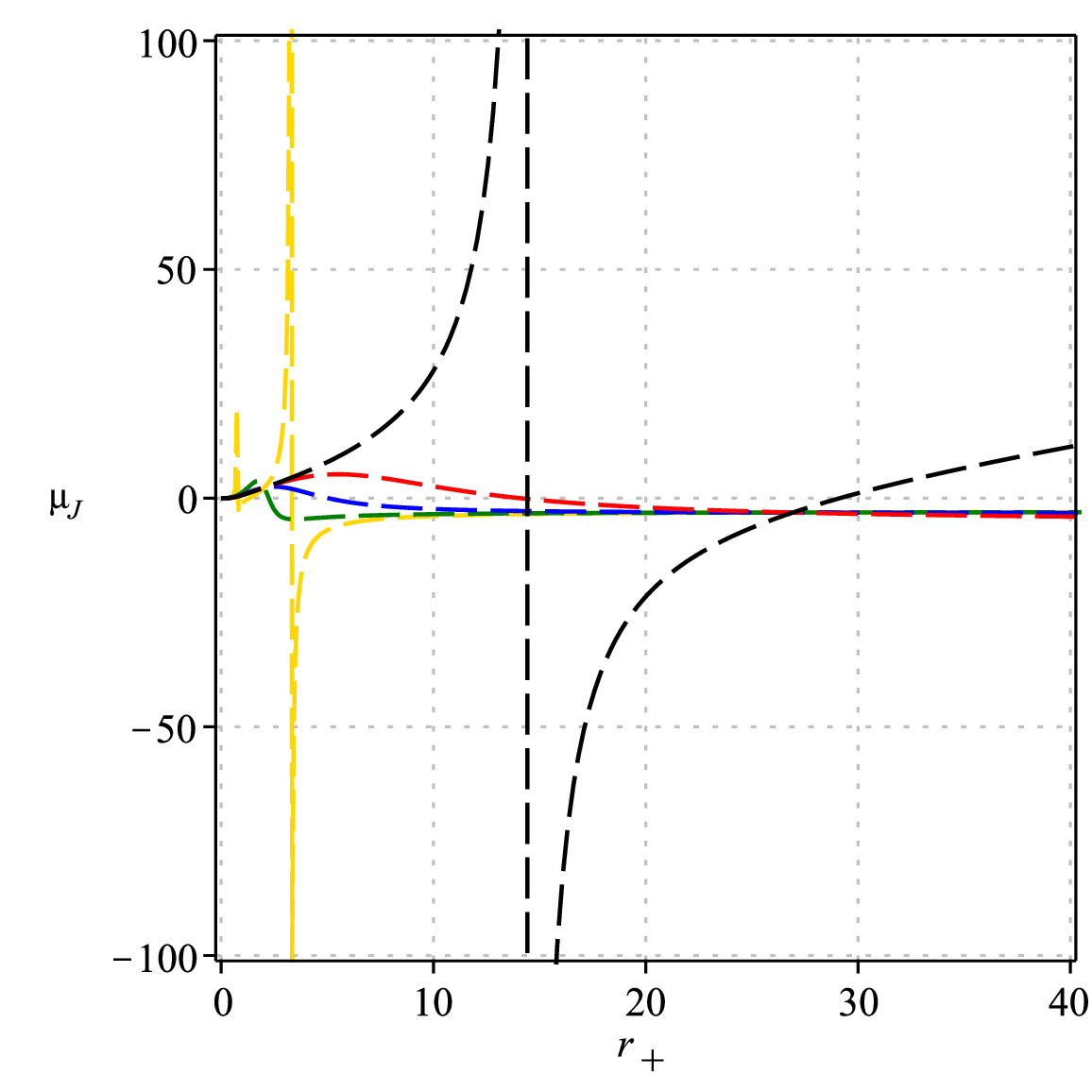}}\hfill
\subfloat[$\alpha=0.4$ \& $\beta=0.5$]{\includegraphics[width=.5\textwidth]{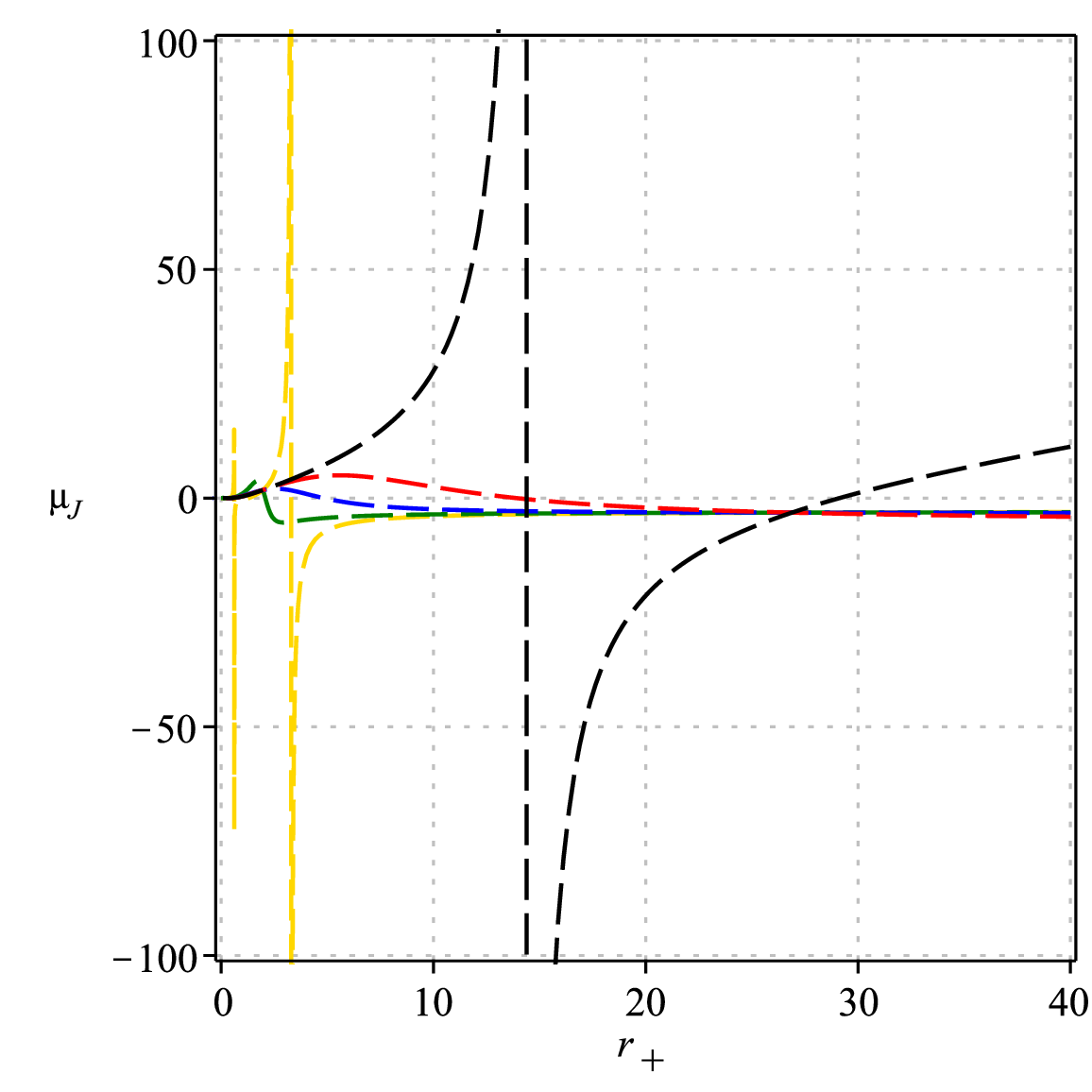}}\hfill
\subfloat[$\alpha=0.8$ \& $\beta=0.5$]{\includegraphics[width=.5\textwidth]{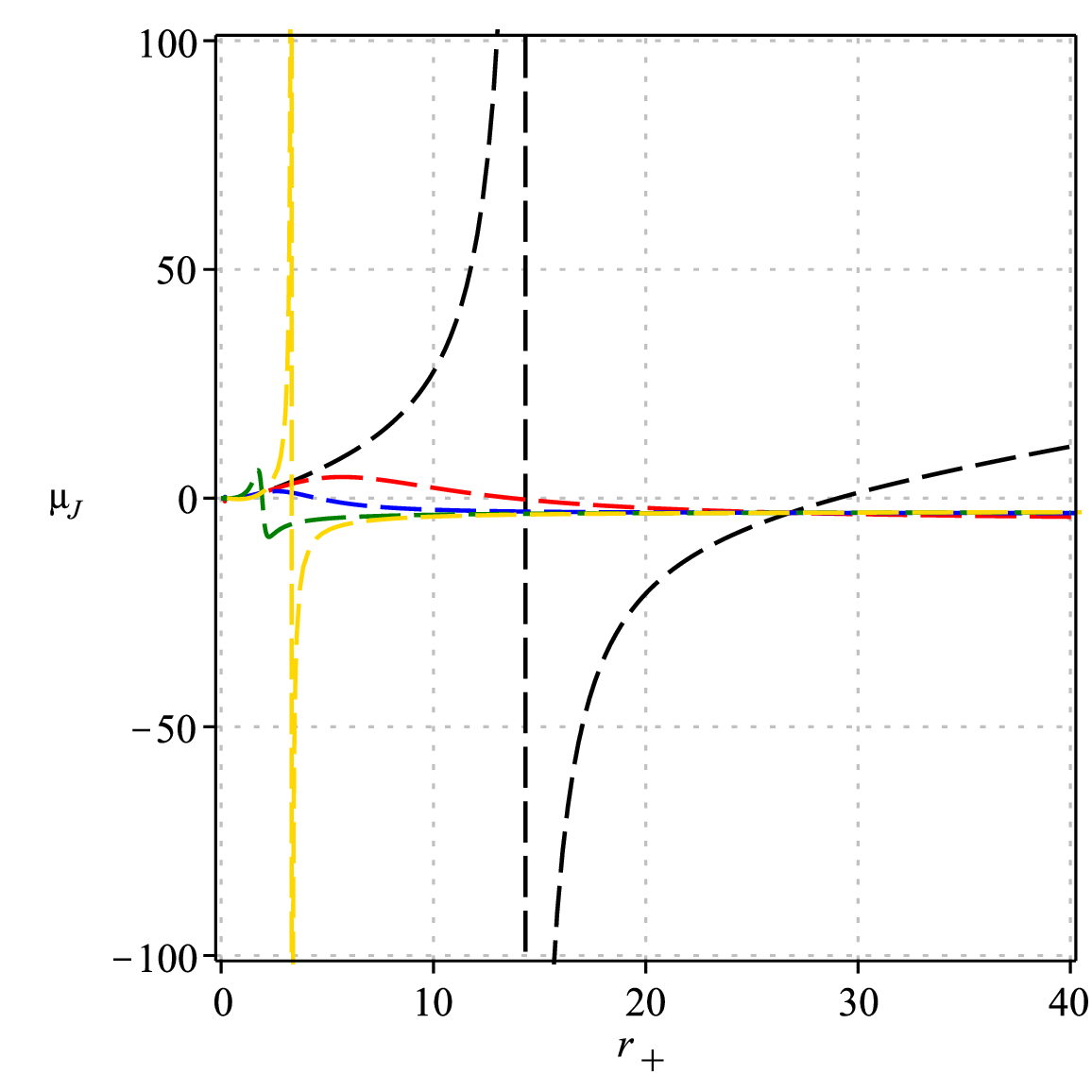}}\hfill
\centering
\begin{subfigure}[b]{0.2\textwidth}
\centering
\includegraphics[width=\textwidth]{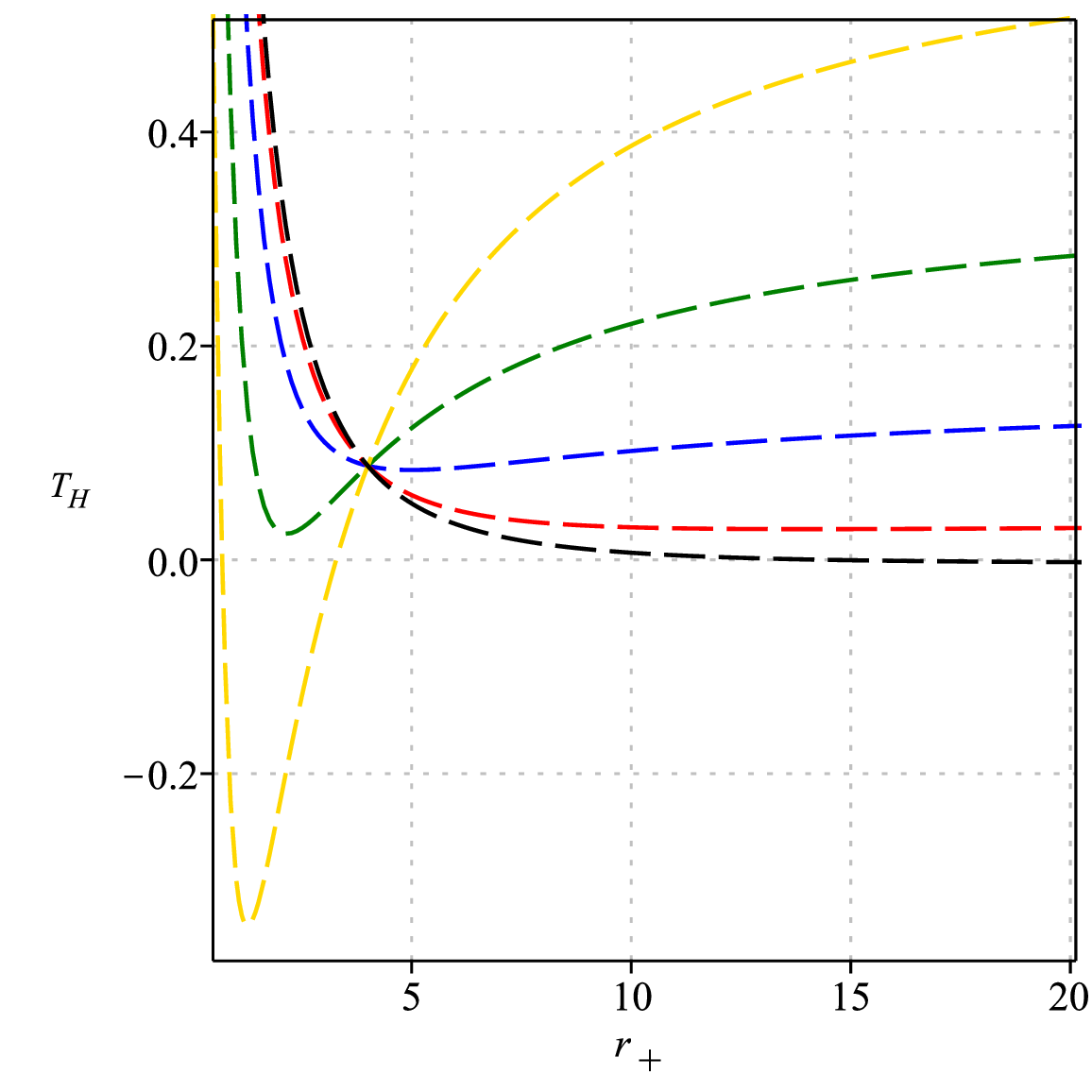}
\caption{$\beta=0.5$ \& $\alpha=0.2$}
\label{fig:47e}
\end{subfigure}
\hfill
\begin{subfigure}[b]{0.2\textwidth}
\centering
\includegraphics[width=\textwidth]{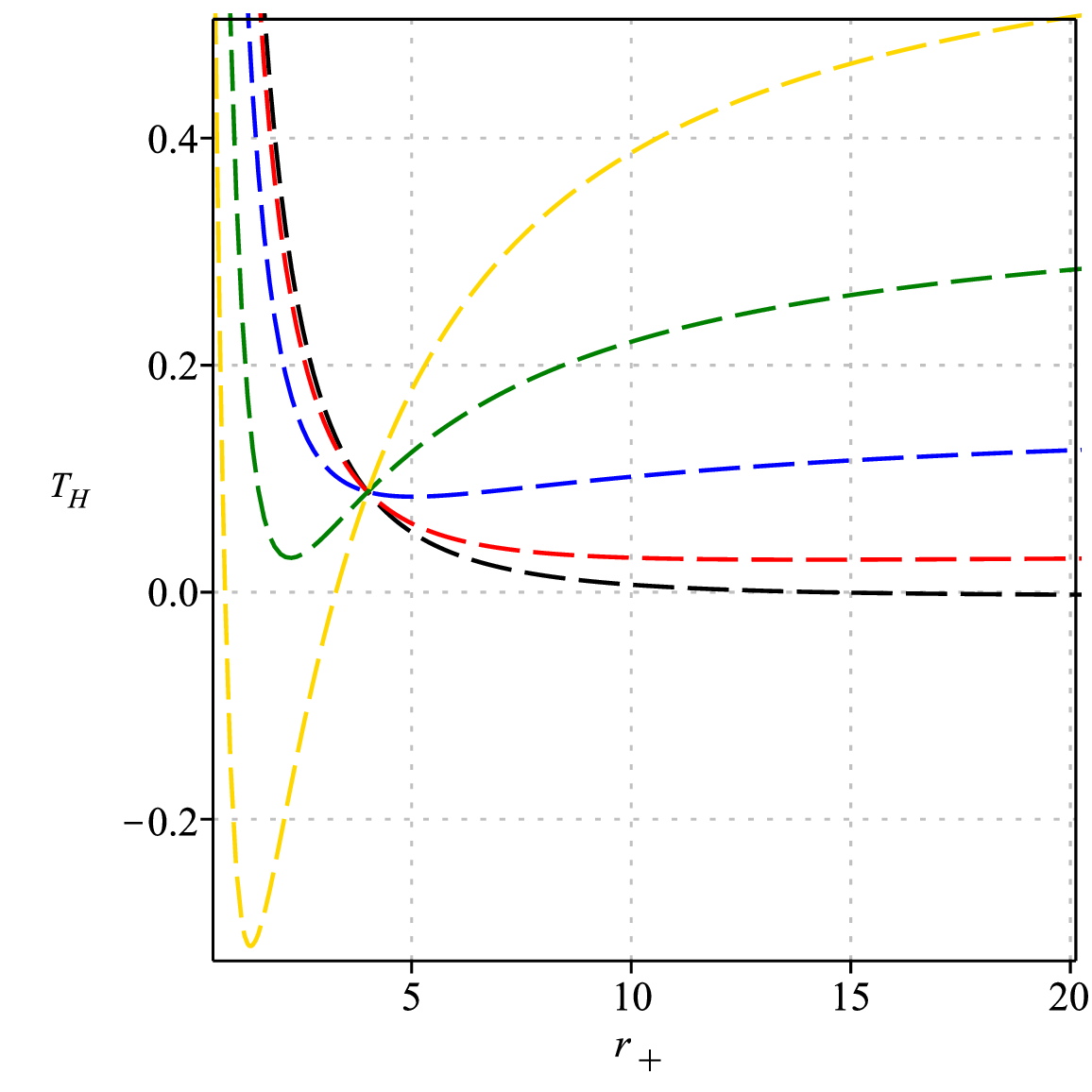}
\caption{$\beta=1.0$ \& $\alpha=0.2$}
\label{fig:47f}
\end{subfigure}
\hfill
\begin{subfigure}[b]{0.2\textwidth}
\centering
\includegraphics[width=\textwidth]{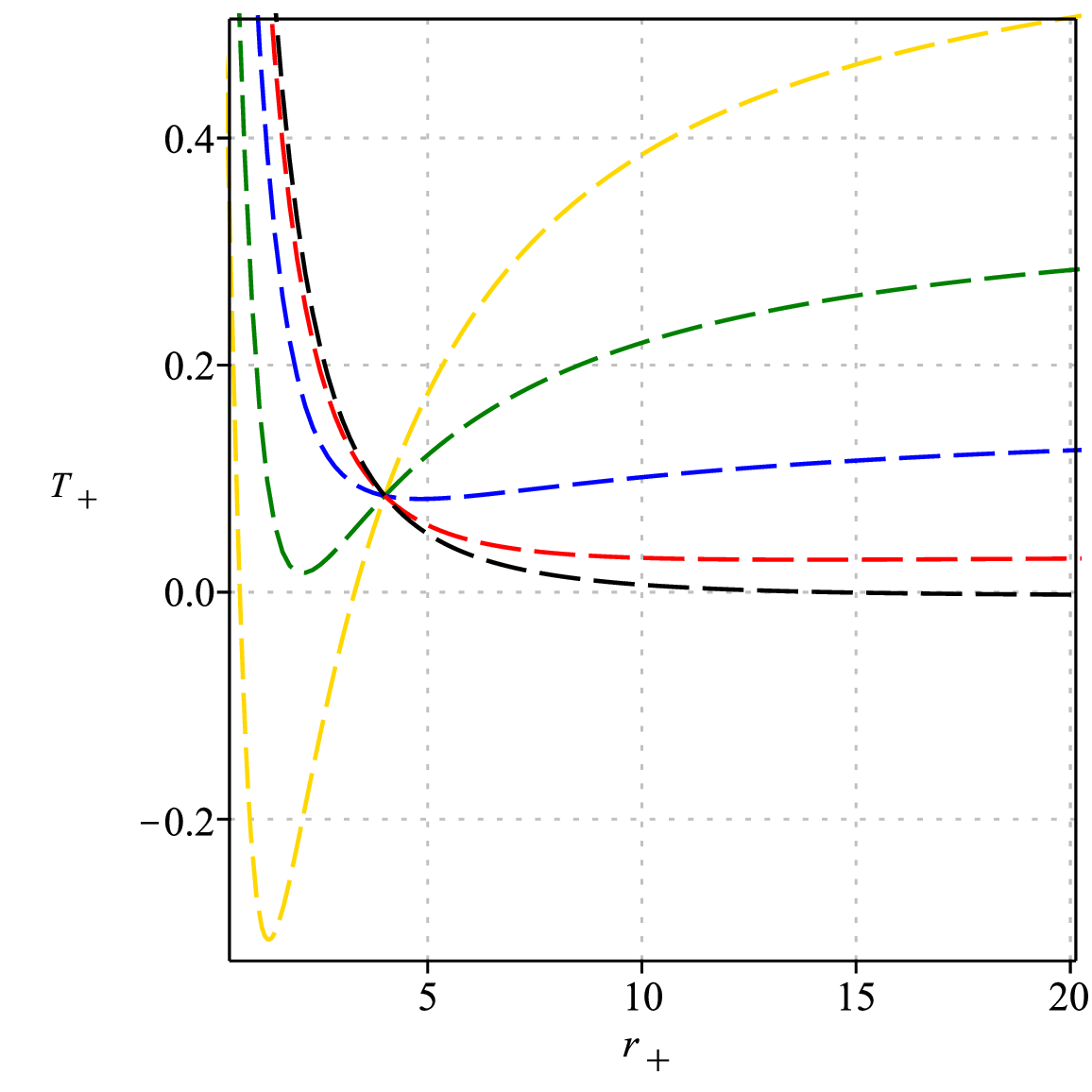}
\caption{$\alpha=0.4$ \& $\beta=0.5$}
\label{fig:47g}
\end{subfigure}
\hfill
\begin{subfigure}[b]{0.2\textwidth}
\centering
\includegraphics[width=\textwidth]{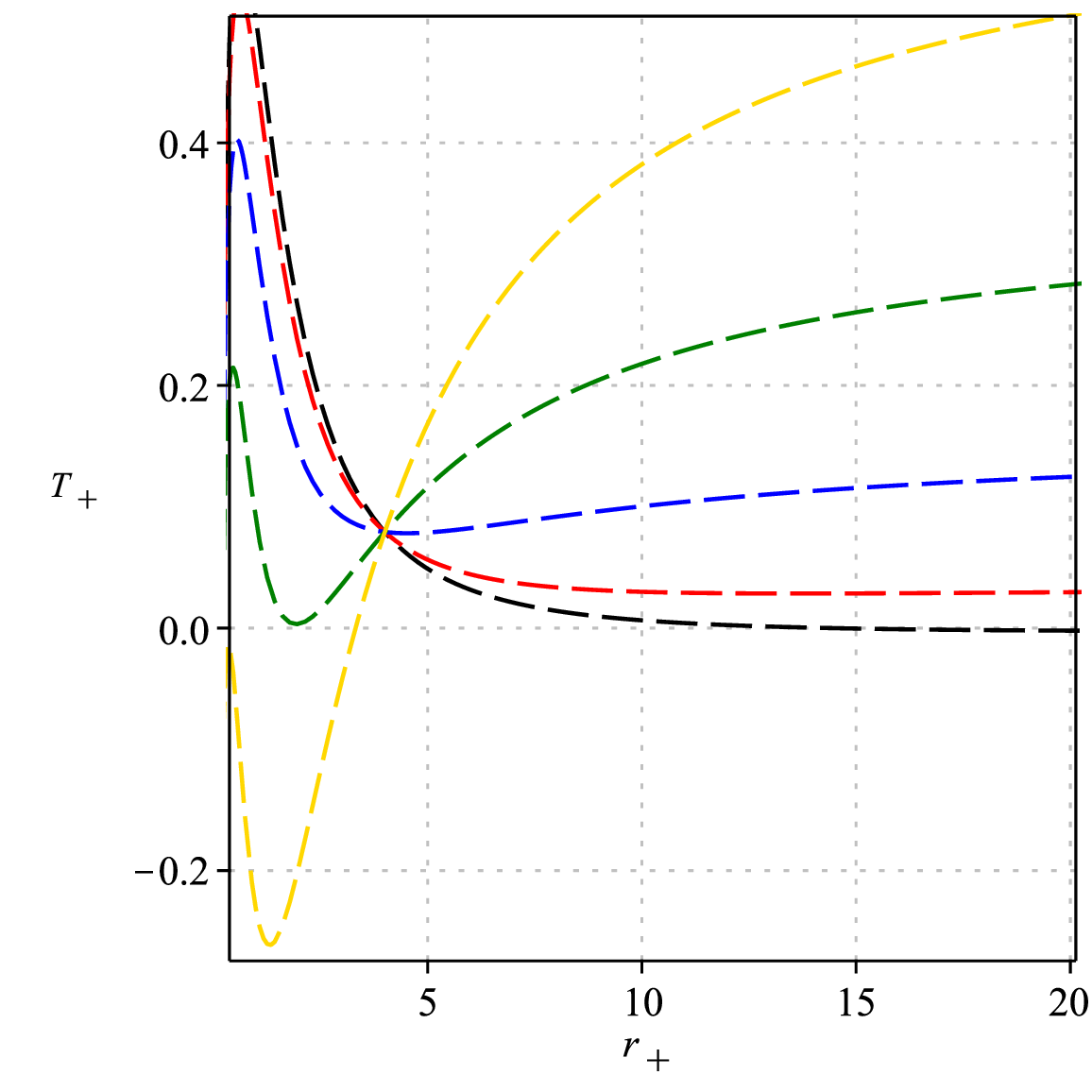}
\caption{$\alpha=0.8$ \& $\beta=0.5$}
\label{fig:47h}
\end{subfigure}
\caption{Black dash line denotes $m=0$, red dash line denotes $m=1.0$, blue dash line denoted $m=2.0$, green dash line denoted $m=3.0$ and gold dash line denoted $m=4.0$ with $M=5$, $Q_m=2$,  $c=1$, $c_1=-1$ and $c_2=1$.}
\label{fig:47}
\end{figure}

In Figs. \ref{fig:47}(a) - \ref{fig:47}(d) we plot the Joule--Thomson coefficient 
of $4D$ EGB massive gravity black hole for different values of NED parameter $\beta$ 
and GB parameter $\alpha$. Figs. \ref{fig:47}(a) \& \ref{fig:47}(b) show 
Joule--Thomson coefficient for different values of $\beta$, where Fig. \ref{fig:47}(c) 
\& \ref{fig:47}(d) show Joule--Thomson coefficient for different value of $\alpha$. 
For $m=4$, $\mu_{J}$ (Fig. \ref{fig:47}a - \ref{fig:47}c) has two diverging points 
except Fig. \ref{fig:47}(d). These diverging points correspond to zero Hawking 
temperature $T_{H}=0$ (Fig. \ref{fig:47}e - \ref{fig:46}f) in $T_{H} - r_{+}$ plane. 
After crossing the second diverging point $\mu_{J}$ is always negative, which 
indicates a heating phase. For $m=0$, $\mu_{J}$ has only one diverging point 
(Fig. \ref{fig:47}a - \ref{fig:46}d). Before the diverging points $\mu_{J}$ are 
positive, which indicates a cooling phase. At the diverging points, a phase 
transition occurs from the cooling phase to the heating phase. After crossing the 
singular point $\mu_{J}$ attains zero ($\mu_{J}$=0) for some particular values of 
horizon radius which is known as an inverse phenomenon and once again $\mu_{J}$ 
changes its sign from negative to positive one. For $m=1$, $2$ \& $3$, $\mu_{J}$ 
is an continuous function of $r_{+}$.

\begin{figure}[H]
\centering
\subfloat[$c_1=-1$ \& $c_2=-1$]{\includegraphics[width=.5\textwidth]{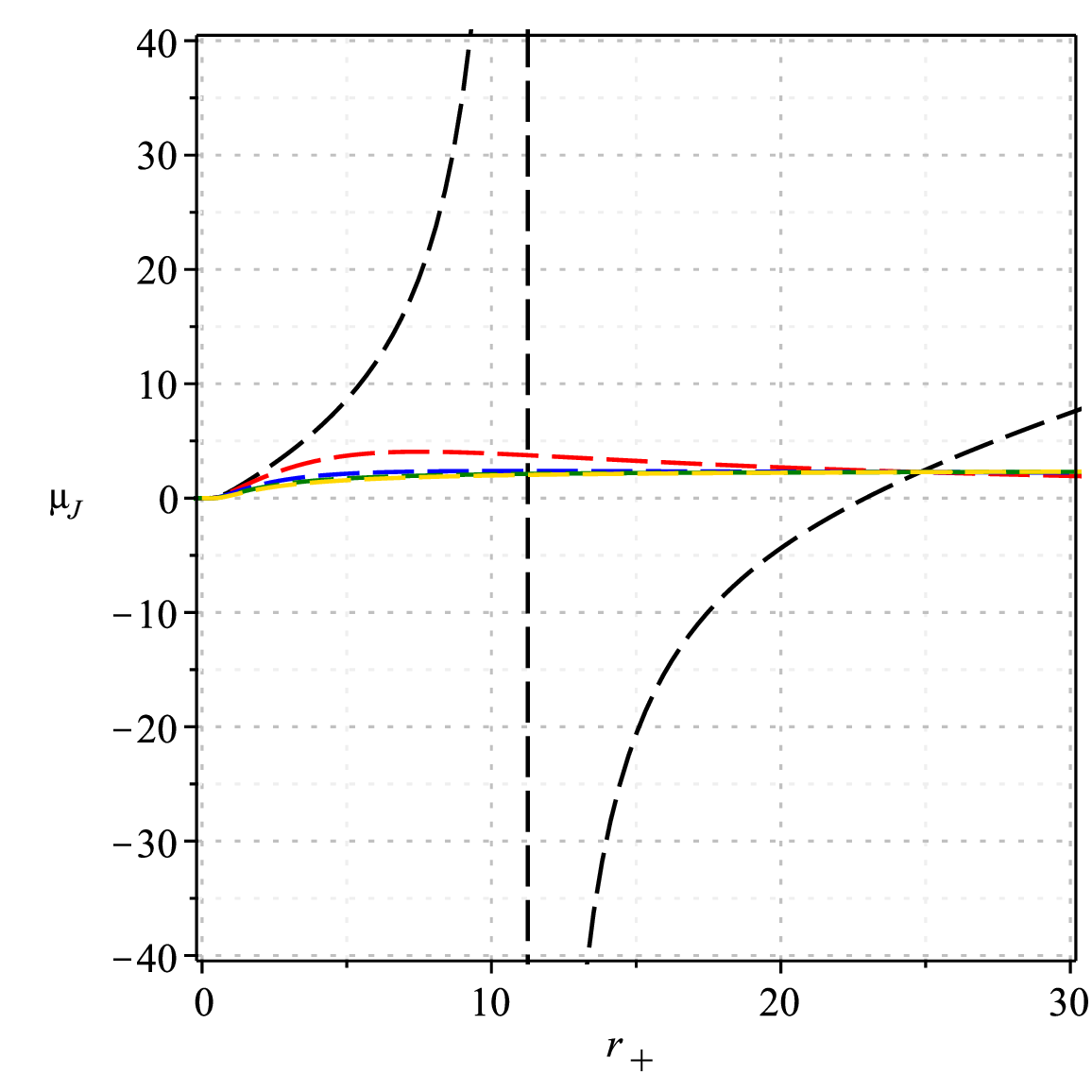}}\hfill
\subfloat[$c_1=-1$ \& $c_2=1$]{\includegraphics[width=.5\textwidth]{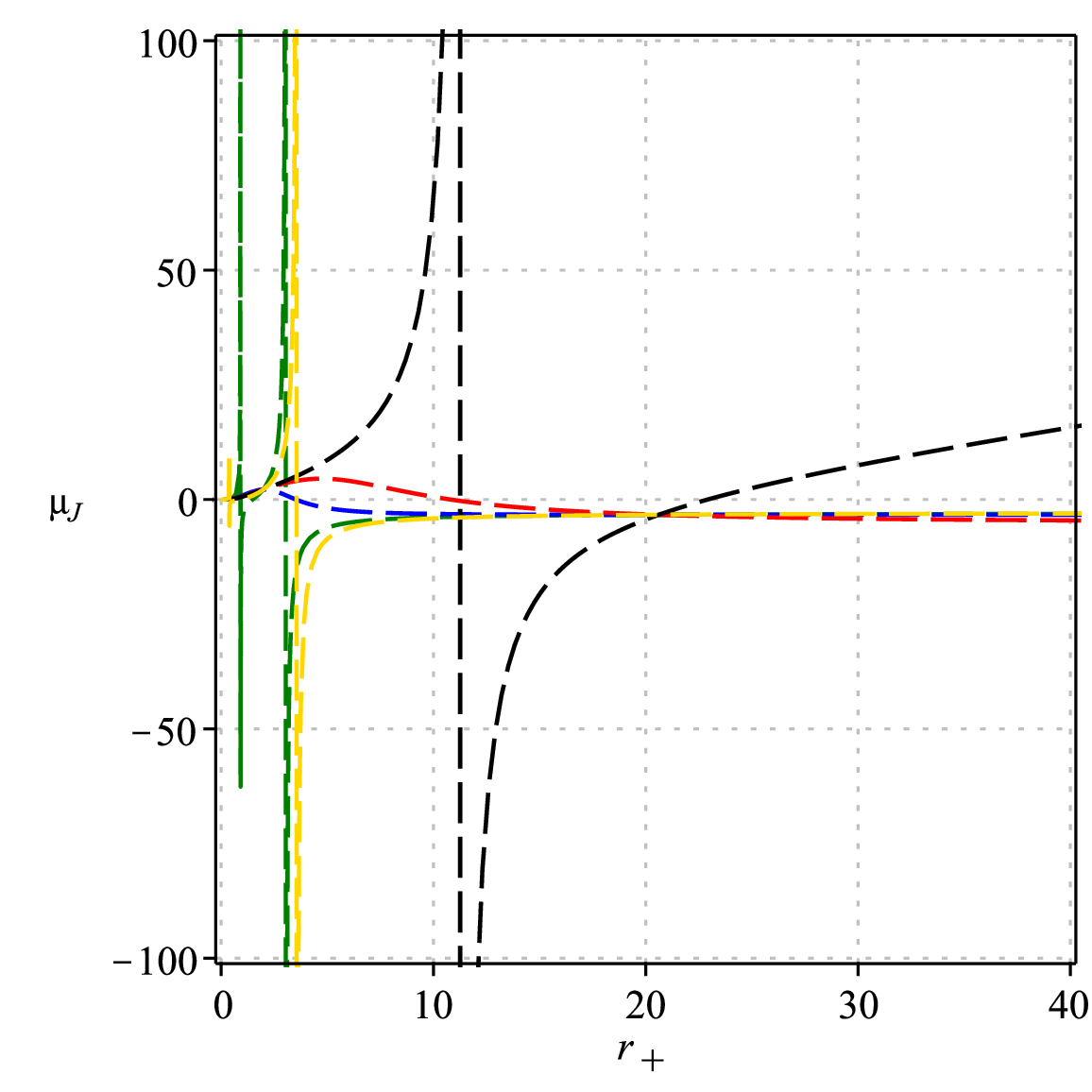}}\hfill
\subfloat[$c_1=1$ \& $c_2=-1$]{\includegraphics[width=.5\textwidth]{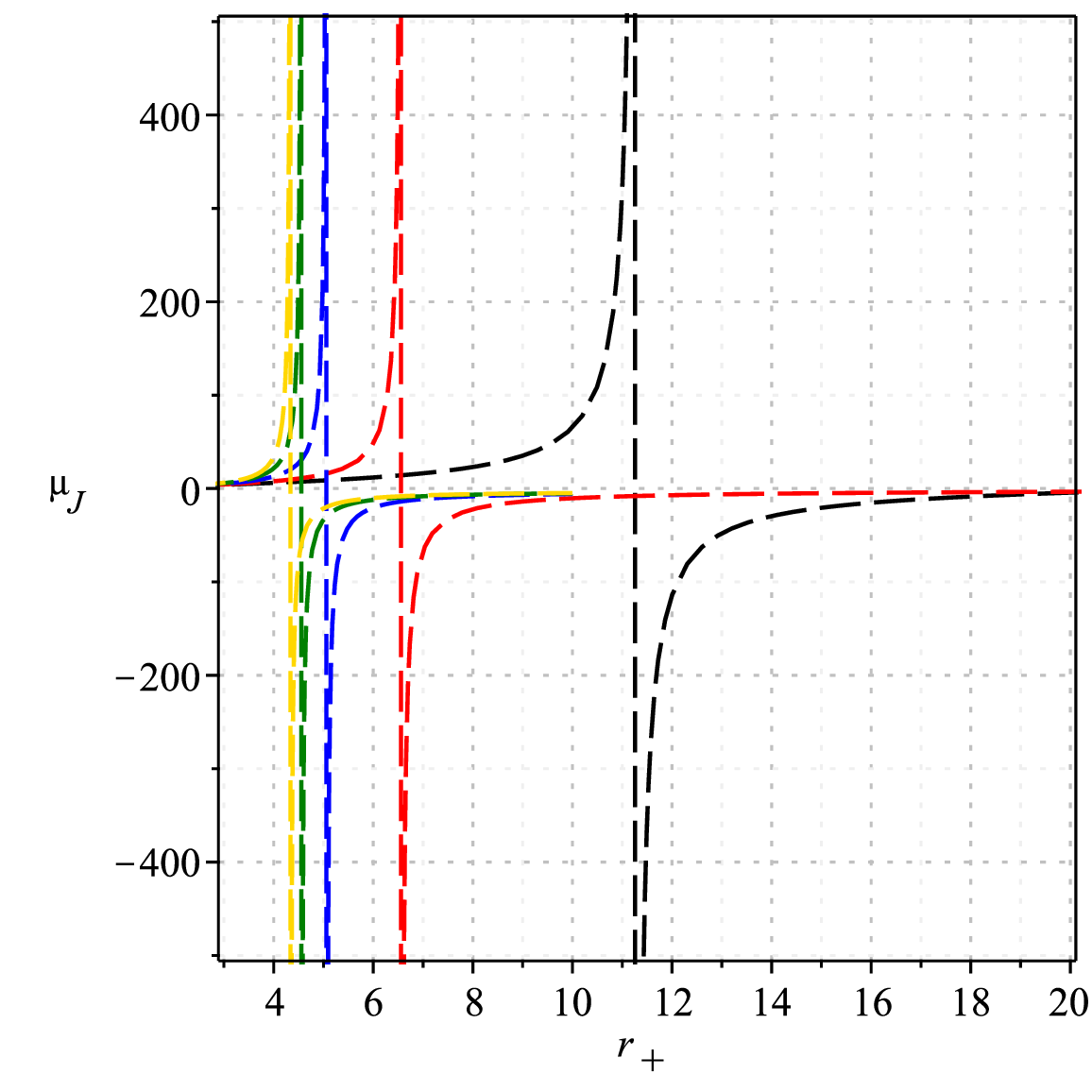}}\hfill
\subfloat[$c_1=1$ \& $c_2=1$]{\includegraphics[width=.5\textwidth]{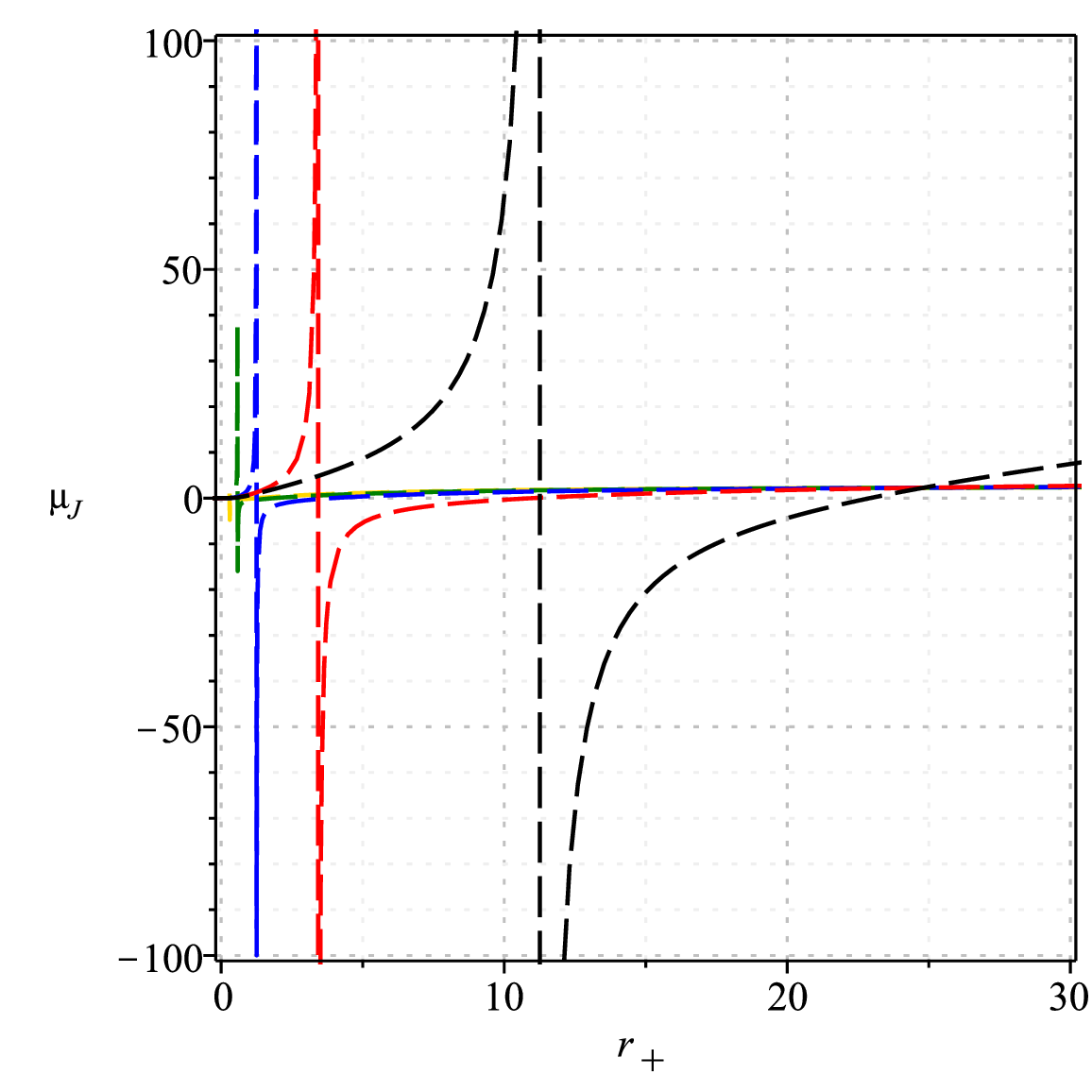}}\hfill
\centering
\begin{subfigure}[b]{0.2\textwidth}
\centering
\includegraphics[width=\textwidth]{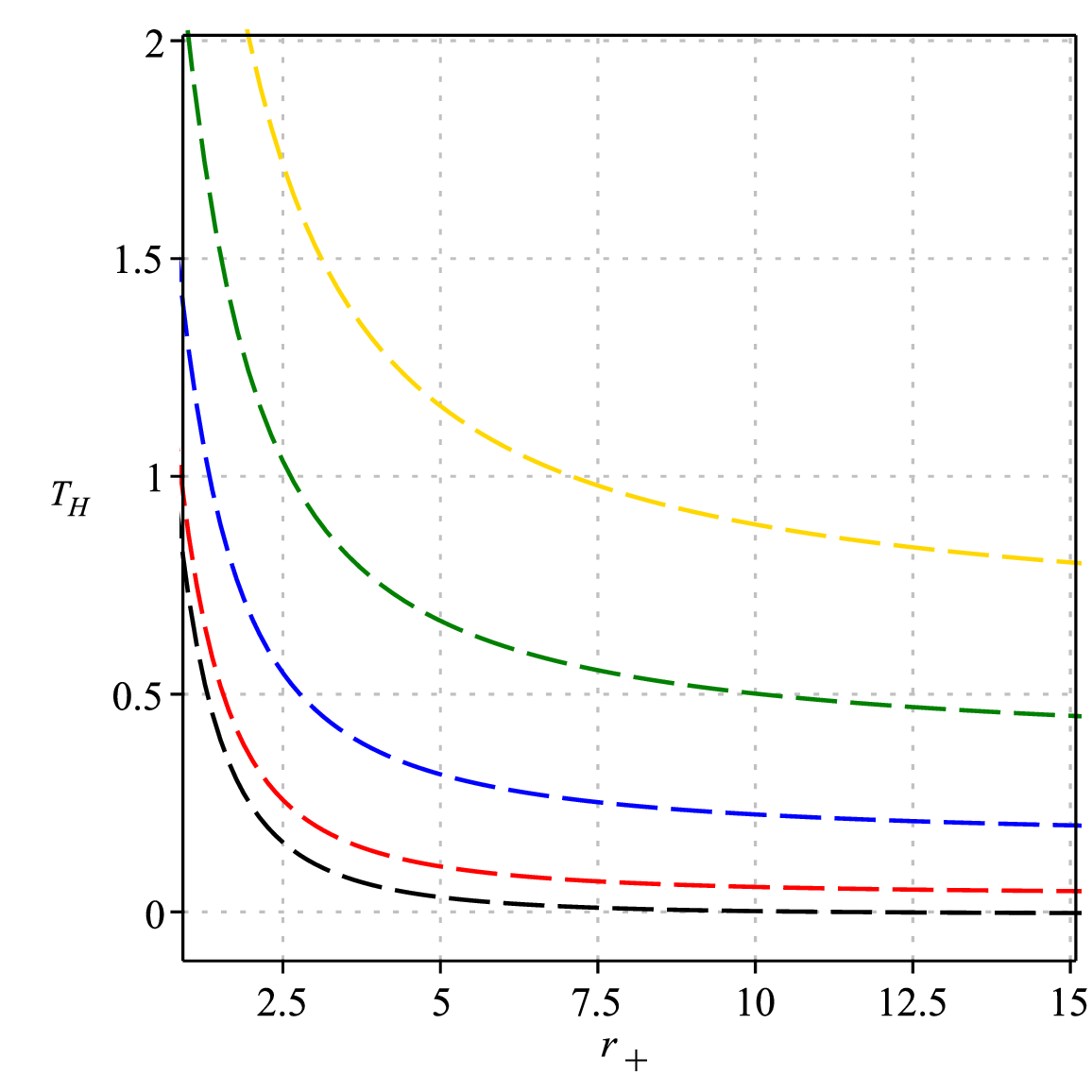}
\caption{$c_1=-1$ \& $c_2=-1$}
\label{fig:48e}
\end{subfigure}
\hfill
\begin{subfigure}[b]{0.2\textwidth}
\centering
\includegraphics[width=\textwidth]{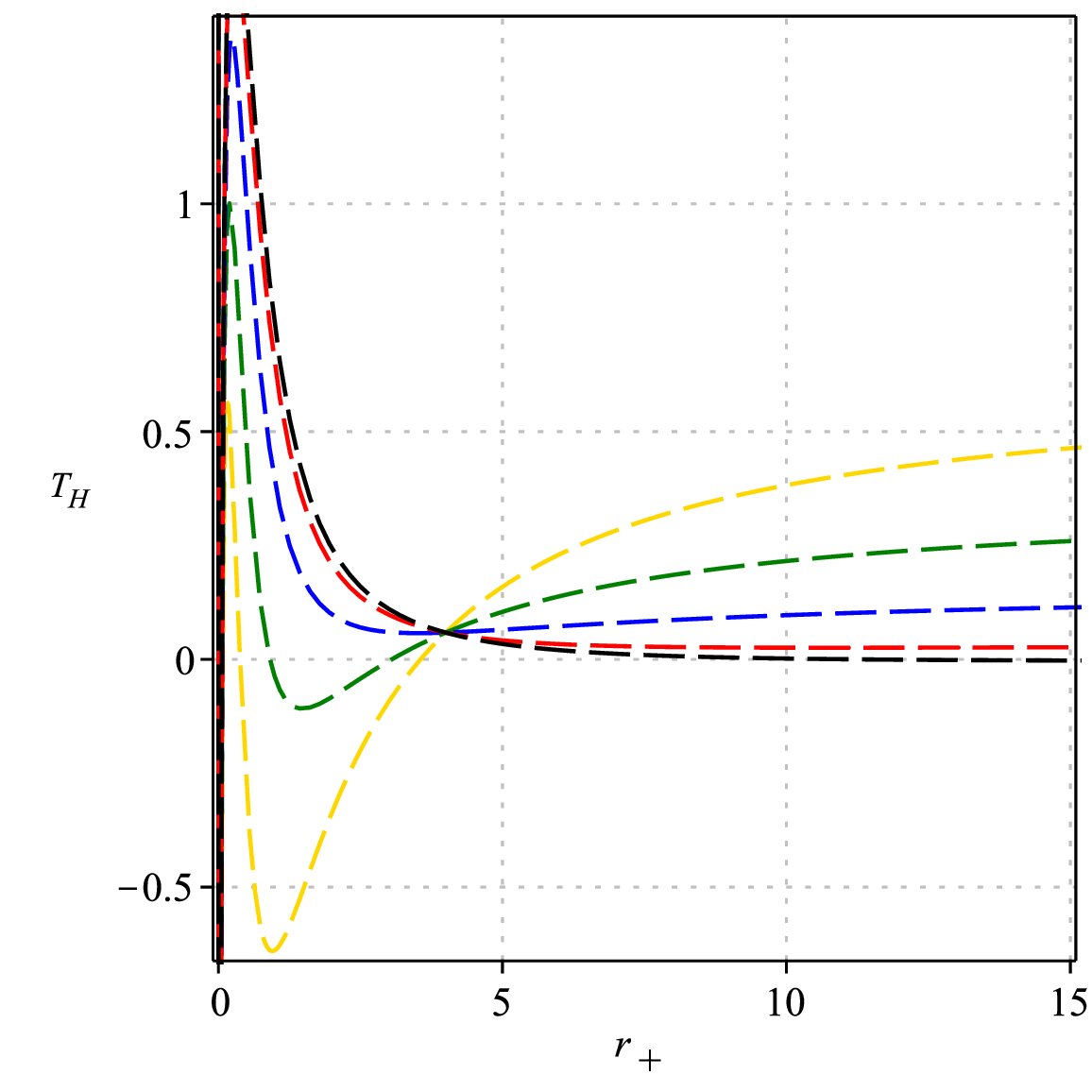}
\caption{$c_1=-1$ \& $c_2=1$}
\label{fig:48f}
\end{subfigure}
\hfill
\begin{subfigure}[b]{0.2\textwidth}
\centering
\includegraphics[width=\textwidth]{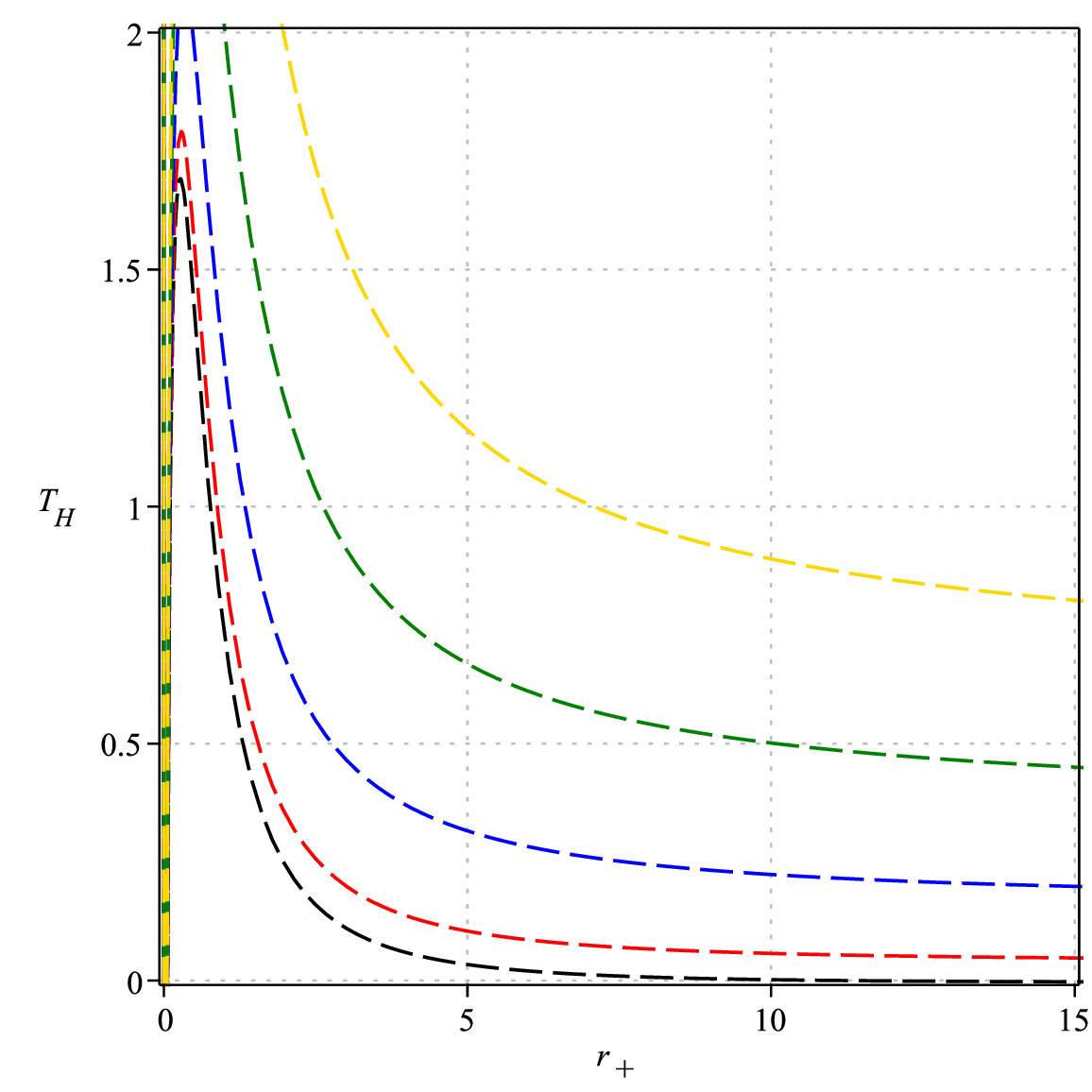}
\caption{$c_1=1$ \& $c_2=-1$}
\label{fig:48g}
\end{subfigure}
\hfill
\begin{subfigure}[b]{0.2\textwidth}
\centering
\includegraphics[width=\textwidth]{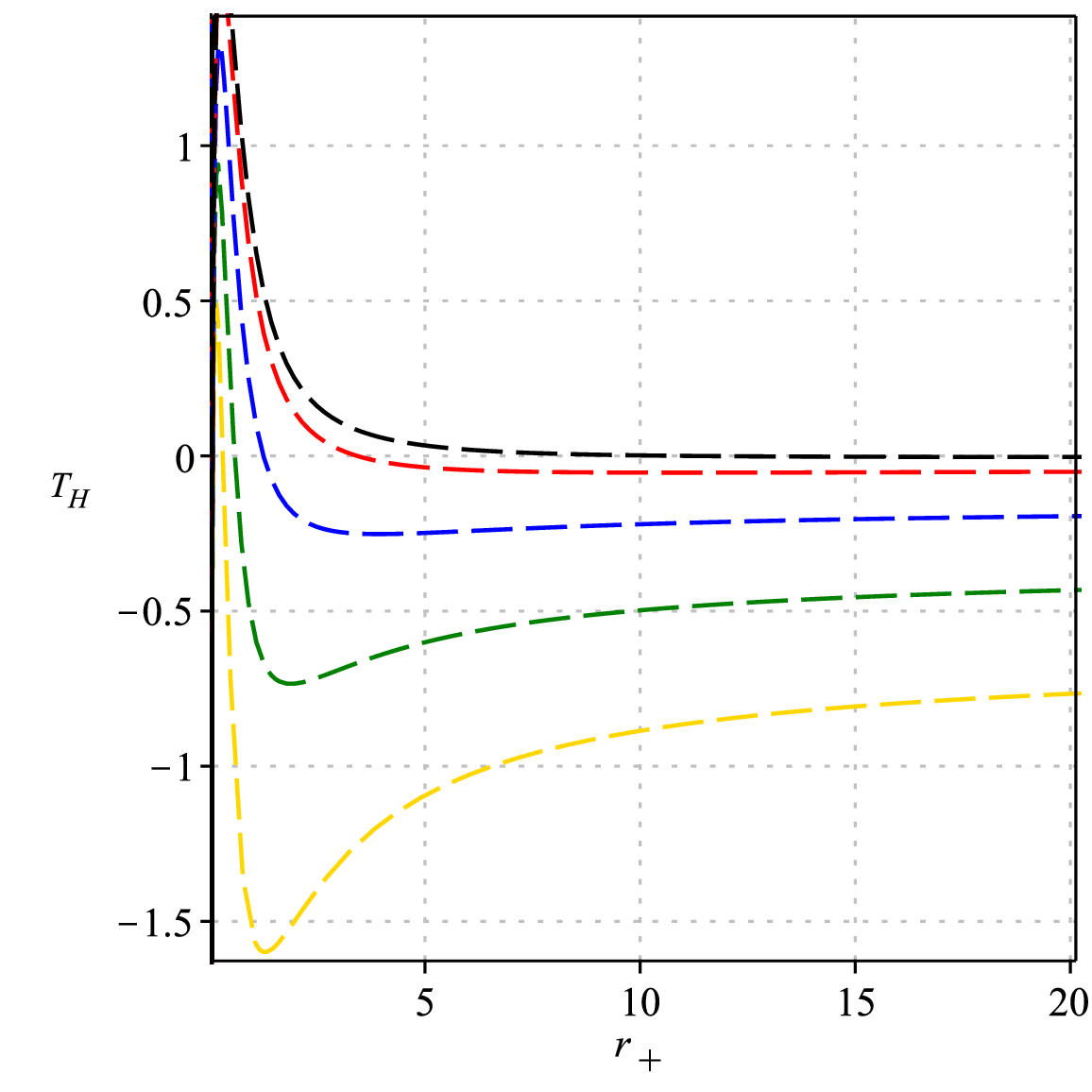}
\caption{$c_1=1$ \& $c_2=1$}
\label{fig:48h}
\end{subfigure}
\caption{Black dash line denotes $m=0$, red dash line denotes $m=1.0$, blue dash line denoted $m=2.0$, green dash line denoted $m=3.0$ and gold dash line denoted $m=4.0$ with $M=4$, $Q_m=2$, $\alpha=0.2$, $\beta=0.5$ and $c=1$.}
\label{fig:48}
\end{figure}

In Fig. \ref{fig:48}(a) - \ref{fig:48}(d) we plot the Joule--Thomson coefficient 
of $4D$ EGB massive gravity black hole for different values of constants $c_{1}$ 
and $c_{2}$. The behaviour of the Hawking temperature is shown in Fig. 
\ref{fig:48}(e) - \ref{fig:48}(f). For $c_{1}=-1$ \& $c_{2}=-1$, Joule--Thomson 
coefficient is depicted in Fig. \ref{fig:48}(a). The Joule-Thomson coefficient 
$\mu_{J}$, has only one single singular point for $m=0$ and at the singular point 
Hawking temperature is zero (Fig. \ref{fig:48}e). For $c_{1}=-1$ \& $c_{2}=1$, 
Joule--Thomson coefficient is depicted in Fig. \ref{fig:48}(b). $\mu_{J}$ has 
two singular points for $m=4$ \& $3$, one singular point for $m=0$ and continuous 
function of $r_{+}$ for $m=1$ \& $2$. The number of singular points in $\mu_{J}$ 
are also evident from the $T_{H} - r_{+}$ plot in Fig. \ref{fig:48}(f). For $m=4$ 
\& $3$, Hawking temperature is zero at two points and at $m=0$ Hawking temperature is 
zero at only one point. For $m=4$ \& $3$, before the first diverging point $\mu_{J}$ 
is positive, at the first diverging point $\mu_{J}(r_{+}^{0})$ changes its phase from 
cooling to heating and between the two diverging points there is an inverse phenomenon 
occurs at which $\mu_{J}(r_{+}^{mid})=0$. Between the regions 
$r_{+}^{0} < r_{+} < r_{+}^{mid}$  and $r_{+}^{mid} < r_{+} < r_{+}^{1}$ 
Joule--Thomson coefficients are negative (heating phase) and positive (cooling phase),
where $r_{+}^{1}$ is position of the second diverging point. At the second singular 
point, once again a phase transition of $\mu_{J}(r_{+}^{1})$ occurs from cooling 
phase (positive) to heating phase (negative). In Fig. \ref{fig:48}(c), Joule--Thomson 
coefficient is shown for $c_{1}=1$ \& $c_{2}=1$. $\mu_{J}$ has only one singular 
point for each value of graviton mass and the position of the singular point 
decreases as graviton mass increases. Fig. \ref{fig:48}(d) ($c_{1}=1$ \& $c_{2}=1$) 
shows similar behaviour as Fig. \ref{fig:48}(c) ($c_{1}=1$ \& $c_{2}=-1$).

\subsection{Black Holes in 4D EGB Massless gravity coupled to NED}

From Hawking temperature \eqref{eq:3.5} one can obtain black hole equation of state 
\begin{equation}\label{eq:5.9}
T_{H}=\frac{8 P \pi  r_{+}^{6}+(8 P \pi  k^{2}+1) r_{+}^{4}+( k^{2}-Q_{m}^{2}-\alpha ) r_{+}^{2}-\alpha  k^{2}}{8 r_{+} (\frac{r_{+}^{2}}{2}+\alpha ) (k^{2}+r_{+}^{2}) \pi}.
\end{equation}
From the mass function \eqref{eq:3.1} pressure of the black hole can be obtained using limit $m \to 0$ as 
\begin{equation}\label{eq:5.10}
    P=\frac{1}{{\pi  r_{+}^{4} k}}\biggr[\frac{3 Q_{m}^{2} r_{+} \arctan (\frac{r_{+}}{k})}{8}-\frac{3 Q_{m}^{2} r_{+} \pi}{16}+\frac{3 M r_{+} k}{4}-\frac{3 k r_{+}^{2}}{8}-\frac{3 \alpha  k}{8} \biggr].
\end{equation}
From equation \eqref{eq:5.9} and equation \eqref{eq:5.2} we obtain inverse pressure as 
\begin{equation*}
P_{i}=\frac{1}{16 \pi  r_{+}^{6} (k^{2}+r_{+}^{2})^{2}} \biggr[ 4 Q_{m}^{2} k^{2} r_{+}^{4}+6 Q_{m}^{2} r_{+}^{6} -4 k^{4} r_{+}^{4}-8 k^{2} r_{+}^{6}-4 r_{+}^{8}+4 Q_{m}^{2} \alpha  k^{2} r_{+}^{2}+8 Q_{m}^{2} \alpha  r_{+}^{4}
\end{equation*}
\begin{equation}\label{eq:5.11}
+2 \alpha  k^{4} r_{+}^{2}+4 \alpha  k^{2} r_{+}^{4}+2 \alpha  r_{+}^{6}+8 \alpha^{2} k^{4}+16 \alpha^{2} k^{2} r_{+}^{2}+8 \alpha^{2} r_{+}^{4} \biggr].
\end{equation}

Using equation \eqref{eq:5.9} and equation \eqref{eq:5.11} we obtain inverse temperature

\begin{equation}\label{eq:5.12}
T_i=\frac{1}{8 \pi  (k^{2}+r_{+}^{2})^{2} r_{+}^{3}} \biggr[ -2 r_{+}^{6}+(-4 k^{2}+4 Q_{m}^{2}+4 \alpha ) r_{+}^{4} -2 ( k^{2}-Q_{m}^{2}-4 \alpha ) k^{2} r_{+}^{2}+4 \alpha  k^{4} \biggr].
\end{equation}
The expression for Joule–Thomson coefficient looks cumbersome, so we will not present it here.

By setting $P_{i}=0$ into equation \eqref{eq:5.11}, one can obtain minimum event horizon radius as 
\begin{equation*}
-4 (r_{{+}}^{min})^{8}+(-8 k^{2}+6 Q_{m}^{2}+2 \alpha ) (r_{{+}}^{min})^{6}
+(-4 k^{4}+(4 Q_{m}^{2}+4 \alpha ) k^{2}+8 Q_{m}^{2} \alpha +8 \alpha^{2}) (r_{{+}}^{min})^{4}
\end{equation*}
\begin{equation}\label{eq:5.13a}
-4 \alpha  ((-{1}/{2}) k^{2}-Q_{m}^{2}-4 \alpha ) k^{2} (r_{{+}}^{min})^{2}+8 \alpha^{2} k^{4} =0.
\end{equation}

With the help of numerical techniques equation \eqref{eq:5.13a} is solved, minimum horizon radius \& 
inverse temperature are obtained. These values are presented in tables \ref{table:11} and \ref{table:12}, 
corresponding to various graviton mass, NED parameter, and GB coupling parameter values.

\begin{figure}[H]
\centering
\subfloat[$\beta=0.5$]{\includegraphics[width=.5\textwidth]{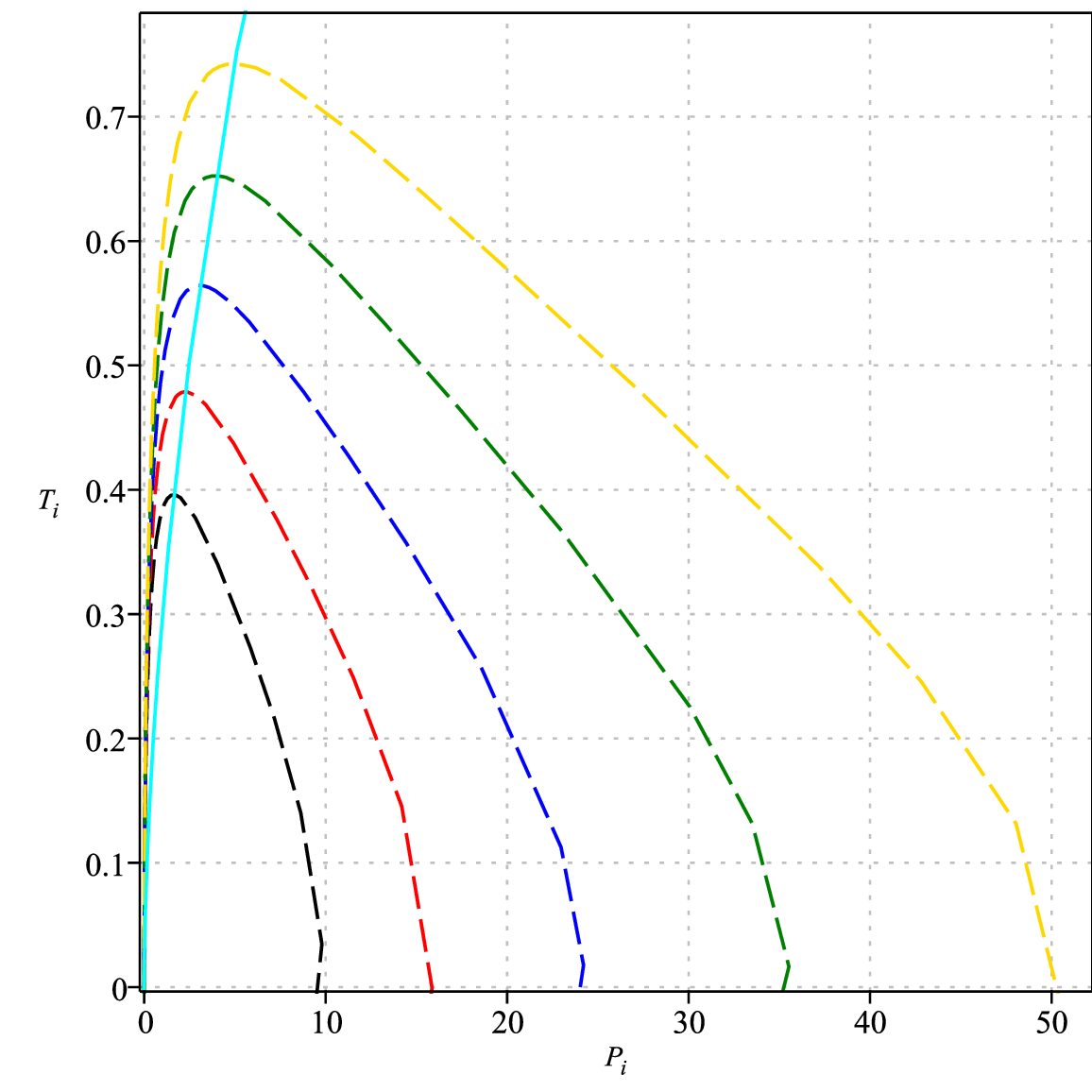}}\hfill
\subfloat[$\beta=1.0$]{\includegraphics[width=.5\textwidth]{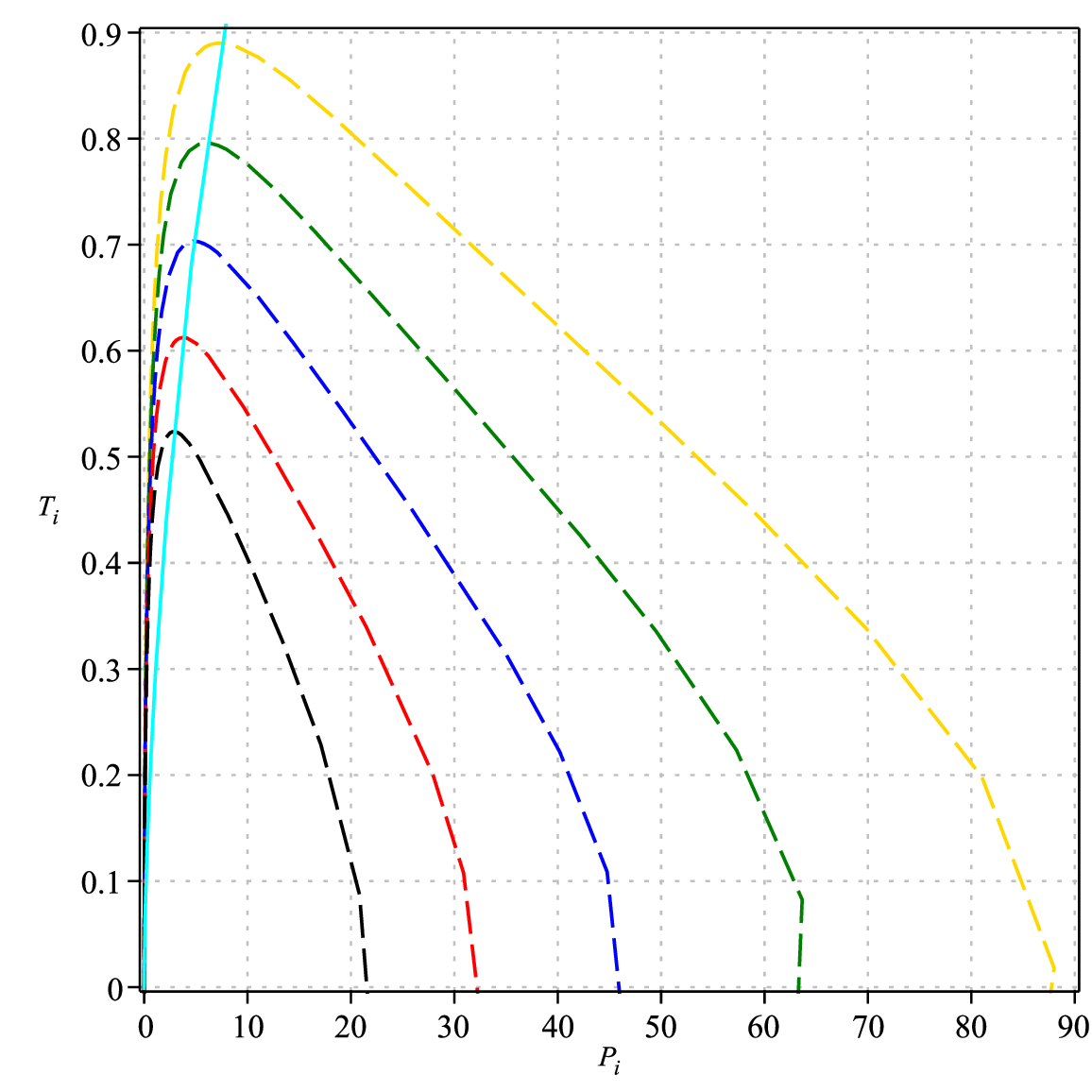}}\hfill
\caption{Black dash line denotes $M=2$, red dash line denotes $M=2.1$, blue dash line denoted $M=2.2$, green dash line denoted $M=2.3$, gold dash line denoted $M=2.4$ and solid cyan line denotes inverse curve with $Q_m=1.5$, $\alpha=0.2$ and $m=0$.}\label{fig:49}
\end{figure}

\begin{figure}[H]
\centering
\subfloat[$\alpha=0.2$]{\includegraphics[width=.5\textwidth]{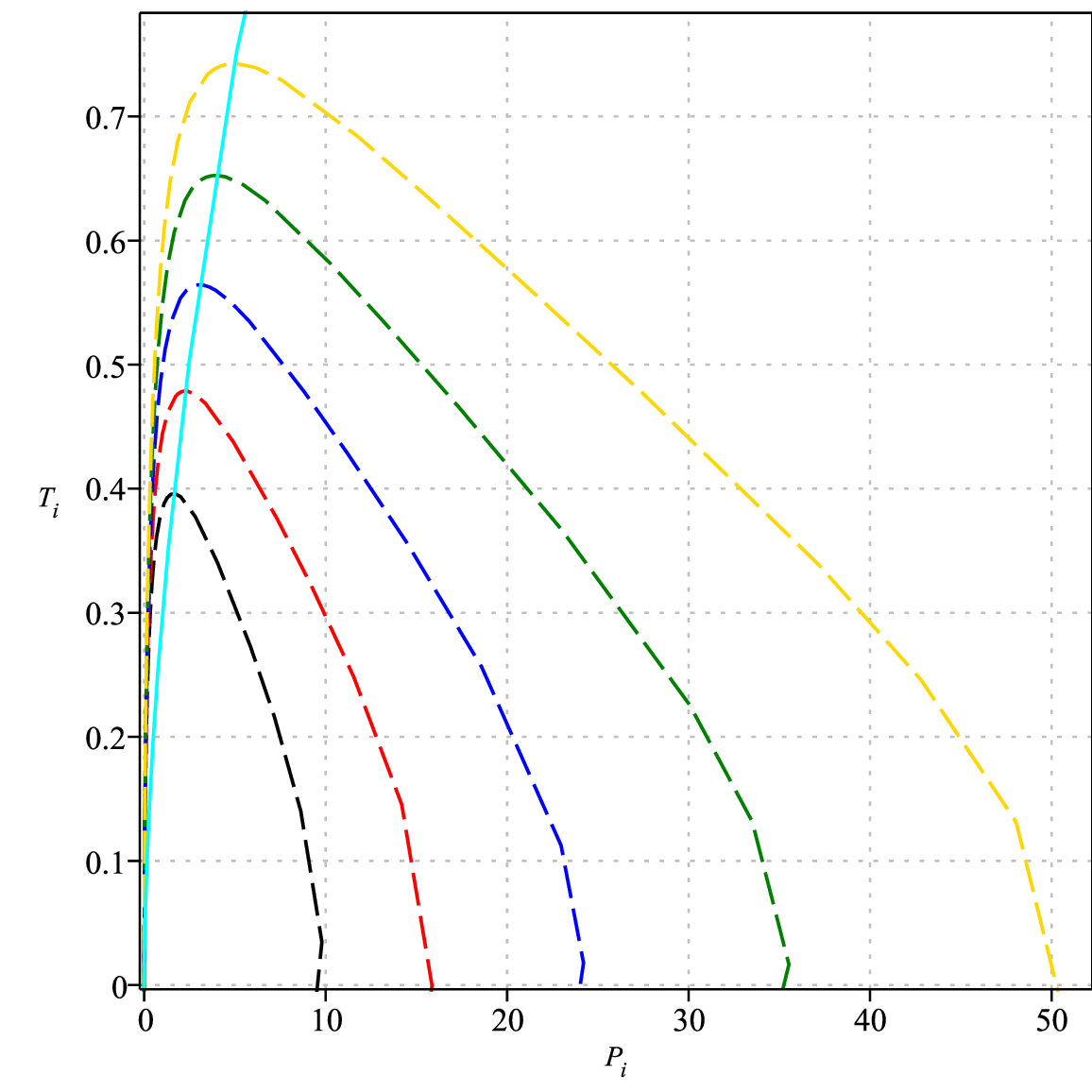}}\hfill
\subfloat[$\alpha=0.8$]{\includegraphics[width=.5\textwidth]{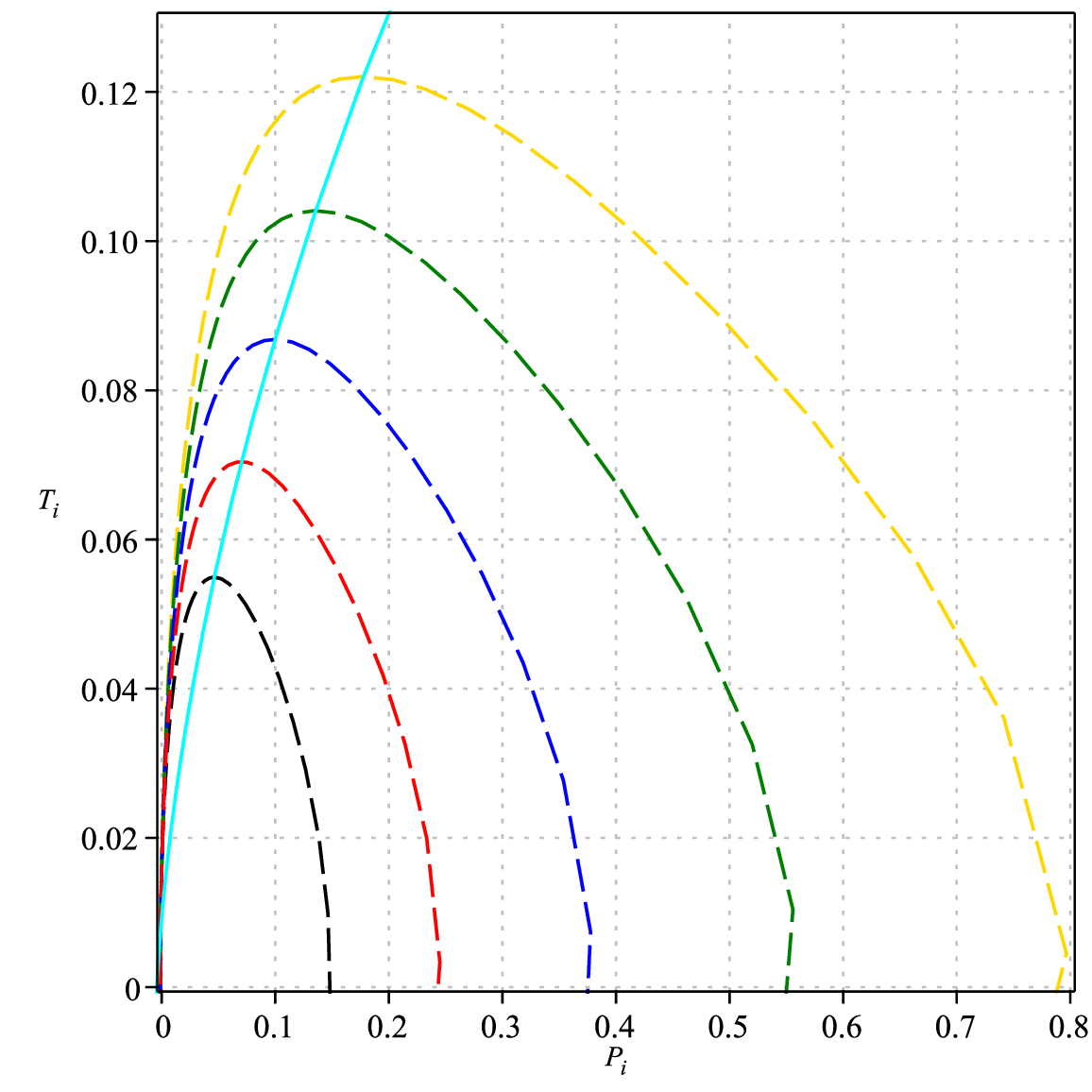}}\hfill
\caption{Black dash line denotes $M=2$, red dash line denotes $M=2.1$, blue dash line denoted $M=2.2$, green dash line denoted $M=2.3$, gold dash line denoted $M=2.4$ and solid cyan line denotes inverse curve with $Q_m=1.5$, $\alpha=0.2$ and $m=0$.}\label{fig:50}
\end{figure}

\begin{table}[H]
\centering
\begin{tabular}{ |p{1.5cm}|p{1.5cm}|p{1.5cm}| } 
\hline
\multicolumn{3}{|c|}{$\alpha=0.2$ } \\
\hline
\textbf{$\beta$} & \textbf{$r_{+}^{min}$} & \textbf{$T_{i}^{min}$}  \\ [0.5ex]  
\hline
0.0 & 2.522 & 0.0100  \\ \hline
0.1 & 2.182 & 0.0099  \\ \hline
0.4 & 1.789 & 0.0094  \\ \hline
0.8 & 1.436 & 0.0090  \\ \hline
1.0 & 1.303 & 0.0089  \\ \hline 
\end{tabular}
\caption{$Q_m=2$, and $m=0$}
\label{table:11}
\end{table}

\begin{table}[H]
\centering
\begin{tabular}{ |p{1.5cm}|p{1.5cm}|p{1.5cm}| } 
\hline
\multicolumn{3}{|c|}{$\beta=0.5$} \\
\hline
\textbf{$\alpha$} & \textbf{$r_{+}^{min}$} & \textbf{$T_{i}^{min}$} \\ [0.5ex]  
\hline
0.0 & 1.287 & 0.0136  \\ \hline
0.1 & 1.570 & 0.0095  \\ \hline
0.2 & 1.688 & 0.0093 \\ \hline
0.3 & 1.787 & 0.0090 \\ \hline
0.4 & 1.874 & 0.0088 \\ \hline
\end{tabular}
\caption{$Q_m=2$, and $m=0$}
\label{table:12}
\end{table}

\begin{figure}[H]
\centering
\subfloat[$\beta=0.5$ \& $\alpha=0.2$]{\includegraphics[width=.5\textwidth]{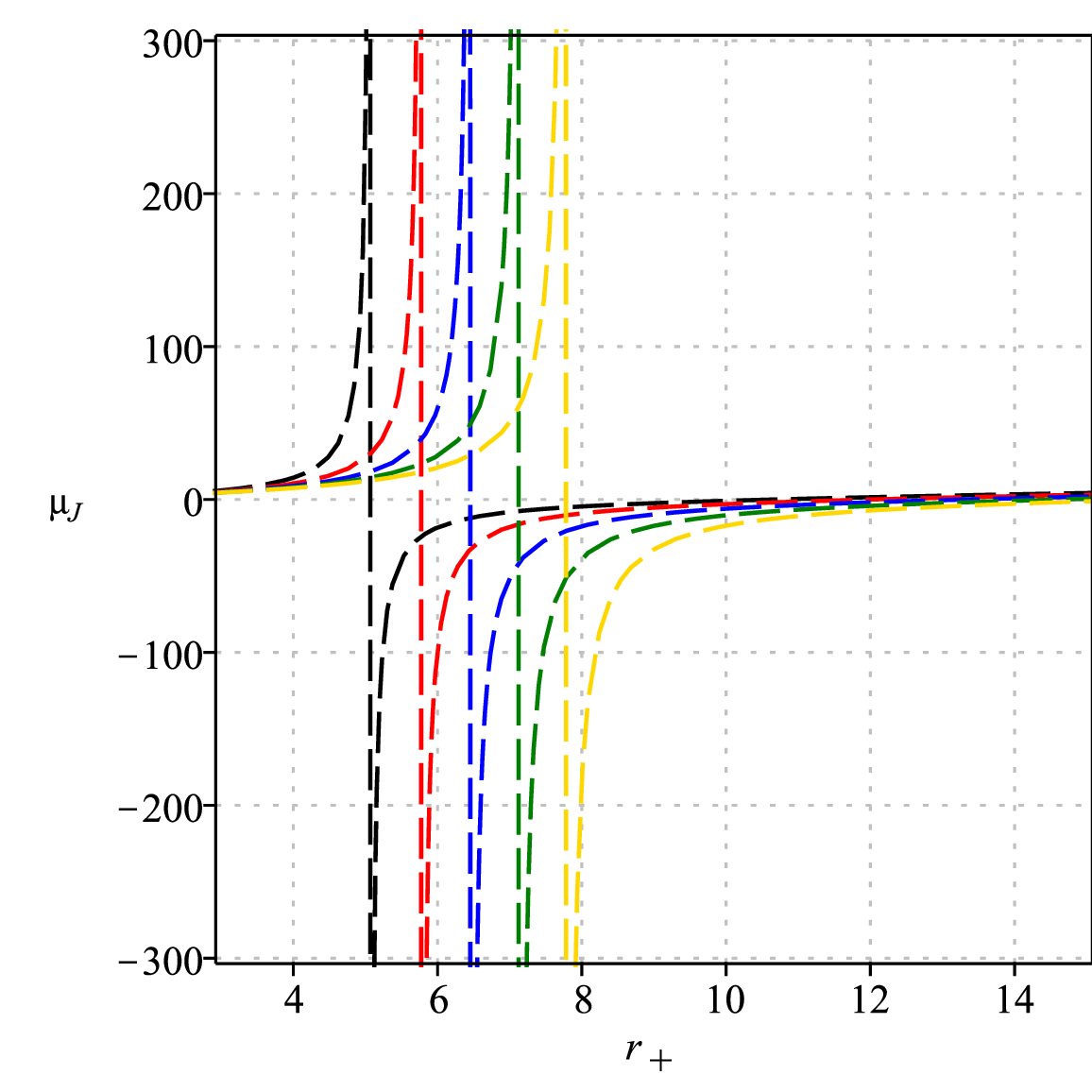}}\hfill
\subfloat[$\beta=1.0$ \& $\alpha=0.2$]{\includegraphics[width=.5\textwidth]{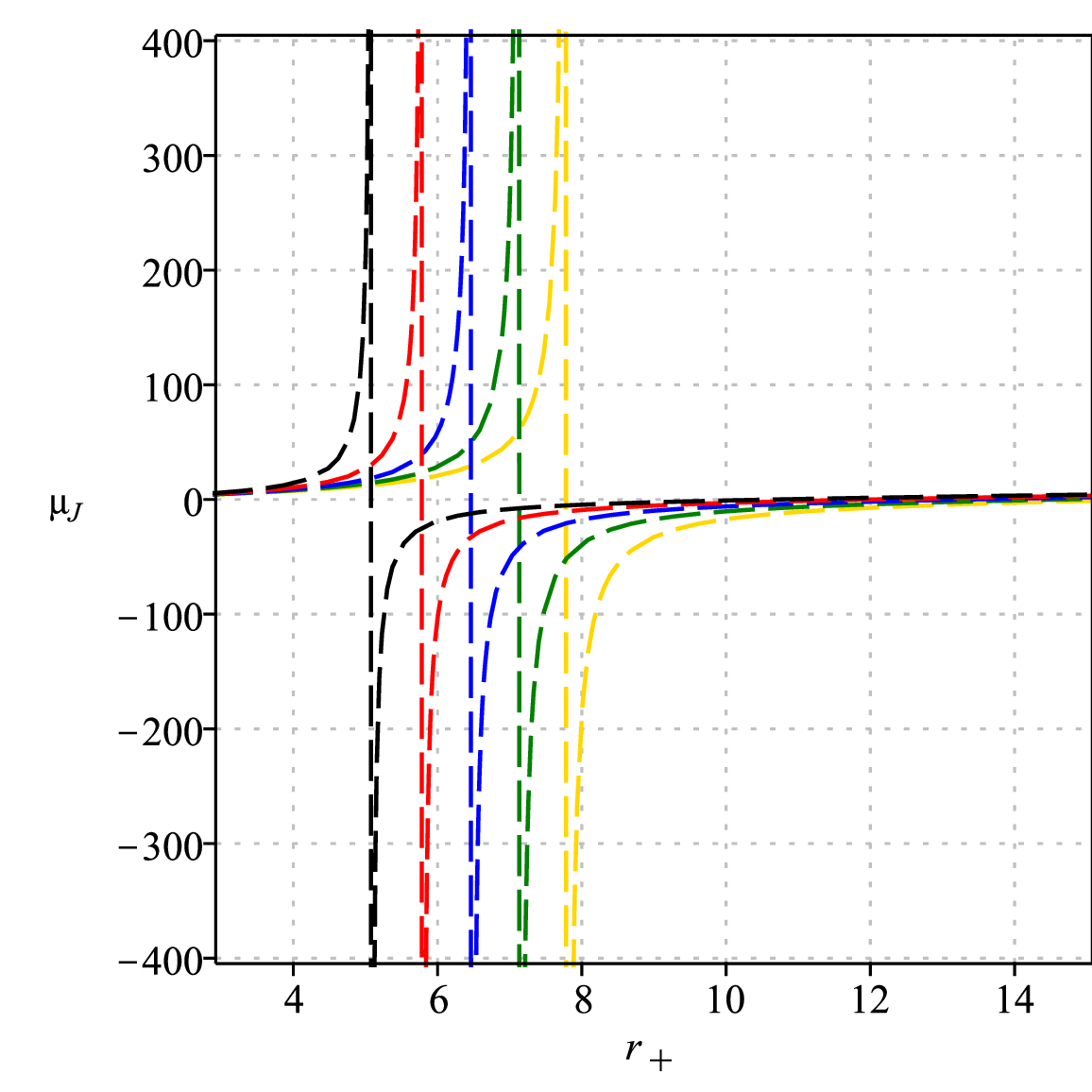}}\hfill
\subfloat[$\alpha=0.2$ \& $\beta=0.5$]{\includegraphics[width=.5\textwidth]{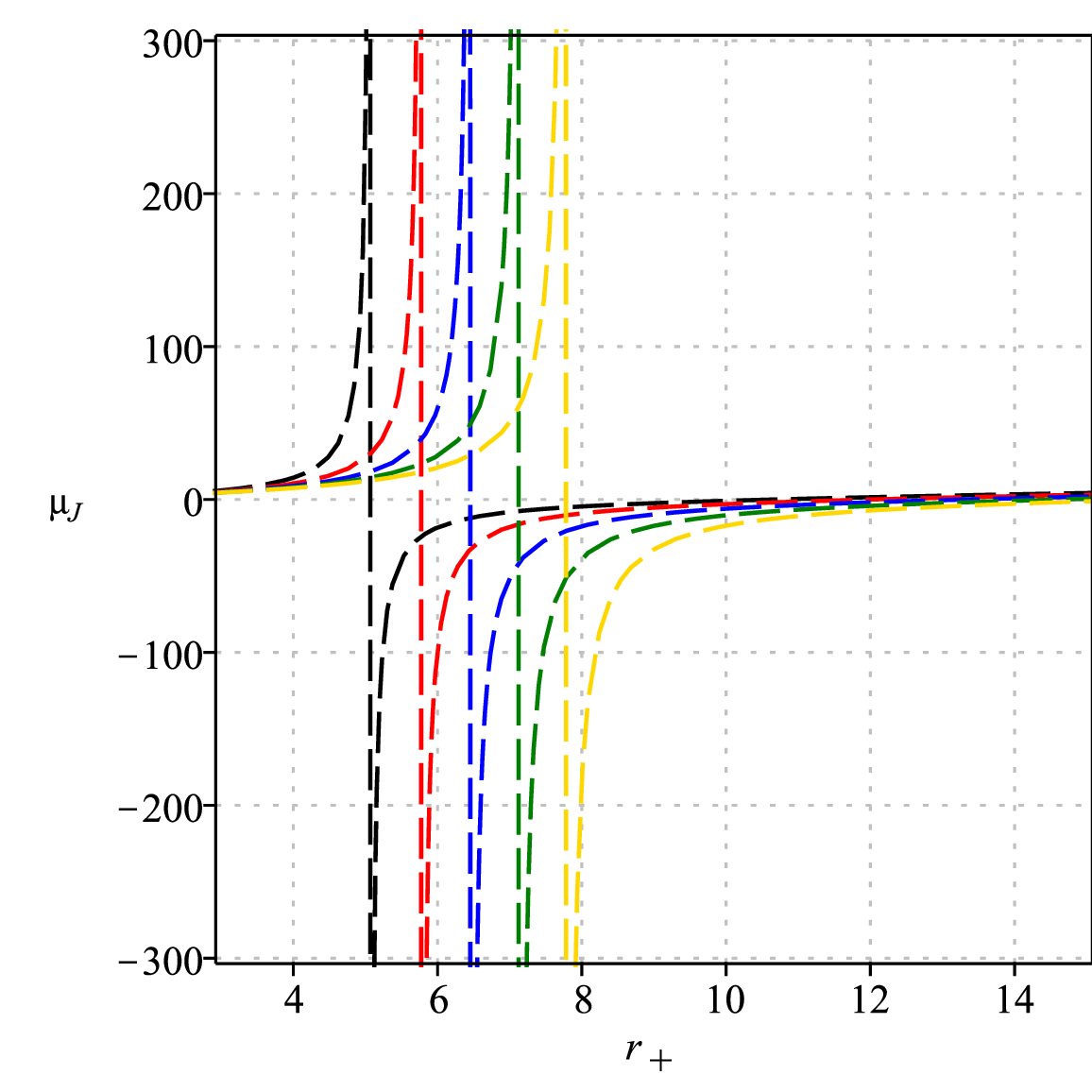}}\hfill
\subfloat[$\alpha=0.8$ \& $\beta=0.5$]{\includegraphics[width=.5\textwidth]{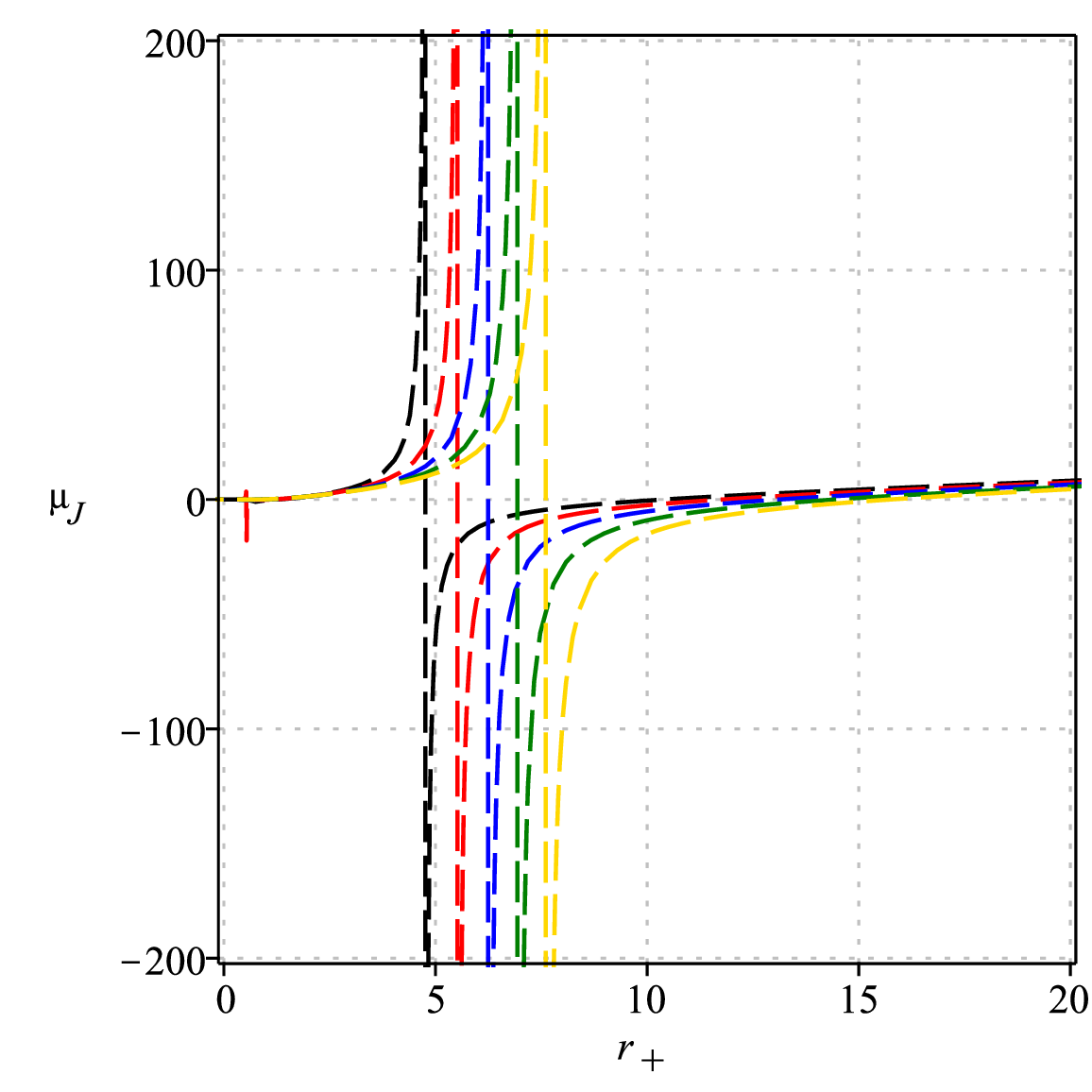}}\hfill
\centering
\begin{subfigure}[b]{0.2\textwidth}
\centering
\includegraphics[width=\textwidth]{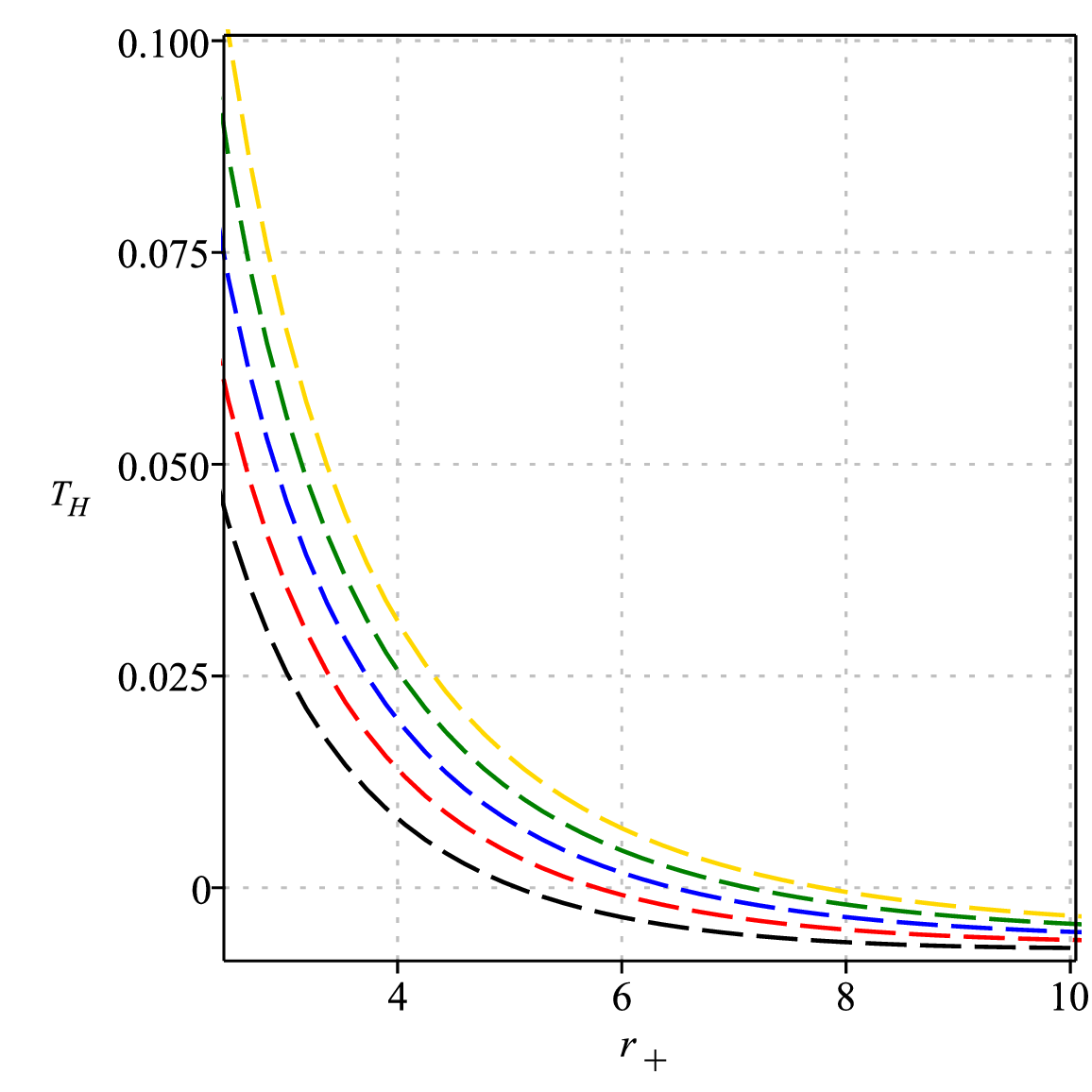}
\caption{$\beta=0.5$ \& $\alpha=0.2$}
\label{fig:51e}
\end{subfigure}
\hfill
\begin{subfigure}[b]{0.2\textwidth}
\centering
\includegraphics[width=\textwidth]{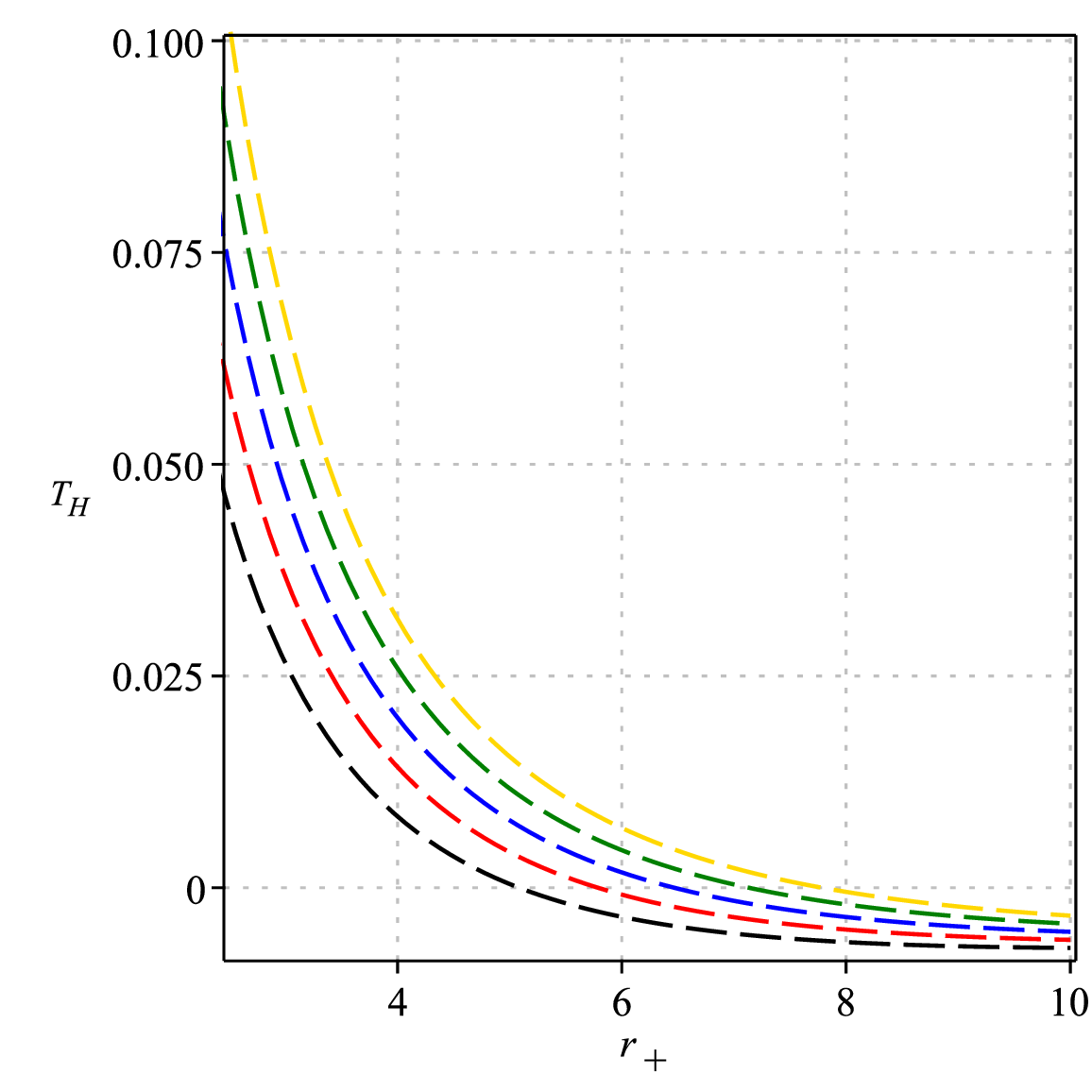}
\caption{$\beta=1.0$ \& $\alpha=0.2$}
\label{fig:51f}
\end{subfigure}
\hfill
\begin{subfigure}[b]{0.2\textwidth}
\centering
\includegraphics[width=\textwidth]{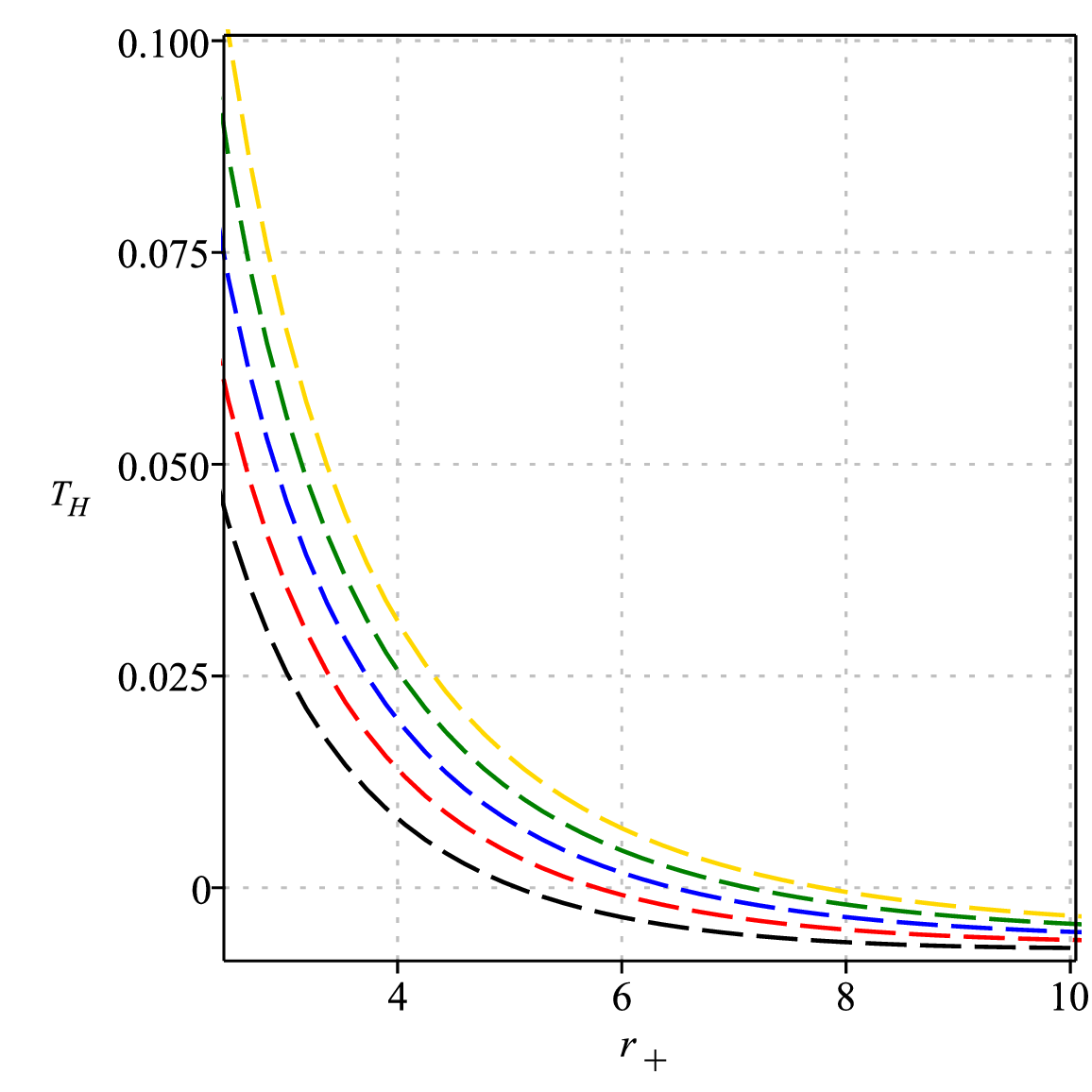}
\caption{$\alpha=0.2$ \& $\beta=0.5$}
\label{fig:51g}
\end{subfigure}
\hfill
\begin{subfigure}[b]{0.2\textwidth}
\centering
\includegraphics[width=\textwidth]{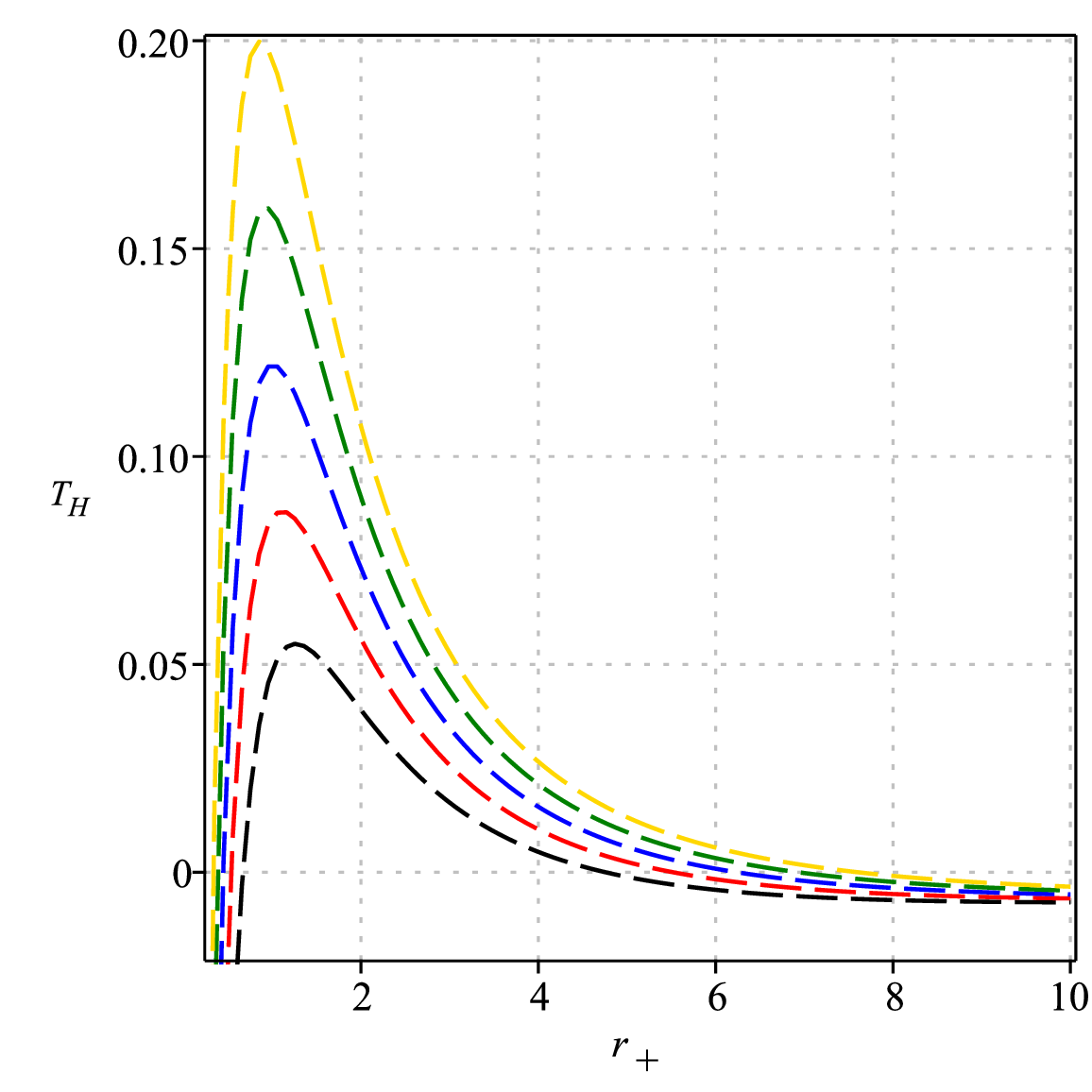}
\caption{$\alpha=0.8$ \& $\beta=0.5$}
\label{fig:51h}
\end{subfigure}
\caption{Black dash line denotes $M=2$, red dash line denotes $M=2.2$, blue dash line denoted $M=2.4$, green dash line denoted $M=2.6$ and gold dash line denoted $M=2.8$ with $Q_m=1.5$, and $m=0$.}
\label{fig:51}
\end{figure}

\begin{figure}[H]
    \centering
    \includegraphics[width=.6\textwidth]{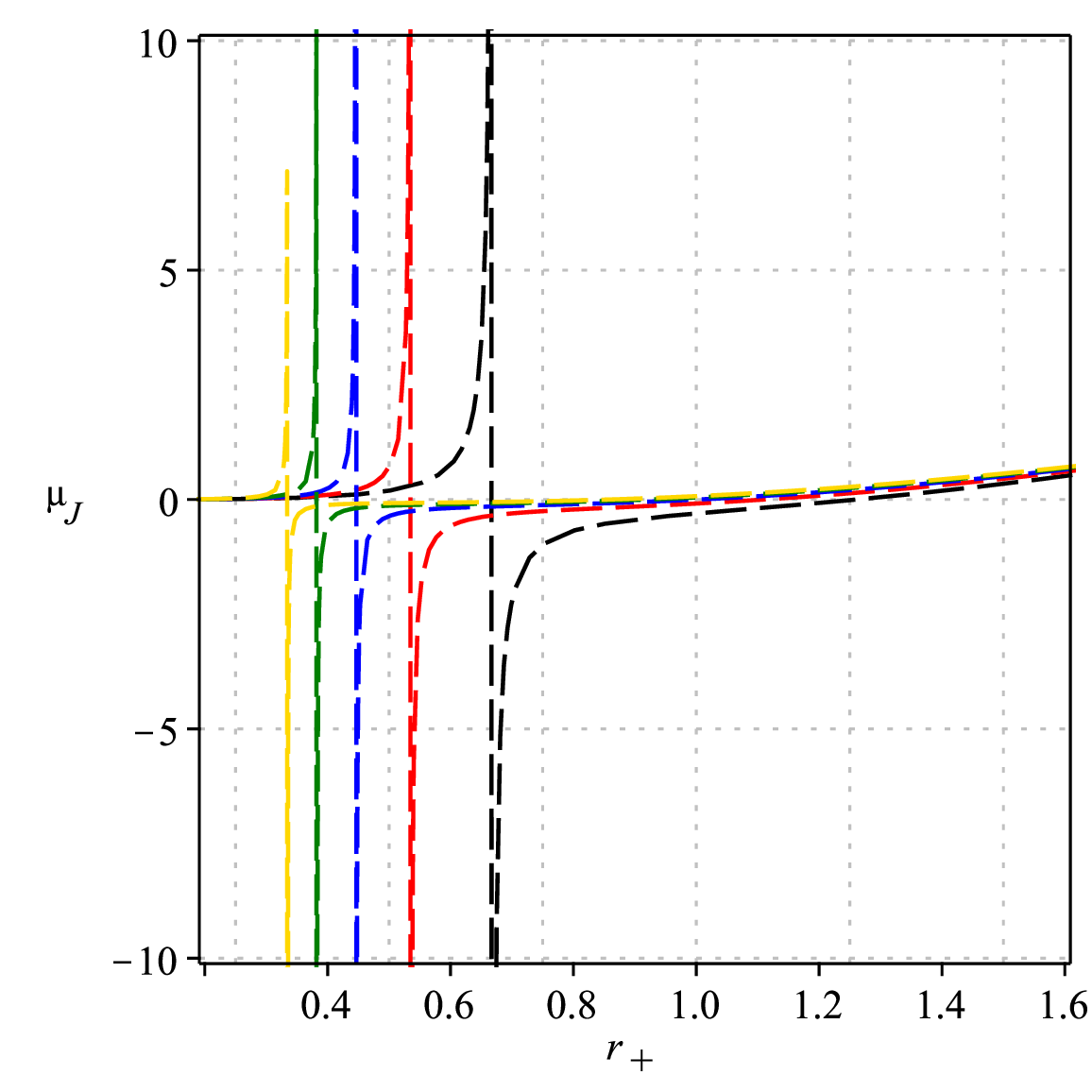}
    \caption{Small scale behaviour of Fig. \ref{fig:51}(d) with $\alpha=0.8$ \& $\beta=0.5$.}
\end{figure}\label{fig:52}

In Figs. \ref{fig:51}(a) \& \ref{fig:51}(b), Joule--Thomson coefficients are shown for 
different values of NED parameter $\beta$. For each value of black hole mass 
(enthalpy), $\mu_{J}$ has only one singular point and the corresponding Hawking 
temperature is zero (Figs. \ref{fig:51}e \& \ref{fig:51}f). Before the singular 
point black hole is in a cooling phase and at the singular point a phase transition 
of the black hole occurs from cooling phase to heating phase. After the singular 
point, an inverse phenomenon occurs at which $\mu_{J}=0$ and Joule--Thomson 
coefficients change its sign from negative to positive one.

In Figs. \ref{fig:51}(c) \& \ref{fig:51}(d), Joule--Thomson coefficients are depicted 
for different values of GB parameter $\alpha$. In Fig. \ref{fig:51}(c), 
$\mu_{J}$ has only one singular point and at the singular point Hawking temperature 
(Fig. \ref{fig:51}g) is zero. But in Fig. \ref{fig:51}(d), Joule--Thomson coefficients 
have two singular points (for small-scale behaviour see Fig. \ref{fig:52}) for each 
value of black hole mass (enthalpy). Therefore, for this black hole, two phase 
transitions occur. At each singular point Hawking temperature goes to zero 
(Fig. \ref{fig:51}h). The two singular points separate three regions. 
Furthermore, between the two singular points, an inverse phenomenon occurs for 
the black hole where the sign of $\mu_{J}$ changes.

\subsection{Black Holes in 4D Massive Einstein gravity coupled to NED}
From Hawking temperature \eqref{eq:3.5} we obtain equation of state 
\begin{equation}\label{eq:5.13}
T_{H}= \frac{1}{4 \pi  r_{+} (k^{2}+r^{2})} \biggr[ {8 r^{4} \pi  P +c c_{1} m^{2} r^{3}+(c^{2} c_{2} m^{2}+8 P \pi  k^{2}+1) r^{2}+c k^{2} m^{2} c_{1} r +(c^{2} c_{2} m^{2}+1) k^{2}-Q_{m}^{2}}
 \biggr]. 
\end{equation}
Using the mass function of the black hole we obtain 
\begin{equation}\label{eq:5.14}
    P=\frac{1}{r^{3} \pi k} \biggr[ -\frac{3 c^{2} c_{2} m^{2} r k}{8}-\frac{3 c m^{2} c_{1} r^{2} k}{16}+\frac{3 Q_{m}^{2} \arctan (\frac{r}{k})}{8}-\frac{3 \pi  Q_{m}^{2}}{16}+\frac{3 M k}{4}-\frac{3 r k}{8} \biggr].
\end{equation}
To obtain inverse pressure we use equation \eqref{eq:5.13} and equation \eqref{eq:5.2}
\begin{equation*}
P_i=\frac{1}{16 r_{{+}}^{2} \pi  (k^{2}+r_{{+}}^{2})^{2}} \biggr[  -4 c^{2} c_{2} k^{4} m^{2}-8 c^{2} c_{2} k^{2} m^{2} r_{{+}}^{2}-4 c^{2} c_{2} m^{2} r_{{+}}^{4}-3 c c_{1} k^{4} m^{2} r_{{+}} -6 c c_{1} k^{2} m^{2} r_{{+}}^{3}
\end{equation*}
\begin{equation}\label{eq:5.15}
-3 c c_{1} m^{2} r_{{+}}^{5}+4 Q_{m}^{2} k^{2}+6 Q_{m}^{2} r_{{+}}^{2}-4 k^{4}-8 k^{2} r_{{+}}^{2}-4 r_{{+}}^{4} \biggr].  
\end{equation}
Using inverse pressure and equation \eqref{eq:5.13} one can obtain inverse temperature as 
\begin{equation*}
T_i= \frac{1}{8 (k^{2}+r_{{+}}^{2})^{2} \pi  r_{{+}}}  \biggr[  -c c_{1} m^{2} r_{{+}}^{5}+(-2 c^{2} c_{2} m^{2}-2) r_{{+}}^{4}-2 c c_{1} k^{2} m^{2} r_{{+}}^{3}+((-4 c^{2} c_{2} m^{2}-4) k^{2}+4 Q_{m}^{2}) r_{{+}}^{2}
\end{equation*}
\begin{equation}\label{eq:5.16}
-c c_{1} k^{4} m^{2} r_{{+}} +(-2 c^{2} c_{2} m^{2}-2) k^{4}+2 Q_{m}^{2} k^{2} \biggr].
\end{equation}
In the limit $m \to 0$, above equations \eqref{eq:5.15} \& \eqref{eq:5.16} are reduced to the inverse pressure \& temperature of black hole in $4D$ massless gravity couples to NED \cite{kruglov2022nonlinearly}
\begin{equation}\label{eq:5.17}
P_i=\frac{1}{16 r_{{+}}^{2} \pi  (k^{2}+r_{{+}}^{2})^{2}} \biggr[ 
4 Q_{m}^{2} k^{2}+6 Q_{m}^{2} r_{{+}}^{2}-4 k^{4}-8 k^{2} r_{{+}}^{2}-4 r_{{+}}^{4} \biggr],     
\end{equation}
\begin{equation}\label{eq:5.18}
T_i= \frac{1}{8 (k^{2}+r_{{+}}^{2})^{2} \pi  r_{{+}}}  \biggr[ -2 r_{{+}}^{4}+(-4 k^{2}+4 Q_{m}^{2}) r_{{+}}^{2}
 -2 k^{4}+2 Q_{m}^{2} k^{2} \biggr].
\end{equation}

By setting $P_{i}=0$ into equation \eqref{eq:5.17}, one can obtain minimum event horizon radius as 
\begin{equation*}
-3 c c_{1} m^{2} (r_{{+}}^{\min})^{9}+(-4 c^{2} c_{2} m^{2}-4) (r_{{+}}^{\min})^{8}-6 c_{1} c m^{2} k^{2} (r_{{+}}^{\min})^{7}+((-8 c^{2} c_{2} m^{2}-8) k^{2}+6 Q_{m}^{2}) (r_{{+}}^{\min})^{6}
\end{equation*}
\begin{equation}\label{eq:5.19}
-3 c_{1} c m^{2} k^{4} (r_{{+}}^{\min})^{5}+((-4 c^{2} c_{2} m^{2}-4) k^{4}+4 Q_{m}^{2} k^{2}) (r_{{+}}^{\min})^{4} =0.
\end{equation}

By performing numerical computations, we derive the smallest possible event horizon radius. 
By applying both the equation \eqref{eq:5.19} and equation \eqref{eq:5.18}, we obtained the minimal 
inverse temperature. These findings are tabulated in table \ref{table:13} and table \ref{table:14} for 
different values of graviton mass and NED parameter. In the limit $m \to 0$, above equations 
\eqref{eq:5.19} and \eqref{eq:5.18} are reduced to the minimum horizon radius \& inversion temperature 
of a magnetic black hole in $4D$ massless Einstein gravity coupled to NED \cite{kruglov2022nonlinearly}

\begin{equation}
r_{+}^{min} =   \frac{\sqrt{3 Q_{m}^{2}-4 k^{2}+\sqrt{9 Q_{m}^{4}-8 Q_{m}^{2} k^{2}}}}{2},
\end{equation}
\begin{equation}
T_{i}^{min} = \frac{\sqrt{3 Q_{m}^{2}-4 k^{2}+Q_{m} \sqrt{9 Q_{m}^{2}-8 k^{2}}}}{\pi  (3 Q_{m} +\sqrt{9 Q_{m}^{2}-8 k^{2}})^{2}}.
\end{equation}

Furthermore, if one takes $\beta \to 0$ limit, into above equations then it's reduced to minimum inversion temperature  for Maxwell--$AdS$ magnetic black holes
\begin{equation}
    r_{+}^{min} = \frac{1}{6 \sqrt{6} \pi Q_m},
\end{equation}
\begin{equation}
    T_{i}^{min} = \frac{\sqrt{6}}{2 Q_m}.
\end{equation}

\begin{table}[H]
\centering
\begin{tabular}{ |p{1.5cm}|p{1.5cm}|p{1.5cm}| } 
\hline
\multicolumn{3}{|c|}{ $\beta=0.5$ } \\
\hline
\textbf{$m$} & \textbf{$r_{+}^{min}$} & \textbf{$T_{i}^{min}$}  \\ [0.5ex]  
\hline
0.0 & 1.414 & 0.0096  \\ \hline
0.1 & 1.415 & 0.0094  \\ \hline
0.2 & 1.420 & 0.0088  \\ \hline
0.3 & 1.428 & 0.0077  \\ \hline
0.4 & 1.444 & 0.0063  \\ \hline 
\end{tabular}
\caption{$Q_m=2$, $c_1=1$, $c_1=-1$ and $c_2=1$}
\label{table:13}
\end{table}

\begin{table}[H]
\centering
\begin{tabular}{ |p{1.5cm}|p{1.5cm}|p{1.5cm}| } 
\hline
\multicolumn{3}{|c|}{$m=0.1$} \\
\hline
\textbf{$\beta$} & \textbf{$r_{+}^{min}$} & \textbf{$T_{i}^{min}$} \\ [0.5ex]  
\hline
0.0 & 2.459 & 0.0104  \\ \hline
0.2 & 1.888 & 0.0102  \\ \hline
0.4 & 1.570 & 0.0098 \\ \hline
0.6 & 1.253 & 0.0089 \\ \hline
0.8 & 0.869 & 0.0071 \\ \hline
\end{tabular}
\caption{$Q_m=2$, $c_1=1$, $c_1=-1$ and $c_2=1$}
\label{table:14}
\end{table}

\begin{figure}[H]
\centering
\subfloat[$\beta=0.3$]{\includegraphics[width=.5\textwidth]{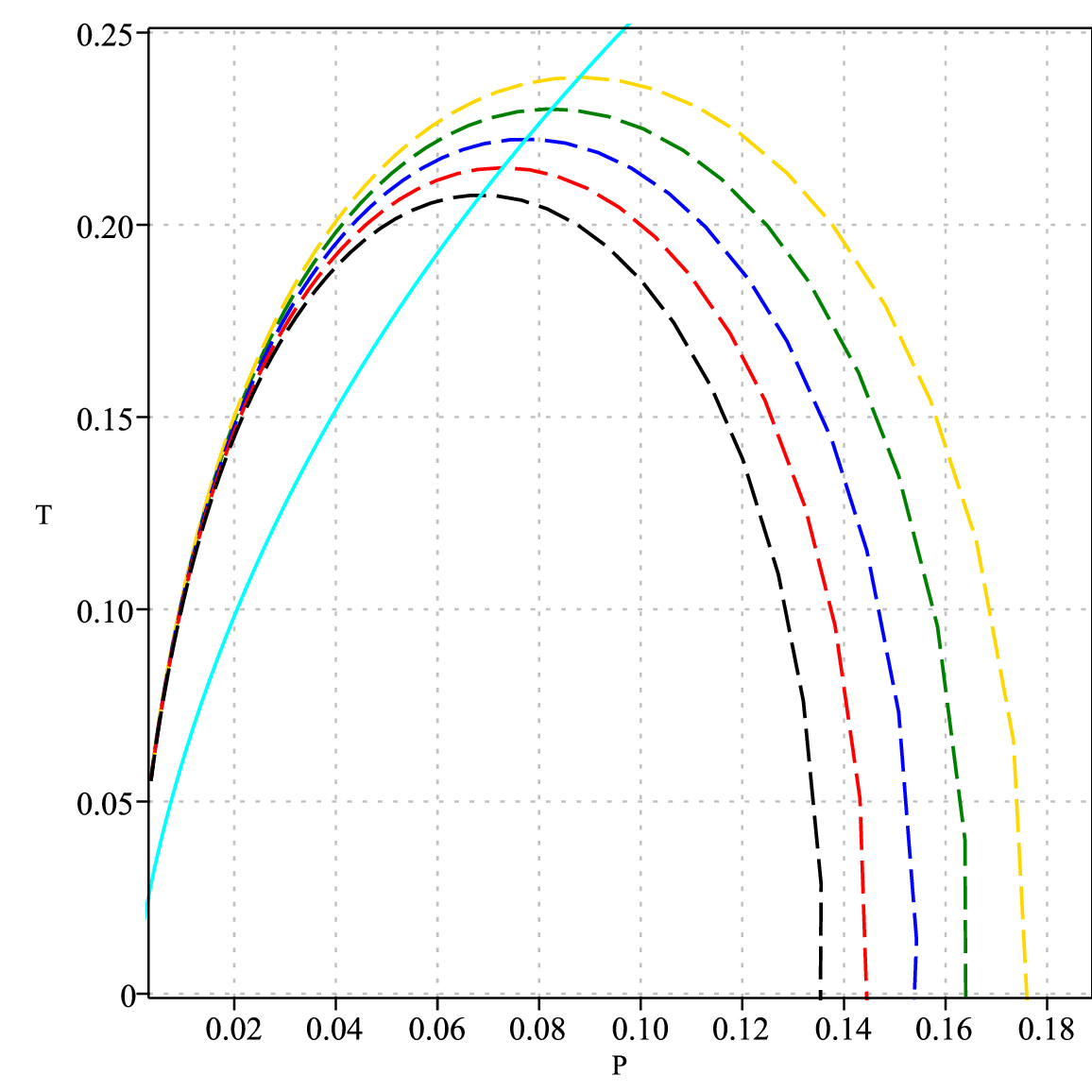}}\hfill
\subfloat[$\beta=0.5$]{\includegraphics[width=.5\textwidth]{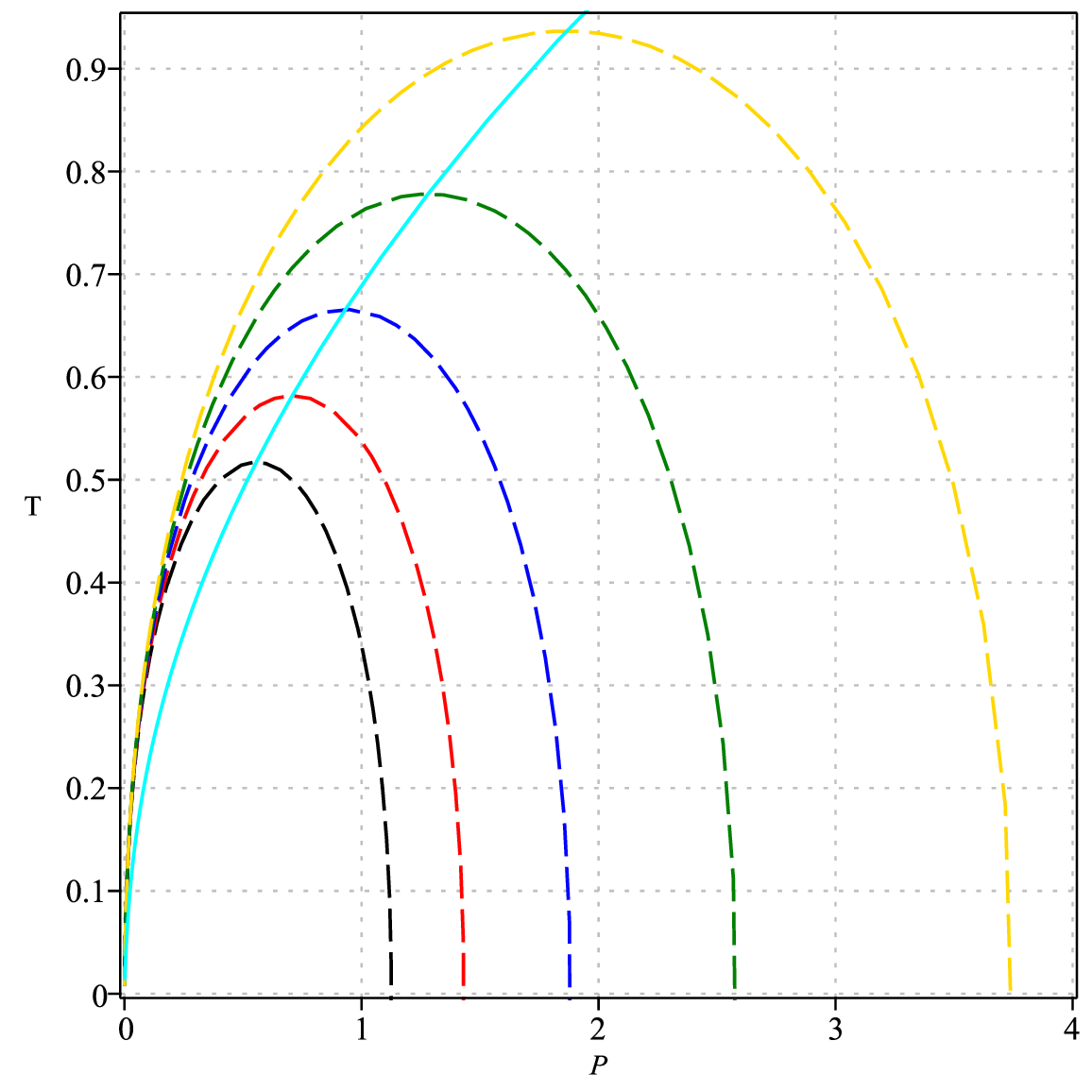}}\hfill
\caption{Black dash line denotes $M=20$, red dash line denotes $M=20.1$, blue dash line denoted $M=20.2$, green dash line denoted $M=20.3$, gold dash line denoted $M=20.4$ and solid cyan line denores inverse curve with $Q_m=10$, $m=0.5$, $c=1$, $c_1=-1$ and $c_2=1$.}\label{fig:53}
\end{figure}

\begin{figure}[H]
\centering
\subfloat[$c_1=-1$ \& $c_2=-1$]{\includegraphics[width=.5\textwidth]{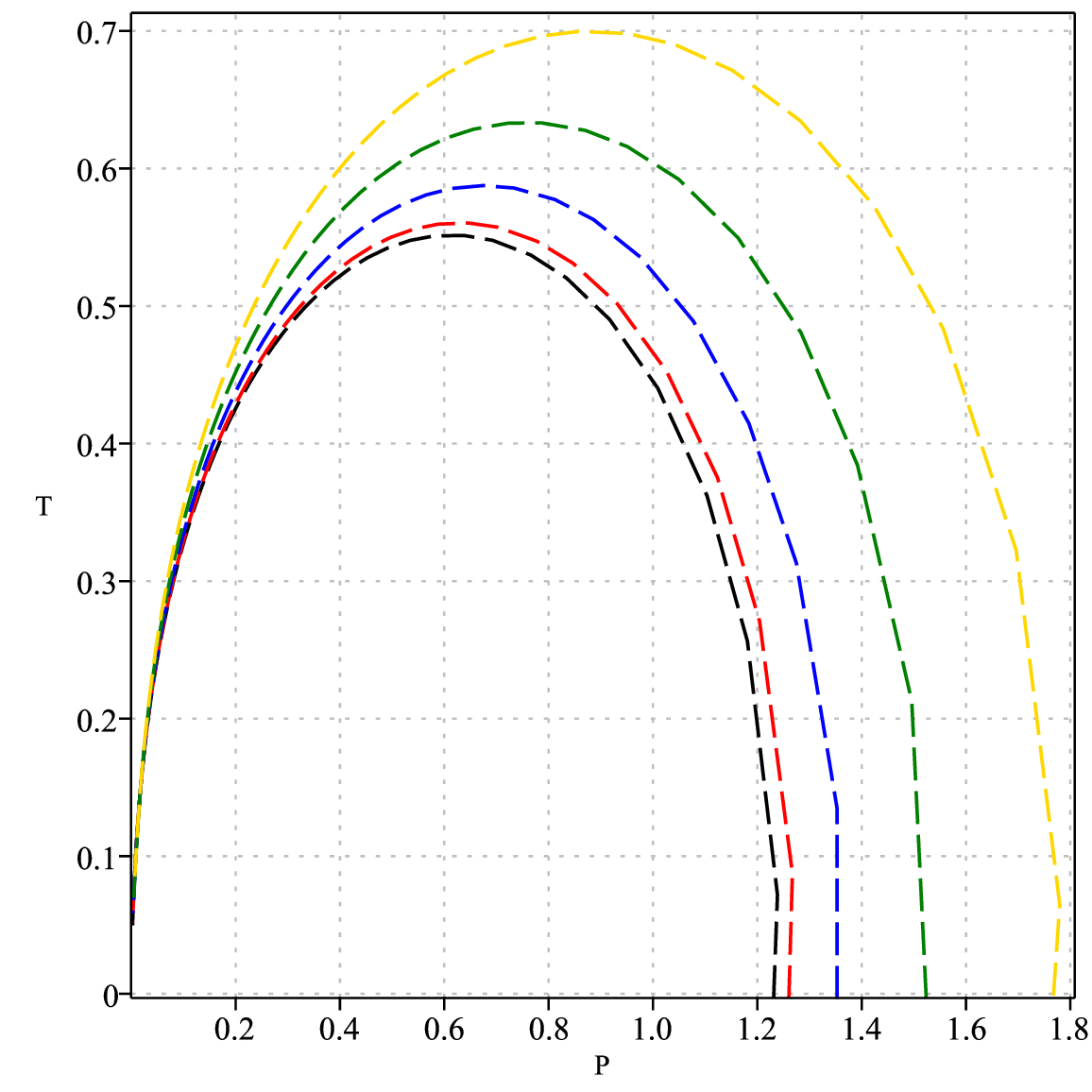}}\hfill
\subfloat[$c_1=-1$ \& $c_2=1$]{\includegraphics[width=.5\textwidth]{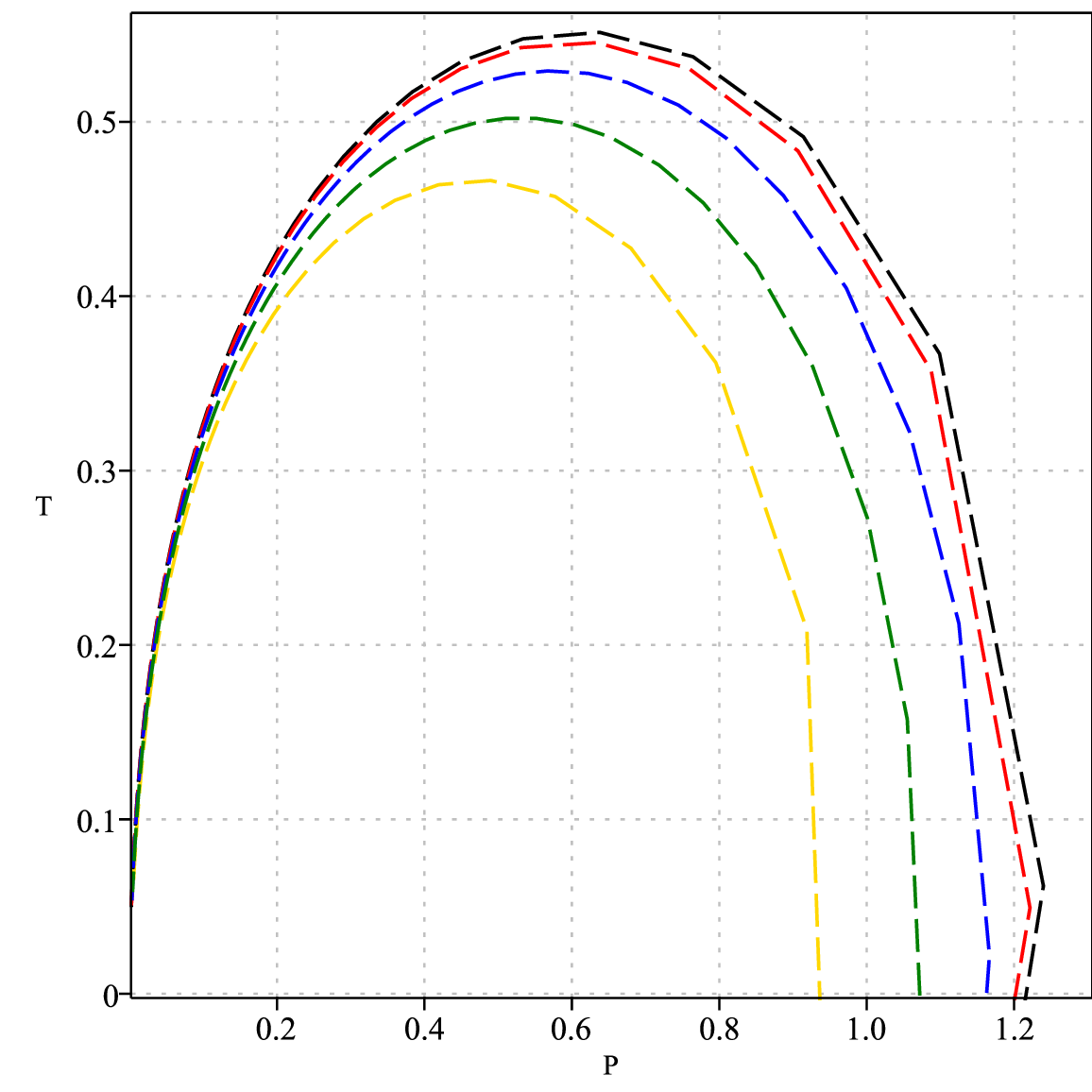}}\hfill
\subfloat[$c_1=1$ \& $c_2=-1$]{\includegraphics[width=.5\textwidth]{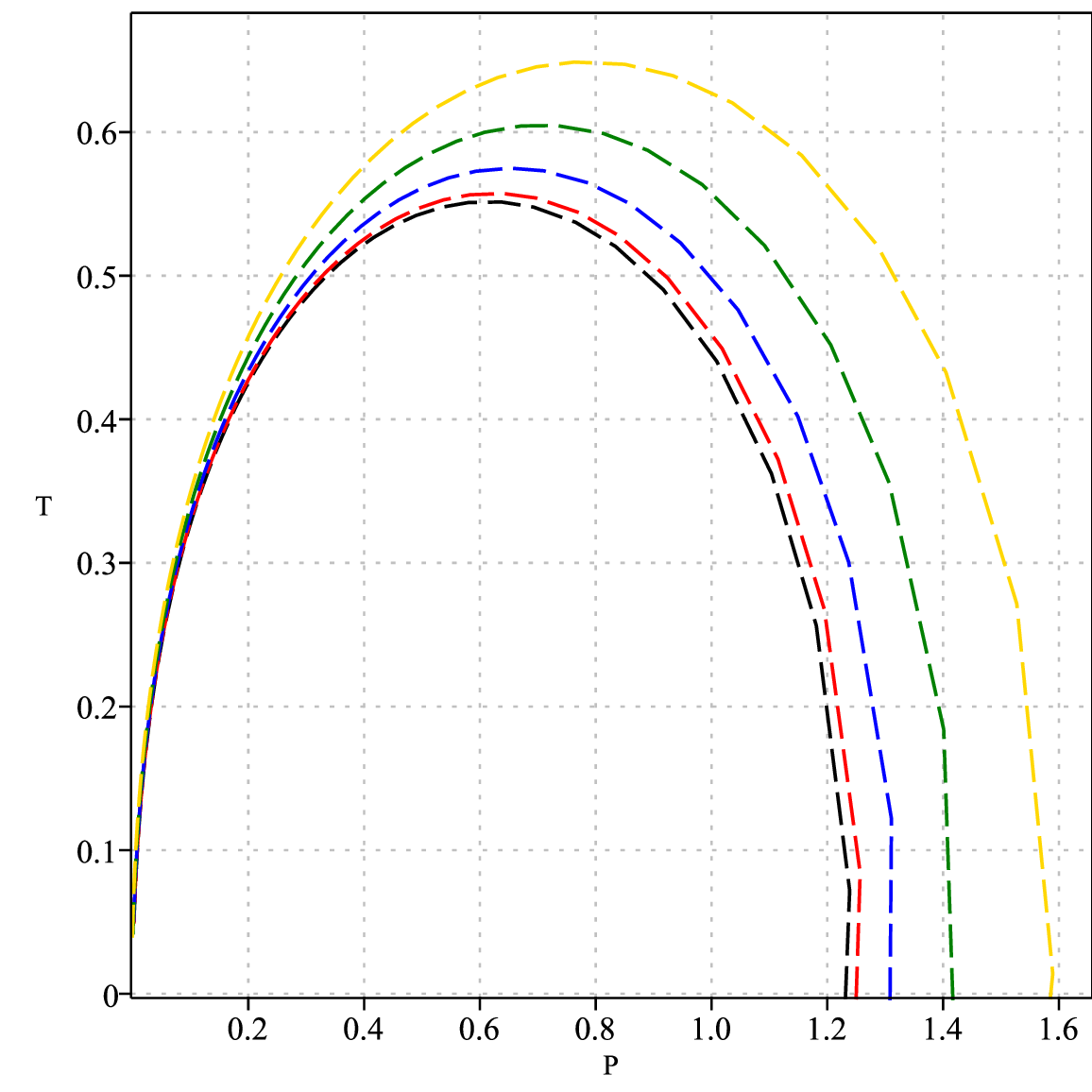}}\hfill
\subfloat[$c_1=1$ \& $c_2=1$]{\includegraphics[width=.5\textwidth]{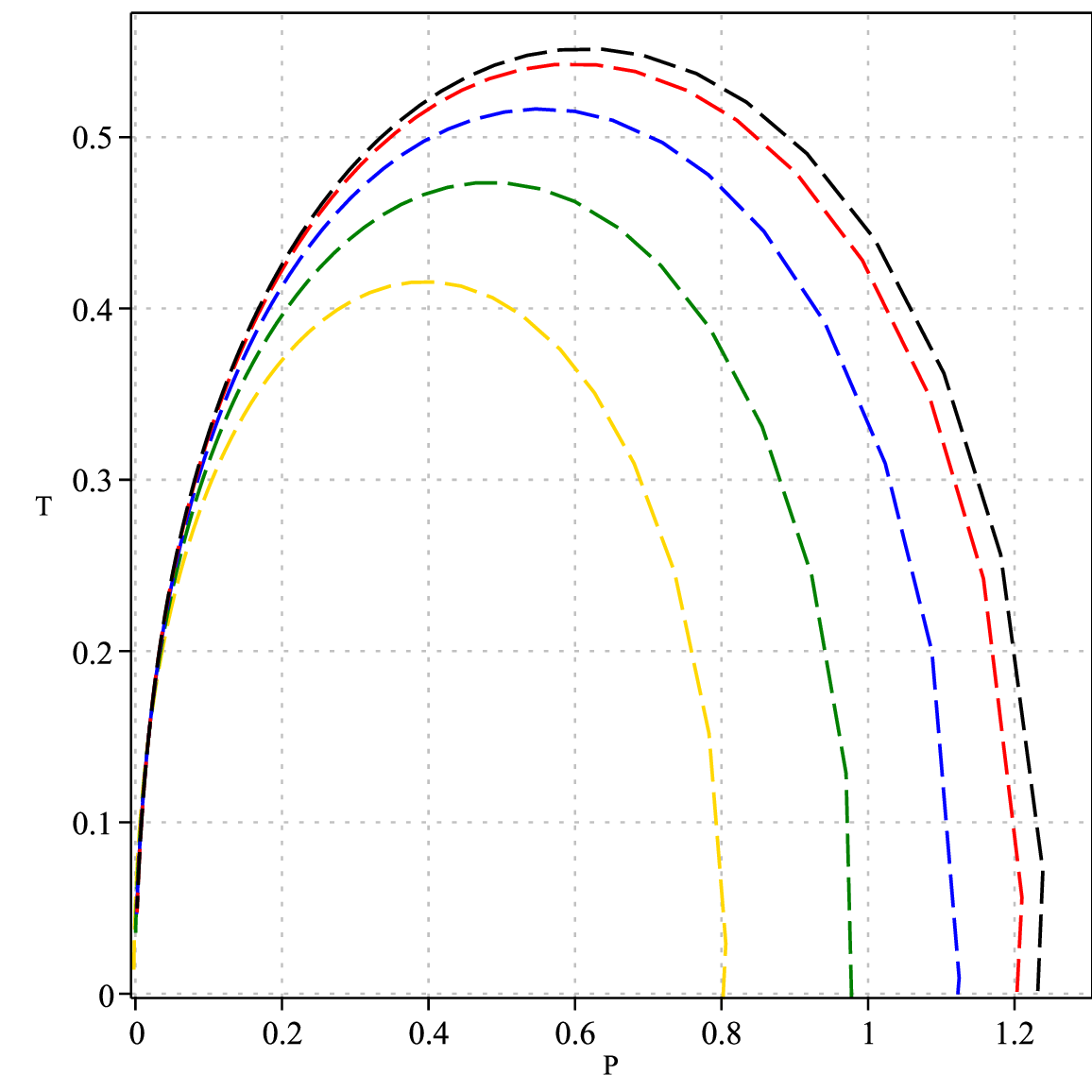}}\hfill
\caption{Black dash line denotes $m=0$, red dash line denotes $m=0.2$, blue dash line denoted $m=0.4$, green dash line denoted $m=0.6$ and gold dash line denoted $m=0.8$ with $Q_m=10$, $\beta=0.5$, $M=20$ and $c=1$.}\label{fig:54}
\end{figure}

\begin{figure}[H]
\centering
\subfloat[$c_1=0$]{\includegraphics[width=.5\textwidth]{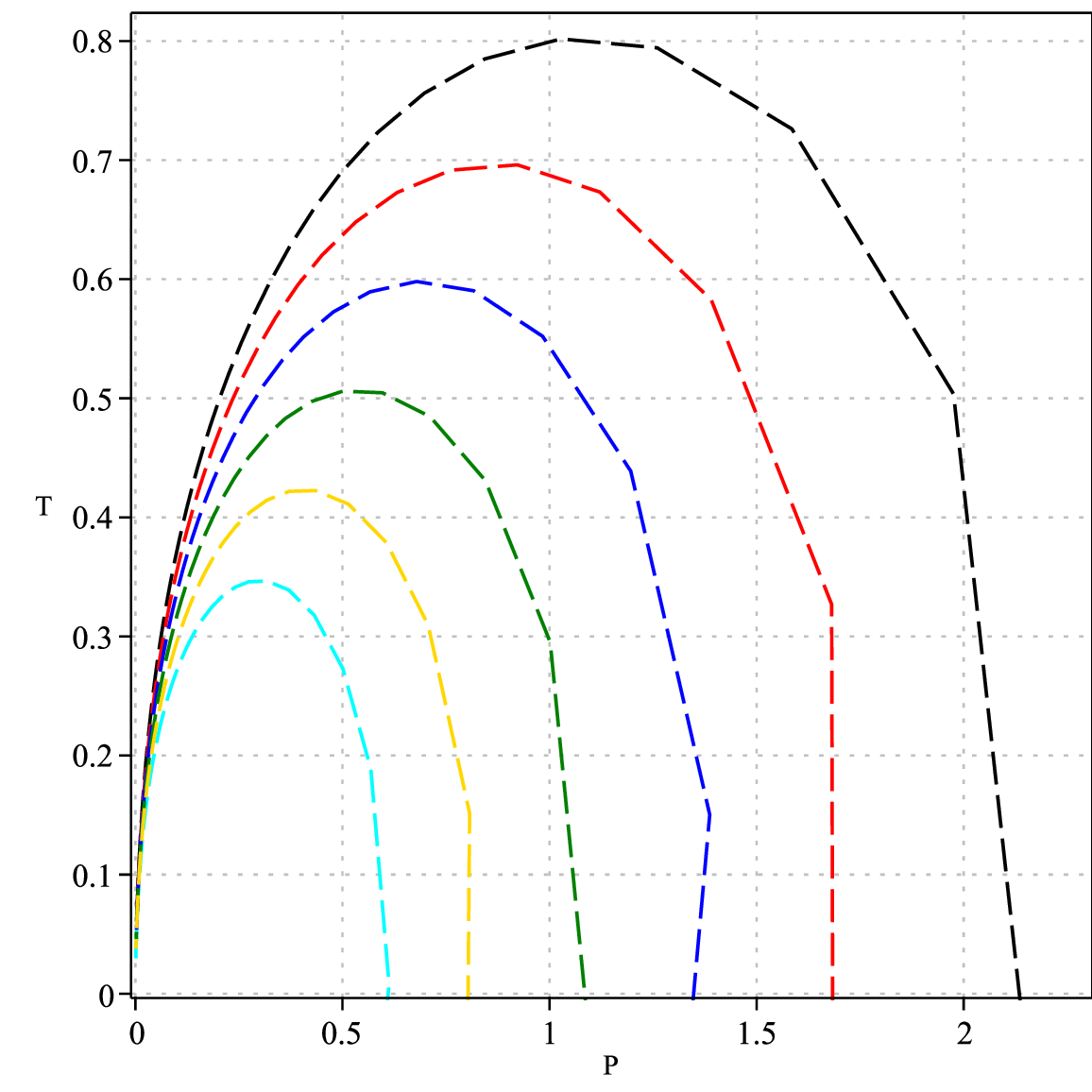}}\hfill
\subfloat[$c_2=0$]{\includegraphics[width=.5\textwidth]{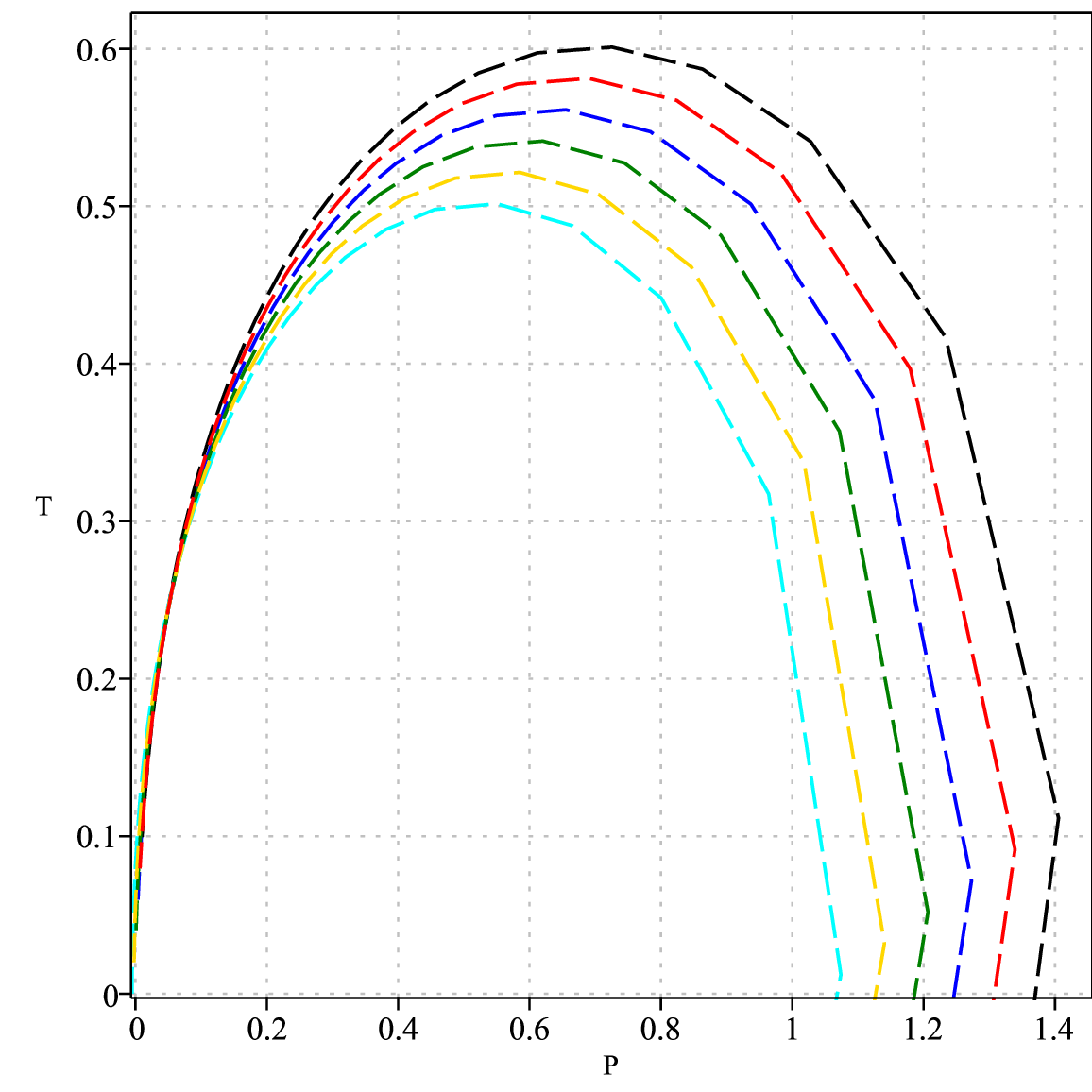}}\hfill
\caption{Left Panel : cyan dash line denotes $c_2=5$, gold dash line denotes 
$c_2=3$, green dash line denotes $c_2=1$, blue dash line denotes $c_2=-1$, red 
dash line denotes $c_2=-3$ and black dash line denotes $c_2=-5$. Right Panel : cyan dash line denotes $c_1=5$, gold dash line denotes $c_1=3$, green dash line denotes $c_1=1$, blue dash line denotes $c_1=-1$, red dash line denotes $c_1=-3$ and black dash line denotes $c_1=-5$. With $M=20$, $Q_m=10$, $\beta=0.5$, $c=1$ and $m=0.5$.}\label{fig:55}
\end{figure}

Using equation \eqref{eq:5.1}, equation \eqref{eq:5.13} and equation \eqref{eq:5.14} one can obtain J-T coefficient 
\begin{equation*}
\biggl(\frac{\partial T_{H}}{\partial r_{+}}\biggl)_{M}= \frac{1}{4 r_{{+}}^{3} k (k^{2}+r_{{+}}^{2})^{2} \pi} \Biggr[-6 (k^{2}+r_{{+}}^{2})^{2} Q_{m}^{2} \arctan  ({r_{{+}}}/{k})+(2 c^{2} c_{2} m^{2} r_{{+}} -12 M +2 r_{{+}} ) k^{5}+3 \pi  Q_{m}^{2} k^{4}
\end{equation*}
\begin{equation}
+((4 c^{2} c_{2} m^{2}+4) r_{{+}}^{3}-24 M r_{{+}}^{2}+4 Q_{m}^{2} r_{{+}} ) k^{3}+6 \pi  Q_{m}^{2} k^{2} r_{{+}}^{2}+((2 c^{2} c_{2} m^{2}+2) r_{{+}}^{5}-12 M r_{{+}}^{4}+6 Q_{m}^{2} r_{{+}}^{3}) k +3 \pi  Q_{m}^{2} r_{{+}}^{4}  \Biggr],
\end{equation}

\begin{equation*}
\biggl(\frac{\partial P}{\partial r_{+}}\biggl)_{M} = \frac{1}{16 (k^{2}+r_{+}^{2}) \pi  r_{+}^{5} k} \Biggr[ -18 r_{{+}} Q_{m}^{2} (k^{2}+r_{{+}}^{2}) \arctan  ({r_{{+}}}/{k})+3 c c_{1} m^{2} r_{{+}}^{5} k +(12 c^{2} c_{2} m^{2}+12) k r_{{+}}^{4}
\end{equation*}
\begin{equation}
+(3 c c_{1} k^{3} m^{2}+9 \pi  Q_{m}^{2}-36 M k ) r_{{+}}^{3}+((12 c^{2} c_{2} m^{2}+12) k^{3}+6 Q_{m}^{2} k ) r_{{+}}^{2}+(9 \pi  Q_{m}^{2} k^{2}-36 M k^{3}) r_{{+}} \Biggr].
\end{equation}

In the limit $m\to 0$, above equation is reduced to the Joule--Thomson coefficients in $4D$ massless GR coupled to NED \cite{kruglov2022nonlinearly}
\begin{equation*}
\mu_{J}= \frac{4 \biggr[ -A+(M -{r_{{+}}}/{6}) k^{5}+(2 M r_{{+}}^{2}-\frac{1}{3} Q_{m}^{2} r_{{+}} - r_{{+}}^{3}/3) k^{3}+r_{{+}}^{3} B k \biggr] r_{{+}}}{3 \biggr[ -A+(M -{r_{{+}}}/{3}) k^{3} (k^{2}+r_{{+}}^{2})+r_{{+}}  k B (k^{2}+r_{{+}}^{2}) \biggr] },
\end{equation*}

where $A$ \& $B$ are given by 
\begin{equation*}
A=\frac{Q_{m}^{2} (k^{2}+r_{{+}}^{2})^2 (\pi -2 \arctan (\frac{r_{{+}}}{k}))}{4},
\end{equation*}
\begin{equation}
B=(M r_{{+}} -\frac{1}{6} Q_{m}^{2}-\frac{1}{3} r_{{+}}^{2}).
\end{equation}

\begin{figure}[H]
\centering
\begin{subfigure}[b]{0.2\textwidth}
\centering
\includegraphics[width=\textwidth]{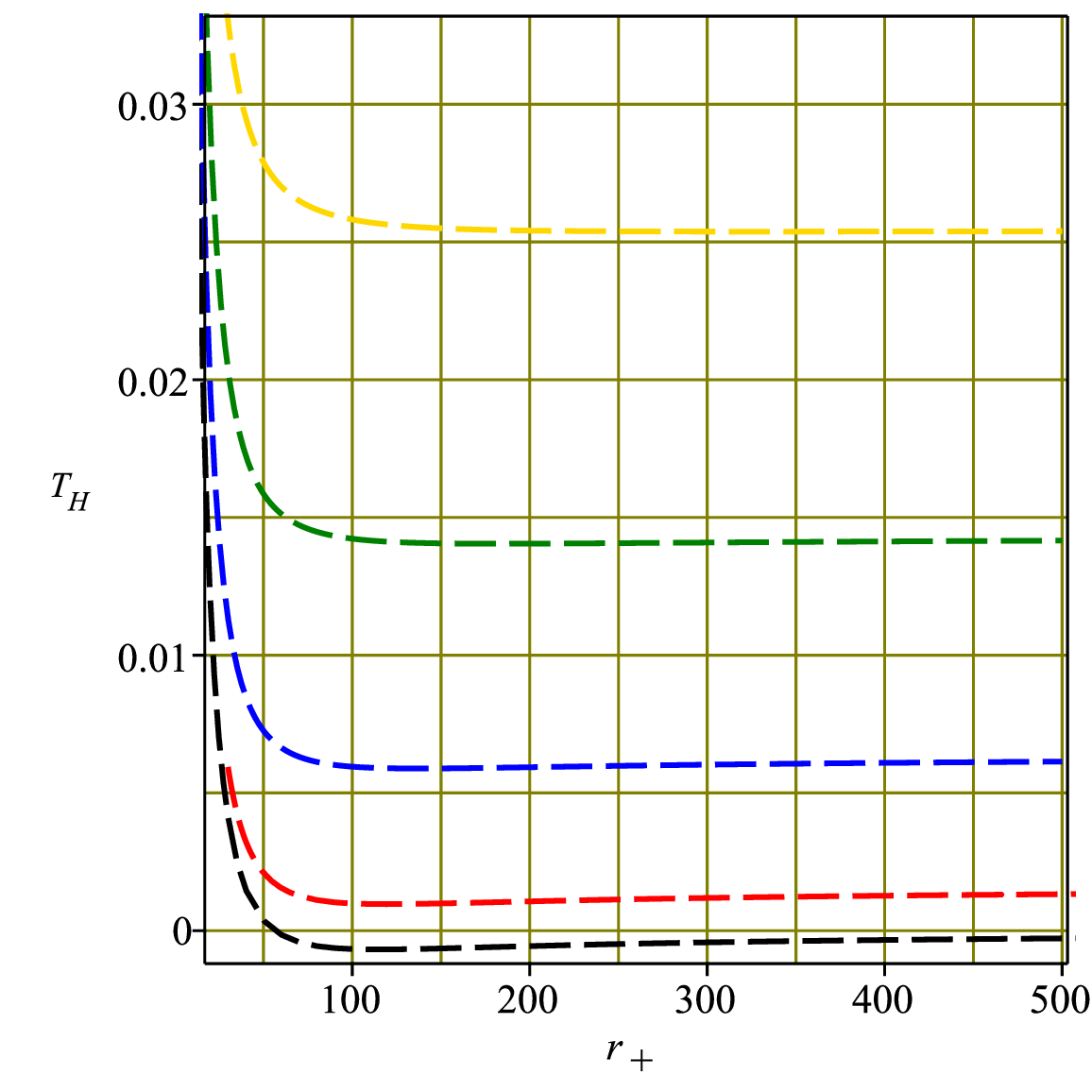}
\caption{Large scale $c_1=-1$ \& $c_2=-1$}
\label{fig:57a}
\end{subfigure}
\hfill
\begin{subfigure}[b]{0.2\textwidth}
\centering
\includegraphics[width=\textwidth]{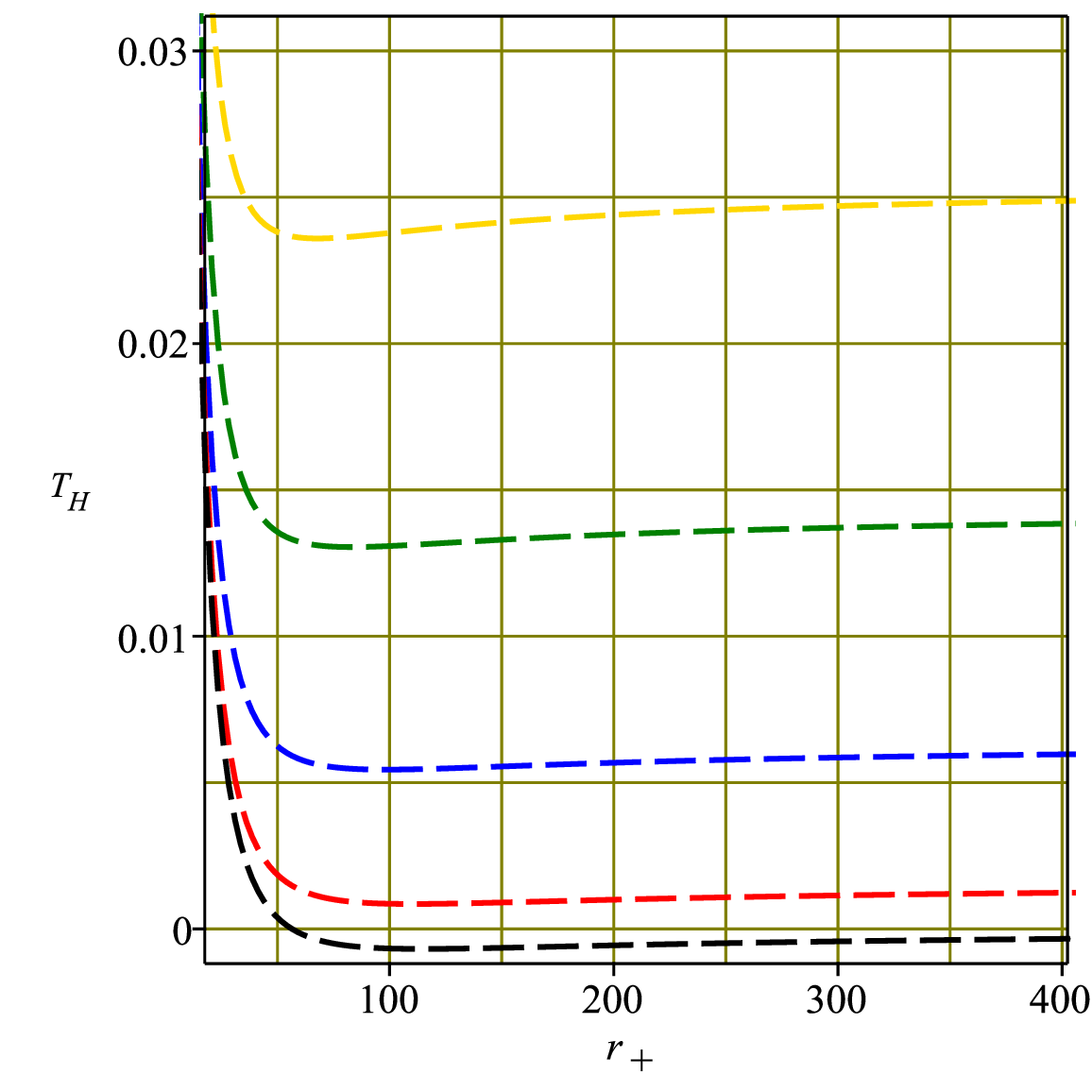}
\caption{Large scale $c_1=-1$ \& $c_2=1$}
\label{fig:57b}
\end{subfigure}
\hfill
\begin{subfigure}[b]{0.2\textwidth}
\centering
\includegraphics[width=\textwidth]{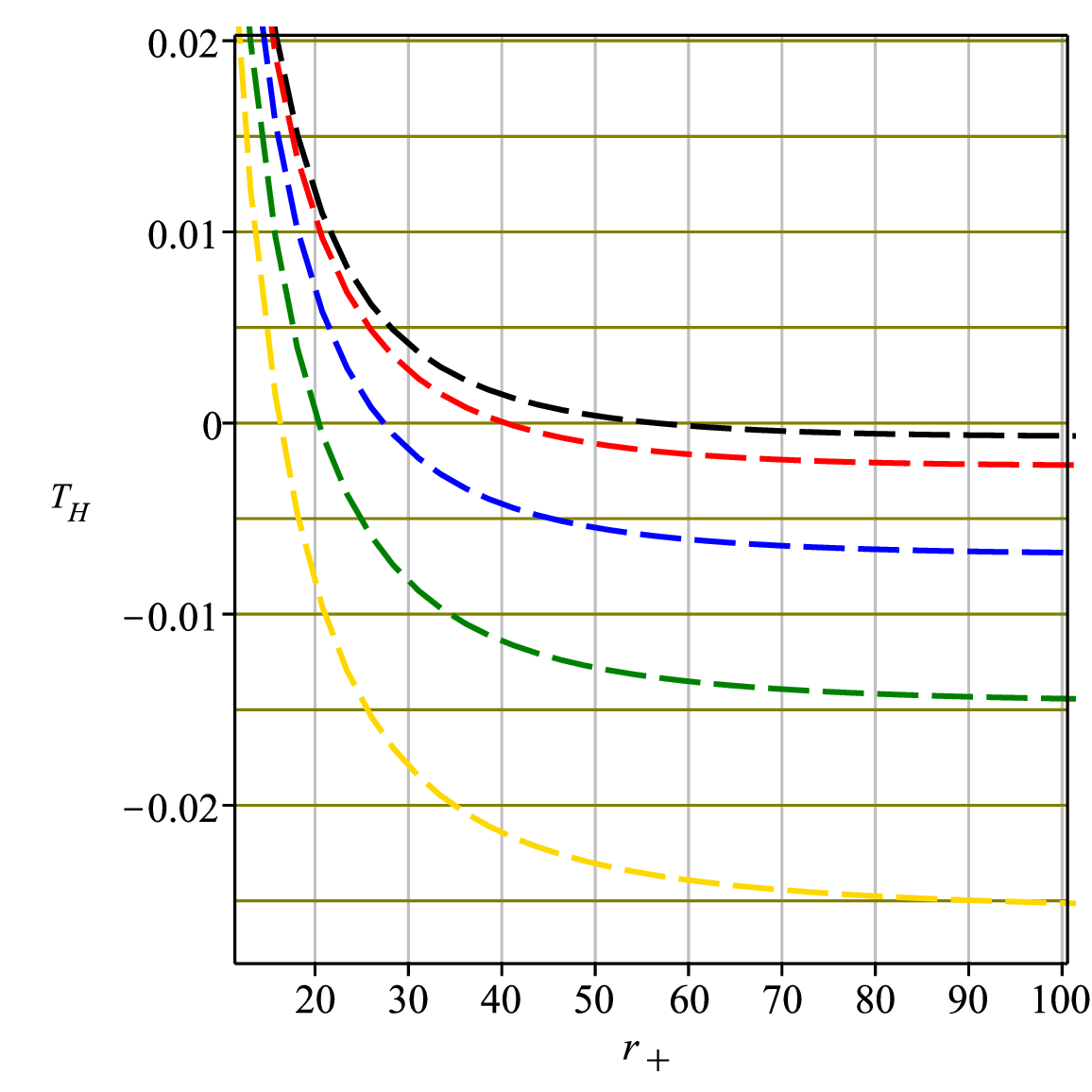}
\caption{Large scale $c_1=1$ \& $c_2=-1$}
\label{fig:57c}
\end{subfigure}
\hfill
\begin{subfigure}[b]{0.2\textwidth}
\centering
\includegraphics[width=\textwidth]{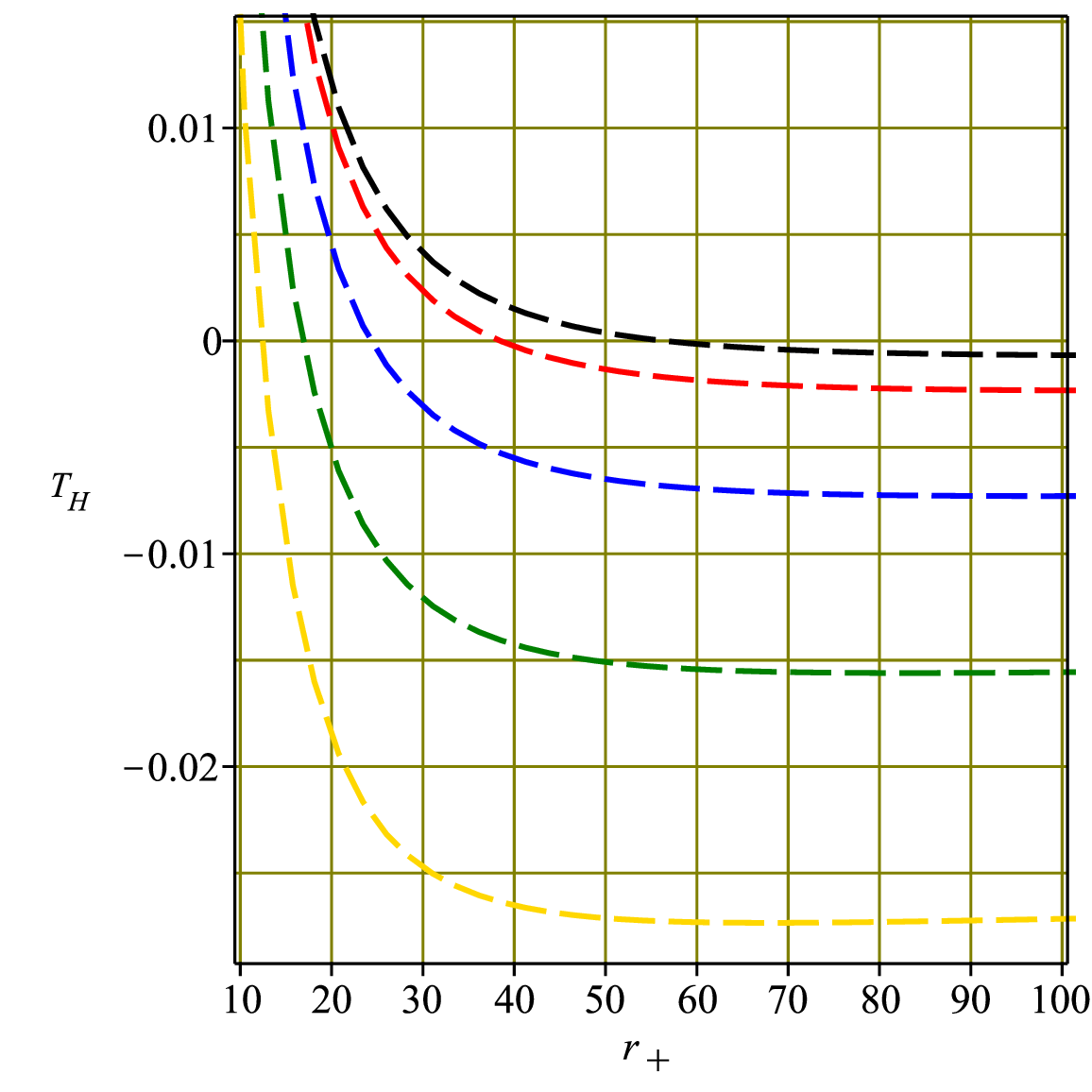}
\caption{Large scale $c_1=1$ \& $c_2=1$}
\label{fig:57d}
\end{subfigure}
\hfill
\begin{subfigure}[b]{0.2\textwidth}
\centering
\includegraphics[width=\textwidth]{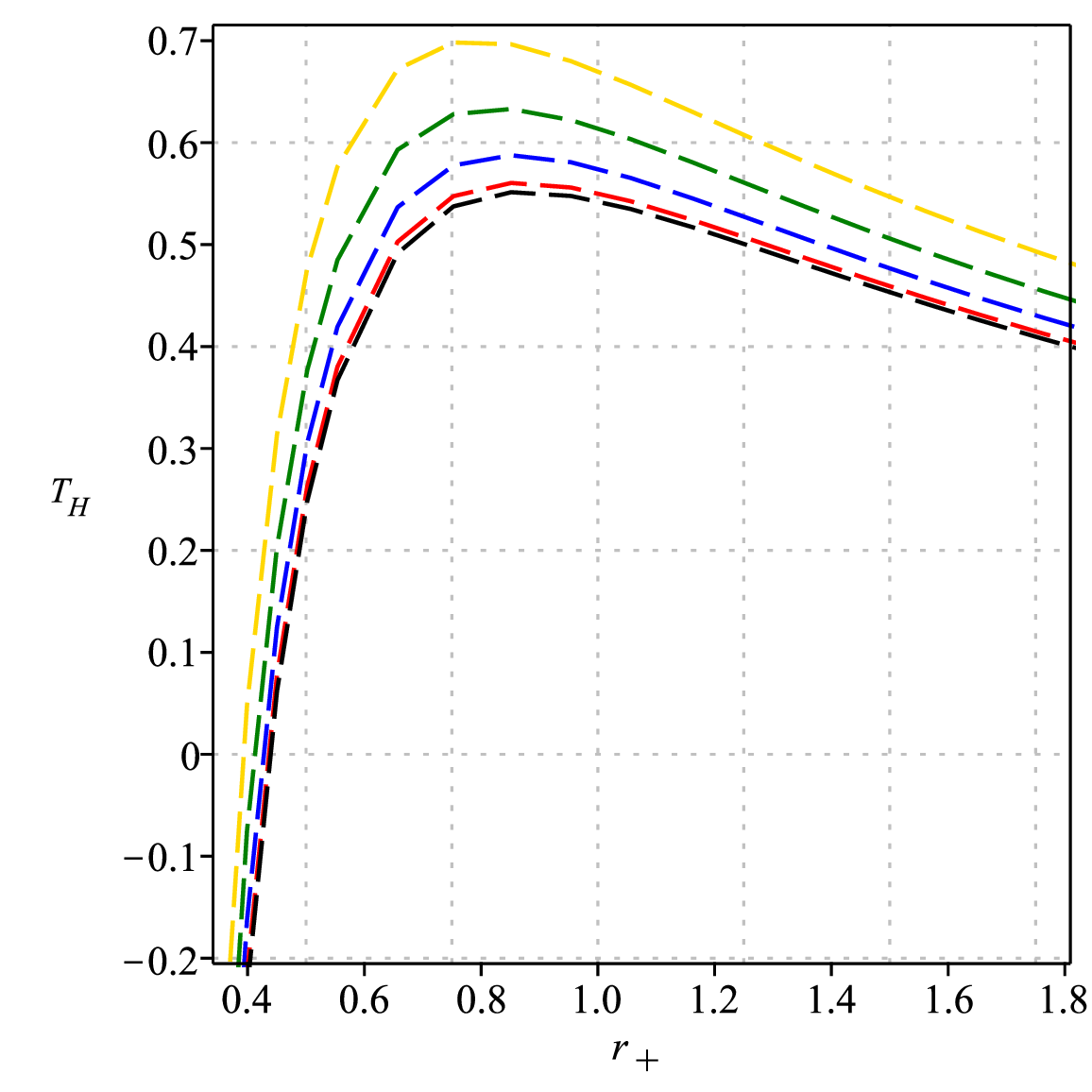}
\caption{Small scale $c_1=-1$ \& $c_2=-1$}
\label{fig:57e}
\end{subfigure}
\hfill
\begin{subfigure}[b]{0.2\textwidth}
\centering
\includegraphics[width=\textwidth]{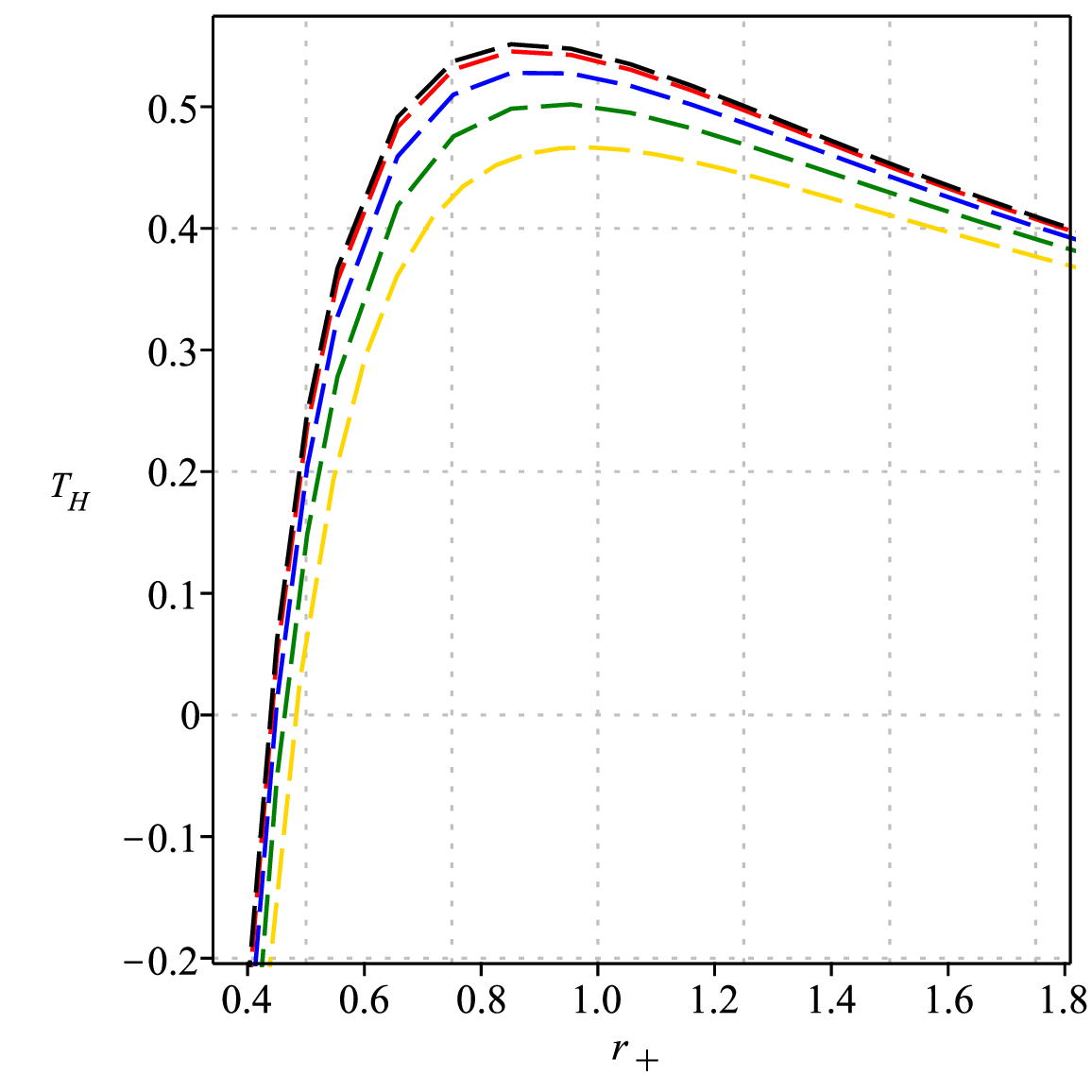}
\caption{Small scale $c_1=-1$ \& $c_2=1$}
\label{fig:57f}
\end{subfigure}
\hfill
\begin{subfigure}[b]{0.2\textwidth}
\centering
\includegraphics[width=\textwidth]{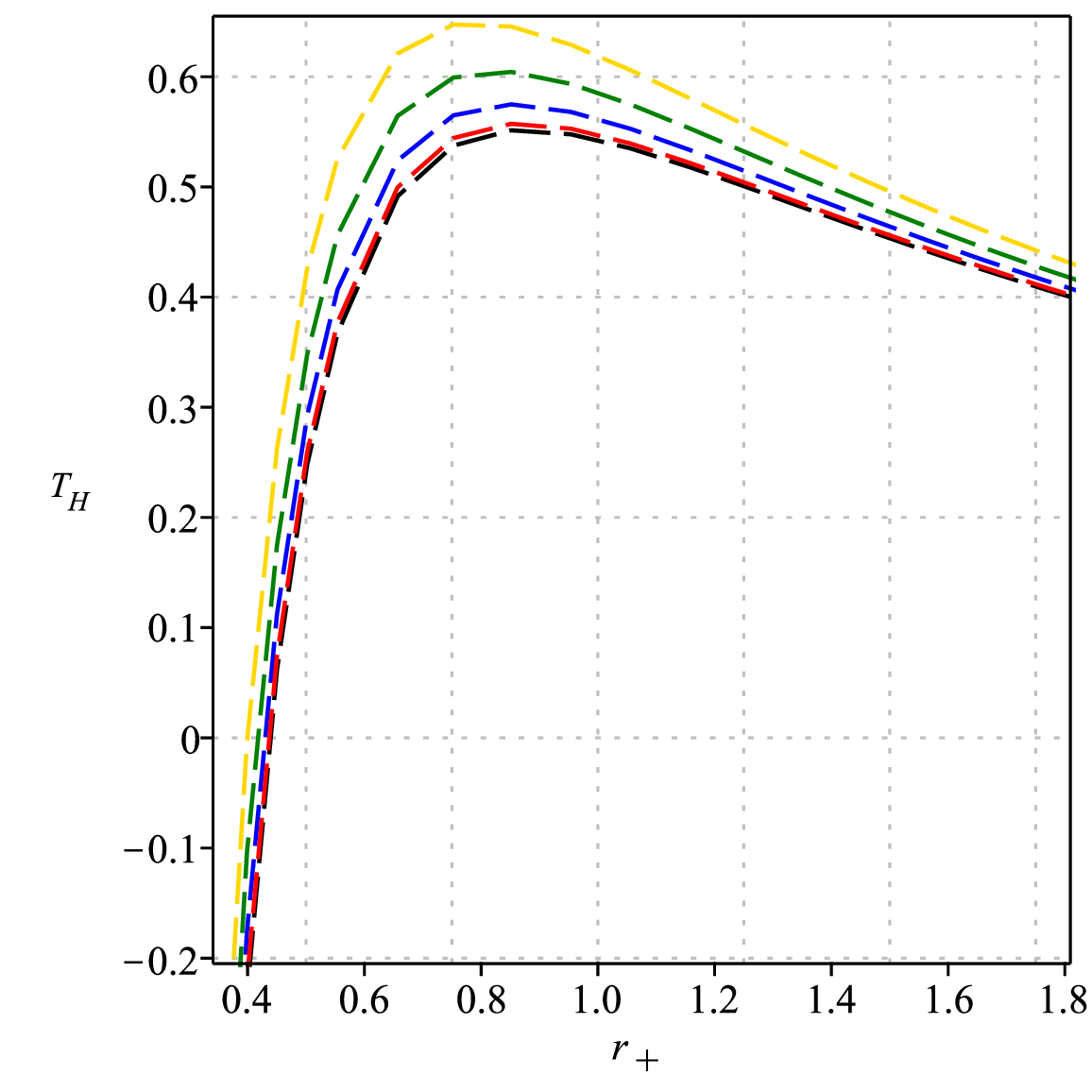}
\caption{Small scale $c_1=1$ \& $c_2=-1$}
\label{fig:57g}
\end{subfigure}
\hfill
\begin{subfigure}[b]{0.2\textwidth}
\centering
\includegraphics[width=\textwidth]{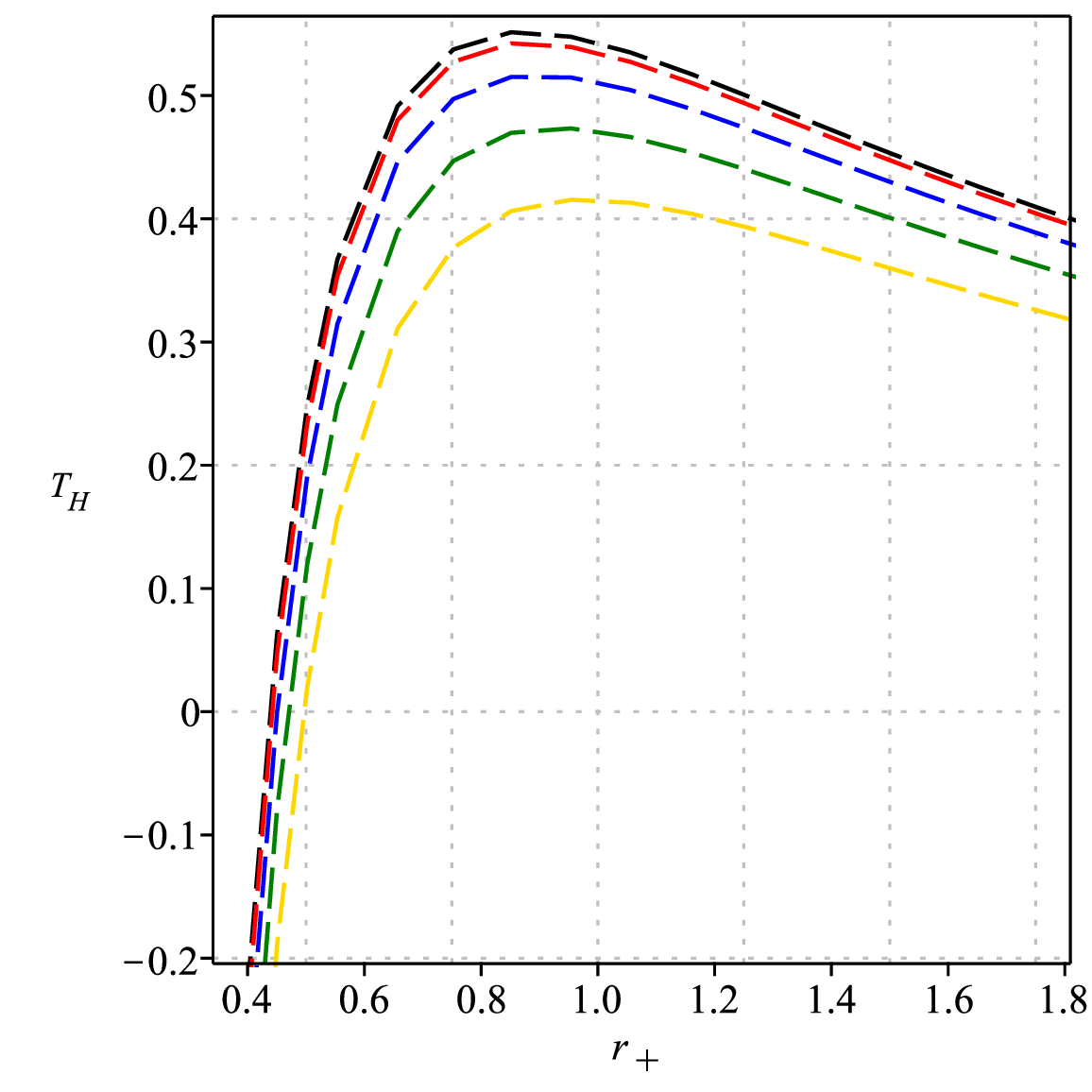}
\caption{Small scale $c_1=1$ \& $c_2=1$}
\label{fig:57h}
\end{subfigure}
\caption{Black dash line denotes $m=0$, red dash line denotes $m=0.2$, blue dash line denoted $m=0.4$, green dash line denoted $m=0.6$ and gold dash line denoted $m=0.8$ with $Q_m=10$, $\beta=0.5$, $M=20$ and $c=1$.}\label{fig:57}
\end{figure}

\begin{figure}[H]
\centering
\subfloat[ $c_1=-1$ \& $c_2=-1$]{\includegraphics[width=.5\textwidth]{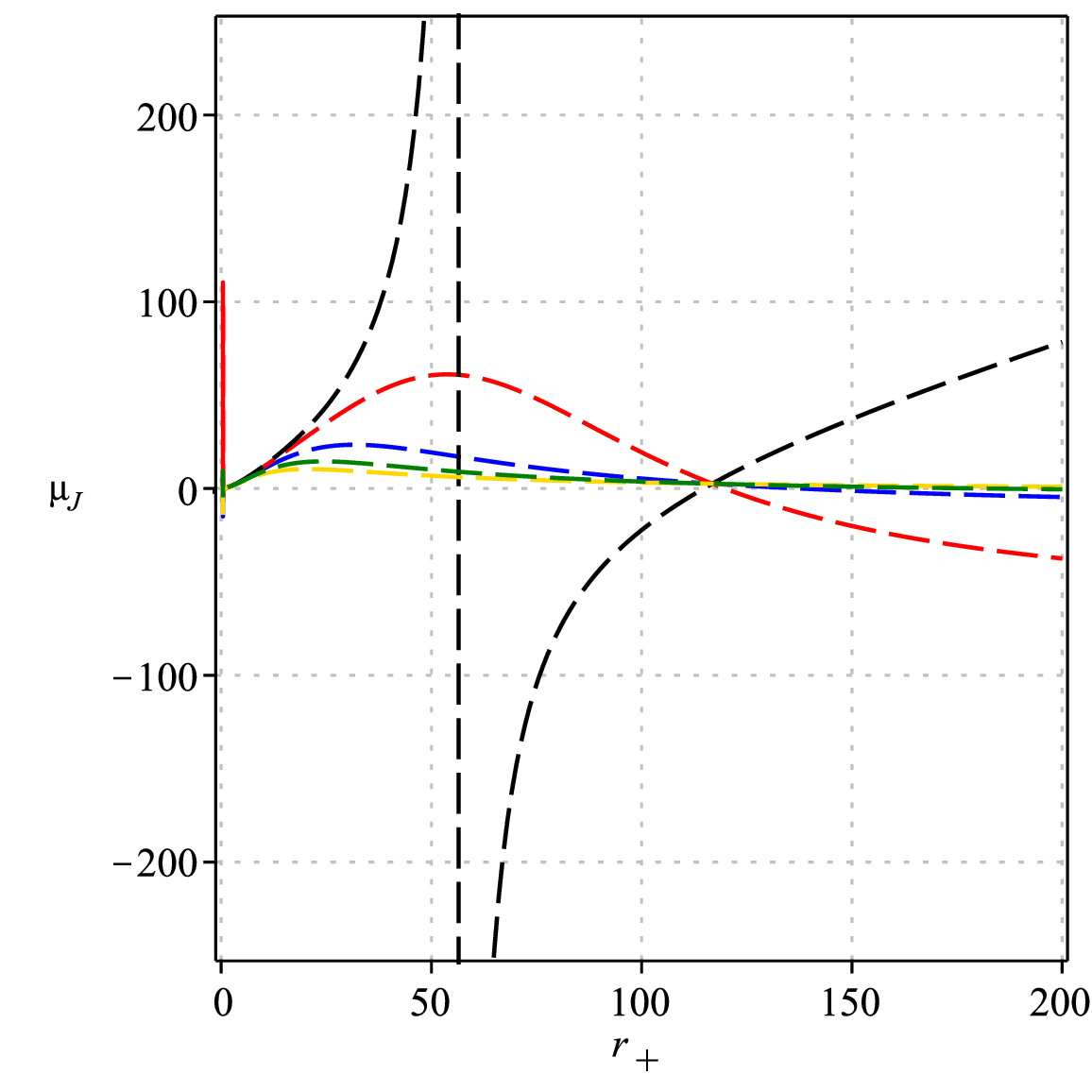}}\hfill
\subfloat[ $c_1=-1$ \& $c_2=1$]{\includegraphics[width=.5\textwidth]{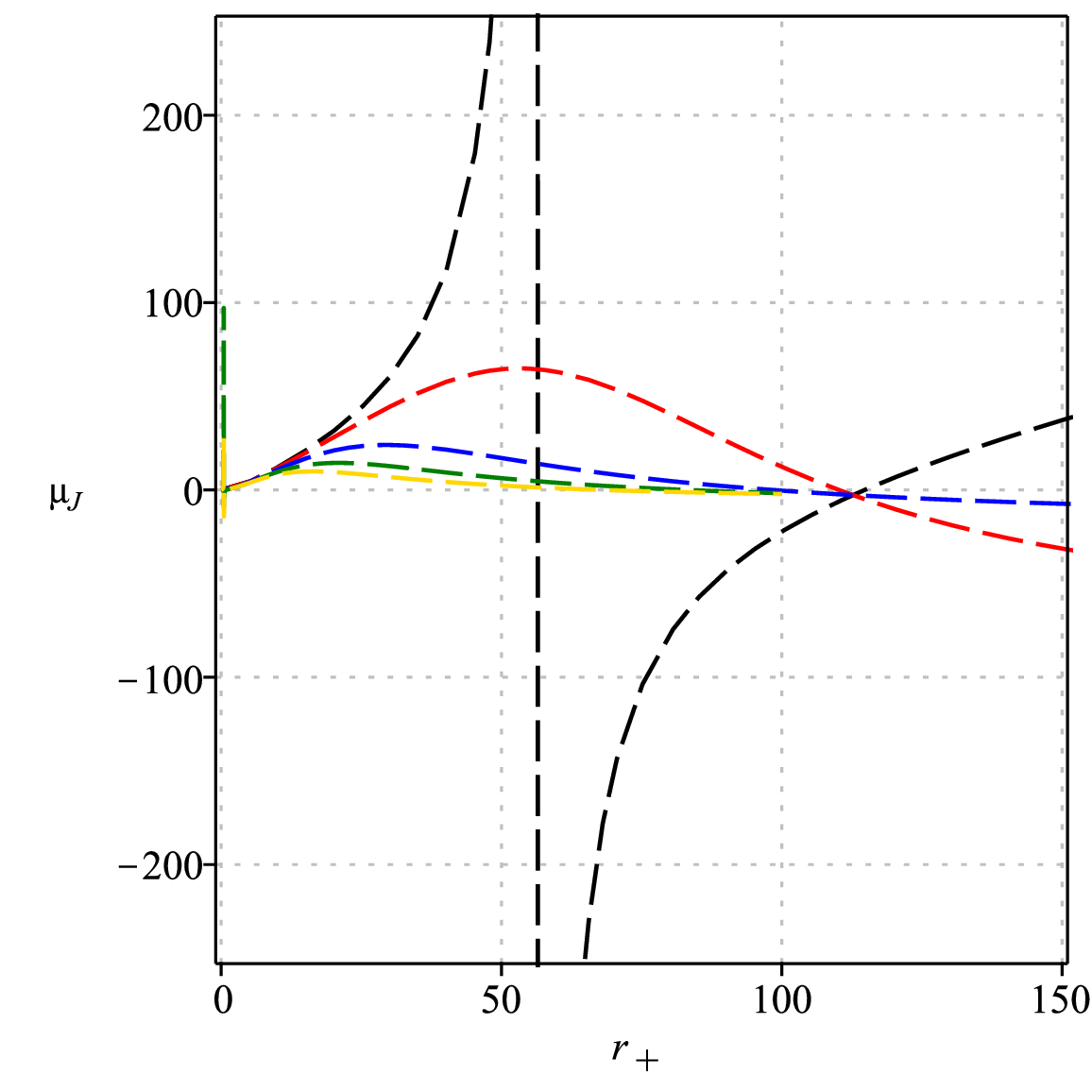}}\hfill
\subfloat[ $c_1=1$ \& $c_2=-1$]{\includegraphics[width=.5\textwidth]{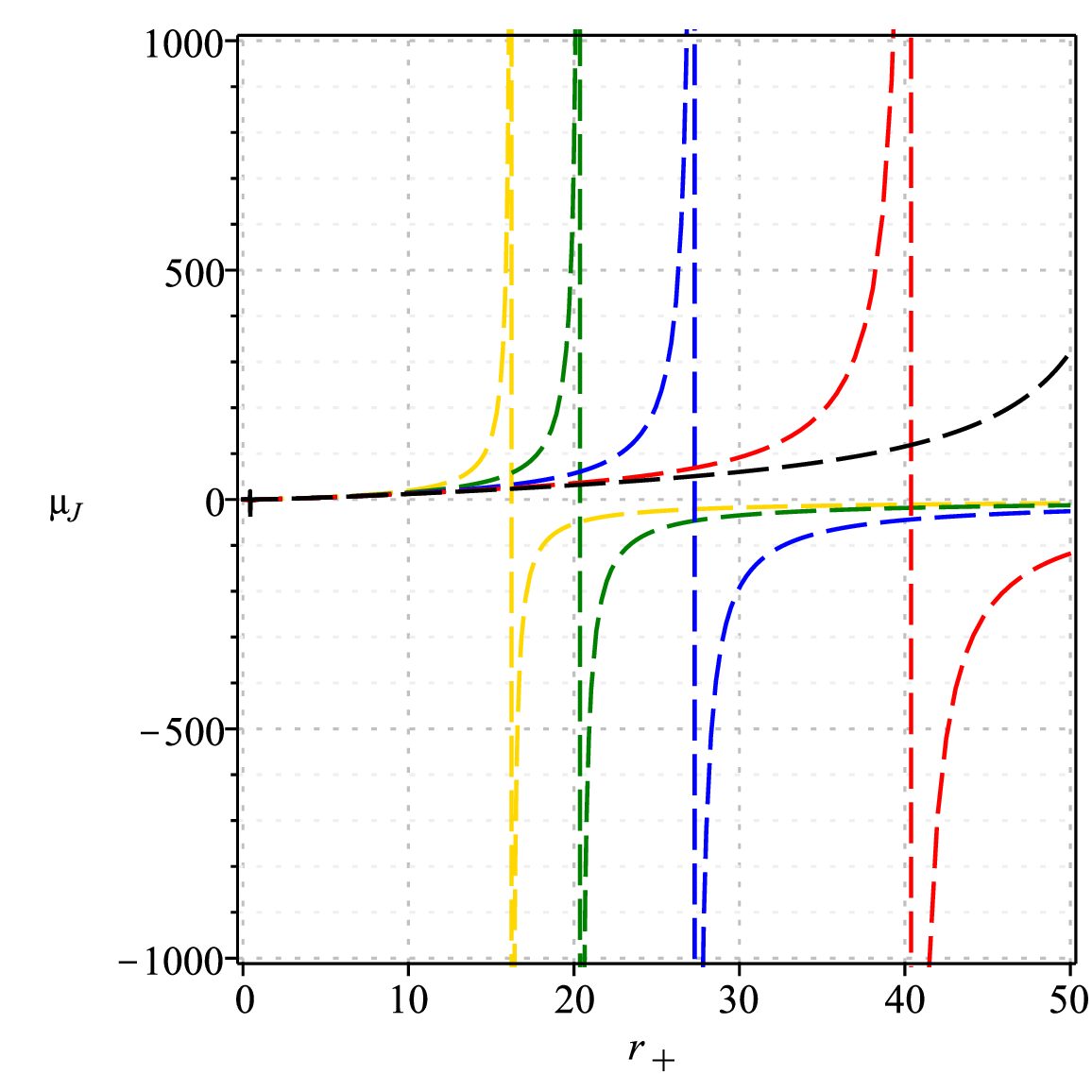}}\hfill
\subfloat[$c_1=1$ \& $c_2=1$]{\includegraphics[width=.5\textwidth]{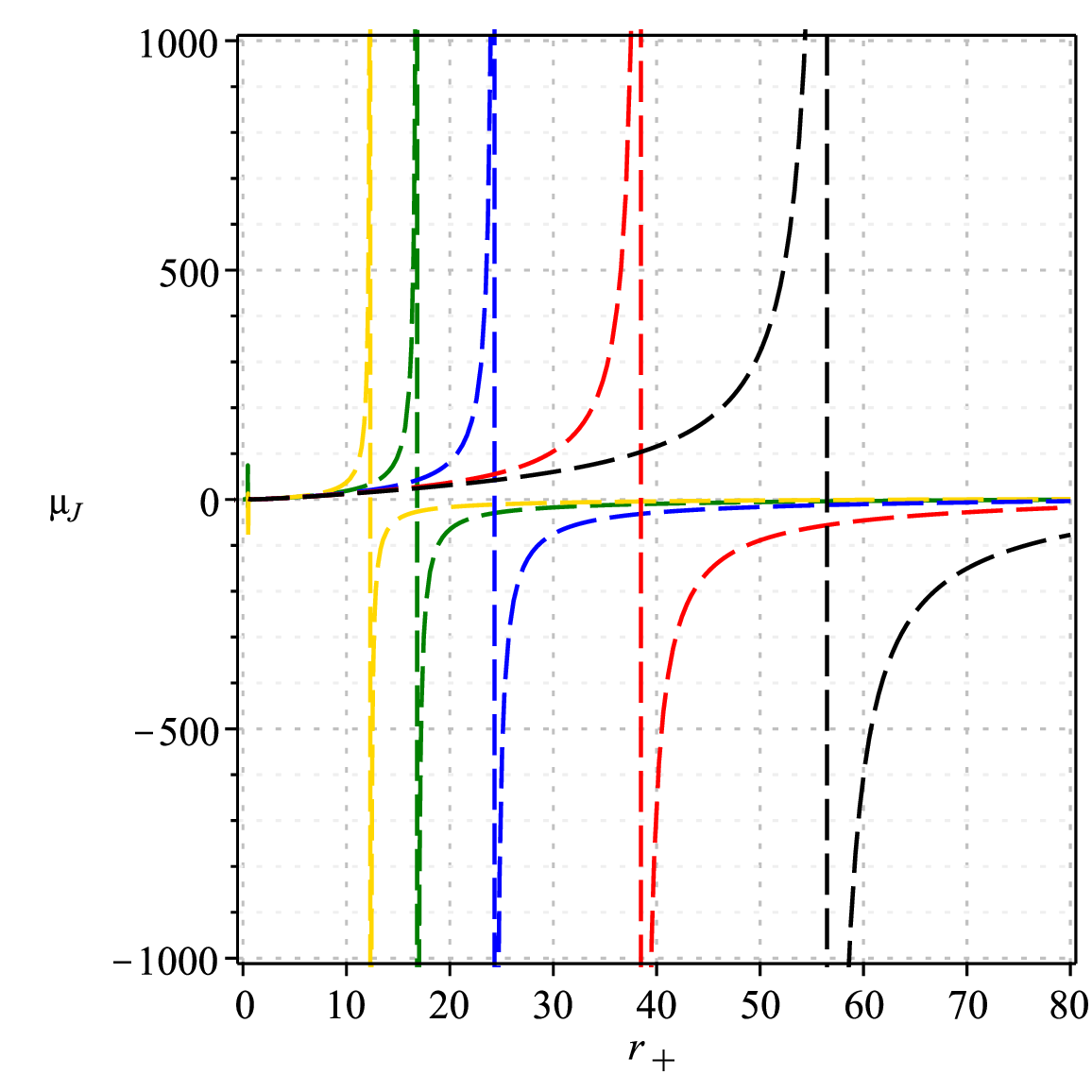}}\hfill
\centering
\begin{subfigure}[b]{0.2\textwidth}
\centering
\includegraphics[width=\textwidth]{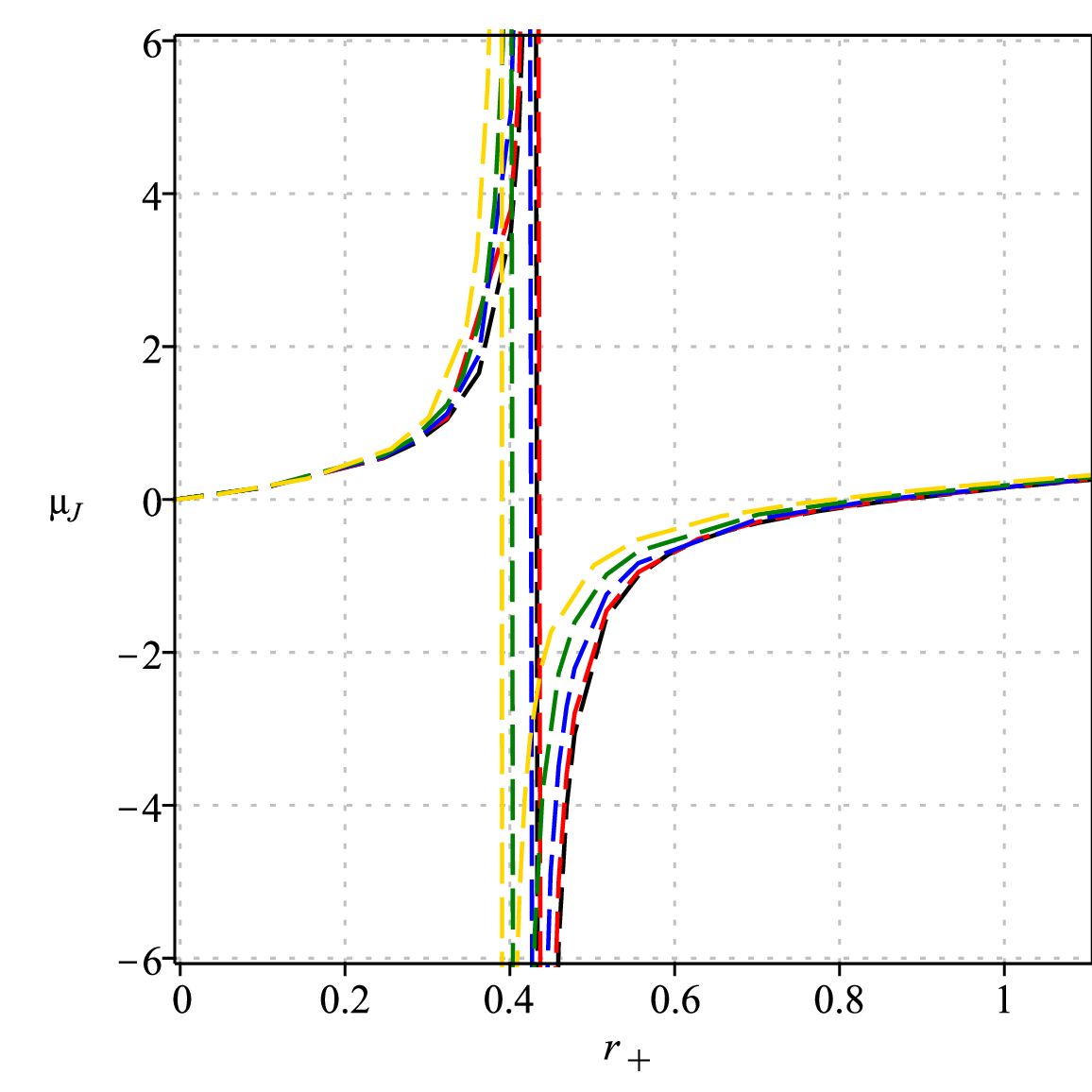}
\caption{Small scale $c_1=-1$ \& $c_2=-1$}
\label{fig:56e}
\end{subfigure}
\hfill
\begin{subfigure}[b]{0.2\textwidth}
\centering
\includegraphics[width=\textwidth]{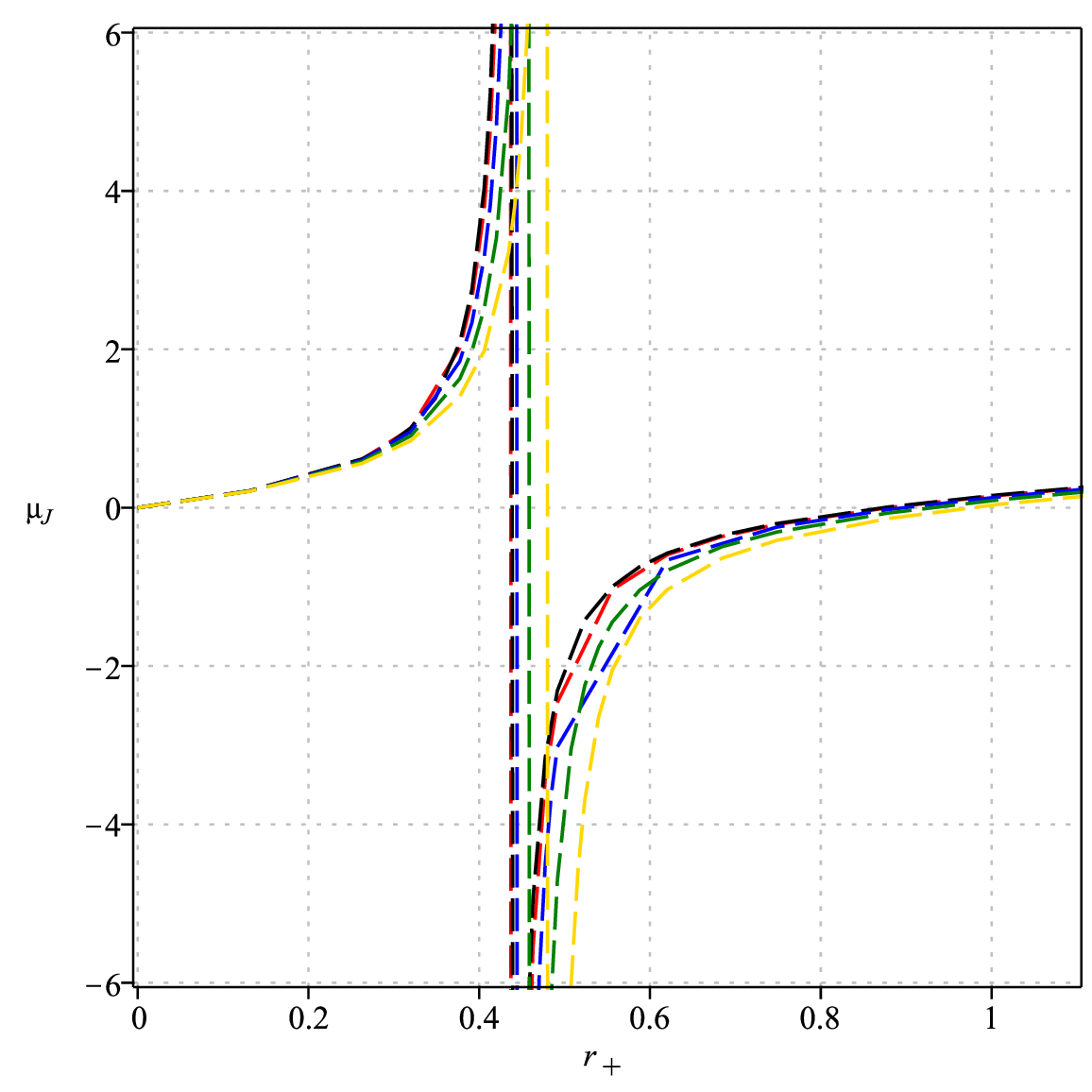}
\caption{Small scale $c_1=-1$ \& $c_2=1$}
\label{fig:56f}
\end{subfigure}
\hfill
\begin{subfigure}[b]{0.2\textwidth}
\centering
\includegraphics[width=\textwidth]{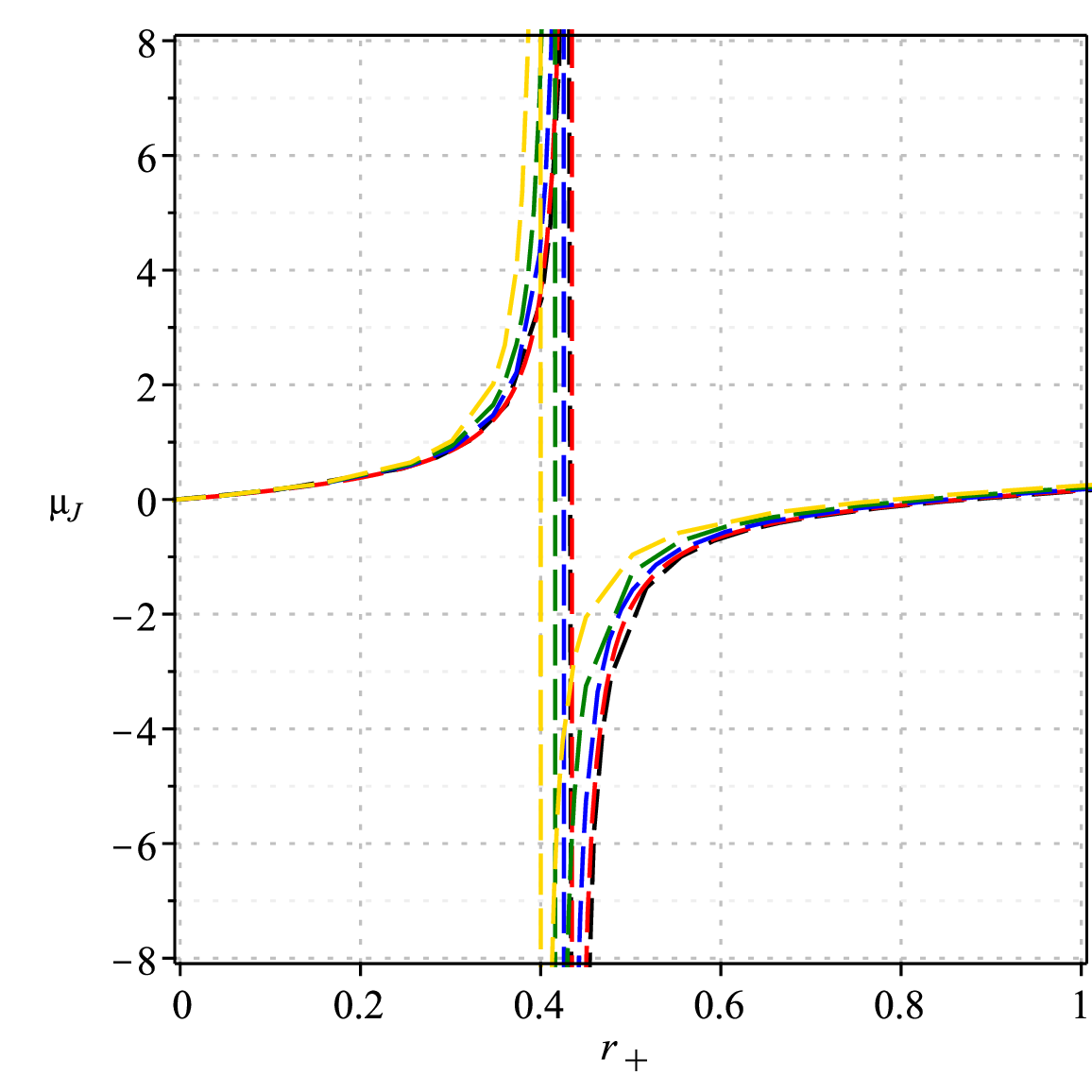}
\caption{Small scale $c_1=1$ \& $c_2=-1$}
\label{fig:56g}
\end{subfigure}
\hfill
\begin{subfigure}[b]{0.2\textwidth}
\centering
\includegraphics[width=\textwidth]{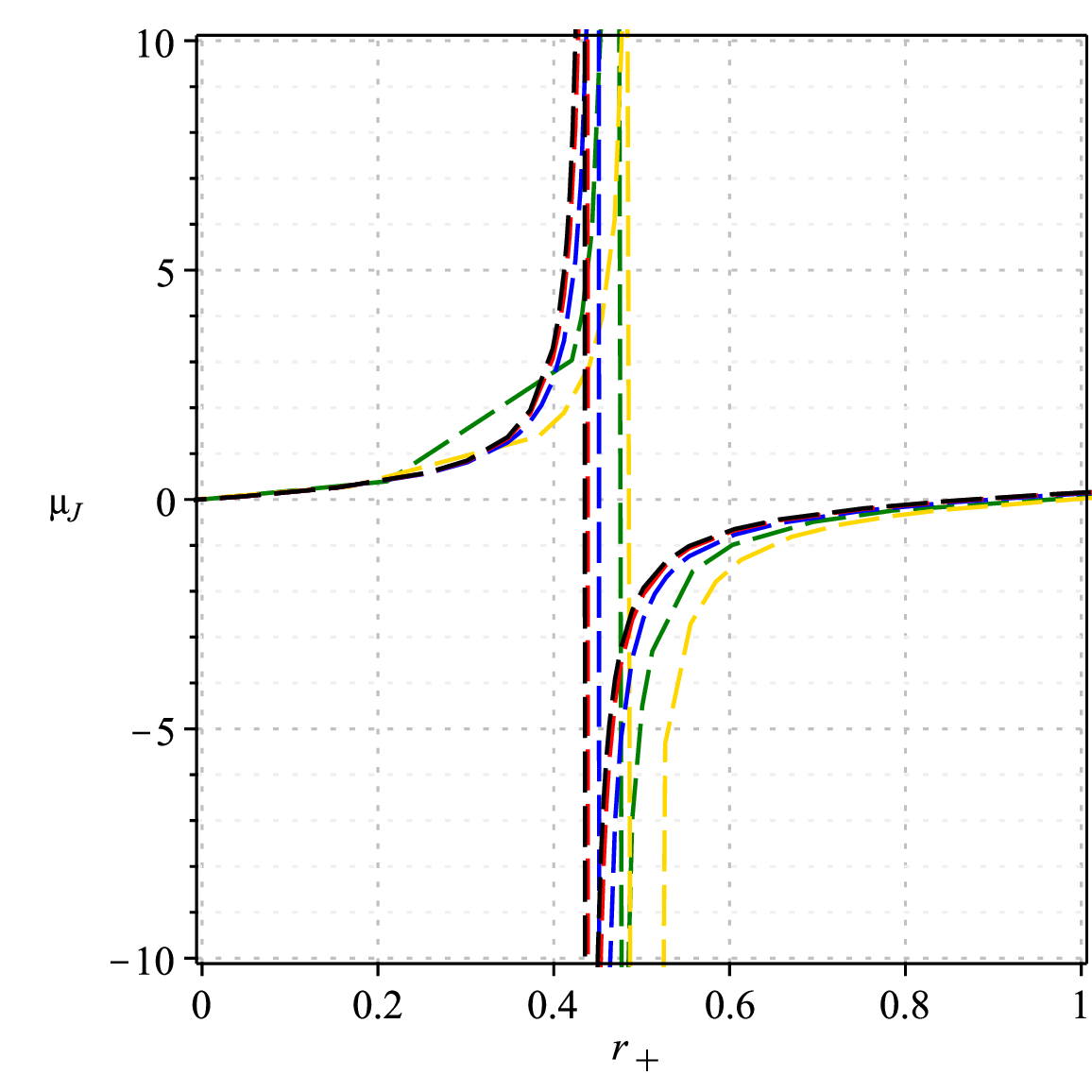}
\caption{Small scale $c_1=1$ \& $c_2=1$}
\label{fig:56h}
\end{subfigure}
\caption{Black dash line denotes $m=0$, red dash line denotes $m=0.2$, blue dash line denoted $m=0.4$, green dash line denoted $m=0.6$ and gold dash line denoted $m=0.8$ with $Q_m=10$, $\beta=0.5$, $M=20$ and $c=1$.}\label{fig:56}
\end{figure}

The behaviour of Joule--Thomson coefficients for different values of constant 
$c_{1,2}$ is shown in Figs. \ref{fig:56}(a) - \ref{fig:56}(h), where Figs. 
\ref{fig:56}(a) - \ref{fig:56}(d) represent the large scale behaviour and 
Figs. \ref{fig:56}(e) - \ref{fig:56}(f) represent the small scale behaviour. 
The corresponding Hawking temperature is shown in Figs. 
\ref{fig:57}(a) - \ref{fig:57}(h), where Figs. \ref{fig:57}(a) - \ref{fig:57}(d) 
represent the large scale behaviour and Figs. \ref{fig:57}(e) - \ref{fig:57}(f) 
represent the small scale behaviour.

In Fig. \ref{fig:56}(a), Joule--Thomson coefficient is depicted for $c_{1}=-1$ 
\& $c_{2}=-1$ (small scale Fig. \ref{fig:56}e). At small scale, $\mu_{J}$ undergoes 
discontinuity for each value of graviton mass and at the singular point Hawking 
temperature goes to zero, which is shown in Fig. \ref{fig:57}(e). At large scale, 
only $m=0$ attains a singular point. For $m=0$, between two singular points $\mu_{J}$ 
attains a zero which is known as inverse phenomenon, where $\mu_{J}$ changes its sign 
from negative to positive. When $m \neq 0$, at large scale $\mu_{J}$ is an continuous 
function of $r_{+}$. Also, at large scale $\mu_{J}$ undergoes an inverse phenomenon for 
$m \neq 0$, where $\mu_{J}$ changes its sign from positive to negative. At large scale, 
an inverse phenomenon occurs for $m=0$, where $\mu_{J}$ changes 
its sign from negative to positive. A similar kind of behaviour is shown in 
Fig. \ref{fig:56}(b).

The Joule--Thomson coefficient for $c_{1}=1$ \& $c_{2}=-1$ is shown in Fig. 
\ref{fig:56}(c) (small scale Fig. \ref{fig:56}g). At small scale, $\mu_{J}$ is 
singular for each value of graviton mass and the corresponding Hawking temperature 
is zero at the singular point, which is shown in Fig. \ref{fig:57}(g). After 
crossing the singular point at small scale $\mu_{J}$ attains zero, where the sign 
of $\mu_{J}$ changes from negative to positive. At large scale $\mu_{J}$ once again 
singular, i.e., a second phase transition occurs for Joule--Thomson coefficients and 
once again Hawking temperature goes to zero, which is shown in Fig. \ref{fig:57}(c). 
After the second singular point, once again $\mu_{J}$ attains zero, where the sign of 
$\mu_{J}$ changes from negative to positive. A similar kind of behaviour is shown in 
Fig. \ref{fig:56}(d).

\begin{figure}[H]
\centering
\begin{subfigure}[b]{0.3\textwidth}
\centering
\includegraphics[width=\textwidth]{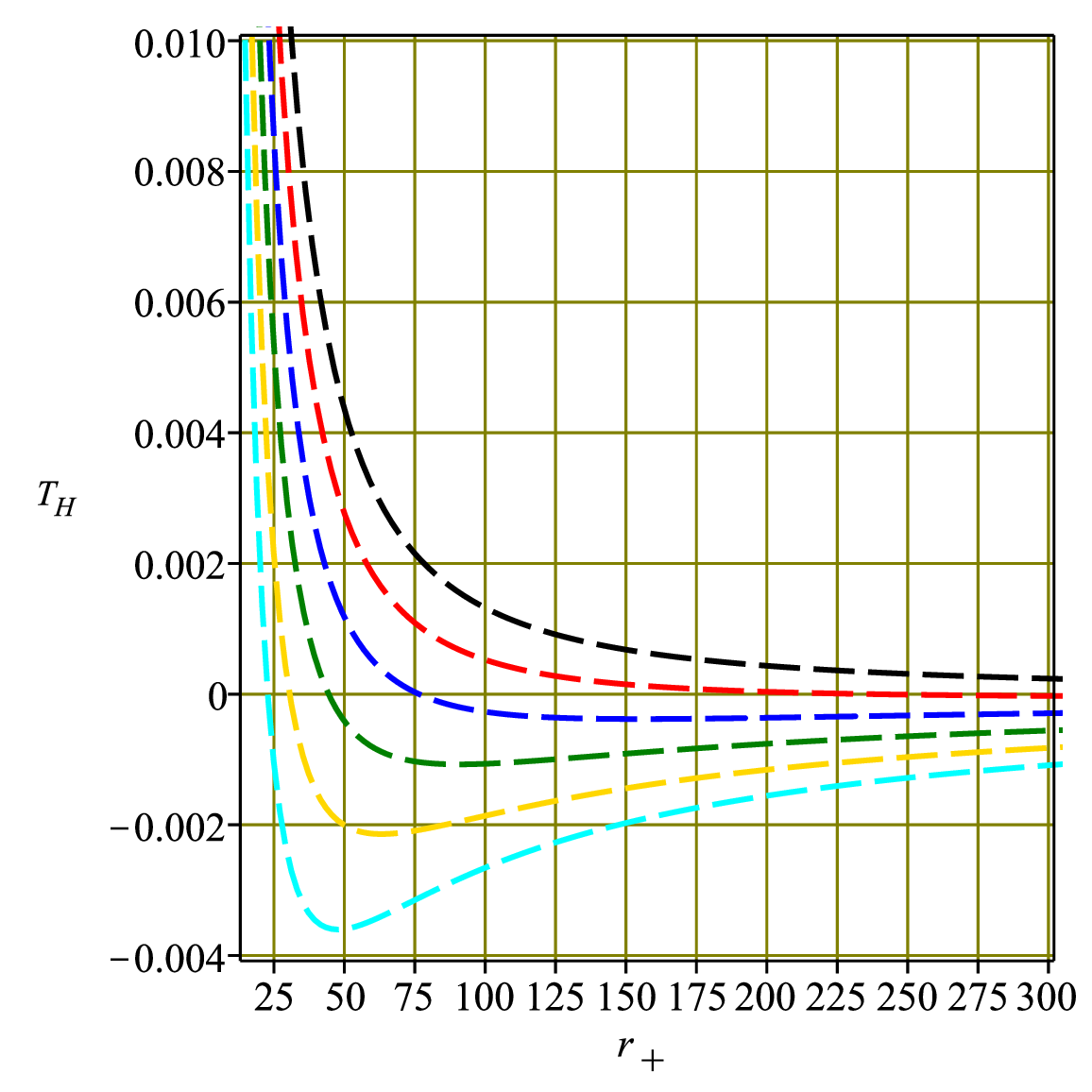}
\caption{Large scale with $c_1=0$}
\label{fig:58a}
\end{subfigure}
\begin{subfigure}[b]{0.3\textwidth}
\centering
\includegraphics[width=\textwidth]{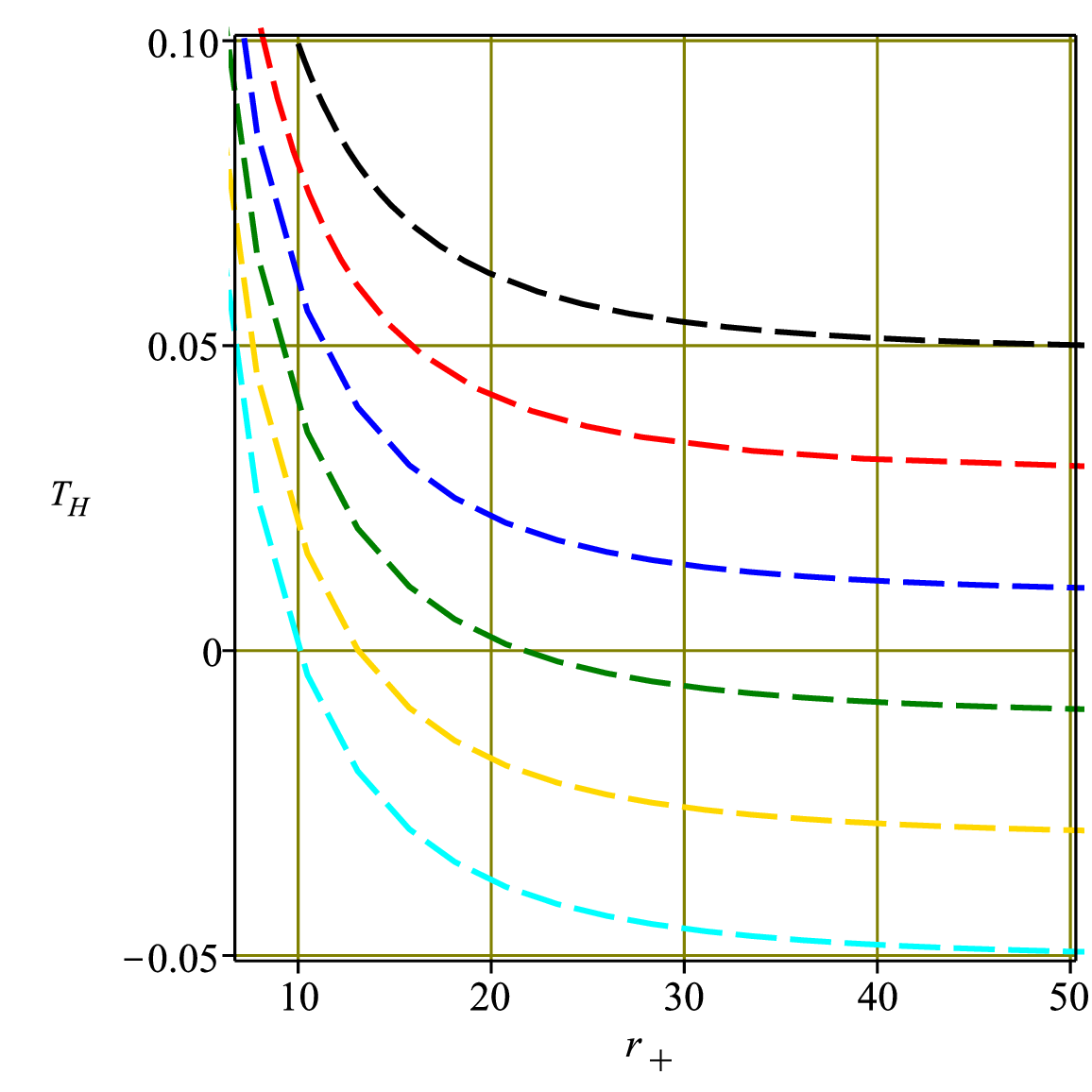}
\caption{Large scale with $c_2=0$}
\label{fig:58b}
\end{subfigure}
\linebreak
\begin{subfigure}[b]{0.3\textwidth}
\centering
\includegraphics[width=\textwidth]{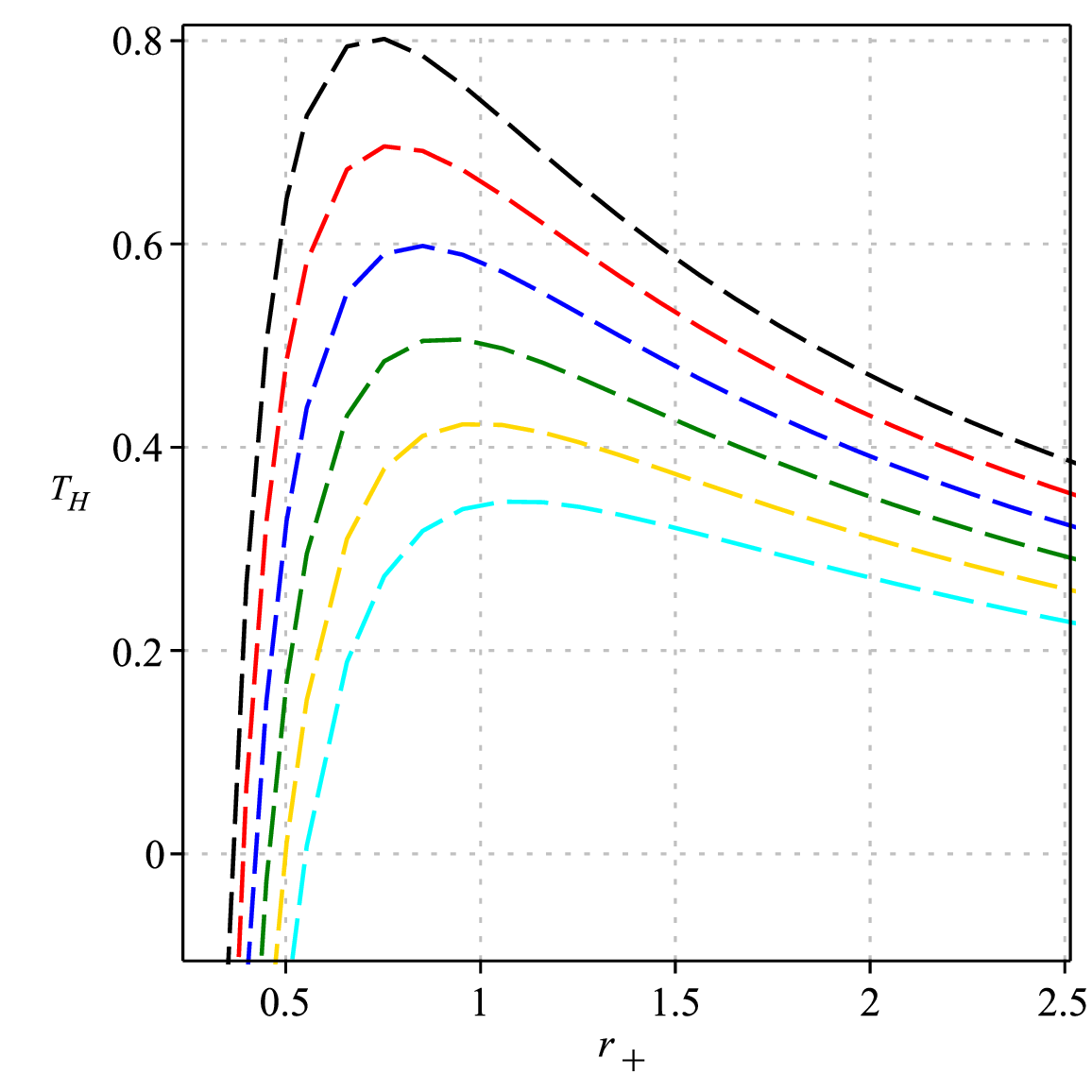}
\caption{Small scale with $c_1=0$}
\label{fig:58c}
\end{subfigure}
\begin{subfigure}[b]{0.3\textwidth}
\centering
\includegraphics[width=\textwidth]{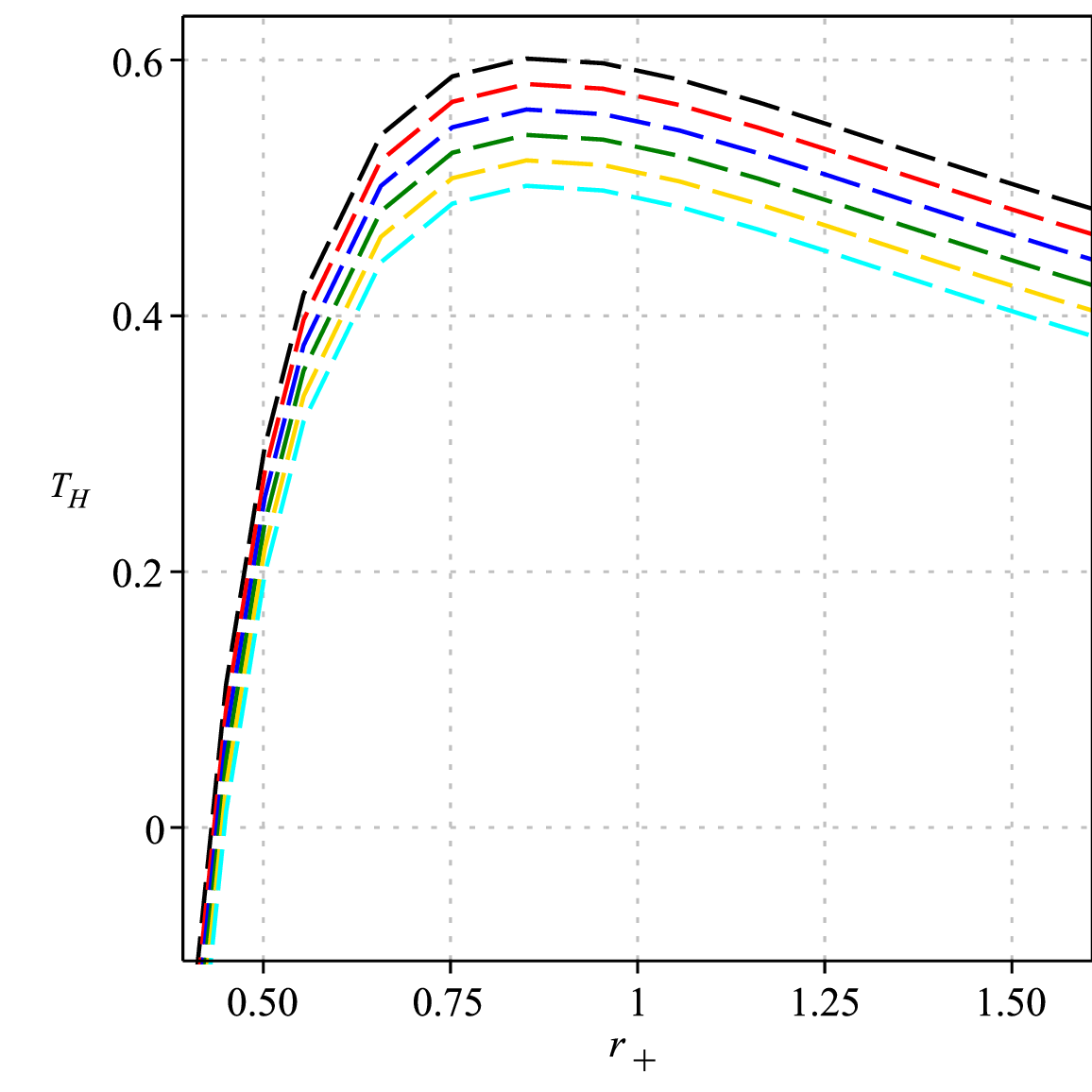}
\caption{Small scale with $c_2=0$}
\label{fig:58d}
\end{subfigure}
\caption{Left panel : cyan dash line denotes $c_2=5$, gold dash line denotes 
$c_2=3$, green dash line denotes $c_2=1$, blue dash line denotes $c_2=-1$, red 
dash line denotes $c_2=-3$ and black dash line denotes $c_2=-5$. Right panel : cyan dash line denotes $c_1=5$, gold dash line denotes $c_1=3$, green dash line denotes $c_1=1$, blue dash line denotes $c_1=-1$, red dash line denotes $c_1=-3$ and black dash line denotes $c_1=-5$. $M=20$, $Q_m=10$, $\beta=0.5$, $c=1$ and $m=0.5$.}\label{fig:58}
\end{figure}

\begin{figure}[H]
\centering
\subfloat[$c_1=0$]{\includegraphics[width=.5\textwidth]{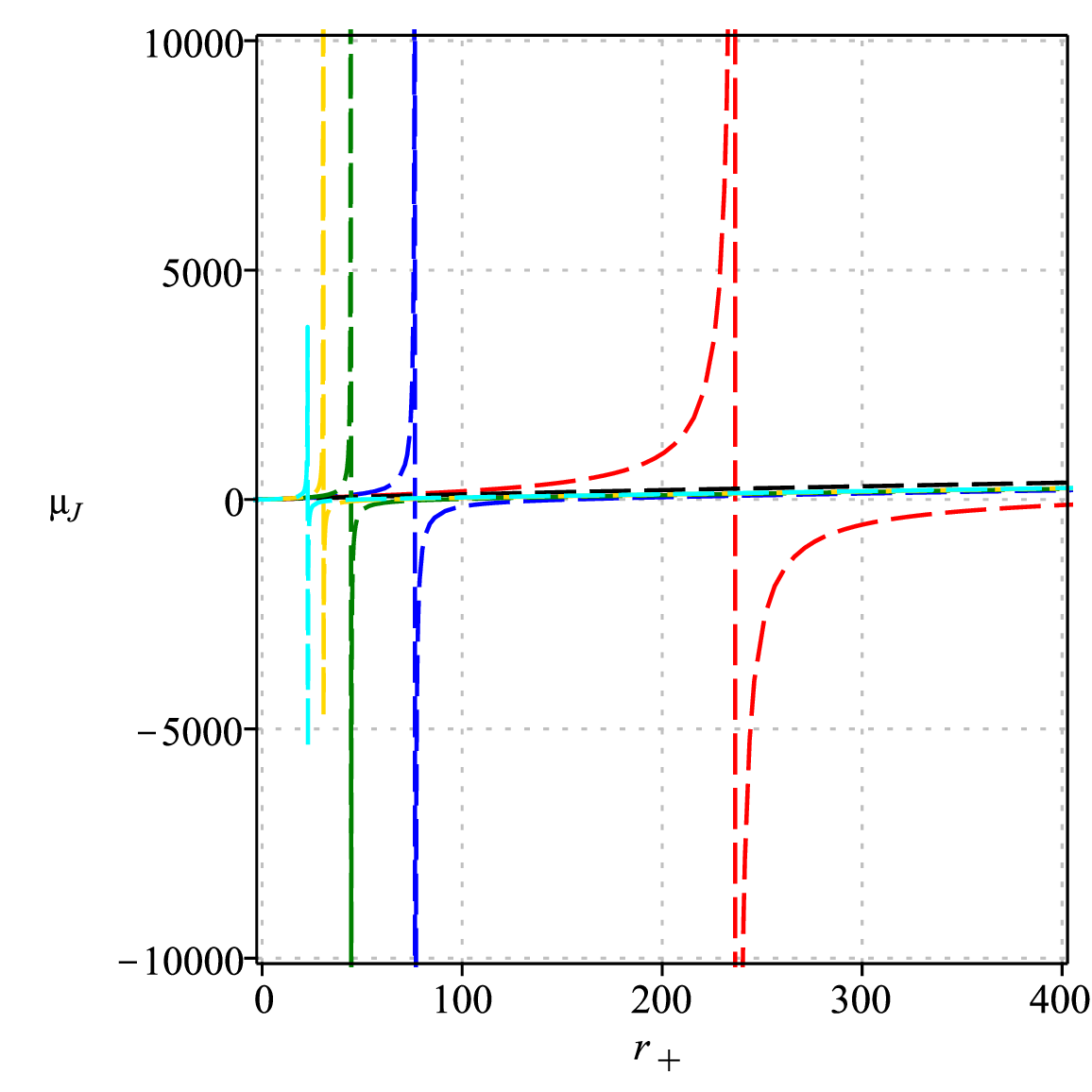}}\hfill
\subfloat[$c_2=0$]{\includegraphics[width=.5\textwidth]{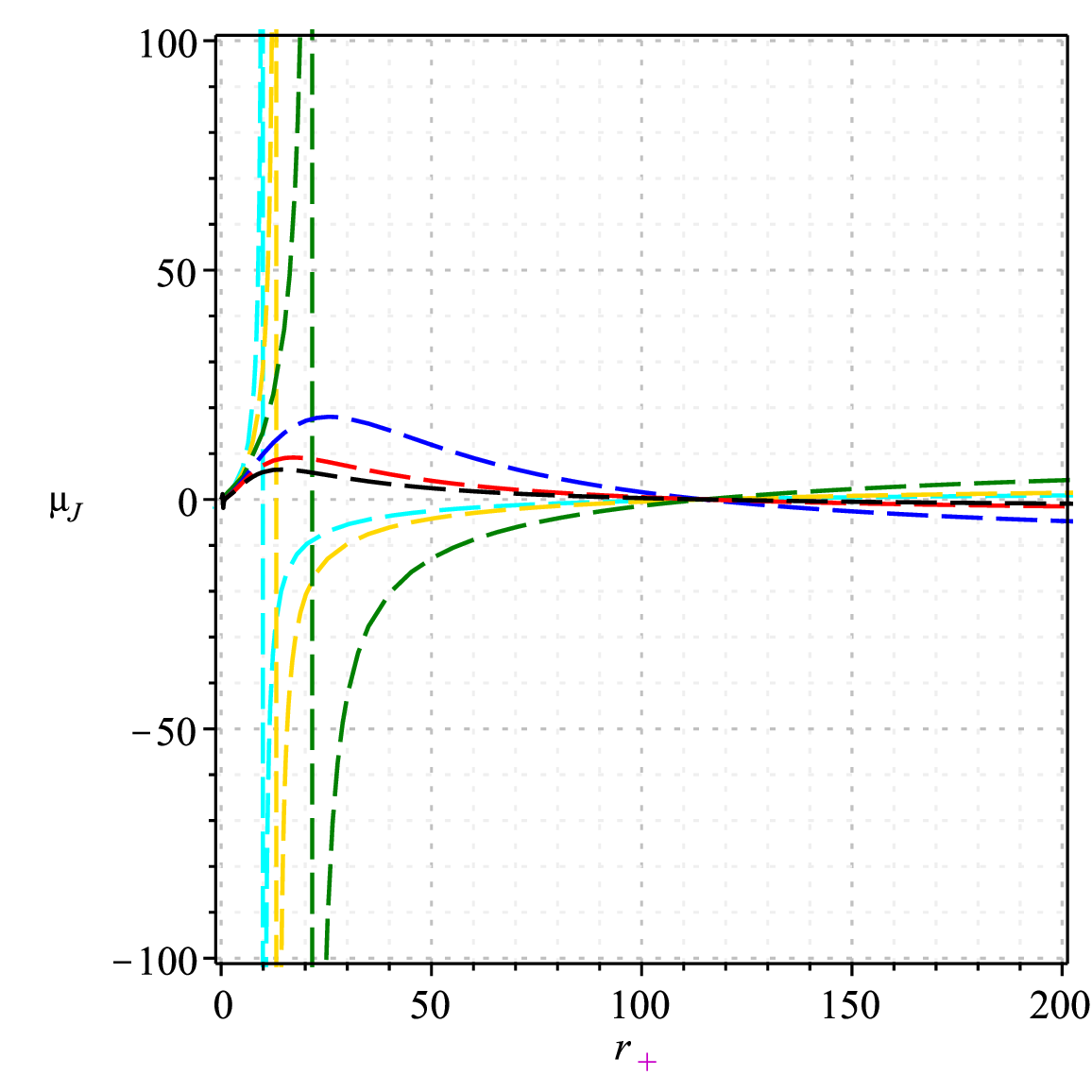}}\hfill
\subfloat[Small scale behaviour of Fig. \ref{fig:59}(a) with $c_1=0$]{\includegraphics[width=.5\textwidth]{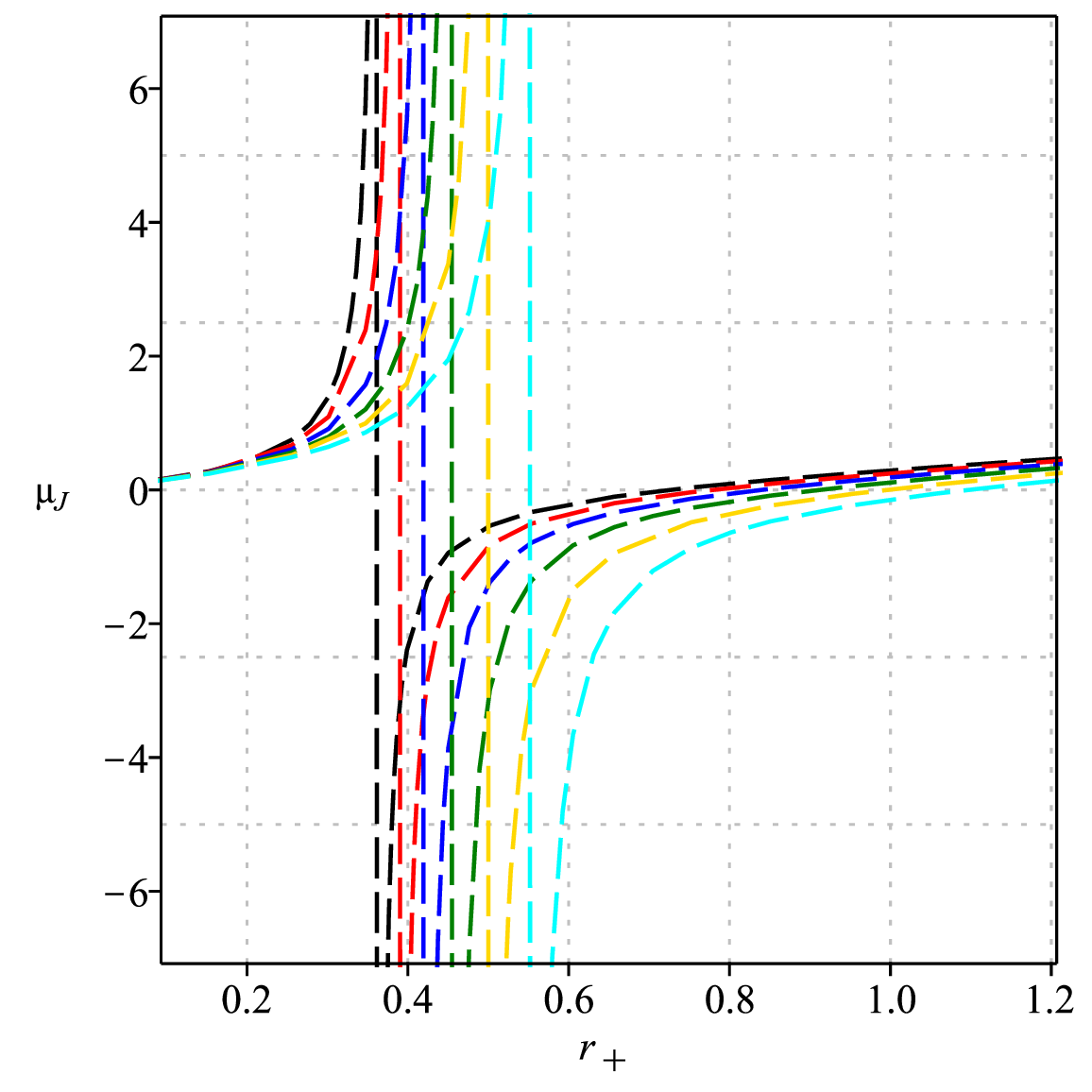}}\hfill
\subfloat[Small scale behaviour of Fig. \ref{fig:59}(b) with $c_2=0$]{\includegraphics[width=.5\textwidth]{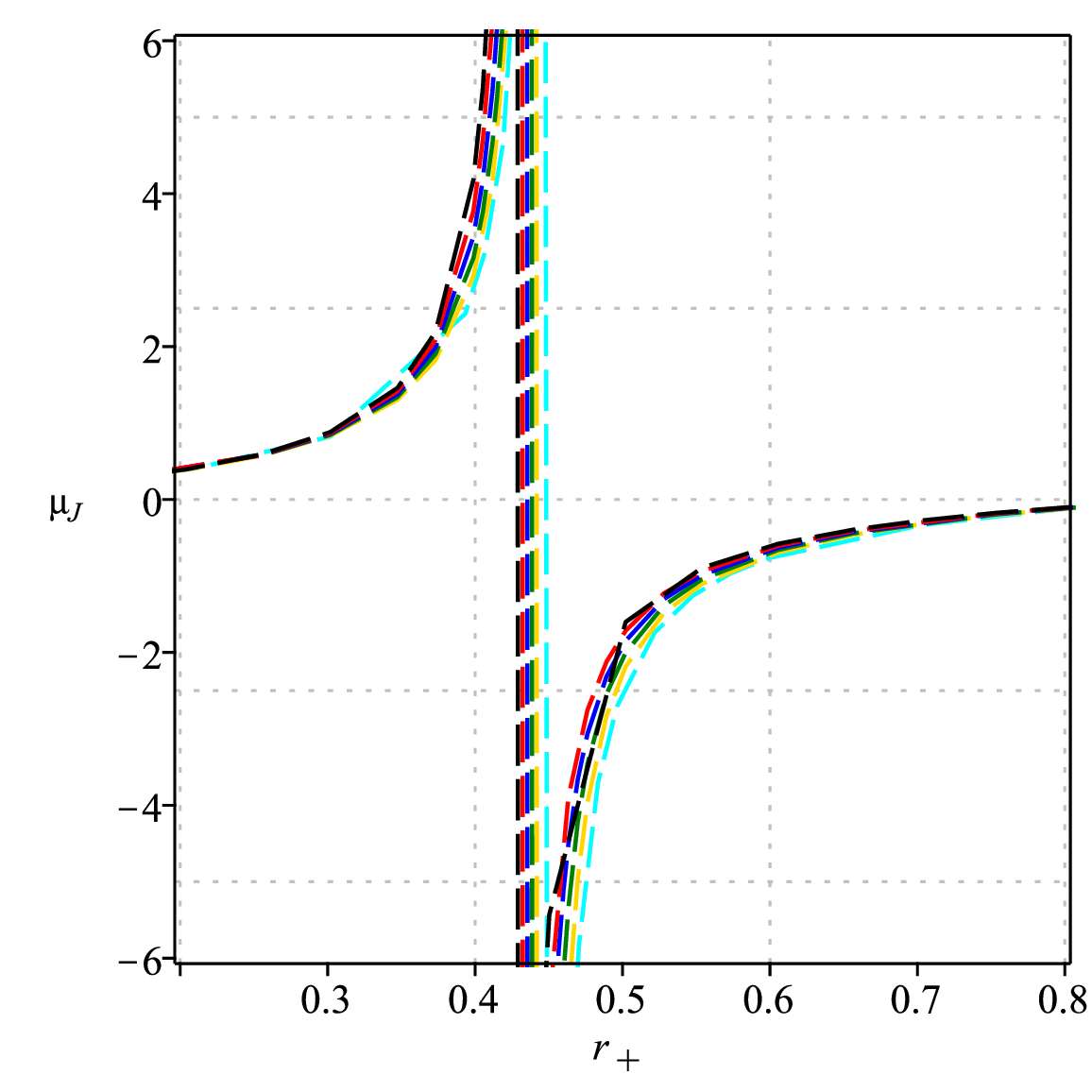}}\hfill
\caption{Left panel : cyan dash line denotes $c_2=5$, gold dash line denotes 
$c_2=3$, green dash line denotes $c_2=1$, blue dash line denotes $c_2=-1$, 
red dash line denotes $c_2=-3$ and black dash line denotes $c_2=-5$. Right panel : cyan dash line denotes $c_1=5$, gold dash line denotes $c_1=3$, green dash line denotes $c_1=1$, blue dash line denotes $c_1=-1$, red dash line denotes $c_1=-3$ and black dash line denotes $c_1=-5$. $M=20$, $Q_m=10$, $\beta=0.5$, $c=1$ and $m=0.5$.}\label{fig:59}
\end{figure}

The Joule--Thomson coefficients for different values of $c_{2}$ $(c_{1}=0)$ \& $c_{1}$ 
$(c_{2}=0)$ are shown in Fig. \ref{fig:59} and corresponding temperature is depicted 
in Fig. \ref{fig:58}. In Fig. \ref{fig:59}(a) (large scale behaviour) and 
\ref{fig:59}(c) (small scale behaviour) we plot $\mu_{J}$ for different values 
of parameter $c_{2}$ ($c_{1}=0$). On small scale, $\mu_{J}$ (Fig. \ref{fig:59}c) is 
singular for each value of constant $c_{2}$ and Hawking temperature 
(Fig. \ref{fig:58}c) goes to zero at the singular point. On large scale, 
$\mu_{J}$ (Fig. \ref{fig:59}a) is singular for each value of constant $c_{2}$ except $c_{2}=-5$ and Hawking temperature (Fig. \ref{fig:58}a) goes to zero at the singular point. Between two singular points, an inverse phenomenon occurs where $\mu_{J}=0$.

In Fig. \ref{fig:59}(b) (large scale behaviour) and \ref{fig:59}(d) (small scale behaviour) we plot $\mu_{J}$ for different values of parameter $c_{1}$ ($c_{2}=0$). On small scale, $\mu_{J}$ (Fig. \ref{fig:59}d) is singular for each value of constant $c_{1}$ and Hawking temperature (Fig. \ref{fig:58}d) goes to zero at the singular point. on large scale, $\mu_{J}$ (Fig. \ref{fig:59}b) is singular for $c_{1}=5,3,1$ only and Hawking temperature (Fig. \ref{fig:58}b) goes to zero at the singular point.

\section{Conclusions}\label{sec:6}
In this paper, we obtained magnetically charged $AdS$ black hole 
solutions in EGB massive gravity coupled to NED. The metric function 
of the black hole is depicted. In the limit, $m \to 0$ and 
$\alpha \to 0$ we obtained the EGB massless gravity black hole 
and massive Einstein gravity black hole. The thermodynamics of 
magnetically charged $AdS$ black holes in extended phase space has 
been studied, where the cosmological constant played the role of 
a thermodynamic pressure. We defined thermodynamic quantity 
$\mathcal{A}$, $\mathcal{C}_{1,2}$, $\mathcal{B}$ conjugate to 
EGB parameter $\alpha$, constant $c_{1,2}$, NED parameter $\beta$ 
and magnetic potential $\Phi_m$ conjugates to magnetic charge $Q_m$. 
We verify first law of black hole thermodynamics and the generalized 
Smarr formula in extended phase space. The local stability of the 
black holes is studied through specific heat. 

The Van der Waals-like 
phase transition of the black holes is analysed. We numerically 
estimated the critical points for EGB/Einstein massive gravity and 
EGB massless gravity black hole. The Gibbs free energy $Vs$ temperature 
showed swallow tail-like behaviour, which indicates that black hole 
undergoes first-order phase transitions. The $P-v$ diagram showed a 
liquid-gas-like phase transition and one inflection point is present 
at $T_{H}=T_{c}$. For a range of $\beta$ values $\beta \in (\beta_1, \beta_2)$
the black hole in massless Einstein gravity coupled to NED undergoes a 
\textbf{LBH}--\textbf{IBH}--\textbf{LBH} phase transitions with two
positive critical pressures. The addition of mass to Einstein's gravity
unchanged the phase transitions of the black hole but it lowers the critical points
$(P_{t},T_{t})$ and $(P_{z},T_{z})$. In the case of EGB black hole, we introduced a 
small value to the GB coupling parameter $(\alpha=0.0001)$, and observed that real tricritical
points occur for a range of $\beta$ values $\beta \in (\beta_1, \beta_2)$. In this 
range a \textbf{SBH}--\textbf{IBH}--\textbf{LBH} phase transitions occur. The 
addition of mass term to the EGB gravity unchanged the phase structure of the 
black hole but it lowers the critical points. 

Finally, we studied the Joule–-Thomson adiabatic 
expansion of the EGB/Einstein massive gravity black hole and massless 
EGB gravity black hole. We plotted the isenthalpic $P - T$ diagrams 
and inversion temperature curve $P_i - T_i$ for each black hole. The 
inversion temperature curve separates the isenthalpic plots into two 
branches corresponding to cooling $(\mu_J > 0)$ region and heating 
$(\mu_J < 0)$ region of the black holes. We numerically estimated 
the minimum inverse temperature and event horizon radius. 
Furthermore, we analysed the effects of EGB parameter $\alpha$, 
massive gravity and NED parameters on the Joule–Thomson coefficients 
as a function of horizon radius.

\begin{appendices}\label{appendicex}
    \setcounter{equation}{0}
    \renewcommand\theequation{\thesection\arabic{equation}} 
  
The NED Lagrangian for Ref. \cite{kruglov2022nonlinearly} is 
\begin{equation}\label{eq:a1}
    \mathcal{L}(\mathcal{F}) = - \frac{\mathcal{F}}{1+\sqrt{2 \beta \mathcal{F}}}
\end{equation}

For the above Lagrangian equations for critical parameters are \cite{kruglov2022nonlinearly}

\begin{align}\label{eq:a2}
\Bigl( v_{c}^{2} +4Q_m \sqrt{\beta} \Bigl)^3 - 
8 Q_m^2 \Bigl( 3v_c^4 +6Q_m \sqrt{\beta}v_c^2+8 \beta Q_m^2 \Bigl) =0, \\  
T_{c} = \frac{1}{ \pi  v_{c}  }  - 
\frac{8 Q_m^2 \bigl( v_c^2 +2Q_m \sqrt{\beta}\bigl)}{\pi v_c (4 Q_m \sqrt{\beta}+{v_{c}^{2}})^{2}}, \\
P_{c} = \frac{1}{2 \pi  v_{c}^2  }  
- \frac{2 Q_m^2 \bigl( 3v_c^2 +4Q_m \sqrt{\beta}\bigl)}{\pi v_c^2 ({4} Q_m \sqrt{\beta}+{v_{c}^{2}})^{2}}. 
\end{align}

Putting $x=v_c^2+4Q_m \sqrt{\beta}$ into the first equation of \eqref{eq:a2} 
 we obtain
\begin{equation}\label{eq:a3}
    x^3-24Q_m^2x^2+144Q^2k^2x-256Q_m^2k^4=0.
\end{equation}

In order to satisfy $v_c \geq 0$, we must have
\begin{equation}\label{eq:a4}
\lvert x \rvert \geq 4Q_m \sqrt{\beta}.
\end{equation}

Three real roots of equation \eqref{eq:a3} occur when 
the discriminant is
\begin{equation}\label{eq:a5}
\Delta= 442368 Q_{m}^{8} \beta  (\sqrt{\beta}-Q_{m} ) (5 Q_{m} -4 \sqrt{\beta}) < 0.
\end{equation}
From above condition $\Delta < 0$ we obtain
\begin{equation}\label{eq:a6}
    {Q_m}=\sqrt{\beta_0} < \sqrt{\beta} < \sqrt{\beta_2}= \frac{5Q_m}{4}.
\end{equation}
To find the solutions of equation \eqref{eq:a3}, 
we will use the Tschirnhaus transformation method.
 Putting $x=t+B$ into equation \eqref{eq:a3}

\begin{equation}\label{eq:a7} 
    t^3 +pt +q=0,
\end{equation}
where we set coefficients of $t^2$ equal to zero \& 
$B=8Q_m^2$. Finally the solutions of equations \eqref{eq:a3} is 
\begin{equation}\label{eq:a8}
    x_{j}=2 \sqrt{\frac{-p}{3}} \cos{\Biggr[\frac{1}{3} \arccos{\biggl(\frac{3q}{2p}\sqrt{\frac{-3}{p} } \biggl)}-\frac{2\pi j}{3} \Biggr]},
\end{equation}
where $j=0$, $1$ \& $2$. The condition in the equation
\eqref{eq:a6} was satisfied for $x_0$ and $x_1$ only, $x_2$ does not
satisfy condition \eqref{eq:a6}. Therefore we have two physical critical points.
The constant $p$ and $q$ are given by
\begin{align}\label{eq:a9}
p &=  144 \sqrt{\beta} Q_{m}^{3}-3 B^{2}, \\ 
q &=  -256 \beta  Q_{m}^{4}+144 B \sqrt{\beta} Q_{m}^{3}-2 B^{3}.
\end{align}
$\beta < \beta_{0}$ admits only one real critical point. For $\beta > \beta_2$ no critical points occur. Finally, the critical radius $v_c$ can be written as
\begin{equation}\label{eq:a10}
v_c=\sqrt{x-4Q_m \sqrt{\beta}} \text{,}  \:    x=
x_0 \; \& \; x_1, \: \text{where} \;  \beta_0  < \beta < \beta_2.
\end{equation}

For two critical pressures to be positive, we must have
\begin{equation}\label{eq:a13}
 \sqrt{\beta} > \sqrt{\beta_1} = \frac{9Q_m}{8}.
\end{equation}
Excluding the range of $\beta$ from $\beta_0$ to $\beta_1$, we can say that 
two critical points with positive critical pressures occur 
for $\beta_1 < \beta <  \beta_2$. For $\beta_0 < \beta <  \beta_1$ two critical 
points occur but with one negative critical pressure. Our table \ref{ta4} is 
consistent with table 1 of Ref. \cite{kruglov2022nonlinearly}.

\begin{table}[H]
    \begin{center}
    \begin{tabular}{ |c|c|c|c|c| } 
    \hline
    Case & CP & CP1 & CP2 \\
    \hline
    \multirow{3}{10em}{$\beta_1 < \beta=1.3 < \beta_2$} & $v_c$ & 1.2996 & 3.0603 \\ 
    & $T_c$ & 0.0458 & 0.0540 \\ 
    & $P_c$ & 0.0013 & 0.0055 \\ 
    \hline
    \multirow{3}{10em}{$\beta_0 < \beta=1.1 < \beta_1 $} & $v_c$ & 0.8319 & 3.3416 \\ 
    & $T_c$ &  0.0251 & 0.0524 \\ 
    & $P_c$ & -0.0115 & 0.0051 \\ 
    \hline
    \multirow{3}{10em}{$0.50=\beta < \beta_0  $} & $v_c$ & $--$ & 4.0206 \\ 
    & $T_c$ & $--$  & 0.0483 \\ 
    & $P_c$ & $--$  & 0.0042 \\ 
    \hline
    \end{tabular}
    \end{center}
    \caption{With $Q_m=1$.}
    \label{ta4}
    \end{table}

\noindent\hrulefill
\end{appendices}

\printbibliography

@article{glavan2020einstein,
  title={Einstein-Gauss-Bonnet gravity in four-dimensional spacetime},
  author={Glavan, Dra{\v{z}}en and Lin, Chunshan},
  journal={Physical review letters},
  volume={124},
  number={8},
  pages={081301},
  year={2020},
  publisher={APS}
}

@article{Fernandes:2020rpa,
    author = "Fernandes, Pedro G. S.",
    title = "{Charged black holes in AdS spaces in 4D Einstein Gauss-Bonnet gravity}",
    journal = "Phys. Lett. B",
    volume = "805",
    pages = "135468",
    year = "2020"
}

@article{Upadhyay:2022axg,
    author = "Upadhyay, Sudhaker and Singh, Dharm Veer",
    title = "{Black hole solution and thermal properties in 4$D$ $AdS$ Gauss\textendash{}Bonnet massive gravity}",
    journal = "Eur. Phys. J. Plus",
    volume = "137",
    number = "3",
    pages = "383",
    year = "2022"
}

@article{Kruglov:2021stm,
    author = "Kruglov, S. I.",
    title = "{Einstein \ensuremath{-} Gauss \ensuremath{-} Bonnet gravity with nonlinear electrodynamics}",
    journal = "Annals Phys.",
    volume = "428",
    pages = "168449",
    year = "2021"
}

@article{kruglov2017nonlinear,
  title={Nonlinear electrodynamics and magnetic black holes},
  author={Kruglov, S.I.},
  journal={Annalen der Physik},
  volume={529},
  number={8},
  pages={1700073},
  year={2017},
  publisher={Wiley Online Library}
}

@article{kruglov2022nonlinearly,
  title={Nonlinearly charged AdS black holes, extended phase space thermodynamics and Joule--Thomson expansion},
  author={Kruglov, S.I.},
  journal={Annals of Physics},
  volume={441},
  pages={168894},
  year={2022},
  publisher={Elsevier}
}

@article{Cai:2014znn,
    author = "Cai, Rong-Gen and Hu, Ya-Peng and Pan, Qi-Yuan and Zhang, Yun-Long",
    title = "{Thermodynamics of Black Holes in Massive Gravity}",
    journal = "Phys. Rev. D",
    volume = "91",
    number = "2",
    pages = "024032",
    year = "2015"
}

@article{Dolan:2010ha,
    author = "Dolan, Brian P.",
    title = "{The cosmological constant and the black hole equation of state}",
    journal = "Class. Quant. Grav.",
    volume = "28",
    pages = "125020",
    year = "2011"
}

@article{Kubiznak:2012wp,
    author = "Kubiznak, David and Mann, Robert B.",
    title = "{P-V criticality of charged AdS black holes}",
    journal = "JHEP",
    volume = "07",
    pages = "033",
    year = "2012"
}

@article{Gunasekaran:2012dq,
    author = "Gunasekaran, Sharmila and Mann, Robert B. and Kubiznak, David",
    title = "{Extended phase space thermodynamics for charged and rotating black holes and Born-Infeld vacuum polarization}",
    journal = "JHEP",
    volume = "11",
    pages = "110",
    year = "2012"
}

@article{Xu:2015rfa,
    author = "Xu, Jianfei and Cao, Li-Ming and Hu, Ya-Peng",
    title = "{P-V criticality in the extended phase space of black holes in massive gravity}",
    journal = "Phys. Rev. D",
    volume = "91",
    number = "12",
    pages = "124033",
    year = "2015"
}

@article{Hawking:1982dh,
    author = "Hawking, S. W. and Page, Don N.",
    title = "{Thermodynamics of Black Holes in anti-De Sitter Space}",
    journal = "Commun. Math. Phys.",
    volume = "87",
    pages = "577",
    year = "1983"
}

@article{Hegde:2020xlv,
    author = "Hegde, Kartheek and Naveena Kumara, A. and Rizwan, C. L. Ahmed and M., Ajith K. and Ali, Md Sabir",
    title = "{Thermodynamics, Phase Transition and Joule Thomson Expansion of novel 4-D Gauss Bonnet AdS Black Hole}",
    month = "3",
    year = "2020"
}

@article{Yang:2020jno,
    author = "Yang, Ke and Gu, Bao-Min and Wei, Shao-Wen and Liu, Yu-Xiao",
    title = "{Born\textendash{}Infeld black holes in 4D Einstein\textendash{}Gauss\textendash{}Bonnet gravity}",
    journal = "Eur. Phys. J. C",
    volume = "80",
    number = "7",
    pages = "662",
    year = "2020"
}

@article{Zhang:2020obn,
    author = "Zhang, Chao-Ming and Zou, De-Cheng and Zhang, Ming",
    title = "{Triple points and phase diagrams of Born-Infeld AdS black holes in 4D Einstein-Gauss-Bonnet gravity}",
    journal = "Phys. Lett. B",
    volume = "811",
    pages = "135955",
    year = "2020"
}

@article{Fernando:2016qhq,
    author = "Fernando, Sharmanthie",
    title = "{Phase transitions of black holes in massive gravity}",
    journal = "Mod. Phys. Lett. A",
    volume = "31",
    number = "16",
    pages = "1650096",
    year = "2016"
}

@article{Hendi:2015pda,
    author = "Hendi, S. H. and Panahiyan, S. and Eslam Panah, B.",
    title = "{Charged Black Hole Solutions in Gauss-Bonnet-Massive Gravity}",
    journal = "JHEP",
    volume = "01",
    pages = "129",
    year = "2016"
}

@article{Zou:2016sab,
    author = "Zou, De-Cheng and Yue, Ruihong and Zhang, Ming",
    title = "{Reentrant phase transitions of higher-dimensional AdS black holes in dRGT massive gravity}",
    journal = "Eur. Phys. J. C",
    volume = "77",
    number = "4",
    pages = "256",
    year = "2017"
}

@article{Zou:2017juz,
    author = "Zou, De-Cheng and Liu, Yunqi and Yue, Rui-Hong",
    title = "{Behavior of quasinormal modes and Van der Waals-like phase transition of charged AdS black holes in massive gravity}",
    journal = "Eur. Phys. J. C",
    volume = "77",
    number = "6",
    pages = "365",
    year = "2017"
}

@article{Hendi:2016yof,
    author = "Hendi, Seyed Hossein and Li, Gu-Qiang and Mo, Jie-Xiong and Panahiyan, Shahram and Eslam Panah, Behzad",
    title = "{New perspective for black hole thermodynamics in Gauss\textendash{}Bonnet\textendash{}Born\textendash{}Infeld massive gravity}",
    journal = "Eur. Phys. J. C",
    volume = "76",
    number = "10",
    pages = "571",
    year = "2016"
}

@article{Okcu:2016tgt,
    author = {\"Okc\"u, \"Ozg\"ur and Ayd\i{}ner, Ekrem},
    title = "{Joule\textendash{}Thomson expansion of the charged AdS black holes}",
    journal = "Eur. Phys. J. C",
    volume = "77",
    number = "1",
    pages = "24",
    year = "2017"
}

@article{Mo:2018rgq,
    author = "Mo, Jie-Xiong and Li, Gu-Qiang and Lan, Shan-Quan and Xu, Xiao-Bao",
    title = "{Joule-Thomson expansion of $d$-dimensional charged AdS black holes}",
    journal = "Phys. Rev. D",
    volume = "98",
    number = "12",
    pages = "124032",
    year = "2018"
}

@article{Okcu:2017qgo,
    author = {\"Okc\"u, \"Ozg\"ur and Ayd\i{}ner, Ekrem},
    title = "{Joule\textendash{}Thomson expansion of Kerr\textendash{}AdS black holes}",
    journal = "Eur. Phys. J. C",
    volume = "78",
    number = "2",
    pages = "123",
    year = "2018"
}

@article{Zhao:2018kpz,
    author = "Zhao, Ze-Wei and Xiu, Yi-Hong and Li, Nan",
    title = "{Throttling process of the Kerr\textendash{}Newman\textendash{}anti-de Sitter black holes in the extended phase space}",
    journal = "Phys. Rev. D",
    volume = "98",
    number = "12",
    pages = "124003",
    year = "2018"
}

@article{Nam:2020gud,
    author = "Nam, Cao H.",
    title = "{Effect of massive gravity on Joule\textendash{}Thomson expansion of the charged AdS black hole}",
    journal = "Eur. Phys. J. Plus",
    volume = "135",
    number = "2",
    pages = "259",
    year = "2020"
}

@article{Zhang:2021kha,
    author = "Zhang, Chao-Ming and Zhang, Ming and Zou, De-Cheng",
    title = "{Joule\textendash{}Thomson expansion of Born\textendash{}Infeld AdS black holes in consistent 4D Einstein\textendash{}Gauss\textendash{}Bonnet gravity}",
    journal = "Mod. Phys. Lett. A",
    volume = "37",
    number = "11",
    pages = "2250063",
    year = "2022"
}

@article{Kruglov:2022sxx,
    author = "Kruglov, S. I.",
    title = "{Magnetic black holes in AdS space with nonlinear electrodynamics, extended phase space thermodynamics and Joule\textendash{}Thomson expansion}",
    journal = "Int. J. Geom. Meth. Mod. Phys.",
    volume = "20",
    number = "01",
    pages = "2350008",
    year = "2023"
}

@article{Kruglov:2022mde,
    author = "Kruglov, S. I.",
    title = "{NED-AdS black holes, extended phase space thermodynamics and Joule\textendash{}Thomson expansion}",
    journal = "Nucl. Phys. B",
    volume = "984",
    pages = "115949",
    year = "2022"
}

@article{Kruglov:2022bhx,
    author = "Kruglov, Sergey Il\textquoteright{}ich",
    title = "{AdS Black Holes in the Framework of Nonlinear Electrodynamics, Thermodynamics, and Joule\textendash{}Thomson Expansion}",
    journal = "Symmetry",
    volume = "14",
    number = "8",
    pages = "1597",
    year = "2022"
}

@book{carroll2019spacetime,
  title={Spacetime and geometry},
  author={Carroll, Sean M.},
  year={2019},
  publisher={Cambridge University Press}
}

@article{weinberg1972gravitation,
  title={Gravitation and cosmology: principles and applications of the general theory of relativity},
  author={Weinberg, Steven},
  year={1972}
}

@article{Barrabes:1997kk,
    author = "Barrabes, C. and Bressange, G. F.",
    title = "{Singular hypersurfaces in scalar - tensor theories of gravity}",
    journal = "Class. Quant. Grav.",
    volume = "14",
    pages = "805--824",
    year = "1997"
}

@article{Cai:1996pj,
    author = "Cai, Rong-Gen and Myung, Y. S.",
    title = "{Black holes in the Brans-Dicke-Maxwell theory}",
    journal = "Phys. Rev. D",
    volume = "56",
    pages = "3466--3470",
    year = "1997"
}

@article{Capozziello:2005bu,
    author = "Capozziello, S. and Troisi, Antonio",
    title = "{PPN-limit of fourth order gravity inspired by scalar-tensor gravity}",
    journal = "Phys. Rev. D",
    volume = "72",
    pages = "044022",
    year = "2005"
}

@article{Sotiriou:2006hs,
    author = "Sotiriou, Thomas P.",
    title = "{f(R) gravity and scalar-tensor theory}",
    journal = "Class. Quant. Grav.",
    volume = "23",
    pages = "5117--5128",
    year = "2006"
}

@article{Moffat:2005si,
    author = "Moffat, J. W.",
    title = "{Scalar-tensor-vector gravity theory}",
    journal = "JCAP",
    volume = "03",
    pages = "004",
    year = "2006"
}

@article{Faraoni:2007yn,
    author = "Faraoni, Valerio",
    title = "{de Sitter space and the equivalence between f(R) and scalar-tensor gravity}",
    journal = "Phys. Rev. D",
    volume = "75",
    pages = "067302",
    year = "2007"
}

@article{Sotiriou:2008ve,
    author = "Sotiriou, Thomas P.",
    editor = "Stergioulas, N. and Tsagas, C.",
    title = "{6+1 lessons from f(R) gravity}",
    journal = "J. Phys. Conf. Ser.",
    volume = "189",
    pages = "012039",
    year = "2009"
}

@article{Sotiriou:2008rp,
    author = "Sotiriou, Thomas P. and Faraoni, Valerio",
    title = "{f(R) Theories Of Gravity}",
    journal = "Rev. Mod. Phys.",
    volume = "82",
    pages = "451--497",
    year = "2010"
}

@article{Capozziello:2011et,
    author = "Capozziello, Salvatore and De Laurentis, Mariafelicia",
    title = "{Extended Theories of Gravity}",
    journal = "Phys. Rept.",
    volume = "509",
    pages = "167--321",
    year = "2011"
}

@article{Berry:2011pb,
    author = "Berry, Christopher P. L. and Gair, Jonathan R.",
    title = "{Linearized f(R) Gravity: Gravitational Radiation and Solar System Tests}",
    journal = "Phys. Rev. D",
    volume = "83",
    pages = "104022",
    year = "2011",
    note = "[Erratum: Phys.Rev.D 85, 089906 (2012)]"
}

@article{Liang:2017ahj,
    author = "Liang, Dicong and Gong, Yungui and Hou, Shaoqi and Liu, Yunqi",
    title = "{Polarizations of gravitational waves in $f(R)$ gravity}",
    journal = "Phys. Rev. D",
    volume = "95",
    number = "10",
    pages = "104034",
    year = "2017"
}

@article{Gogoi:2020ypn,
    author = "Gogoi, Dhruba Jyoti and Dev Goswami, Umananda",
    title = "{A new $f(R)$ gravity model and properties of gravitational waves in it}",
    journal = "Eur. Phys. J. C",
    volume = "80",
    number = "12",
    pages = "1101",
    year = "2020"
}

@article{Horava:2009uw,
    author = "Horava, Petr",
    title = "{Quantum Gravity at a Lifshitz Point}",
    journal = "Phys. Rev. D",
    volume = "79",
    pages = "084008",
    year = "2009"
}

@article{Blas:2009yd,
    author = "Blas, D. and Pujolas, O. and Sibiryakov, S.",
    title = "{On the Extra Mode and Inconsistency of Horava Gravity}",
    journal = "JHEP",
    volume = "10",
    pages = "029",
    year = "2009"
}

@article{Sotiriou:2010wn,
    author = "Sotiriou, Thomas P.",
    editor = "Perivolaropoulos, Leandros and Kanti, Panagiota",
    title = "{Horava-Lifshitz gravity: a status report}",
    journal = "J. Phys. Conf. Ser.",
    volume = "283",
    pages = "012034",
    year = "2011"
}

@article{Wang:2017brl,
    author = "Wang, Anzhong",
    title = "{Ho\v{r}ava gravity at a Lifshitz point: A progress report}",
    journal = "Int. J. Mod. Phys. D",
    volume = "26",
    number = "07",
    pages = "1730014",
    year = "2017"
}

@article{Lovelock:1971yv,
    author = "Lovelock, D.",
    title = "{The Einstein tensor and its generalizations}",
    journal = "J. Math. Phys.",
    volume = "12",
    pages = "498--501",
    year = "1971"
}

@article{Lovelock:1972vz,
    author = "Lovelock, D.",
    title = "{The four-dimensionality of space and the einstein tensor}",
    journal = "J. Math. Phys.",
    volume = "13",
    pages = "874--876",
    year = "1972"
}

@article{Deruelle:1989fj,
    author = "Deruelle, Nathalie and Farina-Busto, Luis",
    title = "{The Lovelock Gravitational Field Equations in Cosmology}",
    journal = "Phys. Rev. D",
    volume = "41",
    pages = "3696",
    year = "1990"
}

@article{Wei:2020poh,
    author = "Wei, Shao-Wen and Liu, Yu-Xiao",
    title = "{Extended thermodynamics and microstructures of four-dimensional charged Gauss-Bonnet black hole in AdS space}",
    journal = "Phys. Rev. D",
    volume = "101",
    number = "10",
    pages = "104018",
    year = "2020"
}

@article{Singh:2020nwo,
    author = "Singh, Dharm Veer and Ghosh, Sushant G. and Maharaj, Sunil D.",
    title = "{Clouds of strings in 4$D$ Einstein-Gauss-Bonnet black holes}",
    journal = "Phys. Dark Univ.",
    volume = "30",
    pages = "100730",
    year = "2020"
}

@article{Singh:2020xju,
    author = "Singh, Dharm Veer and Siwach, Sanjay",
    title = "{Thermodynamics and P-v criticality of Bardeen-AdS Black Hole in 4$D$ Einstein-Gauss-Bonnet Gravity}",
    journal = "Phys. Lett. B",
    volume = "808",
    pages = "135658",
    year = "2020"
}

@article{EslamPanah:2020hoj,
    author = "Eslam Panah, B. and Jafarzade, Kh. and Hendi, S. H.",
    title = "{Charged 4D Einstein-Gauss-Bonnet-AdS black holes: Shadow, energy emission, deflection angle and heat engine}",
    journal = "Nucl. Phys. B",
    volume = "961",
    pages = "115269",
    year = "2020"
}

@article{Singh:2021xbk,
    author = "Singh, Dharm Veer and Singh, Benoy Kumar and Upadhyay, Sudhaker",
    title = "{4D AdS Einstein\textendash{}Gauss\textendash{}Bonnet black hole with Yang\textendash{}Mills field and its thermodynamics}",
    journal = "Annals Phys.",
    volume = "434",
    pages = "168642",
    year = "2021"
}

@article{Godani:2022jwz,
    author = "Godani, Nisha and Singh, Dharm Veer and Samanta, Gauranga C.",
    title = "{Stability of thin-shell wormhole in 4D Einstein\textendash{}Gauss\textendash{}Bonnet gravity}",
    journal = "Phys. Dark Univ.",
    volume = "35",
    pages = "100952",
    year = "2022"
}

@article{Wang:2020pmb,
    author = "Wang, Yuan-Yuan and Su, Bing-Yu and Li, Nan",
    title = "{Hawking\textendash{}Page phase transitions in four-dimensional Einstein\textendash{}Gauss\textendash{}Bonnet gravity}",
    journal = "Phys. Dark Univ.",
    volume = "31",
    pages = "100769",
    year = "2021"
}

@article{Fierz:1939ix,
    author = "Fierz, M. and Pauli, W.",
    title = "{On relativistic wave equations for particles of arbitrary spin in an electromagnetic field}",
    journal = "Proc. Roy. Soc. Lond. A",
    volume = "173",
    pages = "211--232",
    year = "1939"
}

@article{Fierz:1939zz,
    author = "Fierz, M.",
    title = "{Force-free particles with any spin}",
    journal = "Helv. Phys. Acta",
    volume = "12",
    pages = "3--37",
    year = "1939"
}

@article{Boulware:1972yco,
    author = "Boulware, D. G. and Deser, Stanley",
    title = "{Can gravitation have a finite range?}",
    journal = "Phys. Rev. D",
    volume = "6",
    pages = "3368--3382",
    year = "1972"
}

@article{deRham:2010ik,
    author = "de Rham, Claudia and Gabadadze, Gregory",
    title = "{Generalization of the Fierz-Pauli Action}",
    journal = "Phys. Rev. D",
    volume = "82",
    pages = "044020",
    year = "2010"
}

@article{deRham:2010kj,
    author = "de Rham, Claudia and Gabadadze, Gregory and Tolley, Andrew J.",
    title = "{Resummation of Massive Gravity}",
    journal = "Phys. Rev. Lett.",
    volume = "106",
    pages = "231101",
    year = "2011"
}

@article{LIGOScientific:2016lio,
    author = "Abbott, B. P. and others",
    collaboration = "LIGO Scientific, Virgo",
    title = "{Tests of general relativity with GW150914}",
    journal = "Phys. Rev. Lett.",
    volume = "116",
    number = "22",
    pages = "221101",
    year = "2016",
    note = "[Erratum: Phys.Rev.Lett. 121, 129902 (2018)]"
}

@article{Hendi:2015bna,
    author = "Hendi, S. H. and Eslam Panah, B. and Panahiyan, S.",
    title = "{Thermodynamical Structure of AdS Black Holes in Massive Gravity with Stringy Gauge-Gravity Corrections}",
    journal = "Class. Quant. Grav.",
    volume = "33",
    number = "23",
    pages = "235007",
    year = "2016"
}

@article{Hendi:2018hdo,
    author = "Hendi, Seyed Hossein and Momennia, Mehrab",
    title = "{Thermodynamic description and quasinormal modes of adS black holes in Born-lnfeld massive gravity with a non-abelian hair}",
    journal = "JHEP",
    volume = "10",
    pages = "207",
    year = "2019"
}

@article{Hendi:2016hbe,
    author = "Hendi, S. H. and Panahiyan, S. and Upadhyay, S. and Eslam Panah, B.",
    title = "{Charged BTZ black holes in the context of massive gravity\textquoteright{}s rainbow}",
    journal = "Phys. Rev. D",
    volume = "95",
    number = "8",
    pages = "084036",
    year = "2017"
}

@article{Upadhyay:2018vfu,
    author = "Upadhyay, Sudhaker and Hendi, Seyed Hossein and Panahiyan, Shahram and Eslam Panah, Behzad",
    title = "{Thermal fluctuations of charged black holes in gravity\textquoteright{}s rainbow}",
    journal = "PTEP",
    volume = "2018",
    number = "9",
    pages = "093E01",
    year = "2018"
}

@article{Singh:2020rnm,
    author = "Singh, Benoy Kumar and Singh, Raj Pal and Singh, Dharm Veer",
    title = "{Extended phase space thermodynamics of Bardeen black hole in massive gravity}",
    journal = "Eur. Phys. J. Plus",
    volume = "135",
    number = "10",
    pages = "862",
    year = "2020"
}

@article{Born:1934gh,
    author = "Born, M. and Infeld, L.",
    title = "{Foundations of the new field theory}",
    journal = "Proc. Roy. Soc. Lond. A",
    volume = "144",
    number = "852",
    pages = "425--451",
    year = "1934"
}

@article{Heisenberg:1936nmg,
    author = "Heisenberg, W. and Euler, H.",
    title = "{Consequences of Dirac's theory of positrons}",
    journal = "Z. Phys.",
    volume = "98",
    number = "11-12",
    pages = "714--732",
    year = "1936"
}

@article{plebanski1966non,
  title={Non-Linear Electrodynamics--a Study},
  author={Plebanski, J.},
  journal={CIEA del IPN, Mexico City},
  year={1966}
}

@article{Natsuume:1994hd,
    author = "Natsuume, Makoto",
    title = "{Higher order correction to the GHS string black hole}",
    journal = "Phys. Rev. D",
    volume = "50",
    pages = "3949--3953",
    year = "1994"
}

@article{Kats:2006xp,
    author = "Kats, Yevgeny and Motl, Lubos and Padi, Megha",
    title = "{Higher-order corrections to mass-charge relation of extremal black holes}",
    journal = "JHEP",
    volume = "12",
    pages = "068",
    year = "2007"
}

@article{Cai:2008ph,
    author = "Cai, Rong-Gen and Nie, Zhang-Yu and Sun, Ya-Wen",
    title = "{Shear Viscosity from Effective Couplings of Gravitons}",
    journal = "Phys. Rev. D",
    volume = "78",
    pages = "126007",
    year = "2008"
}

@article{Liu:2008kt,
    author = "Liu, James T. and Szepietowski, Phillip",
    title = "{Higher derivative corrections to R-charged AdS(5) black holes and field redefinitions}",
    journal = "Phys. Rev. D",
    volume = "79",
    pages = "084042",
    year = "2009"
}

@article{Anninos:2008sj,
    author = "Anninos, Dionysios and Pastras, Georgios",
    title = "{Thermodynamics of the Maxwell-Gauss-Bonnet anti-de Sitter Black Hole with Higher Derivative Gauge Corrections}",
    journal = "JHEP",
    volume = "07",
    pages = "030",
    year = "2009"
}

@article{Soleng:1995kn,
    author = "Soleng, Harald H.",
    title = "{Charged black points in general relativity coupled to the logarithmic U(1) gauge theory}",
    journal = "Phys. Rev. D",
    volume = "52",
    pages = "6178--6181",
    year = "1995"
}

@article{Hendi:2013dwa,
    author = "Hendi, S. H.",
    title = "{Asymptotic Reissner-Nordstroem black holes}",
    journal = "Annals Phys.",
    volume = "333",
    pages = "282--289",
    year = "2013"
}

@article{Hendi:2014mna,
    author = "Hendi, S. H.",
    title = "{Thermodynamic properties of asymptotically Reissner-Nordstrom black holes}",
    journal = "Annals Phys.",
    volume = "346",
    pages = "42--50",
    year = "2014"
}

@article{Kruglov:2019ybs,
    author = "Kruglov, S. I.",
    title = "{Dyonic Black Holes with Nonlinear Logarithmic Electrodynamics}",
    journal = "Grav. Cosmol.",
    volume = "25",
    number = "2",
    pages = "190--195",
    year = "2019"
}

@article{Kruglov:2023qed,
    author = "Kruglov, Sergey II'ich",
    title = "{4D Einstein--Gauss--Bonnet gravity coupled to modified logarithmic nonlinear electrodynamics}",
    journal = "Universe",
    volume = "9",
    pages = "24",
    year = "2023"
}

@article{Kruglov:2014iqa,
    author = "Kruglov, S. I.",
    title = "{On Generalized Logarithmic Electrodynamics}",
    journal = "Eur. Phys. J. C",
    volume = "75",
    number = "2",
    pages = "88",
    year = "2015"
}

@article{Gullu:2020ant,
    author = "Gullu, Ibrahim and Mazharimousavi, S. Habib",
    title = "{Double-logarithmic nonlinear electrodynamics}",
    journal = "Phys. Scripta",
    volume = "96",
    number = "4",
    pages = "045217",
    year = "2021"
}

@article{Hendi:2012zz,
    author = "Hendi, S. H.",
    title = "{Asymptotic charged BTZ black hole solutions}",
    journal = "JHEP",
    volume = "03",
    pages = "065",
    year = "2012"
}

@article{Hendi:2012um,
    author = "Hendi, S. H. and Vahidinia, M. H.",
    title = "{Extended phase space thermodynamics and P-V criticality of black holes with a nonlinear source}",
    journal = "Phys. Rev. D",
    volume = "88",
    number = "8",
    pages = "084045",
    year = "2013"
}

@article{Hendi:2016usw,
    author = "Hendi, S. H. and Eslam Panah, B. and Panahiyan, S. and Talezadeh, M. S.",
    title = "{Geometrical thermodynamics and P-V criticality of black holes with power-law Maxwell field}",
    journal = "Eur. Phys. J. C",
    volume = "77",
    number = "2",
    pages = "133",
    year = "2017"
}

@article{Kruglov:2017fck,
    author = "Kruglov, S. I.",
    title = "{Black hole as a magnetic monopole within exponential nonlinear electrodynamics}",
    journal = "Annals Phys.",
    volume = "378",
    pages = "59--70",
    year = "2017"
}

@article{kruglov2015nonlinear,
  title={Nonlinear arcsin-electrodynamics},
  author={Kruglov, S. I.},
  journal={Annalen der Physik},
  volume={527},
  number={5-6},
  pages={397--401},
  year={2015},
  publisher={Wiley Online Library}
}

@article{Kruglov:2016ezw,
    author = "Kruglov, S. I.",
    title = {{Nonlinear arcsin-electrodynamics and asymptotic Reissner-Nordstr\"om black holes}},
    journal = "Annalen Phys.",
    volume = "528",
    pages = "588--596",
    year = "2016"
}

@article{Kruglov:2019okd,
    author = "Kruglov, S. I.",
    title = "{Dyonic and magnetic black holes with nonlinear arcsin-electrodynamics}",
    journal = "Annals Phys.",
    volume = "409",
    pages = "167937",
    year = "2019"
}

@article{Kruglov:2018xzs,
    author = "Kruglov, S. I.",
    title = "{Holographic superconductor with nonlinear Born\textendash{}Infeld-type electrodynamics}",
    journal = "Int. J. Mod. Phys. A",
    volume = "34",
    number = "03n04",
    pages = "1950019",
    year = "2019"
}

@article{Kruglov:2021btd,
    author = "Kruglov, S. I.",
    title = "{Einstein-Gauss-Bonnet gravity with rational nonlinear electrodynamics}",
    journal = "EPL",
    volume = "133",
    number = "6",
    pages = "6",
    year = "2021"
}

@article{Kruglov:2021pdp,
    author = "Kruglov, Sergey Il'ich",
    title = "{4D Einstein\textendash{}Gauss\textendash{}Bonnet Gravity Coupled with Nonlinear Electrodynamics}",
    journal = "Symmetry",
    volume = "13",
    number = "2",
    pages = "204",
    year = "2021"
}

@article{Kruglov:2021qzd,
    author = "Kruglov, Sergey Il'ich",
    title = "{Einstein\textendash{}Gauss\textendash{}Bonnet Gravity with Nonlinear Electrodynamics: Entropy, Energy Emission, Quasinormal Modes and Deflection Angle}",
    journal = "Symmetry",
    volume = "13",
    number = "6",
    pages = "944",
    year = "2021"
}

@article{Kruglov:2021rqf,
    author = "Kruglov, Sergey Il\textquoteright{}ich",
    title = "{New model of 4D Einstein--Gauss--Bonnet gravity coupled with nonlinear electrodynamics}",
    journal = "Universe",
    volume = "7",
    pages = "249",
    year = "2021"
}

@article{Kruglov:2016ymq,
    author = "Kruglov, S. I.",
    title = {{Asymptotic Reissner-Nordstr\"om solution within nonlinear electrodynamics}},
    journal = "Phys. Rev. D",
    volume = "94",
    number = "4",
    pages = "044026",
    year = "2016"
}

@article{cataldo1999three,
  title={Three dimensional black hole coupled to the Born-Infeld electrodynamics},
  author={Cataldo, Mauricio and Garcia, Alberto},
  journal={Physics Letters B},
  volume={456},
  number={1},
  pages={28--33},
  year={1999},
  publisher={Elsevier}
}

@article{cai2004born,
  title={Born-Infeld black holes in AdS spaces},
  author={Cai, Rong-Gen and Pang, Da-Wei and Wang, Anzhong},
  journal={Physical Review D},
  volume={70},
  number={12},
  pages={124034},
  year={2004},
  publisher={APS}
}

@article{panahiyan2020nonlinearly,
  title={Nonlinearly charged dyonic black holes},
  author={Panahiyan, Shahram},
  journal={Nuclear Physics B},
  volume={950},
  pages={114831},
  year={2020},
  publisher={Elsevier}
}

@article{kruglov2017born,
  title={Born--Infeld-type electrodynamics and magnetic black holes},
  author={Kruglov, S.I.},
  journal={Annals of Physics},
  volume={383},
  pages={550--559},
  year={2017},
  publisher={Elsevier}
}

@article{Yerra:2022eov,
    author = "Yerra, Pavan Kumar and Bhamidipati, Chandrasekhar",
    title = "{Topology of Born-Infeld AdS black holes in 4D novel Einstein-Gauss-Bonnet gravity}",
    journal = "Phys. Lett. B",
    volume = "835",
    pages = "137591",
    year = "2022"
}

@incollection{bekenstein2020black,
  title={Black holes and the second law},
  author={Bekenstein, Jacob D.},
  booktitle={JACOB BEKENSTEIN: The Conservative Revolutionary},
  pages={303--306},
  year={2020},
  publisher={World Scientific}
}

@article{bekenstein1973extraction,
  title={Extraction of energy and charge from a black hole},
  author={Bekenstein, Jacob D.},
  journal={Physical Review D},
  volume={7},
  number={4},
  pages={949},
  year={1973},
  publisher={APS}
}

@article{bekenstein1974generalized,
  title={Generalized second law of thermodynamics in black-hole physics},
  author={Bekenstein, Jacob D.},
  journal={Physical Review D},
  volume={9},
  number={12},
  pages={3292},
  year={1974},
  publisher={APS}
}

@article{bardeen1973four,
  title={The four laws of black hole mechanics},
  author={Bardeen, James M. and Carter, Brandon and Hawking, Stephen W.},
  journal={Communications in mathematical physics},
  volume={31},
  pages={161--170},
  year={1973},
  publisher={Springer}
}

@article{hawking1975particle,
  title={Particle creation by black holes},
  author={Hawking, Stephen W.},
  journal={Communications in mathematical physics},
  volume={43},
  number={3},
  pages={199--220},
  year={1975},
  publisher={Springer}
}

@article{maldacena1999large,
  title={The large-N limit of superconformal field theories and supergravity},
  author={Maldacena, Juan},
  journal={International journal of theoretical physics},
  volume={38},
  number={4},
  pages={1113--1133},
  year={1999},
  publisher={Springer}
}

@article{Witten:1998qj,
    author = "Witten, Edward",
    title = "{Anti-de Sitter space and holography}",
    journal = "Adv. Theor. Math. Phys.",
    volume = "2",
    pages = "253--291",
    year = "1998"
}

@article{gubser1998gauge,
  title={Gauge theory correlators from non-critical string theory},
  author={Gubser, Steven S. and Klebanov, Igor R. and Polyakov, Alexander M.},
  journal={Physics Letters B},
  volume={428},
  number={1-2},
  pages={105--114},
  year={1998},
  publisher={Elsevier}
}

@article{Aharony:1999ti,
    author = "Aharony, Ofer and Gubser, Steven S. and Maldacena, Juan Martin and Ooguri, Hirosi and Oz, Yaron",
    title = "{Large N field theories, string theory and gravity}",
    journal = "Phys. Rept.",
    volume = "323",
    pages = "183--386",
    year = "2000"
}

@article{kastor2009enthalpy,
  title={Enthalpy and the mechanics of AdS black holes},
  author={Kastor, David and Ray, Sourya and Traschen, Jennie},
  journal={Classical and Quantum Gravity},
  volume={26},
  number={19},
  pages={195011},
  year={2009},
  publisher={IOP Publishing}
}

@article{kastor2010smarr,
  title={Smarr formula and an extended first law for Lovelock gravity},
  author={Kastor, David and Ray, Sourya and Traschen, Jennie},
  journal={Classical and Quantum Gravity},
  volume={27},
  number={23},
  pages={235014},
  year={2010},
  publisher={IOP Publishing}
}

@article{dolan2011cosmological,
  title={The cosmological constant and black-hole thermodynamic potentials},
  author={Dolan, Brian P.},
  journal={Classical and Quantum Gravity},
  volume={28},
  number={12},
  pages={125020},
  year={2011},
  publisher={IOP Publishing}
}

@article{dolan2011pressure,
  title={Pressure and volume in the first law of black hole thermodynamics},
  author={Dolan, Brian P.},
  journal={Classical and Quantum Gravity},
  volume={28},
  number={23},
  pages={235017},
  year={2011},
  publisher={IOP Publishing}
}

@article{dehghani2014p,
  title={P- V criticality of charged dilatonic black holes},
  author={Dehghani, M.H. and Kamrani, S. and Sheykhi, A.},
  journal={Physical Review D},
  volume={90},
  number={10},
  pages={104020},
  year={2014},
  publisher={APS}
}

@article{hennigar2015p,
  title={P- v criticality in quasitopological gravity},
  author={Hennigar, Robie A. and Brenna, Wilson G. and Mann, Robert B.},
  journal={Journal of High Energy Physics},
  volume={2015},
  number={7},
  pages={1--31},
  year={2015},
  publisher={Springer}
}

@article{hendi2016extended,
  title={Extended phase space thermodynamics and P--V criticality: Brans--Dicke--Born--Infeld vs. Einstein--Born--Infeld-dilaton black holes},
  author={Hendi, S.H. and Tad, R. Moradi and Armanfard, Z. and Talezadeh, M.S.},
  journal={The European Physical Journal C},
  volume={76},
  pages={1--15},
  year={2016},
  publisher={Springer}
}

@article{hansen2017universality,
  title={Universality of P- V criticality in horizon thermodynamics},
  author={Hansen, Devin and Kubiz{\v{n}}{\'a}k, David and Mann, Robert B.},
  journal={Journal of High Energy Physics},
  volume={2017},
  number={1},
  pages={1--24},
  year={2017},
  publisher={Springer}
}

@article{majhi2017pv,
  title={PV criticality of AdS black holes in a general framework},
  author={Majhi, Bibhas Ranjan and Samanta, Saurav},
  journal={Physics Letters B},
  volume={773},
  pages={203--207},
  year={2017},
  publisher={Elsevier}
}

@article{hendi2017geometrical,
  title={Geometrical thermodynamics and P--V criticality of the black holes with power-law Maxwell field},
  author={Hendi, S.H. and Panah, B. Eslam and Panahiyan, S. and Talezadeh, M.S.},
  journal={The European Physical Journal C},
  volume={77},
  pages={1--16},
  year={2017},
  publisher={Springer}
}

@article{upadhyay2017p,
  title={P- V criticality of first-order entropy corrected AdS black holes in massive gravity},
  author={Upadhyay, S. and Pourhassan, B. and Farahani, H.},
  journal={Physical Review D},
  volume={95},
  number={10},
  pages={106014},
  year={2017},
  publisher={APS}
}

@article{johnson2014holographic,
  title={Holographic heat engines},
  author={Johnson, Clifford V.},
  journal={Classical and Quantum Gravity},
  volume={31},
  number={20},
  pages={205002},
  year={2014},
  publisher={IOP Publishing}
}

@article{belhaj2015heat,
  title={On heat properties of AdS black holes in higher dimensions},
  author={Belhaj, A. and Chabab, M. and El Moumni, H. and Masmar, K. and Sedra, M.B. and Segui, A.},
  journal={Journal of High Energy Physics},
  volume={2015},
  number={5},
  pages={1--13},
  year={2015},
  publisher={Springer}
}

@article{setare2015polytropic,
  title={Polytropic black hole as a heat engine},
  author={Setare, M.R. and Adami, H.},
  journal={General Relativity and Gravitation},
  volume={47},
  pages={1--5},
  year={2015},
  publisher={Springer}
}

@article{johnson2016gauss,
  title={Gauss--Bonnet black holes and holographic heat engines beyond large N},
  author={Johnson, Clifford V.},
  journal={Classical and Quantum Gravity},
  volume={33},
  number={21},
  pages={215009},
  year={2016},
  publisher={IOP Publishing}
}

@article{johnson2016born,
  title={Born--Infeld AdS black holes as heat engines},
  author={Johnson, Clifford V.},
  journal={Classical and Quantum Gravity},
  volume={33},
  number={13},
  pages={135001},
  year={2016},
  publisher={IOP Publishing}
}

@article{bhamidipati2017heat,
  title={Heat engines for dilatonic Born--Infeld black holes},
  author={Bhamidipati, Chandrasekhar and Yerra, Pavan Kumar},
  journal={The European Physical Journal C},
  volume={77},
  pages={1--15},
  year={2017},
  publisher={Springer}
}

@article{hennigar2017holographic,
  title={Holographic heat engines: general considerations and rotating black holes},
  author={Hennigar, Robie A. and McCarthy, Fiona and Ballon, Alvaro and Mann, Robert B.},
  journal={Classical and Quantum Gravity},
  volume={34},
  number={17},
  pages={175005},
  year={2017},
  publisher={IOP Publishing}
}

@article{mo2017heat,
  title={Heat engine in the three-dimensional spacetime},
  author={Mo, Jie-Xiong and Liang, Feng and Li, Gu-Qiang},
  journal={Journal of High Energy Physics},
  volume={2017},
  number={3},
  pages={1--11},
  year={2017},
  publisher={Springer}
}

@article{hendi2018black,
  title={Black holes in massive gravity as heat engines},
  author={Hendi, S.H. and Panah, B. Eslam and Panahiyan, S. and Liu, H. and Meng, X-H},
  journal={Physics Letters B},
  volume={781},
  pages={40--47},
  year={2018},
  publisher={Elsevier}
}

@article{altamirano2013reentrant,
  title={Reentrant phase transitions in rotating anti--de Sitter black holes},
  author={Altamirano, Natacha and Kubiz{\v{n}}{\'a}k, David and Mann, Robert B.},
  journal={Physical Review D},
  volume={88},
  number={10},
  pages={101502},
  year={2013},
  publisher={APS}
}

@article{frassino2014multiple,
  title={Multiple reentrant phase transitions and triple points in Lovelock thermodynamics},
  author={Frassino, Antonia M. and Kubiz{\v{n}}{\'a}k, David and Mann, Robert B. and Simovic, Fil},
  journal={Journal of High Energy Physics},
  volume={2014},
  number={9},
  pages={1--47},
  year={2014},
  publisher={Springer}
}

@article{hennigar2015reentrant,
  title={Reentrant phase transitions and van der Waals behaviour for hairy black holes},
  author={Hennigar, Robie A. and Mann, Robert B.},
  journal={Entropy},
  volume={17},
  number={12},
  pages={8056--8072},
  year={2015},
  publisher={MDPI}
}

@article{Paul:2023mlh,
    author = "Paul, Prosenjit and Upadhyay, Sudhaker and Singh, Dharm Veer",
    title = "{Charged AdS black holes in 4D Einstein\textendash{}Gauss\textendash{}Bonnet massive gravity}",
    journal = "Eur. Phys. J. Plus",
    volume = "138",
    number = "6",
    pages = "566",
    year = "2023"
}

@article{Nam:2019zyk,
    author = "Nam, Cao Hoang",
    title = "{Heat engine efficiency and Joule\textendash{}Thomson expansion of nonlinear charged AdS black hole in massive gravity}",
    journal = "Gen. Rel. Grav.",
    volume = "53",
    number = "3",
    pages = "30",
    year = "2021"
}

@article{NaveenaKumara:2020biu,
    author = "Naveena Kumara, A. and Ahmed Rizwan, C. L. and Hegde, Kartheek and Ali, Md Sabir and Ajith, K. M.",
    title = "{Ruppeiner geometry, reentrant phase transition, and microstructure of Born-Infeld AdS black hole}",
    journal = "Phys. Rev. D",
    volume = "103",
    number = "4",
    pages = "044025",
    year = "2021"
}

@article{hudson1904gegenseitige,
  title={Die gegenseitige l{\"o}slichkeit von nikotin in wasser},
  author={Hudson, Claude Silbert},
  journal={Zeitschrift f{\"u}r Physikalische Chemie},
  volume={47},
  number={1},
  pages={113--115},
  year={1904},
  publisher={De Gruyter Oldenbourg}
}

@article{narayanan1994reentrant,
  title={Reentrant phase transitions in multicomponent liquid mixtures},
  author={Narayanan, T and Kumar, Anil},
  journal={Physics Reports},
  volume={249},
  number={3},
  pages={135--218},
  year={1994},
  publisher={Elsevier}
}

@article{Altamirano:2013uqa,
    author = "Altamirano, Natacha and Kubiz\v{n}\'ak, David and Mann, Robert B. and Sherkatghanad, Zeinab",
    title = "{Kerr-AdS analogue of triple point and solid/liquid/gas phase transition}",
    journal = "Class. Quant. Grav.",
    volume = "31",
    pages = "042001",
    year = "2014"
}

@article{Altamirano:2014tva,
    author = "Altamirano, Natacha and Kubiznak, David and Mann, Robert B. and Sherkatghanad, Zeinab",
    title = "{Thermodynamics of rotating black holes and black rings: phase transitions and thermodynamic volume}",
    journal = "Galaxies",
    volume = "2",
    pages = "89--159",
    year = "2014"
}

@article{Kubiznak:2015bya,
    author = "Kubiznak, David and Simovic, Fil",
    title = "{Thermodynamics of horizons: de Sitter black holes and reentrant phase transitions}",
    journal = "Class. Quant. Grav.",
    volume = "33",
    number = "24",
    pages = "245001",
    year = "2016"
}

@article{Frassino:2014pha,
    author = "Frassino, Antonia M. and Kubiznak, David and Mann, Robert B. and Simovic, Fil",
    title = "{Multiple Reentrant Phase Transitions and Triple Points in Lovelock Thermodynamics}",
    journal = "JHEP",
    volume = "09",
    pages = "080",
    year = "2014"
}

@article{Wei:2014hba,
    author = "Wei, Shao-Wen and Liu, Yu-Xiao",
    title = "{Triple points and phase diagrams in the extended phase space of charged Gauss-Bonnet black holes in AdS space}",
    journal = "Phys. Rev. D",
    volume = "90",
    number = "4",
    pages = "044057",
    year = "2014"
}

@article{Hennigar:2015wxa,
    author = "Hennigar, Robie A. and Mann, Robert B.",
    title = "{Reentrant phase transitions and van der Waals behaviour for hairy black holes}",
    journal = "Entropy",
    volume = "17",
    number = "12",
    pages = "8056--8072",
    year = "2015"
}

@article{Kumar:2023cmo,
    author = "Kumar, Amit and Singh, Dharm Veer and Myrzakulov, Yerlan and Yergaliyeva, Gulmira and Upadhyay, Sudhaker",
    title = "{Exact solution of Bardeen black hole in Einstein\textendash{}Gauss\textendash{}Bonnet gravity}",
    journal = "Eur. Phys. J. Plus",
    volume = "138",
    number = "12",
    pages = "1071",
    year = "2023"
}

@article{Kruglov:2023ktg,
    author = "Kruglov, Sergey Il\textquoteright{}ich",
    title = "{Einstein-AdS Gravity Coupled to Nonlinear Electrodynamics, Magnetic Black Holes, Thermodynamics in an Extended Phase Space and Joule\textendash{}Thomson Expansion}",
    journal = "Universe",
    volume = "9",
    number = "10",
    pages = "456",
    year = "2023"
}

@article{Myrzakulov:2023rkr,
    author = "Myrzakulov, Yerlan and Myrzakulov, Kairat and Upadhyay, Sudhaker and Singh, Dharm Veer",
    title = "{Quasinormal modes and phase structure of regular AdS Einstein\textendash{}Gauss\textendash{}Bonnet black holes}",
    journal = "Int. J. Geom. Meth. Mod. Phys.",
    volume = "20",
    number = "07",
    pages = "2350121",
    year = "2023"
}

@article{Ghosh:2020ijh,
    author = "Ghosh, Sushant G. and Singh, Dharm Veer and Kumar, Rahul and Maharaj, Sunil D.",
    title = "{Phase transition of AdS black holes in 4D EGB gravity coupled to nonlinear electrodynamics}",
    journal = "Annals Phys.",
    volume = "424",
    pages = "168347",
    year = "2021"
}

@article{Singh:2024jgo,
    author = "Singh, Bhupendra and Veer Singh, Dharm and Kumar Singh, Benoy",
    title = "{Thermodynamics, phase structure and quasinormal modes for AdS Heyward massive black hole}",
    journal = "Phys. Scripta",
    volume = "99",
    number = "2",
    pages = "025305",
    year = "2024"
}
\end{document}